\documentclass[dvips]{article}
\usepackage{amsmath}

\usepackage{graphicx}
\usepackage{longtable}
\usepackage{multirow}
\usepackage{subfigure}

\usepackage{amssymb,mathrsfs}

\usepackage{url}

\usepackage{icrctc07_edit}

\newcommand{\HESSONE}{HESS~J1718$-$385}
\newcommand{\PSRONE}{PSR~J1718$-$3825}
\newcommand{\HMS}[3]{$#1^{\mathrm{h}}#2^{\mathrm{m}}#3^{\mathrm{s}}$}
\newcommand{\DMS}[3]{$#1^\circ #2' #3''$}

\newcommand{\aj}{AJ}
\newcommand{\apj}{ApJ}
\newcommand{\apjl}{ApJL}
\newcommand{\apjs}{ApJS}
\newcommand{\aap}{A\&A}
\newcommand{\aapr}{A\&A Rev.}
\newcommand\araa{ARA\&A}% Annual Review of Astron and Astrophys
\newcommand\aaps{A\&AS}% Astronomy and Astrophysics, Supplement
\newcommand{\nat}{Nature}
\newcommand{\mnras}{MNRAS}
\newcommand{\prd}{Phys. Rev. D}
\newcommand{\pasj}{PASJ}

\def\g{\hbox{$\gamma$}}
\newcommand\arcmin{\mbox{$^\prime$}}%
\newcommand\arcsec{\mbox{$^{\prime\prime}$}}%
\newcommand\degr{\arcdeg}%

\newcommand{\gr}{$\gamma$-ray} \newcommand{\grs}{$\gamma$-rays}
\newcommand{\hess}{H.E.S.S.}  \newcommand{\hgc}{HESS~J1745-290}
\newcommand{\astar}{Sgr~A$^*$} \newcommand{\aeast}{Sgr~A~East}
\newcommand{\pwn}{G359.95-0.04}

\newcommand{\HESSa}{HESS~J1427$-$608}
\newcommand{\HESSb}{HESS~J1626$-$490}
\newcommand{\HESSc}{HESS~J1702$-$420}
\newcommand{\HESSd}{HESS~J1708$-$410}
\newcommand{\HESSe}{HESS~J1731$-$347}
\newcommand{\HESSf}{HESS~J1841$-$055}
\newcommand{\HESSg}{HESS~J1857$+$026}
\newcommand{\HESSh}{HESS~J1858$+$020}

\newcommand{\comment}[1]{}
\def\pks{PKS$\;$2155$-$304}

\newcommand{\rxj}{RX~J1713.7$-$3946}

\newcommand{\lsi}{LS I 61$^{\circ}$+303 }

\newcommand {\sax} {{\it Beppo}SAX }

\newcommand {\gradi} {^{\circ} }

\def\hessns{H.E.S.S.}
\def\mbh{M_{\mathrm{BH}}}
\def\mbhrm{M_{\mathrm{BH}}}

\begin{document}

%%%%%%%%%%%%%%%%
%  Titelseite  %
%%%%%%%%%%%%%%%%

\begin{titlepage}
\pagestyle{empty}
\begin{figure*}[!h]
\vspace*{-3.6cm}
\includegraphics[bb = 42 0 637 842]{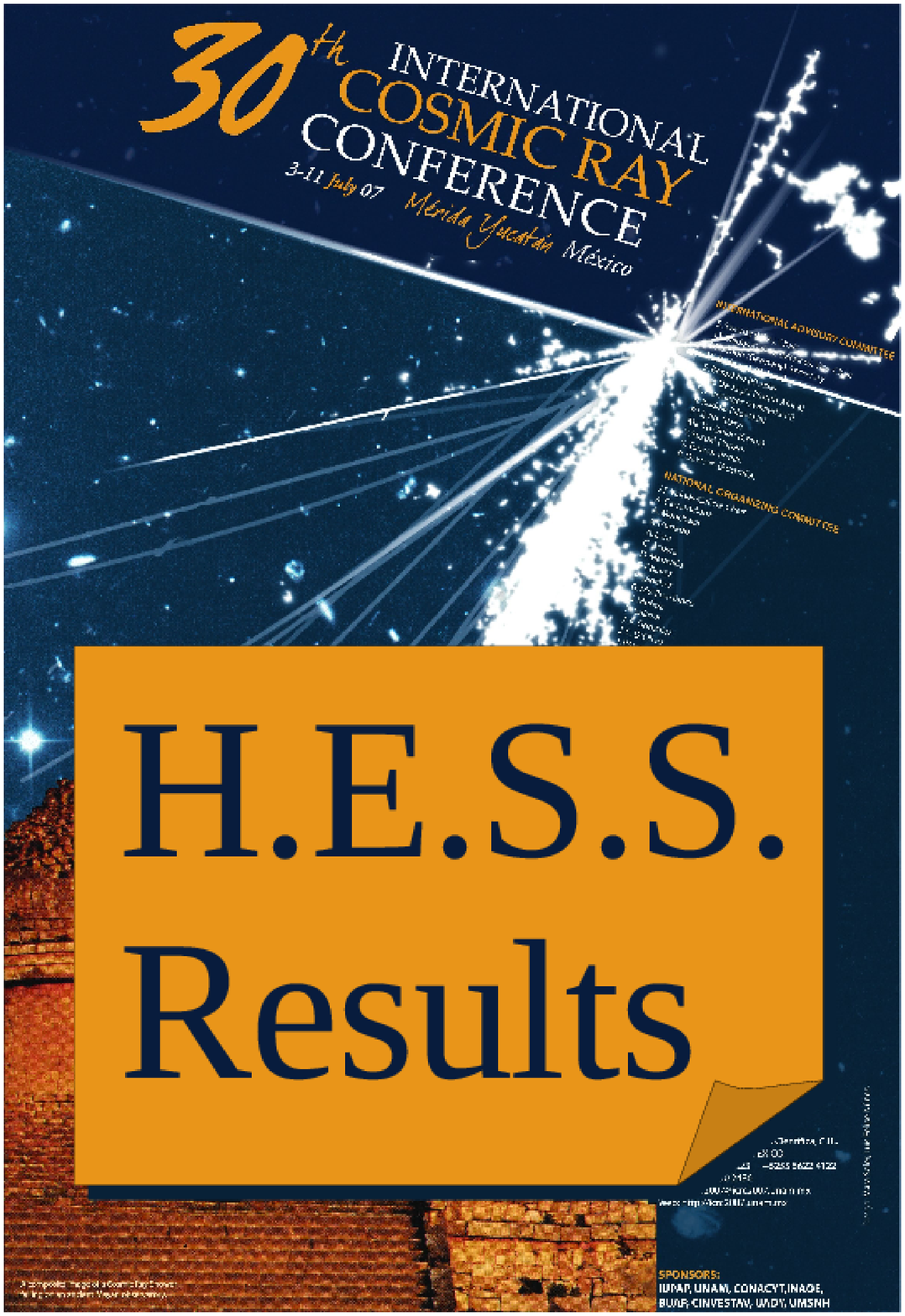}
\end{figure*}
\end{titlepage}

\begin{titlepage}
\pagestyle{empty}
\tableofcontents
\end{titlepage}

%%%%%%%%%%%%%%%%%%%%%%%%%%%%%%
% The H.E.S.S. Collaboration %
%%%%%%%%%%%%%%%%%%%%%%%%%%%%%%
\begin{titlepage}

\title{The H.E.S.S. Collaboration}

\authors{\small{F. Aharonian$^{1,13}$, A.G.~Akhperjanian $^{2}$, U.~Barres de Almeida $^{8}$ %\thanks{supported by CAPES Foundation, Ministry of Education of Brazil}
, A.R.~Bazer-Bachi $^{3}$, B.~Behera $^{14}$, M.~Beilicke $^{4}$, W.~Benbow $^{1}$%, D.~Berge $^{1}$ \thanks{now at CERN, Geneva, Switzerland}$
, K.~Bernl\"ohr $^{1,5}$, C.~Boisson $^{6}$, O.~Bolz $^{1}$
% ...
}
}

\afiliations{
$^{1}$Max-Planck-Institut f\"ur Kernphysik, P.O. Box 103980, D 69029
Heidelberg, Germany, 
$^{2}$Yerevan Physics Institute, 2 Alikhanian Brothers St., 375036 Yerevan,
Armenia, 
$^{3}$Centre d'Etude Spatiale des Rayonnements, CNRS/UPS, 9 av. du Colonel Roche, BP
4346, F-31029 Toulouse Cedex 4, France, 
$^{4}$Universit\"at Hamburg, Institut f\"ur Experimentalphysik, Luruper Chaussee
149, D 22761 Hamburg, Germany, 
$^{5}$Institut f\"ur Physik, Humboldt-Universit\"at zu Berlin, Newtonstr. 15,
D 12489 Berlin, Germany, 
$^{6}$LUTH, Observatoire de Paris, CNRS, Universit\'e Paris Diderot, 5 Place Jules Janssen, 92190 Meudon, 
France, 
% ...
$^{8}$University of Durham, Department of Physics, South Road, Durham DH1 3LE,
U.K., 
% ...
$^{13}$Dublin Institute for Advanced Studies, 5 Merrion Square, Dublin 2,
Ireland, 
$^{14}$Landessternwarte, Universit\"at Heidelberg, K\"onigstuhl, D 69117 Heidelberg, Germany, 
% ...
supported by CNRS and MPG\\
\\
\\
}

\maketitle
\end{titlepage}

%%%%%%%%
%  01  %
%%%%%%%%

%The paper title
\title{Primary particle acceleration above 100 TeV in the shell-type
Supernova Remnant RX J1713.7--3946 with deep H.E.S.S. observations}
%Short title to print in the headers to the final publication (Not showed in this print).
\shorttitle{Particle acceleration above 100 TeV in the SNR RX
J1713.7--3946 with H.E.S.S.}

%All paper authors
\authors{
  D.~Berge$^{1,2}$,
  F.~Aharonian$^{2,3}$,
  W.~Hofmann$^{2}$,
  M.~Lemoine-Goumard$^{4}$,
  O.~Reimer$^{5}$,
  G.~Rowell$^{6}$,
  H.J.~V\"olk$^{2}$,
  for the H.E.S.S.\ Collaboration
}

%Short title to print in the headers to the final publication (Not shown in this print).
\shortauthors{Berge et al.}

%All the affiliations.
\afiliations{
  $^1$ CERN PH Department, CH-1211 Geneva 23, Switzerland\\
  $^2$Max-Planck-Institut f\"ur Kernphysik, P.O. Box 103980, D-69029
  Heidelberg, Germany\\
  $^3$Dublin Institute for Advanced Studies, 5 Merrion Square,
  Dublin 2, Ireland\\
  $^4$Laboratoire Leprince-Ringuet, IN2P3/CNRS, Ecole Polytechnique,
  F-91128 Palaiseau,\\France\\
  $^5$Stanford University, HEPL $\&$ KIPAC, CA 94305-4085, USA\\
  $^6$School of Chemistry \& Physics, University of Adelaide, Adelaide
  5005, Australia
}
\email{berge@cern.ch}

%The abstract.
\abstract{The shell-type supernova remnant RX J1713.7--3946 was
observed during three years with the H.E.S.S. Cherenkov telescope
system. The first observation campaign in 2003 yielded the first-ever
resolved TeV gamma-ray image. Follow-up observations in 2004 and 2005
revealed the very-high-energy gamma-ray morphology with unprecedented
precision and enabled spatially resolved spectral studies. Combining
the data of three years, we obtain significantly increased statistics
and energy coverage of the gamma-ray spectrum as compared to earlier
H.E.S.S. results. We present the analysis of the data of different
years separately for comparison and demonstrate that the telescope
system operates stably over the course of three years. When combining
the data sets, a gamma-ray image is obtained with a superb angular
resolution of 0.06 degrees. The combined spectrum extends over three
orders of magnitude, with significant gamma-ray emission approaching
100 TeV. For realistic scenarios of very-high-energy gamma-ray
production, the measured gamma-ray energies imply efficient particle
acceleration of primary particles, electrons or protons, to energies
exceeding 100 TeV in the shell of RX J1713.7--3946.}

\maketitle

%Begin the section.

\setcounter{page}{6}

\addtocontents{toc}{\protect\contentsline {part}{\protect\large Supernova Remnants (SNR)}{}}
\addcontentsline{toc}{section}{Primary particle acceleration above 100 TeV in the shell-type Supernova Remnant RX J1713.7--3946 with deep H.E.S.S. observations}
\setcounter{figure}{0}
\setcounter{table}{0}
\setcounter{equation}{0}

\section*{Introduction}
\vspace{-0.4cm}
%%%%%%%%%%%%%%%%%%%%%%%%%%%  Fig3   %%%%%%%%%%%%%%%%%%%%%%%%%%%%%%%%%

\begin{figure*}
  \begin{center}
    \includegraphics [width=0.78\textwidth]{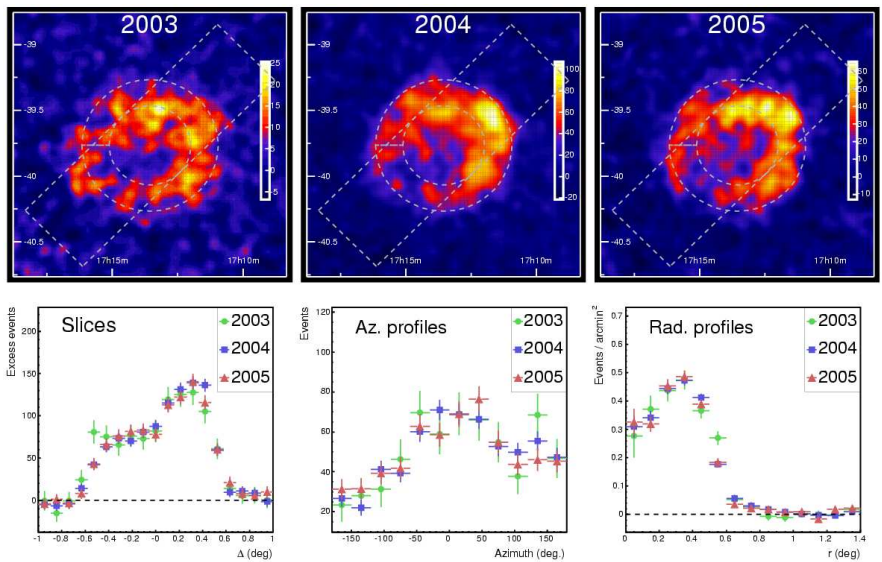}
  \end{center}
  \caption{\underline{\textbf{Upper panel:}} H.E.S.S.\ gamma-ray
    excess images from the region around RX~J1713.7$-$3946\ are shown
    for three years. \underline{\textbf{Lower panel:}} 1D
    distributions generated from the non-smoothed,
    acceptance-corrected gamma-ray excess images.  }
  \label{fig1}
\end{figure*}

 %    The images are corrected for the decline of the
 %    system acceptance with increasing distance to the SNR centre and
 %    smoothed with a Gaussian of $2\arcmin$. The linear colour scale
 %    is in excess counts per smoothing radius. The dashed box
 %    (dimensions $2\degr \times 0.6\degr$) and ring ($r_1 = 0.3\degr$,
 %    $r_2 = 0.5\degr$) are used for obtaining the 1D distributions
 %    shown in the lower panel. 
 %

%     \textbf{Left:} Slices taken within the rotated dashed box
%     running through the SNR region. \textbf{Middle:} Azimuth profiles
%     integrated in a thick ring covering the shell of
%     RX~J1713.7$-$3946. \textbf{Right:} Radial profiles around the
%     centre of the SNR.

%$0\degr$ corresponds to the west part of the shell, $90\degr$ is
%    north or upward, $-90\degr$ is south or downward. 

%%%%%%%%%%%%%%%%%%%%%%%%%%%%%%%%%%%%%%%%%%%%%%%%%%%%%%%%%%%%%%%%%%%%%%%%%

The energy spectrum of cosmic rays measured at Earth exhibits a
power-law dependence over a broad energy range. Starting at a few GeV
$(1~\mathrm{GeV} = 10^9~\mathrm{eV})$ it continues to energies of at
least $10^{20}~\mathrm{eV}$. The power-law index of the spectrum
changes at two characteristics energies: in the region around $3
\times 10^{15}~\mathrm{eV}$ -- the \emph{knee} region -- the spectrum
steepens, and at energies beyond $10^{18}~\mathrm{eV}$ it hardens
again. Up to the knee, cosmic rays are believed to be of Galactic
origin, accelerated in shell-type supernova remnants (SNRs) --
expanding shock waves initiated by supernova
explosions~\cite{HillasReview}. However, the experimental confirmation
of an SNR origin of Galactic cosmic rays is difficult due to the
propagation effects of charged particles in the interstellar
medium. The most promising way of proving the existence of high-energy
particles in SNR shells is the detection of very-high-energy (VHE)
gamma rays ($E > 100~\mathrm{GeV}$), produced in interactions of
cosmic rays close to their acceleration site~\cite{DAV}.

Recently the VHE gamma-ray instrument H.E.S.S. has detected two
shell-type SNRs, RX~J1713.7$-$3946~\cite{Hess1713a,Hess1713b} and
RX~J0852.0--4622~\cite{HessVelaJr_a,HessVelaJr_b}. The two objects
show an extended morphology and exhibit a shell structure, as expected
from the notion of particle acceleration in the expanding shock
fronts.
% Both reveal gamma-ray spectra that can be described by a hard power
% law (with photon index $\Gamma \sim 2.0$) over a broad energy
% range. For RX~J1713.7$-$3946\ significant deviations from a pure power
% law at larger energies are measured~\cite{Hess1713b}.
While it is difficult to attribute the measured VHE gamma rays
unambiguously to nucleonic cosmic rays (rather than to cosmic
electrons), the measured spectral shapes favour indeed in both cases a
nucleonic cosmic-ray origin of the gamma
rays~\cite{Hess1713b,HessVelaJr_b}.

%   In the case of RX~J1713.7$-$3946\ in addition a narrow shock
%   filament seen in X-rays~\cite{HiragaXMM} indicates strong
%   amplification of the magnetic field at least in one region of the
%   rim~\cite{BerezhkoVoelk}. If such an amplified magnetic field exists
%   throughout the main volume of the SNR~--~the region for which VHE
%   gamma-ray data is presented here~--~and if consequently high
%   magnetic field values are found not only in one shock filament, but
%   on a large part of the shock surface, a leptonic origin of the VHE
%   gamma rays becomes increasingly unlikely just based on the absolute
%   level of X-ray and gamma-ray flux of
%   RX~J1713.7$-$3946~\cite{Hess1713b}.
% 
Apart from the first unambiguous proof of multi-TeV particle
acceleration in SNRs, the question of the highest observed energies
remains an important one. Only the detection of gamma rays with
energies of 100~TeV provides experimental proof of acceleration of
primary particles, protons or electrons, to the \emph{knee} region
(1~PeV). Here we present a combined analysis of H.E.S.S.\ data of
RX~J1713.7$-$3946\ of three years, from 2003 to 2005. A comparison of
the three data sets demonstrates the expected steady emission of the
source as well as the stability of the system. Special emphasis is
then devoted to the high-energy end of the combined spectrum.
%\footnote{The analysis
%presented here is already published, where appropriate we condensed
%the information and would like to refer the reader for more details to
%the original publication~\cite{Hess1713c}.}

\section*{\hess\ observations}
\vspace{-0.4cm}
The High Energy Stereoscopic System (H.E.S.S.) consists of four
identical Cherenkov telescopes that are operated in the Khomas
Highland of Namibia. Its large field of view of $\approx 5\degr$ make
H.E.S.S.\ the currently best suited experiment in the field for the
study of extended VHE gamma-ray sources such as young Galactic SNRs.

The H.E.S.S.\ observation campaign of RX~J1713.7$-$3946\ started in
2003. The data were recorded during the commissioning phase of the
telescope system, with 2 out of the 4 telescopes operational. The data
set revealed extended gamma-ray emission resembling a shell
structure. It was actually the first ever resolved image of an
astronomical source obtained with VHE gamma rays. In 2004,
observations were conducted with the full telescope array. The
H.E.S.S.\ data enabled analysis of the gamma-ray morphology and the
spectrum of the remnant with unprecedented precision. A very good
correlation was found between the X-ray and the gamma-ray image. The
differential spectrum showed deviations from a pure power law at high
energies. The 2005 observation campaign was aiming at extending the
energy coverage of the spectrum to as high energies as
possible. Therefore the observations were preferentially pursued at
zenith angles larger than in the two years before to make use of the
drastically increased effective collection area of the experiment at
high energies. The analysis of these data are presented in the
following (a more detailed discussion of this analysis can be found in
\cite{Hess1713c}).

\section*{Analysis results}
\vspace{-0.4cm}

%%%%%%%%%%%%%%%%%%%%%%%%%%%  Fig2   %%%%%%%%%%%%%%%%%%%%%%%%%%%%%%

\begin{figure}
  \begin{center}
    \noindent
    \includegraphics [width=0.38\textwidth]{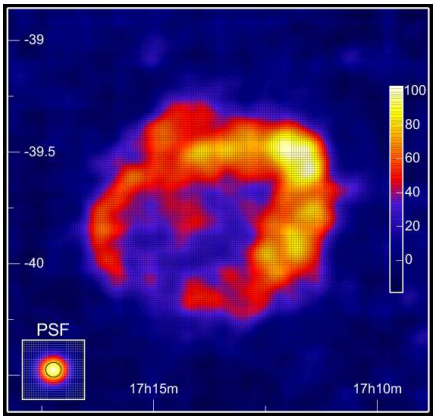}
  \end{center}
  \caption{Combined H.E.S.S.\ image of the SNR \rxj\ from the 2004 and
      2005 data. A simulated point source (\emph{PSF}) is also shown.}
  \label{fig2}
\end{figure}
%The acceptance-corrected gamma-ray excess image.

%  Note that for the 2005 data, only data recorded at zenith angles
%  less than $60\degr$ are taken into account. 

% The image is smoothed as in Fig.~\ref{fig1},

%%%%%%%%%%%%%%%%%%%%%%%%%%%%%%%%%%%%%%%%%%%%%%%%%%%%%%%%%%%%%%%%%%%%%%%%%

%%%%%%%%%%%%%%%%%%%%%%%%%%%  Fig3   %%%%%%%%%%%%%%%%%%%%%%%%%%%%%%%%%

\begin{figure*}
  \begin{center}
    \includegraphics [width=0.8\textwidth]{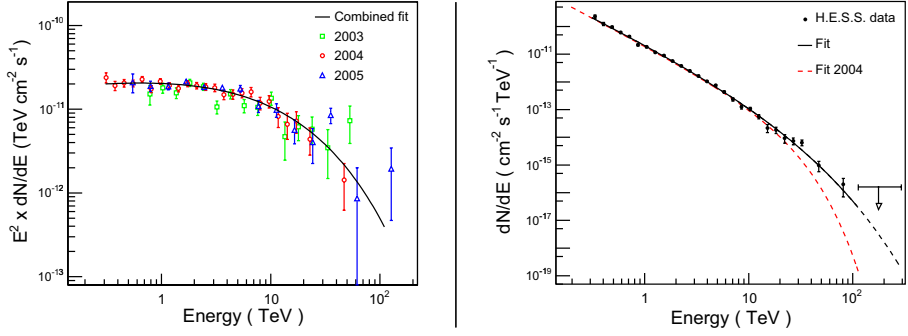}
  \end{center}
  \caption{\textbf{Left:} Comparison of H.E.S.S.\ energy-flux spectra
    of three years. The black curve is the best fit of a power law
    with exponential cutoff to the combined data, as shown on the
    \textbf{right}, where the combined \hess\ \gr\ spectrum of \rxj\
    is shown. The data are well described by the fit function, which
    is continued as dashed line beyond the fit range for
    illustration. The arrow is a model-independent upper limit,
    determined in the energy range from 113 to 300~TeV.}
  \label{fig3}
\end{figure*}

The analysis techniques used here are presented in detail
elsewhere~(\cite{Hess2155,BackgroundPaper}). The gamma-ray
morphology measured in three years is seen in the upper panel of
Fig.~\ref{fig1}. The images are readily comparable. Very similar
angular resolutions are achieved for all years. Good agreement is
achieved, as can also be seen from the 1D distributions shown in the
lower panel, where also the statistical errors are plotted. Shown from
left to right are a slice along a thick box (cf.\ Fig.~\ref{fig1},
upper panel), an azimuthal profile of the shell region, and a radial
profile. All the distributions are generated from the non-smoothed,
acceptance-corrected excess images. Clearly, there is no sign of
disagreement or variability, the H.E.S.S.\ data of three years are
well compatible with each other.
  
The combined H.E.S.S.\ image is shown in Fig.~\ref{fig2}. Data of 2004
and 2005 are used for this Gaussian smoothed~($\sigma=2\arcmin$),
acceptance-corrected gamma-ray excess image. In order to obtain
optimum angular resolution, a special high-resolution analysis is
applied here. Besides choosing only well reconstructed events, the cut
on the minimum event multiplicity is raised to three telescopes,
disregarding the 2003 data. Moreover, an advanced reconstruction
method is chosen, \emph{algorithm~3} of \cite{HofmannShowerReco}. The
image corresponds to 62.7 hours of observation time. 6702 gamma-ray
excess events are measured with a statistical significance of
$48\sigma$. An angular resolution of $0.06\degr$ ($3.6\arcmin$) is
achieved. The image confirms nicely the published H.E.S.S.\
measurements~\cite{Hess1713a,Hess1713b}, with 20\% better angular
resolution and increased statistics. The shell of RX~J1713.7$-$3946,
somewhat thick and asymmetric, is clearly visible and almost
closed. The gamma-ray brightest parts are located in the north and
west of the SNR.

The gamma-ray spectra measured with H.E.S.S.\ in three consecutive
years are compared to each other in Fig.~\ref{fig3}~(left). In order
to compare the data, a correction for the variation of optical
efficiency of the telescopes over the years must be
applied~\cite{HessCrab}. After that correction, very good agreement is
found. The measured spectral shape remains unchanged over time. The
absolute flux levels are well within the systematic uncertainty of
20\%. As expected for an object like RX~J1713.7$-$3946, no flux
variation is seen on yearly timescales. Clearly, the performance of
the telescope system is under good control.

The combined data of three years are shown in
Fig.~\ref{fig3}~(right). This energy spectrum of the whole SNR region
corresponds to $91$~hours of H.E.S.S.\ observations. The combined
spectrum extends over almost three decades in energy beyond 30~TeV,
and is compatible with previous H.E.S.S.\ measurements. Taking all
events with energies above 30~TeV, the cumulative significance is
$4.8~\sigma$. Different spectral models can be fit to the data. A pure
power law is clearly ruled out, alternative spectral models like a
power law with exponential cutoff, a broken power law, or a power law
with energy-dependent index, all provide significantly better
descriptions of the data, but none of these alternative models is
favoured over the other.

\section*{Summary}
\vspace{-0.4cm}
The complete H.E.S.S.\ data set of the SNR RX~J1713.7$-$3946\ recorded
from 2003 to 2005 is presented here. Very good agreement is found for
both the gamma-ray morphology and the differential energy spectra over
the years. The combined analysis confirms the earlier findings nicely:
the gamma-ray image reveals a thick, almost circular shell with
significant brightness variations. The spectrum follows a hard power
law with significant deviations at higher energies (beyond a few TeV).

In the combined image using $\sim63$~hours of H.E.S.S.\ observations
an unprecedented angular resolution of $0.06\degr$ is achieved. The
high-energy end of the combined spectrum approaches 100~TeV with
significant emission $(4.8\sigma)$ beyond 30~TeV. Given the systematic
uncertainties in the spectral determination at these highest energies
and comparable statistical uncertainties despite the long exposure
time, this measurement is presumably close to what can be studied with
the current generation of imaging atmospheric Cherenkov telescopes.

From the largest measured gamma-ray energies one can estimate the
corresponding energy of the primary particles. In case of
$\pi^0$-decay gamma rays, energies of 30~TeV imply that primary
protons are accelerated to $30~\mathrm{TeV} / 0.15 = 200~\mathrm{TeV}$
in the shell of RX~J1713.7$-$3946. On the other hand, if the gamma
rays are due to Inverse Compton scattering of VHE electrons, the
electron energies at the current epoch can be estimated in the
Thompson regime as
$E_\mathrm{e}~\approx~20~\sqrt{E_\gamma}~\mathrm{TeV} \approx
110~\mathrm{TeV}$. At these large energies Klein--Nishina effects
start to be important and reduce the maximum energy slightly such that
$\sim100~\mathrm{TeV}$ is a realistic estimate.

RX~J1713.7$-$3946\ remains an exceptional SNR in respect of its VHE
gamma-ray observability, being at present the remnant with the widest
possible coverage along the electromagnetic spectrum. The H.E.S.S.\
measurement of significant gamma-ray emission beyond 30~TeV without
indication of a termination of the high-energy spectrum provides proof
of particle acceleration in the shell of RX~J1713.7$-$3946\ beyond
$10^{14}$~eV, up to energies which start to approach the region of the
cosmic-ray \emph{knee}.

\bibliographystyle{plain}

%%%%%%%%
%  02  %
%%%%%%%%

%The paper title
\title{H.E.S.S. observations of the supernova remnant RX~J0852.0-4622: 
shell-type morphology and spectrum of a widely extended VHE gamma-ray 
source}
%Short title to print in the headers to the final publication (Not showed in this print).
\shorttitle{H.E.S.S. observations of the SNR RX~J0852.0-4622}

%All paper authors
\authors{M. Lemoine-Goumard$^{1}$, F. Aharonian$^{2,3}$, B. Degrange$^{4}$, L. Drury$^{2}$, U. Schwanke$^{5}$, for the H.E.S.S. Collaboration}
%Short title to print in the headers to the final publication (Not shown in this print).
\shortauthors{M. Lemoine-Goumard (for the H.E.S.S. Collaboration) et al.}
%All the affiliations.
\afiliations{$^1$CENBG, Universit\'e Bordeaux I, CNRS-IN2P3, Le Haut-Vigneau, 33175 Gradignan, France\\ $^2$Dublin Institute for Advanced Studies, 5 Merrion Square, Dublin, Ireland\\$^3$Max-Planck-Institut f\"ur Kernphysik, P.O. Box 103980, Heidelberg, Germany\\$^4$LLR, Ecole Polytechnique, CNRS-IN2P3, 91128 Palaiseau, France\\$^5$Institut f\"ur Physik, Humboldt-Universit\"at zu Berlin, D 12489 Berlin, Germany}
\email{lemoine@cenbg.in2p3.fr}

%The abstract.
\abstract{The shell-type supernova remnant RX~J0852.0-4622 was detected in 2004 and re-observed between December 2004 and May 2005 with the High Energy Stereoscopic System (H.E.S.S.), an array of four Imaging Cherenkov Telescopes located in Namibia and dedicated to the observations of $\gamma$-rays above 100~GeV. The angular resolution of $< 0.1^{\circ}$ and the large field of view of H.E.S.S. ($5^{\circ}$ diameter) are well adapted to studying the morphology of the object in very high energy gamma-rays, which exhibits a remarkably thin shell very similar to the features observed in the radio range and in X-rays. The spectral analysis of the source from 300~GeV to 20~TeV will be presented. Finally, the possible origins of the very high energy gamma-ray emission (Inverse Compton scattering by electrons or the decay of neutral pions produced by proton interactions) will be discussed, on the basis of morphological and spectral features obtained at different wavelengths.}

\maketitle

\addcontentsline{toc}{section}{H.E.S.S. observations of the supernova remnant RX~J0852.0-4622: shell-type morphology and spectrum of a widely extended VHE gamma-ray source}
\setcounter{figure}{0}
\setcounter{table}{0}
\setcounter{equation}{0}

%Begin the section.
\section*{Introduction}

Shell-type supernova remnants (SNR) are widely believed to be the prime candidates for accelerating cosmic rays up to $10^{15}$~eV, but until recently, this statement was only supported by indirect evidence, namely non-thermal X-ray emission interpreted as synchrotron radiation from very high energy electrons from a few objects. A more direct proof is provided by the detection of very high energy $\gamma$-rays, produced in nucleonic interactions with ambient matter or by inverse Compton scattering of accelerated electrons off ambient photons.\\
Here, we present recent data on RX~J0852.0$-$4622 obtained with H.E.S.S. in 2004 and 2005. %With a diameter of $2^{\circ}$, this source is the most extended SNR ever resolved in $\gamma$-rays.

\section*{The H.E.S.S. detector and the analysis technique}

H.E.S.S. is an array of four 13 m diameter imaging Cherenkov telescopes 
located in the Khomas Highlands in Namibia, 1800~m above sea level~\cite{HESS}. Each telescope has a tesselated mirror with an area of 107~m$^{2}$~\cite{HESSOptics}
and is equipped with a camera comprising 960 photomultipliers~\cite{HESSCamera} 
covering a field of view of 5$^{\circ}$ diameter. Due to the 
powerful rejection of hadronic showers provided 
by stereoscopy, the complete system (operational since December 2003) 
can detect point sources at flux levels of about 1\% of 
the Crab nebula flux near zenith with a significance of 5~$\sigma$ in 25 hours of observation. 
This high sensitivity, the angular resolution of a few arc minutes and the large field of 
view make H.E.S.S. ideally suited for 
the study of the $\gamma$-ray morphology of extended sources. 
During the observations, an array level 
hardware trigger required each shower to be observed by at least two telescopes within a coincidence window of 60~ns~\cite{HESSTrigger}. The data were recorded in runs of typical 28 minute 
duration in the so-called ``wobble mode'', where the source is offset from the center of 
the field of view, and were calibrated as described in detail in~\cite{HESSCalib}. In a first stage, 
a standard image cleaning was applied to shower images to remove the pollution due to 
the night sky background. The results presented in 
this paper were obtained using a 3D-modeling of the light-emitting region of an 
electromagnetic air shower, a method referred to as ``the 3D-model analysis''~\cite{HESSModel3D}, and were also cross-checked with the standard H.E.S.S. stereoscopic analysis based on the Hillas parameters of showers images~\cite{aharonian04}. The excess skymap was produced with a background subtraction called the ``Weighting Method''~\cite{WeightingMethod}. In this method, the signal and 
the background are estimated simultaneously in the same portion of the sky. In each sky bin 
(treated independently), the signal and the background are estimated from those events 
originating from this bin exclusively; this is done by means of a likelihood fit 
in which each event is characterized by a discriminating parameter whose distribution is fairly different for $\gamma$-rays and hadrons. In the case of the 3D-Model, this discriminating parameter is the 3D-width of the electromagnetic shower. 

\section*{H.E.S.S. results}
\begin{figure}[htbp]
\centering
\includegraphics[width=0.45\textwidth]{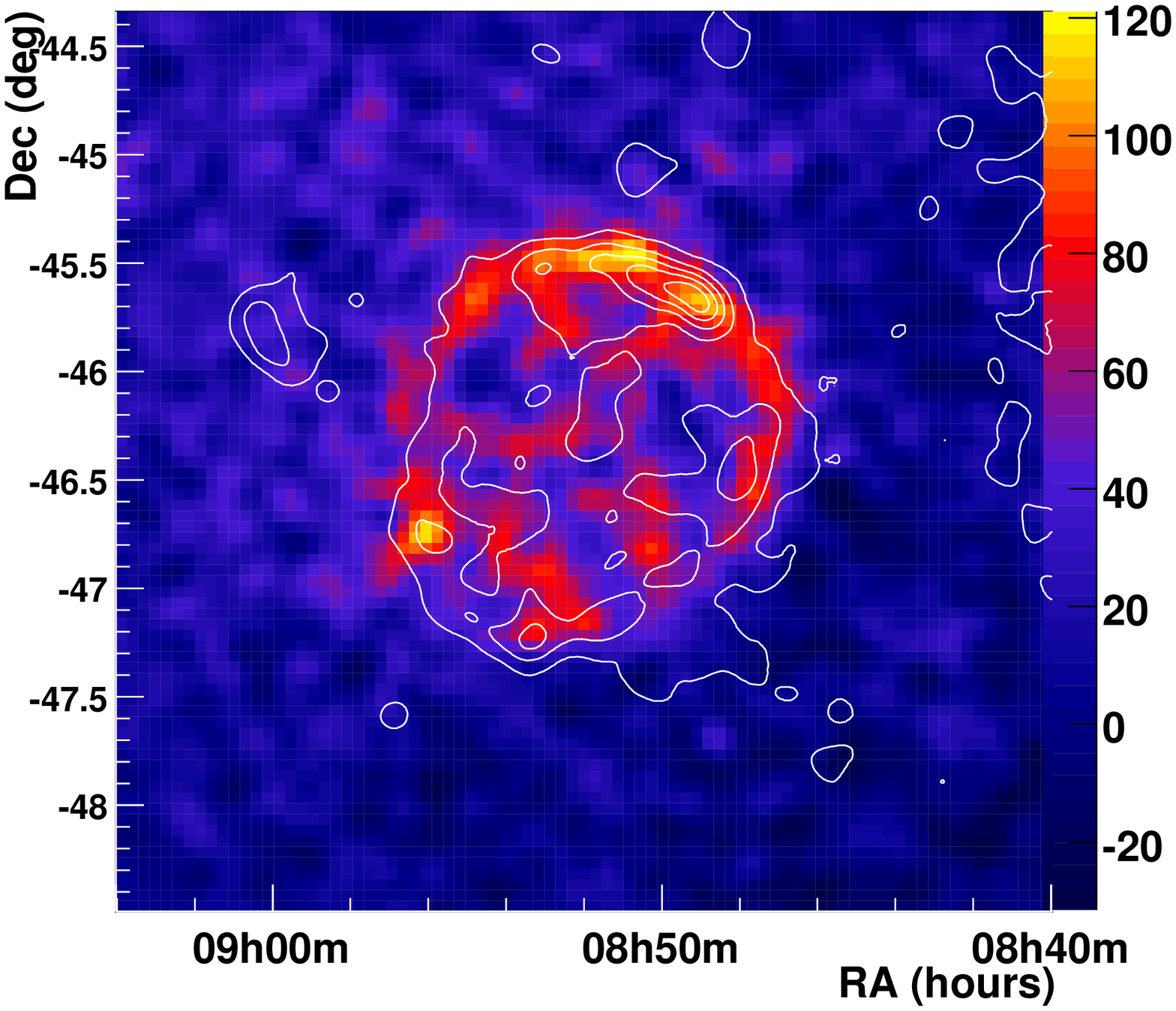}
\caption{Excess skymap of RX~J0852.0$-$4622 smoothed with a Gaussian of 0.06$^\circ$ 
standard deviation, obtained with the 3D-Model. The white lines are the contours of the X-ray data from the 
ROSAT All Sky Survey for energies higher than 1.3 keV (smoothed with a Gaussian 
of 0.06$^\circ$ standard deviation to enable direct comparison of the two images).
\label{fig:velajrmap}}
\end{figure}

\begin{figure}[htbp]
\centering
\includegraphics[width=0.45\textwidth]{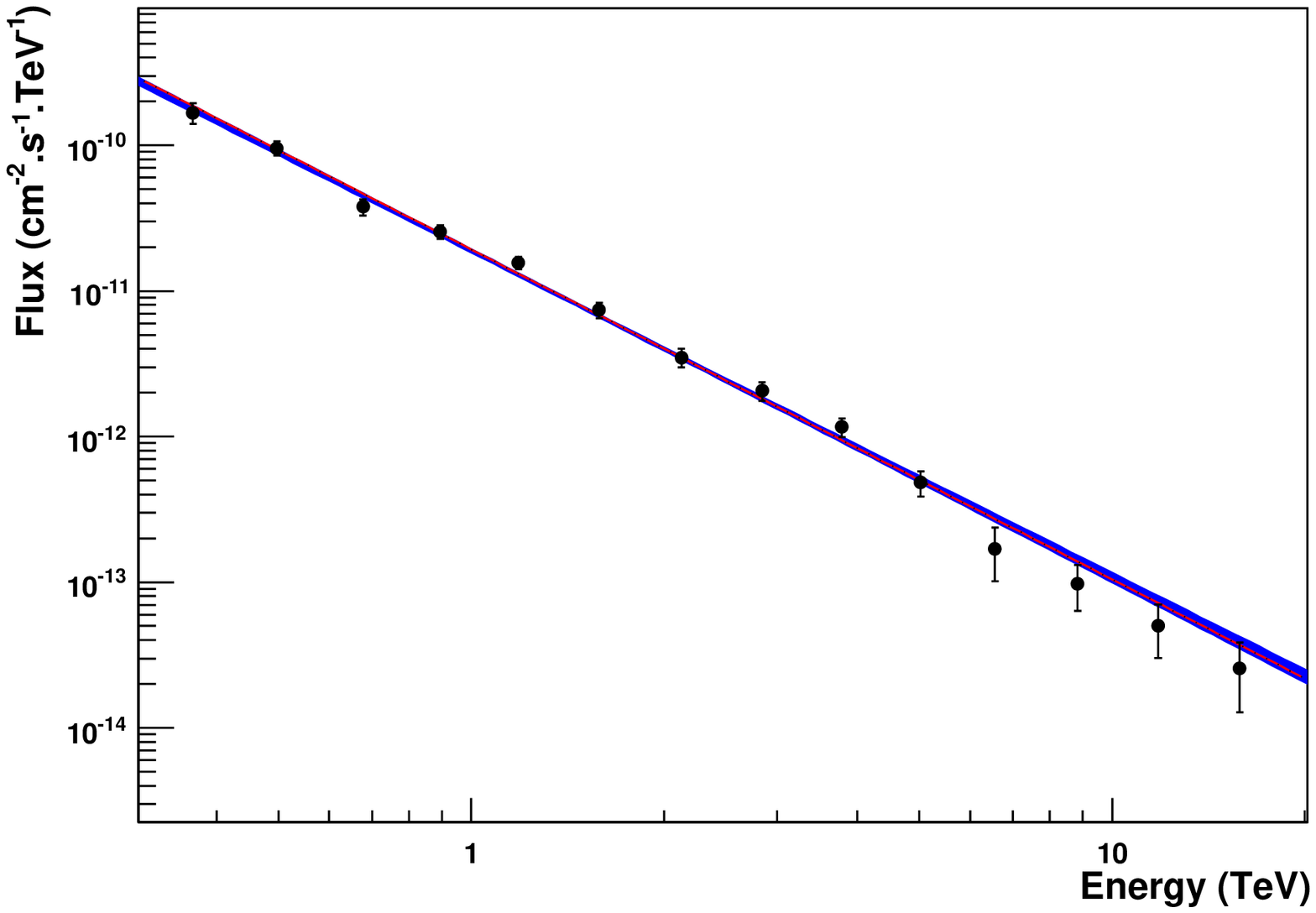}
\caption{Differential energy spectrum of RX~J0852.0$-$4622, for the whole region of 
the SNR. The shaded area gives the 1$\sigma$ confidence region for the spectral 
shape under the assumption of a power law. The spectrum ranges from 300 GeV to 20 TeV.
\label{fig:velajrspec}}   
\end{figure}

RX~J0852.0$-$4622 is a shell-type SNR discovered in the ROSAT all-sky survey. Its X-ray emission is mostly non-thermal~\cite{aschen98}. Indeed, up to now no thermal X-rays were detected from this source, which could imply a limit on the density of the material in the remnant $n_0 < 2.9 \times 10^{-2} (D/ 1 \, \mathrm{kpc})^{-1/2} f^{-1/2} \, \mathrm{cm^{-3}}$, where $f$ is the filling factor of a sphere taken as the emitting volume in the region chosen~\cite{slane}. The X-ray non-thermal spectrum of the whole remnant in the 2-10 keV energy band is well described by a power law with a spectral index of $2.7 \pm 0.2$ and a flux $F_X = 13.8 \times 10^{-11} \, \mathrm{erg \, cm^{-2} \, s^{-1}}$~\cite{velajrApJ}. In the TeV range, the announcement of a signal from the North-Western part of the remnant by CANGAROO was rapidly followed by the publication of a complete $\gamma$-ray map by H.E.S.S. obtained from a short period of observation (3.2~hours)~\cite{HESSVelaJr}. The study of this source is really complex due to several points: its extension (it is the largest extended source ever detected by a Cherenkov telescope), its location at the South-Eastern corner of the Vela remnant and the uncertainty on its distance and age. Indeed, RX~J0852.0$-$4622 could be as close as Vela ($\sim$ 250~pc) and possibly in
interaction with Vela, or as far as the Vela Molecular Ridge ($\sim$
1~kpc). Figure~\ref{fig:velajrmap} presents the $\gamma$-ray image of RX~J0852.0$-$4622 obtained with the 3D-Model from a long observation in 2005 (corresponding to 20~hours live time). The morphology appearing from this skymap reveals a very thin shell of $1^{\circ}$ radius and thickness smaller than $0.22^{\circ}$. Another interesting feature is the remarkably circular shape of this shell, even if the Southern part shows a more diffuse emission. Keeping all events inside a radius of $1^{\circ}$ around the center of the remnant, the cumulative significance is about 19$\sigma$ and the cumulative excess is $\sim 5200$ events. The overall $\gamma$-ray morphology seems to be similar to the one seen 
in the X-ray band, especially in the Northern part of the remnant where a brightening is seen in both wavebands. The correlation coefficient between the $\gamma$-ray 
counts and the X-ray counts in bins of 0.2$^\circ$ $\times$ 0.2$^\circ$ is found 
to be equal to 0.60 and comprised between 0.54 and 0.67 at 95$\%$ confidence level. The differential energy spectrum (Fig.~\ref{fig:velajrspec}) extends from 300~GeV up to 20~TeV. The spectral parameters were obtained from a maximum likelihood fit of a power law 
hypothesis dN/dE = $\mathrm{N_0}$~$\mathrm{(E/1 \, TeV)^{-\Gamma}}$ to the data, resulting 
in an integral flux above 1 TeV of ($15.2 \pm 0.7_{\mathrm{stat}} \pm 3.20_{\mathrm{syst}}$) $\times$ $10^{-12} \mathrm{cm^{-2}} 
\mathrm{s^{-1}}$ and a spectral index of 2.24 $\pm 0.04_{\mathrm{ stat}} 
\pm 0.15_{\mathrm{ syst}}$. An indication of curvature at high energy can be noticed. 

\section*{Emission processes}
One of the key issues is the interpretation of the $\gamma$-ray signal in terms of an electronic or a hadronic scenario. Despite the large uncertainty on the distance and age of the remnant, the multi-wavelength data already give some strong constraints. In a leptonic scenario, where $\gamma$-rays are produced by Inverse Compton scattering of high energy electrons off ambient photons, the ratio of the X-ray flux and the $\gamma$-ray flux determines the magnetic field to be close to $7 \, \mu \rm G$. This value is completely independent of the distance and only assumes a filling factor (fraction of the Inverse Compton emitting electrons containing the magnetic field responsible for the synchrotron emission) of 1; this low magnetic field seems hardly compatible with the amplification suggested by the thin filaments resolved by Chandra~\cite{chandra}. In the nearby case ($\sim 200$~pc), the limit on the width of the shell $\Delta R$ obtained by the morphological analysis of the H.E.S.S. data is $\Delta R < 0.7$~pc, which leads to an escape time by diffusion and by convection lower than both the age of the remnant and the synchrotron cooling time for energies higher than $\sim 10$~TeV. Therefore, one would expect to see a variation of the width of the shell with the energy, which is not observed by H.E.S.S.~\cite{velajrApJ} and disfavours the electronic scenario at this distance. In a hadronic scenario, in which we assume that the $\gamma$-ray flux is entirely due to proton-proton interactions, one can estimate the total energy in accelerated protons in the range $10 - 100$~TeV required to produce the $\gamma$-ray luminosity observed by H.E.S.S.:
\begin{eqnarray}
L_{\gamma} (1-10 \, \mathrm{TeV}) & = & 4 \pi D^2 \int_{1\,\mathrm{TeV}}^{10\,\mathrm{TeV}} E \phi(E) dE \nonumber \\ 
                               & = & 2.6 \times 10^{32} \big(\frac{D}{200 \, \mathrm{pc}} \big)^2 \mathrm{erg \, s^{-1}} \nonumber 
\end{eqnarray}
In this energy range, the characteristic cooling time of protons through the $\pi^0$ production channel is approximately independent of the energy and can be estimated to be: $\tau_{\gamma} = 4.4 \times 10^{15} \left(\frac{n}{1 \, \mathrm{cm^{-3}}} \right)^{-1}$. Assuming that the proton spectrum continues down to $E \approx 1$~GeV with the same spectral slope as that of the photon spectrum, the total energy injected into protons is estimated to be:
$$ W_p^{\rm tot} \approx 10^{49} \left( \frac{D}{200 \, \mathrm{pc}} \right)^2  \left( \frac{n}{1 \, \mathrm{cm^{-3}}} \right)^{-1} \mathrm{erg}$$ 
Therefore, for densities compatible with the absence of thermal X-rays, the only way to explain the entire $\gamma$-ray flux by proton-proton interactions in a homogeneous medium is to assume that RX~J0852.0$-$4622 is a nearby supernova remnant (D $< 600$~pc). Indeed, for larger distances and a typical energy of the supernova explosion of $10^{51}$~erg, the acceleration efficiency would be excessive. Nevertheless, a distance of 1~kpc should also be considered if one assumes that RX~J0852.0$-$4622 is the result of a core collapse supernova which exploded inside a bubble created by the wind of a massive progenitor star~\cite{bervolk06}. According to stellar wind theory, the size of the bubble evolves according to the formula: $R = 45 \left(\frac{n_0}{1 \, \rm{cm^{-3}}}\right)^{-0.2}$~pc. For a density of 1~$\mathrm{cm^{-3}}$, the radius of this bubble would be equal to 45~pc. In the case of a close supernova remnant, its size would be significantly lower than the size of the bubble and the hypothesis of a homogeneous medium would be satisfactory. In the opposite, for larger distances ($D \sim 1$~kpc), the presence of the Vela Molecular Ridge can produce a sudden increase of the density leading to a smaller bubble ($15.6$~pc for a density of $200 \, \mathrm{cm^{-3}}$), which would make the proton-proton interactions efficient at the outer shock.
 
\section*{Summary}
We have firmly established that the shell-type supernova remnant RX~J0852.0$-$4622 is a TeV emitter and for the first time we have resolved its morphology in the $\gamma$-ray range, which is highly correlated with the emission observed in X-rays. Its overall $\gamma$-ray energy spectrum extends over two orders of magnitude, providing the direct proof that particles of $\sim 100$~TeV are accelerated at the shock.\\
The question of the nature of the particles producing the $\gamma$-ray signal observed by H.E.S.S. was also addressed. In the case of a close remnant, the results of the morphological study combined with our spectral modeling highly disfavour the leptonic scenario which is unable to reproduce the thin shell observed by H.E.S.S. and the thin filaments resolved by Chandra. In the case of a medium distance, the explosion energy needed to explain the $\gamma$-ray flux observed by H.E.S.S., taking into account the limit on the density implied by the absence of thermal X-rays, would highly disfavour the hadronic process. At larger distances, both the leptonic and the hadronic scenario are possible, at the expense, for the leptonic process, of a low magnetic field of $\approx 7 \, \mu \rm G$. Such a small magnetic field exceeds typical interstellar values only slightly and is difficult to reconcile with the theory of magnetic field amplification at the region of the shock.\\
However, at present, no firm conclusions can be drawn from the spectral shape. The results which should hopefully be obtained by GLAST or H.E.S.S. II at lower energies will therefore have a great interest for the domain.

\section*{Acknowledgements}
The support of the Namibian authorities and of the University of Namibia
in facilitating the construction and operation of H.E.S.S. is gratefully
acknowledged, as is the support by the German Ministry for Education and
Research (BMBF), the Max Planck Society, the French Ministry for Research,
the CNRS-IN2P3 and the Astroparticle Interdisciplinary Programme of the
CNRS, the U.K. Science and Technology Facilities Council (STFC),
the IPNP of the Charles University, the Polish Ministry of Science and 
Higher Education, the South African Department of
Science and Technology and National Research Foundation, and by the
University of Namibia. We appreciate the excellent work of the technical
support staff in Berlin, Durham, Hamburg, Heidelberg, Palaiseau, Paris,
Saclay, and in Namibia in the construction and operation of the
equipment.

\bibliographystyle{plain}

%%%%%%%%
%  03  %
%%%%%%%%

%The paper title
\title{H.E.S.S. observations of the supernova remnant RCW 86}
%Short title to print in the headers to the final publication (Not showed in this print).
\shorttitle{H.E.S.S. observations of the SNR RCW~86}

%All paper authors
\authors{S. Hoppe$^{1}$ \& M. Lemoine-Goumard$^{2}$ for the
H.E.S.S. Collaboration}
%Short title to print in the headers to the final publication (Not shown in this print).
\shortauthors{S. Hoppe (for the H.E.S.S. Collaboration) et al.}
%All the affiliations.
\afiliations{$^1$Max-Planck-Institut f\"ur Kernphysik, P.O. Box
103980, Heidelberg, Germany\\ $^2$CENBG, Universit\'e Bordeaux I,
CNRS-IN2P3, Chemin du Solarium, 33175 Gradignan, France} \email{
hoppe@mpi-hd.mpg.de, lemoine@cenbg.in2p3.fr}

%The abstract.
\abstract{The shell-type supernova remnant (SNR) RCW 86 - possibly
associated with the historical supernova SN 185 - was observed during
the past three years with the High Energy Stereoscopic System
(H.E.S.S.), an array of four atmospheric-Cherenkov telescopes located
in Namibia. The multi-wavelength properties of RCW 86, e.g. weak radio
emission and North-East X-ray emission almost entirely consisting of
synchroton radiation, resemble those of two very-high energy (VHE;$>$
100 GeV) $\gamma$-ray emitting SNRs RX J1713.7-3946 and RX
J0852-4622. The H.E.S.S. observations reveal a new extended source of
VHE $\gamma$-ray emission.The morphological and spectral properties of
this new source will be presented.}

\maketitle

\addcontentsline{toc}{section}{H.E.S.S. observations of the supernova remnant RCW 86}
\setcounter{figure}{0}
\setcounter{table}{0}
\setcounter{equation}{0}

%Begin the section.
\section*{Introduction}

Shell-type supernova remnants are widely believed to be the prime
candidates for accelerating cosmic rays up to $10^{15}$~eV. The most
promising way of proving the existence of high energy particles
accelerated in SNR shells is the detection of VHE $\gamma$-rays
produced in nucleonic interactions with ambient matter. Recently, the
H.E.S.S. instrument has detected VHE $\gamma$-ray emission from two
shell-type SNRs, RX~J1713.7-3946 \cite{rxj1713} and RX~J0852.0-4622
\cite{velajr}. They both show an extended morphology highly correlated
with the structures seen in X-rays. Although a hadronic origin is
highly probable in the above cases, a leptonic origin can not be ruled
out.\\ Another young shell-type SNR is RCW~86 (also known as
G315.4-2.3 and MSH14-63). It has a complete shell in
radio~\cite{kesteven}, optical~\cite{smith} and
X-rays~\cite{pisarski}, with a nearly circular shape of $40$'
diameter. It received substantial attention because of its possible
association with SN~185, the first historical galactic
supernova. However, strong evidence for this connection is still
missing: using optical observations, Rosado et al.~\cite{rosado} found
an apparent kinematic distance of 2.8~kpc and an age of ~10 000 years,
whereas recent observations of the North-East part of the remnant with
the Chandra and XMM-Newton satellites strengthen the case that the
event recorded by the Chinese was a supernova and that RCW 86 is its
remnant~\cite{vink}. These observations also reveal that RCW 86 has
properties resembling the already established TeV emitting SNRs
mentioned above: weak radio emission and X-ray emission (almost)
entirely consisting of synchrotron radiation, which could be due to
the expansion of the shock in a wind blown bubble. The South-Western
rim seems to be completely different, with hard X-ray emission,
observed by ROSAT~\cite{bocchino}, mainly coming from stellar ejecta
possibly interacting with circumstellar layers ejected before the SN
explosion. The relatively large size of the remnant - about 40' in
diameter - and the observation of non-thermal X-rays make it a
promising target for $\gamma$-ray observations, aiming at increasing the
currently modest number of remnants where the shells are resolved in
VHE $\gamma$-rays. Hints for $\gamma$-ray emission from RCW 86 were
seen with the CANGAROO-II instrument~\cite{cangaroo}, but no firm
detection was claimed. Here, we present recent data on RCW~86 obtained
with the full H.E.S.S. array in 2005 and 2006 .

\section*{The H.E.S.S. detector and the analysis technique}

H.E.S.S. is an array of four imaging Cherenkov telescopes located
1800~m above sea level in the Khomas Highlands in
Namibia~\cite{HESS}. Each telescope has a tesselated mirror with an
area of 107~m$^{2}$~\cite{HESSOptics} and is equipped with a camera
comprising 960 photomultipliers~\cite{HESSCamera} covering a field of
view of 5$^{\circ}$ diameter. Due to the effective rejection of
hadronic showers provided by its stereoscopy, the complete system
(operational since December 2003) can detect point sources at flux
levels of about 1\% of the Crab nebula flux near zenith with a
statistical significance of 5~$\sigma$ in 25 hours of observation
\cite{hess_crab}.  This high sensitivity, the angular resolution of a
few arc minutes and the large field of view make H.E.S.S. ideally
suited for morphology studies of extended VHE $\gamma$-ray sources.\\
The data on RCW~86 were recorded in runs of typically 28 minutes duration in the
so-called ``wobble mode'', where the source is slightly offset from
the center of the field of view. As a cross-check, the obtained data were analysed
using two independent analysis chains, which share only the raw data. 
The first one is based on the combination of a semi-analytical shower model and a
parametrisation based on the moment method of Hillas to yield the
combined likelihood of the event being initiated by a
$\gamma$-ray~\cite{denaurois}. We will call this method the ``Combined
Model analysis'' in the following. The second analysis method is the
standard stereoscopic analysis based on the Hillas parameters of
the shower images~\cite{hillas}.

\section*{H.E.S.S. results}
\begin{figure}[h!]
  \begin{minipage}[t]{9.3cm}
    \includegraphics[width=0.8\textwidth]{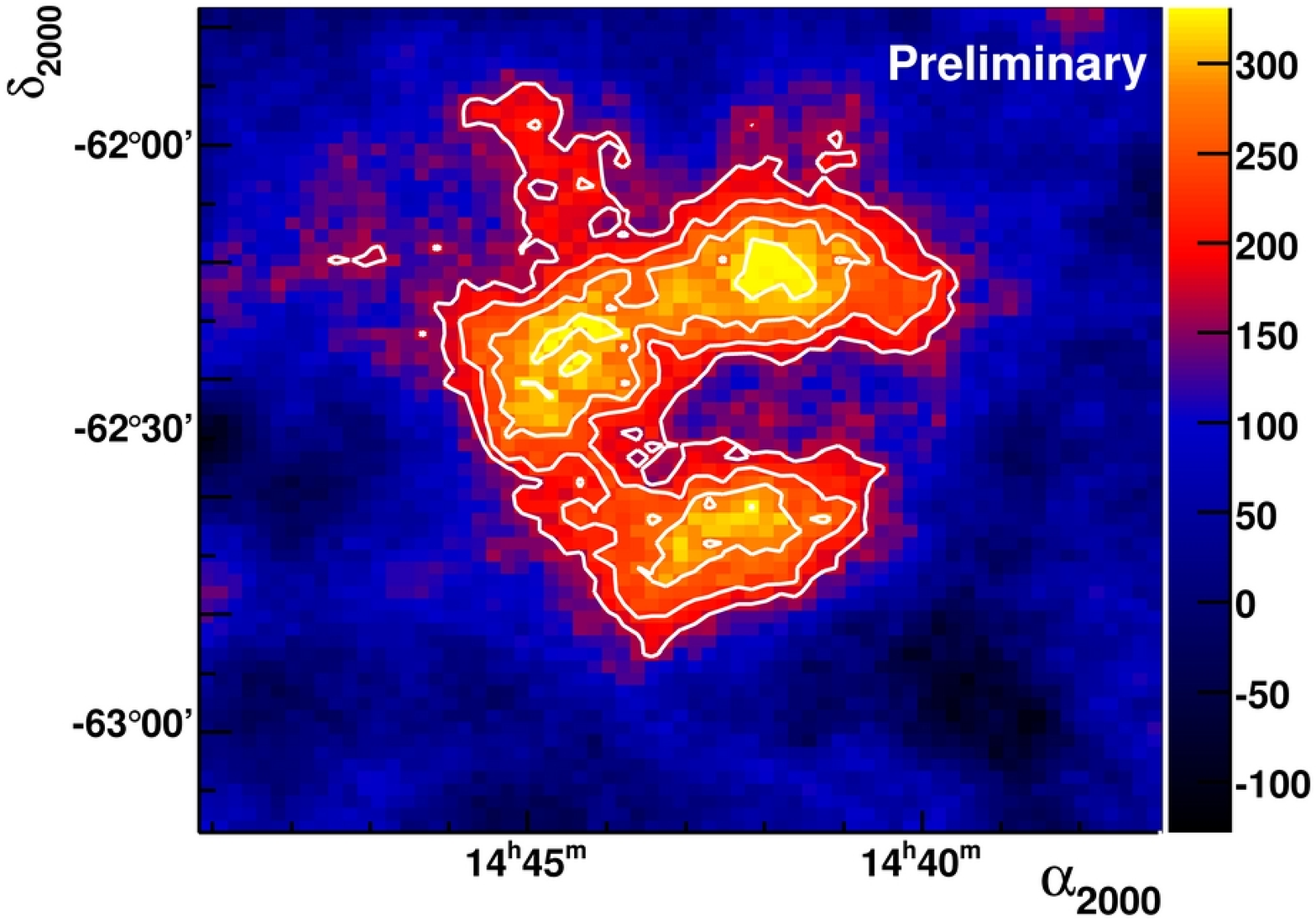}
  \end{minipage}
  \vspace{0.1cm}
  \begin{minipage}[t]{9.3cm}
    \includegraphics[width=0.8\textwidth]{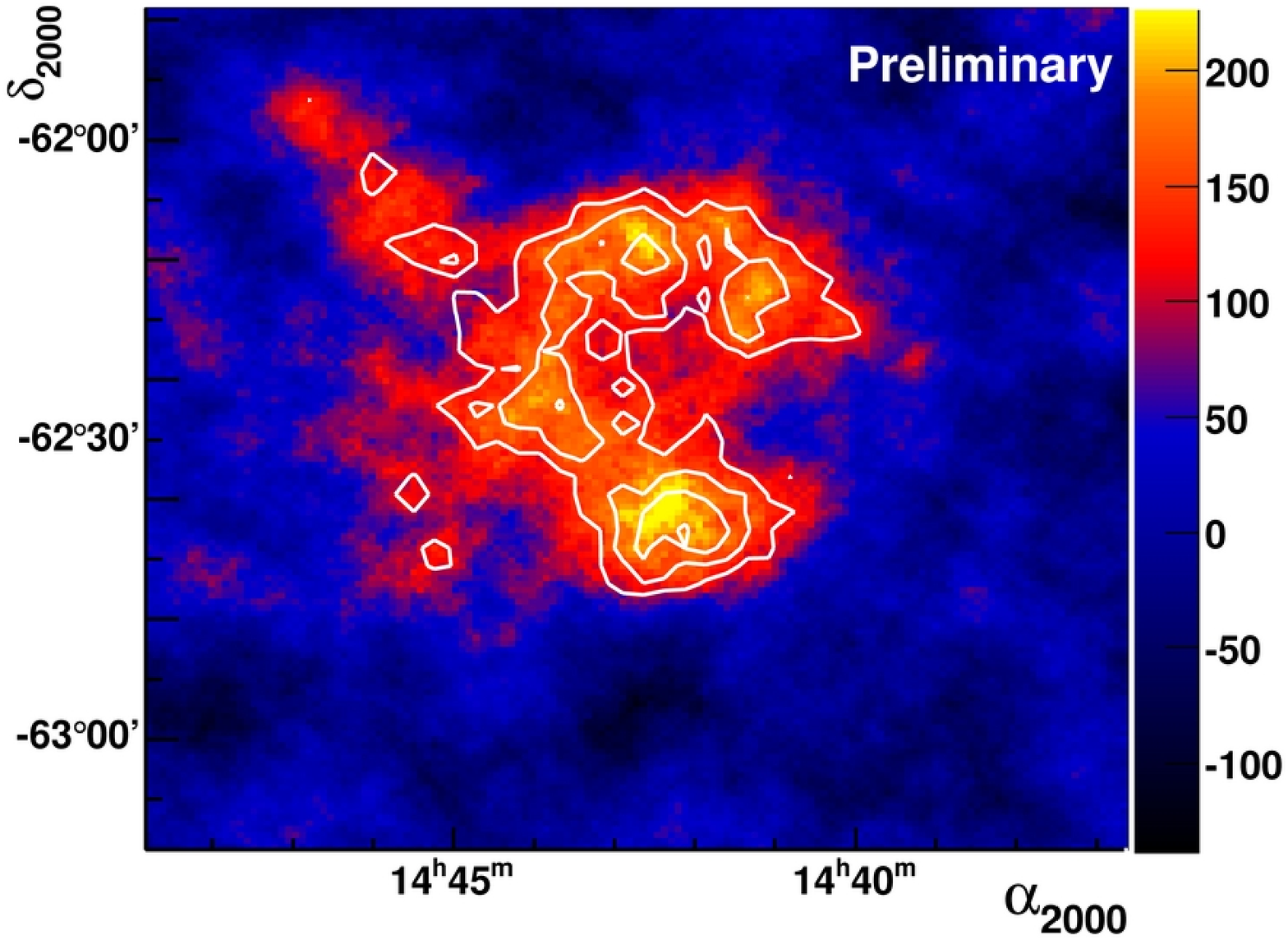}
  \end{minipage}
  \caption{VHE $\gamma$-ray emission from RCW~86, as seen
with H.E.S.S.. The top image shows the excess skymap obtained with the
Combined Model analysis where shower images are matched against image
templates, whereas the bottom image results from the classical,
slightly less sensitive Hillas analysis technique. White contours
correspond to 3, 4, 5, 6 sigma significance, obtained by counting
gamma rays within 0.11$^{\circ}$ from a given location.
\label{fig:rcw86model}}
\end{figure}

\begin{figure}[t!]
\centering
\includegraphics[width=0.4\textwidth]{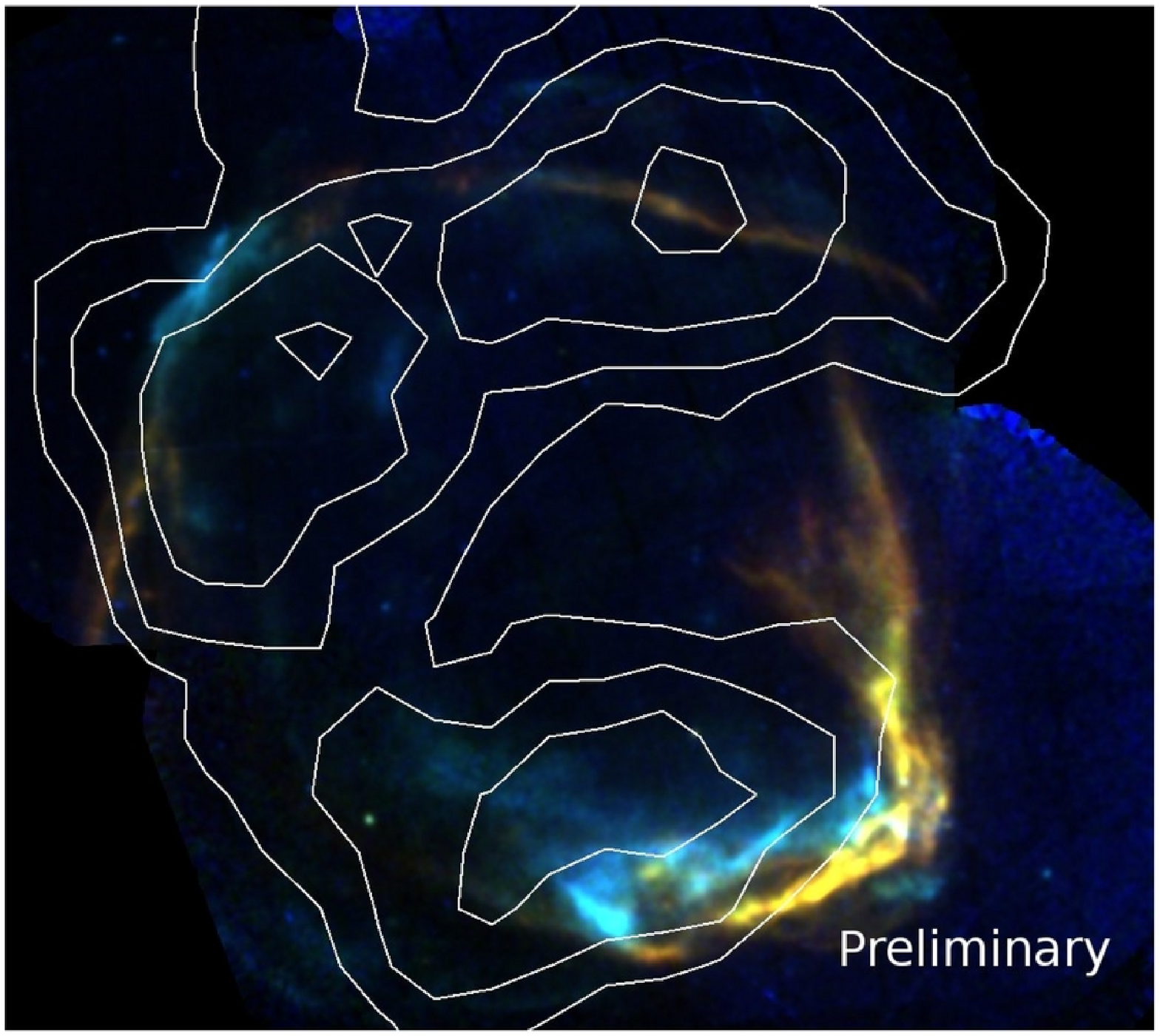}
\caption{Significance contours of gamma-ray emission (from the Combined Model analysis; 3, 4, 5, 6 sigma)
superimposed onto the XMM X-ray image of the remnant \cite{vink}.
\label{fig:xmm}}
\end{figure}

RCW 86 was observed for about 30 hours with the H.E.S.S. instrument
with a mean zenith angle of $41^{\circ}$. Within a circular region
of 27' radius (defined a priori so that it encompasses the
whole remnant) around the centre of the SNR ($\alpha_{J2000}$ =
14$^h$42$^m$43$^s$, $\delta_{J2000}$ = $-62^\circ$29'), a clear VHE
$\gamma$-ray signal with more than $9$ standard deviations is detected
with both analysis methods described above. The exact morphology of the
gamma-ray emission is still under study: whereas one type of data
analysis shows hints of a 3/4 shell resembling the shape of the X-ray
emission (Fig.~\ref{fig:rcw86model} top, Fig~\ref{fig:xmm}), this
morphology is not quite as evident with the other analysis technique
(Fig.~\ref{fig:rcw86model} bottom), and more data may be required to
fully settle this issue. The differential energy spectrum of RCW~86, $\phi(E)$,
was extracted from a circular region of diameter 22' around the position 
$\alpha_{J2000}$ = 14$^h$42$^m$12$^s$, $\delta_{J2000}$ = $-62^\circ$24' 
which is -- different from the region for which the detection significance was
determined -- adjusted to the H.E.S.S. data to include $\sim$ 90 \% of
the $\gamma$-ray excess. It is well described by a power-law with a spectral index of $\Gamma = 2.5 \pm 0.1_{\rm{stat}}$ and a flux normalisation at 1 TeV
of $\phi(1TeV) = (2.71 \pm 0.35_{\rm{stat}}) \times 10^{-12} \mathrm{cm^{-2}}
\mathrm{s^{-1}} \mathrm{TeV^{-1}} $ . The integral flux in the energy
range 1 - 10 TeV is $\sim$ 8\% of the integrated flux
of the Crab nebula within the same range. However, at this level of data
statistics, a power-law with index $\Gamma \sim 1.9$ and an
exponential cut-off at $E_c^{\gamma} \sim 5$ TeV is also a good fit to
the data.

\section*{Discussion}
There are two basic mechanisms for $\gamma$-ray production in young
supernova remnants, inverse Compton scattering of high energy
electrons off ambient photons (leptonic scenario) and $\pi^0$ mesons
produced in inelastic interactions of accelerated protons with ambient
gas decaying into $\gamma$-rays (hadronic scenario). The measured
$\gamma$-ray flux spectrum from RCW~86 translates into an energy flux
between 2 and 10 TeV of $3.4 \times 10^{-12} \, \rm erg \, cm^{-2} \,
s^{-1}$. In a leptonic scenario, the ratio of this energy flux and the
X-ray flux between 2 and 10 keV ($1.7 \times 10^{-10} \, \rm erg \,
cm^{-2} \, s^{-1}$, see Winkler \cite{winkler}) determines the magnetic field to be
close to $22 \, \mu \rm G$. This value, completely independent of the
distance and age of the SNR, is compatible with the calculation made
by J. Vink et al.~\cite{vink} based on the thin filaments resolved by
Chandra for a distance of 2.5~kpc in which he also deduces a high
speed of the blast wave ($\sim 2700 \, \rm km s^{-1}$). However, it is
more than ten times lower than the value proposed by H. J. Voelk and
his colleagues~\cite{volk} for the same distance using a much lower
velocity of the shock of $800 \, \rm km s^{-1}$ as suggested by
optical data in the Southern region of the SNR~\cite{rosado}.\\ In a
hadronic scenario, one can estimate the total energy in accelerated
protons $W_p$ in the range $10 - 100$~TeV required to produce the
$\gamma$-ray luminosity $L_{\gamma}$ observed by H.E.S.S. using the
relation:
\begin{equation}
W_p (10 - 100 \mathrm{TeV}) \approx \tau_{\gamma} \times L_{\gamma}(1
- 10 \mbox{TeV})
\end{equation}
in which $\tau_{\gamma} \approx 4.4 \times 10^{15} \left(\frac{n}{1 \,
\mathrm{cm^{-3}}} \right)^{-1}$ is the characteristic cooling time of
protons through the $\pi^0$ production channel. The correspnding
$L_{\gamma}$ can be calculated using:
\begin{eqnarray}
  L_{\gamma} (1-10 \, \mathrm{TeV}) & = & 4 \pi D^2
  \int_{1\,\mathrm{TeV}}^{10\,\mathrm{TeV}} E \phi(E) dE \nonumber \\
  & = & 2.8 \times 10^{31} \big(\frac{D}{200 \, \mathrm{pc}} \big)^2
  \mathrm{erg \, s^{-1}} \nonumber
\end{eqnarray} 
Finally, the total energy injected in protons is calculated by
extrapolating the proton spectrum with the same spectral shape as the
photon spectrum down to 1 GeV. Therefore, this estimation is highly
dependent on the shape of the $\gamma$-ray spectrum. Assuming that the
proton spectrum is a power-law with index $\Gamma = 2.5$, one would
obtain a total energy injected into protons of $W_p \rm (tot) = 3
\times 10^{51} \left( \frac{D}{2.5 \, \mathrm{kpc}} \right)^2 \left(
\frac{n}{1 \, \mathrm{cm^{-3}}} \right)^{-1} \, \rm erg$. 
For densities of $\sim 1 \rm \, cm^{-3}$, the only way to explain the
entire $\gamma$-ray flux by proton-proton interactions in a
homogeneous medium is to assume that RCW~86 is a close supernova
remnant ($\sim 1$ kpc). Indeed, for larger distances and a typical
energy of the supernova explosion of $10^{51}$~erg, the acceleration
efficiency would be excessive. For an exponential cut-off power-law
with $\Gamma= 1.9$ and $E_c = 10 \times E_c^{\gamma} = 50$ TeV, the
total energy injected into protons would be $10^{50} \left(
\frac{D}{2.5 \, \mathrm{kpc}} \right)^2 \left( \frac{n}{1 \,
\mathrm{cm^{-3}}} \right)^{-1}$ erg which would make the hadronic
scenario possible even at larger distances. However, the observation
of TeV gamma-rays from the remnant up to more than 10 TeV favors
somewhat the scenario of a young -- and therefore close-by -- remnant
with high expansion speed, easing the acceleration of high-energy
particles.

\section*{Summary}

H.E.S.S. observations have led to the discovery of the shell-type SNR
RCW~86 in VHE $\gamma$-rays . The $\gamma$-ray signal is extended but
the exact morphology of the emission region is still under
study. The flux from the remnant is $\sim$8\% of the flux from the
Crab nebula, with a similar spectral index of 2.5, but the spectrum is
also well described by a power law with index 1.9 and a cutoff around
5 TeV. The question of the nature of the particles producing the
$\gamma$-ray signal observed by H.E.S.S. was also addressed. However,
at present, no firm conclusions can be drawn from the spectral shape.

\section*{Acknowledgements}
The support of the Namibian authorities and of the University of Namibia
in facilitating the construction and operation of H.E.S.S. is gratefully
acknowledged, as is the support by the German Ministry for Education and
Research (BMBF), the Max Planck Society, the French Ministry for Research,
the CNRS-IN2P3 and the Astroparticle Interdisciplinary Programme of the
CNRS, the U.K. Science and Technology Facilities Council (STFC),
the IPNP of the Charles University, the Polish Ministry of Science and 
Higher Education, the South African Department of
Science and Technology and National Research Foundation, and by the
University of Namibia. We appreciate the excellent work of the technical
support staff in Berlin, Durham, Hamburg, Heidelberg, Palaiseau, Paris,
Saclay, and in Namibia in the construction and operation of the
equipment.

%This is the reference to .bib file (Without .bib!)

%This in the bibtex style, is ok.
\bibliographystyle{plain}

%%%%%%%%
%  04  %
%%%%%%%%

\newcommand{\etal}{\hbox{et al.} } 
\newcommand{\micro}{\mbox{\usefont{U}{eur}{m}{n}\char22}}

%The paper title
\title{Discovery of very high energy gamma-ray emission in the W~28 (G6.4$-$0.1) region, and multiwavelength comparisons}
%Short title to print in the headers to the final publication (Not showed in this print).
\shorttitle{VHE emission from the W~28 region}
%All paper authors
\authors{G. Rowell$^{1\dagger}$, E. Brion$^{2\dagger}$, O. Reimer$^{3\dagger}$, 
  Y. Moriguchi$^{4}$, Y. Fukui$^{4}$, A. Djannati-Ata\"i$^{5\dagger}$, S. Funk$^{3\dagger}$}
%Short title to print in the headers to the final puplication (Not showed in this print).
\shortauthors{G. Rowell et al.}
%All the affiliations.
\afiliations{$^1$ School of Chemistry \& Physics, University of Adelaide, Adelaide 5005, Australia\\
  $^2$ DAPNIA/DSM/CEA, CE Saclay, F-91191 Gif-sur-Yvette, Cedex, France\\
  $^3$ Stanford University, HEPL \& KIPAC, Stanford, CA 94305-4085, USA\\
  $^4$ Department of Astrophysics, Nagoya University, Chikusa-ku, Nagoya 464-8602, Japan\\
  $^5$ APC, 11 Place Marcelin Berthelot, F-75231 Paris Cedex 05, France \\
  $\dagger$ for the H.E.S.S. Collaboration {\tt www.mpi-hd.mpg.de/hfm/HESS}}

%The abstract.
\abstract{
  H.E.S.S. observations of the old-age ($>$10$^4$~yr; $\sim 0.5^\circ$ diameter) composite supernova remnant (SNR) W~28
  reveal very high energy (VHE) $\gamma$-ray emission situated at its northeastern and southern boundaries. The
  northeastern VHE source (HESS~J1801$-$233) is in an area where W~28 is interacting with a dense molecular cloud, 
  containing OH masers, local radio and X-ray peaks. The southern VHE sources (HESS~J1800$-$240 with components labelled
  A, B and C) are found in a region occupied by several HII regions, including the ultracompact HII region W~28A2. 
  Our analysis of NANTEN CO data reveals a dense molecular cloud enveloping this southern region, and our reanalysis
  of EGRET data reveals MeV/GeV emission centred on HESS~J1801$-$233 and the northeastern interaction region.
}

\email{growell@physics.adelaide.edu.au}

\maketitle

%Begin the section.
\addcontentsline{toc}{section}{Discovery of very high energy gamma-ray emission in the W~28 (G6.4$-$0.1) region, and multiwavelength comparisons}
\setcounter{figure}{0}
\setcounter{table}{0}
\setcounter{equation}{0}

\section*{Introduction \& H.E.S.S. Results}
The study of shell-type SNRs at $\gamma$-ray energies is motivated
by the idea that they are the dominant sites of hadronic Galactic cosmic-ray (CR)
acceleration to energies approaching the \emph{knee} ($\sim 10^{15}$~eV) and beyond, e.g. \cite{Ginzburg:1}. 
CRs are then accelerated via the diffusive shock acceleration (DSA) process 
(eg. \cite{Bell:1,Blandford:2}).
Gamma-ray production from the interaction of these CRs with ambient 
matter and/or electromagnetic fields is a tracer of such particle acceleration, 
and establishing the hadronic or electronic nature of the parent CRs in any $\gamma$-ray source is a key
issue. 
Already, two shell-type SNRs, RX~J1713.7$-$3946 and RX~J0852.0$-$4622, exhibit shell-like morphology in VHE $\gamma$-rays 
\cite{HESS_RXJ1713_II,HESS_VelaJnr_II,HESS_RXJ1713_III} to 20~TeV and above.
Although a hadronic origin of the VHE $\gamma$-ray emission is highly likely in the above cases, an electronic origin  
is not ruled out.

W~28 (G6.4$-$0.1) is a composite or mixed-morphology
SNR, with dimensions 50$^\prime$x45$^\prime$ and an estimated distance between 1.8 and 3.3~kpc 
(eg. \cite{Goudis:1,Lozinskaya:1}). 
It is an old-age SNR (age 3.5$\times 10^4$ to 15$\times 10^4$~yr \cite{Kaspi:1}), thought to have entered its 
radiative phase of evolution \cite{Lozinskaya:1}.
The shell-like radio emission \cite{Long:1,Dubner:1} peaks at the northern and northeastern
boundaries where interaction with a molecular cloud \cite{Wootten:1} is established \cite{Reach:1,Arikawa:1}.
The X-ray emission, which overall is well-explained by a thermal model, peaks in the SNR centre but has local enhancements 
in the northeastern SNR/molecular cloud interaction region \cite{Rho:2}. 
Additional SNRs in the vicinity of W~28 have also been identified: G6.67$-$0.42 and G7.06$-$0.12 \cite{Yusef:1}.
The pulsar PSR~J1801$-$23
with 
spin-down luminosity $\dot{E} \sim 6.2\times 10^{34}$ erg~s$^{-1}$ and distance $d\geq9.4$~kpc \cite{Claussen:3},
is at the northern radio edge.
                                   
Given its interaction with a molecular cloud, W~28 is an ideal target for VHE observations. 
This interaction is traced by the high concentration of 1720~MHz OH masers \cite{Frail:2,Claussen:1,Claussen:2}, 
and also the location of very high-density ($n>10^3$~cm$^{-3}$) shocked gas \cite{Arikawa:1,Reach:1}.  
Previous observations of the W~28 region at VHE energies by the CANGAROO-I telescope revealed no evidence for such emission 
\cite{Rowell:1} from this and nearby regions.

\begin{figure*}[th]
  \centering
  \hbox{
    \begin{minipage}{0.55\textwidth}
      \includegraphics[width=\textwidth]{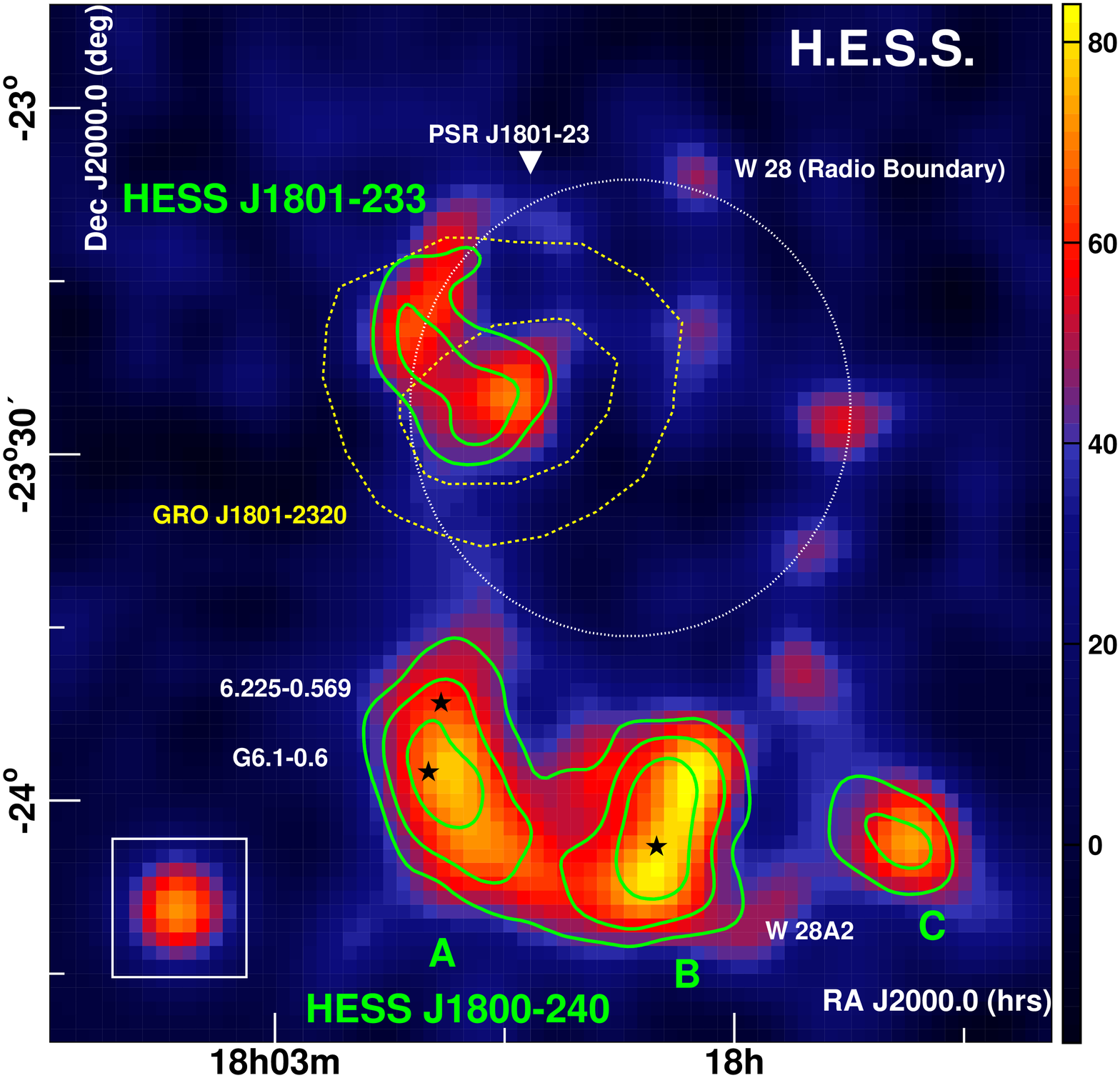}
    \end{minipage}
    \begin{minipage}{0.45\textwidth}    
      \caption{H.E.S.S. VHE $\gamma$-ray excess counts, corrected for 
	exposure and Gaussian smoothed (with 4.2$^\prime$ std. dev.). 
	Solid green contours represent excess  
	significance levels of 4, 5, and 6$\sigma$, for an integrating radius $\theta$=0.1$^\circ$. 
	The VHE sources HESS~J1801$-$233 
	and a complex of sources HESS~J1800-240 (A, B \& C) are indicated.
	The thin-dashed circle depicts the approximate radio boundary of the SNR W~28 \cite{Dubner:1,Brogan:1}.
	Additional objects indicated are: HII regions (black stars); W~28A2, {G6.1$-$0.6} %(Kuchar \& Clark \cite{Kuchar:1}), 
	{6.225$-$0.569}; %(Lockman \cite{Lockman:1});
	The 68\% and 95\% location contours (thick-dashed yellow lines) of the $E>100$~MeV EGRET source {GRO~J1801$-$2320};
	the pulsar {PSR~J1801$-$23} (white triangle). The inset depicts a pointlike source under identical analysis and smoothing 
	as for the main image.}
      \label{fig:tevskymap}
    \end{minipage}
  }
\end{figure*}

The High Energy Stereoscopic System (H.E.S.S.:  see \cite{Hinton:1} for details and performance) has observed 
the W~28 region over the 2004, 2005 and 2006 seasons.
After quality selection, a total of $\sim$42~hr observations were available for analysis. 
Data were analysed using the moment-based Hillas analysis procedure employing {\em hard cuts} (image size $>$200~p.e.), 
the same used in the analysis of the  inner Galactic Plane Scan datasets \cite{HESS_GalScan,HESS_GalScan_II}. An energy threshold of
$\sim 320$~GeV results from this analysis.
The VHE $\gamma$-ray image in Fig.~\ref{fig:tevskymap} shows that two source of 
VHE $\gamma$-ray emission are located at the northeastern and southern boundaries of W~28.
The VHE sources are labelled HESS~J1801$-$233 and HESS~J1801$-$240 where the latter can be further subdivided into
three components A, B, and C. The excess significances of both sources exceed $\sim$8$\sigma$ after integrating events within
their fitted, arcminute-scale sizes. 
Similar results were also obtained using an alternative analysis \cite{Mathieu:1}.

\section*{W~28: The Multiwavelength View}

We have revisited EGRET MeV/GeV data, including data from the CGRO observation cycles (OC) 1 to 6,
which slightly expands on the dataset of the 3rd EGRET catalogue (using OCs 1 to 4; \cite{Hartman:1}, 
revealing the source 3EG~J1800$-$2338. We find a pointlike $E>100$~MeV source in the W~28 region, labelled GRO~J1801$-$2320
in Fig~\ref{fig:tevskymap}. 
The 68\% and 95\% location contours of GRO~J1801$-$2320 match well the location of HESS~J1801$-$233. However we cannot 
rule out a connection to HESS~J1800$-$240 due to the degree-scale EGRET PSF.

\begin{figure*}[h]
  \centering
  \hbox{
      \includegraphics[width=0.5\textwidth]{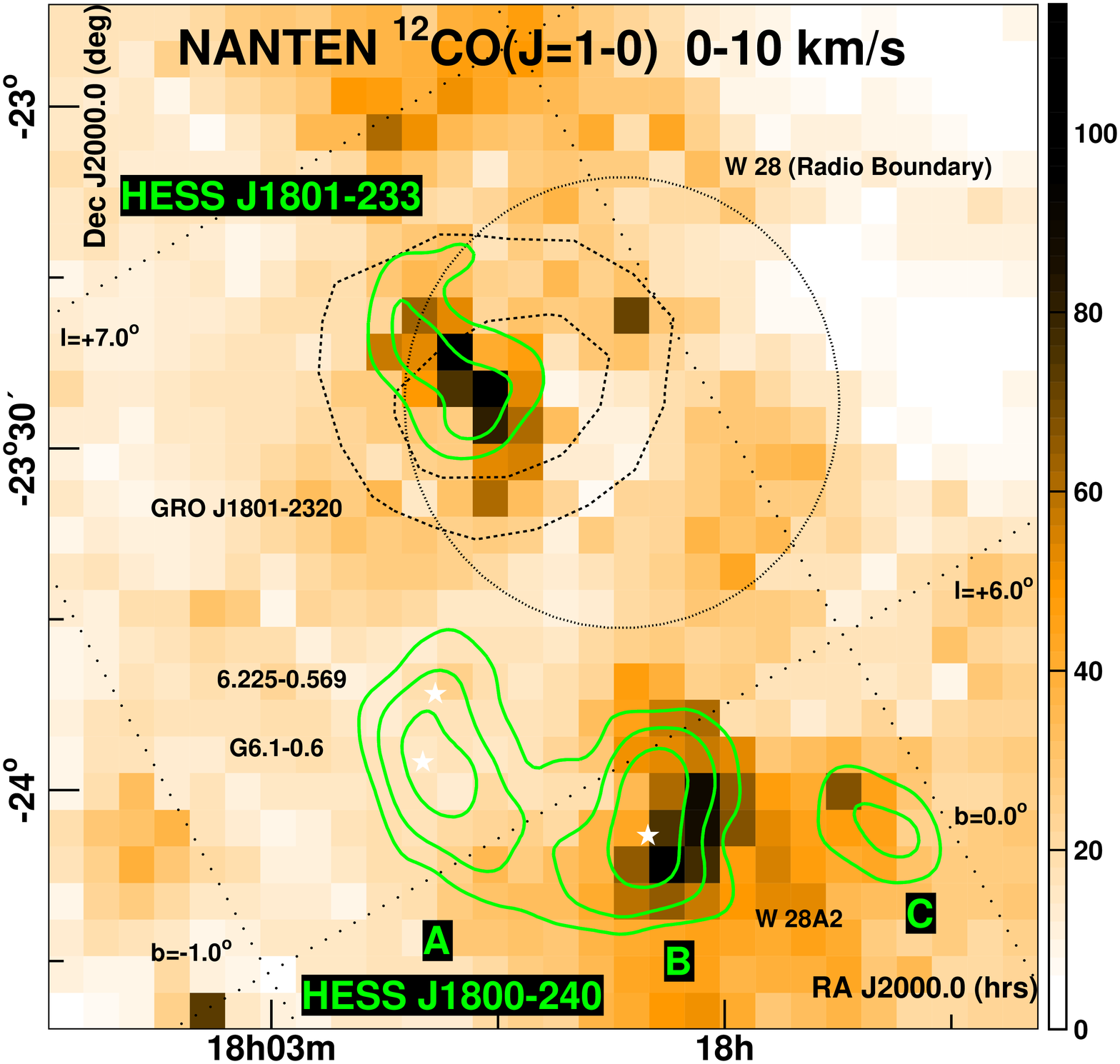}
	\includegraphics[width=0.5\textwidth]{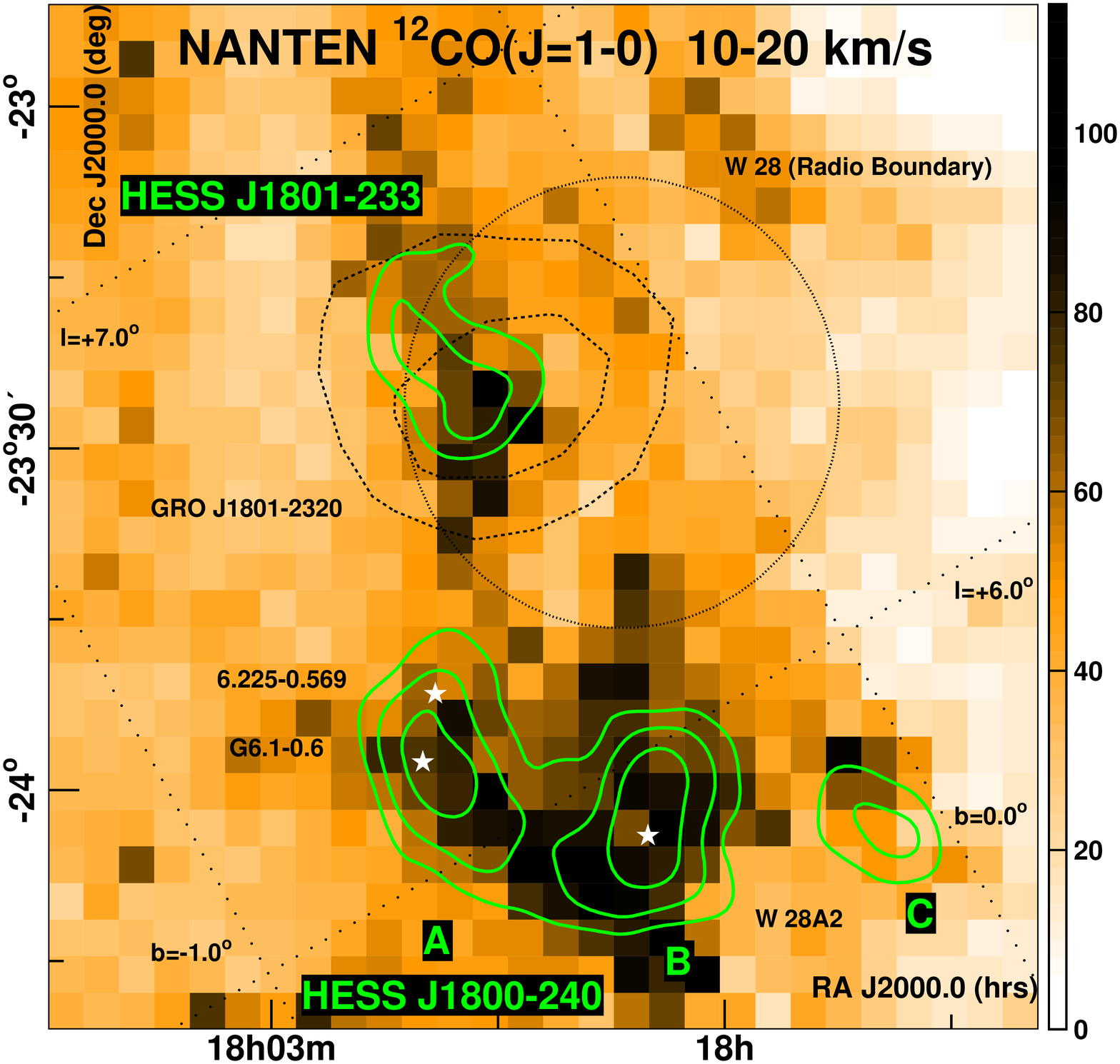}
  }
  \caption{{\bf Left:} NANTEN $^{12}$CO(J=1-0) image 
    (linear scale in K~km~s$^{-1}$) for $V_{\rm LSR}$=0 to 10~km~s$^{-1}$ with VHE $\gamma$-ray significance 
    contours overlaid (green) --- levels 4,5,6$\sigma$ and other features as in Fig.~\ref{fig:tevskymap}. 
    {\bf Right:}  
    NANTEN $^{12}$CO(J=1-0) image for $V_{\rm LSR}$=10 to 20~km~s$^{-1}$ (linear scale and same maximum as for left panel).}
  \label{fig:co_tev}
\end{figure*}

Fig.~\ref{fig:co_tev} presents $^{12}$CO ($J$=1--0) observations from the NANTEN \cite{Mizuno:1} Galactic Plane survey 
\cite{Matsunaga:1} covering the line-of-sight velocity ranges $V_{\rm LSR}$= 0 to 10~km~s$^{-1}$ and 10 to 20~km~s$^{-1}$.
These ranges represent distances 0 to $\sim$2.5~kpc and 2.5 to $\sim$4~kpc respectively and encompass the
distance estimates for W~28. We cannot rule out however, distances $\sim$4~kpc for the  $V_{\rm LSR}>$10~km~s$^{-1}$ cloud components.
It is clear that molecular clouds coincide well with the VHE sources. The northeastern cloud  $V_{\rm LSR}<$10~km~s$^{-1}$ component near HESS~J1801$-$233, is already
well-studied \cite{Reach:1,Arikawa:1}. Contributions from the $V_{\rm LSR}>$10~km~s$^{-1}$ cloud components are also likely.
The molecular cloud to the south of W~28 coincides well with HESS~J1800$-$240 and its three VHE components. 
The $V_{\rm LSR}<$10~km~s$^{-1}$ component of
this cloud coincides well with  HESS~J1800$-$240B, and may represent the dense molecular matter surrounding the ultra-compact HII region W~28A2.
This cloud also extends to $V_{\rm LSR}\sim$20~km~s$^{-1}$ and thus, similar to HESS~J1801$-$233, the total VHE emission in  HESS~J1800$-$240 may result
from several molecular cloud components in projection.  

\begin{figure*}[bh]
  \centering
  \hbox{
    \includegraphics[width=0.5\textwidth]{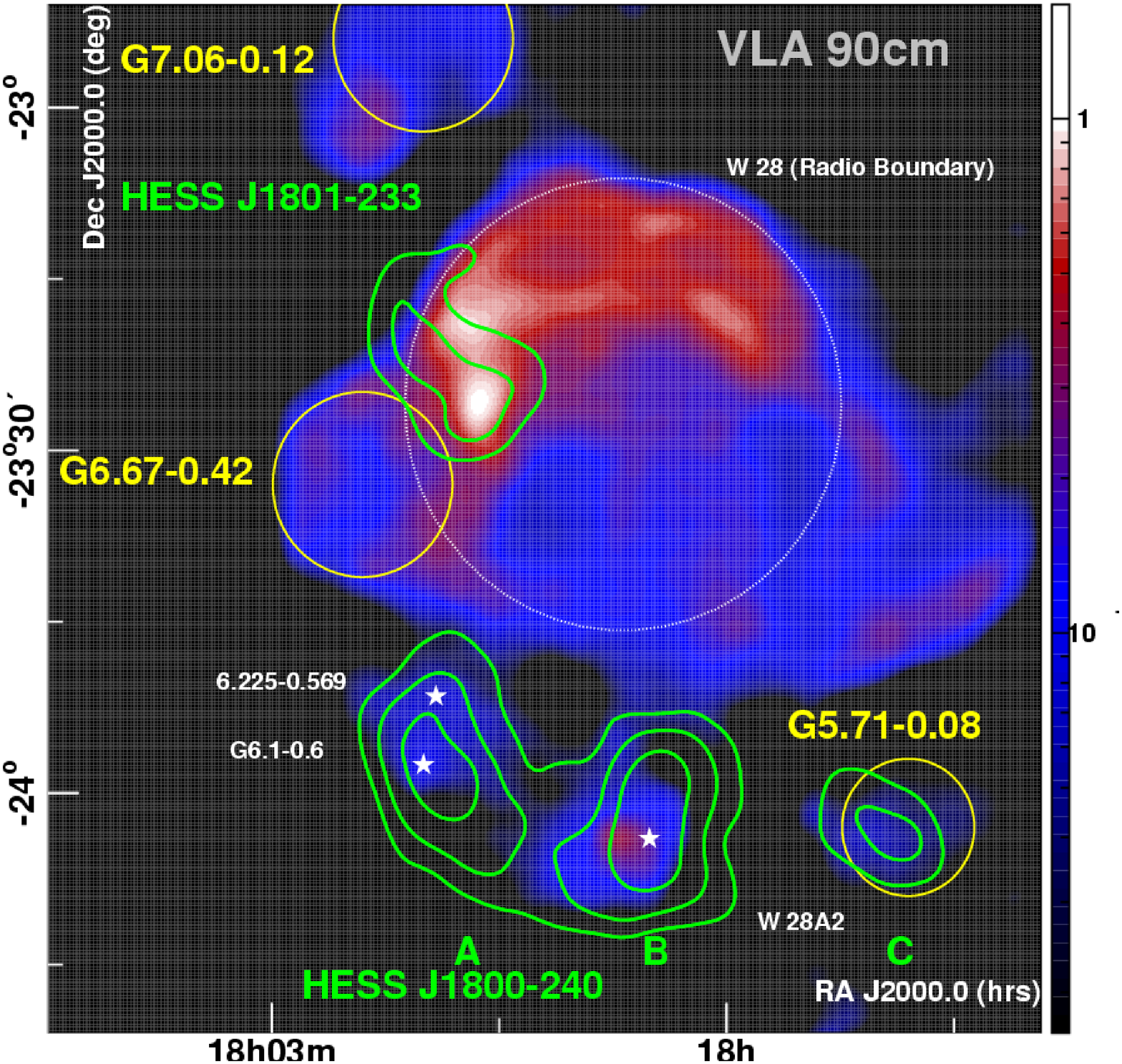}
    \includegraphics[width=0.5\textwidth]{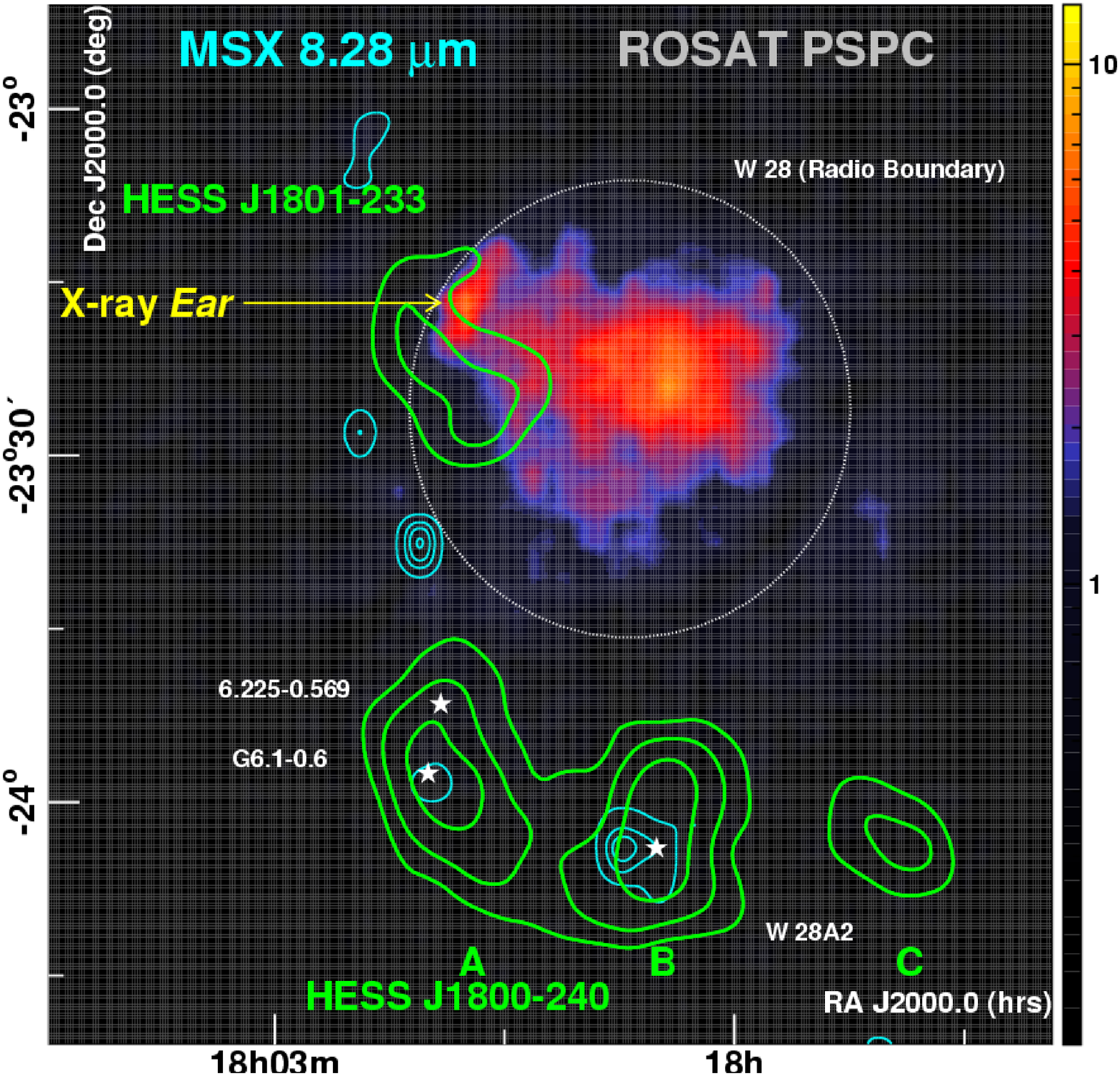}
  }
  \caption{{\bf Left:} VLA 90cm radio image \cite{Brogan:1} in Jy~beam$^{-1}$. 
    %(rebinned by a factor 1.2 compared to the original). 
    The VHE significance contours (green) from 
    Fig.~\ref{fig:tevskymap} are overlaid along with the HII regions (white stars) and the additional SNRs and SNR candidates 
    (with yellow circles indicating their location and approximate dimensions) discussed in text.
    {\bf Right:} 
    ROSAT PSPC image --- 0.5 to 2.4~keV (smoothed counts per bin \cite{Rho:2}). Overlaid are contours 
    (cyan --- 10 linear levels up to 5$\times 10^{-4}$~W~m$^{-2}$~sr$^{-1}$) from the MSX~8.28~\micro m image. Other contours and
    objects are as for the left panel. The X-ray {\em Ear} representing a peak at the northeastern edge is indicated.}
   \label{fig:mwl}
\end{figure*} 

Fig.~\ref{fig:mwl} compares the radio (left panel --- VLA 90~cm \cite{Brogan:1}), infrared and X-ray views (right panel 
MSX 8.28~\micro m and ROSAT PSPC 0.5 to 2.4~keV \cite{Rho:2}) 
with the VHE results. HESS~J1801$-$233 overlaps the northeastern shell of the SNR, coinciding with a strong peak in the 
90~cm continuum emission.
Additional SNRs G6.67$-$0.42 and G7.06$-$0.12 \cite{Yusef:1,Helfand:1} 
are indicated.
The non-thermal radio arc G5.71$-$0.08 is a SNR candidate \cite{Brogan:1}, and is possibly a counterpart to HESS~J1801$-$240C.
The distances to G6.67$-$0.42 and G5.71$-$0.08 are presently unknown. The unusual, ultracompact HII region W~28A2, 
is positioned within $0.1^\circ$ of the centroid of HESS~J1800$-$240B. W~28A2, at a distance $d\sim$2~kpc, 
exhibits energetic bipolar molecular outflows \cite{Harvey:1,Acord:1,Sollins:1} and may therefore be an energy source for 
particle acceleration in the region. The other HII regions 
G6.1$-$0.6 \cite{Kuchar:1} and 6.225$-$0.569 \cite{Lockman:1} are also associated with radio emission.

The X-ray morphology (Fig.~\ref{fig:mwl} right panel) shows the central concentration of
X-ray emission. A local X-ray peak or {\em Ear} is seen at the northeastern W~28 boundary.
The HII regions, W~28A2 and G6.1$-$0.6 %and 6.225$-$0.569 
are prominent in the MSX 8.28~\micro m image (Fig.~\ref{fig:mwl} right panel), indicating that 
a high concentration of heated dust still surrounds these very young stellar objects.

\section*{Discussion and Conclusions}
H.E.S.S. and NANTEN observations reveal VHE emission in the W~28 region spatially coincident with
molecular clouds. The VHE emission and molecular clouds are found at the northeastern boundary, 
and $\sim 0.5^\circ$ south of W~28 respectively. 
%The northeastern VHE source coincides with
%a radio peak and is adjacent to an X-ray feature. The southern VHE source has a component which 
%coincides with the ultra-compact HII region W~28-A2. 
The SNR W~28 may be a source of power for the VHE sources,
although there are additional potential particle accelerators in the region such as other SNR/SNR-candidates,
HII regions and open clusters.   
Further details concerning these results and discussion are presented in \cite{HESS_W28}.

\section*{Acknowledgements}

{\renewcommand{\baselinestretch}{-0.5}
\scriptsize
The support of the Namibian authorities and of the University of Namibia\\[-1.5mm]
in facilitating the construction and operation of H.E.S.S. is gratefully\\[-1.5mm]
acknowledged, as is the support by the German Ministry for Education \\[-1.5mm]
and Research (BMBF), the Max Planck Society, the French Ministry \\[-1.5mm]
for Research, the CNRS-IN2P3 and the Astroparticle Interdisciplinary  \\[-1.5mm]
Programme of the CNRS, the U.K. Particle Physics and Astronomy  \\[-1.5mm]
Research Council (PPARC), the IPNP of the Charles University,  \\[-1.5mm]
the Polish Ministry of Science and Higher Education, the South  \\[-1.5mm]
African Department of Science and Technology and National Research  \\[-1.5mm]
Foundation, and by the University of Namibia. We appreciate the  \\[-1.5mm]
excellent work of the technical support staff in Berlin, Durham,  \\[-1.5mm]
Hamburg, Heidelberg, Palaiseau, Paris, Saclay, and in Namibia in the \\[-1.5mm]
construction and operation of the equipment. The NANTEN project is \\[-1.5mm] 
financially supported from JSPS (Japan Society for the Promotion of \\[-1.5mm]
Science) Core-to-Core Program, MEXT Grant-in-Aid for Scientific \\[-1.5mm] 
Research on Priority Areas, and SORST-JST (Solution Oriented \\[-1.5mm]
Research for Science and Technology: Japan Science and Technology \\[-1.5mm]
Agency). We also thank Crystal Brogan for the VLA 90~cm image.}

%\nocite{ref4}
%\nocite{ref5}
%\nocite{ref6}
%\nocite{ref7}
%This is the reference to .bib file (Whitout .bib!)

{
\scriptsize
\bibliographystyle{plain}

}
\normalsize

%%%%%%%%
%  05  %
%%%%%%%%

%The paper title
\title{The Monoceros very-high-energy gamma-ray source}
%Short title to print in the headers to the final publication (Not showed in this print).
\shorttitle{A point-like $\gamma$-ray source in Monoceros}
%All paper authors
\authors{A. Fiasson$^{1}$, J. A. Hinton$^{2}$, Y. Gallant$^{1}$, 
A. Marcowith$^{1}$, O. Reimer$^{3}$, G. Rowell$^{4}$, for the H.E.S.S. Collaboration}
%Short title to print in the headers to the final puplication (Not showed in this print).
\shortauthors{A. Fiasson et al}
%All the affiliations.
\afiliations{$^1$Laboratoire de Physique Th\'eorique et Astroparticules, IN2P2/CNRS, Universit\'e Montpellier II, CC 70, Place Eug\`ene Bataillon, F-34095 Montpellier Cedex 5, France\\ $^2$School of Physics \& Astronomy, University of Leeds, Leeds LS2 9JT, UK\\$^3$Stanford University, HEPL \& KIPAC, Stanford, CA 94305-4085, USA\\$^4$School of Chemistery \& Physics, University of Adelaide, Adelaide 5005, Australia}
\email{Armand.Fiasson@lpta.in2p3.fr, J.A.Hinton@leeds.ac.uk}

%The abstract.
\abstract{The H.E.S.S. telescope array has observed the complex Monoceros Loop SNR/Rosette Nebula region which contains unidentified high energy EGRET sources and potential very-high-energy (VHE) $\gamma$-ray source. We announce the discovery of a new point-like VHE $\gamma$-ray sources, HESS J0632+057. It is located close to the rim of the Monoceros SNR and has no clear counterpart at other wavelengths. Data from the NANTEN telescope have been used to investigate hadronic interactions with nearby molecular clouds. We found no evidence for a clear association. The VHE $\gamma$-ray emission is possibly associated with the lower energy $\gamma$-ray source 3EG J0634+0521, a weak X-ray source 1RXS J063258.3+054857 and the Be-star MWC 148.}

\maketitle

\addcontentsline{toc}{section}{The Monoceros very-high-energy gamma-ray source}
\setcounter{figure}{0}
\setcounter{table}{0}
\setcounter{equation}{0}

%Begin the section.

\section*{Introduction}
Shell type supernova remnants (SNRs) are believed to be particle accelerator to energy up to a few hundred TeV. Observations of very high energy $\gamma$-ray (VHE; E $\geq$ 100 GeV) from these objects (Aharonian et al. 2006) confirm the presence of particles with energy higher than 10 TeV in these regions. The presence of molecular clouds in the vicinity of SNRs could reveal the nature of such particles as they would interact and produce VHE $\gamma$ rays. The Monoceros SNR (G205.5+0.5), situated at $\sim$1.6 kpc (Graham et al. 1982), apparently interacting with the Rosette Nebula (a young stellar cluster/ molecular cloud complex situated at 1.4 $\pm$ 0.1 kpc(Heinsberger et al. 2000)) is a candidate.\\
In the case of interaction of accelerated particles with interstellar medium producing neutral pions which decays in two $\gamma$ rays, we expect a correlation between $\gamma$-ray emission and matter concentration. We used NANTEN data to trace target material. The NANTEN 4m diameter sub-mm telescope at Las Campanas observatory, Chile, has been conducting a $^{12}$CO (J=1$\rightarrow$0) survey of the galactic plane since 1996, including the Monoceros region (Mizuno \& Fukui 2004).
 
\section*{H.E.S.S. observations and results}
The H.E.S.S. experiment is an array of four Cherenkov telescope installed in Namibia which detects $\gamma$ rays with energy in the 100 GeV to 50 TeV range. A more complete description of the H.E.S.S. experiment is given in Aharonian et al. 2004. The Monoceros loop region has been observed between March 2004 and March 2006 (Aharonian et al. 2007). The dataset includes 13.5~hours of data after quality selection and dead-time correction and was taken at zenith angles ranging between 29$^{\circ}$ and 59$^{\circ}$. It corresponds to a mean energy threshold of 400 GeV with standard cuts used in spectral analysis and 750 GeV with hard cuts used for the source search and position fitting.\\
We made a search for a point-like source on this dataset using a source size of 0.11$^{\circ}$ and a ring of radius 0.5$^{\circ}$ for background estimation. We found an excess corresponding to a statistical significance of 7.1$\sigma$. Fig. \ref{fig1} shows the NANTEN $^{12}$CO map with 4 and 6$\sigma$ levels of statistical confidence contours for the VHE $\gamma$-ray excess (yellow contours). As we made a blind search for a point-like source, the probability we get an excess at this position is increased by the number of positions in the field of view, here $\approx$10$^{5}$. This leads to a post-trials statistical significance of 5.3$\sigma$. The excess is confirmed by an independent analysis based on a fit of camera images to a shower model (\textit{Model Analysis}, see de Naurois 2006), which yields to a significance of 7.3$\sigma$ (5.6$\sigma$ post-trials).\\
\begin{figure}
\begin{center}
\includegraphics [width=0.32\textwidth]{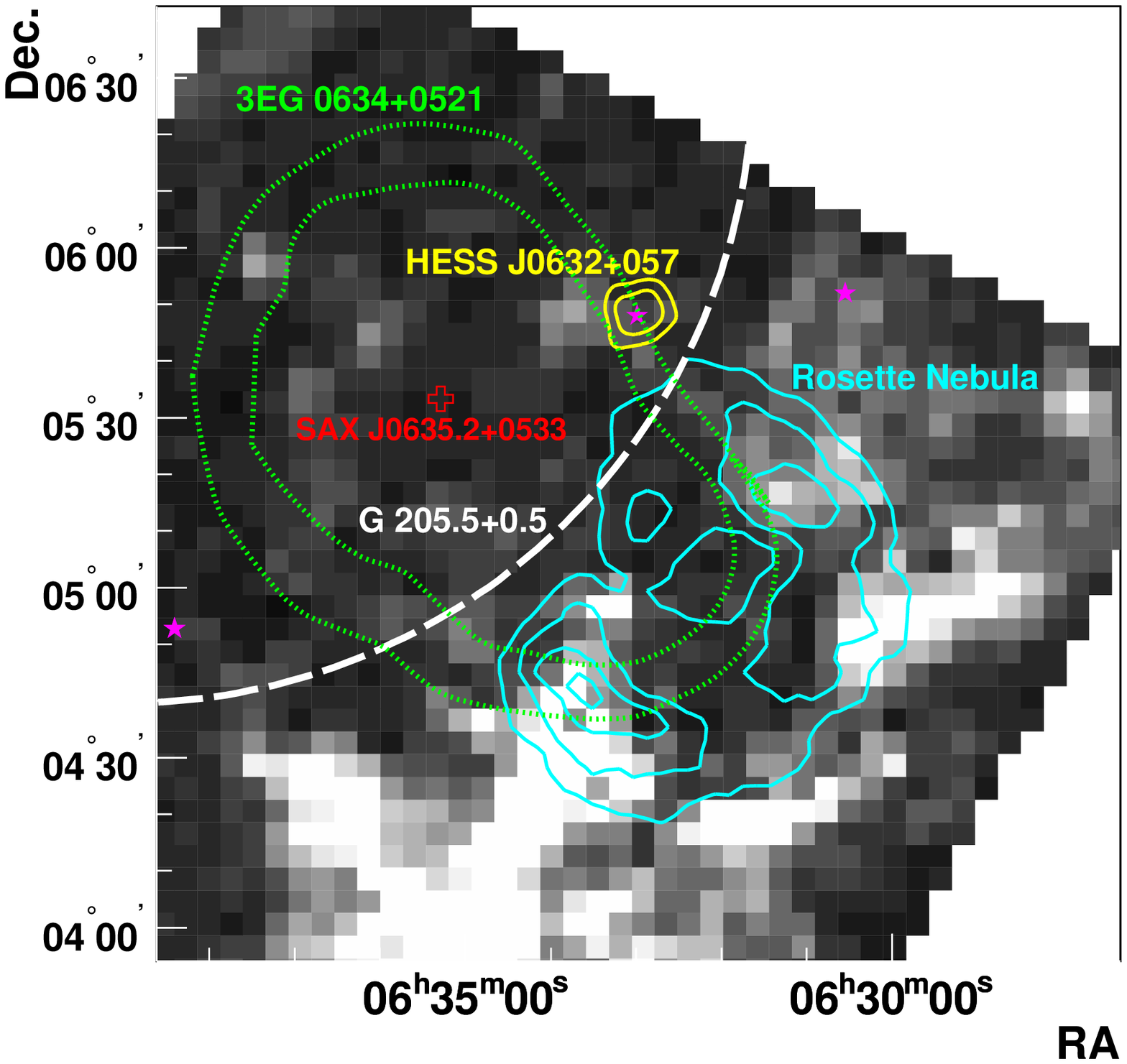}
\end{center}
\caption{$^{12}$CO (J=1$\rightarrow$0) emission from the Monoceros SNR / Rosette Nebula region. The gray-scale corresponds to velocity integrated (0-30 km.s$^{-1}$) emission from the NANTEN Galactic Plane Survey (white areas mean highest flux). The 4$\sigma$ and 6$\sigma$ levels for the statistical significance of a point-like VHE $\gamma$-ray source are shown as yellow contours. Extended cyan contours are radio observations at 8.35 GHz of the Rosette Nebula. The white dashed circle is the Green catalog nominal position and size of the Monoceros SNR. The dotted green contours are 95\% and 99\% confidence level for the position of the EGRET source 3EG J0634+0521. And last, the position of the binary pulsar SAX J0635.2+053 is marked as a red square and the position of Be-stars with pink stars.}
\label{fig1}
\end{figure}
\begin{figure}
\begin{center}
\includegraphics [width=0.33\textwidth]{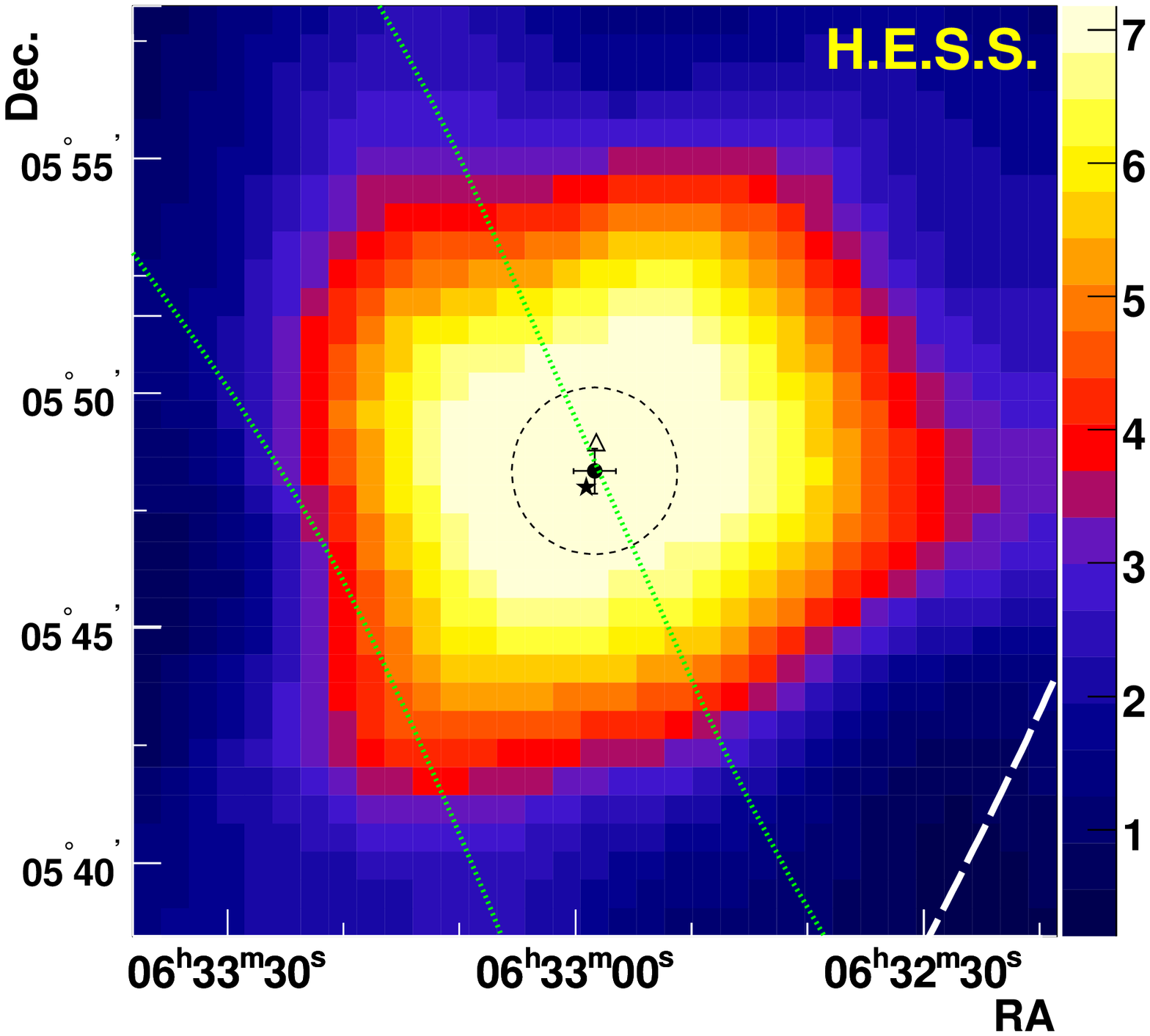}
\end{center}
\caption{Statistical significance map of the H.E.S.S. VHE $\gamma$-ray source. The rms size limit is shown as a dotted circle. Dotted green contours are 95\% and 99\% confidence level for the position of the EGRET source 3EG J0634+0521. The unidentifed X-ray source 1RXS J063258.3+054857 is marked with a triangle and the Be-star MCW 148 with a star.}
\label{fig2}
\end{figure}
The fitted position of this new source HESS J0632+057, is 6$^{h}$32$^{min}$58.3$^{s}$, +5$^{\circ}$48'20" (RA/Dec. J2000) with 28" statistical errors on each axis (fig. \ref{fig2}). We estimated systematics errors at 20" on each axis. The fig. \ref{fig3} represents the distribution of signal in function of the angular distance around the fitted position. The distribution is fully compatible with the point spread function (red curve). We derived an upper limit on the size of the source of 2' at 95\% confidence assuming a Gaussian profile for the source.\\
The reconstructed energy spectrum of the excess is consistent with a power-law of index $\Gamma = 2.53 \pm 0.26 \pm 0.20$ and differential flux at 1 TeV $\Phi_{TeV} = 9.1 \pm 1.7 \pm 3.0 \times 10^{-13} $cm$^{-2}$s$^{-1}$TeV$^{-1}$. The first errors are statistical errors and the second are estimated systematic errors. Fig.\ref{fig4} represents the VHE $\gamma$-ray reconstructed flux together with that for the EGRET sources 3EG0634+0521 and the upper limit derived by the HEGRA telescope array for the EGRET source position (converted to differential flux assuming the spectral shape observed by H.E.S.S.). There is no evidence of flux variability in our dataset but the sparse sampling of data together with the weakness of the source do not permit to constrain strongly intrinsic variability of the source.\\

\begin{figure}
\begin{center}
\includegraphics [width=0.45\textwidth]{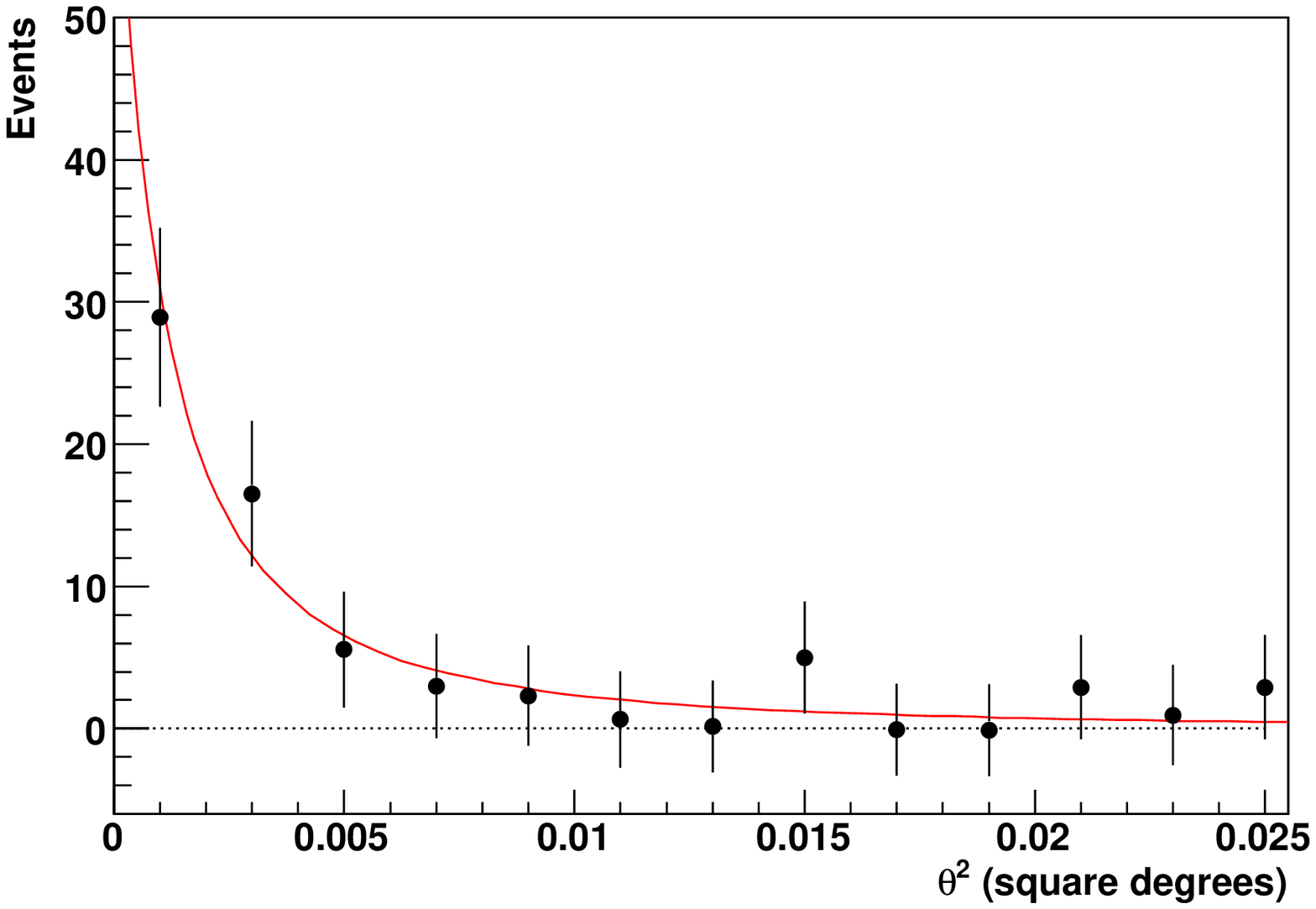}
\end{center}
\caption{Distribution of $\gamma$-ray candidates events as function of squared angular distance from the bes fit position of HESS J0632+057. The red line is the point spread function corresponding to this dataset obtained with Monte-Carlo simulations.}
\label{fig3}
\end{figure}
\begin{figure}
\begin{center}
\includegraphics [width=0.45\textwidth]{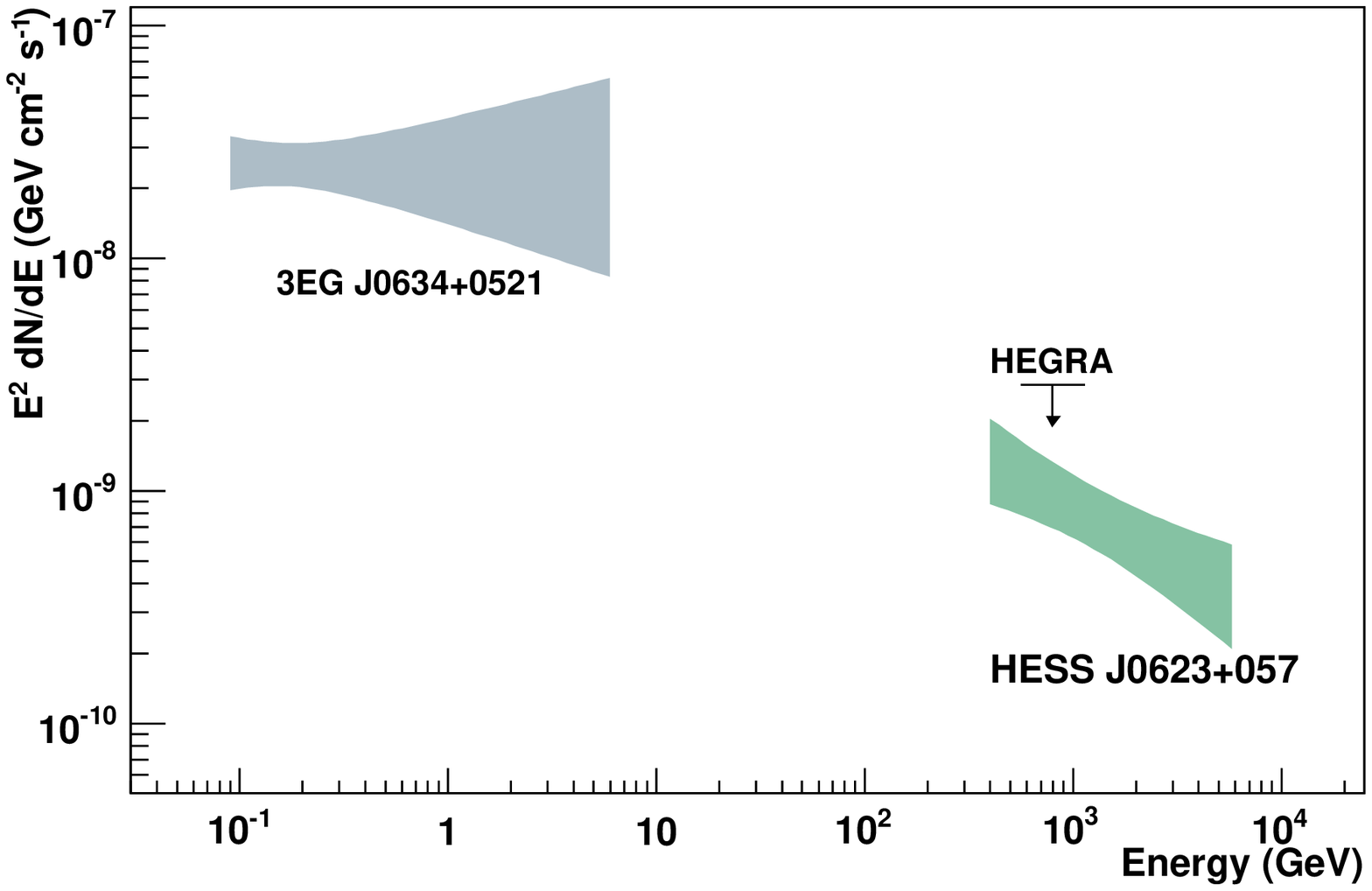}
\end{center}
\caption{Reconstructed VHE $\gamma$-ray spectrum of HESS J0632+057 compared to the EGRET source 3EG J0634+0521. The upper limit obtained using the HEGRA instrument for the EGRET source position is shown.}
\label{fig4}
\end{figure}

\section*{Possible associations}
The region where lies HESS J0632+057 is a complex region and although there is no clear counterpart, it may be associated with various objects known at other wavelengths. 
\subsection*{3EG J0634+0521}
In the same region lies also an EGRET source, 3EG J0634+0521 (Hartman et al., 1999). Considering that the source is flagged as confused and possibly extended, our measurement, which lies between 95\% and 99\% confidence region, is compatible with its position. Furthermore, the reported third EGRET catalogue flux above 100 MeV is consistent with an extrapolation of the H.E.S.S. spectrum. A global fit of the two spectra gives a photon index of 2.41 $\pm$ 0.06 (fig.\ref{fig4}). 

\subsection*{The Monoceros Loop SNR}
The possible association of spectra in the GeV and TeV band is an argument in favor of an hadronic interpretation of the VHE $\gamma$-ray emission. In this case, the Monoceros loop SNR is a good candidate for acceleration of particles. This remnant, which has an age of $\sim$ 3$\times$10$^{4}$ years, is rather old in comparison to known shell type SNRs emitting VHE $\gamma$ rays ($\sim$2000 years). However, cosmic rays acceleration may occur even at later evolutionary phase (late Sedov or Radiative, see Yamazaki et al. 2006). Given the point-like nature of the VHE $\gamma$-ray emission, to explain VHE $\gamma$ rays as a product of accelerated cosmic rays interacting with interstellar medium requires the presence of a dense molecular cloud 
coincident with the emission. An unresolved molecular cloud listed in CO survey at 115 GHz (Oliver et al. 1996) lies rather close to HESS J0632+057. The distance estimate for this cloud (1.6 kpc) is consistent with that for the Monoceros SNR. NANTEN survey shows that the intensity peak of this cloud is significantly shifted to the east of the H.E.S.S. source (fig. \ref{fig1}). There is no evidence of other dense clouds along the line of sight in the NANTEN data.

\subsection*{1 RXS J063258.3+054857}
1 RXS J063258.3+054857 is a faint ROSAT source which is potential counterpart of HESS J0632+057, given the uncertainty of the position of the two objects. Given the number of sources in the field of view, the chance probability of coincidence of the two source is 0.1\%. X rays are useful to discriminate between scenario of VHE $\gamma$-ray emission. If $\gamma$-rays are due to inverse Compton scattering from a population of accelerated electrons, X rays are expected to come from synchrotron emission of the same population. In this case, the weakness of this source ($\sim$10$^{-13}$erg cm$^{-2}$ s$^{-1}$) compared to the TeV flux ($\sim$10$^{-12}$erg cm$^{-2}$ s$^{-1}$) required a very low magnetic field ($\ll$3$\mu$G), unless a strong radiation source exists in the neighbourhood of the emission region. Important absorption of the X-ray emission may also explain weakness of the ROSAT source. In the case of a hadronic scenario, production of pions leads to secondary electrons which produce a weaker X-ray source, probably compatible with the measured ROSAT flux.

\subsection*{MWC 148}
A massive emission-line Be-star lies within the H.E.S.S. error circle. Given the fact that there are only three stars of this type in the field of view, the chance probability of the association is $\approx$10$^{-4}$. Stars of this spectral type have winds with typical velocities and mass loss rates of 1000 km.s$^{-1}$ and 10$^{-7}$\textit{M}$_{\odot}$. Stellar winds may induce internal or external shocks where particles can be accelerated, but no association of VHE $\gamma$-ray emission with similar stars have been already detected and seems unlikely. Another hypothesis is that this star is a part of a binary system with a compact companion not already detected. Further observations are required to constrain this scenario.

\section*{Acknowledgments}
The support of the Namibian authorities and of the University of Namibia in facilitating the construction and operation of H.E.S.S. is gratefully acknowledged, as is the support by the German Ministry for Education and Research, the CNRS-IN2P3 and the Astroparticle Interdisciplinary Programme of the CNRS, the U.K. Particle Physics and Astronomy Research Coucil (PPARC), the IPNP of the Charles University, the South African Department of Science and Technology and National Research Foundation, and by the University of Namibia. We appreciate the excellent work of the technical support staff in Berlin, Durham, Hamburg, Heidelberg, Palaiseau, Paris, Saclay, and in Namibia in the construction and operation of the equipment. The NANTEN project is financially supported from JSPS (Japan Society for the Promotion of Science) Core-to-Core Program, MEXT Grant-in-Aid for Scientific Research on Priority Areas, and SORST-JST (Solution Oriented Research for Science and Technology: Japan science and Technology Agency). We would also like to thank Stan Owocki and James Urquhart for very useful discussions.
\nocite{ref1}
\nocite{ref2}
\nocite{ref3}
\nocite{ref4}
\nocite{ref4b}
\nocite{ref5}
\nocite{ref6}
\nocite{ref7a}
\nocite{ref7}
%This in the bibtex style, is ok.
\bibliographystyle{plain}
%This is the reference to .bib file (Whitout .bib!)

%%%%%%%%
%  06  %
%%%%%%%%

%The paper title
\title{Crab nebula spectrum as seen by H.E.S.S.}
%Short title to print in the headers to the final publication (Not showed in this print).
\shorttitle{Crab nebula spectrum as seen by H.E.S.S.}
%All paper authors
\authors{B. Kh\'elifi$^{1}$, C. Masterson$^{2}$, S. Pita$^{3}$, E. O\~{n}a-Wilhelmi$^{3}$ for the H.E.S.S. collaboration}
%Short title to print in the headers to the final puplication (Not showed in this print).
\shortauthors{B. Kh\'elifi et al.}
%All the affiliations.
\afiliations{$^1$Laboratoire Leprince-Ringuet, Ecole Polytechnique/IN2P3/CNRS, Palaiseau, France\\
 $^2$Dublin Institute for Advanced Studies, 5 Merrion Square, Dublin 2, Ireland \\
 $^3$AstroParticule et Cosmologie, Paris VII/IN2P3/CNRS, Paris, France }
\email{khelifi@llr.in2p3.fr}

%The abstract.
\abstract{The H.E.S.S. stereoscopic Cherenkov telescope system has observed the
Crab nebula since December 2003 with the complete four-telescope array.
The stable signal from this pulsar wind nebula (PWN) has been used to
verify the performance and calibration of the instrument thanks to its
high flux compared to the H.E.S.S sensitivity.
These observations allow us also to study the radiation mechanisms of
this PWN, in particular by focusing on the high energy part of its
energy spectrum, where gamma-ray emission at energies above 30 TeV has
been detected.}

\maketitle

\addtocontents{toc}{\protect\contentsline {part}{\protect\large Pulsar Wind Nebulae (PWN)}{}}
\addcontentsline{toc}{section}{Crab nebula spectrum as seen by H.E.S.S.}
\setcounter{figure}{0}
\setcounter{table}{0}
\setcounter{equation}{0}

%Begin the section.
\section*{Introduction}

The Crab nebula was discovered at very high energies (VHE; $>$100~GeV) in 1989~\cite{whip} and the emission has been confirmed by a number of other experiments (e.g. \cite{heg, cat, magic}). This pulsar wind nebula (PWN) has a high flux relative to other known VHE sources and its emission is expected to be stable. As a result, the Crab nebula is commonly used as a standard `calibration candle' 
for the ground-based gamma-ray detectors, and a particular attention is paid here to the control of the analysis chain accuracy.
Indeed, the detector ageing results from a decrease of the overall optical efficiency (a combination of mirrors, light-cones, and photomultipliers degradation) and from ageing of electronics components of cameras. The detector response is measured, calibrated \cite{calib} and used for the data analysis \cite{crab}.

Important questions on the origin of the non-thermal emission of the Crab nebula remain. 
It is commonly admitted that its spectral energy distribution (SED) can be well-reproduced with a mechanism based 
on a synchrotron self-Compton (SSC) emission of high energy
electrons/positrons (e.g. \cite{horns}) even if a contribution from proton radiation is not excluded at high energies (e.g.
\cite{amato}). However, the acceleration mechanisms of these leptons and hadrons are still under investigation (Cf. \cite{kirk} for a
recent review). Thus, multi-wavelength observations are still necessary to understand the underlying physics, in particular
observations of VHE gamma-rays above 30~TeV.

\section*{H.E.S.S. observations and data analysis}

The Crab nebula has been observed with the complete array for 58.4~hours from December 2003 to December 2006.
After data-quality selection based on good
weather conditions and good detector operation, an exposure of 29.4~hours live-time is obtained. The periods of the Crab observations
suffer sometimes of poor weather conditions in Namibia. All observations were taken in {\it wobble} mode whereby the source is
alternately offset by a fixed distance within the field of view, alternating between 28 minutes runs in positive and negative
declination (or right ascension) directions. 

In table~\ref{tab:obs} we present, for each observation period considered, the live-time (in hours), mean zenith angle (in degrees),
mean position (in degrees) of the Crab pulsar position relative to the centre of the field of view and mean optical efficiency (in percent) of the detection system.\\

\begin{table}[t]
\begin{center}
\small
    \begin{tabular}{ c c c c c }
      \hline
Year                &2004   &2005   &2006   &All\\
\hline                
\hline
Live-time [h]       &20.6   &5.4    &3.4    &29.4 \\
Zenith Angle [deg]  &52.2   &47.7   &49.2   &51.1 \\
Offset [deg]        &0.65   &0.58   &0.70   &0.65 \\
OptEff [\%]         &8.3    &7.8    &7.0    &8.1 \\
      \hline
      \hline
\end{tabular}
\end{center}
\vspace{-0.5cm}
\caption{Summary of the Crab observations. The row descriptions are given in the text.}
\label{tab:obs}
\vspace{-0.5cm}
\end{table}

The data are processed with the HAP (H.E.S.S. Analysis Package) software as follows. In order to reject the overwhelming
background of night-sky diffuse light and hadronic showers, a two-level image cleaning is performed to remove pixels containing only
background noise. After image cleaning, the Hillas parameters \cite{hillas} are computed. For comparison, two methods are used to
reconstruct the characteristics of the atmospheric showers, i.e. the impact parameter ($D$), the shower maximum ($H$) and the shower
direction. The first method \cite{crab}, called hereafter {\it Hillas}, is based on a geometrical reconstruction of the shower
characteristics from the Hillas parameters (tracks of the projected direction of the shower in the field of view). The second one, called {\it Model3D} \cite{model3d}, uses a model of the atmospheric shower as a `Cherenkov ellipsoid' and its parameters are
adjusted to the camera images. Cuts are applied to the parameters derived by these methods to improve the signal to (hadronic) noise
ratio. For the {\it Model3D} analysis, the standard cuts of the {\it Hillas} analysis are applied together with cuts on the `Cherenkov
ellipsoid' size. The remaining background is estimated from regions at same distance from the field of view centre as the Crab pulsar
position for the observations (cf. fig.~9 of \cite{crab}).

The energy of each event is estimated from $D$, $H$ and the
images charges within the Hillas ellipses ($Q$). Look-up tables given the image charges as a function of energy ($E$), $D$ and $H$
($Q=f(E,D,H)$ are derived from gamma-ray simulations made with Kaskade \cite{kaskade} for different fixed energies, zenith angles, offsets and
optical efficiencies. Given the measured $Q$, $D$ and $H$, inverting the tables provides an estimation of the event energy. To determine the energy spectrum, the
instrument response functions (effective areas and energy resolutions) are derived from the same gamma-ray simulations, and 
a forward-folding algorithm developed by the CAT collaboration \cite{sp} is used. A likelihood fit
is used to adjust different spectral shape hypotheses. A test of the hypotheses with a likelihood ratio is made to determine the 
spectrum shape that best adjusts to the data.

\section*{H.E.S.S. results}

The main results of the analysis of the Crab observations are given in table~\ref{tab:res}. For each method of shower reconstruction and for each year, the
number of gamma-rays above the analysis energy threshold, the significance and the integral flux above 1~TeV are listed. A strong signal is detected and,
independently of the year and the analysis method, the integral flux is basically constant, illustrating the good correction for the effects of the detector
ageing.

\begin{figure}[h!]
\begin{center}
\vspace{-0.2cm}
\includegraphics [width=0.43\textwidth]{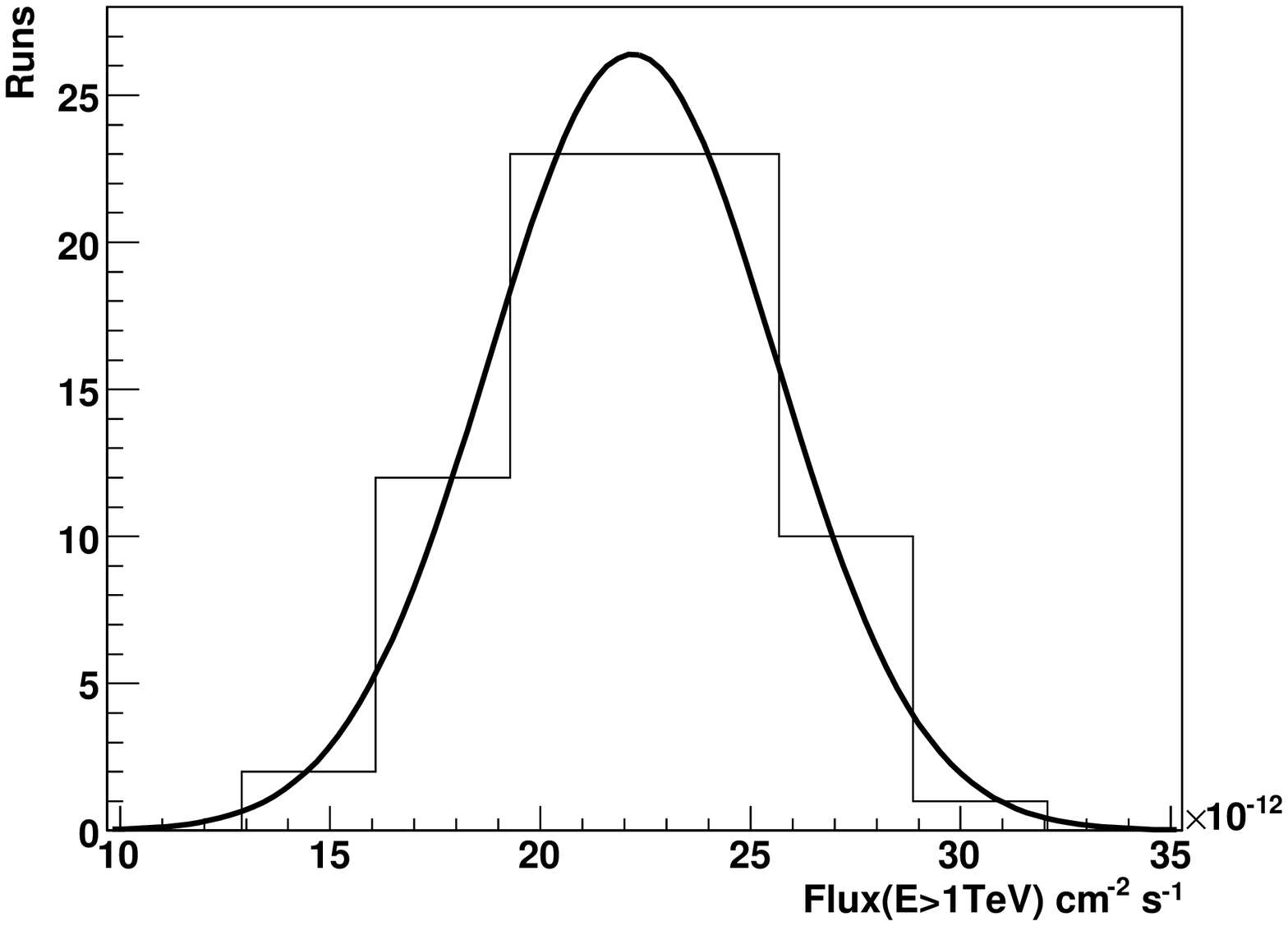}
\end{center}
\vspace{-0.75cm}
\caption{Distribution of the run-wise integral fluxes above 1~TeV for the {\it Hillas} analysis.}
\label{fig:fig1}
\end{figure}

The run-wise fluxes are also computed and their distribution is given in fig.~\ref{fig:fig1} for the {\it Hillas} analysis. It follows
a Gaussian distribution (black line) with a $\chi^2$/dof of $0.50/3$. The best-fit parameters are $\textrm{Mean}=2.22\pm0.04$ and
$\textrm{Sigma}=0.34\pm0.02$ in units of $10^{-11} \textrm{cm}^{-2} \textrm{s}^{-1}$. The flux derived is thus compatible with a
steady flux with Gaussian fluctuations of $\sim$$15\%$.\\

\begin{table*}[!t]
\begin{center}
\small
    \begin{tabular}{ c c c c c }
      \hline
Method              &Year   &Excess   &Significance   &$\textrm{F}_{>1\,{\rm TeV}}$\\
                    &       &[$\gamma$] &[$\sigma$]   &[$\times 10^{-11} \textrm{cm}^{-2} \textrm{s}^{-1}$]\\
\hline                
\hline
{\it Hillas}        &2004   &5788 &122  &$2.22\pm0.07$ \\
                    &2005   &1674 &70   &$2.18\pm0.06$ \\
                    &2006   &1069 &57   &$2.41\pm0.10$ \\
                    &All    &8531 &151  &$2.22\pm0.04$ \\
\hline
{\it Model3D}       &2004   &5208 &130  &$2.20\pm0.06$ \\
                    &2005   &1612 &74   &$2.13\pm0.18$ \\
                    &2006   &1008 &59   &$2.37\pm0.12$ \\
                    &All    &7828 &161  &$2.22\pm0.05$ \\
      \hline
      \hline
\end{tabular}
\end{center}
\vspace{-0.5cm}
\caption{Results of the observations. The column descriptions are given in the text.}
\label{tab:res}
\vspace{-0.5cm}
\end{table*}

\begin{table}[h]
\begin{center}
\vspace{-0.3cm}
\small
    \begin{tabular}{ c c c}
      \hline
                	      &{\it Hillas}   &{\it Model3D}\\
\hline                
\hline
$\Phi_0^{\textrm{PL}}$        &$3.52\pm0.04$  &$3.46\pm0.04$   \\
$\Gamma^{\textrm{PL}}$        &$2.60\pm0.01$  &$2.61\pm0.01$   \\
\hline
$\Phi_0^{\textrm{EC}}$        &$3.53\pm0.04$  &$3.48\pm0.04$   \\
$\Gamma^{\textrm{EC}}$        &$2.40\pm0.03$  &$2.42\pm0.03$   \\
$E_c^{\textrm{EC}}$   	      &$16.7\pm2.5$   &$16.1\pm2.5$   \\
$\lambda^{\textrm{EC}}$       &$74.4$         &$66.6$ 	\\
\hline
\hline
\end{tabular}
\end{center}
\vspace{-0.5cm}
\caption{Summary of spectrum fits. The row descriptions are given in the text.}
\label{tab:fit}
\vspace{-0.2cm}
%\vspace{+0.5cm}
\end{table}

For both analyses, the energy spectrum is computed for two different spectral hypotheses: a pure power-law $\mathscr{H}_0$ ($\textrm{d}N/\textrm{d}E_{\textrm{PL}} = \Phi_0
\times E^{-\Gamma}$) and a power-law with an exponential cut-off $\mathscr{H}_1$ ($\textrm{d}N/\textrm{d}E_{\textrm{EC}} = \Phi_0 \times
E^{-\Gamma}\times\textrm{e}^{-E/E_c}$). The fit results are listed in table~\ref{tab:fit}. The parameter $\Phi_0$ is in units of $10^{-11} \textrm{cm}^{-2} \textrm{s}^{-1}
\textrm{TeV}^{-1}$, $E_c$ in TeV. $\lambda$ is the ratio between the maximum likelihood of the $\mathscr{H}_1$ fit over the $\mathscr{H}_0$ fit and its distribution follows
asymptotically a $\chi^2$ law with one degree of freedom. From this parameter and independently of the analysis method used, it can clearly be seen that the fitted spectrum
shape is not compatible with a pure power-law  with a probability less than $10^{-6}$. The use of a `parabolic' spectrum shape ($E^{-\alpha-\beta \log(E)}$) fits the data
equally well as a power-law with an exponential cut-off. Note that the fit results are compatible between the different analyses.

\begin{figure}[h!]
\begin{center}
\vspace{-0.4cm}
\includegraphics [width=0.48\textwidth]{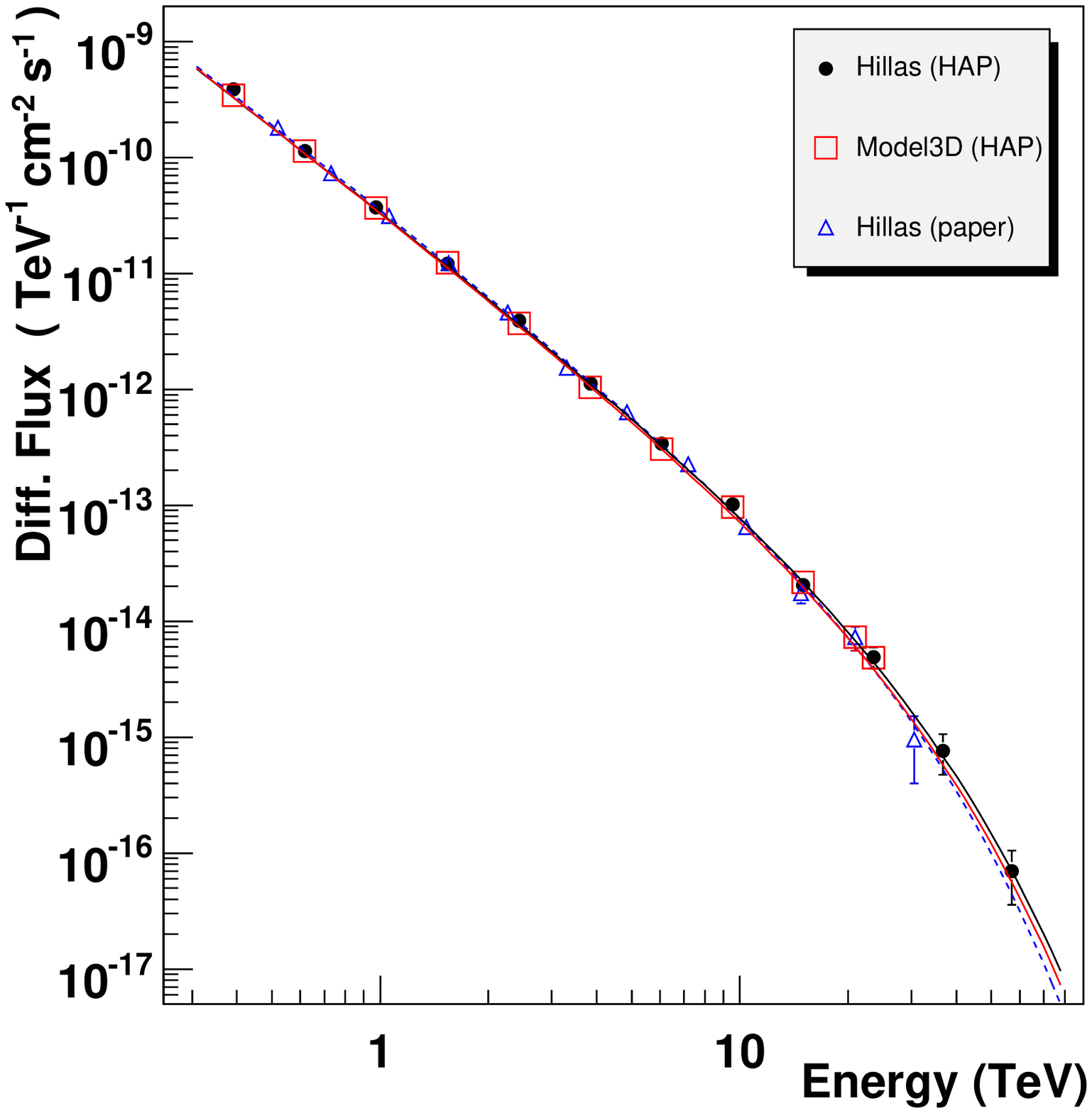}
\end{center}
\vspace{-0.75cm}
\caption{Comparison of the Crab spectrum fits between this analysis and that published in \cite{crab}. The lines are the best-fit
shapes.}
\label{fig:fig2}
\end{figure}

\begin{figure}[h!]
\begin{center}
\vspace{0.2cm}
\includegraphics [angle=-90, width=0.45\textwidth]{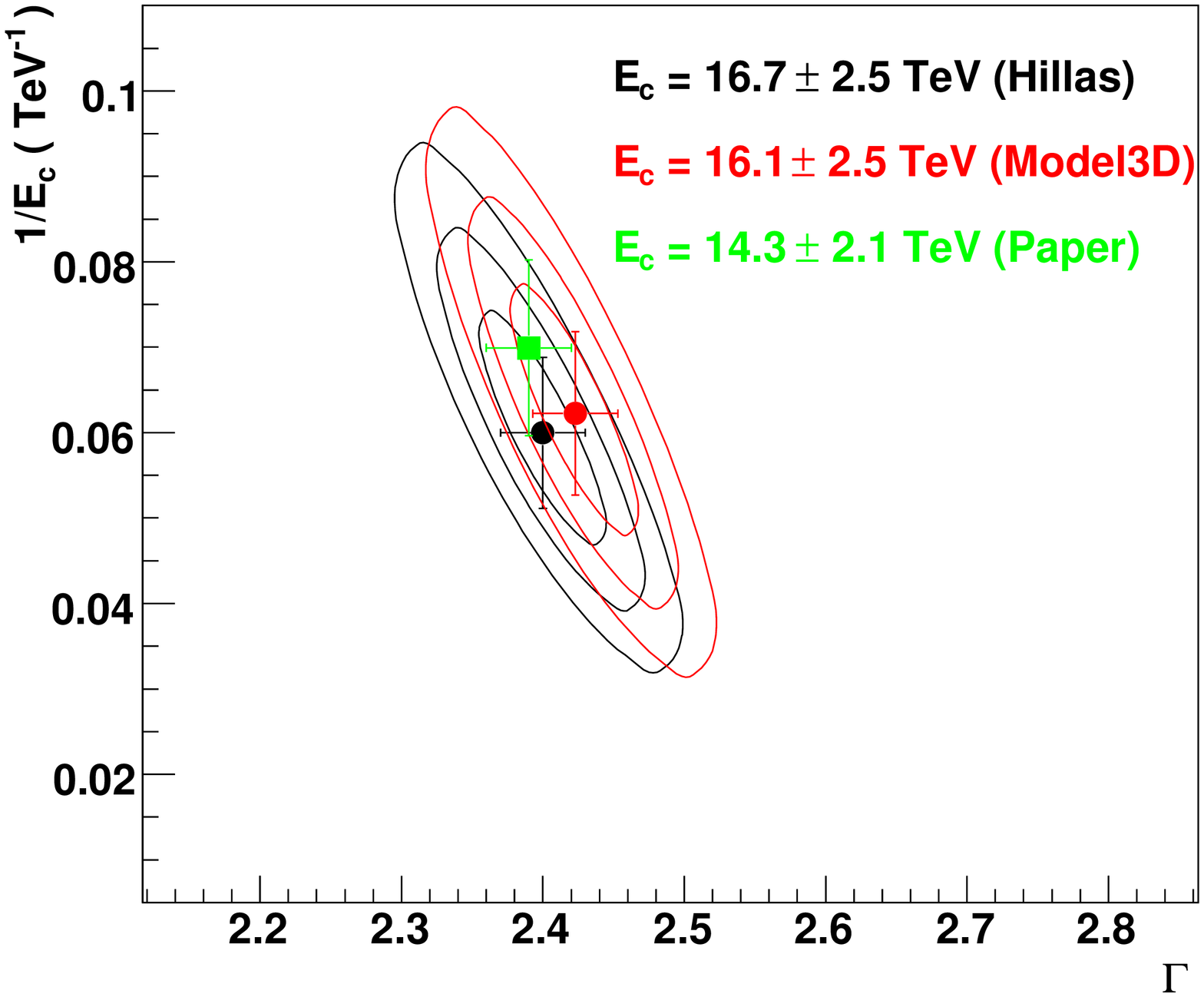}
\end{center}
\vspace{-0.65cm}
\caption{Comparison of the best-fit parameters between this analysis and those published in \cite{crab}.}
\label{fig:fig3}
\vspace{-0.5cm}
\end{figure}

Figure~\ref{fig:fig2} shows the Crab spectrum
derived with these two analyses carried out with the HAP software, together with the H.E.S.S. spectrum published in \cite{crab}. In the
following, the results of the $\mathscr{H}_1$ fit for the {\it Hillas} analysis are used and the flux measurements for each energy bin
(differential flux) are given in table~\ref{tab:sp}. Here, the measurements on high energy bins above 30~TeV should be emphasised in which a signal is
detected at the level of $\sim$$6$$\sigma$. 
A signal is detected significantly at the highest energies which allows the spectrum curvature to be measured more accurately . Figure~\ref{fig:fig3}
shows the comparison of the best-fit parameters $\Gamma$ and $1/E_c$ between these analyses and the results from~\cite{crab}.
The parameters are quite compatible between these and the exponential cut-off energy, $E_c$, is compatible with $\sim$15~TeV.

\begin{table*}[!t]
\begin{center}
\small
    \begin{tabular}{ c c c}
      \hline
Mean Energy         &Significance   &$\frac{\textrm{d}N}{\textrm{d}E}$\\
                    &[$\sigma$]     &[$\textrm{cm}^{-2} \textrm{s}^{-1} \textrm{TeV}^{-1}$]\\
\hline                
\hline
0.39           &16.7        &$(3.87\pm0.38) \times 10^{-10}$ \\
0.62           &73.4        &$(1.14\pm0.03) \times 10^{-10}$ \\
0.97           &78.4        &$(3.72\pm0.08) \times 10^{-11}$ \\
1.54           &67.7        &$(1.21\pm0.03) \times 10^{-11}$ \\
2.43           &55.9        &$(3.93\pm0.12) \times 10^{-12}$ \\
3.84           &41.3        &$(1.13\pm0.05) \times 10^{-12}$ \\
6.06           &30.2        &$(3.40\pm0.19) \times 10^{-13}$ \\
9.54           &22.4        &$(1.01\pm0.07) \times 10^{-13}$ \\
15.0           &12.4        &$(2.05\pm0.27) \times 10^{-14}$ \\
23.5           &7.9         &$(4.91\pm1.00) \times 10^{-15}$ \\
36.7           &4.1         &$(7.56\pm2.84) \times 10^{-16}$ \\
57.0           &3.8         &$(7.22\pm3.40) \times 10^{-17}$ \\
\hline                           
\hline
\end{tabular}
\end{center}
\vspace{-0.5cm}
\caption{Flux measurements for each energy bin for the {\it Hillas} analysis.}
\label{tab:sp}
\vspace{-0.5cm}
\end{table*}

\section*{Conclusions}

Analysis of Crab data carried out with the new software framework HAP yields results which are consistent with those published previously by
H.E.S.S. in~\cite{crab}. The measured Crab flux is compatible with a steady flux between December 2003 and December 2006, indicating that all effects of the
detector ageing are correctly taken into account. The integral flux above 1~TeV is
$F(>1\,\textrm{TeV}) = (2.22\pm0.07) \times 10^{-11} \textrm{ cm}^{-2} \textrm{s}^{-1}$. Its energy spectrum is not compatible with a
pure power-law shape and is well-represented by a power-law with an exponential cut-off ($E_c=16.7\pm2.5\,\textrm{TeV}$).
%$\textrm{dN/dE}$$=$$(2.16\pm0.05)$$\times$$\textrm{E}^{-2.41\pm0.03}$$\textrm{e}^{-\textrm{E}/(30.3\pm8.7 \textrm{TeV})}$$\times$$10^{-11}
%\textrm{cm}^{-2} \textrm{s}^{-1} \textrm{TeV}^{-1}$

Comparing the results of different analyses presented here, one finds that the differences of flux and spectrum index estimated
are well within the systematics detailed in \cite{crab}.

A clear signal is detected above 30~TeV which allows the curved nature of the Crab nebula spectrum to be clearly confirmed. This measured spectrum seems to
be still compatible with a SSC scenario in the Klein-Nishina regime as described in~\cite{horns}. An adjustment of the fit parameters of this radiation model
on our data is still necessary to confirm this scenario.

\section*{Acknowledgements}
The support of the Namibian authorities and of the University of Namibia in facilitating the
construction and operation of H.E.S.S. is gratefully acknowledged, as is the support by the German
Ministry for Education and Research (BMBF), the Max Planck Society, the French Ministry for Research,
the CNRS-IN2P3 and the Astroparticle Interdisciplinary Programme of the CNRS, the U.K. Science and
Technology Facilities Council (STFC), the IPNP of the Charles University, the Polish Ministry of
Science and  Higher Education, the South African Department of Science and Technology and National
Research Foundation, and by the University of Namibia. We appreciate the excellent work of the
technical support staff in Berlin, Durham, Hamburg, Heidelberg, Palaiseau, Paris, Saclay, and in
Namibia in the construction and operation of the equipment.

\bibliographystyle{plain}

%%%%%%%%
%  07  %
%%%%%%%%

%The paper title
\title{Energy Dependent Morphology in the PWN candidate HESS\,J1825--137}
%Short title to print in the headers to the final publication (Not showed in this print).
\shorttitle{Energy Dependent Morphology in HESS\,J1825--137}

%All paper authors
\authors{S. Funk$^{1}$, J.~A. Hinton$^{2}$,O.~C. deJager$^{3}$ for the
H.E.S.S.\ collaboration}
%Short title to print in the headers to the final publication (Not shown in this print).
\shortauthors{Funk et al.}
%All the affiliations.
\afiliations{
  $^1$Kavli Institute for Particle Astrophysics and
  Cosmology, SLAC, Menlo Park, CA-94025, USA\\ 
  $^2$ School of Physics and Astronomy, University of Leeds, Leeds LS2
  9JT, UK\\
  $^3$ Unit for Space Physics, North-West University, Potchefstroom
  2520, South Africa 
}

\email{Stefan.Funk@slac.stanford.edu}

%The abstract.
\abstract{Observations with H.E.S.S.\ revealed a new source of very
high-energy (VHE) gamma-rays above 100 GeV -- HESS\,J1825--137 --
extending mainly to the south of the energetic pulsar
PSR\,B1823--13. A detailed spectral and morphological analysis of
HESS\,J1825--137 reveals for the first time in VHE gamma-ray astronomy
a steepening of the energy spectrum with increasing distance from the
pulsar. This behaviour can be understood by invoking radiative cooling
of the IC-Compton gamma-ray emitting electrons during their
propagation. In this scenario the vastly different sizes between the
VHE gamma-ray emitting region and the X-ray PWN associated with
PSR\,B1823--13 can be naturally explained by different cooling
timescales for the radiating electron populations. If this scenario is
correct, HESS\,J1825--137 can serve as a prototype for a whole class
of asymmetric PWN in which the X-rays are extended over a much smaller
angular scales than the gamma-rays and can help understanding recent
detections of X-ray PWN in systems such as HESS\,J1640--465 and
HESS\,J1813--178. The future GLAST satellite will probe lower electron
energies shedding further light on cooling and diffusion processes in
this source.}

\maketitle

\addcontentsline{toc}{section}{Energy Dependent Morphology in the PWN candidate HESS\,J1825--137}
\setcounter{figure}{0}
\setcounter{table}{0}
\setcounter{equation}{0}

%Begin the section.
\section*{Introduction}

The pulsar PSR\,B1823--13 and its surrounding X-ray pulsar wind nebula
(PWN) G18.0--0.2 is a system that has been studied by H.E.S.S.\ in
very high-energy gamma-rays above 200~GeV in unprecedented
detail~\cite{HESSJ1825II}. PWNe seem to constitute a significant
fraction of the population of identified Galactic VHE gamma-ray
sources detected by H.E.S.S.~\cite{FunkBarcelona} and as also
suggested by a statistical assessment of the correlation between
Galactic VHE $\gamma$-ray sources and energetic pulsars (see Carrigan
et al., these proceedings).  The gamma-ray emission in these objects
is typically thought to be generated by Inverse Compton scattering of
relativistic electrons accelerated in the termination shock of the
PWN.

Considering the population of VHE gamma-ray PWNe, HESS\,J1825--137 is
probably thus far the best example of the emerging class of so-called
\emph{offset Pulsar Wind nebulae} in which an extended VHE gamma-ray
emission surrounding an energetic pulsar is offset into one direction
of the pulsar. This offset is generally thought to arise from dense
molecular material in one direction of the pulsar that prevents an
symmetric expansion of the PWN (see e.g.~\cite{Blondin} for a
hydro-dynamical simulation and discussion of this effect).

As one of the best studied objects in VHE gamma-rays with an
observation time of nearly 70~hours, HESS\,J1825--137 has been used as
a template for the association of asymmetric PWN in VHE $\gamma$-rays
and X-rays~\cite{FunkBarcelona, YvesBarcelona}. In HESS\,J1825--137
the claimed association between the VHE $\gamma$-ray source and the
X-ray PWN rests on the following properties of the source:
\begin{itemize}
\item Same morphology (i.e.\ asymmetric extension to the south) in both
  bands but X-ray nebula much smaller ($\sim$ 5'') than $\gamma$-ray
  ($\sim$ 0.5$^{\circ}$) emission region
\item Spectral steepening of the VHE gamma-ray source away from the
  pulsar (i.e.\ decrease of gamma-ray extension with increasing
  energy). Interestingly the maximum of the VHE $\gamma$-ray emission
  is not coincident with the pulsar position but is shifted $\sim 17'$
  to the south-west.
\end{itemize}

The vastly different sizes of the emission region in the two wavebands
prevents at first glance a direct identification as a counterpart,
since the morphology can not be matched between X-rays and
gamma-rays. As will be explained in the following, the different sizes
can be explained in a time-dependent leptonic model by different
cooling timescales of the X-ray and of the VHE gamma-ray emitting
regions. Caution should however be used, if such an association serves
as a template for other unidentified H.E.S.S. VHE gamma-ray sources
with an energetic pulsar in the vicinity, in cases in which no X-ray
PWN has been detected so far.

\section*{Observational data}

CO-Observations performed in the composite survey~\cite{Dame} show a
dense molecular cloud in the distance band between 3.5 and 4~kpc to
the north of PSR\,B1823--13 (located at $\sim 4$
kpc)~\cite{Lemiere}. This cloud seems to support the picture of an
offset PWN and could explain why the X-ray and VHE emission is shifted
to the south of the pulsar. Given the relatively high gamma-ray flux
and the rather large distance of the system of 4~kpc (in comparison to
the Crab), the required gamma-ray luminosity $L_\gamma \sim 3 \times
10^{35} \ \rm erg/s$ is comparable to the Crab luminosity. The
spin-down luminosity of the pulsar is, however, two orders of
magnitude lower than the Crab spin-down luminosity. Assuming the
distance of $\sim 4$~kpc is correct this shows that the efficiency of
converting spin-down power to gamma-ray luminosity must be much higher
than in the Crab Nebula, not unexpected, given the large magnetic
field in the Crab Nebula. Detailed time-dependent modelling of the
source shows indeed that (especially below $\sim 1$TeV) the energy
injection into the system must have been about an order of magnitude
higher in the past. Potentially the spin-down power of the pulsar was
significantly higher in the early stage of the pulsar evolution. For
the lower energy end of the H.E.S.S.\ spectrum and for modest magnetic
fields of a few $\mu$G as suggested by the large VHE gamma-ray flux,
the electron lifetimes become comparable to the pulsar age and
therefore ``relic'' electrons released in the early history of the
pulsar can survive until today and provide the required luminosity. It
should be noted that to this date no sensitive X-ray observation of
the region coinciding with the peak of the VHE gamma-ray emission has
been performed and a low surface-brightness extension to the south of
the X-ray PWN found by Gaensler et al.~\cite{Gaensler1825} remains an
interesting possiblility that should eventually be tested.

\begin{figure}
\begin{center}
\noindent
\fbox{\hbox{\vbox{\hsize=60mm \includegraphics
      [width=60mm]{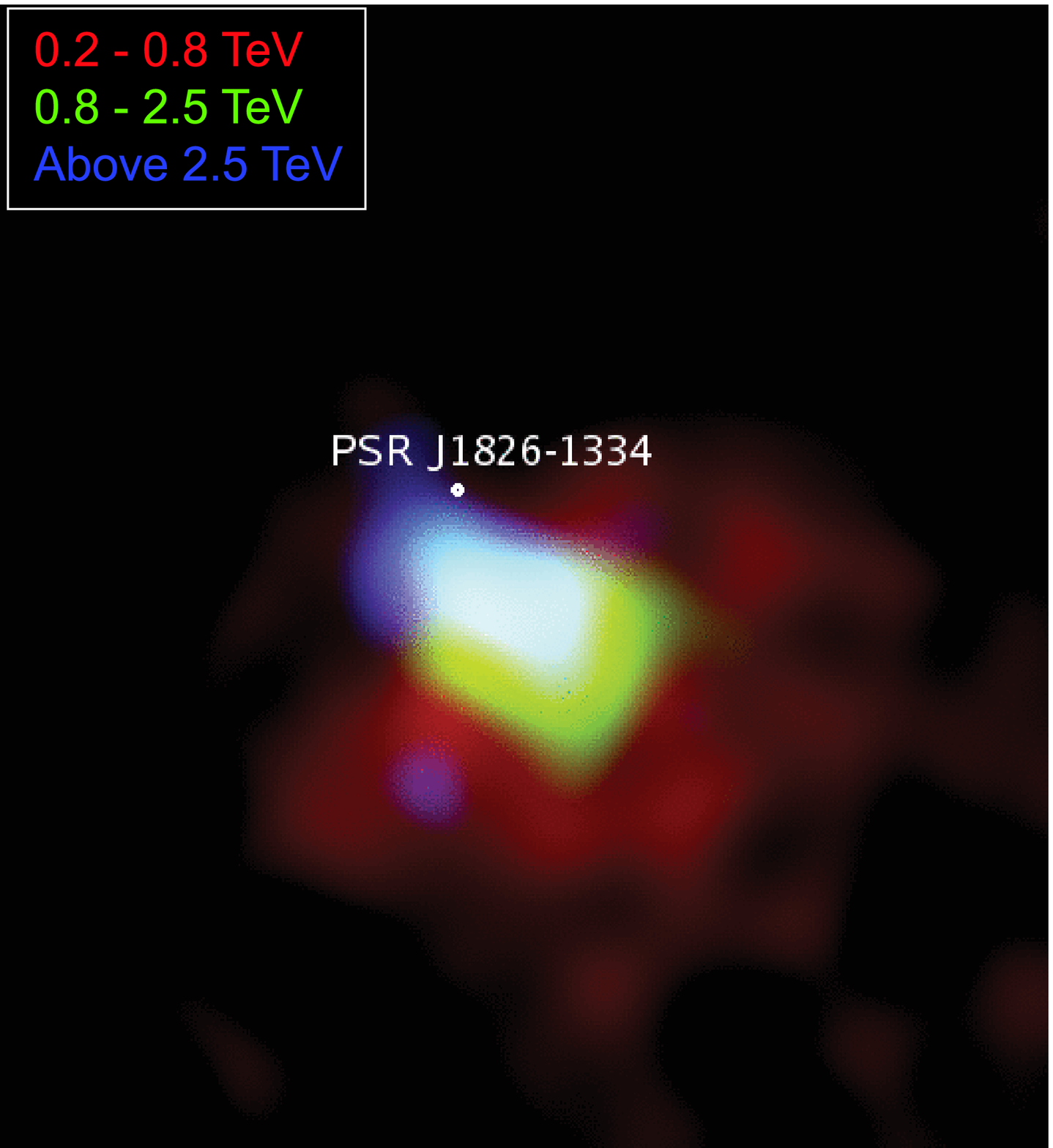} }}}
\end{center}
\caption{Three-colour image showing the gamma-ray emission in
  different energy bands (red: 0.2-0.8 TeV, green 0.8-2.5 TeV and
  blue: above 2.5 TeV). The different gamma-ray energy bands show a
  shrinking with increasing energy away from the pulsar
  PSR\,B1823--13.}\label{fig::HESS}
\end{figure}

\section*{Energy dependent morphology}

\begin{figure*}[ht]
\begin{center}
\noindent 
%\fbox{\hbox{\vbox{\hsize=130mm \hfill \vspace{50mm}}}}
\fbox{\hbox{\vbox{\hsize=150mm 
\includegraphics
    [width=150mm]{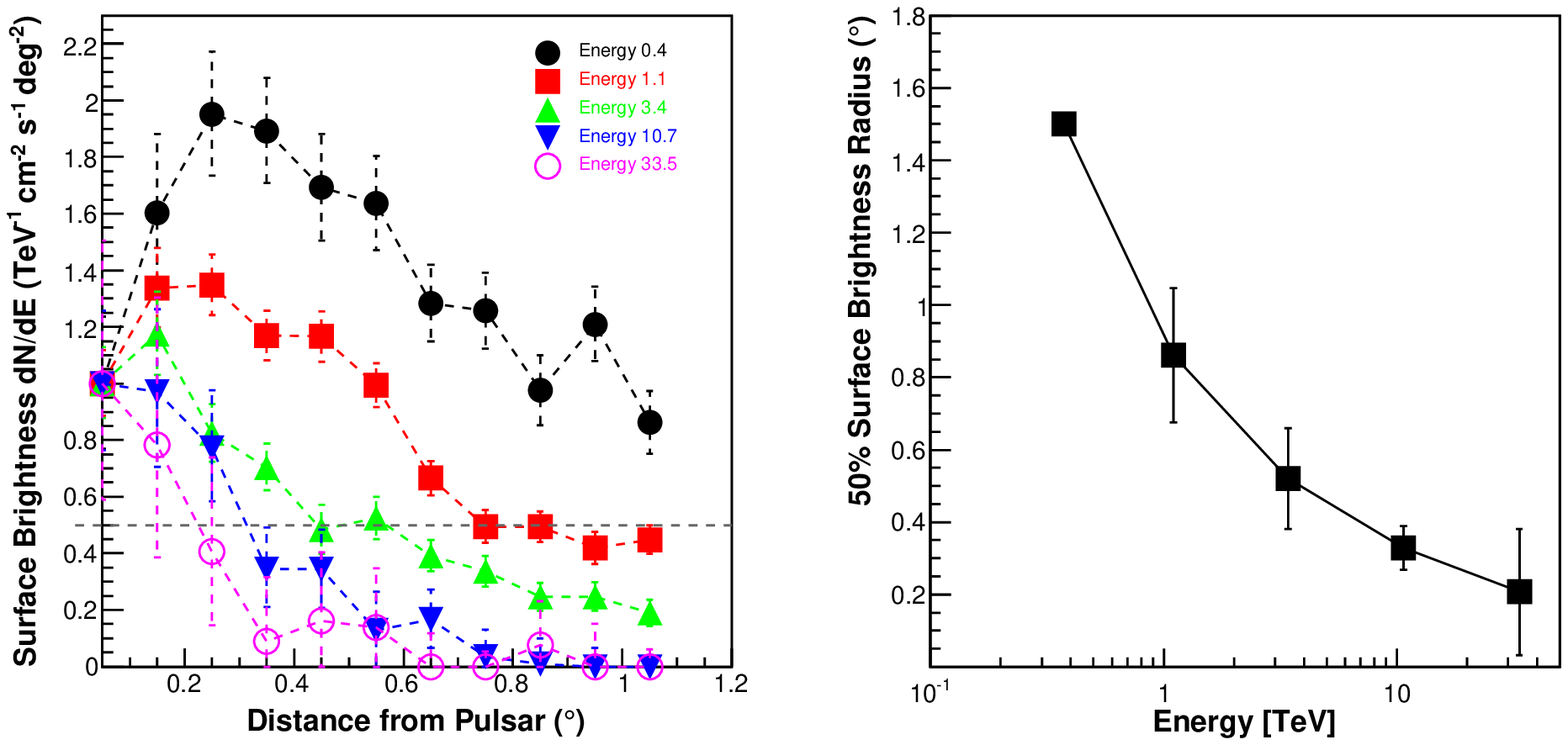}}}}
\end{center}
\caption{{\bf{Left:}} Surface brightness as a function of distance from
  the pulsar for different energy bands (derived from Figure 4 in
  Aharonian et al~\cite{HESSJ1825II}. The surface brightness is
  defined as the differential gamma-ray flux at a given energy scaled
  by the area of the extraction region and normalised by the flux for
  that energy at the pulsar position r = 0. {\bf{Right:}} Distance from
  the pulsar at which the surface brightness drops to 50\% of the flux
  at the pulsar position. The error bars are derived by fitting the
  falling points of the left plot, varying the fit parameters within
  the errors and recalculating the 50\% containment
  radius. }\label{fig::HESS2}
\end{figure*}

Given the large data set with nearly 20,000 $\gamma$-ray excess
events, a spatially resolved spectral analysis of HESS\,J1825--137
could be performed. For the the first time VHE $\gamma$-ray astronomy
an energy dependent morphology (see Figure~\ref{fig::HESS}) was
established~\cite{HESSJ1825II} in which the size of the emission
region decreases with increasing energy. This shrinking size with
increasing energy is equivalent to the statement of a steepening of
the spectral index away from the pulsar. The spectrum in
HESS\,J1825--137 changes from a rather hard photon index $\sim 2$
close to the pulsar to a softer value of $\sim 2.5$ at a distance of
$1^{\circ}$ away from the pulsar. Figure~\ref{fig::HESS2} shows the
surface brightness as a function of the distance from the pulsar for
different energy bands. Two clear trends are apparent in this figure:
a) the peak of the surface brightness shifts to lower energies (as
already suggested by the steepening of the energy spectrum away from
the pulsar) b) at low energies the surface brightness is nearly
independent of the distance whereas at the higher energies the surface
brightness drops rapidly with increasing distance from the pulsar. The
right panel of Figure~\ref{fig::HESS2} shows the derived radius
$R_{50}$ corresponding to the 50\% containment of the surface
brightness. This radius $R_{50}$ drops with increasing energy as
already apparent in Figure~\ref{fig::HESS}.

The steepening of the energy spectrum away from a central pulsar is a
property commonly observed in X-ray studies of PWNe other than the
Crab. For most of these system the total change in the photon index is
close to $\sim 0.5$ similar to what is seen in HESS\,J1825--137. It
should be noted that the results shown here represent the first
unambiguous detection of a spectral steepening at a fixed electron
energy (since the synchrotron emission seen in X-rays depends on the
magnetic field) in a PWN system. Spectral variation with distance from
the pulsar could result from (1) energy loss of particles during
propagation, with radiative cooling of electrons as the main loss
mechanism, from (2) energy dependent diffusion or convection speeds,
and from (3) variation of the shape of the injection spectrum with age
of the pulsar. Concerning (1): Loss mechanisms include amongst others
adiabatic expansion, ionisation loss, bremsstrahlung, synchrotron
losses and inverse Compton losses. Only synchrotron and IC losses can
result in a electron lifetime that decreases with increasing energy. A
source decreasing source size with increasing energy is therefore
generally seen as indicative of electrons as the radiating particles.

For continuous injection and short radiative lifetimes of the
electrons (in comparison to the age of the source), the spectral index
of the electrons changes by one unit as a result of the cooling,
yielding in a change of 0.5 in the the photon index. This matches
roughly what is seen in HESS\,J1825--137 when comparing the inner and
the outer nebula.  The lower energy gamma-rays (i.e. below $\sim 0.6$
TeV) correspond to mostly un-cooled low energy electrons (i.e. the
spectral index consistent with the injection spectral index). At these
low energies the electron lifetime becomes comparable to the age of
the source and the size is rather independent of the energy. At higher
energies the cooling break takes effect and the source size shrinks
with increasing energy as expected from electron cooling. The
XMM-Newton X-ray emitting electrons typically have much higher
energies ($\sim 100$~TeV) than the $\gamma$-ray emitting electrons
($\sim 10$~TeV), assuming a typical magnetic field of 5~$\mu$G. The
synchrotron cooling lifetime of X-ray emitting electrons is therefore
expected to be much smaller, resulting in a smaller spatial extension
in X-rays.

For systems like HESS\,J1825--137 a detailed study in X-rays trying to
detect the low surface brightness nebula in the soft X-ray band would
be very beneficial, is however very hard to achieve given the
absorption of soft X-rays.  The upcoming GLAST-satellite will observe
this object in a thus far rather unexplored energy regime especially
above $\sim 10$~GeV, where the angular resolution of the instrument
becomes comparable to the angular resolution of the ground-based
instrument. In this energy range GLAST will probe even lower energy
electrons and it will be interesting to compare the sizes of the GLAST
and the H.E.S.S.\ emission region. The H.E.S.S.\ results have shown
that a wealth of detail exists in gamma-rays at an angular scale of
$\sim 0.1^{\circ}$. Future instruments like CTA or AGIS might improve
this angular resolution even further.

\section*{Acknowledgements}
"The support of the Namibian authorities and of the University of
Namibia in facilitating the construction and operation of H.E.S.S. is
gratefully acknowledged, as is the support by the German Ministry for
Education and Research (BMBF), the Max Planck Society, the French
Ministry for Research, the CNRS-IN2P3 and the Astroparticle
Interdisciplinary Programme of the CNRS, the U.K. Science and
Technology Facilities Council (STFC), the IPNP of the Charles
University, the Polish Ministry of Science and Higher Education, the
South African Department of Science and Technology and National
Research Foundation, and by the University of Namibia. We appreciate
the excellent work of the technical support staff in Berlin, Durham,
Hamburg, Heidelberg, Palaiseau, Paris, Saclay, and in Namibia in the
construction and operation of the equipment."

\bibliographystyle{plain}

%%%%%%%%
%  08  %
%%%%%%%%

%The paper title
\title{Discovery of the candidate pulsar wind nebula HESS J1718-385 in very-high-energy gamma-rays}
%Short title to print in the headers to the final publication (Not showed in this print).
\shorttitle{Discovery of the candidate pulsar wind nebula HESS J1718-385} 
%All paper authors
\authors{S.~Carrigan$^{1}$, Y.A.~Gallant$^{2}$, J.A.~Hinton$^{1,3}$,
  Nu.~Komin$^{2}$, K.~Kosack$^{1}$ and C.~Stegmann$^{4}$ for the
  H.E.S.S. collaboration}
%Short title to print in the headers to the final puplication (Not showed in this print).
\shortauthors{S. Carrigan et al.}
%All the affiliations.
\afiliations{
\small
  $^1$Max-Planck-Institut f\"ur Kernphysik, P.O. Box 103980, D 69029 Heidelberg, Germany  \\
  $^2$Laboratoire de Physique Th\'eorique et Astroparticules, IN2P3/CNRS, Universit\'e Montpellier II, CC 70, Place Eug\`ene Bataillon, F-34095 Montpellier Cedex 5, France  \\
  $^3$Landessternwarte, Universit\"at Heidelberg, K\"onigstuhl, D 69117 Heidelberg, Germany  \\
  $^4$Universit\"at Erlangen-N\"urnberg, Physikalisches Institut, Erwin-Rommel-Str. 1, D 91058 Erlangen, Germany}
\email{svenja.carrigan@mpi-hd.mpg.de}

%The abstract.
\abstract{Motivated by recent detections of pulsar wind nebulae in
  very-high-energy (VHE) gamma rays, a systematic search for VHE
  gamma-ray sources associated with energetic pulsars was performed,
  using data obtained with the H.E.S.S. (High Energy Stereoscopic
  System) instrument. The search for VHE gamma-ray sources near the
  pulsar PSR J1718-3825 revealed the new VHE gamma-ray source HESS
  J1718-385. We report on the results from the HESS data analysis of
  this source and on possible associations with the pulsar and at
  other wavelengths. We investigate the energy spectrum of HESS
  J1718-385 that shows a clear peak. This is only the second time a
  VHE gamma-ray spectral maximum from a cosmic source was observed,
  the first being the Vela X pulsar wind nebula.}

%\email{aastex-help@aas.org}
\maketitle

\addcontentsline{toc}{section}{Discovery of the candidate pulsar wind nebula HESS J1718-385 in very-high-energy gamma-rays}
\setcounter{figure}{0}
\setcounter{table}{0}
\setcounter{equation}{0}
%Begin the section.

\section*{Introduction}

It has long been known that pulsars can drive powerful winds of highly
relativistic particles. Confinement of these winds leads to the
formation of strong shocks, which may accelerate particles to
$\sim$PeV energies.

The best studied example of a pulsar wind nebula (PWN) is the Crab
nebula, which exhibits strong non-thermal emission across most of the
electromagnetic spectrum from radio to $>$50~TeV $\gamma$-rays
\cite{WHIPPLE:crab}.  More recently, VHE $\gamma$-ray emission has
been detected from the Vela\,X PWN \cite{HESS:velax}, which is an
order of magnitude older ($\sim$11\,kyr) than the Crab nebula, and its
nebula is significantly offset from the pulsar position, both in
X-rays and VHE $\gamma$-rays.  Offset nebulae in both X-rays and VHE
$\gamma$-rays have also been observed in the Kookaburra Complex
\cite{HESS:kookaburra} and for the PWN associated with the
$\gamma$-ray source HESS~J1825$-$137 \cite{HESS:J1825a,HESS:J1825b}.
The latter source appears much brighter and more extended in VHE
$\gamma$-rays than in keV X-rays.  This suggests that searches at TeV
energies are a powerful tool for detecting PWNe.

Motivated by these detections, a systematic search for VHE
$\gamma$-ray sources associated with high spin-down energy loss rate
pulsars was performed, using data obtained with the
H.E.S.S. instrument. The VHE $\gamma$-ray data set used in the search
includes all data used in the H.E.S.S. Galactic plane survey
\cite{HESS:scanpaper2}, an extension of the survey to $-60^{\circ} < l
< -30^{\circ}$, dedicated observations of Galactic targets and
re-observations of H.E.S.S. survey sources. It spans Galactic
longitudes $-60^{\circ} < l < 30^{\circ}$ and Galactic latitudes
$-2^{\circ} < b < 2^{\circ}$, a region covered with high sensitivity
in the survey. These data are being searched for VHE emission from
pulsars from the Parkes Multibeam Pulsar Survey~\cite{Parkes1}. The
search for a possible $\gamma$-ray excess is done in a circular region
with radius $\theta = 0.22^{\circ}$ (as in \cite{HESS:scanpaper2})
around each pulsar position, sufficient to encompass a large fraction
of a possible PWN.  The statistical significance of the resulting
associations of the VHE $\gamma$-ray source with the pulsar is
evaluated by repeating the procedure for randomly generated pulsar
samples, modelled after the above-mentioned parent population.

In this search, it is found that pulsars with high spin-down energy
loss rates are on a statistical basis accompanied by VHE emission.
The search for VHE $\gamma$-ray emission near the pulsar \PSRONE\
revealed the new VHE $\gamma$-ray source \HESSONE. This paper deals
with the results from the HESS data analysis of \HESSONE\ and with its
possible associations with \PSRONE\ and other objects seen in radio
and X-ray wavelengths.

\section*{H.E.S.S. Observations and Analysis}

The data on \HESSONE\ are composed primarily from dedicated
observations of the supernova remnant RX~J1713.7$-$3946
\cite{HESS:RXJ1713}, which is located at about 1.6$^\circ$ south-west
of \HESSONE. After passing the H.E.S.S.\ standard data quality
criteria based on hardware and weather conditions, the data set for
\HESSONE\ has a total live time of $\sim$82 hours. The standard
H.E.S.S.\ analysis scheme \cite{HESS:crab} is applied to the data,
including optical efficiency corrections. In this analysis, \emph{hard
  cuts} are applied, which include a rather tight cut on the shower
image brightness of 200 photo-electrons and are suitable for extended,
hard-spectrum sources such as PWN. These cuts also improve the angular
resolution and therefore suppress contamination from the nearby
RX~J1713.7$-$3946. To produce a sky map, the background at each test
position in the sky is derived from a ring surrounding this position
with a mean radius of 1$^\circ$ and a width scaled to provide a
background area that is about 7 times larger than the area of the
on-source region.

For spectral studies, only observations in which the camera centre is
offset by less than 2$^\circ$ from the best-fit source position are
used to reduce systematic effects due to reconstructed $\gamma$-ray
directions falling close to edge of the field of view. The remaining
live time of the data sample is $\sim$73 hours. The spectral
significance is calculated by counting events within a circle of
radius 0.2$^\circ$ from the best-fit position, chosen to enclose the
whole emission region while reducing systematic effects arising from
morphology assumptions. The proximity of the strong source makes it
necessary to choose the background data from off-source observations
(matched to the zenith angle and offset distribution of the on-source
data) instead of from areas in the same field of view. For a more
detailed description of methods for background estimation, see
\cite{HESS:bg}.

\section*{Results}

The detection significance from the search for VHE $\gamma$-ray
emission within 0.22$^\circ$ of the location of \PSRONE\ is
$7.9\sigma$. A very conservative estimate of the number of trials
involved (\cite{HESS:scanpaper2}) leads to a corrected significance of
$6.2\sigma$.

\begin{figure}[!hbt!]
  \centering
  \includegraphics[width=0.48\textwidth]{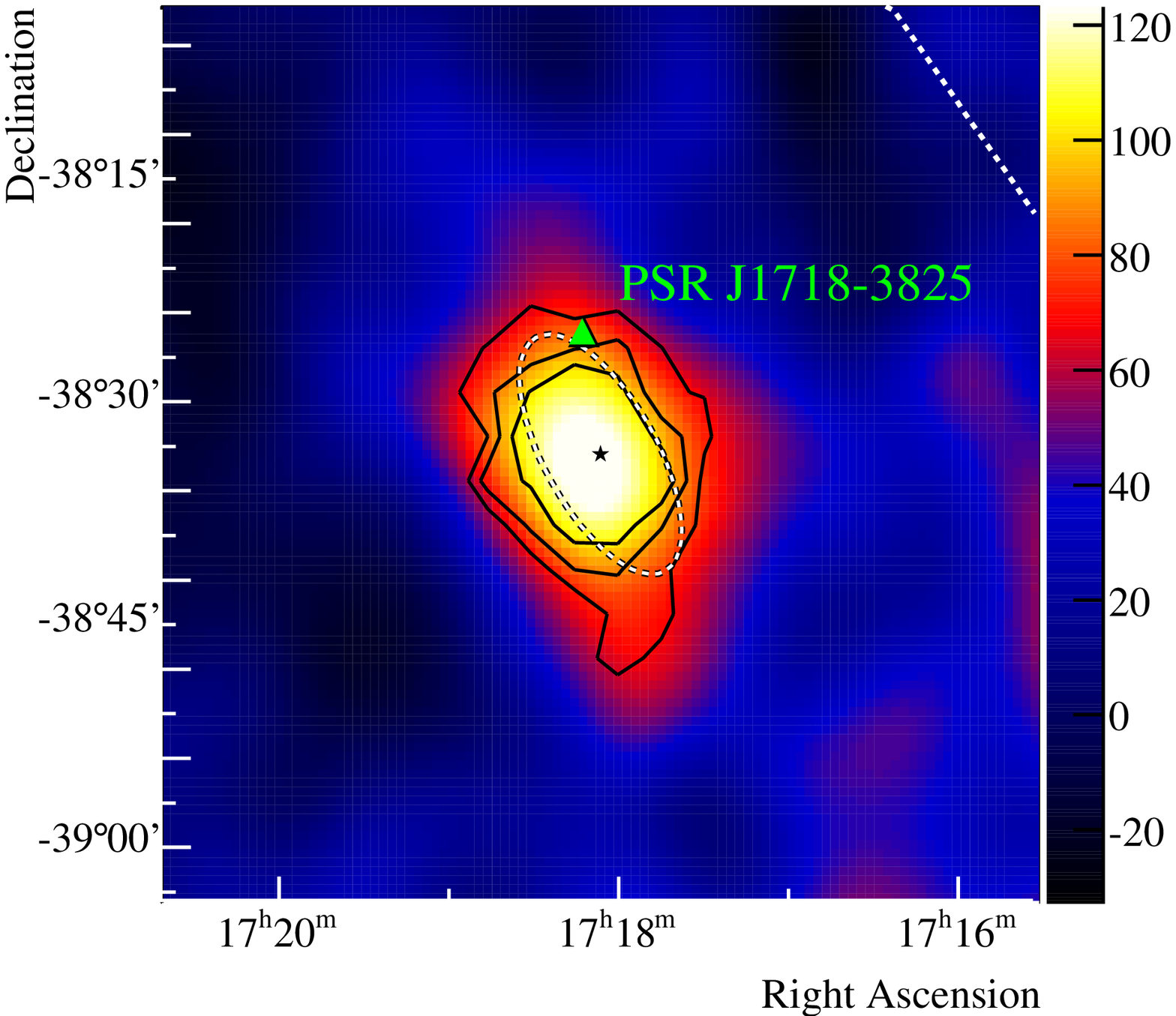}
  \caption{\label{PP_1718_figA}\small An image of the VHE $\gamma$-ray
    excess counts of \HESSONE, smoothed with a Gaussian of width
    0.06$^\circ$. The colour scale is set such that the blue/red
    transition occurs at approximately the 3$\sigma$ significance
    level. The black contours are the 4, 5 and 6$\sigma$ significance
    contours. The position of the pulsar \PSRONE\ is marked with a
    green triangle and the Galactic plane is shown as a white dotted
    line.  The best-fit position for the $\gamma$-ray source is marked
    with a black star and the fit ellipse with a dashed line.}
\end{figure}
  
Figure \ref{PP_1718_figA} shows the smoothed excess count map of the
1$^\circ$~$\times$~1$^\circ$ region around \HESSONE. A two-dimensional
Gaussian brightness profile, folded with the H.E.S.S. point-spread
function, is fit to the distribution before smoothing. Its parameters
are the width in two dimensions and the orientation angle, defined
counter-clockwise from North. The intrinsic widths (with the effect of
the point-spread function removed) for the fit are $9' \pm 2'$ and $4'
\pm 1'$ and the orientation angle is $\sim$33$^{\circ}$. The best-fit
position for the centre of the excess is RA~=~\HMS{17}{18}{7}\,$\pm
5^{s}$, Dec~=~$-38^{\circ}33'\pm2'$ (epoch J2000). H.E.S.S. has a
systematic pointing error of $\sim 20''$.

\begin{figure}[!hbt!]
  \centering
  \vspace{3ex}
  \includegraphics[width=0.48\textwidth]{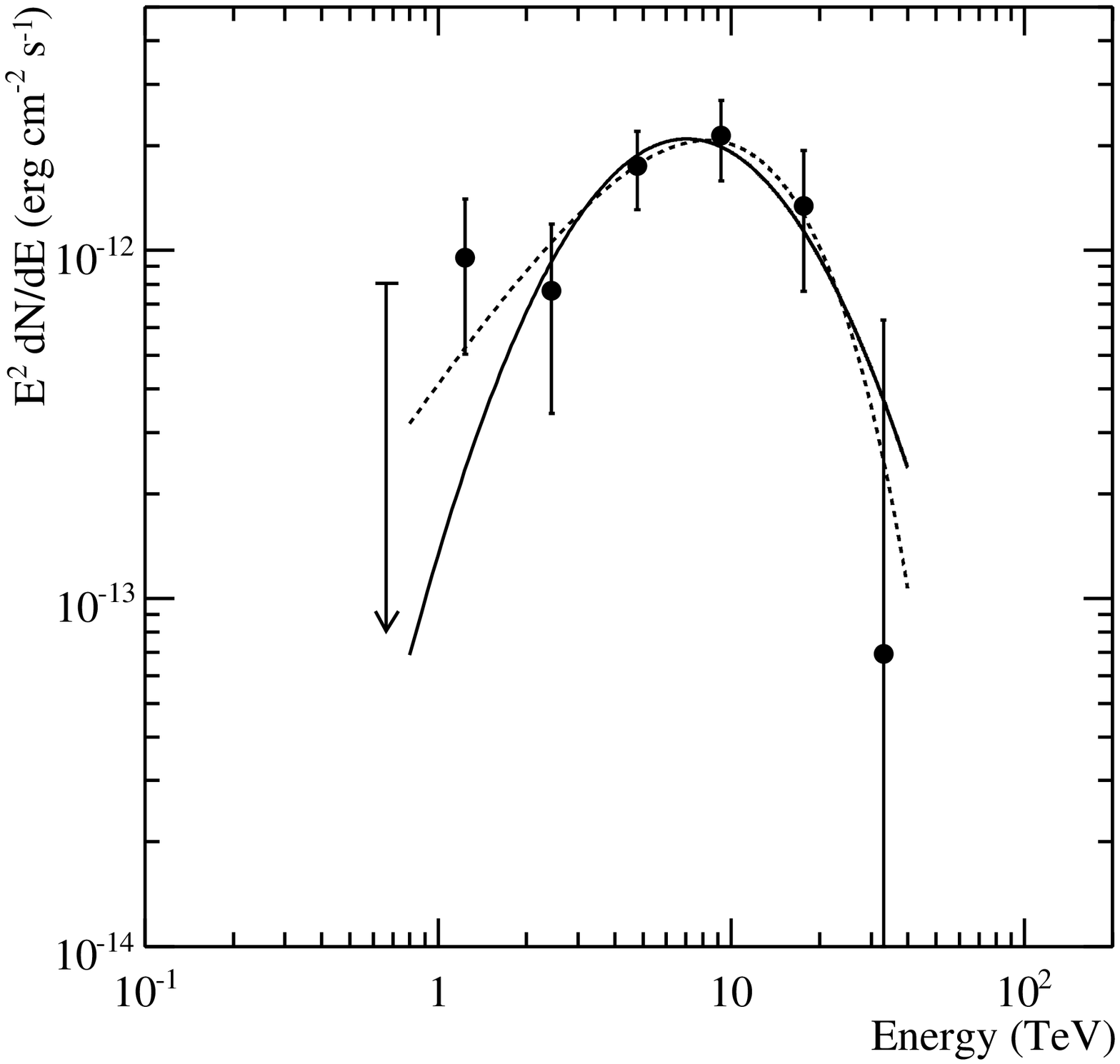}
  \caption{\label{J1718-both}\small The energy spectrum of \HESSONE, which is
    fit by a curved profile (solid line).  Alternatively, the fit of
    an exponentially cut-off power law is shown (dashed line, refer to
    the text for details on both fits). The first point in the
    spectrum lacks statistics due to lower exposure at small zenith
    angles and is plotted as an upper limit with at a confidence level
    of $2\sigma$.}
\end{figure}

For the spectral analysis, a statistical significance of $6.8\sigma$
(with 343 excess counts) is derived. Figure \ref{J1718-both} shows the
measured spectral energy distribution for \HESSONE\ (in $E^{2}\,dN/dE$
representation).

The spectrum is fit by a curved profile (shown as the solid line):
\begin{equation}
  \frac{dN}{dE} = N_0 \left(\frac{E_{\mathrm{peak}}}{1\,\mathrm{TeV}}\right)^{-2} \left(\frac{E}{E_{\mathrm{peak}}}\right)^{\beta \cdot \mathrm{ln}(E/E_{\mathrm{peak}}) - 2}
\end{equation}
The peak energy $E_{\mathrm{peak}}$ is $(7 \pm 1_\mathrm{stat} \pm
1_\mathrm{sys})$\,TeV, the differential flux normalisation $N_0 = (1.3
\pm 0.3_\mathrm{stat} \pm 0.5_\mathrm{sys}) \times
10^{-12}$\,TeV$^{-1}$\,cm$^{-2}$\,s$^{-1}$ and $\beta = -0.7 \pm
0.3_\mathrm{stat} \pm 0.4_\mathrm{sys}$. This fit has a
$\chi^2/{d.o.f.}$ of $3.2/3$. The integral flux between $1-10$~TeV is
about 2\,\% of the flux of the Crab nebula in the same energy
range~\cite{HESS:crab}. 

Alternatively, fitting the spectrum by an exponentially cut-off power
law ($dN/dE = N_0 E^{-\Gamma} e^{-E/E_{\mathrm{cut}}}$) gives $N_0 =
(3.0 \pm 1.9_\mathrm{stat} \pm 0.9_\mathrm{sys}) \times
10^{-13}$\,TeV$^{-1}$\,cm$^{-2}$\,s$^{-1}$, photon index $\Gamma = 0.7
\pm 0.6_\mathrm{stat}\pm 0.2_\mathrm{sys}$ and a cut-off in the
spectrum at an energy of $E_{\mathrm{cut}}=(6 \pm 3_\mathrm{stat} \pm
1_\mathrm{sys})$\,TeV. This fit, which is shown as a dashed line in
Figure \ref{J1718-both}, has a $\chi^2/{d.o.f.}$ of $1.6/3$.

Both the curved and exponentially cut-off power law profiles fit the
data well; the former has the advantage of showing explicitly the
peak energy of the spectrum, which has to date only been resolved in
one other VHE source, Vela\,X \cite{HESS:velax}.

\section*{Possible Associations}

The $\gamma$-ray source \HESSONE\ is located $\sim$0.14$^\circ$ south
of the pulsar \PSRONE. \PSRONE\ appears to be a Vela-like pulsar, as
it is of comparable age, 90~kyr, and has a similar spin period,
75\,ms.  From the spectral fit of a curved profile, the energy flux of
\HESSONE\ between (1 -- 10)\,TeV is estimated to
$2.9\times10^{-12}$~erg cm$^{-2}$s$^{-1}$.  With a distance of
$\sim$4\,kpc and a spin-down luminosity of $\dot{E} =
1.3\times10^{36}$\,erg\,s$^{-1}$, \PSRONE\ is energetic enough to
power \HESSONE, with an implied efficiency of $\epsilon_\gamma \equiv
L_\gamma/\dot{E} = 0.5\,\%$.

\begin{figure}[!hbt!]
  \centering
  \vspace{3ex}
  \includegraphics[width=0.48\textwidth]{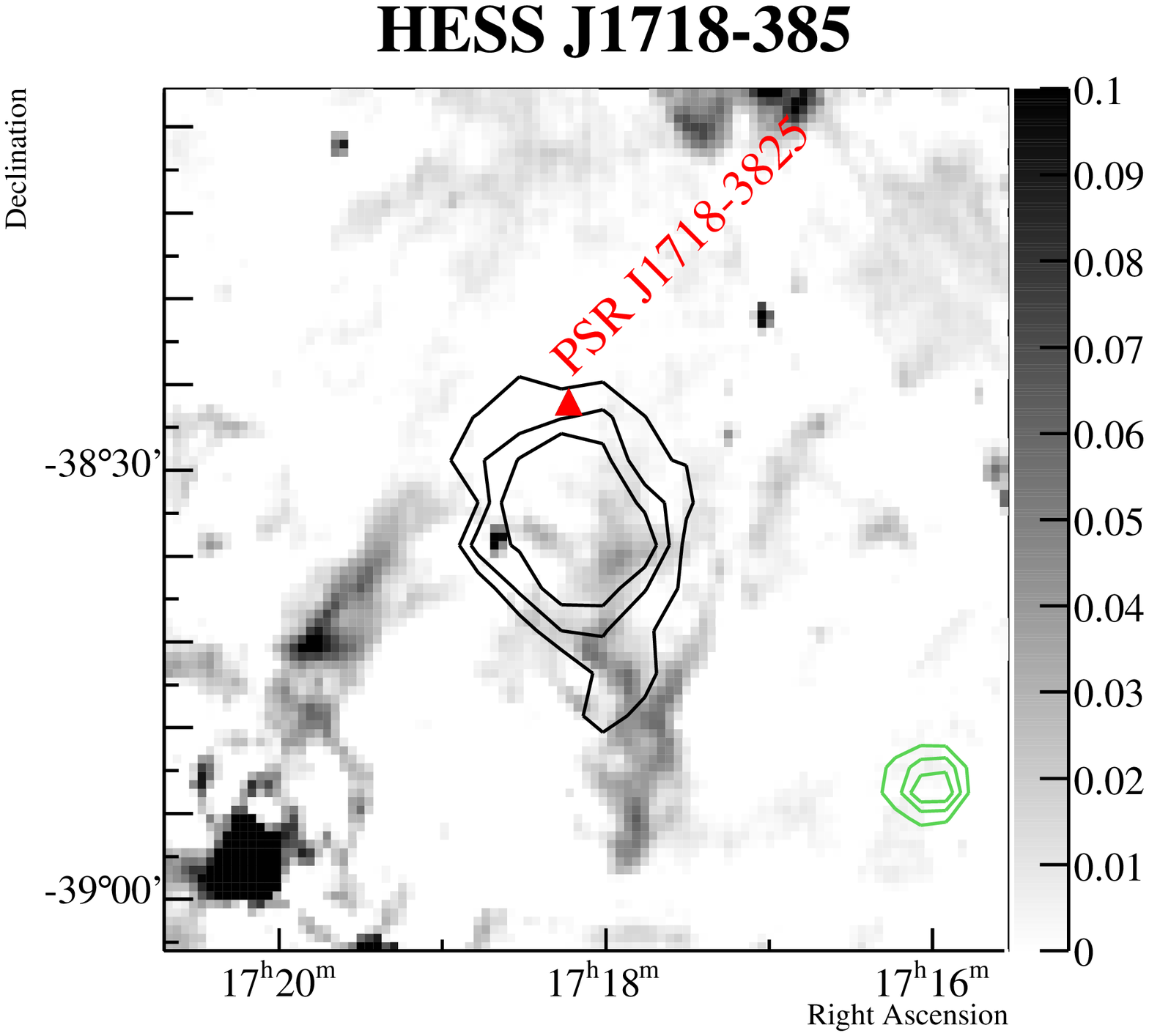}
  \caption{\label{PP_1718_figB}\small Radio image from the Molonglo Galactic
    Plane Survey at 843\,MHz~\cite{Molonglo} (in Jy/beam). The
    H.E.S.S. significance contours are overlaid in black and the
    pulsar position is marked with a red triangle. Adaptively smoothed
    ROSAT hard-band X-ray contours are shown in green~\cite{ROSAT}.}
\end{figure}

As can be seen in Figure \ref{PP_1718_figB}, no obvious X-ray
counterpart is visible for \HESSONE. There is diffuse extended radio
emission, which is partially coincident with the VHE emission.
However, this emission seems to be correlated with thermal dust
emission visible in the IRAS Sky Survey Atlas~\cite{IRAS}, suggesting
that the radio emission is thermal and is thus not likely associated
with a possible PWN. The brightest part of this diffuse feature is
catalogued as PMN~J1717$-$3846~\cite{PMN:1}. From the point of view of
positional coincidence, energetics, and lack of other counterparts,
the association of \HESSONE\ with \PSRONE\ seems plausible. To confirm
this, additional evidence from spectral and morphological studies in
VHE $\gamma$-rays and from data at other wavelengths is needed.

\HESSONE\ may well represent the first VHE $\gamma$-ray PWN found in a
systematic search for pulsar associations, despite the present lack of
a PWN detection in other wave bands. The remarkable similarity between
\HESSONE\ and other known VHE PWNe, together with the lack of other
probable counterparts, gives additional confidence. The detection of
an X-ray PWN would provide confirmation.

\section*{Acknowledgements}
\small The support of the Namibian authorities and of the University
of Namibia in facilitating the construction and operation of H.E.S.S.
is gratefully acknowledged, as is the support by the German Ministry
for Education and Research (BMBF), the Max Planck Society, the French
Ministry for Research, the CNRS-IN2P3 and the Astroparticle
Interdisciplinary Programme of the CNRS, the U.K. Science and
Technology Facilities Council (STFC), the IPNP of the Charles
University, the Polish Ministry of Science and Higher Education, the
South African Department of Science and Technology and National
Research Foundation, and by the University of Namibia. We appreciate
the excellent work of the technical support staff in Berlin, Durham,
Hamburg, Heidelberg, Palaiseau, Paris, Saclay, and in Namibia in the
construction and operation of the equipment.\\We have made use of the
ROSAT Data Archive of the Max-Planck-Institut fuer extraterrestrische
Physik (MPE) at Garching, Germany.

%This is the reference to .bib file (Whitout .bib!)

%This in the bibtex style, is ok.
\bibliographystyle{plain}
\normalsize

%%%%%%%%
%  09  %
%%%%%%%%

%\pdfoutput=1 

\def\degr{\hbox{$^\circ$}}
%\newcommand{\HMS}[3]{$#1^{\mathrm{h}}#2^{\mathrm{m}}#3^{\mathrm{s}}$}
% Bibliography and bibfile
\def\aj{AJ}%
          % Astronomical Journal
\def\actaa{Acta Astron.}%
          % Acta Astronomica
\def\araa{ARA\&A}%
          % Annual Review of Astron and Astrophys
\def\apj{ApJ}%
          % Astrophysical Journal
\def\apjl{ApJ}%
          % Astrophysical Journal, Letters
\def\apjs{ApJS}%
          % Astrophysical Journal, Supplement
\def\ao{Appl.~Opt.}%
          % Applied Optics
\def\apss{Ap\&SS}%
          % Astrophysics and Space Science
\def\aap{A\&A}%
          % Astronomy and Astrophysics
\def\aapr{A\&A~Rev.}%
          % Astronomy and Astrophysics Reviews
\def\aaps{A\&AS}%
          % Astronomy and Astrophysics, Supplement
\def\azh{AZh}%
          % Astronomicheskii Zhurnal
\def\baas{BAAS}%
          % Bulletin of the AAS
\def\bac{Bull. astr. Inst. Czechosl.}%
          % Bulletin of the Astronomical Institutes of Czechoslovakia 
\def\caa{Chinese Astron. Astrophys.}%
          % Chinese Astronomy and Astrophysics
\def\cjaa{Chinese J. Astron. Astrophys.}%
          % Chinese Journal of Astronomy and Astrophysics
\def\icarus{Icarus}%
          % Icarus
\def\jcap{J. Cosmology Astropart. Phys.}%
          % Journal of Cosmology and Astroparticle Physics
\def\jrasc{JRASC}%
          % Journal of the RAS of Canada
\def\mnras{MNRAS}%
          % Monthly Notices of the RAS
\def\memras{MmRAS}%
          % Memoirs of the RAS
\def\na{New A}%
          % New Astronomy
\def\nar{New A Rev.}%
          % New Astronomy Review
\def\pasa{PASA}%
          % Publications of the Astron. Soc. of Australia
\def\pra{Phys.~Rev.~A}%
          % Physical Review A: General Physics
\def\prb{Phys.~Rev.~B}%
          % Physical Review B: Solid State
\def\prc{Phys.~Rev.~C}%
          % Physical Review C
\def\prd{Phys.~Rev.~D}%
          % Physical Review D
\def\pre{Phys.~Rev.~E}%
          % Physical Review E
\def\prl{Phys.~Rev.~Lett.}%
          % Physical Review Letters
\def\pasp{PASP}%
          % Publications of the ASP
\def\pasj{PASJ}%
          % Publications of the ASJ
\def\qjras{QJRAS}%
          % Quarterly Journal of the RAS
\def\rmxaa{Rev. Mexicana Astron. Astrofis.}%
          % Revista Mexicana de Astronomia y Astrofisica
\def\skytel{S\&T}%
          % Sky and Telescope
\def\solphys{Sol.~Phys.}%
          % Solar Physics
\def\sovast{Soviet~Ast.}%
          % Soviet Astronomy
\def\ssr{Space~Sci.~Rev.}%
          % Space Science Reviews
\def\zap{ZAp}%
          % Zeitschrift fuer Astrophysik
\def\nat{Nature}%
          % Nature
\def\iaucirc{IAU~Circ.}%
          % IAU Cirulars
\def\aplett{Astrophys.~Lett.}%
          % Astrophysics Letters
\def\apspr{Astrophys.~Space~Phys.~Res.}%
          % Astrophysics Space Physics Research
\def\bain{Bull.~Astron.~Inst.~Netherlands}%
          % Bulletin Astronomical Institute of the Netherlands
\def\fcp{Fund.~Cosmic~Phys.}%
          % Fundamental Cosmic Physics
\def\gca{Geochim.~Cosmochim.~Acta}%
          % Geochimica Cosmochimica Acta
\def\grl{Geophys.~Res.~Lett.}%
          % Geophysics Research Letters
\def\jcp{J.~Chem.~Phys.}%
          % Journal of Chemical Physics
\def\jgr{J.~Geophys.~Res.}%
          % Journal of Geophysics Research
\def\jqsrt{J.~Quant.~Spec.~Radiat.~Transf.}%
          % Journal of Quantitiative Spectroscopy and Radiative Trasfer
\def\memsai{Mem.~Soc.~Astron.~Italiana}%
          % Mem. Societa Astronomica Italiana
\def\nphysa{Nucl.~Phys.~A}%
          % Nuclear Physics A
\def\physrep{Phys.~Rep.}%
          % Physics Reports
\def\physscr{Phys.~Scr}%
          % Physica Scripta
\def\planss{Planet.~Space~Sci.}%
          % Planetary Space Science
\def\procspie{Proc.~SPIE}%
          % Proceedings of the SPIE
\let\astap=\aap
\let\apjlett=\apjl
\let\apjsupp=\apjs
\let\applopt=\ao

%The paper title
\title{Morphological Studies of the PWN Candidate HESS J1809-193}
%Short title to print in the headers to the final publication (Not showed in this print).
\shorttitle{The PWN candidate HESS J1809-193}
%All paper authors
\authors{Nu. Komin$^{1}$,
  S. Carrigan$^{2}$,
   A. Djannati-Ata\"i$^{3}$,
   Y.A. Gallant$^{1}$,\\
   K. Kosack$^{2}$,
   G. Puehlhofer$^{4}$,
   S. Schwemmer$^{4}$,
   for the H.E.S.S. Collaboration$^{5}$}
%Short title to print in the headers to the final puplication (Not showed in this print).
\shortauthors{Komin and et al}
%All the affiliations.
\afiliations{$^1$ LPTA, Universit\'e Montpellier 2, CNRS/IN2P3, Montpellier, France\\ 
$^2$ Max-Planck-Institut f\"ur Kernphysik, Heidelberg, Germany\\
$^3$ APC (CNRS, Universit\'e Paris VII, CEA, Observatoire de Paris), Paris, France\\
$^4$ Landessternwarte, Universit\"at Heidelberg, Germany\\
$^5$ {\tt www.mpi-hd.mpg.de/HESS }}
\email{komin@lpta.in2p3.fr}

%The abstract.
\abstract{The source HESS J1809$-$193 was discovered in 2006 in data
of the Galactic Plane survey, followed by several re-observations. It
shows a hard gamma-ray spectrum and the emission is clearly
extended. Its vicinity to PSR\,J1809-1917, a high spin-down luminosity
pulsar powerful enough to drive the observed gamma-ray emission, makes
it a plausible candidate for a TeV Pulsar Wind Nebula (PWN). On the
other hand, in this region of the sky a number of faint,
radio-emitting supernova remnants can be found, making a firm
conclusion on the source type difficult.

Here we present a detailed morphological study of recent H.E.S.S. data
and compare the result with X-ray measurements taken with
\emph{Chandra} and radio data. The association with a PWN is likely,
but contributions from supernova remnants cannot be ruled out.

}

%\email{aastex-help@aas.org}
\maketitle

\addcontentsline{toc}{section}{Morphological Studies of the PWN Candidate HESS J1809-193}
\setcounter{figure}{0}
\setcounter{table}{0}
\setcounter{equation}{0}

%Begin the section.

\section*{Introduction}

Since the beginning of observations with H.E.S.S. (High Energy
Stereoscopic System) in 2003 the number of known TeV gamma-ray
emitting sources has increased drastically. The ongoing scan of the
Galactic plane revealed several bright and extended sources for which
no clear association with objects in other wavelength could be found
\cite{Scan1,Scan2}.

Pulsars, rapidly rotating neutron stars, are widely believed to be
able to accelerate particles up to PeV energies. Those objects loose
their rotational energy in winds of relativistic particles.  The
confinement of the wind in the interaction with the ambient
interstellar material forms shocks; the particles accelerated there
are visible as a Pulsar Wind Nebula (PWN) (see \cite{GaenslerSlane}
for a review). Synchrotron radiation seen in radio and X-rays prove
the existence of relativistic electrons in the PWN. These electrons
undergo inverse Compton (IC) scattering off ambient radiation fields,
like the Cosmic Microwave Background, Galactic infrared background and
optical star light, leading to the production of TeV gamma-rays.

Here we present the observation of one TeV source, HESS\,J1809$-$193,
which is located close to a powerful pulsar and thus a good PWN
candidate. X-ray emission from the direction of the pulsar support the
theory of being a PWN. However, confusion with other sources cannot be
ruled out.

\section*{TeV observations of  HESS\,J1809$-$193}

H.E.S.S. is a system of four Imaging Atmospheric Cherenkov telescopes
(IACTs) dedicated to the observation of TeV gamma-rays. Its high
sensitivity allows the detection of point sources with a flux of $1\%$
of that of the Crab nebula within 25\,h \cite{Crab}. Its large field
of view and an angular resolution of better than $0.1\degr$ makes it
an ideal tool for observations of extended objects and for the
conduction of sky surveys.

In the original Galactic plane survey conducted with H.E.S.S., TeV
emission from HESS\,J1809$-$193 was only marginally detected. Further
re-observations confirmed the existence of gamma-ray
emission \cite{paper}. Further observations were performed in autumn
2006; in total data with a live time of 32\,h is available.

\begin{figure}
\begin{center}
\includegraphics [width=0.44\textwidth]{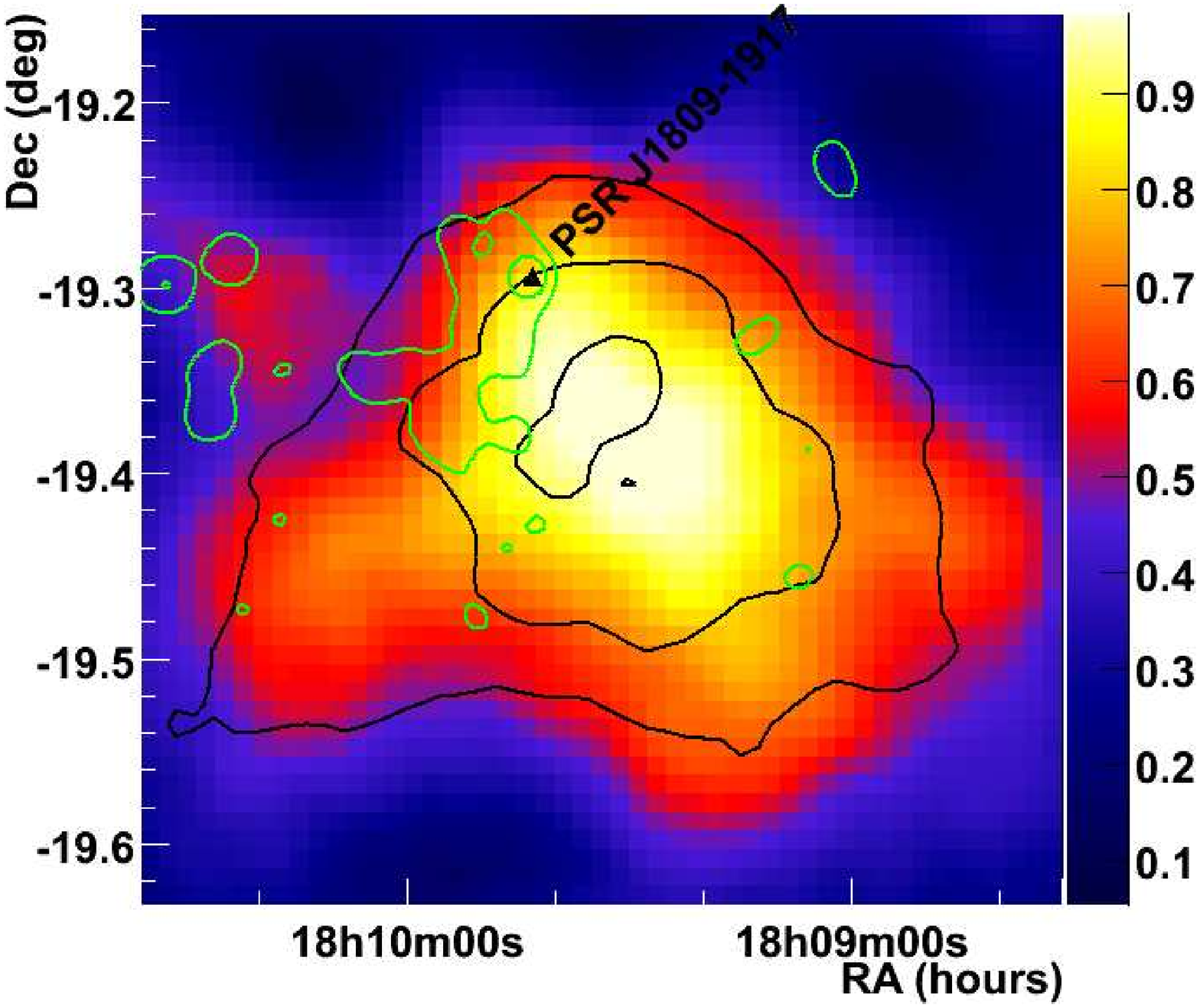}
\end{center}
\caption{TeV gamma-ray excess counts from the direction of
HESS\,J1809$-$193 (colour scale). The image is smoothed with the
point-spread function. Overlaid are 5, 7 and 9\,$\sigma$ significance
contours, oversampled with a circle with radius $0.1\degr$. The
position of the pulsar is marked with a black triangle. The green
contours denote the diffuse X-ray emission from
Fig.~\ref{fig:chandra}.}
\label{fig:skymap}
\end{figure}

The gamma-ray excess map of the source HESS\,J1809$-$193 is shown in
Fig.~\ref{fig:skymap}. The emission is clearly extended lying
south-west of the pulsar. In addition faint emission can be seen to
the south-east. In total, an excess of 3600 events with a significance
of $19\,\sigma$ was detected. The source shows an energy spectrum
consistent with a power law with an index of $2.2 \pm
0.1_\mathrm{stat} \pm 0.2_\mathrm{syst}$ and an energy flux between 1
and 10\,TeV of roughly $1.3 \times 10^{-11} \,
\mathrm{erg\,cm}^{-2}\,\mathrm{s}^{-1}$ \cite{paper}. If this energy
flux is projected to the distance of the pulsar, only $1.2\%$ of the
pulsar's spin down luminosity of
$1.8\times10^{36}\,\mathrm{erg\,s}^{-1}$ is needed to power the
H.E.S.S. source. Therefore it seems to be plausible that
HESS~J1809$-$193 is indeed a Pulsar Wind Nebula.

\section*{X-Ray Observations}

In the data of the Galactic plane scan performed with the ASCA
satellite diffuse emission was detected \cite{Bamba}, which turned out
to be coincident with the TeV source. The X-ray source G$11.0+0.0$ has
been discussed to be either a young shell-type supernova remnant (SNR)
or a plerionic SNR.

\begin{figure}
\begin{center}
\includegraphics [width=0.48\textwidth]{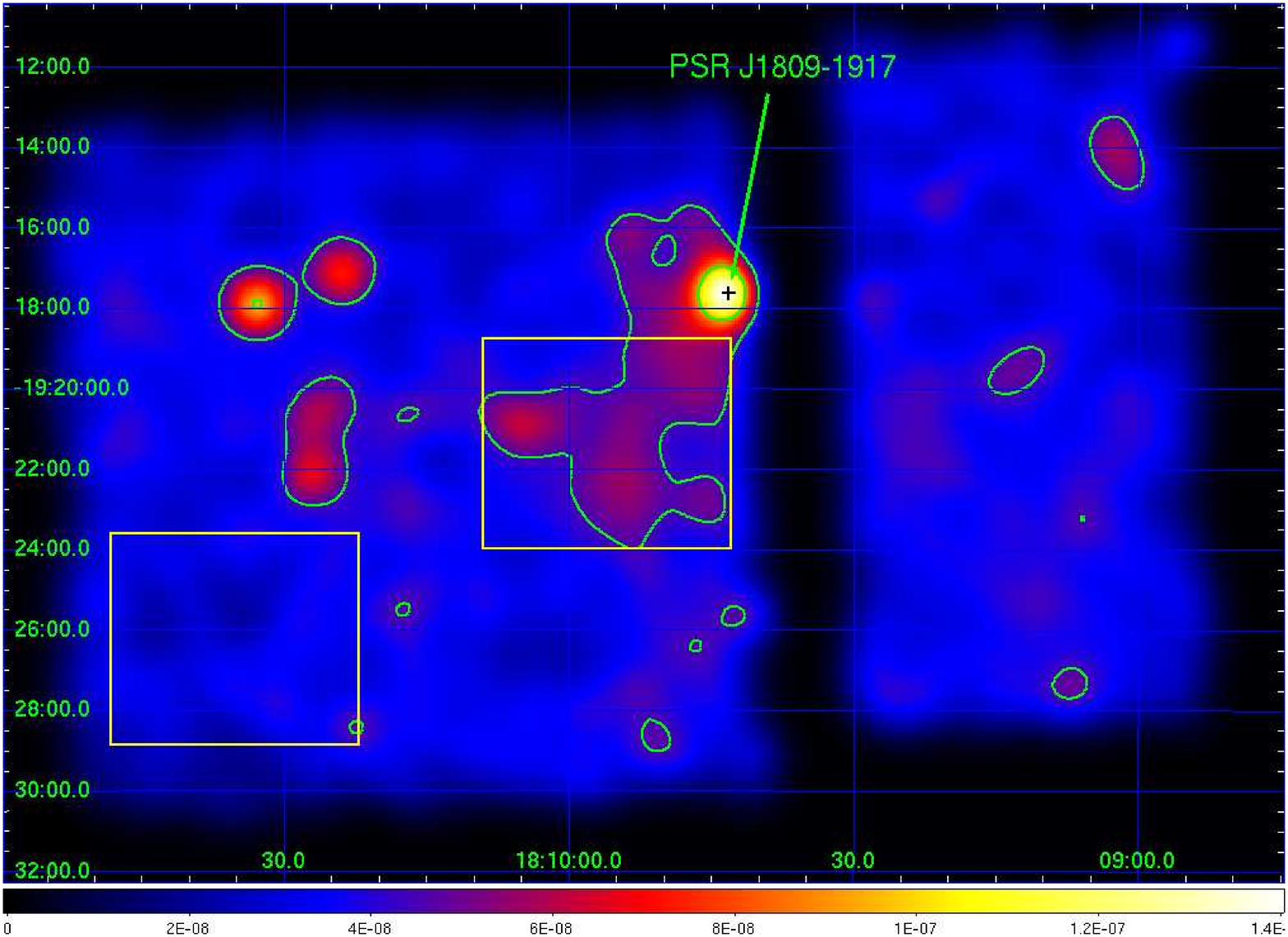}
\end{center}
\caption{Chandra X-ray excess of the field of view of HESS
  J1809$-$193. The map is exposure-corrected and smoothed with a
  Gaussian with $32''$.}
\label{fig:chandra}
\end{figure}

High-resolution observations with the \emph{Chandra} satellite revealed
a compact X-ray nebula north of the pulsar and additional faint
emission south \cite{PSR}. Here we present \emph{Chandra} data which
was taken in February 2007 (ObsID 6720).  Figure~\ref{fig:chandra}
shows the exposure-corrected and smoothed X-ray excess map. It shows a
strong X-ray nebula, high resolution images show its extension to the
north of the pulsar \cite{PSR}. Further faint emission can be seen to
the south.

The significance of the diffuse emission was tested by comparing the
on-source region with an off-source region in the same field of view
(these regions are indicated by the yellow rectangles in
Fig.~\ref{fig:chandra}). Taking into account the small acceptance
difference of 4\% (estimated from the exposure map at 2 keV), the
source region shows an excess of about 900 events with a statistical
significance of $10\,\sigma$.

The contour of the diffuse X-ray emission is overlaid in the TeV
excess map in Fig.~\ref{fig:skymap}. It should be noted that due to
the gap between the chips of the X-ray detector and the different
nature of the chips on the right hand side, no conclusions can be
drawn on the existence of diffuse emission to the west. However, it
can be seen that the X-ray nebula's extension to the south is far
smaller than the extension of the TeV emission.

\section*{Radio Data of the Field of View}

\begin{figure}
\begin{center}
\includegraphics [width=0.48\textwidth]{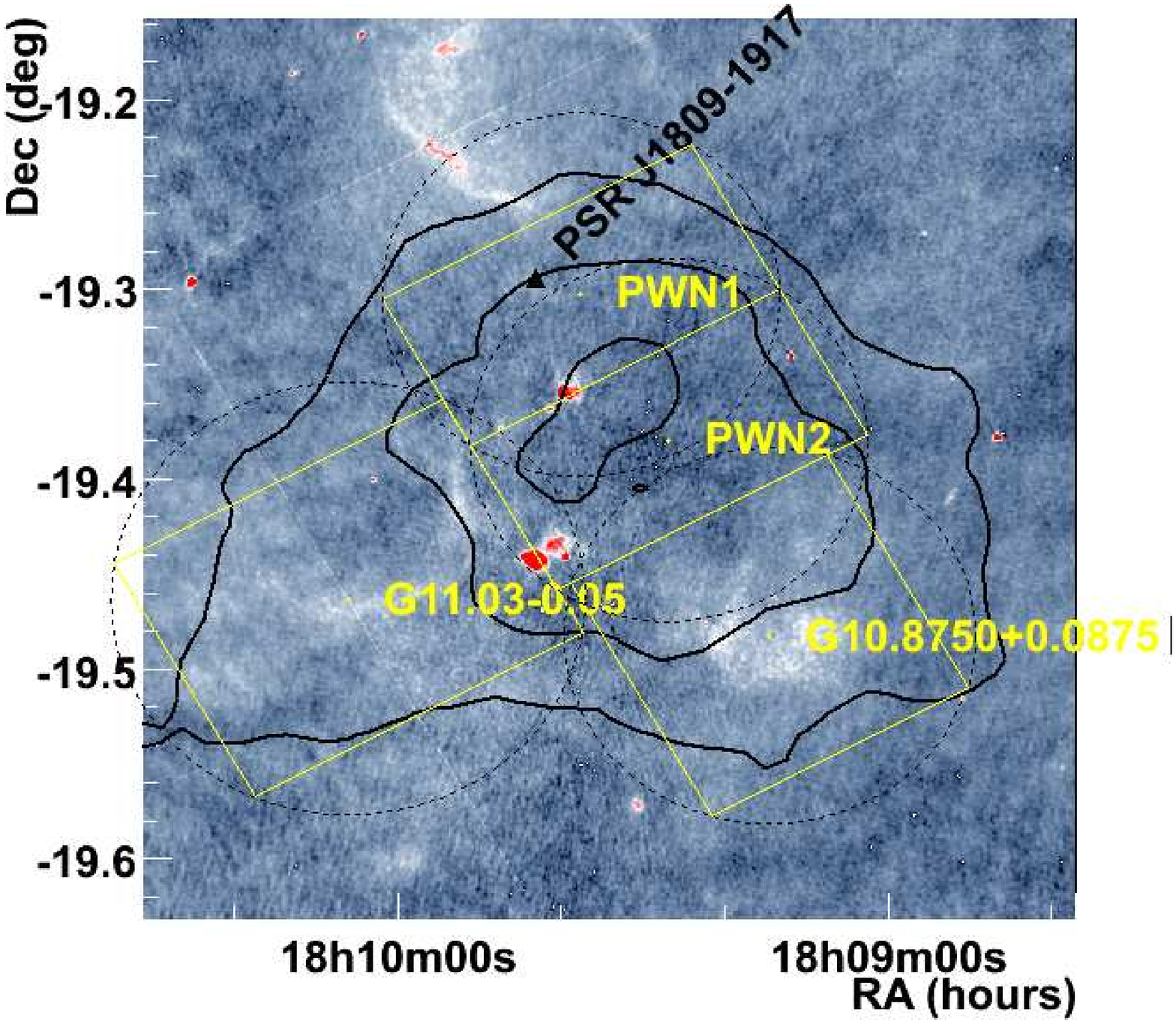}
\end{center}
\caption{MAGPIS radio image. Overlaid are the 5, 7 and 9\,$\sigma$
  H.E.S.S. significance contours. The yellow rectangles indicate the
  test regions for the spectral analysis of the H.E.S.S. data.}
\label{fig:radio}
\end{figure}

We compared the region of HESS~J1809$-$193 with radio data of the
Multi-Array Galactic Plane Imaging Survey \cite{MAGPIS} shown in
Fig.~\ref{fig:radio}. North of the pulsar is the SNR G$11.18+0.11$
\cite{Brogan,Green}, not coincident with the H.E.S.S. excess. Located
south-east of the pulsar and coincident with the H.E.S.S. excess is
the SNR G$11.03-0.05$ \cite{Brogan,Green}. Further south-west of the
pulsar is the supernova remnant candidate $10.8750+0.0875$
\cite{MAGPIS}. In the region between the latter two SNRs and the
pulsar no diffuse radio emission can be found.

For a spectral analysis different regions according to the radio data
have been chosen. Two regions for the SNRs G$11.03-0.05$ and
$10.8750+0.0875$ and another two regions for the possible pulsar wind
nebula (PWN1, PWN2). These regions are indicated in
Fig.~\ref{fig:radio}.

\section*{Spectral Analysis}

\begin{figure}
\begin{center}
\includegraphics [width=0.48\textwidth]{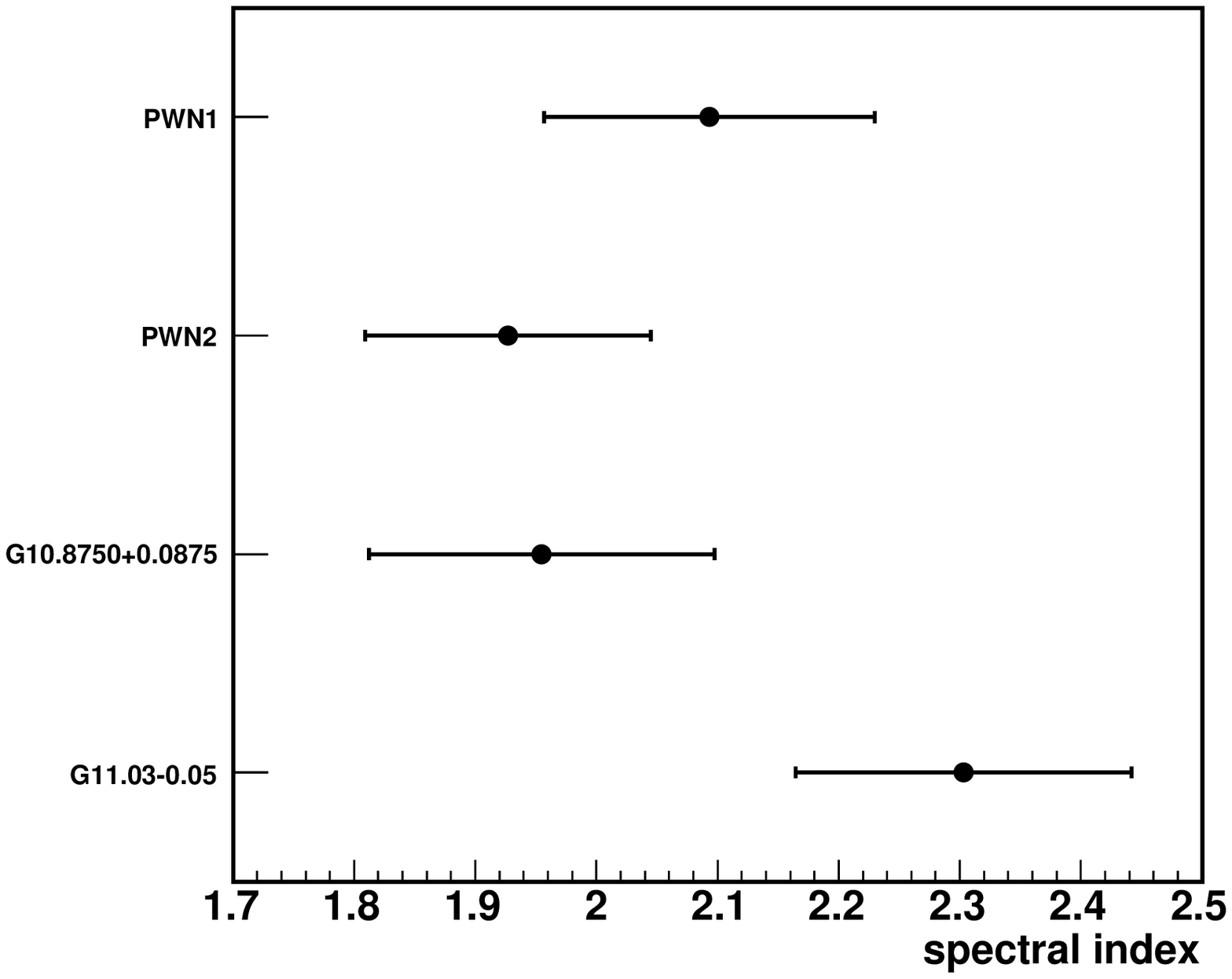}
\end{center}
\caption{Spectral indices for the test regions indicated in
Fig.~\ref{fig:radio}.}
\label{fig:spectrum}
\end{figure}

The TeV energy spectra of each region were fitted with a power
law. The spectral indices are shown in Fig.~\ref{fig:spectrum}.  The
regions PWN1, PWN2 and $10.8750+0.0875$ are with increasing distance
to the pulsar. A steepening of the spectrum with increasing distance
would be a clear indication for a PWN (see \cite{J1825_2}). A
significant different energy spectrum between the region of the SNR
G$11.03-0.05$ and the rest of the emission region would be a hint for
different source associations. However, due to the large statistic
uncertainties no conclusion on spectral variations over the extension
of the source can be drawn.

\section*{Discussion}

The existence of the powerful pulsar PSR~J$1809-1917$ which can easily
power the TeV emission suggests that the TeV emission is a PWN
associated with the pulsar. The association with a PWN is further
supported by a compact X-ray nebula and diffuse X-ray emission
coincident with the H.E.S.S. source. The X-ray emission is
significantly smaller than the TeV source. This has been already seen
for the PWN HESS~J$1825-137$ \cite{J1825}.

On the other hand, two supernova remnants coincide with the TeV
emission. They do not show X-ray emission, however, TeV emission can
still be expected, in particular if if they are associated with dense
molecular clouds \cite{W28,Yamazaki}. Further studies will include the
search for dense molecular clouds in the region.

\section*{Conclusion}

Detailed studies of the source HESS~J$1809-193$ and comparison with
objects in other wavelength show that this source is likely a PWN
powered by the pulsar PSR~J$1809-1917$.  The number of TeV emitting
PWNe is increasing, showing that TeV PWN constitute a significant
fraction of the Galactic TeV gamma-ray source population.

Contribution of gamma-ray emission from faint radio supernova remnants
cannot be ruled out. Radio-emitting, X-ray quiet SNRs, possibly in
connection with dense molecular clouds, remain interesting targets for
gamma-ray observations.

\section*{Acknowledgements}

The support of the Namibian authorities and of the University of
Namibia in facilitating the construction and operation of H.E.S.S. is
gratefully acknowledged, as is the support by the German Ministry for
Education and Research (BMBF), the Max Planck Society, the French
Ministry for Research, the CNRS-IN2P3 and the Astroparticle
Interdisciplinary Programme of the CNRS, the U.K. Science and
Technology Facilities Council (STFC), the IPNP of the Charles
University, the Polish Ministry of Science and Higher Education, the
South African Department of Science and Technology and National
Research Foundation, and by the University of Namibia. We appreciate
the excellent work of the technical support staff in Berlin, Durham,
Hamburg, Heidelberg, Palaiseau, Paris, Saclay, and in Namibia in the
construction and operation of the equipment.

\bibliographystyle{plain}

%%%%%%%%
%  10  %
%%%%%%%%

%The paper title
\title{New VHE emitting middle-age pulsar wind nebula candidates in
the extended H.E.S.S. Galactic plane survey data}
%Short title to print in the headers to the final publication (Not showed in this print).
\shorttitle{Dependence of the energy}
%All paper authors
\authors{A. Lemiere$^{1},^{3}$, A.Djannati$^{1}$, O. deJager$^{2}$,
R. Terrier$^{1}$} 
\shortauthors{A. Lemière et al.}
%All the affiliations.
\afiliations{$(^1)$ APC - Astroparticule et Cosmologie - CNRS
Universite Paris VII, Paris.\\
$(^2)$ Unit for Space Physics, North-West University, Potchefstroom 2520,
South Africa. \\
$(^3)$ Center for Astrophysics, Smithonian-Harvard Observatory 
60 Garden street, Cambridge MA , USA.}
\email{alemiere@head.cfa.harvard.edu}

%The abstract.
\abstract{The H.E.S.S. 2004-2005 survey of the Galactic Plane at energies above 200
 GeV had revealed a number of pulsar wind nebulae candidates, including the
 remarkable source HESS J1825-137. Spatially resolved spectral measurements
 of this source gave the first evidence of an energy-dependent morphology
 which was interpreted as being due to the cooling of relic electrons
 cumulated throughout pulsar's history. Also for a few number of
 sources the asymmetry of the pulsar with respect to the nebula could be
 attributed to an asymmetric reverse shock from the associated supernova
 remnant due to inhomogeneities in the interstellar matter. Subsequently a
 class of large offset and relic nebulae emerged as an outstanding new type
 of VHE $\gamma$-ray source.\\
 We discuss here the cases of such nebulae in the extended H.E.S.S.
 Galactic Plane survey data through an energetic criterion taking into
 account earlier epochs of pulsar injection as well as through
 investigation of CO data to search for
 inhomogeneities.}

\maketitle

%Begin the section.
\addcontentsline{toc}{section}{New VHE emitting middle-age pulsar wind nebula candidates in the extended H.E.S.S. Galactic plane survey data}
\setcounter{figure}{0}
\setcounter{table}{0}
\setcounter{equation}{0}

\section*{Introduction}

%-----------------------------------------------------------------------------------------
% * Les donnees du scan 
% * les nouvelles PWN resolues 
% * HESSJ1825-137 :  - La difference de taille X/gamma : le cooling ?
%                    - La spectro-morphologie : le cooling... 
%                    - le model : qui confirme le cooling 
%                    - L'asymetrie : en X et gamma
%                    - l'asymetrie : Blondin et al : l'hypothese
%                    - sondage au CO : resultat
%                   
% * beaucoup de source sont des caracteristqiues similaire a J1825 : 
% grande extension, asymetrie, faibles X, pulsat middel-age
%
% * HESSJ1825 : un prototype pour les autres sources : extension de son etude aux autres !                      
%------------------------------------------------------------------------------------------

%le scan
During 2004-2006 H.E.S.S. (High energy stereoscopic system) performed 
a Survey of the inner part of the Galaxy \cite{HESSScanII} where its excellent
capability allowed to mark a breakthrough in the field of Pulsar wind nebula 
(PWN) study : for the first time the morphological structure of many PWN was resolved 
in the $\gamma$-ray band and many of them appears to belong 
to the middle-age Vela-like pulsar class.\\ 
HESS$~$J1825-137 is the archetypal example of this population. 
For this source, a detailed spectral and morphological analysis \cite{HESS1825II} revealed 
for the first time in $\gamma$-ray a steepening of the energy spectrum 
with increasing distance from the pulsar. This fact was interpreted as due to 
the radiative cooling of the emitting electrons during their propagation. 
%and confirmed by showing that one can reproduce the data by cumulating the 
%electrons and making them cooling during all the pulsar life.
Indeed very high energy (VHE) Inverse Compton (IC) flux typically provide information 
on lower energy electrons than those of the keV synchrotron flux and 
imply a larger life-time for IC emittors. 
%than for keV emittors. 
%and causes very larger TeV size than keV. 
For the middle-age PWN, electrons emitting at TeV energies consist 
of cooled particles cumulated during few tenth of kyrs (called relic electrons thereafter). 
%and as shown by \cite{J1825model},  
%gives access to the magnetic field evolution.\\
%a one zone time-dependent model taking 
%into account the particles injected during the pulsar life and the cooling of electrons 
%in the nebula could successfully reproduce the TeV total spectrum as well as the X-ray 
%spectrum and gave strong constraints on the magnetic field evolution.\\
HESS~J1825-137 can also be taken as a prototype by its asymmetrical configuration with 
respect to the pulsar, a morphological characteristic observed in many of the middle-age 
candidates. The similar morphology of HESS~J1825-137 to that of the X-ray
nebula G18.0-0.7 ~\cite{Gaensler03} suggested an asymmetric reverse shock to have happened 
in the 10 first kyrs of the pulsar life, consisting of the interaction of a supernova 
remnant (SNR) with an inhomogeneous surrounding
material implying different colliding times of the reverse shock with
the PWN, and resulting in a one-sided morphology ~\cite{HESS1825I}. 
To examine this hypothesis, \cite{CO1825} probed the
interstellar matter density near G18.0-0.7 through $^{12}$CO line emissions. Two 
molecular structures were discovered and their characteristics were found compatible with the
observed offset of the VHE nebula HESS~J1825$-$137 within the
framework of the simulations by \cite{bcf01}.\\
Here, we propose to extend this study to the large set of asymmetric middle-age 
PWN candidates found in the 2004-2007 H.E.S.S. data.   
In one first step, the association between the TeV sources and the pulsars will be 
tested by the calculation of a new energy criterion. It will result in the first catalogue of middle-age PWN 
at VHE $\gamma$-ray. For the selected candidates which exhibit an asymmetric shape,  the pulsar configuration 
with respect to the H$_{2}$ density profile and the high energy
emissions will be considered and compared to the numerical model of \cite{bcf01}, 
as it has been done in \cite{CO1825}.
%%%%%%%%%%%%%%%%%%%%%%%%%%%%%%%%%%%%%%%%%%%%%%%%%%%%%%%%%%%%%%%%%%%%%%%%%%%%%%%%%%%%%%%%%%%%%%%%%%%%
%ou annoncer les nouveaux candidats ??? au moins les 4 derniers...
%%%%%%%%%%%%%%%%%%%%%%%%%%%%%%%%%%%%%%%%%%%%%%%%%%%%%%%%%%%%%%%%%%%%%%%%%%%%%%%%%%%%%%%%%%%%%%%%%%%%
%\section*{Test the association between TeV sources and pulsars}
\section*{Build a catalogue of PWN candidates}
\subsection*{Method}

During the first part of the H.E.S.S. scan of the Galactic Plane, we buit a list of six VHE sources 
for which the spectral and morphological characteristics together with the proximity of an energetic 
pulsar placed them as potential PWN candidates. 
With the extended scan performed during 2006-2007, 3 new sources 
have been added to the list : HESS$~$J1718-385 and HESS$~$J1809-193 \cite{TwoPWN}, 
%HESS$~$J1357-645 
and HESS$~$J1912+101 \cite{month}. 
All the sources of this list have ages between few to few hundred of kyrs (middle-age class) and 
the high energetic pulsars are closer that 7 kpc from us.\\
%We know that some fraction of the rotational power of pulsars goes in to driving a relativistic wind. 
%When this wind is confined by the surrounding medium it decelerates and the terminal shock form,
% then some fraction of the power available in the wind is converted into non thermal electrons 
%radiating by Synchroton and inverse Compton processes and forming what we call the pulsar wind nebula 
%(PWN).\\
To evaluate the likelihood of an association between the TeV sources and the nearby pulsars, 
the power available for $\gamma$-ray production must be assessed.
Since pulsar's rotational energy $\dot{\rm E}$ is the source for most of the emission seen 
from PWNe, it is used to consider the actual spin evolution of pulsars with respect to the 
PWN flux.  %Indeed, despite the great differences in their rotational paramters, magnetic flied,
%the rotation-powered X-ray sources revealed a correlation between their high-energy luminosity 
%and their spin-down luminosity (Possenti et al 2001).ICRC2007_outPWN.aux
%then for a 100\% conversion efficiency, the expected gamma-ray luminosity is just 
%$\frac{\dot{E}}{4 \pi d^2}$. 
%To power a typical gamma-ray PWN, 
%the efficiency of conversion of spin down power to gamma-ray emission must be ~ 1\% 
%(in the 0.1-10 TeV band) (de l'ordre de quelques pourcents).
But since the sources we consider have middle-ages, and given the fact that TeV electrons emittors 
are cooled particles in the nebula, we must consider also the earlier 
stages of the system if we want to get a true estimation of the TeV flux (contraty the X-ray nebulae 
for which the actual $\dot{\rm E}$ is relevant). \\
Our solution is to take into account the electrons injected by the pulsar in the nebula 
along the pulsar life-time, and the radiative losses dominated by the synchrotron component.
The model assumes a power law electron injection spectrum with index of 2.02 and a fixed minimum 
energy $E_{\rm min} $= 1 TeV.
The maximum energy is constrained by the acceleration condition that the gyroradius must not exceed the 
terminal shock radius  and the normalization is determined by $\dot{\rm E}$ at each time and by 
the ratio between particles and field $\sigma$ defined in \cite{KennelCoro} and fixed at 0.003 
(similar to Crab Nebula's). 
The time evolution of the spin down power of the pulsar is written 
%\begin{equation}
$\dot{E}(t) = \frac{\dot{E_{o}}}{( 1 + t/\tau_{o} )^2}$, 
%\end{equation}
 with $\dot{E_{o}}$, the initial spin down power of the pulsar, $\tau_{o}$ the characteristic pulsar 
time scale (we fix it at the standard value of 400 years) 
and a breaking index of 3.   
The total injected energy by the pulsar during the time dt is: 
\begin{equation}
\rm dW = \frac{1}{\sigma +1}~ \dot{\rm E}\rm (t) \rm dt
\end{equation}
The synchrotron losses are calculated with a typical average magnetic field of 5 $\mu$G :
\begin{equation}
\rm dP_{\rm losses} = \rm k.\rm E^{2}.\rm B^{2} \rm dt
\end{equation}
%If electrons are injected between 10 GeV and 100 TeV, the total equivalent losses are :
%\begin{equation}
% - P_{pertes} = \int_{10 GeV}^{100 TeV} \Big( \int_{T} k.E^{2}.B^{2} dt \Big) dE
%\end{equation}
The energetic balance for the electrons in the nebula can then be written at each time :
\begin{equation}
\rm dU = \rm dW -  \rm dP_{\rm losses}
\end{equation} 
After integration of this equation over the pulsar life-time, we obtain the actual 
total electron spectrum in the nebula. These electrons produce Inverse Compton photons that can 
be seen in the $\gamma$-ray band. We use an average value for the targets density of 0.25 $\rm eV.cm^{-3}$ for the CMB 
and 0.5 $\rm eV.cm^{-3}$ for the star and dust light.  
%The calculation applied to all the middle-age PWN candidates gives us a new energetic criteria 
%and efficiency epsilon (see Table 1).
%We then calculate the efficiency implied by the comparison between the seasured flux and 
%the estimated flux taking into account the pulsar characteristics. We then can use this quantities to 
%estimate credibility of such an association.
We define as $\epsilon$ an equivalent percentage of the ratio between 
the measured and predicted (this calculation) flux and propose to use it as an energetic criterion 
to estimate the credibility of an association between the VHE sources and corresponding pulsars. 
%%%%%%%%%%%%%%%%%%%%%%%%%%%%%%%%%%%%%%%%%%%%%%%%%%%%%%%%%%%%%%%%%%%%%%%
\subsection*{Results}

Table \ref{FinalTable} shows the complete selected sources list, with the corresponding pulsars 
characteristics, the measured and predicted $\gamma$-ray luminosity and the $\gamma$-ray efficiency. 
A 30$\%$ efficiency criterion of acceptance looks reasonable, since it gives the insurance that 
the pulsar generated enough power to reproduce the observed flux, even if the conversion efficiency at the 
shock radius is low. Using this criterion, a first class of 9 
good candidates appears clearly in the upper part of the table, whereas the two last sources are definitely 
rejected. All the selected candidates are extended and have a spectral index between 2 and 2.5, 
in good agreement with the known X-ray nebula.  Six of them have an X-ray nebula detected 
but only one has a counterpart in radio (mostly due to the faint radio emission of such extended object). \\
One of the most remarkable characteristic of these sources is that almost all of them show an asymmetric 
shape. We now propose to investigate possible reasons of this asymmetry by exploring the Interstellar medium (ISM).
%%%%%%%%%%%%%%%%%%%%%%%%%%%%%%%%%%%%%%%%%%%%%%%%%%%%%%%%%%%%%%%%%%%%%%%%%%%%%%%%%%%%%%%%%%%
\section*{Search for inhomogeneity in the ISM}
%possiblly explaning the asymetry morphology of the source
\subsection*{Data and Method}
The $^{12}\rm CO$ observations used here are taken from the
Composite CO Survey from \cite{Dame}. This
survey compiles observations from 37 individual surveys with a
resolution of 0.2$^\circ$ and a FWHM velocity resolution of 2
km.$\rm s^{-1}$~\cite{Dame}.\\
As a first step, clouds are searched for through the scan
of the CO intensity in the (l, b) plane as a function of the radial
velocity in the vicinity of the associated pulsar. 
If some structures are detected, the average CO velocity profiles are made 
in the line of sight of these structures and they are extracted by 
detecting CO pics.  We take the fit (gaussian) centroid of the $^{12} \rm CO$ peak 
to establish the kinematic distance to the clouds using the empirical Galactic 
rotation curve models \cite{Clemens85}. Only molecular clouds with a compatible 
distance with the pulsar are selected. 
%Dire un mot sur la morphologie ici mettre un plot en exemple ?... 
%The CO emission contours have
%been obtained through integration over the velocity-selected region $\delta V$ of
%$^{12} \rm CO$ peaks and subsequent conversion to $\rm H_2$ column density
%($\rm cm^{-2}$). 
In order to compare the compatibility of the interstellar matter 
distribution with the one by \cite{bcf01}, we analyze the average gradient of 
molecular matter along the vector defined from the pulsar position, 
in the direction of the TeV emission's center of gravity. 
The CO emission is spatially averaged for each
band and integrated over the cloud's velocity range. 
These values (usually called $\langle \rm W(CO) \rangle $) are converted into average 
$\rm H_{2}$ column density using the conversion coefficient by \cite{Hunter97}.  
We consider here the total average gradient of each distribution by
defining the contrast as the ratio between the maximum and minimum
values of the density and the characteristic scale as the distance
needed for the density to decrease by a factor two. 
These numbers are given for each sources in Table \ref{FinalCO}. 
%%%%%%%%%%%%%%%%%%%%%%%%%%%%%%%%%%%%%%%%%%%%%%%%%%%%%%%%
\subsection*{Results}
Half of the cases seems to have a configuration compatible with the simulations 
of \cite{bcf01}. Only one case is rejected (HESSJ$~$1420-607) due to the 
geometric configuration incompatible with the reverse shock hypothesis.
%Indeed, the correct estimation of the gradient density began very difficult 
%when the cloud and the pulsar are not exectily at the same distance. 
The major conclusion of this study is that the asymmetric reverse shock 
is a credible hypothesis if we consider the ISM distribution around most of the 
middle-age PWN VHE candidates.
However it remains difficult to get a strong conclusion on this study, 
firstly because of the distance near-far ambiguity and secondly because of 
the effect of projection in the line of sight.
\section*{Conclusion}
We established a list of potential PWNe candidates in the H.E.S.S. data and calculated an energetic criterion for each of them, taking into account the 
pulsar spindown power evolution over time together with average synchrotron losses. 
This study allowed to build the first VHE catalogue of middle-age PWN candidates 
consisting of 9 sources. 
Many of these candidates show an asymmetric shape around the putative associated pulsar.
By probing the ISM through CO data, we showed that for many of them, the hypothesis 
of an aymmetric reverse shock is not to exclude, given the ISM density distribution.   

%\section*{Acknowledgements}
%This work is partially supported by the Brazilian agencies FAPESP and CNP, 
%and the Argentinian agencies
%CNEA and ANPCyT.
%\nocite{ref4}
%\nocite{ref5}
%\nocite{ref6}
%\nocite{ref7}
%This is the reference to .bib file (Whitout .bib!)

%This in the bibtex style, is ok.

\bibliographystyle{plain}

\begin{table*}[htb]%htb
%\centering  
\label{table:4}      % is used to refer this table in the text                       
\begin{tabular}{l c c c c c c c c c c l}        % centered columns (4 columns)
\hline\hline                 %
 Name  & Pulsar & Age   & Dist & $\dot{\rm E}$         & $ \rm L_{\gamma_{\rm mes}}(\theta)$ & $\rm L_{\gamma_{\rm theo}}$ & $\epsilon$ & RX & A\\
 HESS &  PSR   & (yrs) & (kpc)& $10^{36}$ &          $10^{35}$            &       $10^{35}$                 &(\%)        &    & \\
      &        &       &      & (erg/s)      &           (erg/s)(deg)               &         (erg/s)                  &            &    & \\
\hline
\hline  
 J1303-631 & J1301-6305  & 11 000  & 6.6  &  1.7  &  1.80(0.32) & 6.78  & 27.00   & x & Y\\
 %J1357-645 & J1357-6429  &  7 310  & 4.09 &  3.1  &  0.50(0.34) & 7.9   & 6.34 & x  & N\\
 J1420-607 & J1420-6048  & 13 000  & 5.6  &  10   & 0.61(0.16)  & 61.37 & 1.01  & rx & Y\\
 J1616-508 & J1617-5055  &  8 100  & 6.5  &  18   & 1.59(0.4)   & 5.04  & 3.16 & -- & Y\\
 J1702-420 & J1702-4128  & 55 000  & 5.2  &  0.34 & 0.57(0.35)  & 10.8  & 5.30  & -- & Y\\
 J1718-385 & J1718-3825  & 89 500  & 4.2  &  1.3  & 0.14(0.18)  & 62.75 & 0.22 & -- & Y\\
 J1804-216 & J1803-2137  & 15 800  & 3.9  &  2.2  & 0.70(0.49)  & 16.79 & 4.20  & x  & Y\\
 J1809-193 & J1809-1917  & 51 000  & 3.7  &  1.8  & 0.33(0.5)   & 54.78 & 0.59 & x  & Y\\
 J1825-137 & J1823-13    & 21 000  & 3.9  &  2.8  & 1.96(0.8)   & 31.74 & 6.10  & x  & Y\\
 J1912+101 & J1913+1011  &170 000  & 4.48 &  2.9  & 0.87(0.5)   & 199.2 & 0.43 & -- & Y \\
           &             &         & &        &   &             &       &      &    &  \\
 J1632-478 & J1632-4818  & 19 800  & 7    &  0.05 & 2.67(0.36)  & 0.42  & 534  & -- & --\\
% J1834-087 & J1833-0827  & 147 000 & 5.7  &  0.58 &  (0.36)    &       & 1.8  & -- & --\\	
 J1745-303 & B1742-30    & 546 000 & 2.1  & 0.008 &  0.12(0.4)  & 0.32  & 32.77  & -- & --\\
% J1626-480 & J1627-4845  &2666 000 & 6.9  &0.00063&0.78(0.38)   & 0.044 & 178  & -- &-- &\\
 %VelaX    & J0835-4510  & 11 300  &...   &...    &...     &...  & & &Y\\ 
\hline \hline                         
\end{tabular}
\caption{This table shows the complete selected sources list, with the corresponding pulsars characteristics, 
the measured (with the associated radius of intergation) and predicted $\gamma$-ray luminosity in the 
[200 GeV - 10 TeV] energy band and the $\gamma$-ray efficiency. The detection of a radio (r) or X-ray(x) nebula 
around the pulsar and an asymmetric shape of the VHE nebula (A) are also indicated.
%Liste finale des candidats nébuleuses de pulsars dans les sources de très haute énergie non identifiées 
%détectées par HESS entre 2004 et 2007. Les tailles apparentes maximales des sources ainsi que le décalage du pulsar associé 
%par rapport au centre de gravité de la source sont indiquées pour les candidats possédant un pulsar. Pour cette même liste 
%de candidats, le nom et les principales caractéristiques du pulsar sont rappelés. Les flux mesurés et prédits ainsi que 
%les efficacités associées sont indiqués.
}
\label{FinalTable}
\end{table*}

\begin{table*}[htb]
 \begin{center}
  \begin{tabular}{l c c c c c c c c c c c c c l}
     \hline\hline
       Source  & Cloud  & $\rm M_{\rm CO}$   & density            &  $\rm D_{Cl}$       & Offset        & H              & Cont  &  D & G & Gr & RC\\
       name & name       & 10$^{3}$    & mol     &  kpc & $\% $ & 10$^{19}$   &  pc       &    &   &    &   \\
      HESS &  MC      &   $(\rm M_{sol})$        & $\rm cm^{-3}$             &     &  $ \rm R_{SNR}$ &  cm      &           &    &   &    &   \\     

     \hline\hline
      J1303-631  & 304.1+0.3  & 62  & 18   &  6.2         & 19.2     &  4    & 2   & Y & Y & Y & Y\\
      J1420-607  & 314+0.0    & 225 & 75   &  3.5/8       &  14      &  ---  &--   & Y & N & N & N\\ 
      J1616-508  & 332.7-0.6  & 222 & 164  &  3.9         & 10       & 12.2  & 2   & Y & Y & Y & Y\\
     % J1632-478  & 336.0-0.2 & 37.82  & 8      & 7/8.5        & 9.2      &   2   &     & Y & Y & Y & Y\\
      J1718-385  & 348.8-0.4  & 11  & 16     &  4           &  20      & ---   & --  & Y & ? & ? & ?\\
                 & 348.8-0.5  & 13  & 40     & 3.5          &  20      & ---   &--   & Y & ? & ? & ?\\ 
      J1804-216  & 8.3+0.25   & 71  & 129  & 3            &  23      & 11.2  &1.3  & Y & Y &  ? & ?\\ 
                 & 8.7-0.6    & 132 & 107 & 4            &  23      & ---   & --  & N & -- & -- & --\\
      J1809-193  & 11.1+0.12  & 80   & 67     & 3.7          &   25     & 13    & 2   & Y & Y & Y & Y\\ 	
      J1825-137  & 18.15-0.32 & 80 000 & 150    & 4            &   20     & 4.2   & 3.6 & Y & Y & Y & Y\\ 
                 & 18.32-0.75 & 16 000 & 200    & 4            &  20      & 4.2   & 3.6 & Y & Y & Y & Y\\  
     \hline\hline
   \end{tabular}
\caption{Summary of the asymmetric PWNe candidates observed by H.E.S.S. for which we have detected 
one or several molecular clouds with distance compatible with the associated pulsar. The name, 
mass, density and approximate distance of the clouds are summarized. The offset between the pulsar and 
center of gravity of the source in percentage of 
the SNR radius, the contrast and the scale of the gradient of matter are calculated. Finally the distance (D), 
geometric configuration (G) and gradient (Gr) compatibility are compared with the simulations of \cite{bcf01}(Y:compatible,
N:not compatible).The final conclusion of the study is given in the last column (RC : reverse shock)
(Y:compatible with the RC hypothesis, N:not compatible with the RC hypothesis).}
\end{center}
\label{FinalCO}
\end{table*}

%%%%%%%%
%  11  %
%%%%%%%%

%The paper title
\title{New Companions for the lonely Crab?\\
VHE emission from young pulsar wind nebul{\ae} revealed by H.E.S.S.
}
%Short title to print in the headers to the final publication (Not showed in this print).
\shorttitle{New Companions for the lonely Crab?}
%All paper authors
\authors{  A.~Djannati-Ata\"i$^{1}$, O.C.~de~Jager$^{2}, $
R.~Terrier$^{1}$,  Y.A.~Gallant$^{3}$ \& S.~Hoppe$^{4}$   
for the H.E.S.S. collaboration$^{5}$}
%Short title to print in the headers to the final puplication (Not showed in this print).
\shortauthors{A.~Djannati-Ata\"i et al}
%All the affiliations.
\afiliations{$^1$ APC (CNRS, Universit\'e Paris VII, CEA, Observatoire de Paris), Paris, France \\ $^2$ Unit for Space Physics, North-West University, Potchefstroom 2520, South Africa\\ $^3$ LPTA,  Universit\'e Montpellier 2, IN2P3/CNRS,, Montpellier, France \\ $^4$Max-Planck-Instituit f\"ur Kernphysik, PO box 103980, D69029 Heidelberg, Germany\\ $^5$ \rm{ \texttt{www.mpi-hd.mpg.de/HESS}}
}
\email{djannati@apc.univ-paris7.fr}

%The abstract.
\abstract{The deeper and more extended survey of the central parts of the 
Galactic Plane by H.E.S.S. during 2005-2007 has revealed a number 
of new point-like, as well as, extended sources. Two point-like sources
can be associated to two remarkable objects around ``Crab-like'' young
and energetic pulsars in our Galaxy : G21.5-0.9 and Kes~75. The
characteristics of each of the sources are presented and possible
interpretations are briefly discussed.}

\maketitle

\addcontentsline{toc}{section}{New Companions for the lonely Crab?\\
VHE emission from young pulsar wind nebul{\ae} revealed by H.E.S.S.}
\setcounter{figure}{0}
\setcounter{table}{0}
\setcounter{equation}{0}

%Begin the section.
\section*{Introduction}

%Crab first TeV source and for longtime archetype of pulsar wind nebulae
%variety of nebulae in X-rays
%HESS reveals middle-aged PWN
\begin{figure*}[!t]% [!tp]
\begin{center}
\includegraphics*[width=0.46\textwidth,angle=0,clip]{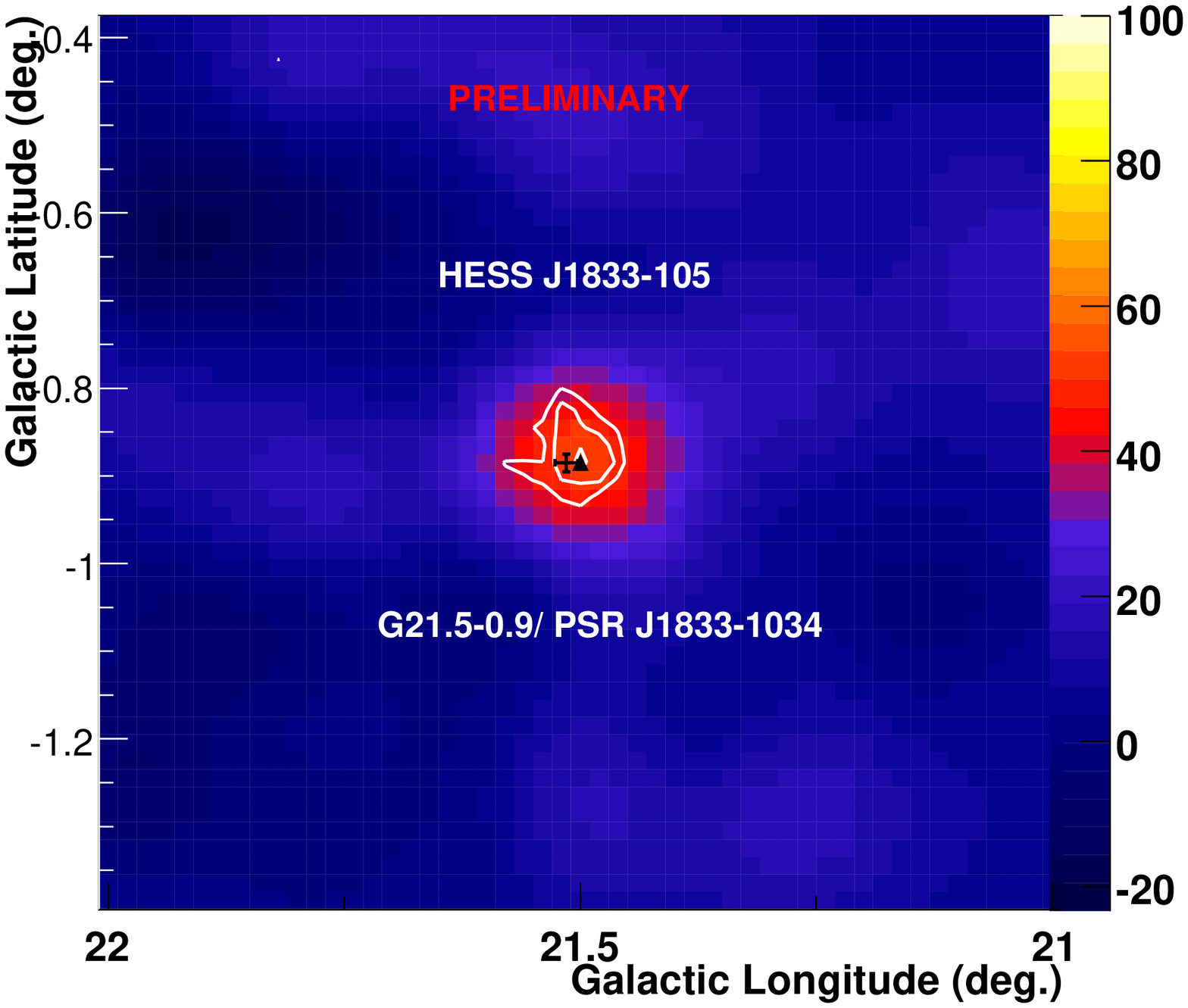}
\includegraphics*[width=0.46\textwidth,angle=0,clip]{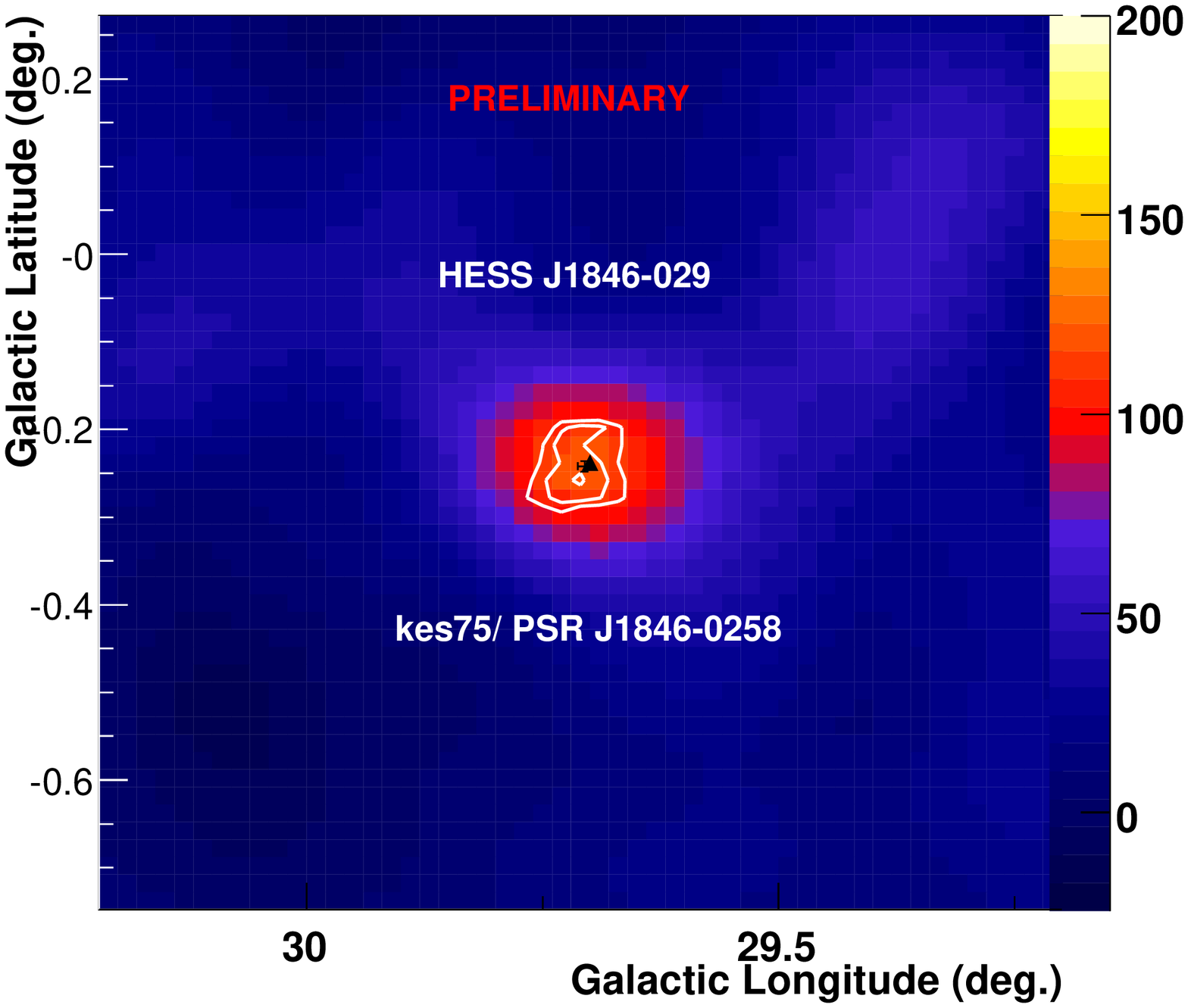}
\end{center}
\caption{ Smoothed excess maps
($\sigma=0.08^{\circ}$) of the 0.5$^{\circ}\times0.5^{\circ}$ field of
view around the positions of HESS~J1833-105 (left) and HESS~J1846-029
(right). The white contours show the pre-trials significance levels for 4, 5,
6 $\sigma$, and 7, 8, 9 $\sigma$, respectively. The black triangle
marks the position of the pulsars. The best-fit positions of the
  two sources are marked with an error cross (for HESS~J1846-029 the
latter overlaps with the triangle).}
\label{skymaps}
\end{figure*}

The standard candle of VHE astronomy, the
Crab Nebula, has served for decades as a yardstick in almost all
wavelengths, and yet it is a very peculiar object, harbouring the most
energetic and one of the youngest pulsars of our Galaxy.
Since the early days, where the similarities of the historical trio Crab/3C~58/G~21.5-0.9
were under debate~\cite{WilsonWeiler76}, radio and X-ray astronomy have
provided a wealth of information by detecting and characterizing
nebulae around rotation-powered pulsars.
In the VHE domain, H.E.S.S. has revealed
more than a dozen pulsar wind nebulae (PWN), either firmly established
as such or compelling candidates~\cite{HESSpwnICRC07}, almost all of
which are middle-aged (at least few kyrs up to $\sim$100 kyrs, except
MSH~15-52) and exhibit an offset between the pulsar position and the
nebula center.
We report here on the VHE emission discovery of two remarkable
objects, G~21.5-0.9 and Kes75, which also harbor very young and
energetic pulsars and which on some aspects, especially their
plerionic nebular emission due to an energetic pulsar, can be
considered as Crab-like.

\begin{figure*}[tb] 
\begin{center}
\includegraphics[width=0.48\textwidth]{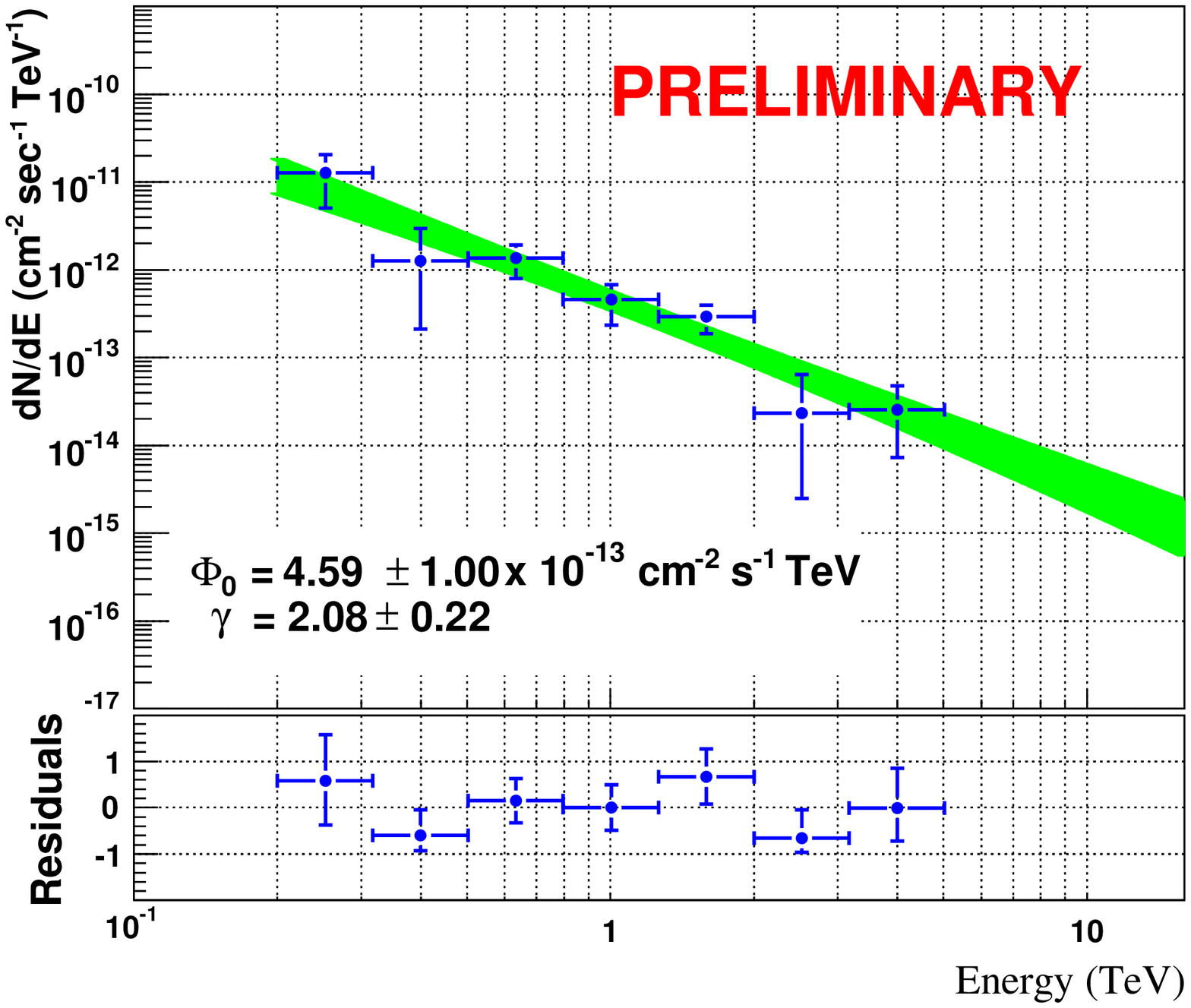}
\includegraphics[width=0.48\textwidth]{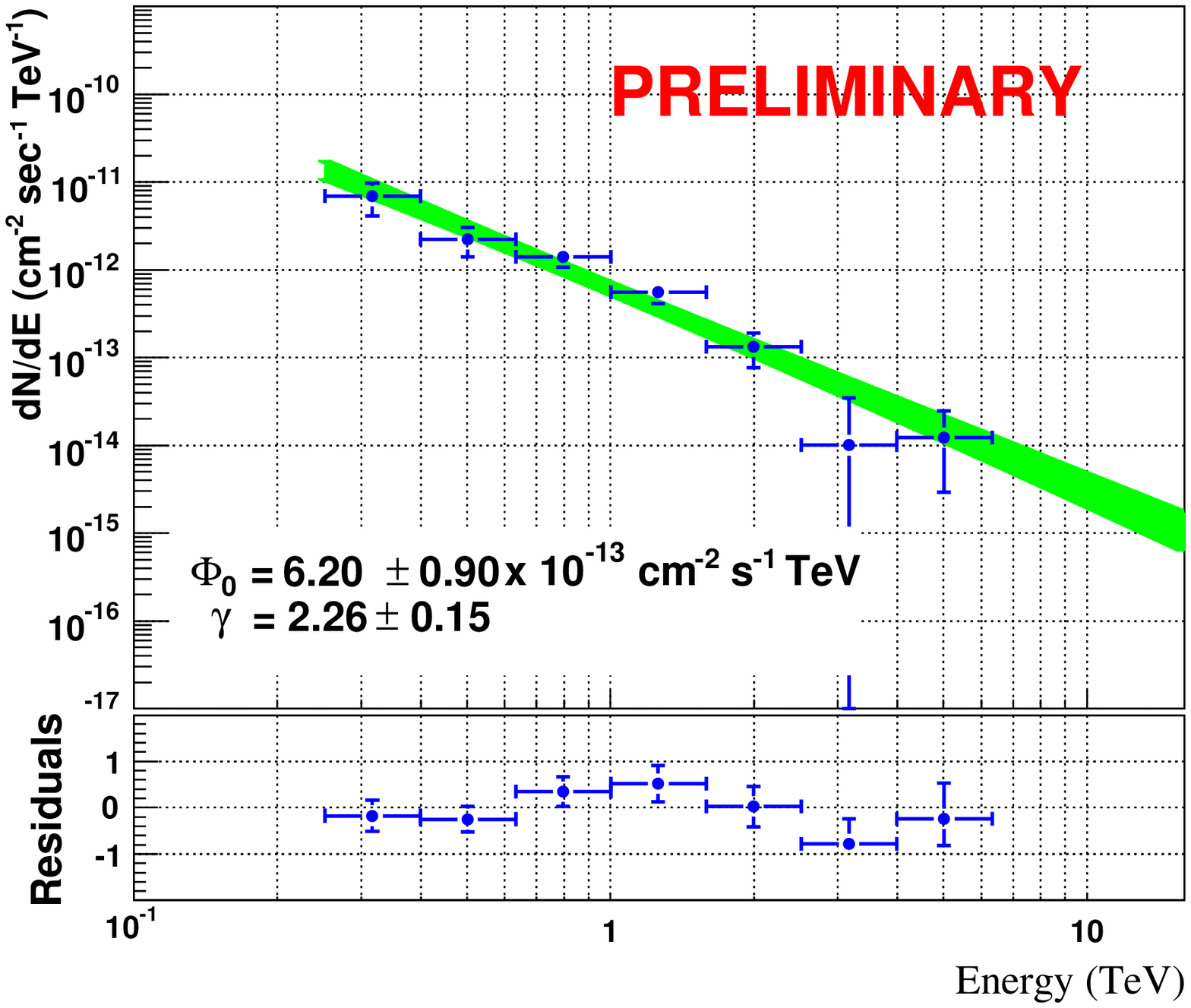}
\end{center}
\caption{Differential energy spectra above for HESS~J1833-105 (left) and HESS~J1846-029
(right). The shaded area shows the 1 $\sigma$
confidence region for the fit parameters.} 
\label{spectra} 
\end{figure*} 

G21.5-0.9 \cite{Altenhoff70}, recently revealed as a composite SNR
consisting of a centrally peaked PWN and a 4{\arcmin}
shell~\cite{Bocchino2005,MathesonSafi-Harb2005}, was 
previously classified as one
of the about ten Crab-like SNR~\cite{Green2004}. Its flat spectrum
PWN is polarised in radio~\cite{BeckerSzymkowiak1981} with a spectral
break above 500 GHz~\cite{GallantTuffs1998}. The non-thermal X-ray PWN 
with radius $\sim$40{\arcsec} shows significant evidence of cooling
~\cite{Slane2000}, with the power-law photon index steepening from
1.43$\pm$0.02 near the pulsar to 2.13$\pm$0.06 at the edge of the 
PWN. 
There appears to be a synchrotron X-ray halo at a radius of 140{\arcsec} from the
pulsar which could originate in the shell~\cite{Bocchino2005,MathesonSafi-Harb2005},
with a contribution of scattering off dust grains
as proposed by Bocchino et
al.~\cite{Bocchino2005}. 
The 61.8 ms pulsar PSR~J1833-1034, with a spin-down power
of ${\dot E} = 3.3\times 10^{37}$erg/s and a characteristic age of
4.9 kyr was discovered only recently through its faint radio pulsed
emission~\cite{Gupta2005,Camilo2006}. 
Given the derived distance of 4.7$\pm$0.4 kpc, the age of G~21.5-0.9
was revised downwards by a factor of $\sim$10 
to force consistency with the freely expanding
SNR shell~\cite{Camilo2006}.  PSR~J1833-1034 in G~21.5-0.9 is the second most energetic
pulsar known in the Galaxy.

Kes 75 (SNR G29.70.3) is also a prototypical example of a composite
remnant for which the distance of 19 kpc was estimated
through neutral hydrogen absorption measurements~\cite{BeckerHelfand1984}.
Its 3.5{\arcmin} radio shell surrounds a flat-spectrum highly
polarized radio core, 
and harbors, at its
center, the 325 ms X-ray Pulsar, PSR~J1846-258~\cite{Gotthelf2000}. 
The latter has the shortest known characteristic age $\tau_c= 723$ yr
and a large inferred magnetar-like magnetic field of B$=4.9\times
10^{13}$G. The pulsar lies within a 25{\arcsec}$\times$20{\arcsec} X-ray
nebula which exhibits an 
photon index of 1.92$\pm$0.04, but no evidence of cooling as a function
of the distance to the pulsar.
Like in G~21.5-0.9 there is an X-ray halo, in this case due mostly to dust
scattering, but a non-thermal contribution 
from electrons accelerated in the shell remains possible~\cite{Helfand2003}.

%%%%%%%%%%%%%%%%%%%%%%%%%%%%%%%%%%%%%%%%%%%%%%%%%%%%%%%%%%%%%%%%%%%%%%%%%%%%%%%%%
\section*{Observations, Analysis \& Results}
\label{results}
%%%%%%%%%%%%%%%%%%%%%%%%%%%%%%%%%%%%%%%%%%%%%%%%%%%%%%%%%%%%%%%%%%%%%%%%%%%%%%%%%

Results presented in this section should be
considered as preliminary.  

The first H.E.S.S. observations of G~21.5-0.9 and Kes~75 were
performed during 2004 and 2005 as part of the systematic survey of the
inner Galactic plane within the longitude range
$ l\in$[$-30^{\circ}$,$+30^{\circ}$] and latitude band 
$b\in$[$-3^{\circ}$,$+3^{\circ}$]. Kes~75, at the edge of the first survey,
was covered in the extension to $l\in$[$+30^{\circ}$,$+60^{\circ}$] of
the survey in the years 2005-2007. The data obtained through the systematic 
survey was completed with followup observations of promising
candidates in wobble mode, hence the two sources are offset at various angular distances with
respect to the center of the field of view. 
The total quality-selected and dead-time corrected data-set includes
19.7 hours of data on G~21.5-0.9 and 24.1 hours on Kes~75, with average offsets of
1.33$^{\circ}$ and 1.1$^{\circ}$, for each source, respectively.

The standard scheme for the reconstruction of events was applied to
the data \cite{HESSCrab}. Cuts on the scaled width and length of images (optimised 
on $\gamma$-ray simulations and off-source data) were used to suppress the
hadronic background. As described e.g. in
\cite{HESSKooka}, sky-maps and morphological analyses are made with a
tight cut on the image size of 200 p.e. (photoelectrons) to achieve a
maximum signal-to-noise ratio and a narrow PSF (point spread
function). For the spectral analysis,
the image size cut is loosened to 80 p.e. in order to cover the
maximum energy range. The background estimation for each position in the
two-dimensional sky map is made in the same way as for search of
extended sources \cite{HESSJ1908ICRC07}, i.e. computed from a ring
with a radius of $1.0^{\circ}$. For a point-like source this radius yields  
seven times a larger area for the background estimation
than the on-source region. 
The background used for the derivation of the spectrum, is evaluated
from circular regions in the field of view with the 
same radius and same offset from the pointing direction as that of the source region.
Finally, to avoid contamination of the background, events coming from
known sources were excluded.

Fig.~\ref{skymaps} shows the Gaussian-smoothed excess maps for
HESS~J1833-105 and HESS~J1846-029 where the white contours mark the
pre-trials significance levels. Both sources were first discovered as 
hot-spots within the analysis scheme described above and then
confirmed through additional followup data at pre-trials significance
of 6.4 and 9.9 standard deviations, respectively. 
A conservative estimate of the trials yields post-trials
significance of 4.0 $\sigma$ and 8.3 $\sigma$ for HESS~J1833-105 and
HESS~J1846-029, respectively.

%Position fit and agreemnt with the pulsars.
 
The extension and the position of the sources were evaluated by adjusting to the images
a symmetrical two-dimensional Gaussian
function, convolved with the instrument PSF (5{\arcmin} for this
analysis). The best-fit positions lie at 
18$^{\rm h}$33$^{\rm m}$32.5$^{\rm s}$$\pm$0.9$^{\rm s}$,$-$10d33\arcmin 19\arcsec$\pm$55\arcsec and 
18$^{\rm h}$46$^{\rm m}$24.1$^{\rm s}$$\pm$0.5$^{\rm s}$,$-$02d58\arcmin 53\arcsec$\pm$34\arcsec. 
The intrinsic extensions are compatible with a point-like source for both
sources and their positions are in a quite good agreement with the pulsars
associated to each supernova remnant, i.e. PSR~J1833-1036
(18$^{\rm h}$33$^{\rm m}$33.57$^{\rm s}$,$-$10d34\arcmin 7.5\arcsec) 
and PSR~J1846-0258 (18$^{\rm h}$46$^{\rm m}$24.5$^{\rm s}$,$-$02d58\arcmin 28\arcsec).

%spectra measurements
The energy spectra of the two sources are derived using the
forward-folding maximum likelihood 
fit of a power-law~\cite{CATSpectrum}. The fluxes are at a level of
$\sim$2\% of that of the Crab Nebula and the spectra are rather hard
(Fig.~\ref{spectra}):  the photon indices are $2.08 \pm
0.22_{\rm stat}$ and $2.26 \pm0.15_{\rm stat}$ for  HESS~J1833-105 and
HESS~J1846-029, respectively, with a systematic error of $\pm 0.1$.

\section*{Discussion}

It is remarkable that de Jager et al.~\cite{deJager1995} predicted 
that plerionic VHE $\gamma$-rays from G21.5-0.9 would be detectable
at a level of $4\times10^{-13}$ cm$^{-2}$~s$^{-1}$ at 1~TeV with an electron spectral index
of $\sim$2.8, which would give a photon index near 2.0 at VHE energies (after including KN
effects given the contributions from dust and CMBR). 
Their prediction was based on an assumed equipartition field strength
of 22 $\mu$G which is close to the value of $\sim$15$\mu$G implied
from $\gamma$-ray observations reported here (assuming IC scattering on CMB photons
only, and using the ratio of the X-ray to the $\gamma$-ray
luminosities: $L_{\rm X}/L_{\gamma}\sim30$). The equipartition field strength was afterwards
increased to 0.3 mG following the revision of the maximum spectral range of the radio PWN
to 500 GHz \cite{GallantTuffs1998, Camilo2006}. However, the detection of VHE
$\gamma$-rays by H.E.S.S. from PWN tends to confirm the suggestion of
Chevalier \cite{Chevalier2004} that some PWN may be particle
dominated, so that the true PWN field strength may be significantly
lower than equipartition for some objects. 
In the case of Kes~75, $L_{\rm X}/L_{\gamma}\sim$10
yields also a lower than equipartition nebular magnetic field
strength of $\sim$10 $\mu$G. It should be noted that Kes~75 shows the
highest conversion efficiency in X-rays ($\sim15$\%) as compared to other
``Crab-like'' pulsars ($\sim3$\% and $\sim0.6$\% for the Crab and
G~21.5-0.9, respectively) and a 100 times larger $\gamma$-ray efficiency
($\sim$2\%) than the Crab and G~21.5-0.9 which are similar in that
respect ($\sim$0.02\%).
However, the latter object's $L_{\rm X}/L_{\gamma}\sim30$ is 4 times
smaller than that of the Crab Nebula $L_{\rm X}/L_{\gamma}\sim120$. 
These numbers together with the spin
parameters and high surface magnetic field in the case of
PSR~J1846-0258, show that these objects, although ``Crab-like'' in
some aspects, do possess peculiar properities.

Given the evidence for synchrotron emission in the SNR
shell, an alternative interpretation of the VHE
emissions of G~21.5-0.9 and Kes75 would 
be radiation from particles accelerated at the non-relativistic forward
shock of the freely expanding SNR. However
the required field strength in the shell to explain the
H.E.S.S. detection in terms of IC scattering should be 
much lower than 10 $\mu$G, value which may be unreasonably low for
typical expanding SNR shells. 
Deeper observations of both sources
could help to constrain the size of the VHE emission region and to 
ascertain whether it is compatible with this scenario.

\section**{Acknowledgments}
The support of the Namibia authorities and of the University of Namibia
in facilitating the construction and operation of H.E.S.S. is gratefully
acknowledged, as is the support by the German Ministry for Education and
Research (BMBF), the Max Planck Society, the French Ministry for Research,
the CNRS-IN2P3 and the Astroparticle Interdisciplinary Programme of the
CNRS, the U.K. Particle Physics and Astronomy Research Council (PPARC),
the IPNP of the Charles University, the Polish Ministry of Science and
Higher Education, the South African Department of
Science and Technology and National Research Foundation, and by the
University of Namibia. We appreciate the excellent work of the technical
support staff in Berlin, Durham, Hamburg, Heidelberg, Palaiseau, Paris,
Saclay, and in Namibia in the construction and operation of the
equipment.

%This in the bibtex style, is ok.
\bibliographystyle{plain}

%%%%%%%%
%  12  %
%%%%%%%%

%The paper title
\title{Search for Pulsed VHE Gamma-Ray Emission from Young Pulsars with H.E.S.S.}
%Short title to print in the headers to the final publication (Not showed in this print).
\shorttitle{Pulsed Emission with H.E.S.S.}
%All paper authors
\authors{M. F\"ussling$^{1}$, S. Schlenker$^{2}$ and C. Venter$^{3}$ for the H.E.S.S.\ Collaboration,
T. Eifert$^{4}$, R. Manchester$^{5}$ and F. Schmidt$^{6}$}
%Short title to print in the headers to the final puplication (Not showed in this print).
\shortauthors{M. Fuessling and et al}
%All the affiliations.
\afiliations{$^1$Institut f\"ur Physik, Humboldt-Universit\"at zu Berlin, Newtonstr.~~15, D~~12489 Berlin, Germany\\ 
$^2$ CERN PH Department, CH-1211 Geneva 23, Switzerland\\
$^3$ Unit for Space Physics, North-West University, Potchefstroom 2520, South Africa\\
$^4$ Universit\'e de Gen\`eve, Section de physique, 24 Quai Ernest-Ansermet, CH-1211 Geneva 4, Switzerland\\
$^5$ Australia Telescope National Facility, CSIRO, PO Box 76, Epping NSW~~1710, Australia\\
$^6$ Department of Astronomy and Astrophysics, The University of Chicago, IL~~60637, USA}
\email{fuessling@physik.hu-berlin.de}

%The abstract.
\abstract{We present the results of a search for pulsed very-high-energy (VHE)
\g-ray emission from young pulsars using data taken with the H.E.S.S.
imaging atmospheric Cherenkov telescope system. Data on eleven pulsars, selected according to their spin-down
luminosity relative to distance, are searched for \g-ray signals with
periodicity at the respective pulsar spin period. Special analysis
efforts were made to improve the sensitivity in the 100\,GeV \g-ray
energy domain in an attempt to reduce the gap between satellite and
ground-based \g-ray instruments. No significant evidence for pulsed emission is found in any data set.
Differential upper limits on pulsed energy flux are determined for all
selected pulsars in the approximate \g-ray energy range between 100\,GeV and
$50$\,TeV, using different limit determination methods, testing a
wide range of possible pulsar light curves and energy spectra. The limits derived here imply that the magnetospheric VHE \g-ray production efficiency in young pulsars is less than $10^{-4}$ of the
pulsar spin-down luminosity, requiring spectral turnovers for the
high-energy emission of four established \g-ray pulsars, and constrain
the inverse Compton radiation component predicted by several outer gap
models.}

\maketitle

\addcontentsline{toc}{section}{Search for Pulsed VHE Gamma-Ray Emission from Young Pulsars with H.E.S.S.}
\setcounter{figure}{0}
\setcounter{table}{0}
\setcounter{equation}{0}

%Begin the section.

\begin{table*}[t]
\caption{\label{table_candidates}
{\footnotesize
The characteristics of the selected pulsars taken from
\cite{atnfcat}. Period, $P$, distance, $D$,
% number of observedglitches, $N_{\rm GLT}$, 
spin-down age, spin-down luminosity, $\dot{E}$, and the corresponding
value for $\dot{E}/D^2$, and calculated magnetic field strength at the
neutron star surface, $B_{\rm surf}$, and the light cylinder, $B_{\rm
LC}$, are listed. The last column shows the rank in $\dot{E}/D^2$
within the ATNF catalogue.}
}
\centering
\resizebox{\hsize}{!}{
\begin{tabular}{ccccccccccc}
\hline\hline
\multicolumn{2}{c}{Pulsar name} & $P$ & $D$ & Age & \multirow{2}*{$\displaystyle\log_{10}\left(\frac{\dot{E}}{\rm erg\,s^{-1}}\right)$} & \multirow{2}*{$\displaystyle\log_{10}\left(\frac{\dot{E}/{\rm erg\,s^{-1}}}{D^2/{\rm kpc^2}}\right)$} & $B_{\rm surf}$ & $B_{\rm LC}$ & Rank\\
\multicolumn{2}{c}{PSR} & $[\rm ms]$ & $[\rm kpc]$ & $[\rm kyears]$ & & & $[\rm 10^{11}\,G]$ & $[\rm 10^4\,G]$ & $\dot{E}/D^2$\\
\hline
B0531+21$^\star$   & J0534+2200   & 33.1 & 2 & 1.24 & 38.7 &
38.1 & 37.8 & 98.0 & 1\\
B0833$-$45$^\star$ & J0835$-$4510 & 89.3 & 0.29 & 11.3 & 36.8 &
37.9 & 33.8 & 4.45 & 2\\
B1706$-$44$^\star$ & J1709$-$4429 & 102  & 1.8 & 17.5 & 36.5 &
36.0 & 31.2 & 2.72 & 6\\
B1509$-$58$^\star$ & J1513$-$5908 & 151  & 4.4 & 1.55 & 37.3 &
36.0 & 154  & 4.22 & 7\\
& J1747$-$2958 & 98.8 & 2.5 & 25.5 & 36.4 &
35.6 & 24.9 & 2.42 & 13\\
B1259$-$63& J1302$-$6350 & 47.8 & 1.5 & 332  & 35.9 &
35.5 &  3.3 & 2.87 & 15\\
& J1811$-$1925 & 64.7 & 5  & 23.3 & 36.8 &
35.4 & 17.1 & 5.92 & 18\\
& J1524$-$5625 & 78.2 & 3.8 & 31.8 & 36.5 &
35.3 & 17.7 & 3.46 & 19\\
& J1420$-$6048 & 68.2 & 7.7 & 13   & 37.0 &
35.3 & 24.1 & 7.13 & 22\\
& J1826$-$1334 & 101  & 4.1 & 21.4 & 36.4 &
35.2 & 27.9 & 2.51 & 23\\
& J1801$-$2451 & 125  & 4.6 & 15.5 & 36.4 &
35.1 & 40.4 & 1.95 & 30\\
\hline
\multicolumn{7}{l}
{
\begin{minipage}[t]{9.cm} 
$^\star$ established as \g-ray pulsars below $\sim$\,$10$\,GeV by EGRET\\
\end{minipage}
}\\
\end{tabular}
}
\end{table*}

\section*{Introduction}
Rotating neutron stars are known to convert a significant part of
their rotational energy into radiation that originates from within the
magnetosphere. This emission is observable as a periodic signal at the
neutron star rotation frequency (the \emph{pulsar} phenomenon). 
For many of the known young and energetic pulsars, the emitted luminosity
peaks at X-ray or \g-ray energies \cite{egret_limits} and is usually attributed to curvature radiation of
accelerated electrons in the strong magnetic fields pervading the
pulsar magnetosphere. The luminosity of the pulsed high-energy
emission was found to correlate significantly with the energy loss
rate of the pulsar, i.e.\ its spin-down power $\dot{E}$
\cite{gamma_energetics, Cheng}. For most of the pulsars with established \g-ray emission
\cite{egret_crabvela%,1509_comptel
}, there is evidence for a turnover
in the pulsed spectrum at a critical energy $E_{\rm c}$ in the sub-GeV
to 10\,GeV range.

The two most commonly discussed scenarios for magnetospheric \g-ray emission place the emission regions either near the magnetic
poles of the neutron star ({\em polar cap} model), or near the null surface in the outer
magnetosphere of the pulsar ({\em outer gap} model)
%resulting in different predicted light-curves,
%spectral shapes, and cutoff energies $E_{\rm c}$ of the emitted high-energy radiation
.
Both models predict a cutoff in the curvature radiation spectrum at
\g-ray energies of the order of GeV up to several tens of
GeV. Additionally, in some outer gap model calculations, a spectral
component in the TeV range due to inverse Compton (IC) up-scattering
of soft ambient seed photons by the accelerated electrons is predicted
\cite{og:2,og2d}.

The prime candidates for the search for very-high-energy (VHE,
energies above $\sim$\,$100$\,GeV) \g-ray emission are the
pulsars with established \g-ray emission at energies below
$\sim$\,$10$\,GeV which have been detected by CGRO instruments.
Some of them have been subject to intensive searches for pulsed VHE \g-ray emission by ground-based
instruments. Up to now, no evidence for pulsed emission has been found
in these observations, and upper limits on the pulsed VHE \g-ray
flux have been derived under various assumptions on the characteristics of
the pulsed emission. However, the IC component predicted by outer gap
models has not yet been significantly constrained.

\section*{H.E.S.S. Observations and Analysis}
The High Energy Stereoscopic System (H.E.S.S.) \cite{status}, an
array of imaging atmospheric Cherenkov telescopes located in the Khomas Highland of Namibia, detects cosmic VHE \g-rays by imaging the Cherenkov
emission of their air showers in the atmosphere using optical
telescopes. The superior sensitivity of H.E.S.S.\ with respect to previous ground-based instruments puts
the predicted pulsed IC component from outer gap models within reach of testability. 

Apart from the known \g-ray pulsars, other candidates for which
H.E.S.S.\ data were available were selected from the ATNF pulsar
catalogue \cite{atnfcat} if their spin-down flux $\dot{E}/D^2$ was greater than $10^{35}\,\rm
erg\,s^{-1}\,kpc^{-2}$. Table~\ref{table_candidates} lists all candidates chosen along with
selected measured and derived characteristics collected from the
literature. 

The data used in this search for pulsed VHE emission were either obtained in pointed observations
or accumulated in the Galactic Plane survey and were analysed using the standard method as described in detail in \cite{CrabMF}. 
Since observational data indicate steep cut-offs in high-energy (GeV) \g-rays, special {\em low energy} cuts have 
been applied in addition to the standard cuts to reduce the gap in observational coverage between satellite and ground-based
\g-ray observations of young pulsars.

\section*{Search for Pulsed Emission}
In order to test for pulsed \g-ray emission at the pulsar position,
the timestamps of each recorded shower passing selection cuts were
transformed from the observer's frame into the pulsar frame and then
folded with the pulsar spin period taken from observations in other energy domains. 
The resulting distribution of pulsar phases corresponding to each shower event was
tested for variability using several statistical tests ($\chi^2$-, $Z^2_m$-, $H$- and Kuiper-test).

\begin{figure}[t]
\centering
\resizebox{\hsize}{!}{\includegraphics{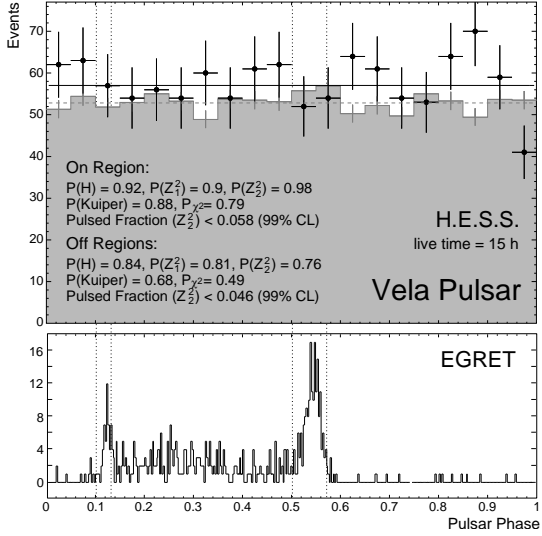}}
\caption{{\footnotesize
{\em Upper plot}: Distribution of event phases for the Vela pulsar
(PSR\,B0833$-$45). The points represent the events in the on-region
at the pulsar position and the histogram the normalised off-region
events. For both regions
the probabilities for being consistent with a uniform distribution
according to the statistical tests on pulsations are listed. No significant deviation from uniformity was found
within any of the statistical tests for pulsations. {\em Lower plot}: Phase distribution for
\g-rays with energies between 2 and 10\,GeV as measured by EGRET
\cite{egret_crabvela}.}}
\label{fig_phaso}
\end{figure}

As an example, the distribution of event phases from observations of
the Vela Pulsar (PSR\,B0833$-$45) is shown in Fig.~\ref{fig_phaso} for the signal ({\em on})
and background ({\em off}) region, obtained using the standard cuts. The difference between on and off
results from the known \g-ray excess from HESS\,J0835$-$456 at the
position of the pulsar.

\begin{figure}[t]
\centering
\resizebox{0.95\hsize}{!}{\includegraphics{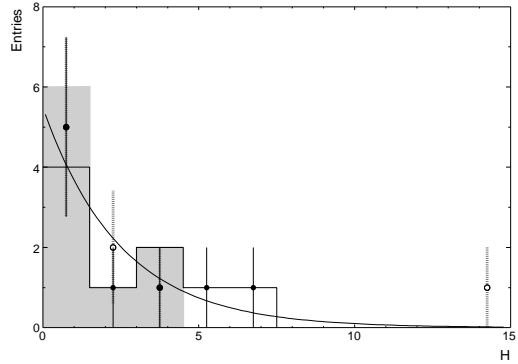}}
\caption{{\footnotesize 
Ensemble distributions of the H-test statistic for the selected
pulsars and their corresponding background control samples. 
The results for the on-region are shown as open and closed circles
for the low energy and standard cut analysis, respectively.
The distributions for the off-regions are displayed as grey filled
and outlined histograms, respectively. The
solid curve shows the expected distribution $N_{\rm H}(H) = N_0 \exp(-\lambda H)\mid_{\lambda = 0.4}$
if no pulsed signals are present.}}
\label{fig_hdist}
\end{figure}

No conclusive evidence for pulsed emission has been found with any
of the statistical tests for pulsations for any data set of the complete sample of pulsars. 
The distribution of the test statistic $H$ of the H-test is compatible with the
expected distribution of $H$ for the case when no pulsed signal is present in any
data set, see Fig.~\ref{fig_hdist}.

\begin{figure}[t]
  \begin{center}
    \begin{minipage}[t]{0.49\textwidth}
      \resizebox{\hsize}{!}{\includegraphics{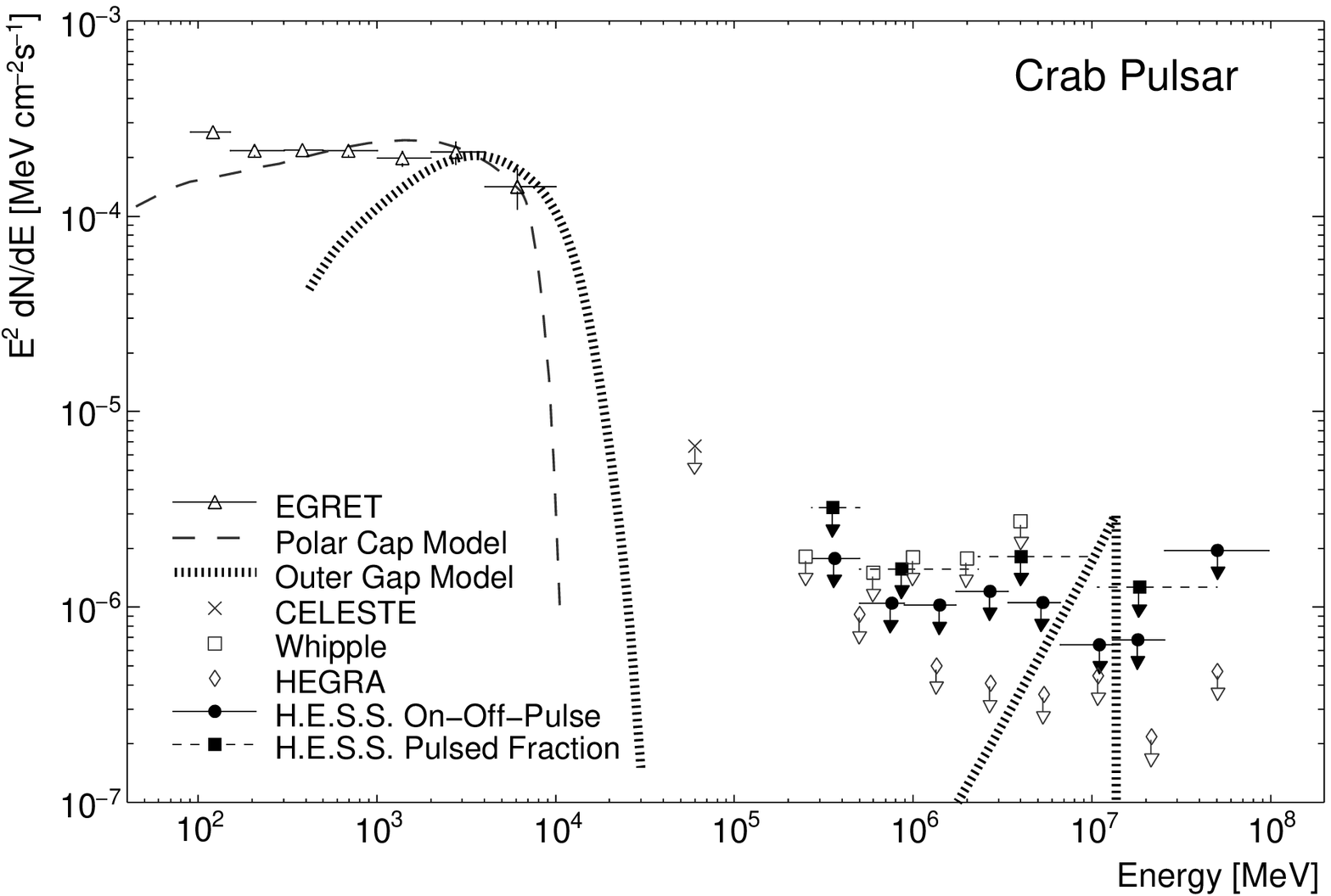}}
    \end{minipage}
    \caption{{\footnotesize 
      H.E.S.S.\ energy flux limits (99\% c.l.) for pulsed emission of the
Crab pulsar. The full circles and full squares correspond to the
on-off-pulse and pulsed fraction limit determination methods,
respectively. Below energies of 0.5\,TeV the results were obtained
with the low energy selection cuts, otherwise the standard cuts were
used. The indicated polar cap curve was generated
according to \cite{pc:3} and the outer gap model curve taken from
\cite{og_crab}. Note that the southern location of H.E.S.S.\
allows only observations at rather high zenith angles for Crab, prohibiting a
deep exposure especially at low energy thresholds.}}
    \label{fig_Crab_spectrum}
  \end{center}
\end{figure}

\begin{figure}[t]
  \begin{center}
    \begin{minipage}[t]{0.49\textwidth}
      \resizebox{\hsize}{!}{\includegraphics{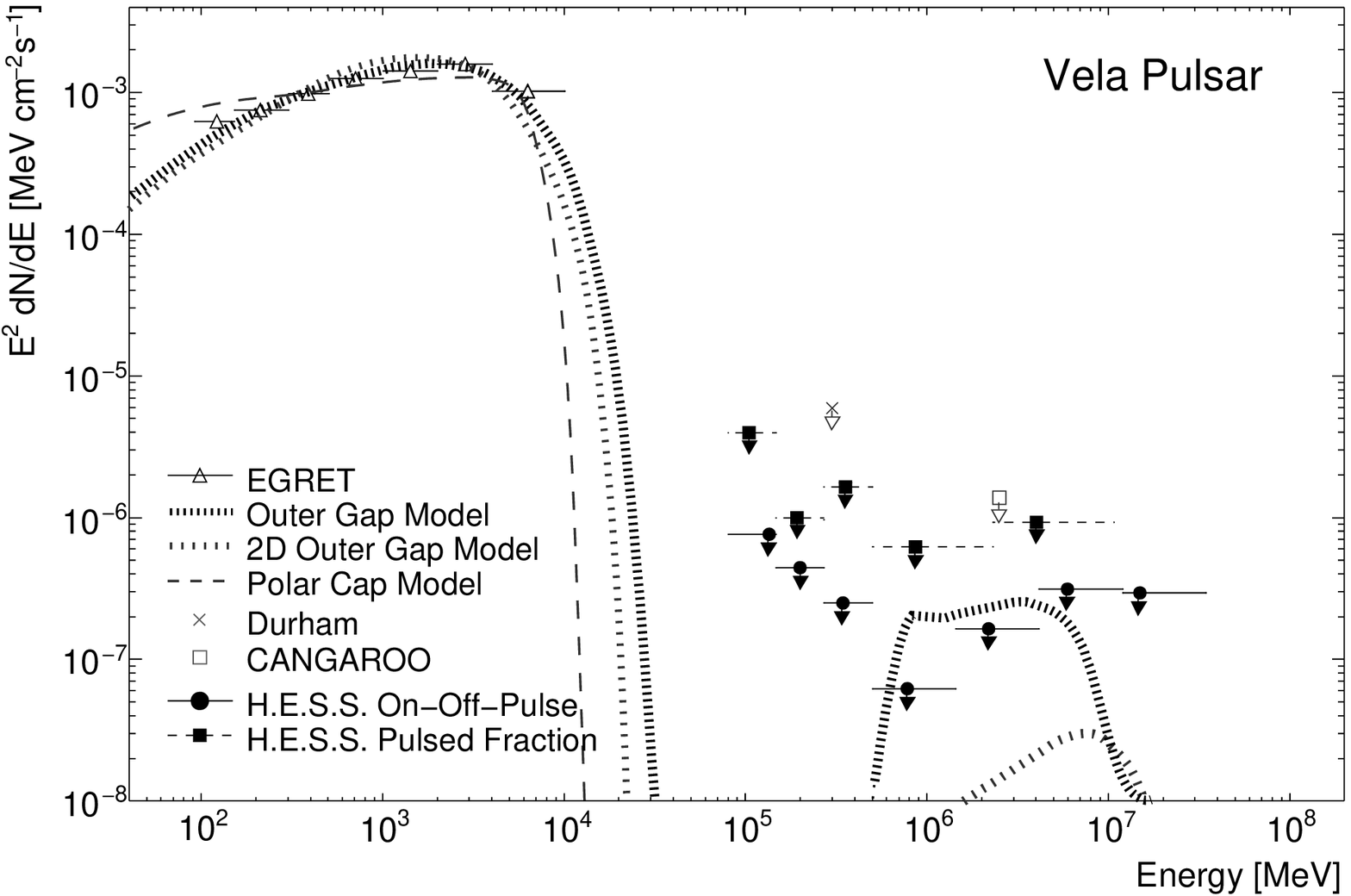}}
    \end{minipage}
    \caption{{\footnotesize 
      H.E.S.S.\ energy flux limits (99\% c.l.) for pulsed emission of the
Vela pulsar (see Fig.~\ref{fig_Crab_spectrum} for point
descriptions). The polar cap (black dotted) and outer gap (dashed) model curves are
generated according to the model of \cite{pc:3} and \cite{og:2}, respectively. The
dotted grey outer gap model curve is taken from \cite{og2d}.}}
    \label{fig_Vela_spectrum}
  \end{center}
\end{figure}

Several methods were applied to obtain upper limits on the \g-ray flux from
the selected pulsars. They differ in the assumptions made concerning
the characteristics of the pulsed emission. The {\em on-off-pulse} 
method assumes similar characteristics of the pulse position and shape as
measured in other energy domains for the hypothetical VHE \g-ray emission whereas 
the {\em pulsed fraction} method only assumes a certain pulse shape.
As an example, the calculated differential flux limits are shown for Crab and Vela, two of the four observed \g-ray pulsars,
in Fig.~\ref{fig_Crab_spectrum} and~\ref{fig_Vela_spectrum}, respectively. 
For more details on the analysis and the limits derived for all selected pulsars refer to \cite{pulsed}. 

For the complete sample of pulsars, the absence of pulsed VHE \g-ray emission
already disfavours a significant contribution of the IC component to the energy loss mechanism
of these pulsars. The flux limits shown here for Crab and Vela significantly constrain the IC component
of selected outer gap models for flux predictions in the TeV range.

\section*{Conclusions}

No conclusive evidence for pulsed emission has been found and differential upper
limits on the pulsed flux were derived, constraining the pulsed flux for a wide range of
possible pulse shapes and spectra in the VHE \g-ray range. 

Although in several cases there is spatial coincidence with extended
TeV \g-ray emission, pulsed emission is not detected in VHE \g-rays.
In particular, the flux upper limits derived are of the order of
$10^{-4}$ to $10^{-6}$ of the pulsar spin-down flux, underlining the
non-magnetospheric origin of the TeV radiation component and
supporting the widely accepted scenario of an effective energy
transport mechanism to, and strong particle acceleration in, the
pulsar wind nebula.

The upper limits imply a steep turnover of the 
pulsed high-energy spectrum at energies of a few tens of GeV.
As the pulsar models differ significantly in their predictions of the exact shape
and energy of the turnover, the search for pulsed \g-ray
emission from pulsars provides interesting prospects for future
satellite-based and ground-based \g-ray instruments.

\section*{Acknowledgements}
{\footnotesize The support of the Namibian authorities and of the University of
Namibia in facilitating the construction and operation of H.E.S.S.\ is
gratefully acknowledged, as is the support by the German Ministry for
Education and Research (BMBF), the Max Planck Society, the French
Ministry for Research, the CNRS-IN2P3 and the Astroparticle
Interdisciplinary Programme of the CNRS, the U.K. Particle Physics and
Astronomy Research Council (PPARC), the IPNP of the Charles
University, the South African Department of Science and Technology and
National Research Foundation, and by the University of Namibia. We
appreciate the excellent work of the technical support staff in
Berlin, Durham, Hamburg, Heidelberg, Palaiseau, Paris, Saclay, and in
Namibia in the construction and operation of the equipment.}

%This in the bibtex style, is ok.
\bibliographystyle{plain}
%This is the reference to .bib file (Whitout .bib!)

%%%%%%%%
%  13  %
%%%%%%%%

%The paper title
\title{Observation of orbital modulation of the VHE emission from the binary system LS~5039 with H.E.S.S.}
%Short title to print in the headers to the final publication (Not showed in this print).
\shorttitle{Observation of LS~5039}

%All paper authors
\authors{Mathieu de Naurois$^{1}$, Gavin Rowell$^{2}$ for the H.E.S.S. Collaboration }
%Short title to print in the headers to the final publication (Not shown in this print).
\shortauthors{de Naurois and et al.}
%All the affiliations.
\afiliations{$^1$LPNHE IN2P3 - CNRS - UniversitÃ©s Paris VI et Paris VII, France\\ $^2$School of Chemistry \& Physics, University of Adelaide, Adelaide 5005, Australia }
\email{denauroi@in2p3.fr}

%The abstract.
\abstract{The binary system LS 5039 was serendipitously discovered with the High Energy
Stereoscopic system (H.E.S.S.)
during the scan of the inner galactic plane in 2004. Deeper observations
were carried out in 2005, and brought clear evidence for TeV emission
perodicity. This is the highest energy periodic source known so far.
The observed flux modulation is attributed to a modulated absorption
of the VHE  gamma-ray emission of the compact object through pair
creation on the stellar photosphere.
Spectral modulation is also observed in this system; this might have
several origins such as modulation of particle acceleration or reprocessing
of high energy photons towards lower energy through cascading.

We will present detailed studies of the source variability (flux and
spectral shape), the timescales compared to other wavelengths, and
briefly review the implications for existing
emission models.}

\maketitle

%%% Begin Binaries %%%%%
\addtocontents{toc}{\protect\contentsline {part}{\protect\large Binaries}{}}
\addcontentsline{toc}{section}{Observation of orbital modulation of the VHE emission from the binary system LS~5039 with H.E.S.S.}
\setcounter{figure}{0}
\setcounter{table}{0}
\setcounter{equation}{0}

%Begin the section.
\section*{Introduction}

\begin{figure}
\begin{center}
\noindent
%\fbox{\hbox{\vbox{\hsize=50mm \hfill \vspace{50mm}}}}
\includegraphics [width=0.45\textwidth]{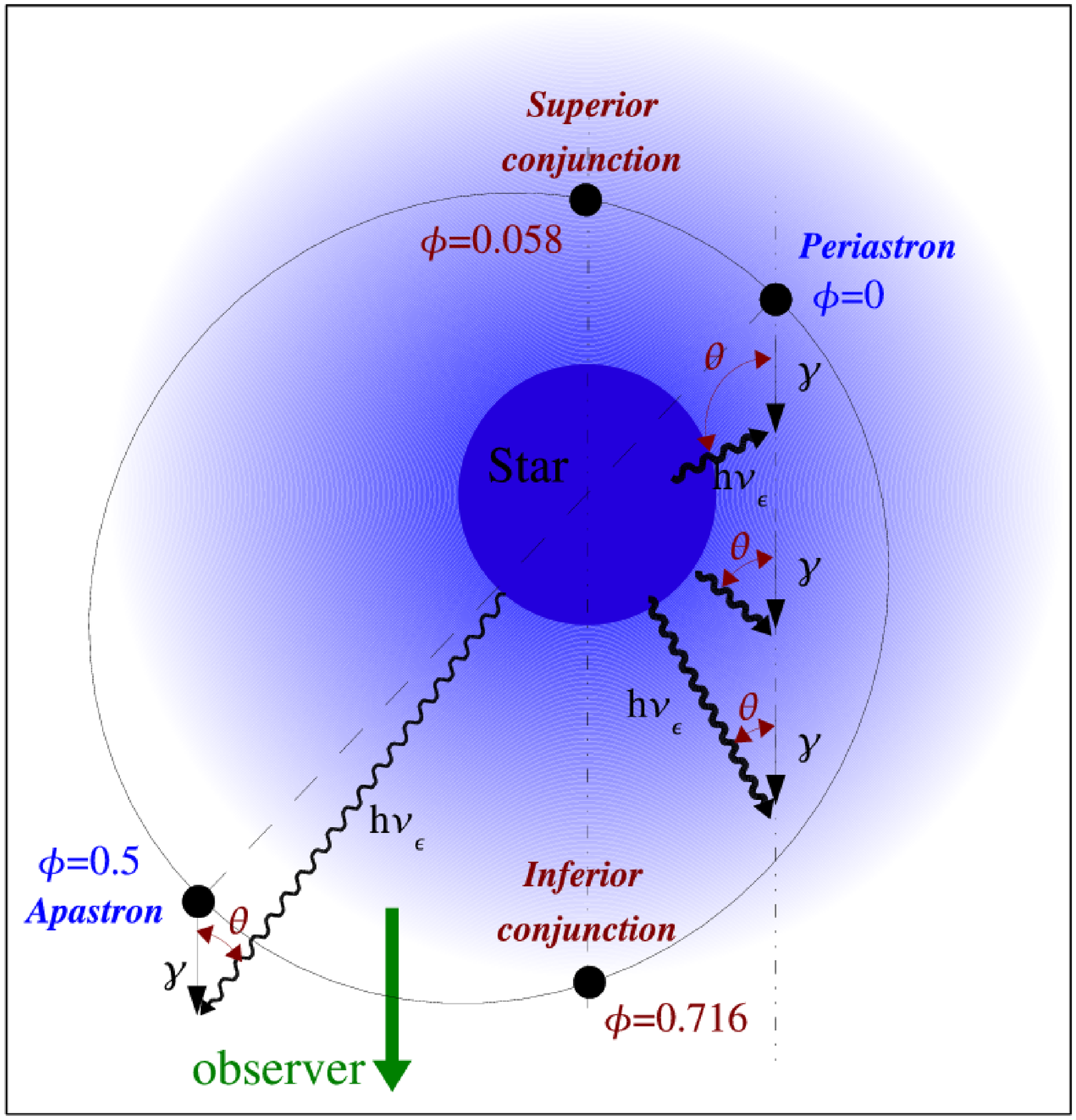}
\end{center}
\vspace{-1.5em}
\caption{Orbital geometry of the binary system LS~5039 viewed from above and using the orbital
parameters derived by Casares et al.\cite{Casares:2005}. Shown are:
phases ($\phi$) of minimum ({\it periastron}) and maximum ({\it apastron}) binary separation;
epoch of superior and inferior conjunctions occurring when the compact object and the star are
aligned along the observer's light-of-sight.}\label{fig:Cartoon}
\end{figure}

In the commonly accepted paradigm, microquasars consist of a compact object (black hole or neutron star) fed by a massive star. 
They can exhibit superluminal radio jets\cite{Mirabel:Nature94}, and emission from the accretion disk.
LS~5039, identified in 1997\cite{Motch:1997} as a massive X-ray binary system with faint radio emission\cite{Marti:1997},
was resolved by Paredes et al.\cite{Paredes:2000} into a bipolar radio outflow emanating from a central core,
thus possibly placing it into the {\it microquasar class}. The detection of radio and variable X-ray 
emission\cite{Bosch:2005} and its possible association with the EGRET source 3EG~J1824-1514 suggested 
the presence of multi-GeV particles accelerated in jets. This binary system (Fig \ref{fig:Cartoon}) consists of 
a massive O6.5V star in a $\sim 3.9$ day mildly  eccentric orbit ($e = 0.35$)\cite{Casares:2005} around a compact object 
whose exact nature (black hole or neutron star) is still under debate.

\section*{H.E.S.S. Observations}

The High Energy Stereoscopic System (H.E.S.S.) is an array of four 
Atmospheric Cherenkov Telescopes (ACT)\cite{Aharonian:HESS_Crab}
located in the Southern Hemisphere (Namibia, 1800 m a.s.l.) and sensitive to
$\gamma$ rays above 100~GeV. 
LS~5039 was serendipitously detected in 2004 during the H.E.S.S. galactic scan\cite{Aharonian:HESS_LS5039}
The 2004 observations have been followed up by a deeper observation campaign\cite{Aharonian:HESS_LS5039:2}
in 2005, leading to a total dataset of 69.2 hours of observation after data quality 
selection. 
%Data were analysed using two separate calibrations\cite{Aharonian:HESS_Calib}
%and analysis pipelines. 
%The results presented here are based on the log-likelihood comparison of the
%shower images with a precalculated semi-analytical model\cite{deNaurois:Model}.

After selection cuts, a total of 1969 $\gamma$-ray events were found within $0.1^\circ$ of the VLBA radio position
of LS~5039 (statistical significance of $40\sigma$). %% (Fig. \ref{fig:LSSkyMap}).
The best fit position, $l=16.879^\circ$, $b=-1.285^\circ$ is compatible with the VLBA position
within uncertainties $\pm 12''$ (stat.) and $\pm 20''$ (syst.).
We obtain an upper limit of $28''$ (at $1\sigma$) on the source size.

\subsection*{Timing Analysis}

\begin{figure}[b!]
\begin{center}
\noindent
%\fbox{\hbox{\vbox{\hsize=50mm \hfill \vspace{50mm}}}}
\includegraphics [width=0.45\textwidth]{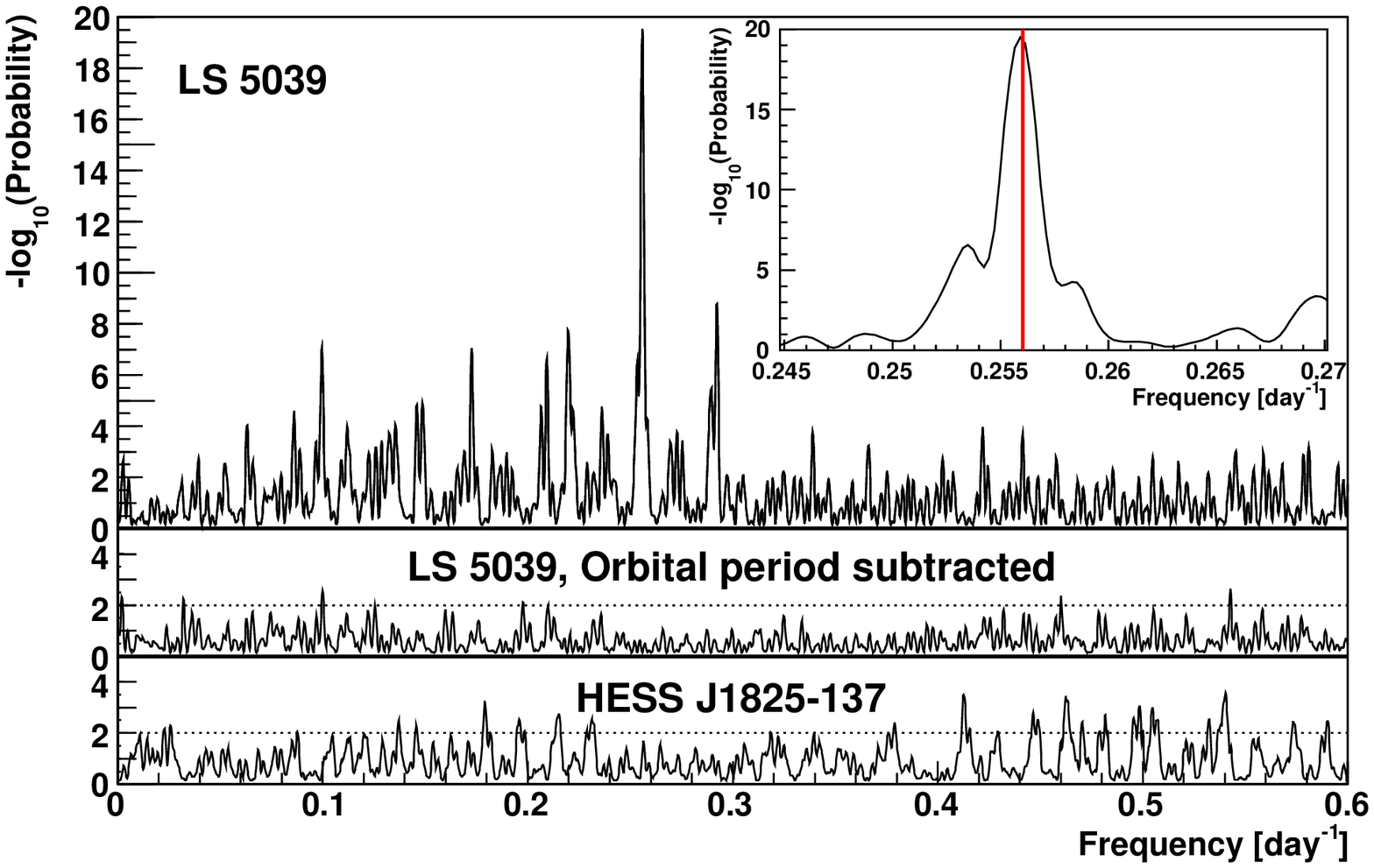}
\end{center}
\vspace{-1.5em}
\caption{Lomb-Scargle (LS) periodogram of the VHE runwise flux of LS~5039 above 1 TeV (Chance probability
to obtain the LS power vs. frequency). From \cite{Aharonian:HESS_LS5039:2}. Zoom: inset around
the highest peak, which corresponds to a period of $3.9078\pm 0.0015$ days, compatible with the optical orbital period\cite{Casares:2005} 
denoted as a red line. Middle: LS periodogram of the same data after subtraction of a pure 
sinusoidal component (see text).
Bottom: LS periodogram obtained on HESS~J1825-137 observed in the same field of view.}\label{fig:Lomb}
\end{figure}

%The high collected photon statistics allows for accurate timing analysis, 
%which was not possible for the 2004 data alone. 
The runwise VHE $\gamma$-ray flux at energies $\geq 1\ \mathrm{TeV}$ was decomposed into its frequency components
using the Lomb-Scargle periodogram\cite{Scargle:1982} (Fig. \ref{fig:Lomb}).
% which is appropriate for
% unevenly sampled datasets such as those collected by H.E.S.S.
A very significant peak (chance probability of $\sim 10^{-20}$ before trials) occurs in the Lomb-Scargle periodogram  
at the period $3.9078\pm 0.0015$ days, consistent with the most recent optical orbital period\cite{Casares:2005}
($3.90603 \pm 0.00017$ days). 
%The effect of subtracting a pure sinusoid at the orbital 
%period is shown in Fig. \ref{fig:Lomb}, middle panel. 
After substraction of a pure sinusoid at the orbital 
period, the orbital peak disappears as expected (Fig. \ref{fig:Lomb}, middle), but also the numerous satellite 
peaks % present in the original periodogram % with chance probabilities less than $10^{-7}$-$10^{-8}$ 
% that were present in the original periodogram
which correspond to beat periods of the orbital period with observation gaps (day-night cycle, moon period, annual period).
%These peaks are beat periods of the orbital period with the various gaps present in the H.E.S.S. dataset
%(day-night cycle, moon period, annual period). 
The bottom panel of the same figure shows the result obtained
on the neighbouring source HESS J1825-137 observed in the same field of view as LS~5039. The absence of any 
significant peak demonstrates that the observed periodicity is genuinely associated with LS~5039.

\subsection*{Flux Modulation}

\begin{figure}[b!]
\begin{center}
\noindent
%\fbox{\hbox{\vbox{\hsize=50mm \hfill \vspace{50mm}}}}
\includegraphics [width=0.45\textwidth]{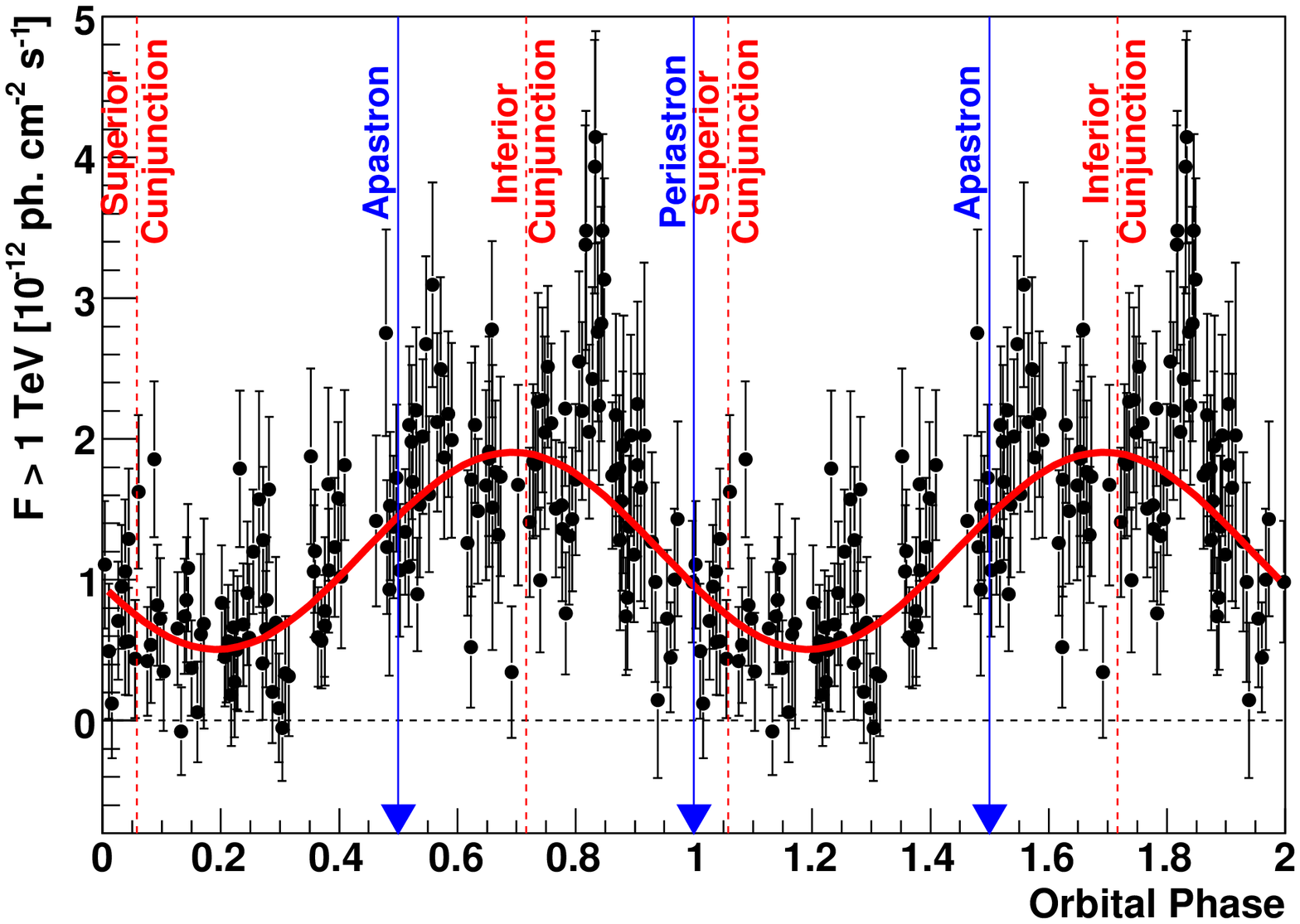}
\end{center}
\vspace{-1.5em}
\caption{Phasogram (Integral run-by-run $\gamma$-ray flux above 1 TeV as function of orbital phase) of LS~5039 
from H.E.S.S. data from 2004 to 2005. %, using the orbital ephemeris\cite{Casares:2005}. 
Each run is $\sim 28$ minutes.
Two full phase periods are shown for clarity. The vertical blue arrows depict the respective phases of minimum ({\it periastron}) 
and maximum ({\it apastron}) binary separation. The vertical dashed red lines show the respective phases of inferior
and superior conjunction, when the star and the compact object are aligned along the observer's line of sight.
From \cite{Aharonian:HESS_LS5039:2}.}\label{fig:Lightcurve}
\end{figure}

The runwise Phasogram (Fig \ref{fig:Lightcurve}) of integral flux at energies $\geq 1\ \mathrm{TeV}$
vs. orbital phase ($\phi$) shows an almost sinusoidal behaviour, with the bulk of the
emission largely confined in a phase interval $\phi$ $\sim 0.45$ to $0.9$, covering about half
of the orbital period.
%The thick red line in Fig \ref{fig:Lightcurve} represents the component at the orbital frequency
%determined with the Lomb-Scargle coefficients. 
The emission maximum ($\phi\sim 0.7$) appear to lag behind  the apastron epoch and to align better with the {\it inferior conjunction} ($\phi=0.716$), 
when the compact object lies in front of the massive star (see Fig. \ref{fig:Cartoon}). 
The VHE flux minimum occurs at phase ($\phi \sim 0.2$), slightly further
along the orbit than {\it superior conjunction} ($\phi=0.058$).
Neither evidence for long-term secular variations nor
any other modulation period are found in the presented H.E.S.S. data.

\subsection*{Spectral Modulation}

Due to the changing environment with orbital phase (magnetic field strength, stellar photon field, relative position
of compact object and star with respect to observer, \dots), the VHE $\gamma$-ray emission spectrum is expected 
to vary along the orbit. 

\begin{figure}[h!]
\begin{center}
\noindent
%\fbox{\hbox{\vbox{\hsize=50mm \hfill \vspace{50mm}}}}
\includegraphics [width=0.45\textwidth]{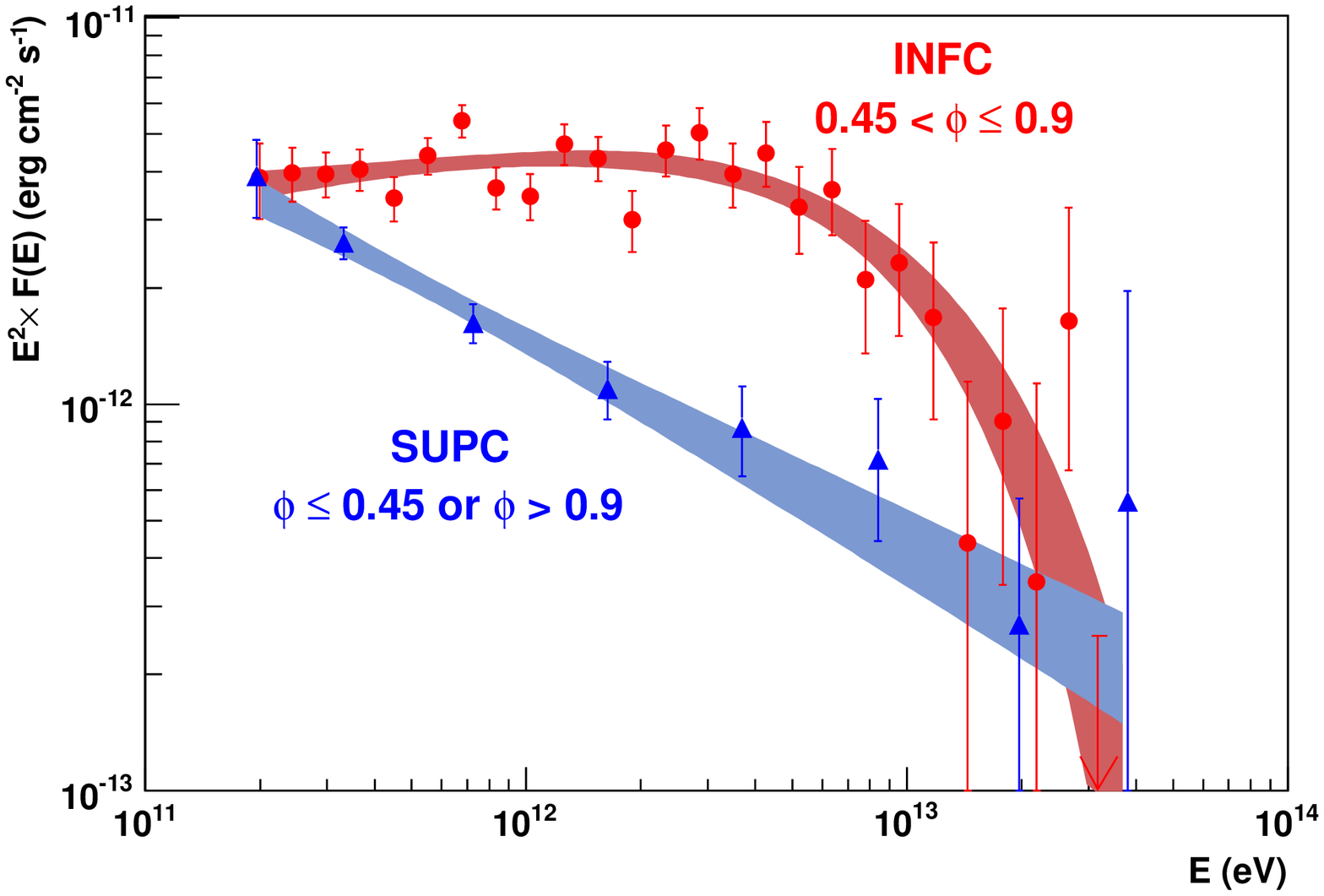}
\end{center}
\vspace{-2em}
\caption{Very high energy spectral energy distribution of LS~5039 for the two broad orbital phase
intervals defines in the text, {\bf INFC} (red circles) and {\bf SUPC} (blue triangles). 
The shaded regions represent the $1\sigma$ confidence bands on the fitted functions.
A clear spectral hardening is occurring in the $200\ \mathrm{GeV}$ to a few TeV range during the {\bf INFC} phase interval.
From \cite{Aharonian:HESS_LS5039:2}.}\label{fig:SED}
\end{figure}

We first define two broad phase intervals: {\bf INFC} centered on the inferior conjunction ($0.45 < \phi \leq 0.9$)
and its complementary {\bf SUPC} centered on the superior conjunction, corresponding respectively to high
and low flux states.
The high state VHE spectral energy distribution (Fig \ref{fig:SED}) is consistent
with a hard power law with index $\Gamma= 1.85 \pm 0.06_{\mathrm{stat}} \pm 0.1_{\mathrm{syst}}$ and exponential 
cutoff at $E_0 = 8.7 \pm 2.0\ \mathrm{TeV}$. In contrast, the spectrum for low state is compatible with a relatively steep
($\Gamma = 2.53 \pm 0.06_{\mathrm{stat}} \pm 0.1_{\mathrm{syst}}$) pure power law extending from $200\ \mathrm{GeV}$ to
$\sim 20\ \mathrm{TeV}$. 
% The spectral shapes of these two states are mutually incompatible at the level of $\sim 2\times 10^{-6}$.
Interestingly, the flux appears to be almost unmodulated at $200\ \mathrm{GeV}$ as well as around $20\ \mathrm{TeV}$,
whereas the modulation is maximum around a few ($\sim 5$) TeV.

\begin{figure}
\begin{center}
\noindent
%\fbox{\hbox{\vbox{\hsize=50mm \hfill \vspace{50mm}}}}
\includegraphics [width=0.45\textwidth]{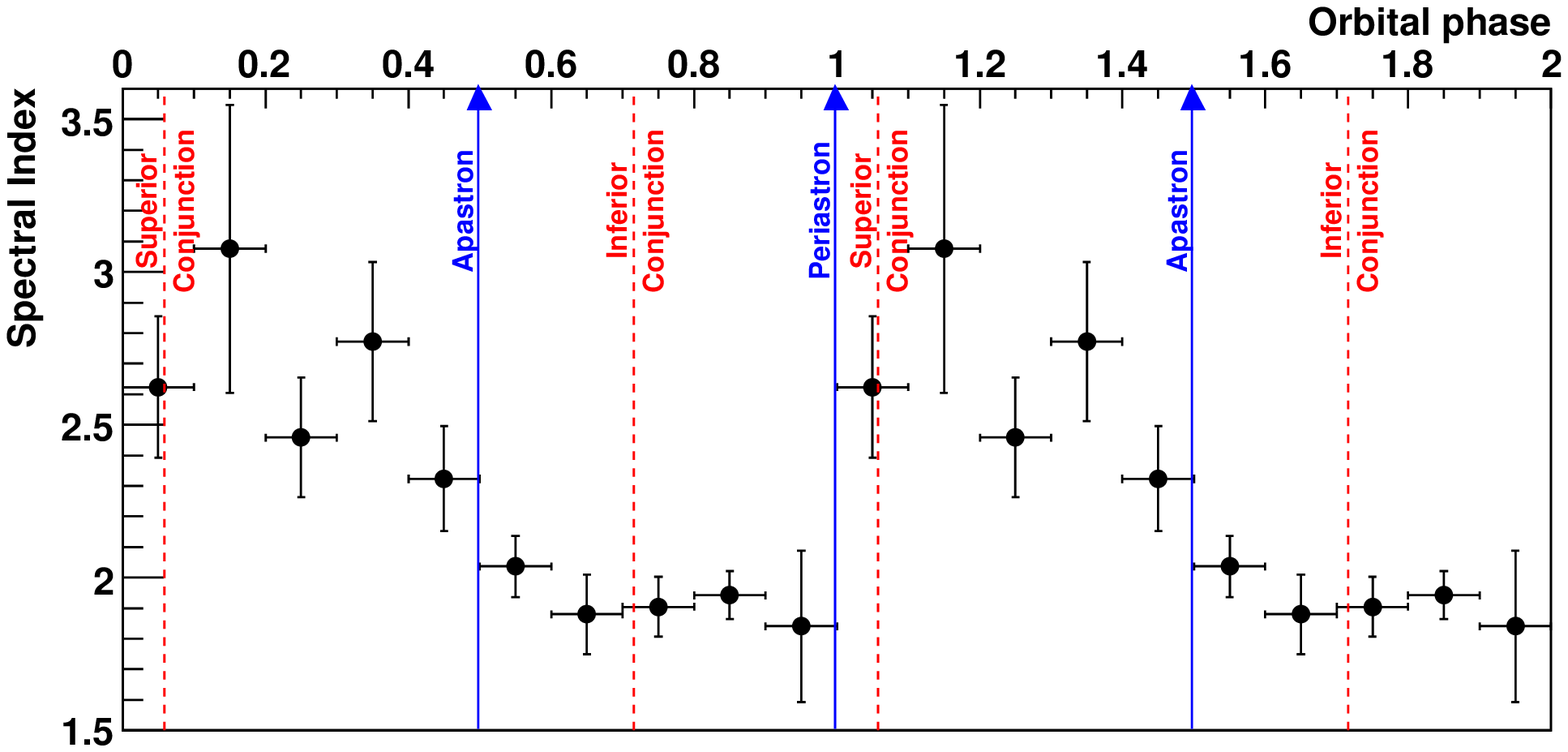}
\vspace{-0.5em}
\includegraphics [width=0.45\textwidth]{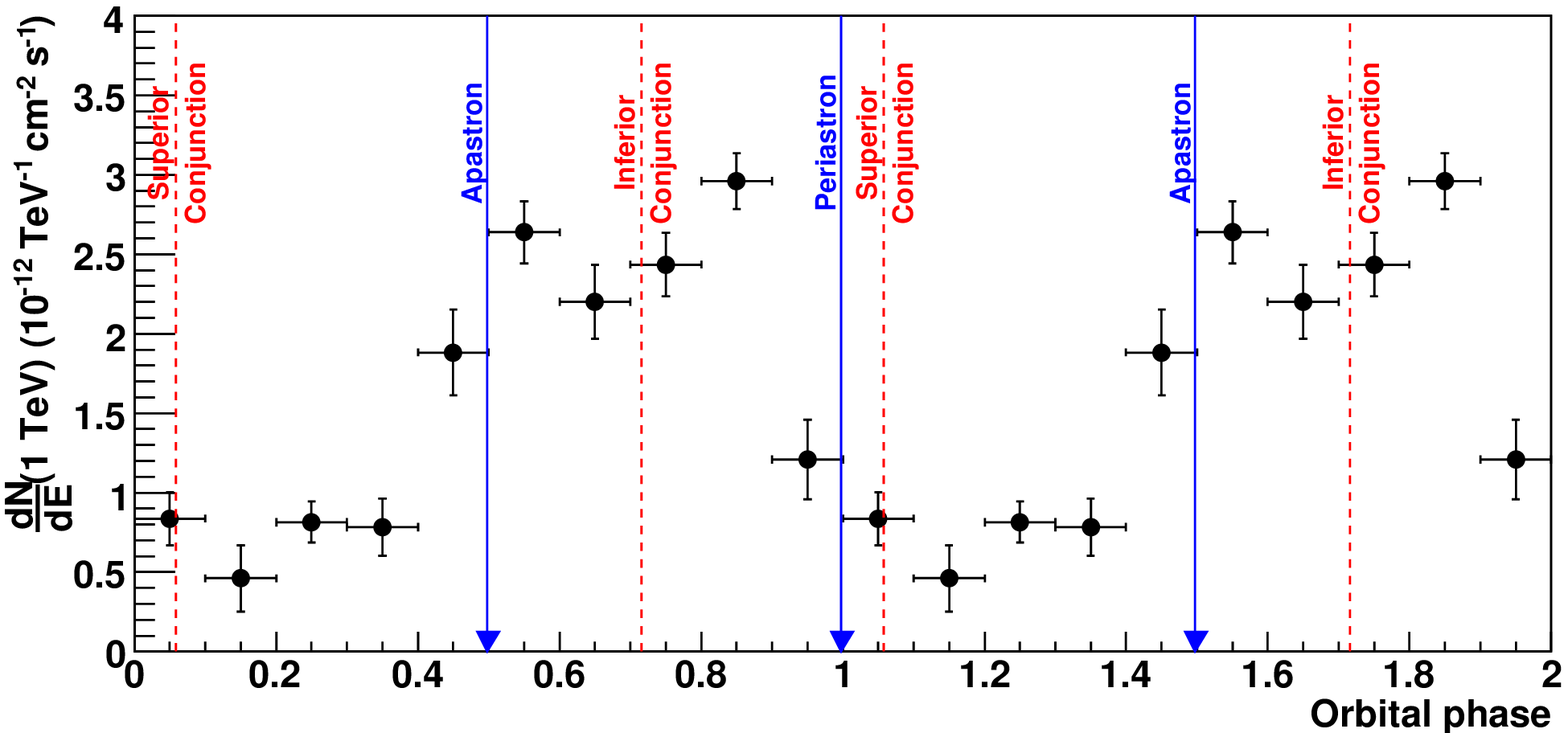}
\end{center}
\vspace{-1em}
\caption{Top: Fitted pure power-law photon index vs. phase interval of width $\Delta \phi = 0.1$.
Bottom: Differential flux at $1\, \mathrm{TeV}$ for the same phase interval. From \cite{Aharonian:HESS_LS5039:2}.}
\vspace{-0.5em}
\label{fig:PhaseSpectra}
\end{figure}

Looking at smaller phase intervals, Fig \ref{fig:PhaseSpectra}
shows the results of a pure power-law fit of 
the high energy spectra in $0.1$ orbital phase bins (restricted to energies below $5\ \mathrm{TeV}$ to avoid
systematic effect introduced by the high state cutoff).
The flux normalisation at $1\ \mathrm{TeV}$ (bottom) and photon index (top) are strongly correlated, the flux being higher
when the spectrum is harder and vice-versa. 
% Interestingly, a similar effect, however in a smaller variation range and a different phasogram,
% was found in X rays\cite{Bosch:2005}.
%However, the X-ray phasogram exhibited a different picture than the VHE one,
%with a flux maximum at $\phi \sim 0.2$  (close to the VHE flux minimum) and a second peak 
%around $\phi\sim 0.8$ better aligned with the VHE flux.

\section*{Interpretation and Conclusion}

The basic paradigm of VHE $\gamma$-ray production requires the presence of particles
accelerated to multi-TeV energies and a target comprising photons (for $\gamma$-ray production
through the Inverse Compton effect) and/or matter of sufficient density (for $\gamma$-ray production
through pion decay in hadronic processes). Several model classes are available to explain VHE emission from 
gamma-ray binaries, 
differentiating one from the other by the nature of accelerated particles and/or the location of the 
acceleration region. In jet-based models, particle acceleration could take place directly inside 
and along the jet, e.g. \cite[and references therein]{Bosch:2004}, and also in the jet termination 
shock regions\cite{Heinz:2002}. Non-jet scenarios are also available, e.g. \cite{Maraschi:1981,Dubus:2006b},
where the emission arises from the interaction of a pulsar wind with the stellar companion 
equatorial wind.

New observations by HESS have established orbital modulation of the VHE $\gamma$-ray flux 
and energy spectrum from the XRB LS~5039. 
The observed VHE modulation indicates that the emission most probably takes place
close (within $\sim 1\,\mathrm{AU}$) to the massive stellar companion, where 
modulated $\gamma$-ray absorption via pair  production ($e^+ e^-$) on the intense stellar
photon field is unavoidable (e.g. \cite{Dubus:2006b}).
The observed spectral modulation is however incompatible with a pure absorption scenario,
which in particular predicts a maximum variability around 300~GeV and a VHE spectral hardening
in the low flux state, inconsistent with observations.

Modulation could also arise from a modulation of the acceleration and cooling timescales 
along the orbit due to varying magnetic field and photon field densities (e.g. \cite[and references therein]{Aharonian:HESS_LS5039:2})
which could modify the maximum electron energy and therefore induce a phase-dependent energy break
in the $\gamma$-ray spectrum.
Modulation of the accretion rate due to interaction of the stellar wind with the compact object 
in the microquasar scenario (e.g. \cite{Paredes:2006}) could
be another ingredient of the observed modulation.

A detailed study is now required to fully explain these new observations and understand the 
complex relationship  between $\gamma$-ray absorption and production processes within these 
binary systems.

\section*{Acknowledgements}

{
\tiny The support of the Namibian authorities and of
the University of Namibia in facilitating the con-
struction and operation of H.E.S.S. is gratefully
acknowledged, as is the support by the German
Ministry for Education and Research (BMBF), the
Max Planck Society, the French Ministry for Re-
search, the CNRS-IN2P3 and the Astroparticle In-
terdisciplinary Programme of the CNRS, the U.K.
Particle Physics and Astronomy Research Council
(PPARC), the IPNP of the Charles University, the
South African Department of Science and Technol-
ogy and National Research Foundation, and by the
University of Namibia. We appreciate the excel-
lent work of the technical support staff in Berlin,
Durham, Hamburg, Heidelberg, Palaiseau, Paris,
Saclay, and in Namibia in the construction and op-
eration of the equipment.\par
}

\bibliographystyle{plain}

%%%%%%%%
%  14  %
%%%%%%%%

%
% Bibliography and bibfile
\def\aj{AJ}%
          % Astronomical Journal
\def\actaa{Acta Astron.}%
          % Acta Astronomica
\def\araa{ARA\&A}%
          % Annual Review of Astron and Astrophys
\def\apj{ApJ}%
          % Astrophysical Journal
\def\apjl{ApJ}%
          % Astrophysical Journal, Letters
\def\apjs{ApJS}%
          % Astrophysical Journal, Supplement
\def\ao{Appl.~Opt.}%
          % Applied Optics
\def\apss{Ap\&SS}%
          % Astrophysics and Space Science
\def\aap{A\&A}%
          % Astronomy and Astrophysics
\def\aapr{A\&A~Rev.}%
          % Astronomy and Astrophysics Reviews
\def\aaps{A\&AS}%
          % Astronomy and Astrophysics, Supplement
\def\azh{AZh}%
          % Astronomicheskii Zhurnal
\def\baas{BAAS}%
          % Bulletin of the AAS
\def\bac{Bull. astr. Inst. Czechosl.}%
          % Bulletin of the Astronomical Institutes of Czechoslovakia 
\def\caa{Chinese Astron. Astrophys.}%
          % Chinese Astronomy and Astrophysics
\def\cjaa{Chinese J. Astron. Astrophys.}%
          % Chinese Journal of Astronomy and Astrophysics
\def\icarus{Icarus}%
          % Icarus
\def\jcap{J. Cosmology Astropart. Phys.}%
          % Journal of Cosmology and Astroparticle Physics
\def\jrasc{JRASC}%
          % Journal of the RAS of Canada
\def\mnras{MNRAS}%
          % Monthly Notices of the RAS
\def\memras{MmRAS}%
          % Memoirs of the RAS
\def\na{New A}%
          % New Astronomy
\def\nar{New A Rev.}%
          % New Astronomy Review
\def\pasa{PASA}%
          % Publications of the Astron. Soc. of Australia
\def\pra{Phys.~Rev.~A}%
          % Physical Review A: General Physics
\def\prb{Phys.~Rev.~B}%
          % Physical Review B: Solid State
\def\prc{Phys.~Rev.~C}%
          % Physical Review C
\def\prd{Phys.~Rev.~D}%
          % Physical Review D
\def\pre{Phys.~Rev.~E}%
          % Physical Review E
\def\prl{Phys.~Rev.~Lett.}%
          % Physical Review Letters
\def\pasp{PASP}%
          % Publications of the ASP
\def\pasj{PASJ}%
          % Publications of the ASJ
\def\qjras{QJRAS}%
          % Quarterly Journal of the RAS
\def\rmxaa{Rev. Mexicana Astron. Astrofis.}%
          % Revista Mexicana de Astronomia y Astrofisica
\def\skytel{S\&T}%
          % Sky and Telescope
\def\solphys{Sol.~Phys.}%
          % Solar Physics
\def\sovast{Soviet~Ast.}%
          % Soviet Astronomy
\def\ssr{Space~Sci.~Rev.}%
          % Space Science Reviews
\def\zap{ZAp}%
          % Zeitschrift fuer Astrophysik
\def\nat{Nature}%
          % Nature
\def\iaucirc{IAU~Circ.}%
          % IAU Cirulars
\def\aplett{Astrophys.~Lett.}%
          % Astrophysics Letters
\def\apspr{Astrophys.~Space~Phys.~Res.}%
          % Astrophysics Space Physics Research
\def\bain{Bull.~Astron.~Inst.~Netherlands}%
          % Bulletin Astronomical Institute of the Netherlands
\def\fcp{Fund.~Cosmic~Phys.}%
          % Fundamental Cosmic Physics
\def\gca{Geochim.~Cosmochim.~Acta}%
          % Geochimica Cosmochimica Acta
\def\grl{Geophys.~Res.~Lett.}%
          % Geophysics Research Letters
\def\jcp{J.~Chem.~Phys.}%
          % Journal of Chemical Physics
\def\jgr{J.~Geophys.~Res.}%
          % Journal of Geophysics Research
\def\jqsrt{J.~Quant.~Spec.~Radiat.~Transf.}%
          % Journal of Quantitiative Spectroscopy and Radiative Trasfer
\def\memsai{Mem.~Soc.~Astron.~Italiana}%
          % Mem. Societa Astronomica Italiana
\def\nphysa{Nucl.~Phys.~A}%
          % Nuclear Physics A
\def\physrep{Phys.~Rep.}%
          % Physics Reports
\def\physscr{Phys.~Scr}%
          % Physica Scripta
\def\planss{Planet.~Space~Sci.}%
          % Planetary Space Science
\def\procspie{Proc.~SPIE}%
          % Proceedings of the SPIE
\let\astap=\aap
\let\apjlett=\apjl
\let\apjsupp=\apjs
\let\applopt=\ao

\def\degr{\hbox{$^\circ$}}
\def\arcmin{\hbox{$^\prime$}}
\def\arcsec{\hbox{$^{\prime\prime}$}}
\def\utw{\smash{\rlap{\lower5pt\hbox{$\sim$}}}}
\def\udtw{\smash{\rlap{\lower6pt\hbox{$\approx$}}}}
\def\fd{\hbox{$.\!\!^{\rm d}$}}
\def\fh{\hbox{$.\!\!^{\rm h}$}}
\def\fm{\hbox{$.\!\!^{\rm m}$}}
\def\fs{\hbox{$.\!\!\;\!^{\rm s}$}}
\def\fdg{\hbox{$.\!\!^\circ$}}
\def\farcm{\hbox{$.\mkern-4mu^\prime$}}
\def\farcs{\hbox{$.\!\!\;\!^{\prime\prime}$}}
\def\sun{{\lower-2pt\hbox{$_\odot$}}}

\def\la{\mathrel{\mathchoice {\vcenter{\offinterlineskip\halign{\hfil
$\displaystyle##$\hfil\cr<\cr\sim\cr}}}
{\vcenter{\offinterlineskip\halign{\hfil$\textstyle##$\hfil\cr
<\cr\sim\cr}}}
{\vcenter{\offinterlineskip\halign{\hfil$\scriptstyle##$\hfil\cr
<\cr\sim\cr}}}
{\vcenter{\offinterlineskip\halign{\hfil$\scriptscriptstyle##$\hfil\cr
<\cr\sim\cr}}}}}
\def\ga{\mathrel{\mathchoice {\vcenter{\offinterlineskip\halign{\hfil
$\displaystyle##$\hfil\cr>\cr\sim\cr}}}
{\vcenter{\offinterlineskip\halign{\hfil$\textstyle##$\hfil\cr
>\cr\sim\cr}}}
{\vcenter{\offinterlineskip\halign{\hfil$\scriptstyle##$\hfil\cr
>\cr\sim\cr}}}
{\vcenter{\offinterlineskip\halign{\hfil$\scriptscriptstyle##$\hfil\cr
>\cr\sim\cr}}}}}

%
%%%upright Greek letters (example below: upright "mu")
\newcommand{\greeksym}[1]{{\usefont{U}{psy}{m}{n}#1}}
\newcommand{\umu}{\mbox{\greeksym{m}}}
\newcommand{\udelta}{\mbox{\greeksym{d}}}
\newcommand{\uDelta}{\mbox{\greeksym{D}}}
\newcommand{\uPi}{\mbox{\greeksym{P}}}
\def\g{$\gamma$}
\def\psr{\hbox{PSR\,B1259$-$63}}
\def\ss{\hbox{SS\,2883}}
\def\sys{\hbox{PSR\,B1259$-$63\,/\,SS\,2883}}
\def\degr{\hbox{$^\circ$}}
\def\arcmin{\hbox{$^\prime$}}
\def\arcsec{\hbox{$^{\prime\prime}$}}

\title{Report on TeV $\gamma$-Ray Observations of PSR\,B1259$-$63/SS2883 near the 2007 Periastron with the H.E.S.S.~Stereoscopic System of Cherenkov Telescopes}

\shorttitle{$\gamma$-Ray Observations of PSR\,B1259$-$63/SS2883 in 2007}

\authors{M.~Kerschhaggl$^{1}$, F.~Aharonian$^{2}$, M.~Fuessling$^{1}$, U.~Schwanke$^{1}$ \\for the H.E.S.S.\ collaboration\\}

\shortauthors{F.~Aharonian et al.}

\afiliations{
	 $^1$Institut f\"ur Physik, Humboldt-Universit\"at zu Berlin,
	Newtonstr. 15, D 12489 Berlin, Germany\\
	$^2$ Dublin Institute for Advanced Studies, 5 Merrion Square, Dublin 2, Ireland
	}
\email{mkersch@physik.hu-berlin.de}

\abstract{PSR\,B1259$-$63\,/ SS\,2883 is a binary system consisting of a 48~ms radio pulsar orbiting 
a Be star with a period of 3.4~y in a highly eccentric orbit (e = 0.87). The 
system was first detected in TeV $\gamma$-rays by H.E.S.S.~around the last periastron passage in March 2004. These observations established 
PSR\,B1259$-$63\,/ SS\,2883 as the first variable galactic source in the very high energy (VHE) regime. A lightcurve for the system, covering mainly the post periastron part, could be deduced, clearly showing a variable flux in VHE photons. New data have been taken this year from April to June with the system approaching its next periastron (July 27, 2007). The status and outcome so far of the corresponding campaign will be discussed.}

\maketitle

\addcontentsline{toc}{section}{Report on TeV $\gamma$-Ray Observations of PSR\,B1259$-$63/SS2883 near the 2007 Periastron with the H.E.S.S.~Stereoscopic System of Cherenkov Telescopes}
\setcounter{figure}{0}
\setcounter{table}{0}
\setcounter{equation}{0}

\section*{Introduction}
The binary system PSR\,B1259$-$63\,/ SS\,2883 was first observed in TeV $\gamma$-rays during its periastron passage between February and June 2004 \cite{1259}, establishing it as the first variable galactic TeV $\gamma$-ray source (see Fig.~\ref{skymap}).   
\begin{figure}[h]
\begin{center}
\resizebox{1.0\hsize}{!}{\includegraphics{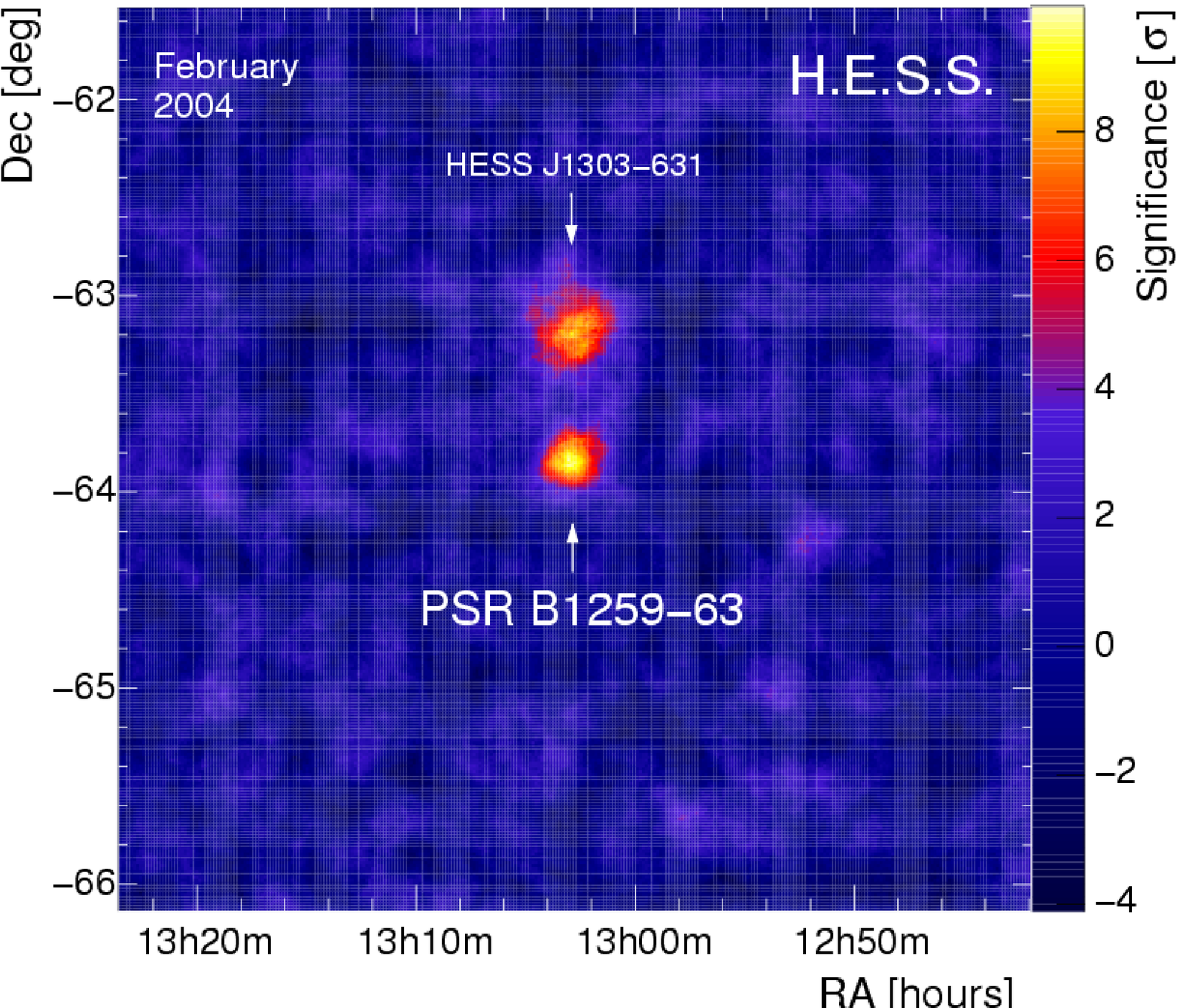}}
\caption{\label{skymap}\em Significance skymap of the PSR\,B1259$-$63 field of view as seen in February 2004. The unidentified $\gamma$-ray source HESS\,J1303$-$631 is located $0.6^{\circ}$ to the north of the source.}
\end{center}
\end{figure}The corresponding data showed a clear pointlike signal with a statistical significance of 13 standard deviations at the position of PSR\,B1259$-$63. A time averaged spectrum as well as a lightcurve for the integrated flux above 380 GeV from this object could be extracted (see Fig.~\ref{spectrum}~\&~\ref{lightcurve}). Moreover, it was the first time in the history of TeV $\gamma$-ray astronomy where two sources have been discovered within the same field of view as this campaign lead to the serendipitous discovery of HESS J1303-631 \cite{1303} (see Fig.~\ref{skymap}).\\
The peculiar shape of the PSR\,B1259$-$63 lightcurve has been object of various model descriptions trying to explain the underlying physical processes causing the VHE emission. Mechanisms like Inverse Compton (IC) scattering of ultrarelativistic electrons on the stellar photons or hadronic scenarios (e.g. $pp\rightarrow\pi^{0}\rightarrow\gamma\gamma$) have been suggested as possible origins of TeV photons in the interactions of the pulsar wind with the stellar outflow and radiation field of the companion Be star (e.g. \cite{Khangulyan}). Some of them take into account the influence of the dense stellar disc that might play a crucial role in the generation mechanism of VHE $\gamma$-rays (see \cite{Neronov}). In order to constrain the various parameters used in the model predictions as well as to be able to discriminate between the models in question, data for the up to now unknown pre-periastron (lightcurve) part of PSR\,B1259$-$63\ are needed.\\As the next periastron takes place on July 27, 2007, a campaign of 60~h of intended exposure during the pre-periastron phase from April to July 2007 has been started.
\begin{figure}[h]
\begin{center}
\resizebox{1.0\hsize}{!}{\includegraphics{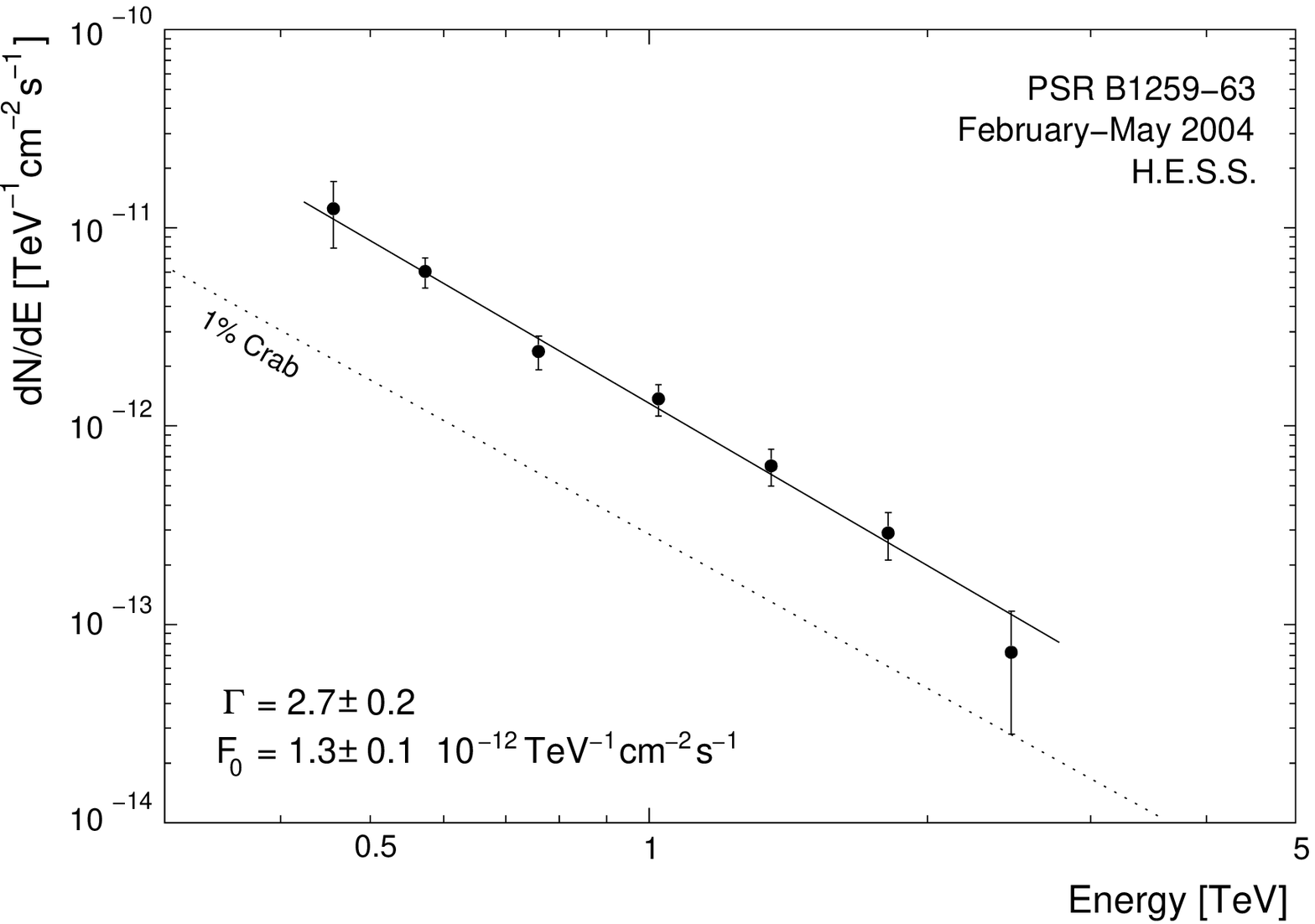}}
\caption{\label{spectrum}\em Energy spectrum $dN/dE$ of $\gamma$-rays from PSR\,B1259$-$63
determined from the H.E.S.S.\ 2004 data. The solid line indicates the
power-law fit $F(E)=F_0E^{-\Gamma}$.}
\end{center}
\end{figure}

\begin{figure}[h]
\begin{center}
\resizebox{1.0\hsize}{0.5\vsize}{\includegraphics{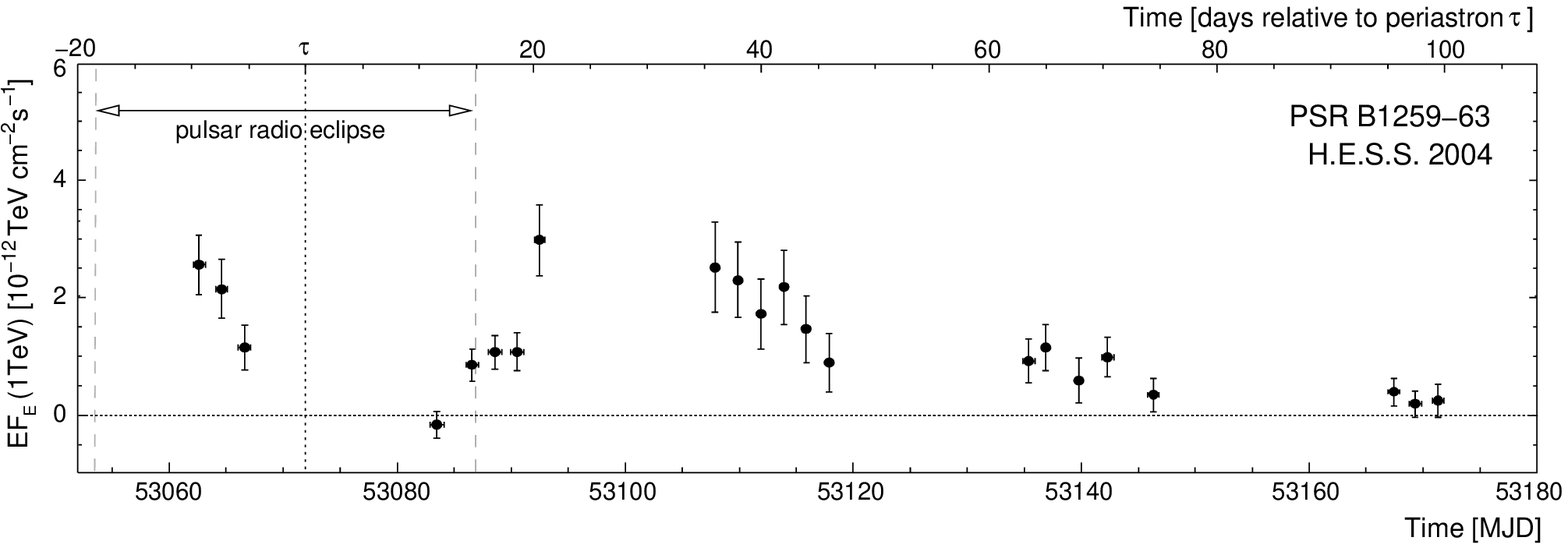}}
\caption{\label{lightcurve}\em The PSR\,B1259$-$63 lightcurve around periastron in 2004. The vertical dashed black line indicates the position of the periastron. The data clearly indicate a variable flux.}
\end{center}
\end{figure}

\section*{The H.E.S.S.~PSR\,B1259$-$63\ 2007 Campaign}
\begin{figure}[h]
\begin{center}
\resizebox{1.1\hsize}{!}{\includegraphics{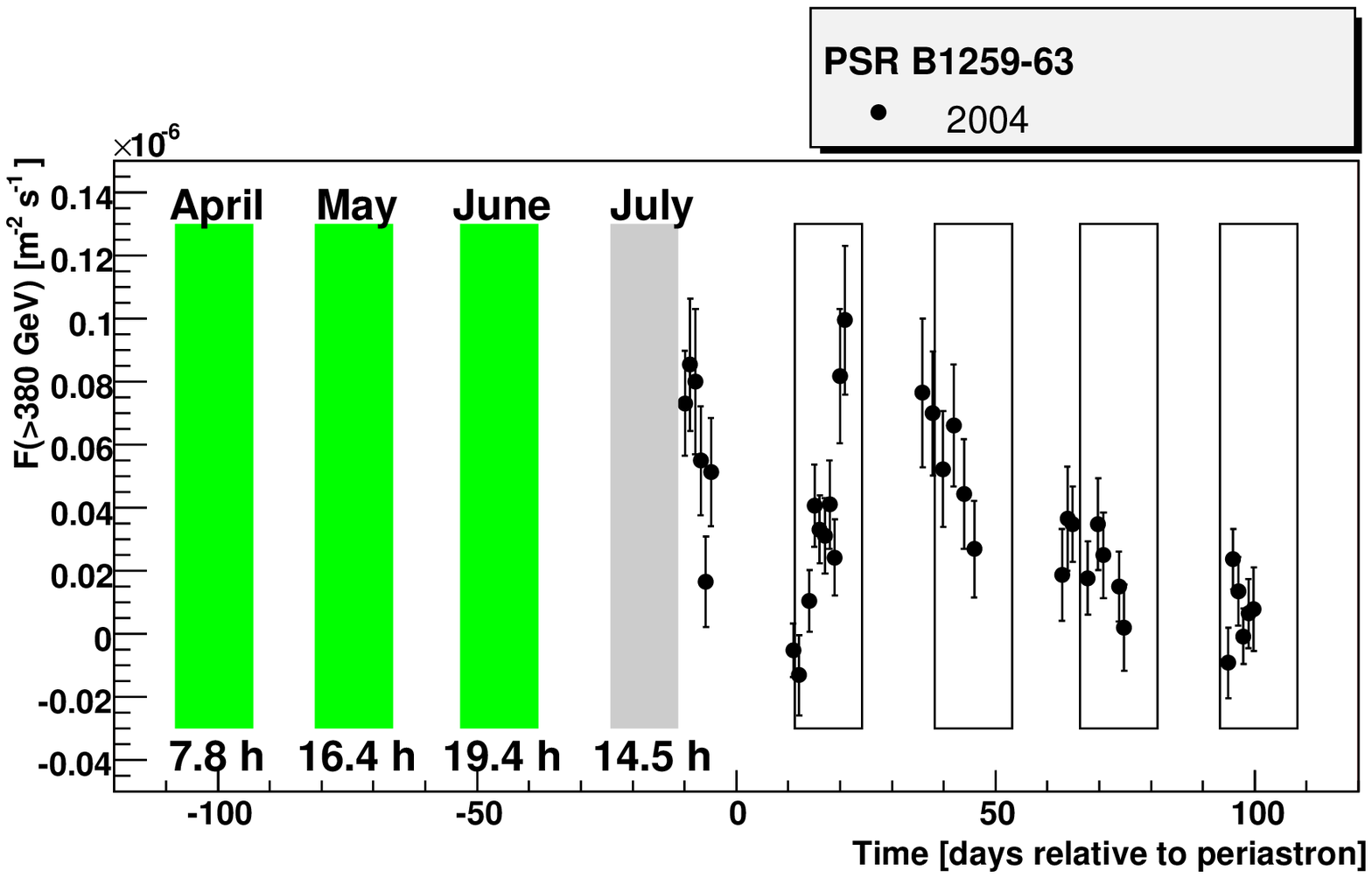}}
\caption{\label{visibility}\em The PSR\,B1259$-$63 observation windows in 2007 for zenith angles $<45^{\circ}$ with respect to the system's time relative to periastron. Green boxes indicate that data taking has already been accomplished during the corresponding month. The empty boxes are the 2007 observation windows mirrored with respect to periastron, overlayed with the 2004 data for comparison.}
\end{center}
\end{figure}

Figure \ref{visibility} shows the H.E.S.S.~visibility windows of PSR\,B1259$-$63 in 2007 for zenith angles below 45$^{\circ}$ together with data from 2004 with respect to periastron. The numbers underneath each observation slot indicate the amount of intended exposure time for this month. Observations will cover the pre-periastron orbital phase until 14 days prior to the periastron passage. Green boxes refer to time windows where data taking already has been accomplished. The overall exposure time of roughly 60~h was chosen to match the dataset from 2004, in terms of good quality data, in order to have a comparable amount of data for the pre-periastron part of the lightcurve.\\
So far, data from April to June with an overall livetime of ~33~h have been taken (see Tab.~\ref{campaign}). The livetime for the overall 2004 dataset was $\sim$ 45~h.

\begin{table}[h]
\begin{center}
\begin{tabular}{cccc}
	\hline\hline
%\multicolumn{4}{|c|}
	%{\textbf{\rule[-3mm]{0mm}{8mm}PSR\,B1259$-$63\ data April-June 2007}} \\
Period	& $\tau~[h]$ & $S~[\sigma]$ & Calibration\\
2007    &            &              & Status \\ \hline
April   &        5.3 & 1.0          &  final \\
May     &       14.6 & 3.2          & final \\
June    &       13.2 & 6.4          & preliminary\\
\hline
\end{tabular}
\caption{\label{campaign}\em PSR\,B1259$-$63\ April-June 2007: Livetime $\tau$, Significance $S$ for a preliminary point source analysis. The June data has been calibrated preliminarily on site in Namibia.}
\end{center}
\end{table}
A preliminary standard point source analysis of these data has been carried out (for details on the H.E.S.S. analysis chain see \cite{1259}). The significance for a $\gamma$-ray excess from the source in April and May is 1.0 and 3.2 standard deviations respectively, showing no strong detection for these months. However, when analyzing data from June on site, using a preliminary on site calibration, a clear signal with a significance of 6.4$\sigma$ can be seen from PSR\,B1259$-$63 (see Tab.~\ref{campaign}).

\section*{Multi Wavelength Coverage}
In order to also have Multi Wavelength (MWL) data available coincident with the TeV data provided by H.E.S.S.~a cooperation with the \emph{SUZAKU} \cite{Mitsuda} satellite project has been established. Corresponding observation schedules have been optimized for a maximum MWL coverage. \emph{SUZAKU} will observe PSR\,B1259$-$63\ from July to September 2007 in eight pointings of 20~ks duration each. Four pointings will take place in July coincident with the H.E.S.S observations covering the assumed first entrance of the pulsar into the dense circumstellar disc. This interesting phase of the system's orbit is therefore covered in the keV and TeV energy bands.

\subsection*{Summary \& Outlook}
PSR\,B1259$-$63 is currently re-observed by the H.E.S.S.~Cherenkov Telescope System in Namibia in a 60~h exposure campaign lasting from April to July 2007. 33~h of livetime data have already been taken during the April to June period. A preliminary point source analysis yields no significant excess from the system for April and May. The preliminarily calibrated June data (on site calibration) however shows a clear signal at 6.4$\sigma$.\\ The campaign is carried on in July with a planned exposure of 14.5~h coinciding with 4 pointings on the target done by the X-ray satellite \emph{SUZAKU} covering the crucial first disc crossing of the pulsar.\\
Further isolated post-periastron H.E.S.S.~observations are planned during August when \emph{SUZAKU} as well as \emph{Chandra} will observe this unique HE \& VHE accelerator.

\bibliographystyle{plain}

%%%%%%%%
%  15  %
%%%%%%%%

\title{A search for VHE $\gamma$-ray binaries in the H.E.S.S. Galactic Plane Scan}
\shorttitle{Gamma-ray binaries in the galaxy}
\authors{Hugh Dickinson$^{1}$, Ian Latham$^{1}$, Paula Chadwick$^{1}$ for the H.E.S.S. Collaboration}
\shortauthors{H.~J.~Dickinson et al.}
\afiliations{$^1$ Physics Department, University of Durham, South Road, Durham, County Durham, DH1 3LE, United Kingdom}
\email{h.j.dickinson@dur.ac.uk}

\abstract{Utilising the unprecedented TeV sky coverage of the H.E.S.S. galactic plane scan, we present the results of a search for Very High Energy gamma-ray sources coincident with the positions of  known X-ray binaries. Although no significant detections were obtained, upper limits to the TeV flux from 18 X-ray binaries were derived.}

\maketitle

\addcontentsline{toc}{section}{A search for VHE $\gamma$-ray binaries in the H.E.S.S. Galactic Plane Scan}
\setcounter{figure}{0}
\setcounter{table}{0}
\setcounter{equation}{0}

\section*{Introduction}\label{sec:intro}

The H.E.S.S. galactic plane scan \cite{aharonian06} is an extensive survey of the inner part of the galaxy in TeV $\gamma$-rays. It consists of  230 hours of observation comprising 500 pointings between $\pm30^{\circ}$ in galactic longitude and  $\pm3^{\circ}$ in galactic latitude. The average flux sensitivity of the scan is $\sim2\%$ of the Crab nebula flux at photon energies above 200 GeV.

X-ray binaries are galactic systems containing a normal donor star which is in the process of transferring mass onto a compact object, such as a neutron star or black hole, to which it is gravitationally bound. A significant fraction of the gravitational potential liberated by this mass transfer is emitted as X-radiation, hence the nomenclature. 

Some X-ray binaries are observed to eject a proportion of the accreted matter in collimated and often highly relativistic jets (See e.g. \cite{mirabel94,fender04}). These `microquasars' are named for their structural similarities with Active Galactic Nuclei (AGN), several of which are known VHE $\gamma$-ray sources (See e.g. \cite{aharonian05_3, aharonian03}). Indeed, the extremes of temperature, pressure and radiation density likely to be generated close to the compact object in X-ray binary systems provide excellent conditions for the acceleration of charged particles to multi-TeV energies. Such acceleration, although not synonymous with VHE $\gamma$-ray emission, has nonetheless been identified as an apparent prerequisite for the production of photons with energies $E_{\gamma} > 200$ GeV.

Several plausible mechanisms exist which permit the generation of VHE $\gamma$-rays in X-ray binary systems. Indeed, both the necessary particle acceleration and the actual photon production can proceed via a number of potential avenues.  In general, particle acceleration is thought to be accomplished at shock fronts either at the interface between the in-falling matter and an outflow or jet , or within the outflow itself. 

In systems where the companion is a low-mass star, the mass transfer proceeds mainly via Roche lobe overflow. In such systems particle acceleration can occur across internal shocks within the jet structures. In systems containing high-mass donor, accretion is often wind-fed, with high energy particles from the stellar wind being accreted onto the compact primary. If the primary is a pulsar, shocks can occur at the pulsar standoff distance, where the ram pressures due to the stellar and pulsar winds equilibrate \cite{dubus06_2}. 

Given a population of energetic charged particles, the mechanisms of $\gamma$-ray production may be broadly segregated into two categories: Those in which the emitting particles are hadronic, and those where they are leptons. Leptonic models closely resemble those used to explain the continuum spectra of AGN, relying as they do on the Synchrotron Self-Compton (SSC) and External Compton (EC) processes \cite{bosch-ramon06, dubus06_2, atoyan99}. The SSC process involves the up-scattering of low energy synchrotron photons by the high energy electrons which generated them, while in the EC process, the target photons are generated elsewhere. In relativistic jets aligned close to the observer line-of-sight, the Comptonised radiation is both beamed and Doppler boosted, producing a measurable flux of photons in the TeV band. 

While protonic SSC and EC processes are possible \cite{aharonian02, bosch-ramon05}, most hadronic emission models rely on the production and subsequent decay of neutral pions.
\begin{equation}
pp\rightarrow pp\pi^{0}\rightarrow e^{+}e^{-}\gamma\notag
\end{equation}
For example, \cite{romero03} and \cite{orellana05} show that pions and consequently TeV photons can be produced via the interaction of stellar wind protons with those in a microquasar jet.

Only three X-ray binary systems are known to emit VHE $\gamma$-rays. LS 5039 \cite{aharonian05} and PSR B1259-63 \cite{aharonian05_2} were detected by the H.E.S.S. Telescope array while in the northern hemisphere LS I +61$^{\circ}$303 \cite{albert06} has been observed by the MAGIC collaboration. While their detections by the MAGIC and H.E.S.S. collaborations confirm the existence of  $\gamma$-ray binaries , the catalogue of such objects remains rather small. In terms of morphology, PSR B1259-63 is a Be star-pulsar binary with a 3.4 yr orbital period and while both LS 5039 and LS I +61$^{\circ}$303 are high-mass X-ray binaries with donor masses of 23 and 12 $M_{\odot}$ and orbital periods of 3.9 and 26.5 days respectively. The nature of the compact primaries in both systems is uncertain, although the LS I +61$^{\circ}$303 system seems likely to contain a pulsar \cite{dhawan06}. \cite{dubus06_2} argues that despite observed milliarsecond radio structure \cite{paredes02,massi04}, the observed TeV emission of both LS I +61$^{\circ}$303 and LS 5039 is rotationally derived from pulsars.

The aim of this work was to expand the catalogue of known $\gamma$-ray binaries and identify candidates for further observation. As galactic objects,  X-ray binaries are concentrated in the plane of the galaxy. Consequently, the H.E.S.S. galactic plane scan is an excellent dataset for our purposes. We make no selections on the basis of target morphology and test the positions of all known X-ray binaries with sufficient exposure.

\begin{table*}
\begin{center}
\begin{tabular}{|l||c|c|c|c|c|}
\hline
~Target Name~ 	& ~Significance~ 	& ~Excess~	 & ~Livetime~ 	& ~Mass~  & ~Flux Upper Limit ~ \\
				& [$\sigma$]		& [counts]		& [hours]	& &($E_{\gamma} > $ 1 TeV) [ph cm$^{-2}$s$^{-1}$] \\			
\hline\hline
RX J1744.7-2713	& 1.367	& 114.874 	& 7.71613 	&H& $5.472\times10^{-12}$\\
AX J1749.2-2725	& 3.021 & 320.949 & 13.46 	&H& $6.3659\times10^{-12}$\\
&&&&&(HESS J1747-281)\\
GRO J1750-27		& 1.521 & 94.006 & 4.715 	&H& $1.050\times10^{-11}$\\
AX J1820.5-1434	&  -1.582 & -22.290 & 4.339 	&H& $1.507\times10^{-12}$\\
H 1833-076		& 1.074 & 23.038& 8.075 	&H& $1.882\times10^{-12}$\\
&&&&&(HESS J1837-069)\\
GS 1839-04		& 0.178	& 2.059	& 1.701 	&H& $2.530\times10^{-12}$\\
AX 1845.0-0433	& 0.639	& 6.574	& 1.711 	&H& $5.142\times10^{-12}$ \\
2S 1845-024		& 0.805& 7.111& 1.711 	&H& $5.142\times10^{-12}$\\
J1744-28		& 4.652	& 461.619 & 10.874 &L& $1.569\times10^{-11}$\\
&&&&&(HESS J1745-290)\\
1742.8-2853	& 11.202	& 1060.91	& 10.0287 &L& $2.564\times10^{-11}$\\
&&&&&(HESS J1745-290)\\
1742.9-2852	& 7.397	& 214.67	& 11.309 	&L& $4.700\times10^{-12}$\\
&&&&&(HESS J1745-290)\\
1743.1-2852	& 6.064	& 177.667	& 11.741 	&L& $4.013\times10^{-12}$\\
&&&&&(HESS J1745-290)\\
1742.9-2849	& 5.553	& 162.729	& 11.741 	&L& $3.637\times10^{-12}$\\
&&&&&(HESS J1745-290)\\
1742.5-2845	& 2.364	& 67.242	& 11.966 	&L& $2.253\times10^{-12}$\\
&&&&&(HESS J1745-290)\\
1743-288		& 3.521	& 101.948 	& 11.7405 	&L& $2.910\times10^{-12}$\\
&&&&&(HESS J1745-290)\\
1743.1-2843	& 3.102	& 93.775 & 12.594 	&L& $2.918\times10^{-12}$\\
&&&&&(HESS J1745-290)\\
J1750.8-2900	& 0.346	& 10.159	& 12.392 	&L& $1.479\times10^{-12}$\\
1739-278		& 2.563	& 65.670	& 5.068 	&L& $2.556\times10^{-11}$\\
J1748-288	& 3.100	& 260.002 & 13.656 	&L& $1.254\times10^{-11}$\\
&&&&&(HESS J1747-281)\\
1735-269		& -0.661	& -5.023	& 1.288 	&L& $2.755\times10^{-12}$\\
1749-285		& -1.570	& -117.739 & 9.444 	&L& $6.684\times10^{-12}$\\
1744-265		& -0.238	& -4.268	& 5.148 	&L& $1.964\times10^{-12}$\\
1745-248		& -1.736	& -36	& 0.435 	&L& $1.372\times10^{-11}$\\
1758-258		& -1.124 & -36.520 & 1.727 	&L& $4.395\times10^{-12}$\\
1758-250		& -0.270 & -14.736 & 3.013 	&L& $6.083\times10^{-12}$\\
J1806-246	& -0.079	& -1.921	& 5.491 	&L& $9.490\times10^{-12}$\\
1758-205		& -0.501	& -8.545	& 3.890	 	&L& $1.534\times10^{-12}$\\
1811-171		& 1.777	& 50.112	& 10.626	 	&L&$1.609\times10^{-12}$\\
1813-140		& -1.088	& -7.373	& 0.855 	&L&$2.451\times10^{-12}$\\
\hline
\end{tabular}
\caption{The results of the X-ray binary search. Where obtainable, upper limits to the $\gamma$-ray flux at energies $>$ 1 TeV are given. Where the target region is contaminated by the flux from a known TeV source, the derivation of an upper limit is not possible, but there is no way to safely associate the observed flux with the X-ray binary system. In this case the contaminating object is indicated in the \textit{Flux Upper Limit} column. Negative excesses and significances result purely from fluctuations in the $\gamma$-ray background and should not be interpreted as a genuine deficit in the photon flux. In the \textit{Mass} column, `H' indicates a high-mass system and `L' a low mass system.}\label{tab:results}
\end{center}
\end{table*}

\section*{The Sample}\label{sec:sample}
As a source of targets for the search we utilise the catalogues of \cite{liu06} for high-mass X-ray binaries and \cite{liu01} for low-mass systems. The selection criteria required only that the target object was within 1.5$^{\circ}$ of the H.E.S.S. camera centre in at least one good observation run in the galactic plane scan. Run quality selection was carried out as discussed in \cite{aharonian06_2} in an effort to minimise the systematic uncertainties introduced inherently by the telescope system itself and also by the atmosphere which, in effect, forms the scintilliating medium of the detector.

The resulting sample consists of 29 X-ray binaries comprising 8 high-mass systems with the remaining 21 having low-mass donors.

\section*{Analysis and Results}\label{sec:analysis}

Data reduction and analysis were carried out using the standard H.E.S.S. analysis procedure outlined in \cite{aharonian06_2}. The event selection cuts placed on image size, $\theta^{2}$ and the mean reduced scaled parameters are identical to those described as standard in \cite{aharonian06_2}, and are consistent with the expected point-like nature of the target objects. The $\gamma$-ray background was estimated using a `reflected' background model with several run dependent \textit{off} regions defined the same distance from the camera camera as the on region. Areas of the sky containing known TeV $\gamma$-ray sources are precluded from being chosen as \textit{off} regions to ensure that the background estimate remains as uncontaminated as possible. Nonetheless, contamination can occur when the \textit{on} region coincides with a known $\gamma$-ray source. Despite having excellent angular resolution for an instrument of its type, the H.E.S.S. point spread function is somewhat extended, with a 68\% containment radius of $\sim0.1^{\circ}$. For this reason it can be impossible to disentangle the signals from nearby objects. This is particularly difficult when the expected target spectrum and flux are unknown.

As reported in \cite{aharonian06} the region exposed by the galactic plane scan is somewhat crowded with VHE $\gamma$-ray sources, and it is therefore unsurprising that some contamination of our targets did indeed occur. Table \ref{tab:results} outlines the results of the search. Upper limits to the photon flux above 1 TeV have been derived for 18 of the 29 targets. These upper limits represent 99\% confidence intervals derived using the unified Feldman-Cousins method \cite{feldman98}. The remaining 11 targets were too close to known TeV emitters for a reliable upper limit or flux estimate to be obtained. In particular, those lying along a line of sight to the galactic centre are subject to heavy contamination from HESS J1745-290.

\section*{Conclusions}\label{sec:conclusions}

 99\% confidence upper limits have been derived for 18 X-ray binaries. The absence of a conclusive detection by H.E.S.S. could be explained in a number of ways. The flux from all three known $\gamma$-ray binaries is highly variable, and their detection would be somewhat dependent upon the timing of observations. Additional variability can occur due to $\gamma\gamma\rightarrow e^{+}e^{-}$ interactions with near infra-red photons absorbing the intrinsic $\gamma$-ray flux. \cite{dubus06} shows that the angular dependence of this process leads to an orbital modulation of the $\gamma$-ray flux similar to that observed in LS 5039 \cite{aharonian06_3}. In some cases the intrinsic flux may be low enough, or the near IR radiation density high enough that no $\gamma$-rays are detected. Finally it may be the case that highly specific conditions are required for the emission of VHE $\gamma$-rays to occur. 
 
Nonetheless, with the increasing sensitivity of ground based Cherenkov telescopes and the advent of experiments such as \textit{GLAST} and H.E.S.S. Phase II to bridge the gap between soft and VHE $\gamma$-rays, it seems unlikely that the $\gamma$-ray binary catalogue of three will remain so small for long.

\bibliographystyle{plain}

%%%%%%%%
%  16  %
%%%%%%%%

\title{A Very High Energy $\gamma$-ray Survey of X-ray Binaries with H.E.S.S.}
\shorttitle{Survey of X-ray Binaries with H.E.S.S.}
\authors{Hugh Dickinson$^{1}$, Ian Latham$^{1}$, Paula Chadwick$^{1}$ for the H.E.S.S. Collaboration}
\shortauthors{H.~J.~Dickinson et al.}
\afiliations{$^1$ Physics Department, University of Durham, South Road, Durham, County Durham, DH1 3LE, United Kingdom}
\email{h.j.dickinson@dur.ac.uk}

\abstract{Since the discovery of TeV emission from the LS 5039/RX J1826.2-1450 binary system, microquasars are an established class of Very High Energy \gr sources. Nonetheless, the current catalogue of \gr binaries remains somewhat limited, with only four examples known. We present the results of a systematic search for TeV emission from known X-ray binaries with similar properties to LS 5039/RX J1826.2-1450 using the H.E.S.S. atmospheric Cherenkov telescope array.}

\maketitle

\addcontentsline{toc}{section}{A Very High Energy $\gamma$-ray Survey of X-ray Binaries with H.E.S.S.}
\setcounter{figure}{0}
\setcounter{table}{0}
\setcounter{equation}{0}

\section*{Introduction}\label{sec:intro}
Galactic binary systems were established as a new class of TeV \gr sources when the pulsar-Be star binary PSR B1259-63 was detected by the \hess Collaboration \cite{aharonian05_2}. In this system very high energy \grs are likely generated as a result of the interaction of a strong pulsar wind with the dense equatorial wind of the Be star companion. 

Subsequent \hess observations of the microquasar LS5039/RX J1826.2-1450 provided the first known example of an orbitally modulated TeV \gr signal \cite{aharonian06_3}. Furthermore, the association of bipolar milliarsecond radio structures with the LS 5039 system \cite{paredes02} permits consideration of jet-powered scenarios of VHE \gr emission \cite{bosch-ramon05_2} suggesting possible parallels with the supermassive cousins of microquasars - the Active Galactic Nuclei (AGNs) (See however \cite{dubus06_2}). The companion star in the LS 5039 system has been spectroscopically identified as a massive O6.5V((f)) star \cite{clark01}. The nature of the compact primary is somewhat ambiguous, with the recent ephemeris of \cite{casares05_2} indicative of a black hole, but only when combined with the assumption of pseudo-synchronicity of the companion star. Neglecting this assumption, the derived lower mass limit of $\approx1.5M_{\odot}$ is consistent with a neutron star primary. \cite{casares05_2} also derive an somewhat short orbital period for the system of 3.9 days, combined with a rather low eccentricity of 0.35. In the X-ray band, LS 5039 is an unremarkable source albeit with a somewhat hard spectrum \cite{martocchia05}. The system was also associated with the soft \gr EGRET source 3EG J1824-1514 \cite{paredes00}.

The most recently discovered \gr binary is the northern hemisphere object LS I +61$^{\circ}$303, detected by the MAGIC collaboration \cite{albert06}. Like LS 5039, this system exhibits a variable VHE \gr flux, although the existence of any orbital modulation is yet to be established. The orbital period of $\approx26.5$ days \cite{gregory02,casares05} is somewhat longer than that of LS 5039 but a higher eccentricity of 0.72 gives a similar periastron distance of $\sim0.1$ AU. \cite{hutchings81} identified the optical counterpart of LS I +61$^{\circ}$303 as a B0 V star with an equatorial disk. At radio frequencies \lsi also exhibits jet-like structure on milliarcsecond scales \cite{massi93}, although recent observations by \cite{dhawan06} suggest that these ``jets'' may in fact result from interactions of a pulsar wind with the stellar wind of the companion. The X-ray charcteristics of \lsi are remarkably similar to those of LS 5039 \cite{harrison00}, and also reminiscent is the tentative association with the EGRET source 2CG 135+01 \cite{tavani98}.

These three systems constitute the entire VHE \gr binary catalogue. \cite{dickinson07} performed a search for TeV signals coincident with the positions of known X-ray binaries using a sample containing a wide range of donor masses, compact primary types, radio, X-ray and soft \gr behaviours. Nonetheless no significant detections were obtained. 

Instead of this blind search approach, a more sensible methodology might be to use the characteristics of the known \gr binaries as selection criteria for a more targeted survey. Indeed it is apparent that the objects in the existing catalogue share several physical and observational characteristics, some or all of which may be prerequisites for detectable TeV emission. In deriving our selection criteria we focus on the shared characteristics of LS 5039 and \lsi, since PSR B1259-63 is only detectable during a very small fraction of its 1237 day orbit and the probability of detecting a similar long period system is consequently rather low. We then use the generated criteria to construct a sample of likely VHE \gr binaries.

\section*{Source Selection}\label{sec:selection}

The final sample of 11 X-ray binaries is shown in Table \ref{tab:sample}, together with an outline of the selection criteria employed, and the degree to which each object in the sample fulfils these criteria. Five characteristics common to both LS 5039 and \lsi were chosen as selection criteria. %Firstly, both systems have a high mass donor star implying the presence of strong stellar winds. The compact primary is likely to be a pulsar in \lsi, but is somewhat ambiguous in nature in the case of LS 5039 although a pulsar is not discounted \cite{dubus06_2}. Both systems exhibit radio structure on millarcsecond scales, indicative of powerful particle acceleration, be it in jets or at a pulsar bow shock. Additionally, both systems have somewhat short orbital periods and comparable periastron distances. Finally we recognise the dual associations with soft \gr sources. 
Based upon the observed similarities we should select short period ($P\sim3-20$ days) systems with high mass donors, feeding neutron star or pulsar primaries, displaying extended milliarcsecond radio structure and carrying associations with known soft \gr sources. 

Unfortunately, there are only two known systems which fulfil all of these criteria, and these are LS 5039 and LS I +61$^{\circ}$303. In fact, choosing targets which do not  precisely match the \gr binary template gives a useful diagnostic of which system properties or combinations thereof are important for the generation of a detectable TeV flux. Suitable targets were identified using  the X-ray binary catalogues of \cite{liu06} and \cite{liu01} together with references therein.
Ultimately, seven of the targets in our sample were chosen because they share at least some of the characteristics of our idealised \gr binary. The remaining four systems, GRS 1915+105, Circinus X-1, GX 339-4 and V4641 Sgr are known superluminal sources. The possibility of observing a transient VHE \gr flare during a superluminal outburst event was seen as sufficient justification for their inclusion in the survey. 

 \begin{table*}
 \begin{center}
 \begin{tabular}{|l|c|c|c|c|c|c|}
 \hline
 \textbf{Name}	&\textbf{Companion}	&\textbf{Compact}	&\textbf{Radio}&\textbf{Orbital}&\textbf{$\gamma$-ray}\\
			&\textbf{Type}			&\textbf{Object}	&\textbf{Structure}&\textbf{Period (d)}&\textbf{Emission}\\
\hline\hline
Vela X-1	& OB &NS &No& 8.96& TeV? \\
Cen X-3	& OB & NS&No& 2.09&GeV/TeV?\\
GX339-4	& LM & BH&Yes& ?&No\\
Cir X-1	& LM &NS &Yes& 16.6&No\\
GRO J1665-40	& LM & BH&Yes& 2.62&GeV\\
OAO 1657-415	& B0-B6 Supergiant& NS&No&10.4& No\\
4U 1700-37	& O6.5Iaf$^{+}$ & NS?&No&3.96& No\\
4U 1538-52	& B0I & NS&No& 4&GeV?\\
V4641 Sgr	& LM & BH&Yes& 2.81&No\\
4U 1907+097	& OB/Be & NS&No& 8.38&No \\
GRS 1915+105	& LM & BH&Yes& 35&No\\
\hline
\end{tabular}
\caption{The targets for our X-ray binary survey are shown in the table below. In the Companion Type column the spectral type of the donor is listed unless LM is specified, indicating a low-mass companion. The compact object type is either a black hole (BH) or neutron star (NS). Question marks indicate ambiguity in the quoted values or an absence of data.
}\label{tab:sample}
\end{center}
\end{table*}

 \begin{table*}
 \begin{center}
\begin{tabular}{|l|c|c|c|c|}
\hline
~\textbf{Target Name}~ 	& ~\textbf{Significance}~ 	& ~\textbf{Excess}~	 & ~\textbf{Livetime}~ 	& ~\textbf{Flux Upper Limit} ~ \\
				& \textbf{[$\sigma$]}	& \textbf{[counts]}		& \textbf{[hours]}	&\textbf{($E_{\gamma} > $ 1 TeV) [ph cm$^{-2}$s$^{-1}$]} \\			
\hline\hline
Vela X-1	&-1.407	& -73.000	&4.360 & $3.550\times10^{-12}$\\
Cen X-3	&0.778	& 21.589	& 5.283& $5.845\times10^{-12}$\\
GX339-4 	&1.349	& 106.640	&8.495 & $6.657\times10^{-12}$\\
Cir X-1	&-0.942	&-106.704 	&27.906 & $1.406\times10^{-13}$\\
GRO J1665-40 & 0.939	& 46.390	& 9.536& $1.625\times10^{-11}$\\
OAO 1657-415 &7.197	&604.597	&26.744& $1.440\times10^{-11}$\\
&&&&(\textbf{Contaminated by HESS J1702-420})\\
4U 1700-37	&2.929	&289.137 	&38.708 & $1.058\times10^{-11}$\\
4U 1538-52 &-0.4256	&-22.339	&7.524 &$2.741\times10^{-12}$\\
V4641 Sgr 	&1.215	&62.175 	& 2.554& $1.271\times10^{-12}$\\
4U 1907+097	&-0.567	& -13.490	&14.997 & $1.372\times10^{-12}$\\
GRS 1915+105 &0.156&4.566&19.692&$7.803\times10^{-13}$\\
\hline
\end{tabular}

\caption{The preliminary results of the survey are shown below. The region exposed by the galactic plane scan is crowded with VHE $\gamma$-ray sources, and it is therefore unsurprising that some contamination of our targets occurred due to overlap with known VHE sources. Where the target region is contaminated by the flux from a known TeV source, the derivation of an upper limit is not possible, but there is no way to safely associate the observed flux with the X-ray binary system. In this case the contaminating object is indicated in the Flux Upper Limit column. Negative excesses and significances result purely from fluctuations in the $\gamma$-ray background and should not be interpreted as a genuine deficit in the photon flux.
}\label{tab:results}
\end{center}
\end{table*}

\section*{Analysis and Results}\label{sec:analysis}

Data reduction and analysis were carried out using the standard H.E.S.S. analysis procedure outlined in \cite{aharonian06_2}. The event selection cuts placed on image size, $\theta^{2}$ and the mean reduced scaled parameters are identical to those described as standard in \cite{aharonian06_2}, and are consistent with the expected point-like nature of the target objects. The $\gamma$-ray background was estimated using a `reflected' background model with several run dependent \textit{off} regions defined the same distance from the camera centre as the on region. Areas of the sky containing known TeV $\gamma$-ray sources are precluded from being chosen as \textit{off} regions to ensure that the background estimate remains as uncontaminated as possible. Nonetheless, contamination can occur when the \textit{on} region coincides with a known $\gamma$-ray source. Despite having excellent angular resolution for an instrument of its type, the H.E.S.S. point spread function is somewhat extended, with a 68\% containment radius of $\sim0.1^{\circ}$. For this reason it can be impossible to disentangle the signals from nearby objects. This is particularly difficult when the expected target spectrum and flux are unknown.

As reported in \cite{aharonian06} the region exposed by the galactic plane scan is somewhat crowded with VHE $\gamma$-ray sources, and it is therefore unsurprising that some contamination of our targets did indeed occur. 

Table \ref{tab:results} outlines the results of the survey. Upper limits to the photon flux above 1 TeV have been derived for 10 of the 11 targets. These upper limits represent 99\% confidence intervals derived using the unified Feldman-Cousins method \cite{feldman98}. The remaining target OAO 1657-415, was too close to the known TeV emitter HESS J1702-420 for a reliable upper limit or flux estimate to be obtained. 

\section*{Conclusions}\label{sec:conclusions}

99\% confidence upper limits to the VHE $\gamma$-ray flux above 1TeV have been derived for seven X-ray binaries with properties similar to LS 5039 and four ÔsuperluminalÕ microquasars. No significant detections were obtained. For the LS5039-like systems this could be due to orbital modulation of the TeV flux and observations contemporaneous with a low flux state. In some cases a rather short exposure time might also explain the lack of a detection. However, it may be that all the specific conditions found in the LS 5039 and \lsi are required to produce a detectable VHE $\gamma$-ray signal. For the superluminal sources, failure to observe during a flaring event is the most likely explanation for a non-detection.

In the absence of a significant detection, it seems conspicuous, given that nearly 300 X-ray binaries are known, that only three should be detectable in the VHE \gr band. It may be that LS 5039, \lsi and PSR B1259-63 are unique systems in our galaxy, or perhaps with the advent of more sensitive instruments such as \hess II combined with the high energy \gr coverage of \textit{GLAST} will reveal a much larger population of faint \gr binaries.

\textit{Acknowledgements}
The support of the Namibian authorities and of the University of Namibia
in facilitating the construction and operation of H.E.S.S. is gratefully
acknowledged, as is the support by the German Ministry for Education and
Research (BMBF), the Max Planck Society, the French Ministry for Research,
the CNRS-IN2P3 and the Astroparticle Interdisciplinary Programme of the
CNRS, the U.K. Science and Technology Facilities Council (STFC),
the IPNP of the Charles University, the Polish Ministry of Science and 
Higher Education, the South African Department of
Science and Technology and National Research Foundation, and by the
University of Namibia. We appreciate the excellent work of the technical
support staff in Berlin, Durham, Hamburg, Heidelberg, Palaiseau, Paris,
Saclay, and in Namibia in the construction and operation of the
equipment.

\bibliographystyle{plain}

%%%%%%%%
%  17  %
%%%%%%%%

%The paper title 
\title{Localising the H.E.S.S. Galactic Center point source} 

%Short title to print in the headers to the final publication (Not
%showed in this print).  
\shorttitle{H.E.S.S. Galactic Center location} 

%All paper authors 
\authors{C. van Eldik$^{1}$,
O.Bolz$^{1}$, I. Braun$^{1}$, G. Hermann$^{1}$, J. Hinton$^{2}$,
W. Hofmann$^{1}$.}  

\shortauthors{C. van Eldik et al.} 

\afiliations{$^1$Max-Planck-Institut f\"ur
Kernphysik, Saupfercheckweg 1, 69117 Heidelberg, Germany\\ $^2$School
of Physics \& Astronomy, University of Leeds, Leeds LS2 9JT, UK}
\email{Christopher.van.Eldik@mpi-hd.mpg.de}

%The abstract.  
\abstract{Observations by the H.E.S.S. system of
imaging atmospheric Cherenkov telescopes provide the most sensitive
measurements of the Galactic Centre region in the energy range 150 GeV
- 30 TeV. The vicinity of the kinetic centre of our galaxy harbours
numerous objects which could potentially accelerate particles to very
high energies (VHE, $> 100$~GeV) and thus produce the $\gamma$-ray flux 
observed. Within statistical and systematic errors, the centroid of
the point-like emission measured by H.E.S.S. was found 
\cite{Aharonian:2006wh} to be in good agreement with the position of the
supermassive black hole Sgr~A* and the recently discovered
PWN candidate G359.95-0.04 \cite{Wang:2005ya}. Given a 
systematic pointing error of about 30'', a possible association with
the SNR Sgr~A~East could not be ruled out with the 2004 \hess\ data.
In this contribution an update is given on the position of the
H.E.S.S. Galactic Centre source using 2005/2006 data. The systematic
pointing error is reduced to 6'' per axis using guiding telescopes for
pointing corrections, making it possible to exclude with high
significance Sgr A East as the source of the VHE \grs.}

\maketitle

%%%%% Begin Galactic Center %%%%%%
\addtocontents{toc}{\protect\contentsline {part}{\protect\large Galactic Center}{}}
\addcontentsline{toc}{section}{Localising the H.E.S.S. Galactic Center point source}
\setcounter{figure}{0}
\setcounter{table}{0}
\setcounter{equation}{0}

%Begin the section.
\section*{Introduction} The centre of the Milky-Way is the most violent
and active region in our galaxy. Dust along the line of sight prevents
observations of the Galactic Centre (GC) by optical telescopes, but
precise data from this region have been obtained at radio, infrared,
X-ray, and hard X-ray/soft \gr\ ($<200$~keV) energies. These data have
established the 
existence of a $2.6\times 10^6\ M_{\odot}$ black hole at the kinematic
centre of our galaxy, commonly identified with the bright compact
radio source \astar, surrounded by a massive star cluster, a bright
supernova remnant shell, and giant molecular clouds (see, e.g., 
\cite{Melia07,Genzel:2007aa} for recent reviews).

VHE \gr\ emission from the direction of the Galactic Centre was
reported by several ground-based \gr\ observatories
\cite{Kosack:2004ri, Tsuchiya:2004wv, Aharonian:2004wa,
Albert:2005kh}. A recent deep exposure by \hess\
\cite{Aharonian:2006au} revealed the existence of two discrete VHE
\gr\ sources, on top of diffuse emission along the inner 300~pc of the
Galactic Centre ridge. One of the sources, HESS~J1747-281
\cite{Aharonian:2005br}, is identified with the pulsar wind nebula
(PWN) associated with the supernova remnant (SNR) G0.9+0.1. However,
no unique identification is possible for \hgc, the position of which
is within errors coincident with the kinematic centre of our galaxy.

A firm identification of \hgc\ is difficult because the GC region is
densely packed with sources of non-thermal radiation -- possibly
emitting at
VHE energies. In direct vicinity of the \hess\ source, at least three
different objects are discussed as possible counterparts of \hgc.
First, various models predict VHE \gr\ production near the
super-massive black hole itself (see, e.g., \cite{Aharonian:2005ti}).
\astar\ is partially 
surrounded by the bright, shell-like radio emission of the SNR \aeast\ 
\cite{Maeda:2005}, which is the second favoured candidate counterpart
of the VHE \gr\ emission. Finally, in a deep Chandra survey, \pwn, a
candidate pulsar wind nebula, was recently discovered
\cite{Wang:2005ya} only $8.7''$ away from \astar. Despite its faint
X-ray flux, models \cite{Hinton:2006zk} predict a TeV \gr\ flux that is
compatible with \hess\ observations. 

A precise localisation of \hgc\ is essential for shedding
light onto this source confusion. In this paper preliminary results
concerning a refined
position measurement of \hgc\ are reported using an improved telescope
pointing strategy, for which the systematic error on the observation
position is reduced by a factor of three compared to previous results.

\section*{\hess\ observations of the Galactic Centre region} 

The most precise published results on the position of \hgc\ are
based on a 
50~h exposure carried out with the \hess\ array in 2004.  Within a
statistical error of 14'' the best-fit position of \hgc\ was found
\cite{Aharonian:2006wh} to coincide with the position of \astar. The
systematic pointing error of the \hess\ telescope system for this data
set is about 28'', already the most precise pointing in the field of
ground-based \gr\ astronomy.  

The results reported here are based on data recorded between May 14th
and July 27th, 2005, and between April 4th and September 24th, 2006.
The total good-quality exposure of the dataset is 73.2~h (live
time). Most of the 
data (66.1 h) were taken in ``wobble mode'' around \astar, i.e. the
observation direction was offset from the source direction by
typically $0.5^\circ-0.7^\circ$ in either right ascension or
declination. The remaining data were taken at various offsets, within
$1.4^\circ$ from \astar. The zenith angle distribution ranges from
$6^\circ-60^\circ$, and the mean zenith angle of observation is
$21.6^\circ$.

Data were analysed with the standard \hess\ calibration and
reconstruction chain \cite{crab}. \emph{Hard cuts}
\cite{Benbow:2005wj} were used for \gr\ selection, resulting in a
sample of well-reconstructed showers with an average angular
resolution of $0.07^\circ$ (68\% containment radius). 
The data show a strong excess of \grs\ from the direction of the GC
source \hgc, accompanied by diffuse \gr\ emission along the Galactic
Plane. An excess of 1300 $\gamma$~events is found within $0.1^\circ$
from the GC, corresponding to a statistical significance of 44.3 standard
deviations above background. The integral \gr\ flux above 1~TeV is in
agreement with published results based on 2004 data
\cite{Aharonian:2006wh}.

\section*{Precision pointing} For an exact localisation of the centroid
of the VHE \gr\ emission, precise knowledge of the telescope pointing
direction is mandatory. The pointing deviation of individual telescopes is
typically of the order of 2-3'. Various causes have been identified,
with the most important ones being
small misalignments of azimuth and altitude axes during
construction,
sagging of telescope foundations over time,
(mostly) elastic deformations of the masts connecting the camera
body to the mirror dish,
gravitational bending of the mirror dish, and
inelastic deformations of the whole structure leading to
hysteresis effects.
The amount these effects contribute to the mispointing
strongly depends on the observation direction. It should however be
noted that - due to the rigidity of the steel construction - the overall
pointing deviation is very small given the size and weight of the \hess\
telescopes. 

\begin{figure}[htbp]
  \includegraphics[width=0.48\textwidth]{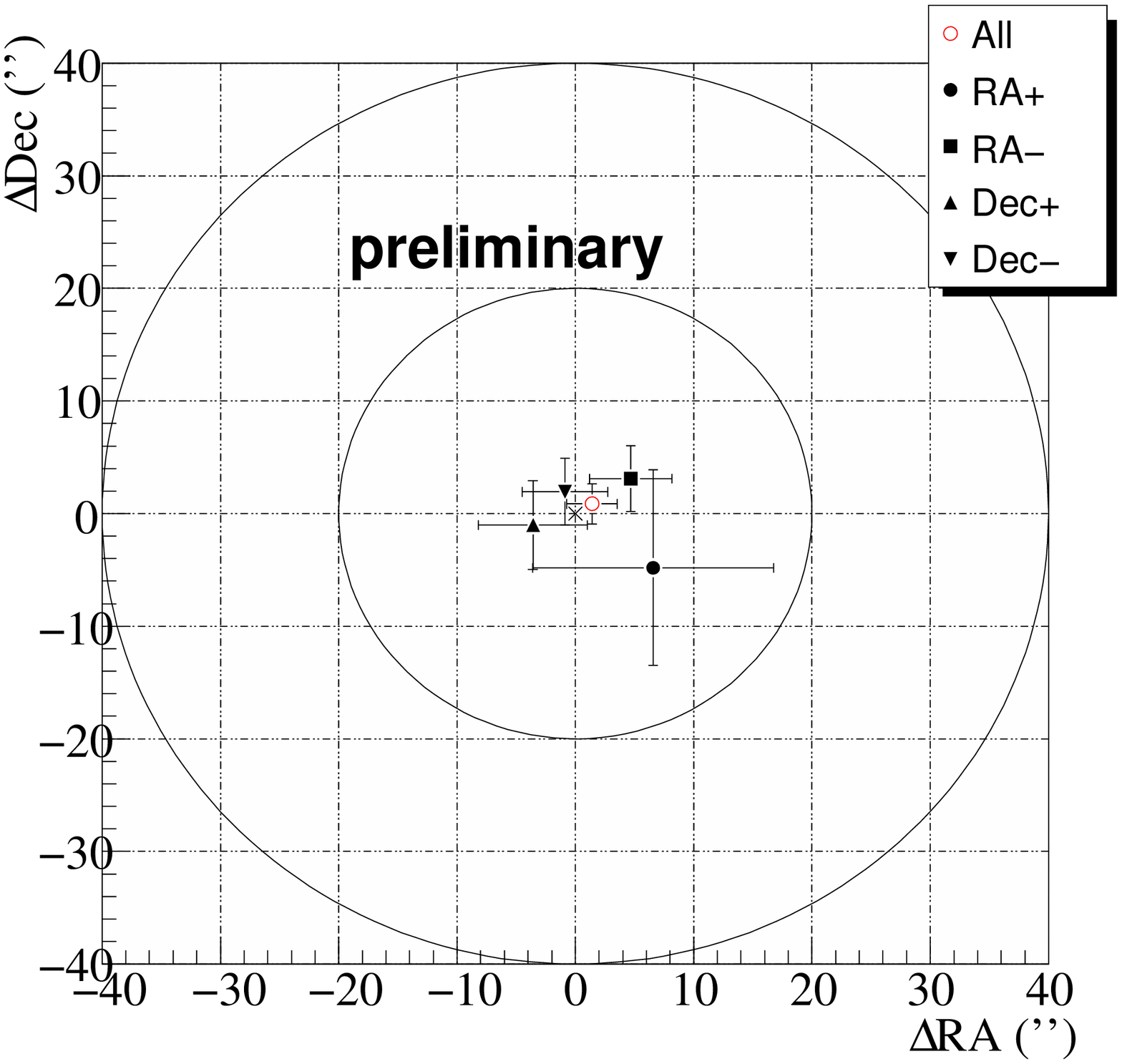}
  \caption{Position of the centroid of VHE \gr\ emission from
    PKS~2155-304 relative to its nominal position. Data were taken in 2006
    during an exceptional VHE \gr\ flare of this source
    \cite{benbow07}. The \gr\ excess was fit by a two-dimensional
    multi-gaussian profile representing the point spread function of the
    \hess\ instrument. The red data point shows the position derived from
    the full data set. When subdividing the data into the four wobble
    offsets, the positions shown by the black symbols are obtained. Note
    that for most of the RA+ wobble data, no bright stars were found in
    the field of view of the guiding telescopes, reducing the available
    live time for this analysis.}
  \label{fig:PointSourceTest}
\end{figure}

Most pointing deviations can be corrected for by taking calibration data at
regular intervals. Each telescope is pointed at typically 50 bright
stars uniformly distributed in the sky. The star is imaged by the
telescope mirror onto a screen in front of the Cherenkov camera, and
an image of the spot is recorded by a central CCD camera mounted at
the centre of the mirror dish. The position of each spot is then
compared to the nominal centre of the Cherenkov camera as determined
from eight positioning LEDs mounted on the camera body. The data are
fit with a 17 parameter model which accounts for elastic deformations
of the telescope structure. In the analysis of \gr\ data, this model
is then used to correct the position of the shower images in the focal
plane of the Cherenkov cameras. The precision achieved on the
observation direction of the \hess\ array is about 20'' per axis
\cite{Gillessen:2004tc}.

For the 2005-2006 data set presented here, the systematic error is
reduced 
further using guiding cameras mounted at each telescope. During
$\gamma$-ray observations, stars in the field of view
($0.3^\circ\times 0.5^\circ$) of these cameras are recorded at a
typical rate of 1~min$^{-1}$, and their reconstructed positions
matched to the Hipparcos and Tycho star catalogues. From this
information position-dependent corrections in right ascension and
declination are calculated for the individual \hess\
telescopes. Additionally, the position of the Cherenkov camera is
monitored by the central CCD camera. With this method, the systematic
error on the telescope orientation is reduced to 6'' per axis for
observations with the full \hess\ array (\cite{braun07}, details will
be published elsewhere).

The procedure was extensively tested on VHE $\gamma$-ray point sources
of known position. Fig. \ref{fig:PointSourceTest} shows a representative
study on the position of the high-frequency peaked BL Lac
PKS~2155-304. Excellent agreement with the nominal position of the
source is found even when splitting the data into different wobble
offsets.

\section*{Position of \hgc} 
The position of \hgc\ is determined by
fitting, in a window of $\pm 0.2^\circ$ around the maximum excess, the
acceptance corrected and background subtracted \gr\ count map. 
Diffuse \gr\ emission is subtracted prior to the fit using the model
presented in \cite{Aharonian:2006au}.  The width of the 2-dimensional
gaussian fit to these data is composed of a fixed term
describing the mean angular resolution of the data set, and a
parameter left free to fit the intrinsic size of the source. The count
map is divided into sky bins of $0.04^\circ\times 0.04^\circ$, and the
fit function is integrated over the bin area for best
accuracy. $\chi^2$-minimisation is used to obtain the best-fit
position.

\begin{figure}[htbp]
  \includegraphics[width=0.52\textwidth]{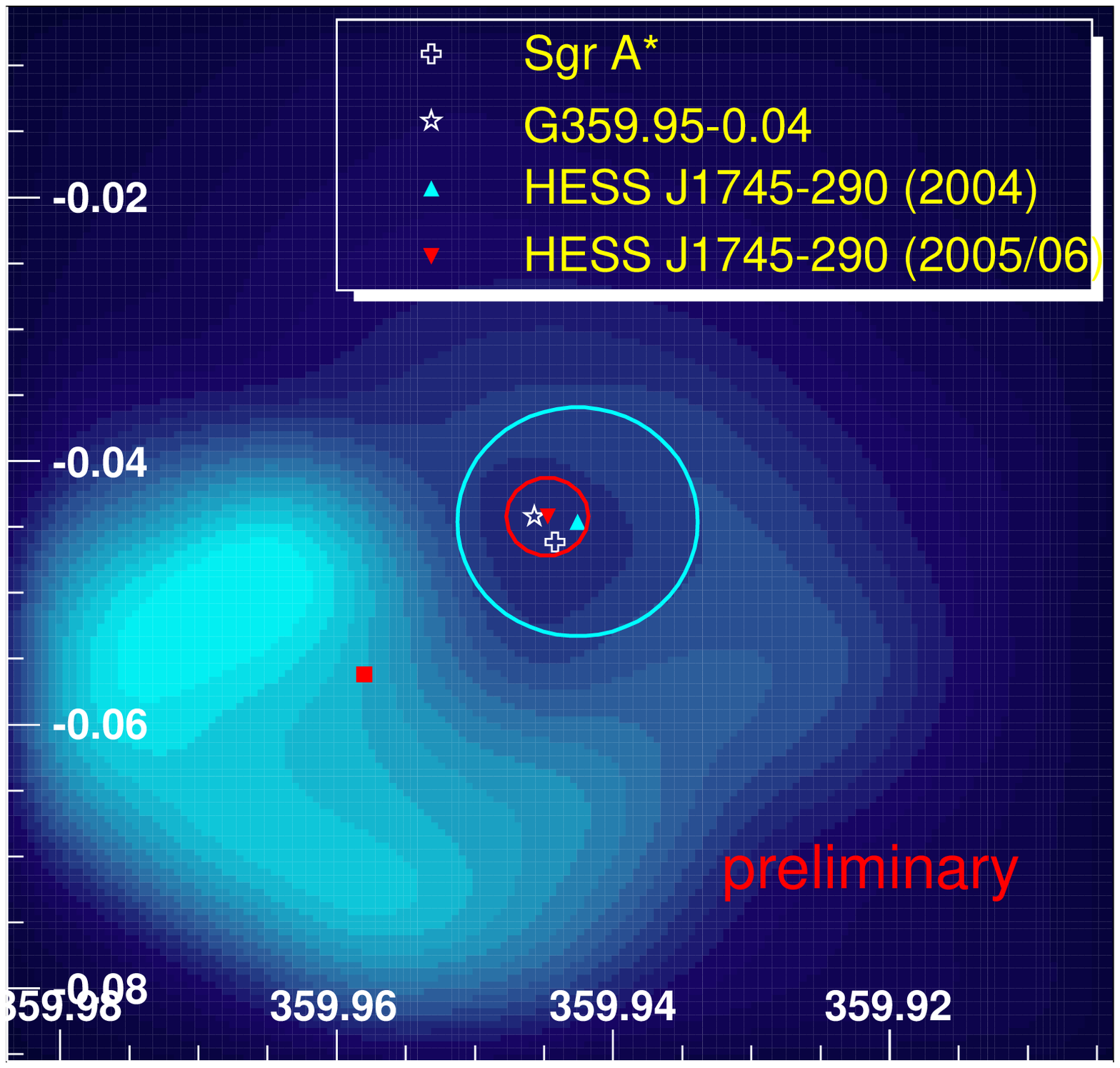}
  \caption{Smoothed 90~cm VLA radio image (reproduced from \cite{LaRosa00})
    of the SNR \aeast\ in Galactic coordinates. The
    position of \astar\ and \pwn\ are marked with a cross and a star,
    respectively. The blue triangle and circle mark the best fit position
    and total error (68\% CL) from the 2004 data set
    \cite{Aharonian:2006wh}. The best fit result of this analysis is shown
    by the red triangle and red circle. The red square marks the expected
    position of the centroid of the VHE \gr\ emission if it followed the
    observed radio flux of \aeast.}
  \label{fig:GCPosition}
\end{figure}

The best-fit position of \hgc\ in Galactic coordinates is $l=359^\circ
56'41.1''\pm 6.4''$~(stat.), $b=-0^\circ 2'39.2''\pm 5.9''$~(stat.).
These results are preliminary and subject to final checks.
Fig. \ref{fig:GCPosition} shows the new \hess\ position measurement on
top of a 90~cm VLA radio image of the inner 10~pc region of the
GC. The shell-like structure of the SNR \aeast\ is clearly visible.
The position of \hgc\ is coincident within $7.3'' \pm 8.7''$~(stat.)
$\pm8.5''$~(syst.) with the radio position of \astar\
\cite{saga_radio}, and is also consistent with the position reported from
the 2004 data set \cite{Aharonian:2006wh}. While the latter 
was marginally consistent with the radio emission from \aeast, the
result obtained in this analysis does rule out \aeast\ as the
counterpart of \hgc\ with high significance. Due to the improved
pointing accuracy of the \hess\ array, the probability that the
observed \gr\ flux is produced near the radio maximum of \aeast\ is
about $10^{-11}$. Assuming that the VHE \gr\ flux follows the radio
morphology of \aeast\ (corresponding to the red square in
Fig. \ref{fig:GCPosition}), the chance probability of finding the  
centroid of the emission at the reported position is $10^{-7}$. 

The position of \hgc\ agrees well with the location of the other two
counterpart candidates, \astar\ and \pwn, which are separated by only
$8.7''$. Since the pointing precision 
obtained in this work is at the limit of what can be achieved with
an instrument such as \hess, other measures have to be taken to
disentangle the remaining source confusion. The most promising method is
to search for variability in the VHE \gr\ flux, which would hint at
a connection between the VHE flux and \astar. The most convincing
signature would
be the detection of correlated flaring in X-rays and VHE \grs.
Such searches have been presented at this conference
\cite{vivier07, hinton07}.

\section*{Acknowledgements} \small The support of the Namibian
authorities and of the University of Namibia in facilitating the
construction and operation of H.E.S.S. is gratefully acknowledged, as
is the support by the German Ministry for Education and Research
(BMBF), the Max Planck Society, the French Ministry for Research, the
CNRS-IN2P3 and the Astroparticle Interdisciplinary Programme of the
CNRS, the U.K. Science and Technology Facilities Council (STFC), the
IPNP of the Charles University, the Polish Ministry of Science and
Higher Education, the South African Department of Science and
Technology and National Research Foundation, and by the University of
Namibia. We appreciate the excellent work of the technical support
staff in Berlin, Durham, Hamburg, Heidelberg, Palaiseau, Paris,
Saclay, and in Namibia in the construction and operation of the
equipment.

\bibliographystyle{plain}
\normalsize

%%%%%%%%
%  18  %
%%%%%%%%

%The paper title
\title{Simultaneous H.E.S.S. and Chandra observations of Sgr A$^{\star}$ during an X-ray flare}
%Short title to print in the headers to the final publication (Not showed in this print).
\shorttitle{Observations of Sgr A$^{\star}$ during an X-ray flare}

%All paper authors
\authors{Jim Hinton$^{1}$, Matthieu Vivier$^{2}$, 
Rolf B\"uhler$^{3}$, 
Gerd P\"uhlhofer$^{4}$, Stefan Wagner$^{4}$ \\ 
for the H.E.S.S. Collaboration}
%Short title to print in the headers to the final publication (Not shown in this print).
\shortauthors{H.E.S.S. Collaboration}
%All the affiliations.
\afiliations{
$^1$School of Physics \& Astronomy, University of Leeds, Leeds LS2 9JT, UK\\
$^2$DAPNIA/DSM/CEA, CE Saclay, F-91191 Gif-sur-Yvette, Cedex, France\\
$^3$Max-Planck-Institut f\"ur Kernphysik, P.O. Box 103980, D 69029 Heidelberg, Germany\\
$^4$Landessternwarte, Universit\"at Heidelberg, K\"onigstuhl, D 69117 Heidelberg, Germany
}

\email{j.a.hinton@leeds.ac.uk}

%VIVIER, Matthieu (DAPNIA/DSM/CEA, CE Saclay)
%BUEHLER, Rolf (Max-Planck-Institut fuer Kernphysik, Heidelberg)
%PUEHLHOFER, Gerd (Landessternwarte, Universitaet Heidelberg)
%WAGNER, Stefan ()

%The abstract.
\abstract{The rapidly varying non-thermal X-ray emission observed from Sgr A$^{\star}$ points to
particle acceleration taking place close to the supermassive black hole. The TeV
$\gamma$-ray source HESS\,J1745$-$290 is coincident with Sgr A$^{\star}$ and may be closely related
to the X-ray emission. Simultaneous X-ray and TeV observations are required to
elucidate the relationship between these two objects. Here we report on joint
H.E.S.S./Chandra observations in July 2005, during which an X-ray flare was detected.
Despite a factor $>10$ increase in the X-ray flux of Sgr~A$^{\star}$, no evidence
is found for an increase in the TeV $\gamma$-ray flux. We find that an increase
of the $\gamma$-ray flux of a factor 2 or greater can be excluded at a 
confidence level of 99\%. This finding disfavours scenarios in which the bulk
of the $\gamma$-ray emission observed is produced close to Sgr A$^{\star}$.
}

\maketitle

\addcontentsline{toc}{section}{Simultaneous H.E.S.S. and Chandra observations of Sgr A$^{\star}$ during an X-ray flare}
\setcounter{figure}{0}
\setcounter{table}{0}
\setcounter{equation}{0}

\section*{Introduction}

The existence of a supermassive ($3.6\pm0.3 \times
10^{6}$ solar mass) black hole at the centre of our galaxy has
been inferred using measurements of stellar orbits in the central parsec
(see e.g. ~\cite{GC:Eisenhauer05}). The supermassive black hole (SMBH)
is coincident with the faint radio source: Sgr~A$^{\star}$.  
The compact nature of Sgr~A$^{\star}$ has been demonstrated both by direct VLBI
measurements~\cite{GC:Shen05} and by the observation of X-ray and near
IR flares with timescales as short as a few minutes (see for
example~\cite{GC:Eckart06,GC:Porquet03}).  Variability on such short
timescales limits the emission region (via causality arguments) to within $<10$ Schwarzchild
radii of the black hole. X-ray flares from \astar\ have reached fluxes of $4\times10^{35}$ erg
s$^{-1}$, two orders of magnitude brighter than the quiescent
flux~\cite{GC:Porquet03, GC:Baganoff03}, and exhibit a range of
spectral shapes~\cite{GC:Porquet03}.  Several models exist for the origin of
this variable emission, all of which invoke non-thermal processes
close to the event horizon of the central black hole to produce a
population of relativistic particles.

Model independent evidence for the existence of ultra-relativistic 
particles close to Sgr~A$^{\star}$ can be provided by the observation
of TeV $\gamma$-rays from this source. Indeed, TeV 
$\gamma$-ray emission has been detected from the Sgr~A region 
by several ground-based instruments~\cite{GC:Whipple04,GC:CANGAROO,HESS:gc04,
GC:MAGIC06}. The most precise measurement of this source, HESS\,J1745$-$290,
are those made using the H.E.S.S. telescope array. The centroid
of the source is located $7'' \pm 14_{\mathrm{stat}}'' \pm 28_{\mathrm{sys}}''$ 
from Sgr~A$^{\star}$, and has an rms extension of  $<1.2'$~\cite{HESS:gcprl}.

TeV emission from \astar\ is expected in several models of 
particle acceleration in the environment of the black hole.
In some of these scenarios \cite{Levinson, AhNer1} TeV 
emission is produced in the immediate vicinity of the  
SMBH and variability is expected. In alternative scenarios
particles are accelerated at \astar\ but radiate in within
the central $\sim$10~parsec region~\cite{AhNer2}, 
or are accelerated at the termination shock of a wind 
driven by the SMBH \cite{AtDer}.
However, several additional candidate objects exist for the origin
of the observed $\gamma$-ray emission. The radio centroid of the 
supernova remnant (SNR) Sgr~A East lies $\sim 1'$ from
Sgr~A$^{\star}$, only marginally inconsistent with the position
of the TeV source given in \cite{HESS:gcprl}. Shell-type SNR are now well established TeV
$\gamma$-ray sources~\cite{HESS:rxj1713,HESS:velajnr}
and several authors have suggested Sgr A East as the origin of
the TeV emission (see for example \cite{Melia}). However,
recent improvements in the statistical and systematic 
uncertainties of the centroid of HESS\,J1745$-$290
effectively exclude Sgr~A East as the dominant 
$\gamma$-ray source in the region~\cite{VanEldik}.
The recently discovered pulsar wind nebula candidate G\,359.95-0.04~\cite{GC:Wang06}
lies only 9 arcseconds from \astar\ and can plausibly explain
the TeV emission~\cite{GC:Hinton07}. Particle
acceleration at stellar wind collision shocks within the central
young stellar cluster has also been hypothesised to explain
the $\gamma$-ray source~\cite{GC:Quataert05}. Finally, an origin of this source
in the annihilation of WIMPs in a central dark matter cusp has 
been extensively discussed~\cite{GC:Hooper04,GC:Profumo05,HESS:gcprl}.
%%% Ballantyne07 (diffuse?), Liu06a (flares), Liu06b (HESS), Liu06c (x-ray), 
%%% Lu06 (red giant), Atoyan (BH wind), Loeb (stellar winds)

Given the limited angular resolution of current VHE $\gamma$-ray 
telescopes, the most promising tool for identification of the
TeV source is the detection of \emph{correlated variability}
between the $\gamma$-ray and X-ray and/or NIR regimes.
A significant increase of the flux of HESS\,J1745$-$290 
simultaneous with a flare in wavebands with sufficient 
angular resolution to isolate \astar, would provide an 
unambiguous identification of the $\gamma$-ray source.
Therefore, whilst not all models for TeV emission from 
Sgr~A$^{\star}$ predict variability of the VHE source, 
coordinated IR/keV/TeV observations can be seen as a key
aspect of the ongoing program to understand the nature
of this enigmatic source.

\section*{Observations \& Results}

A coordinated multi-wavelength campaign on \astar\ 
took place during July/August 2005. As part of this campaign 
observations with H.E.S.S. occurred
for 4-5 hours each night from the 27th of July to the 1st of August
(MJD  53578-53584). Four Chandra observations with IDs 5950-5954 took 
place between the 24th of July and the 2nd of August.
A search for flaring events in the X-ray data yielded
two significant events during the Chandra campaign, both during 
observation ID 5953 on the 30th of July. The second of these flares 
occurred during a period of H.E.S.S. coverage, at MJD 53581.94.

\begin{figure}[t]
\begin{center}
\noindent
%\fbox{\hbox{\vbox{\hsize=50mm \hfill \vspace{130mm}}}}
\includegraphics [width=0.52\textwidth]{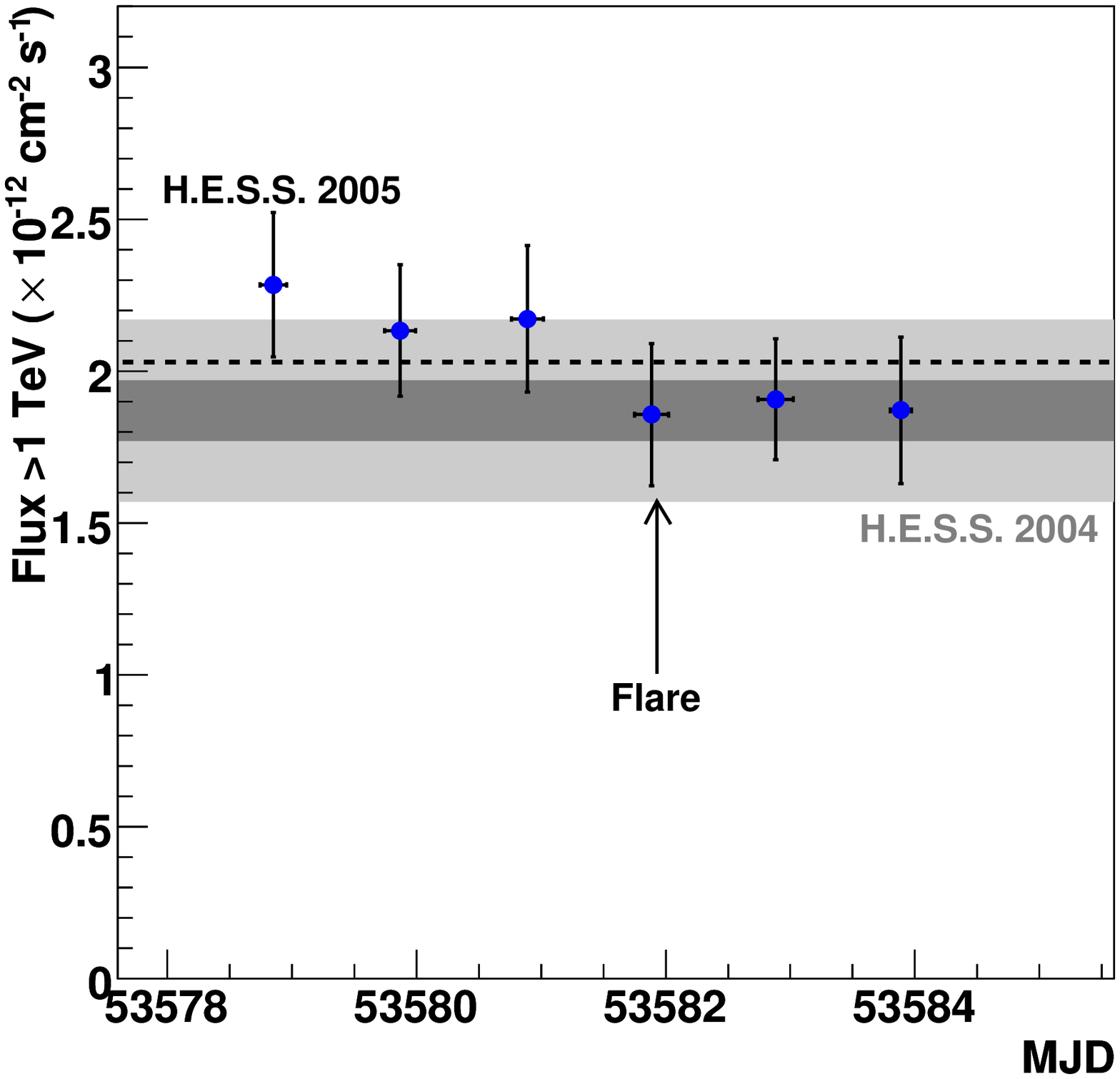}
\caption{Nightly $\gamma$-ray flux light-curve of HESS\,J1745$-$290 from the 
27th of July to the 1st of August 2005.
The H.E.S.S. data have typical thresholds of 150-300 GeV. 
The grey band shows the mean flux from 2004 observations as
published in \cite{HESS:gcprl}. Statistical (dark grey) and
systematic (light grey) errors are shown. The dashed line 
is a fit to the MJD 53578-53584 data.
}\label{fig0}
\end{center}
\end{figure}

The $\gamma$-ray data consist of 72 twenty-eight minute runs,
66 of which pass all the quality selection cuts described by 
\cite{HESS:crab}. All runs on the night of the X-ray flare pass
these cuts and in addition we find no evidence for cloud cover in the 
simultaneous sky temperature (radiometer) measurements 
(see \cite{HESS:crab,HESS:atmosphere}). 
These data were analysed using the H.E.S.S. standard 
\emph{Hillas parameter} based method 
described in \cite{HESS:crab}. An independent analysis based
on the \emph{Model Analysis} method described in \cite{deNaurois:Model} 
produced consistent results. Figure~\ref{fig0} 
shows a night-by-night TeV flux light-curve for this period.
There is no evidence for variations of the flux on day 
timescales and the mean $>1$ TeV $\gamma$-ray flux for this 
week of observations 
was $2.03 \pm 0.09_{\mathrm{stat}} \times 10^{-12}$ cm$^{-2}$ s$^{-1}$, 
consistent with the
average value for H.E.S.S. observations in 2004, 
$1.87 \pm 0.1_{\mathrm{stat}} \pm 0.3_{\mathrm{sys}} \times 10^{-12}$ 
cm$^{-2}$ s$^{-1}$~\cite{HESS:gcprl}.

%demonstrate
%the absence of cloud cover during these observations.

%\section*{Results}

Figure~\ref{fig1} shows the X-ray and $\gamma$-ray light curves 
for the night MJD 53581-2. There is a clear increase in the X-ray
flux of \astar\ with an excess of $103\pm10$ events with respect 
to the quiescent level. The time-profile of this excess is 
consistent with a Gaussian of rms $13.1\pm2.5$ minutes. The time
window for the $\gamma$-ray analysis is defined as the region within
$\pm1.3\sigma$ of the X-ray flare (containing 80\% of the signal).
The lower panel of Figure~\ref{fig1} shows the mean TeV flux
within this time window (grey shaded region) of 
$2.05\pm 0.76\,\times\,10^{-12}$ cm$^{-2}$ s$^{-1}$
as a short dashed line. This flux level is almost identical
to the mean flux level for this week of observations.
There is therefore no evidence for an increase in
$\gamma$-ray flux of HESS\,J1745-290 during the flare and a limit on the relative 
flux increase of $<$ a factor 2 is derived at the 99\% confidence 
level. In principle a (positive or negative) 
time lag might be expected between the X-ray and any associated 
$\gamma$-ray flare. The existence of a counterpart $\gamma$-flare with a 
flux increase $\gg2$ requires a lag of at least 100 minutes.

\begin{figure*}
\begin{center}
\noindent
%\fbox{\hbox{\vbox{\hsize=50mm \hfill \vspace{130mm}}}}
\includegraphics [width=1.0\textwidth]{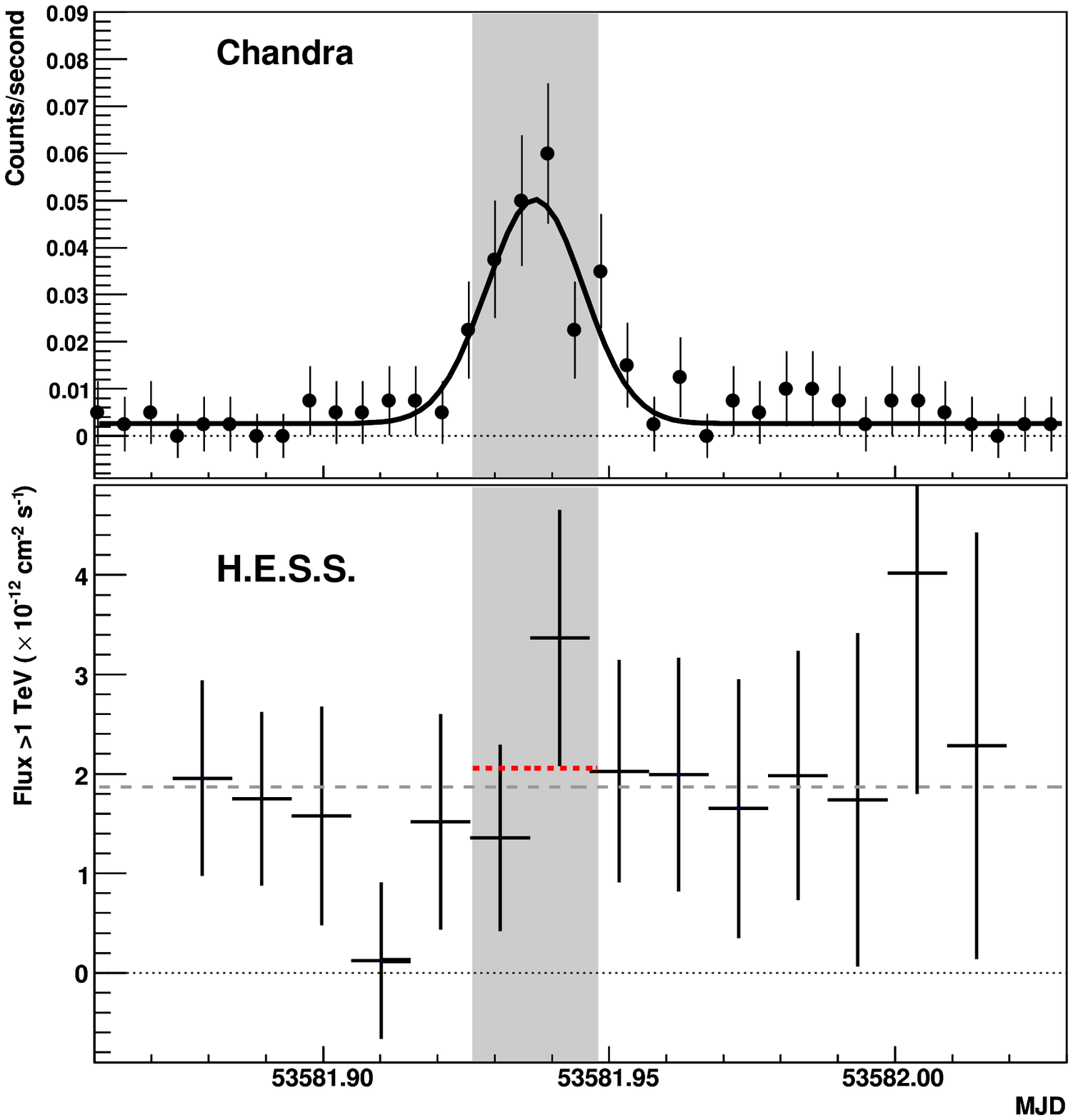}
\caption{X-ray and $\gamma$-ray light curves for the Galactic Centre on MJD 53581. 
  Top: Chandra 1-10 keV count rate in 400 second bins. The X-ray flare is well 
described by Gaussian (solid curve), the shaded region shows $\pm1.3\sigma$ of the
flare position.
% within an $X$ arcsecond  region of \astar. 
Bottom: Very High Energy $\gamma$-ray light curve from H.E.S.S. in
15 minute bins. The long dashed line shows the historical flux
level \cite{HESS:gcprl}. The short dashed line indicates the mean 
TeV flux during the X-ray flare.
}	
\label{fig1}
\end{center}
\end{figure*}

%%\bibliography{flare_icrc}
%This in the bibtex style, is ok.
%\bibliographystyle{plain}

\section*{Summary}

For the first time simultaneous TeV $\gamma$-ray observations 
have been presented for a period of X-ray activity of \astar.
The non-detection of an increase in the TeV flux provides an
important constraint on scenarios in which the source 
HESS\,J1745-290 is associated with the supermassive black hole.

%%%%%%%%
%  19  %
%%%%%%%%

%The paper title
\title{Search for variability and QPO activity from SgrA* from H.E.S.S observations}
%Short title to print in the headers to the final publication (Not showed in this print).
\shorttitle{GC variability}

%All paper authors
\authors{M.Vivier$^{1}$, O.de Jager$^{2}$, J. Hinton$^{3}$ for the H.E.S.S. collaboration}
%Short title to print in the headers to the final publication (Not shown in this print).
\shortauthors{M.Vivier et al.}
%All the affiliations.
\afiliations{$^1$DAPNIA/DSM/CEA, CE Saclay, F-91191 Gif-sur-Yvette, Cedex France\\ $^2$Unit for Space Physics, North-West University, Potchefstroom 2520, South Africa\\$^3$School of Physics and Astronomy, University of Leeds, Leeds LS2 9JT, UK }
\email{matthieu.vivier@cea.fr}

%The abstract.
\abstract{One interesting possibility is that the galactic center (GC) source HESS J1745-290 is associated with the galactic center source Sgr A*, the galactic center black hole, in which case we may expect variability as seen in IR and X-rays, with QPO frequencies predicted by Aschenbach et al. (2006). We will present the results of a search for such variable signatures using HESS observations of this source.}

\maketitle

\addcontentsline{toc}{section}{Search for variability and QPO activity from SgrA* from H.E.S.S observations}
\setcounter{figure}{0}
\setcounter{table}{0}
\setcounter{equation}{0}

%Begin the section.
\section*{Introduction}
H.E.S.S (High Energy Stereoscopic System) \cite{Hofmann} consists of four imaging atmospheric-Cherenkov telescopes (IACT's) located in Namibia. The Cherenkov light emitted by $\gamma$-induced air showers is detected by a camera of 960 photomultiplier tubes, located at the focus of the four 12 m diameter telescopes. The field of view of a camera is 5$^{\circ}$ in diameter. A description of the H.E.S.S. instrument and operation can be found in \cite{Aharonian1, Funk}.\\ 
A strong signal was detected by H.E.S.S. toward the GC \cite{Aharonian2,PRL}. A refined analysis of the HESS J1745-290 source position is presented in \cite{VanEldik}. Given the HESS J1745-290 source position uncertainties, several sources are plausible candidates. Among them, the black hole SgrA* located at the center of our galaxy \cite{NeronovMV,Atoyan} is one of the most popular candidate. The dark matter origin of the H.E.S.S. signal has also been strongly constrained in \cite{PRL}. Thus, it is important to study the flux variability of the source to constrain the emission mechanisms of the detected TeV signal. X-ray flares of SgrA* were detected with periods ranging from ~100 s to 2250 s \cite{Aschenbach}. Here we report a search for similar variabilities in the TeV energy range using H.E.S.S. GC data.

The observations presented here were obtained during the 2004, 2005 and 2006 observing seasons of the GC. Photons were selected and reconstructed with the so-called ''model analysis'' \cite{deNaurois1,deNaurois2}. Spectra are presented elsewhere \cite{Ripken}. Results were checked with a different reconstruction method based on Hillas statistical moment of the images \cite{Hillas}.
The total live time of the whole data set is 97 h. The total significance corresponds to a 59.8 $\sigma$ deviation above the background using a point-like source analysis (i.e. using a cut on the angular distance $\theta$ between the reconstructed direction of the $\gamma$-rays and the pointed direction: $\theta < 0.14^{\circ}$.).\\

\begin{figure}[h]	
\begin{center}
\noindent
%\fbox{\hbox{\vbox{\hsize=130mm \hfill \vspace{50mm}}}}
\vspace{-0.5cm}
\mbox{\hspace{-0.3cm}\includegraphics [width=0.5\textwidth]{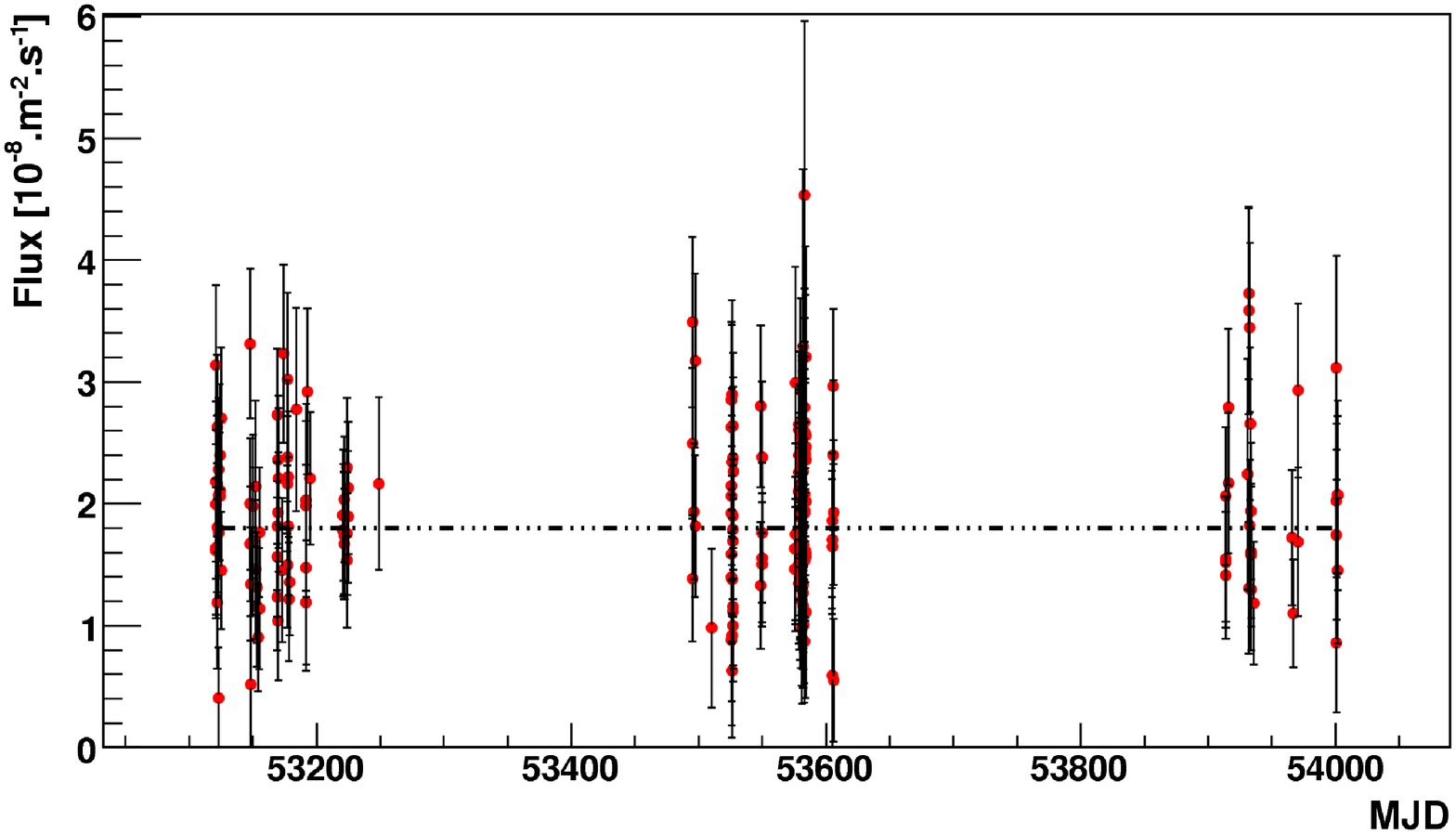}}
\end{center}
\vspace{-0.6cm}\caption{ Run by run light curve of HESS J1745-290. Data cover the 2004-2006 time period.}\label{fig1MV}
\end{figure}

The run by run integral flux of $\gamma$-rays above 1 TeV is shown on Fig.\ref{fig1MV}. A run corresponds to a 28 min observation time. The $\chi^{2}$ of a constant fit to data is 228/211~d.o.f. This is consistent with a constant flux of: 
\begin{equation*}
\Phi(>\mbox{1 TeV})= 2.14 \ 10^{-12} \mbox{cm}^{-2} \mbox{s}^{-1}
\end{equation*}
 No significant variations are detected on timescales longer than 28 min.

\section*{Flare sensitivity}
As mentioned in the introduction, the flaring of SgrA* has been detected in various passbands such as IR or X-rays. Simultaneous H.E.S.S./Chandra observations were carried out during an X-ray flare. No flare was detected by H.E.S.S. as reported in \cite{Hinton}.
Because of the large error bars implied by low statistics, the H.E.S.S. signal is sensitive to relatively larger amplitude flares. The flare sensitivity was estimated by adding a fake gaussian with variable duration $\sigma_t$ and maximum amplification time t$_0$ to the H.E.S.S. light curve (LC). The modified LC is thus represented by:

\begin{equation}
\mbox{LC}_{mod}\mbox{(t)} = \mbox{LC(t)}\times(1+A\times e^{\frac{(t-t_0)^{2}}{\sigma_t^{2}}})
\end{equation}

\begin{figure}[h]	
\begin{center}
\noindent
%\fbox{\hbox{\vbox{\hsize=130mm \hfill \vspace{50mm}}}}
\vspace{-0.3cm}
\mbox{\hspace{-0.2cm}}\includegraphics [width=0.5\textwidth]{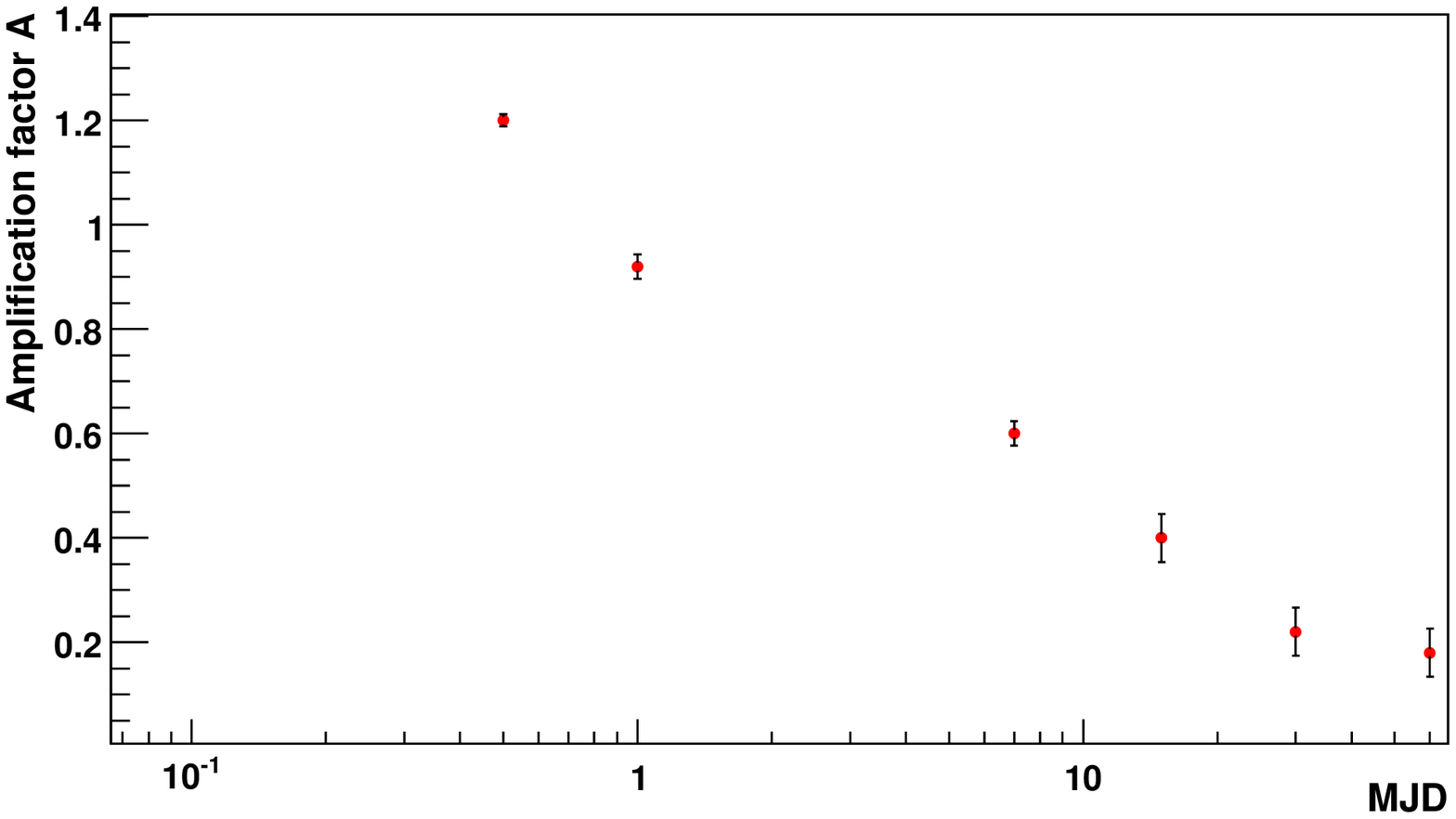}
\end{center}
\vspace{-0.5cm}\caption{Mean sensitivity to a 3-$\sigma$ flare detection as a function of the flare duration. The mean sensitivity decreases with the flare duration.}\label{fig2MV}
\end{figure}

Fig.\ref{fig2MV} shows the maximum amplification factor $A$ compatible with no flare detection at the 3-$\sigma$ confidence level as function of the flare duration. Typical numbers for $A$ are of order unity. $A$ decreases with the flare duration as expected.

\section*{Search for QPO's at the X-ray periods} 

\begin{figure}[h]	
\begin{center}
\noindent
%\fbox{\hbox{\vbox{\hsize=130mm \hfill \vspace{50mm}}}}
\vspace{-0.3cm}
\mbox{\hspace{-0.2cm}}\includegraphics [width=0.5\textwidth]{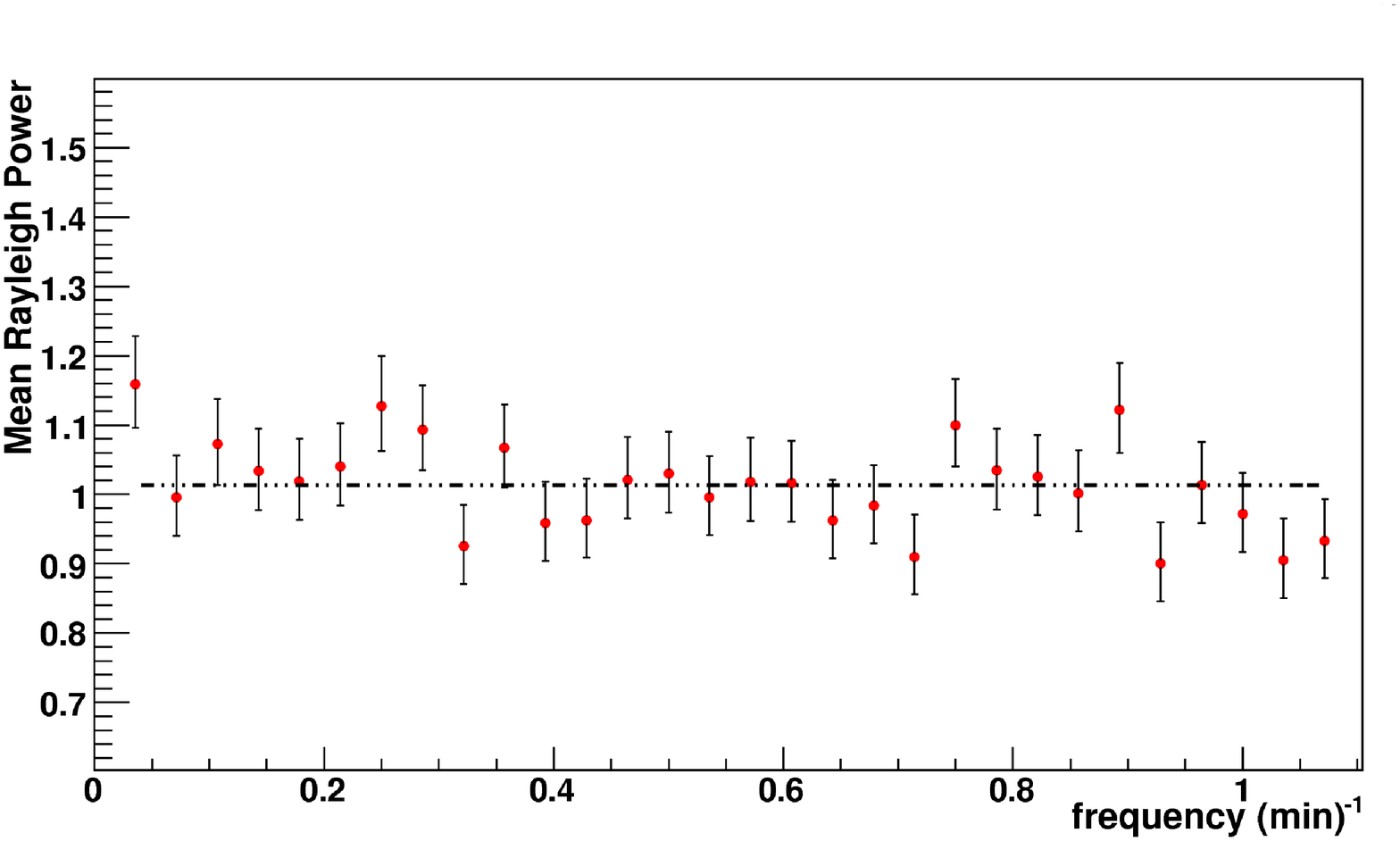}
\end{center}
\vspace{-0.5cm}\caption{Rayleigh power plotted as a function of the frequency. A fit to a constant power gives a $\chi^{2}$ of 35/29 d.o.f. compatible with a flat distribution. }\label{fig3MV}
\end{figure}

Four oscillation frequencies ranging from 100 s to 2250 s have been observed in the X-ray light curve of SgrA* \cite{Aschenbach}. These frequencies are likely to correspond to gravitationnal cyclic modes associated with the accretion disk of SgrA*. We searched for the occurence of these frequencies in our data. First, we assume that the coherence time of oscillations is less than 28 min. We then perform a Rayleigh test \cite{deJager} on photon time arrival distributions for continuous observations of 28 min. The Rayleigh power averaged over 2004-2006 data is shown on Fig.\ref{fig3MV} as a function of the frequency. The probed frequencies range from 1/28 min$^{-1}$ to the inverse of the average time spacing between two consecutive events of 1.2 min$^{-1}$. The Rayleigh power is compatible with a flat function of frequency. No significant peaks are seen at the 100 s, 219 s, 700 s and 1150 s periods observed in X-rays.\\

\begin{figure}[h]	
\begin{center}
\noindent
%\fbox{\hbox{\vbox{\hsize=130mm \hfill \vspace{50mm}}}}
\vspace{-0.3cm}
\mbox{\hspace{-0.2cm}}\includegraphics [width=0.5\textwidth]{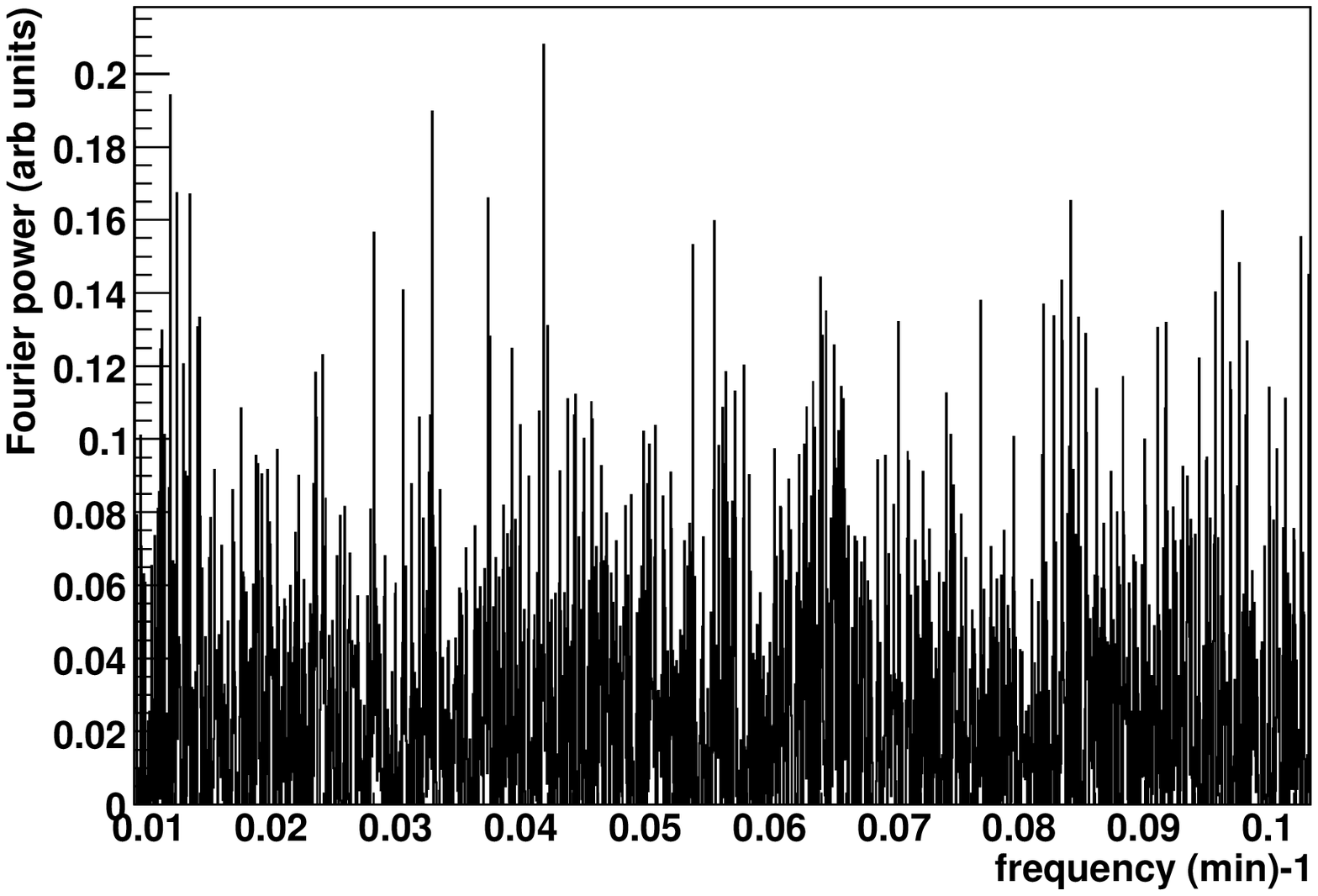}
\end{center}
\vspace{-0.5cm}\caption{[10$^{-2}$ min$^{-1}$ - 0.1 min$^{-1}$] Lomb-Scargle periodogram of the H.E.S.S. SgrA* light curve averaged over the 2004-2006 nights of observation. No significant peak is visible.}\label{fig4MV}
\end{figure}

\begin{figure}[h]	
\begin{center}
\noindent
%\fbox{\hbox{\vbox{\hsize=130mm \hfill \vspace{50mm}}}}
\vspace{-0.3cm}
\mbox{\hspace{-0.2cm}}\includegraphics [width=0.5\textwidth]{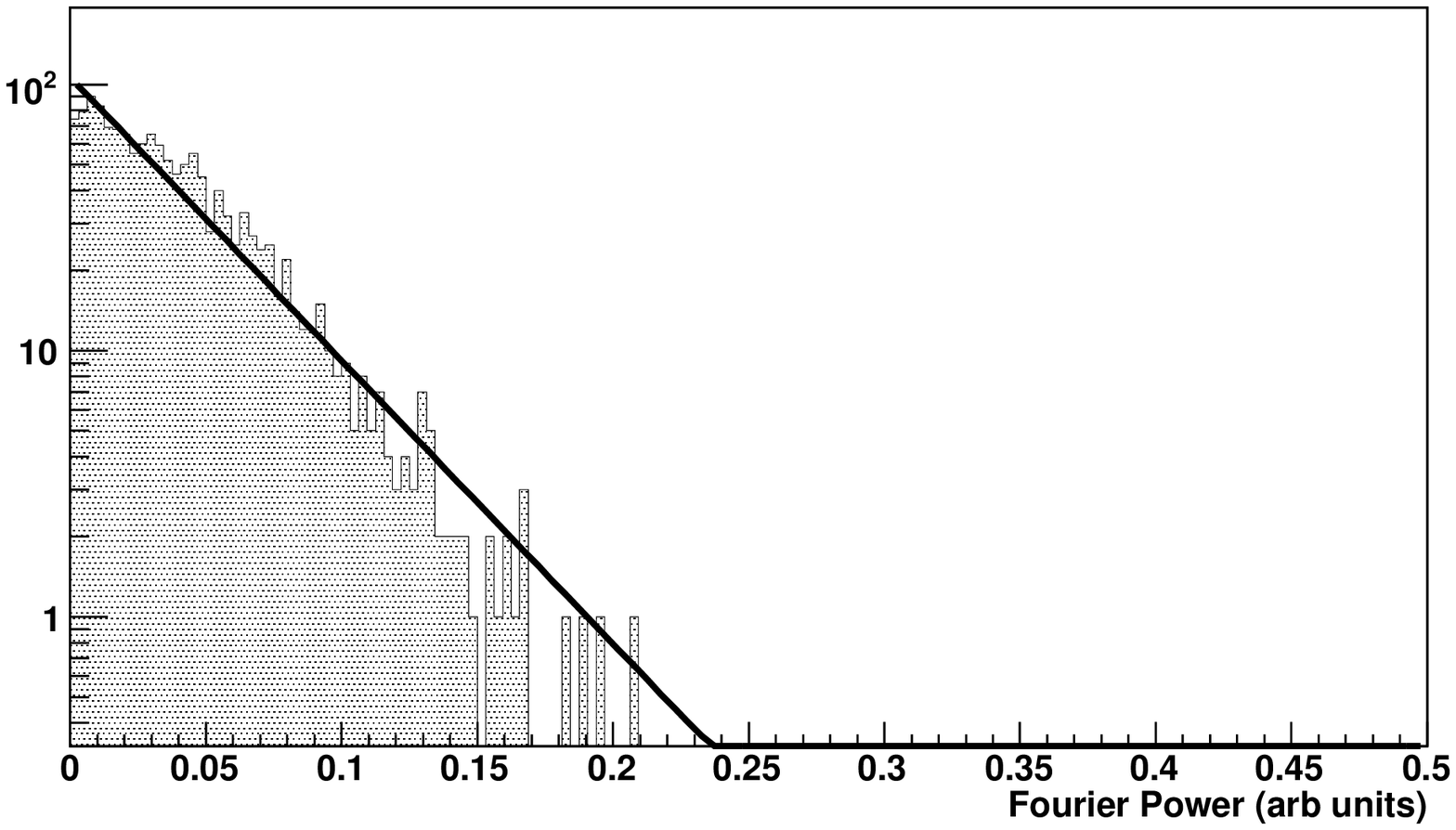}
\end{center}
\vspace{-0.5cm}\caption{Fourier power distribution derived from the average Lomb-Scargle periodogram. The $\chi^{2}$ of the exponential fit to data is 72/55 d.o.f.}\label{fig5MV}
\end{figure}

Next, we assume that the coherence time of oscillations is of the order of a few hours. We then construct the Fourier power distribution using a Lomb-Scargle periodogram \cite{Scargle} for each night of our dataset. Data are binned into 5 min points. The Fourier power averaged over 2004-2006 data is displayed on Fig.\ref{fig4MV} as a function of the frequency. Frequencies tested range from 10$^{-2}$ min$^{-1}$ to 0.1 min$^{-1}$. No significant oscillation frequencies are detected, as shown on Fig.\ref{fig5MV}.

\section*{Conclusions}
The light curve of the very high energy source HESS J1745-290 observed towards the GC is compatible with a constant integrated flux above 1 TeV of \mbox{$\Phi(>1 TeV)= 2.14 \ \mathrm{10^{-12} cm^{-2} s^{-1}}$}. The source appears to be steady in time for timescales from a year down to 1 min. The flare sensitivity study shows that H.E.S.S. was sensitive to flux increases with amplification factor of order of the unity. As the flare duration increases, lower amplification factors are needed for a 3-$\sigma$ detection. Searches for QPO's in the H.E.S.S. GC signal did not uncover any periodicities, in particular none of those observed by \cite{Aschenbach}. The constraints on the variability of HESS J1745-290 will be improved by new observations taken in 2007.

\section*{Acknowlegdments}
The support of the Namibian authorities and of the University of Namibia
in facilitating the construction and operation of H.E.S.S. is gratefully
acknowledged, as is the support by the German Ministry for Education and
Research (BMBF), the Max Planck Society, the French Ministry for Research,
the CNRS-IN2P3 and the Astroparticle Interdisciplinary Programme of the
CNRS, the U.K. Particle Physics and Astronomy Research Council (PPARC),
the IPNP of the Charles University, the Polish Ministry of Science and 
Higher Education, the South African Department of
Science and Technology and National Research Foundation, and by the
University of Namibia. We appreciate the excellent work of the technical
support staff in Berlin, Durham, Hamburg, Heidelberg, Palaiseau, Paris,
Saclay, and in Namibia in the construction and operation of the
equipment.

%\bibliography{biblio.bib}

%%%%%%%%
%  20  %
%%%%%%%%

%The paper title
\title{The H.E.S.S. survey of the inner Galactic plane}
%Short title to print in the headers to the final publication (Not showed in this print).
\shorttitle{The H.E.S.S. survey of the inner Galactic plane}

%All paper authors
\authors{S.~Hoppe$^{1}$,  for the H.E.S.S. collaboration}
%Short title to print in the headers to the final publication (Not shown in this print).
\shortauthors{S.~Hoppe}
%All the affiliations.
\afiliations{$^1$Max-Planck-Institut f\"ur Kernphysik, P.O. Box
  103980, D 69029 Heidelberg, Germany}

\email{stefan.hoppe@mpi-hd.mpg.de}

%The abstract.
\abstract{The High Energy Stereoscopic System (H.E.S.S.), located in
  the Khomas Highlands of Namibia, is an array of four imaging
  atmospheric-Cherenkov telescopes designed to detect $\gamma$-rays in
  the very high energy (VHE; $>$ 100 GeV) domain. Its high sensitivity
  and large field-of-view (5$^{\mbox{\tiny o}}$) make it an ideal
  instrument to perform a survey within the Galactic plane for new VHE
  sources. Previous observations in 2004/2005 resulted in numerous
  detections of VHE gamma-ray emitters in the region l =
  330$^{\mbox{\tiny o}}$ - 30$^{\mbox{\tiny o}}$ Galactic
  longitude. Recently the survey was extended, covering the
  regions l = 280$^{\mbox{\tiny o}}$ - 330$^{\mbox{\tiny o}}$ and l =
  30$^{\mbox{\tiny o}}$ - 60$^{\mbox{\tiny o}}$, leading to the
  discovery of several previously unknown sources with high
  statistical significance. The current status of the survey will be
  presented.}

\maketitle

%%%%% Begin Survey %%%%%%
\addtocontents{toc}{\protect\contentsline {part}{\protect\large Survey}{}}
\addcontentsline{toc}{section}{The H.E.S.S. survey of the inner Galactic plane}
\setcounter{figure}{0}
\setcounter{table}{0}
\setcounter{equation}{0}

\section*{Introduction}
The majority of the newly discovered sources of very high energy (VHE;
$>$ 100 GeV) $\gamma$-rays are related to late phases of stellar
evolution, either directly to massive stars or to the compact objects
they form after their collapse. The possible associations include
pulsar wind nebulae (PWN) of high spin-down luminosity pulsars such as
G\,18.0$-$0.7 \cite{hess_j1825}, supernova remnants like
RX\,J1713.7$-$3946 \cite{RXJ1713}, and open star clusters like
Westerlund\,2 \cite{westerlund2}. As these objects cluster closely
along the Galactic plane, a survey of this region is an effective
approach to discover new sources and source classes of VHE
$\gamma$-ray emission.
\section*{The H.E.S.S.  experiment and its Galactic plane survey}
\begin{figure}[t!]
\centering
\includegraphics[width=0.47\textwidth]{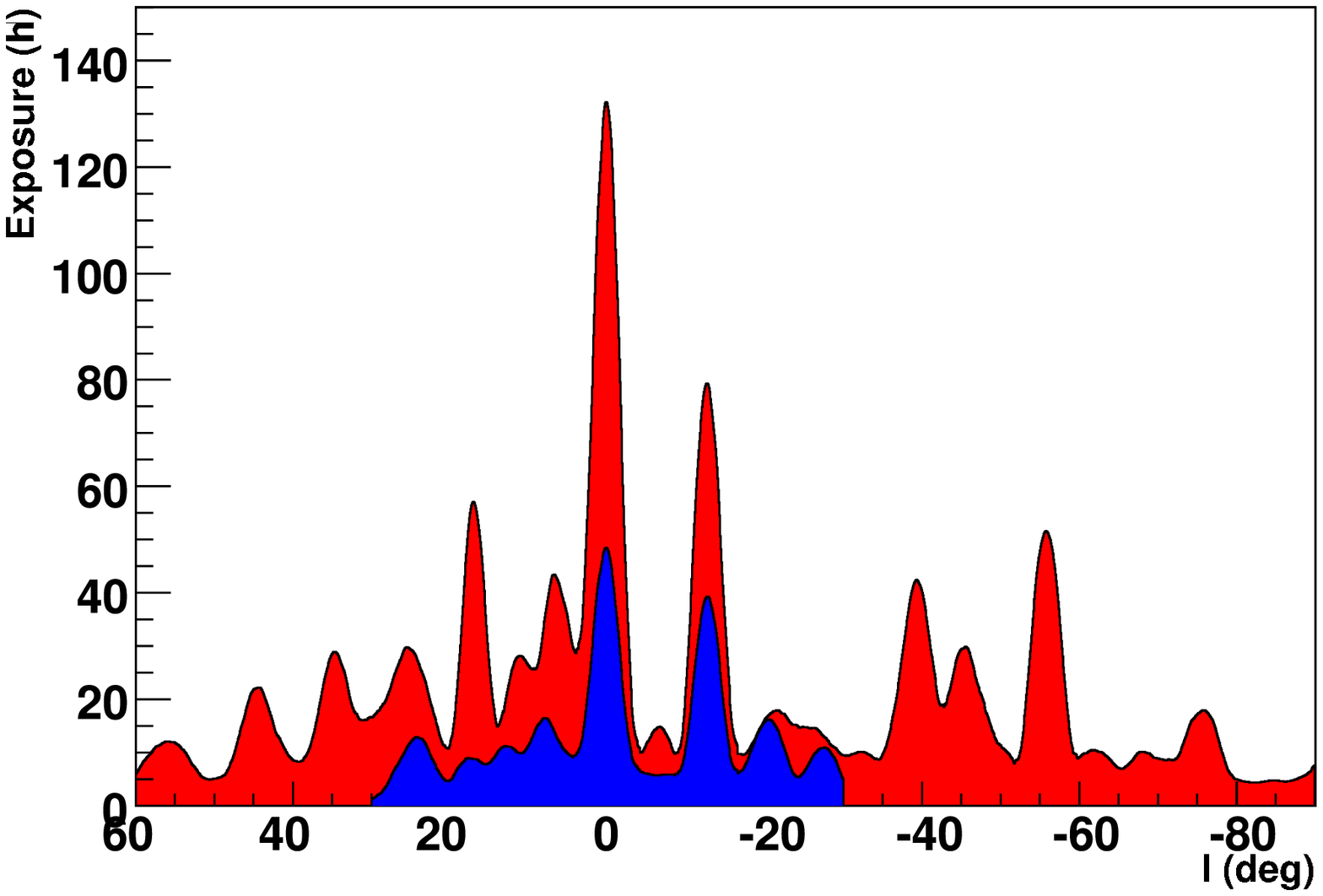}
\caption{Acceptance corrected livetime (equivalent time spent at an
  offset of 0.5$^{\circ}$) along the Galactic plane. All observations
  passing quality selection are considered, including survey-mode
  observations, re-observations of promising source candidates, and
  dedicated observations of known or expected VHE $\gamma$-ray
  sources. {\it Blue}: Observations taken in 2004/2005, published in
  \cite{hess_surveyI}. {\it Red}: Present status of data taking near
  the Galactic plane.
\label{fig:livetime}}
\end{figure}

The High Energy Stereoscopic System (H.E.S.S.) is an array of four
imaging atmospheric-Cherenkov telescopes located 1800~m above sea
level in the Khomas Highlands in Namibia \cite{hess_crab}. Each of the
telescopes is equiped with a camera comprising 960 photomultipliers
and a tesselated mirror with an area of 107\,m$^2$, resulting in a
comparatively large field-of-view of 5$^{\circ}$ in diameter. The H.E.S.S.
array can detect point sources at flux levels of about 1\% of the Crab
nebula flux near zenith with a statistical significance of 5$\sigma$
in 25 hours of observation. This high sensitivity and the large
field-of-view enable H.E.S.S. to survey large celestial areas -- such as
the Galactic plane -- within a reasonable time.\\
The H.E.S.S. Galactic plane survey began 2004 and has been a major
part of the observation program since. In the years 2004/2005 the
survey was conducted in the Galactic longitude band $\pm$ 30$^{\circ}$
around l = 0$^{\circ}$, covering most of the inner part of the
Galactic plane from the tangent of the Norma arm to the tangent of the
Scutum arm. Observations of 28 minutes duration each were taken at
pointings with a spacing of 0.7$^{\circ}$ in longitude in three strips
in Galactic latitude, covering an approximately 6$^{\circ}$ wide
region along the Galactic plane. 95\,h of data were taken in pure
survey mode. Promising source candidates were re-observed in dedicated
observations, comprising 30\,h of data. In addition, dedicated
observations in this region were taken on known or assumed VHE
$\gamma$-ray sources. The total amount of good quality data in this
region was 230 hours, Fig. \ref{fig:livetime} (blue). This first stage
of the H.E.S.S. Galactic plane survey resulted in the discovery of
eight previously unknown sources of VHE $\gamma$-rays at a
significance level greater than 6$\sigma$ after accounting for all
trials involved in the search (post-trials)
\cite{hess_survey_science}. Additionally, six likely sources were
found with significances above 4$\sigma$ \cite{hess_surveyI}.\\
In the years 2005-2007, the survey region was extended further along
the Galactic Plane. The scan region now covers $-$85$^{\circ}$ $<$ l $<$
60$^{\circ}$, $-$3$^{\circ}$ $<$ b $<$ 3$^{\circ}$, containing the
Carina-Sagittarius arm and part of the Perseus arm. In total,
$\sim$325\,h of data were taken in survey mode within this region,
together with 625\,h of pointed observations, which include
re-observations of source candidates and dedicated observations of
known or assumed VHE $\gamma$-ray emitters. Figure \ref{fig:livetime}
shows the present (red) and past (blue) exposure of the H.E.S.S. Galactic
plane scan.
\section*{New sources of VHE $\gamma$-rays}
In the continuation of the H.E.S.S. Galactic plane survey, $>$14 new VHE
$\gamma$-ray sources were discovered so far at statistical significances
larger than 5$\sigma$ post-trials. The possible associations range
from young pulsars such as PSR\,J1846$-$258 (Kes 75), over middle-aged
pulsars (PSR\,J1913+1011) to a source first discovered at TeV energies
by the Milagro collaboration (MGRO\,J1908+06). A non-negligible fraction
of the sources, however, have no obvious counterparts.
\subsection*{PSR\,J1846$-$0258 and Kes\,75}
\begin{figure}[t!]
\centering
\includegraphics[width=0.45\textwidth]{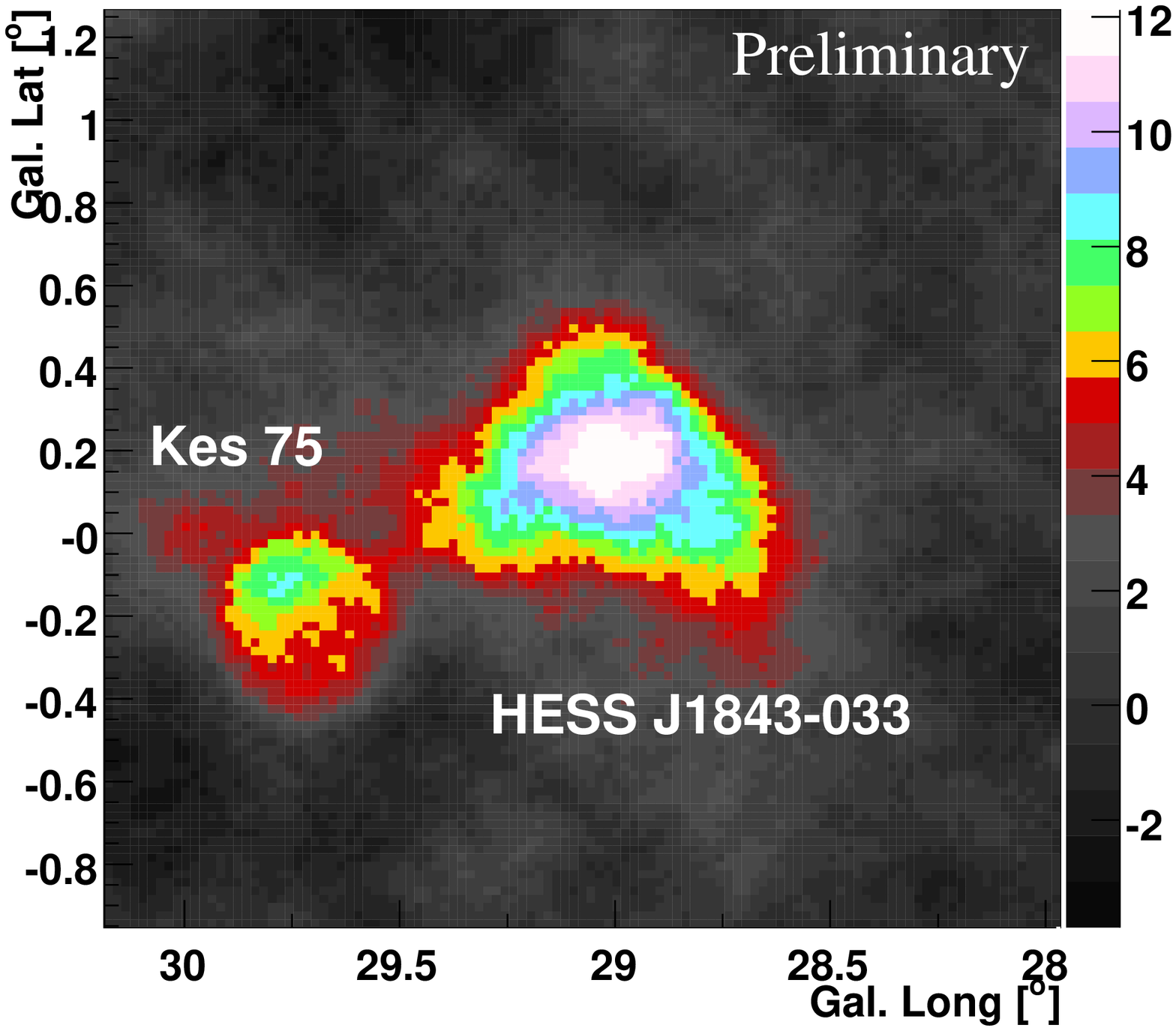}
\caption{$\gamma$-ray significance map of the region containing
  Kes\,75, obtained by counting $\gamma$-rays within 0.22$^{\circ}$
  from a given location. The integration radius is part of the
  standard survey analysis, chosen a-priori and therefore not adjusted
  to the individual source sizes. Significance values shown do not
  take the statistical trials involved in the survey into account.
  \label{fig:kes75}}
\end{figure}
The young shell-type supernova remnant (SNR) Kes\,75 is in many ways
similar to the well-studied Crab SNR. It contains the central pulsar
PSR\,J1846$-$0258, which powers an extended radio and X-ray core, and is
therefore another example of a centre-filled SNR, or plerion. Its
distance is estimated as $\sim$19~kpc \cite{kes75_extension}.
PSR\,J1846$-$0258 has a rotation period of 325~ms and a spin-down age
of 728~yrs \cite{atnf}, which apparently makes it the youngest
rotation-powered pulsar yet discovered \cite{kes75_age}. The
extensions of the core and the shell are 30'' and 3.5', respectively
\cite{kes75_extension}. Like from the Crab nebula, a point-like source
of VHE $\gamma$-ray emission is detected, coincident with the position
of Kes\,75, at a significance level of more than 8$\sigma$
post-trials. For details concerning the H.E.S.S. detection of this
object see \cite{proc_kes75}.
%
%The young shell-type supernova remnant (SNR) Kes~75 contains a central
%pulsar PSR~J1846-0258, which powers an extended radio and X-ray
%core. The distance is approximated to $\sim$ 19~kpc, the age estimates
%range from 980 to 1770 yrs. The spin-down age of the pulsar is 723
%yrs, which makes it the apperently youngest pulsar yet discovered.
%The extensions of the core and the shell are 30'' and 3.5'
%respectively. H.E.S.S. discovered a point-like source of VHE
%$\gamma$-ray emission coincident with the position of Kes~75 at a
%significance level of more than 5 $\sigma$ post-trials. For details
%concerning the corresponding H.E.S.S. source (HESS~J) see.\\
%
In the same field of view, an extended source HESS\,J1843$-$303 was
discovered with a statistical significance of more than 11$\sigma$
post-trials. In contrast to Kes\,75, no obvious counterpart for this
source was found yet, but a detailed archival search is still ongoing.
\subsection*{HESS\,J1912+101}
Another possible example of VHE $\gamma$-ray emission from a PWN of a
high spin-down luminosity pulsar is HESS\,J1912+101, located at l =
44.4$^{\circ}$ and b = $-$0.1$^{\circ}$, detected at a post-trials
significance of more than 5$\sigma$. The corresponding pulsar,
PSR\,J1913+1011, is rather old, with a spin-down age of $t_{c} = 1.7
\times 10^{5}$~yrs, and nearby, at a distance of $\sim$4.5 kpc
\cite{atnf}. In contrast to PSR\,J1846$-$0258 mentioned earlier, no
PWN was detected during a dedicated Chandra observation of the
pulsar. The H.E.S.S. source is offset from the position of
PSR\,J1913+1011, which can be explained by either the proper motion of
the pulsar, or by an expansion of the PWN in an inhomogeous medium
\cite{pwn_asym}. The latter explanation seems plausible as clumpy
molecular material was found close to the pulsar position in
$^{13}$CO(J=1$\rightarrow$0) line measurements \cite{co13_survey}. The
scenario of an asymmetric PWN would make HESS\,J1912+101 similar to
HESS\,J1825$-$137 \cite{hess_j1825}.
\begin{figure}[t!]
\centering
\includegraphics[width=0.47\textwidth]{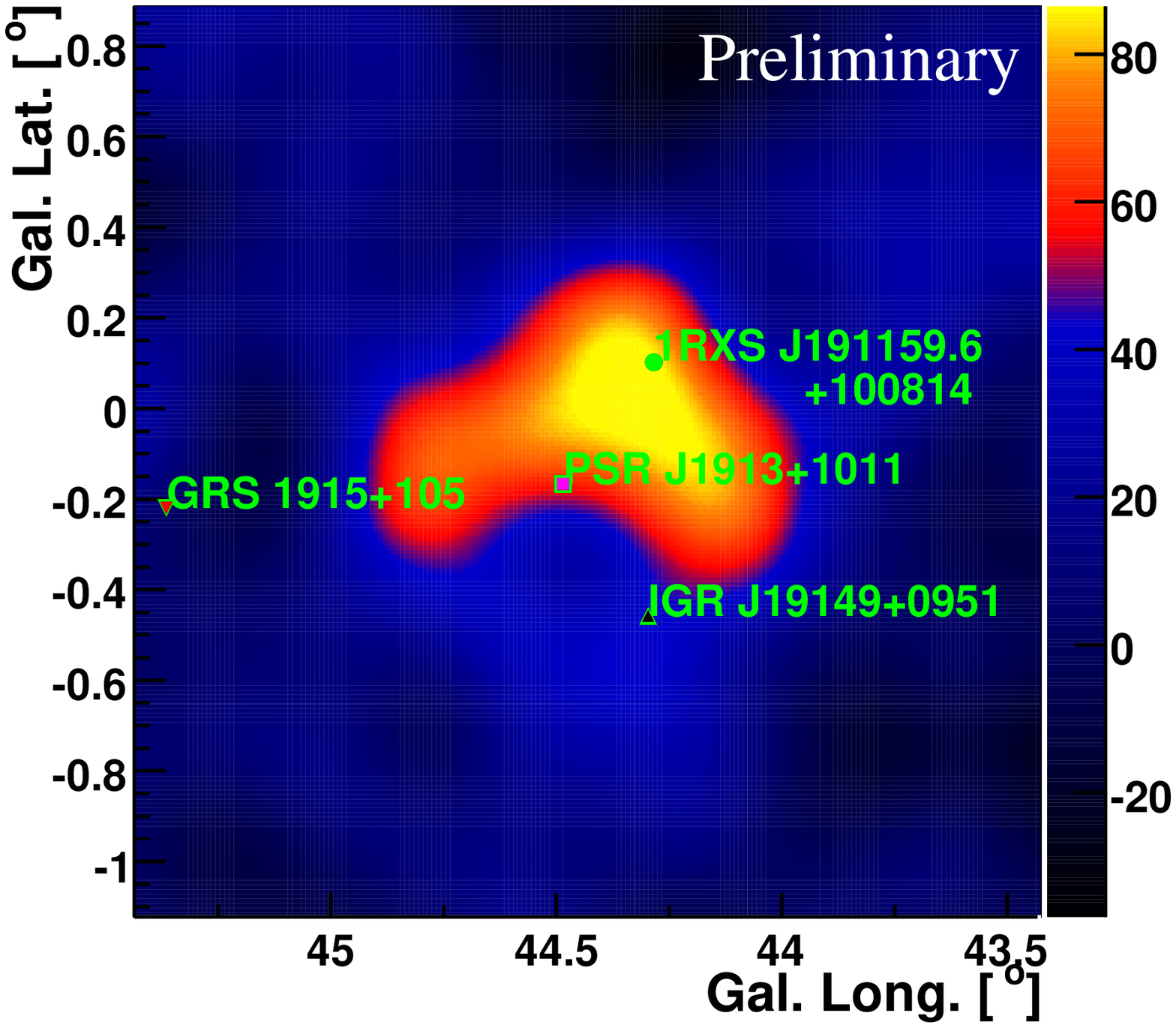}
\caption{Image of the VHE $\gamma$-ray excess from HESS\,J1912+101,
smoothed with a Gaussian profile of width 0.13$^{\circ}$. The
positions of the pulsar PSR\,J1913+1011, the ROSAT
source 1RXS\,J191159.6+100814, the INTEGRAL source
IGR\,J19149+0951 and the microquasar GRS\,1915+105 are marked. 
\label{fig:1912}}
\end{figure}
\subsection*{MGRO\,J1908+06}
The Milagro collaboration, operating a ground-based air shower
detector near Los Alamos, announced the detection of several new
candidate emitters of TeV $\gamma$-rays in the Galactic plane
\cite{milagro}. Compared to the H.E.S.S. array, Milagro has a higher
energy threshold - the median energy of detected events is about 20
TeV - and a reduced angular resolution of about 1$^{\circ}$. The
Milagro coverage of the Galactic plane extends from about 30$^{\circ}$
longitude towards higher longitudes. Four sources are detected at
sufficient significance, the Crab Nebula and the new sources
MGRO\,2019+37, MGRO\,1908+06 and MGRO\,2031+41. Of these, only
MGRO\,1908+06, with a flux of 80\% of the Crab flux and a diameter of
up to 2.6$^{\circ}$, located at around 40$^{\circ}$ longitude is also
contained within the H.E.S.S. Galactic plane survey. Confirming the
Milagro result, a $\gamma$-ray source is detected with a significance
of more than 5$\sigma$ post-trials. The H.E.S.S. source is located at
l = 40.45$^{\circ}$ and b = $-$0.80$^{\circ}$, consistent with the
Milagro position of l = 40.4$^{\circ}$ and b = $-$1.0$^{\circ}$, with
an error radius on the Milagro position of 0.24$^{\circ}$. The rms
size of the H.E.S.S. source is about 0.2$^{\circ}$. For more details
on the H.E.S.S. result see \cite{proc_1908}.
%The Milagro Galactic plane survey resulted in the detection of 8
%candidate sources of TeV $\gamma$-ray emission in the region 30 < l <
%220 and -10 < b < 10. Only the region 30 < l < 65 is covered in both
%surveys and only one candidate source MGRO\,J1908+06 is located within
%this region. H.E.S.S. has detected a VHE $\gamma$-ray source
%coincident with MGRO\,J1908+06 at a siginficance level of more than 5
%$\sigma$ post-trials.
\subsection*{Unidentified sources}
A significant fraction of the recently discovered sources of VHE
$\gamma$-rays within the Galactic plane lack obvious counterparts. For
seven of these sources extensive archival searches in multi-wavelength
data and standard catalogues were performed to search for associated
objects in the radio, X-ray and GeV $\gamma$-ray domains.  While some
of them are partially coincident with known or unidentified X-ray
sources, none provide a clear counterpart which matches all of the
observed characteristics of the VHE emission. The lack of a
lower-energy counterpart challenges VHE emission scenarios, both
leptonic and hadronic. More details are given in a seperate contribution
to this conference \cite{proc_darksources} .
\section*{Summary}
The H.E.S.S. Galactic plane survey, which started in the year 2004,
now reaches from $-$85$^{\circ}$ longitude to 60$^{\circ}$ longitude,
and covers an approximately 6$^{\circ}$ broad band around latitude b =
0$^{\circ}$.  In total, more than 950\,hours of data were taken in this
region, including survey mode observations, re-observations of source
candidates and dedicated observations of known or suspected
$\gamma$-ray sources. The first stage of the survey, covering the
inner 60$^{\circ}$ of the Galactic plane, has increased the number of
known VHE $\gamma$-sources within this region from three at the
beginning of 2004 to seventeen. Further follow-up observations within this
region and the extension of the survey along the Galactic
plane resulted in the discovery of even more additional VHE
$\gamma$-ray emitters. Most of them were presented during this
conference. Multi-wavelength follow-up observations and archival
searches have already begun, and will be crucial for understanding the
underlying processes at work in these astrophysical objects. 
\section*{Acknowledgments}
The support of the Namibian authorities and of the University of Namibia
in facilitating the construction and operation of H.E.S.S. is gratefully
acknowledged, as is the support by the German Ministry for Education and
Research (BMBF), the Max Planck Society, the French Ministry for Research,
the CNRS-IN2P3 and the Astroparticle Interdisciplinary Programme of the
CNRS, the U.K. Science and Technology Facilities Council (STFC),
the IPNP of the Charles University, the Polish Ministry of Science and 
Higher Education, the South African Department of
Science and Technology and National Research Foundation, and by the
University of Namibia. We appreciate the excellent work of the technical
support staff in Berlin, Durham, Hamburg, Heidelberg, Palaiseau, Paris,
Saclay, and in Namibia in the construction and operation of the
equipment.\\

\bibliographystyle{plain}

%%%%%%%%
%  21  %
%%%%%%%%

%The paper title
\title{Establishing a connection between high-power pulsars and
  very-high-energy gamma-ray sources}
%Short title to print in the headers to the final publication (Not showed in this print).
\shorttitle{High-power pulsars and
  very-high-energy gamma-ray sources}
%All paper authors
\authors{S. Carrigan$^{1}$, J.A.~Hinton$^{1,2}$, W. Hofmann$^{1}$, K. Kosack$^{1}$, T.
  Lohse$^{3}$ and O. Reimer$^{4}$ for the H.E.S.S. Collaboration}
%Short title to print in the headers to the final puplication (Not showed in this print).
\shortauthors{S. Carrigan for H.E.S.S.}
%All the affiliations.
\afiliations{ \small
  $^1$Max-Planck-Institut f\"ur Kernphysik, P.O. Box 103980, D 69029 Heidelberg, Germany\\
  $^2$Landessternwarte, Universit\"at Heidelberg, K\"onigstuhl, D 69117 Heidelberg, Germany\\
  $^3$Institut f\"ur Physik, Humboldt-Universit\"at zu Berlin, Newtonstr. 15, D 12489 Berlin, Germany\\
  $^4$Stanford University, HEPL \& KIPAC, Stanford, CA 94305-4085, USA}
\email{svenja.carrigan@mpi-hd.mpg.de}

%The abstract.
\abstract{In the very-high-energy (VHE) gamma-ray wave band, pulsar
  wind nebulae (PWNe) represent to date the most populous class of
  Galactic sources. Nevertheless, the details of the energy conversion
  mechanisms in the vicinity of pulsars are not well understood, nor
  is it known which pulsars are able to drive PWNe and emit
  high-energy radiation. In this paper we present a systematic study
  of a connection between pulsars and VHE $\gamma$-ray sources based
  on a deep survey of the inner Galactic plane conducted with the High
  Energy Stereoscopic System (H.E.S.S.). We find clear evidence that
  pulsars with large spin-down energy flux are associated with VHE
  $\gamma$-ray sources. This implies that these pulsars emit on the
  order of 1\% of their spin-down energy as TeV $\gamma$-rays.
}

\maketitle

\addcontentsline{toc}{section}{Establishing a connection between high-power pulsars and very-high-energy gamma-ray sources}
\setcounter{figure}{0}
\setcounter{table}{0}
\setcounter{equation}{0}

In 1989, the Crab Nebula was discovered as the first celestial source
of VHE $\gamma$-radiation \cite{WhippleCrabDisc}. The pulsar inside the nebula drives a
powerful wind of highly relativistic particles that ends in a
termination shock from which high-energy particles with a wide
spectrum of energies emerge \cite{Gaensler06}. High-energy
electrons\footnote{here and in the following, `electrons' refers to both
electrons and positrons} among these particles
can give rise to two components of electromagnetic radiation: a
low-energy component from synchrotron radiation and a high-energy
component from inverse Compton (IC) up-scattering of ambient photons.

Recently, advances in VHE instrumentation have made the discovery of
many new, predominantly Galactic, sources possible. Of these, a
significant number can be identified as PWNe. Prominent examples are
the PWN of the energetic pulsar PSR~B1509$-$58 in the supernova
remnant MSH~15$-$5$2$ \cite{HESS:MSH1552}, and HESS~J0835$-$455 \cite{HESS:velaxSC},
associated with Vela~X, the nebula of the Vela pulsar. 

These $\gamma$-ray PWNe are extended objects with an angular size of a fraction of a
degree, translating into a size of some 10\,pc for typical distances
of a few kpc.  In addition to the open puzzle of pulsar spin-down
power conversion, a surprising observation is that the centroids of
these $\gamma$-ray PWNe are often displaced from their pulsars by
distances similar to the nebular size. Such displacements, although
usually at smaller scales, are also seen in some X-ray PWNe. The
origin of the displacement remains unknown. It might be attributed to
pulsar motion (e.g. \cite{Swaluw04}), causing the pulsar to leave its nebula
behind, or to a density gradient in the ambient medium \cite{blondin01:PWN}.

The aforementioned examples of coincidences between VHE $\gamma$-ray
sources and radio pulsars motivated a systematic search for VHE
counterparts of energetic pulsars using the H.E.S.S. system of imaging
Cherenkov telescopes located in Namibia \cite{Hinton:2004}. To be
detectable by H.E.S.S., a source at distance $d$ has to provide a
$\gamma$-ray luminosity in the 1~TeV to 10~TeV range of $L_\gamma \sim
10^{32}~d^2$~erg\,s$^{-1}$kpc$^{-2}$. Assuming a conversion efficiency
of ~1\% of pulsar spin-down energy loss $\dot{E}$ into TeV
$\gamma$-rays (where $\dot{E}$ is determined from the measurement of
the rotation period $\Omega$ and the rate at which the rotation slows
down $\dot{\Omega}$), PWNe of pulsars with $\dot{E}$ around
$10^{34}~d^2$~erg\,s$^{-1}$kpc$^{-2}$ might be detectable. We note
that for typical electron spectra, only a small fraction of the total
energy in electrons is carried by the multi-TeV electrons, that are
responsible for TeV $\gamma$-rays by IC scattering off ambient photons
(including those from the cosmic microwave background) and for keV
$\gamma$-rays by synchrotron radiation. Even a 1\% energy output in
TeV $\gamma$-rays already implies a large fraction of spin-down energy
loss going into relativistic electrons.

Here we investigate how the probability to detect in VHE $\gamma$-rays
PWNe surrounding known pulsars varies with the spin-down energy loss
of the pulsar, testing the plausible assumption that the $\gamma$-ray
output of a PWN correlates in some fashion with the power of the
pulsar feeding it.

The VHE $\gamma$-ray data set used to search for $\gamma$-ray emission
near the location of known radio pulsars comprises all data used in
the H.E.S.S. Galactic plane survey
\cite{HESS:scanpaper1,HESS:scanpaper2SC}, including an extension of the
survey to Galactic longitudes $-60^{\circ} < l < -30^{\circ}$,
dedicated observations of Galactic targets and re-observations of
H.E.S.S. survey sources. The search covers a range in Galactic
longitude from $-60^{\circ}$ to $30^{\circ}$ while the range in Galactic
latitude is restricted to $\pm$2\,deg, a region well covered in the
survey. A total of 435 pulsar locations are tested, taken from the
Parkes Multibeam Pulsar Survey (PMPS, \cite{Parkes4} and references
therein), as recorded in the ATNF pulsar catalogue. Pulsars without
measured period derivatives are ignored.  Over the range of the
H.E.S.S. survey, the PMPS provides reasonably uniform sensitivity
\cite{ATNF}, enabling a reliable estimate of the frequency of chance
coincidences between a $\gamma$-ray source and a pulsar. The analysis
of the $\gamma$-ray data follows the standard H.E.S.S. analysis
\cite{HESS:crabSC}. Initially, a sky map is generated providing the
significance of a $\gamma$-ray excess for a given position. Taking
into account the properties of known $\gamma$-ray PWNe, the search is
optimised for slightly extended sources -- on the scale of the angular
resolution ($\approx$~0.1\,deg) of the H.E.S.S. telescopes -- and
allows for small offsets from the pulsar positions. Each excess is
determined by counting $\gamma$-ray candidate events within $\theta
\leq 0.22$\,deg ($\theta^2 \leq 0.05$\,deg$^2$) of a given position
and subtracting a background estimated from areas in the same field of
view. The sky map is used to look up the significance of a
$\gamma$-ray excess at the position of the radio pulsars, as well as
for randomly generated test positions used to evaluate the statistical
significance of the association (details are given below). We require
an excess significance of at least 5 standard deviations above the
background as a signature of a VHE $\gamma$-ray signal.  Given the
modest number of trials - the 435 pulsar locations - the number of
false detections is negligible with this requirement and in any case
small compared to the probability for chance coincidences between
radio pulsars and VHE $\gamma$-ray sources.

\begin{figure*}[ht]
  \centering
  \resizebox{0.8\hsize}{!}{\includegraphics{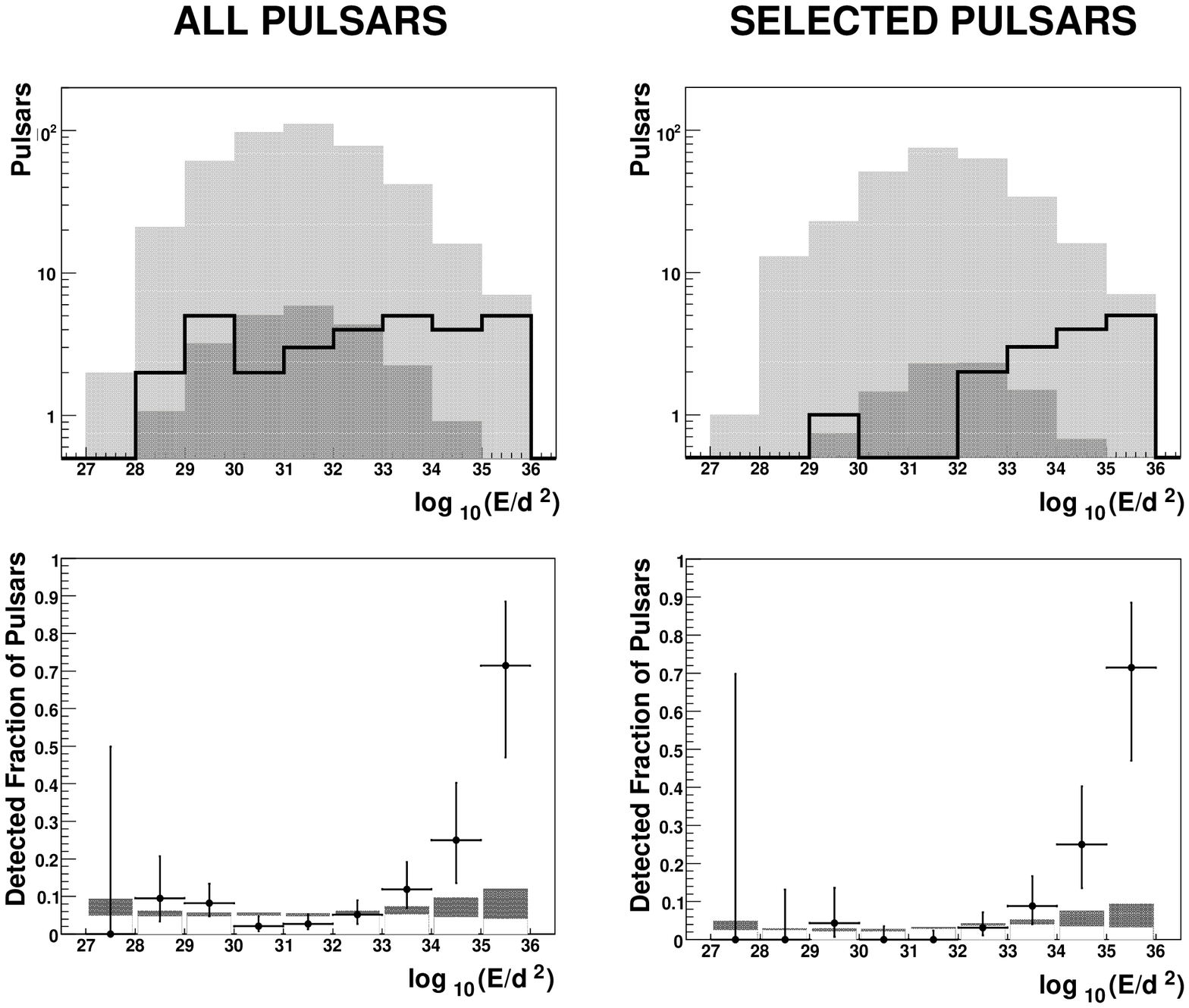}}
\caption{\emph{\bf{Top row:}} Distribution in
log$_{10}(\dot{E}/d^2)$ of all PMPS pulsars in the H.E.S.S. scan range
(shaded in light grey), of chance coincidences (shaded in dark grey)
and of detected pulsars (black line). Here, $\dot{E}/d^2$ is measured
in erg\,s$^{-1}$kpc$^{-2}$. \emph{\bf{Bottom row:}} The points
show the fraction of pulsars with significant $\gamma$-ray excess at
the pulsar position, as a function of log$_{10}(\dot{E}/d^2)$.  
%The
%two-sided binomial errors give the 68\% confidence range for the
%probability to find a $\gamma$-ray source at the pulsar location. 
The
shaded band represents the probability for a chance coincidence. The
width of the band accounts for the uncertainty in the width of the
latitude distribution of pulsars.
%; since $\gamma$-ray sources tend to
%cluster at very low latitude, the rate of chance coincidences is
%sensitive to this width. 
\emph{\bf{Left:}} all pulsars; \emph{\bf{right:}} double occurrences
of gamma-ray sources removed by omitting pulsars which overlap with
stronger pulsars or known non-pulsar sources.
%To
%derive the spin-down flux $\dot{E}/d^2$, the pulsar distances
%resulting from Taylor and Cordes (1993) were used. A comparison with
%pulsar distances updated according to Cordes and Lazio (2002) gives
%consistent results.
}
\end{figure*}

Of the 435 pulsars, 30 are found with significant $\gamma$-ray
emission at the pulsar location (Fig.~1, top left panel).  The lower
left panel of Fig.~1 displays the fraction of pulsars with such
$\gamma$-ray emission for different intervals in spin-down flux
$\dot{E}/d^2$. The fraction is about 5\% for pulsars with spin-down
flux below $10^{33}$~erg\,s$^{-1}$kpc$^{-2}$ and increases to about
70\% for pulsars with $\dot{E}/d^2$ above
$10^{35}$~erg\,s$^{-1}$kpc$^{-2}$. Not all of these associations are
necessarily genuine. The rate of chance coincidences is estimated by
generating $10^6$ realisations of random pulsar samples (each
consisting on average of 435 ``pulsars'') following the distribution
in longitude and latitude of the PMPS pulsars and taking into account
the narrowing of the distribution in latitude with increasing
spin-down flux. The expected fraction of chance coincidences is shown
as dark shaded areas in Fig.~1 and varies between 4\% to 12\%. All
associations with pulsars with $\dot{E}/d^2 <
10^{33}$~erg\,s$^{-1}$kpc$^{-2}$ are within statistical errors
consistent with chance coincidences. Indeed for plausible values of
the ratio between the $\gamma$-ray luminosity and the pulsar spin-down
energy loss, $L_\gamma/\dot{E}$, no detectable emission would be
expected from such pulsars.  On the other hand, the detection of
emission from high-power pulsars is statistically significant.  The
probability that the detection of VHE sources coincident with 9 or
more of the total of 23 pulsars above $\dot{E}/d^2 >
10^{34}$~erg\,s$^{-1}$kpc$^{-2}$ results from a statistical
fluctuation is $\sim 3.4 \times 10^{-4}$. For detection of 5 or more
of the total of 7 pulsars above $10^{35}$~erg\,s$^{-1}$kpc$^{-2}$, the
chance probability is $\sim 4.2 \times 10^{-4}$.

Given the high density of pulsars, a single $\gamma$-ray source may
even coincide with more than a single pulsar, and thus appear more
than once amongst the ``detections'' in the upper left panel of
Fig.~1. Removal of such double occurrences (Right panels of Fig.~1)
does not change the conclusion, and none of the high-luminosity
pulsars is affected. Details will be given elsewhere.

The results shown in Fig.~1 demonstrate that a large fraction of
high-luminosity pulsars correlate with sources of VHE $\gamma$-rays,
emitting with a $\gamma$-ray luminosity of order 1\% of the pulsar
spin-down power. The positive correlation does not necessarily imply
that the pulsar or PWN itself is responsible for the $\gamma$-ray
flux. It could also result from some other mechanism correlated with
the pulsar or its creation, such as a supernova shock wave. The
correlation found between $\gamma$-ray detectability and spin-down
flux $\dot{E}/d^2$ argues in favour of a pulsar-related origin of the
$\gamma$-ray signal. On the other hand, for the PMPS pulsar sample,
$\dot{E}/d^2$ also correlates closely with the spin-down age $T$ of
the pulsar, $\dot{E}/d^2 \sim T^{-3/2}$, and obviously with distance
$d$, both parameters relevant for determining the $\gamma$-ray flux
from shock-wave driven supernova remnants.
 
The exact relation between pulsar parameters and $\gamma$-ray
luminosity is an interesting issue. Variations in exposure and hence
in detection threshold over the survey range, as well as the
uncertainty in pulsar distance will smear out the turn-on curve of
detectability versus $\dot{E}/d^2$ shown in Fig.~1, but cannot fully
account for the rather slow turn-on over a range of more than one
order of magnitude in $\dot{E}/d^2$, combined with a detection
probability below unity for even the highest-power pulsars.  This
indicates that $\dot{E}/d^2$ cannot be the only parameter relevant for
the $\gamma$-ray flux.  The same conclusion is obtained from the
observed variation of $L_\gamma/\dot{E}$ of about an order of
magnitude among the detected pulsars. However, this present pulsar
sample is too small to investigate the dependence of $L_\gamma$ on
multiple pulsar parameters, e.g. including pulsar age.

A constant $L_\gamma/\dot{E}$ is also not necessarily expected. For a
given age, the integral energy fed by the pulsar into the PWN
increases with $\dot{E}$.  Apart from expansion losses, pulsar
spin-down power is shared between particle energy and magnetic field
energy. If equipartition between the two energy densities is assumed
\cite{Reynolds84,rees74}, the magnetic field in the PWN will increase
with $\dot{E}$ and hence the energy loss by synchrotron radiation will
increase relative to and at the expense of inverse Compton
$\gamma$-ray production.  Indeed, un-pulsed X-ray luminosity of
pulsars is observed to increase faster then $\dot{E}$, $L_X \propto
\dot{E}^{1.4\pm0.1}$ \cite{cheng04}. In such scenarios, magnetic field
values and therefore the balance between X-ray and $\gamma$-ray
emission will also depend on volume, i.e. on the expansion speed of
the nebula and hence on the ambient medium. In addition, the current
spin-down luminosity $\dot{E}$ may not be the only relevant scale; if
the pulsar age is shorter than or comparable to the electron cooling
time, relic electrons injected in early epochs with higher spin-down
power will still contribute and may enhance $L_\gamma$ significantly
compared to the quasi-steady state achieved for old pulsars.

\small

\section*{Acknowledgements}
The support of the Namibian authorities and of the University of Namibia
in facilitating the construction and operation of H.E.S.S. is gratefully
acknowledged, as is the support by the German Ministry for Education and
Research (BMBF), the Max Planck Society, the French Ministry for Research,
the CNRS-IN2P3 and the Astroparticle Interdisciplinary Programme of the
CNRS, the U.K. Science and Technology Facilities Council (STFC),
the IPNP of the Charles University, the Polish Ministry of Science and 
Higher Education, the South African Department of
Science and Technology and National Research Foundation, and by the
University of Namibia. We appreciate the excellent work of the technical
support staff in Berlin, Durham, Hamburg, Heidelberg, Palaiseau, Paris,
Saclay, and in Namibia in the construction and operation of the
equipment.

%This is the reference to .bib file (Whitout .bib!)

%This in the bibtex style, is ok.
\bibliographystyle{plain}

\normalsize

%%%%%%%%
%  22  %
%%%%%%%%

\title{H.E.S.S. VHE Gamma-ray sources without identified counterparts}
\shorttitle{H.E.S.S. Unidentified}

\authors{K. Kosack $^1$, A. Djannati-Atai$^2$, A. Lemiere$^2$,
  E. Moulin$^3$, G. P\"uhlhofer$^4$, E.~de O\~{n}a Wilhelmi$^2$ for
  the H.E.S.S. Collaboration}

\shortauthors{K. Kosack et al.}

\afiliations{$^1$Max-Planck-Institut f\"ur Kernphysik, P.O. Box
  103980, D 69029 Heidelberg, Germany , $^2$APC, 11 Place Marcelin
  Berthelot, F-75231 Paris Cedex 05, France,$^3$ Laboratoire de
  Physique Th\'eorique et Astroparticules, IN2P3/CNRS, Universit\'e
  Montpellier II, CC 70, Place Eug\`ene Bataillon, F-34095 Montpellier
  Cedex 5, France. $^4$Landessternwarte, Universit\"at Heidelberg,
  K\"onigstuhl, D 69117 Heidelberg, Germany }
\email{Karl.Kosack@mpi-hd.mpg.de}

%The abstract.
\abstract{Scan-based observations of the Galactic plane and continuing
re-observations of known very-high-energy (VHE) gamma-ray sources with
the H.E.S.S. system of imaging atmospheric Cherenkov telescopes have
revealed a wide variety of new VHE objects.  While in many cases these
objects can be associated with known sources in the X-ray, radio, or optical
wavebands, a subset of them currently have no obvious cataloged lower-energy
counterpart. An analysis of 8 such unidentified sources is
presented here.}

\maketitle

\addcontentsline{toc}{section}{H.E.S.S. VHE Gamma-ray sources without identified counterparts}
\setcounter{figure}{0}
\setcounter{table}{0}
\setcounter{equation}{0}

%Begin the section.

\section*{Introduction}

The current generation of Imaging Atmospheric Cherenkov Telescopes
(IACTs) have provided an unprecedented level of sensitivity to the
field of VHE ($E=$100 GeV--100 TeV) astronomy. In particular, the
H.E.S.S. instrument and the ongoing H.E.S.S. Galactic Plane Survey of
the inner Galaxy \cite{HESS:scanpaper2}, has increased the number of
known VHE sources by nearly an order of magnitude. While many of the
new VHE sources discovered in the survey can be associated through
multi-wavelength data with previously identified objects
(e.g. shell-type supernova remnants, pulsar-wind nebulae, or X-ray
binaries), a growing population of VHE sources have yet to be
identified.  Since at least weak X-ray and radio emission is predicted
by most common VHE emission models, the lack of lower-energy
detections may provide substantial model constraints and may even
point to a new class of objects which emit primarily in the VHE energy
range.

The results presented here should be considered preliminary; further
details (sky maps and spectra) and final results will be available
shortly in a refereed article. 

\section*{Technique}

H.E.S.S. (the High Energy Stereoscopic System) is an array of four
atmospheric Cherenkov telescopes located in the Khomas highland of
Namibia at an altitude of $1800\:\mathrm{m}$ above sea-level. Each
telescope consists of a $107 \mathrm{m^2}$ optical reflector made up
of segmented mirrors that focus light into a camera of 960
photo-multiplier tube pixels \cite{HESS:optics}. The telescopes image
the UV/blue flashes of Cherenkov light emitted by the secondary
particles produced in gamma-ray-induced air-showers. Stereoscopic
shower observations using the \emph{imaging atmospheric Cherenkov
technique} (e.g. \cite{hillas96:technique,weekes96:acts,HEGRA:acts})
allow for accurate reconstruction of the direction and energy of the
primary gamma rays as well as for the rejection of background events
from air showers of cosmic ray origin. H.E.S.S. is sensitive to gamma
rays above a post-cuts threshold energy of approximately 150 GeV and
has an average energy resolution of $\sim16\%$
\cite{HESS:crab}.

The data discussed here were taken as part of the H.E.S.S. Galactic
Plane Survey, which covers the band $-50^\circ < l < 60^\circ$ in
galactic longitude and $-3^\circ < b < 3^\circ$ in latitude. Data are
taken as a series of 28-minute ``runs'', each centered on regular grid
points along the survey region, or in case of re-observed sources, in
\emph{wobble-mode}, where the runs are taken at alternating offsets
from the source position (typically $0.7^\circ$). The data are
analyzed using the standard H.E.S.S. analysis and calibration
techniques described in \cite{HESS:crab}. The predefined \emph{hard}
gamma-ray selection cuts were applied to the data, which provide
better gamma-hadron separation (and are thus better for source
detection) at the expense of a higher analysis energy threshold.  For
source detection and morphology studies, the \emph{ring-background}
technique \cite{HESS:background} was used for background subtraction
with an on-source integration radius of 0.22$^\circ$ and an off-source
annulus with typical radius $0.7^\circ$ (standard for H.E.S.S. scan
sources).

% For the spectral analysis, the integration
% region was taken to entirely enclose the observed VHE emission
% (reducing systematic errors, but also the signal-to-noise ratio) and
% background was estimated using the \emph{reflected-region} technique.

\section*{Source Selection}

The VHE sources discussed here include all new sources discovered
(with post-trials significances over $6\sigma$) in the
H.E.S.S. Galactic Plane Survey for which there is no obvious cataloged
counterpart at lower wavelengths, according to the criteria cited
below. Additionally, two sources meeting these criteria that
were previously published in \cite{HESS:scanpaper2} are included due
to a substantial increase in exposure time. A search for counterparts
to the VHE emission was made by first looking in source catalogs for
objects which are of a type known to produce VHE photons, including
the ATNF pulsar catalog \cite{ATNF}, the Green's supernova remnant
catalog \cite{green04:SNRs}, and the High-Mass X-ray binary (HMXB)
catalog by \cite{liu06:HMXB}. We also checked the Low-Mass X-ray
binary (LMXB) catalog by \cite{liu07:LMXB}, the INTEGRAL source
catalog \cite{INTEGRAL:3IBIS}, and the SIMBAD database. Additionally,
publicly available images for lower-wavelength survey data in the
radio and X-ray wavebands, from the Molonglo \cite{Molonglo,SUMSS},
NRAO VLA \cite{NVSS}, ROSAT \cite{ROSAT}, ASCA \cite{ASCA} Galactic
plane surveys, were compared with the H.E.S.S. excess maps.

To reduce the possibility of chance coincidences, some minimal
selection criteria were applied to the possible candidates. For pulsar
wind nebulae, only pulsars with spin-down fluxes $\dot E/D^2 >
10^{33}\:\mathrm{erg\:sec^{-1}\:kpc^{-2}}$ (e.g. ones which would
require $<100\%$ spin-down power to gamma ray conversion
efficiency) were considered.  For shell-type SNRs, only those that
reasonably match the morphology (size and position) of the VHE
emission, and for XRBs (which have so far not been observed to produce
extended emission), only those with small offsets from the VHE source
were considered plausible candidates.

\section*{Results}

The details of the six new and two updated unidentified H.E.S.S. VHE
sources are presented in Table \ref{tab:sources}, while previously
published unidentified VHE sources are listed in
\ref{tab:other_sources} for reference. The results of a simple
two-dimensional Gaussian function convolved with the
H.E.S.S. point-spread function to the uncorrelated excess event maps
is given in Table \ref{tab:morphology}. This gives a rough impression
of the size of each object, however as the emission is in most cases
not Gaussian.

\begin{table*}
  \begin{center}
    \begin{tabular}{l c c c c c c c}
      \hline\hline
      Source & R. A. & Dec & $l(^\circ)$ & $b(^\circ)$ & T (hrs) & $S$ ($\sigma$) & Counts\\
      \hline
      \HESSa                   &  \HMS{14}{27}{2}  &  \DMS{-60}{51}{00}  & 314.409 & -0.145 & 21 & 7.3  & 197 \\
      \HESSb                   &  \HMS{16}{26}{04}  &  \DMS{-49}{05}{13}  & 334.772 & 0.045  & 12 & 7.5  & 153 \\
      \HESSc$\dagger$          &  \HMS{17}{02}{44}  &  \DMS{-42}{00}{57}  & 344.304 & -0.184 & 9 & 12.8 & 412 \\
      \HESSd$\dagger$          &  \HMS{17}{08}{24}  &  \DMS{-41}{05}{24}  & 345.683 & -0.469 & 39 & 10.7 & 513 \\
      \HESSe                   &  \HMS{17}{31}{55}  &  \DMS{-34}{42}{36}  & 353.565 & -0.622 & 14 & 8.1  & 218 \\
      \HESSf                   &  \HMS{18}{40}{55}  & \DMS{-05}{33}{00}    & 26.795  & -0.197 & 26 & 10.6 & 346 \\
      \HESSg                   &  \HMS{18}{57}{11}  &  \DMS{02}{40}{00}   & 35.972  & -0.056 & 21 & 8.7  & 223 \\
      \HESSh                   &  \HMS{18}{58}{20}  &  \DMS{02}{05}{24}   & 35.578  & -0.581 & 25 & 7.0  & 168 \\
      \hline\hline
      
    \end{tabular}
    \caption{ \label{tab:sources} Positions in equatorial (J2000
      epoch) and Galactic ($l$,$b$) coordinates along with the
      detection significances of unidentified sources in the
      H.E.S.S. Galactic Plane scan discussed in this proceeding.  $S$ is
      the significance (number of standard deviations above the
      background level) of the source using a fixed integration radius
      of $0.22^\circ$, which was used for selecting the sources from
      the scan data. The position of each source is based on a model
      fit to the background-subtracted gamma-ray maps. The fit
      positions have an average statistical error of 0.05
      degrees. Sources marked with a $\dagger$ are previously
      published in \cite{HESS:scanpaper2} and have been updated with
      new data. The exposure time is corrected for the off-axis
      sensitivity of the telescope system and accounts for instrumental readout
      dead-time. }
  \end{center}
\end{table*}

\begin{table}
  \begin{center}
    \begin{tabular}{l c c c}
      \hline\hline
      Source & R.A. & Dec \\
      \hline
      HESS~J1303-631$\ddag$ & \HMS{13}{03}{00}  & \DMS{-63}{11}{55}  \\
      HESS~J1614-518$\ddag$ & \HMS{16}{14}{19}  & \DMS{-51}{49}{12}  \\
      HESS~J1632-478 & \HMS{16}{32}{09}  & \DMS{-47}{49}{12} \\ %HESS HMXB/INTEGRAL counterpart, but
                                %HESS source is extended
      HESS~J1634-472 & \HMS{16}{34}{58}  & \DMS{-47}{16}{12}  \\ % Has IGR unid counterpart and
                                % possible SNR (neither match
                                % morphologically)
      HESS~J1745-303 & \HMS{17}{45}{02} & \DMS{30}{22}{12} \\ % EGRET counterpart, unid
      HESS~J1837-069 & \HMS{18}{37}{38} & \DMS{-6}{57}{00} \\ % AXJ 1838.0-0655 is counterpart,
                                % possibly HMXB, but could be PWN -
                                % unid so far
      TeV~J2032+4130$\ddag$ & \HMS{20}{32}{57}  & \DMS{41}{29}{57}  \\
      \hline\hline
    \end{tabular}    
    \caption{ \label{tab:other_sources} Previously published
      unidentified VHE sources, not discussed here. Coordinates are in
      J2000 epoch. Sources with $\ddag$ have no obvious
      lower-wavelength counterpart. For other sources, possible
      counterpars exist, which are however unidentified themselves or
      did not yet permit an identification of the VHE source. Results
      are from \cite{HESS:J1303}, \cite{HESS:scanpaper2}, and
      \cite{HEGRA:TeVJ2032_Final}. }
  \end{center}
\end{table}

\begin{table}
  \begin{center}
    \begin{tabular}{l c@{$\:\pm\:$}c c@{$\:\pm\:$}c c@{$\:\pm\:$}c }
      \hline\hline
      Source    
      & \multicolumn{2}{c}{$\sigma_1$ ($^\circ$)} 
      & \multicolumn{2}{c}{$\sigma_2$ ($^\circ$)}  
      & \multicolumn{2}{c}{$\phi$ ($^\circ$)} \\
      \hline
      \scriptsize \HESSa & \scriptsize  0.04  & \scriptsize  0.02  & \scriptsize  0.08   & \scriptsize  0.03  &\scriptsize  80  &\scriptsize  17 \\
      \scriptsize \HESSb & \scriptsize  0.07  & \scriptsize  0.02  & \scriptsize  0.10   & \scriptsize  0.05  &\scriptsize   3  &\scriptsize  40 \\
      \scriptsize \HESSc & \scriptsize  0.30  & \scriptsize  0.02  & \scriptsize  0.15   & \scriptsize  0.01  &\scriptsize  68  &\scriptsize  7\\
      \scriptsize \HESSd & \scriptsize  0.06  & \scriptsize  0.01  & \scriptsize  0.08   & \scriptsize  0.01  &\scriptsize  -20 &\scriptsize  23\\
      \scriptsize \HESSe & \scriptsize  0.18  & \scriptsize  0.07  & \scriptsize  0.11   & \scriptsize  0.03  &\scriptsize  -89 &\scriptsize  21\\
      \scriptsize \HESSf & \scriptsize  0.41  & \scriptsize  0.04  & \scriptsize 0.25    & \scriptsize  0.02  &\scriptsize  39  &\scriptsize  6\\
      \scriptsize \HESSg & \scriptsize  0.11  & \scriptsize  0.08  & \scriptsize  0.08   & \scriptsize  0.03  &\scriptsize  -3  &\scriptsize  49\\
      \scriptsize \HESSh & \scriptsize  0.08  & \scriptsize  0.02  & \scriptsize  0.02   & \scriptsize  0.04  &\scriptsize  4   &\scriptsize  17\\
      \hline\hline
    \end{tabular}
    \caption{\label{tab:morphology} Results from an elongated 2-D
      Gaussian model fit to the gamma-ray excess for each source. $\sigma_1$ and $\sigma_2$ are the
      intrinsic semi-major and semi-minor axes (in degrees on the
      sky), with the effect of the point-spread function removed. The
      errors are statistical. The position angle ($\phi$) is measured
      counter-clockwise in degrees relative to the RA axis.}
  \end{center}
\end{table}

\section*{Discussion}

Though the general characteristics (size, location, flux) of the eight
unidentified sources described here are similar to previously
identified galactic VHE sources (e.g. PWNe), they have so far no clear
counterpart in lower wavebands and further multi-wavelength study is
required to understand the emission mechanisms powering
them. Therefore, follow-up observations with higher-sensitivity X-ray
and GeV gamma-ray telescopes will be beneficial.  Since most VHE
sources are predicted to emit X-ray and radio emission, a
non-detection of lower-wavelength emission with current-generation
experiments for some of these objects may indicate a new VHE
source class (as suggested in \cite{HESS:scanpaper1}), and may provide
new insight into high-energy processes within our Galaxy.

\section*{Acknowledgements}
{\small
  The support of the Namibian authorities and of the University of
  Namibia in facilitating the construction and operation of
  H.E.S.S. is gratefully acknowledged, as is the support by the German
  Ministry for Education and Research (BMBF), the Max Planck Society,
  the French Ministry for Research, the CNRS-IN2P3 and the
  Astroparticle Interdisciplinary Programme of the CNRS, the
  U.K. Science and Technology Facilities Council (STFC), the IPNP of
  the Charles University, the Polish Ministry of Science and Higher
  Education, the South African Department of Science and Technology
  and National Research Foundation, and by the University of
  Namibia. We appreciate the excellent work of the technical support
  staff in Berlin, Durham, Hamburg, Heidelberg, Palaiseau, Paris,
  Saclay, and in Namibia in the construction and operation of the
  equipment.

  This research has made use of the SIMBAD database, operated at CDS,
  Strasbourg, France and the ROSAT Data Archive of the
  Max-Planck-Institut f\"ur extraterrestrische Physik (MPE) at
  Garching, Germany.

\bibliographystyle{plain}
}

%%%%%%%%
%  23  %
%%%%%%%%

%The paper title
\title{HESS~J1023--575: Non-thermal particle acceleration associated \\ with the young stellar cluster Westerlund~2}
%Short title to print in the headers to the final publication (Not showed in this print).
\shorttitle{VHE gamma-rays from Westerlund~2}
%All paper authors
\authors{O.~Reimer$^{1}$, J.~Hinton$^{2}$, W.~Hofmann$^{3}$, S.~Hoppe$^{3}$, C.~Masterson$^{4}$, M.~Raue$^{5}$,\\
 for the H.E.S.S. Collaboration$^{6}$.}
%Short title to print in the headers to the final puplication (Not showed in this print).
\shortauthors{O. Reimer et al}
%All the affiliations.
\afiliations{
$^1$ Stanford University, HEPL \& KIPAC, Stanford, CA 94305-4085, USA\\ 
$^2$ School of Physics \& Astronomy, University of Leeds, Leeds LS2 9JT, UK\\
$^3$ Max-Planck-Institut f\"ur Kernphysik, P.O. Box 103980, 69029 Heidelberg, Germany\\
$^4$ Dublin Institute for Advanced Studies, 5 Merrion Square, Dublin 2, Ireland\\
$^5$ Universit\"at Hamburg, Institut f\"ur Experimentalphysik, Luruper Chaussee 149, 22761 Hamburg, Germany\\
$^6$ http://www.mpi-hd.mpg.de/hfm/HESS/public/hn\_hesscollab.html
}
\email{olr@stanford.edu; martin.raue@desy.de}

%The abstract.
\abstract{The results from H.E.S.S. observations towards Westerlund~2 are presented. 
The detection of very-high-energy gamma-ray emission towards the young stellar cluster
Westerlund 2 in the HII complex RCW49 by H.E.S.S. provides ample evidence that
particle acceleration to extreme energies is associated with this region. A 
variety of possible emission scenarios is mentioned, ranging from high-energy 
gamma-ray production in the colliding wind zone of the massive Wolf-Rayet 
binary WR~20a, collective wind scenarios, diffusive shock acceleration at the 
boundaries of wind-blown bubbles in the stellar cluster, and outbreak phenomena 
from hot stellar winds into the interstellar medium. These scenarios are briefly 
compared to the characteristics of the associated new VHE gamma-ray source HESS~J1023--575, 
and conclusions on the validity of the respective emission scenarios for 
high-energy gamma-ray production in the Westerlund~2 system are drawn.}

\maketitle

\addcontentsline{toc}{section}{HESS~J1023--575: Non-thermal particle acceleration associated with the young stellar cluster Westerlund~2}
\setcounter{figure}{0}
\setcounter{table}{0}
\setcounter{equation}{0}

%Begin the section.

\section*{The young stellar cluster Westerlund~2 in the HII region RCW~49}
The prominent giant HII region RCW 49, and its ionizing young stellar cluster Westerlund 2, are located 
towards the outer edge of the Carina arm of our Milky Way. RCW 49 is a luminous, massive star formation region, 
and has been extensively studied at various wavelengths. Recent mid-infrared measurements with SPITZER 
revealed still ongoing massive star formation \cite{ref1}. The regions surrounding Westerlund 2 appear evacuated 
by stellar winds and radiation, and dust is distributed in fine filaments, knots, pillars, bubbles, 
and bow shocks throughout the rest of the HII complex \cite{ref2, ref3}. Radio continuum observations revealed 
two wind-blown shells in the core of RCW~49 \cite{ref4}, surrounding the central region of Westerlund~2, 
and the prominent Wolf-Rayet star WR~20b. A long-standing distance ambiguity has been recently \cite{ref5} 
revised in a determination of the distance to Westerlund~2 by spectro-photometric measurements 
of 12 cluster member O-type stars of $(8.3 \pm 1.6)$ kpc. This value is in good agreement with the 
measurements from the light curve of the eclipsing binary WR~20a \cite{ref6}, associating WR 20a as a cluster 
member of Westerlund 2 (Note, however the 2.8 kpc as of \cite{ref_fn}). The stellar cluster contains an extraordinary ensemble of hot and massive stars, 
presumably at least a dozen early-type O-stars, and two remarkable WR stars. Only recently WR20a was 
established to be a binary \cite{ref7, ref8} by presenting a solutions for a circular orbit with a period of 3.675, and 3.686 days, 
respectively. Based on the orbital period, the minimum masses have been found to be $(83 \pm 5)$\,M$_{\odot}$ and $(82 \pm 5)$\,M$_{\odot}$ 
for the binary components \cite{ref6}. At that time, it classified the WR binary WR 20a as the most massive of all confidently 
measured binary systems in our Galaxy. The supersonic stellar winds of both WR stars collide, and a wind-wind 
interaction zone forms at the stagnation point with a reverse and forward shock. In a detached binary system 
like WR~20a, the colliding wind zone lies between the two stars, and is heavily skewed by Coriolis forces. 
The winds of WR~20a can only be accelerated to a fraction of their expected wind speed $v_\infty\sim 2800$ km/s, and a 
comparatively low pre-shock wind velocity of $\sim 500$ km/s follows. Synchrotron emission has not yet been detected 
from the WR~20a system, presumably because of free-free-absorption in the optically thick stellar winds along 
the line of sight. WR~20a has been detected in X-rays \cite{ref9}, but non-thermal and thermal components of the X-ray 
emission remain currently indistinguishable. Detectable VHE gamma-radiation from the WR~20a binary system was 
only predicted in a pair cascade model \cite{ref10}, although detailed modeling of the WR~20a system in other scenarios 
(e.g. as of \cite{ref19} when produced either by optically-thin inverse Compton scattering of relativistic electrons with the dense photospheric 
stellar radiation fields in the wind-wind collision zone or in neutral pion decays, with the mesons produced by inelastic 
interactions of relativistic nucleons with the wind material) is still pending. 
At VHE gamma-rays, photon-photon absorption will diminish the observable flux from a close binary system such as WR~20a 
\cite{ref11}.

\section*{H.E.S.S. observations of Westerlund~2}

The H.E.S.S. (High Energy Stereoscopic System) collaboration observed the Westerlund~2 region between March and July 2006, 
and obtained 14 h (12.9 h live time) of data, either on the nominal source location of WR~20a or overlapping data from the 
ongoing Galactic plane survey. Standard quality selections were imposed on the data. The data have been obtained in 
wobble-mode observations to allow for simultaneous background estimation. The wobble offsets for these observations 
range from 0.5$^\circ$ to 2$^\circ$, with the majority of data taken with wobble offset less than 0.8$^\circ$. The zenith angles range 
between 36$^\circ$ and 53$^\circ$, resulting in an energy threshold of 380 GeV for the analysis. The data have been analyzed using 
the H.E.S.S. standard Hillas analysis with standard cuts ($>$ 80 p.e.). A point source analysis on the nominal position 
of WR~20a resulted in a clear signal with a significance of 6.8$\sigma$. Further investigations revealed an extended excess 
with a peak significance exceeding 9$\sigma$ (Fig.3 left). The center of the excess was derived by fitting the two-dimensional 
point spread function (PSF) of the instrument folded with a Gaussian to the uncorrelated excess map:  
$\alpha_{2000}$ = $10^{\rm h}23^{\rm m}18^{\rm s} \pm 12^{\rm s}$, $\delta_{2000}$ = -57$^\circ$45'50'' $\pm$ 1'30''. 
The systematic error in the source location is 20'' in both coordinates. The source is clearly extended beyond the nominal 
extension of the PSF (Fig.~1). A fit of a Gaussian folded with the PSF of the H.E.S.S. instruments gives an extension of $0.18^\circ \pm 0.02^\circ$. 

\begin{figure}[ht]
   \centering
   \includegraphics[width=0.48\textwidth]{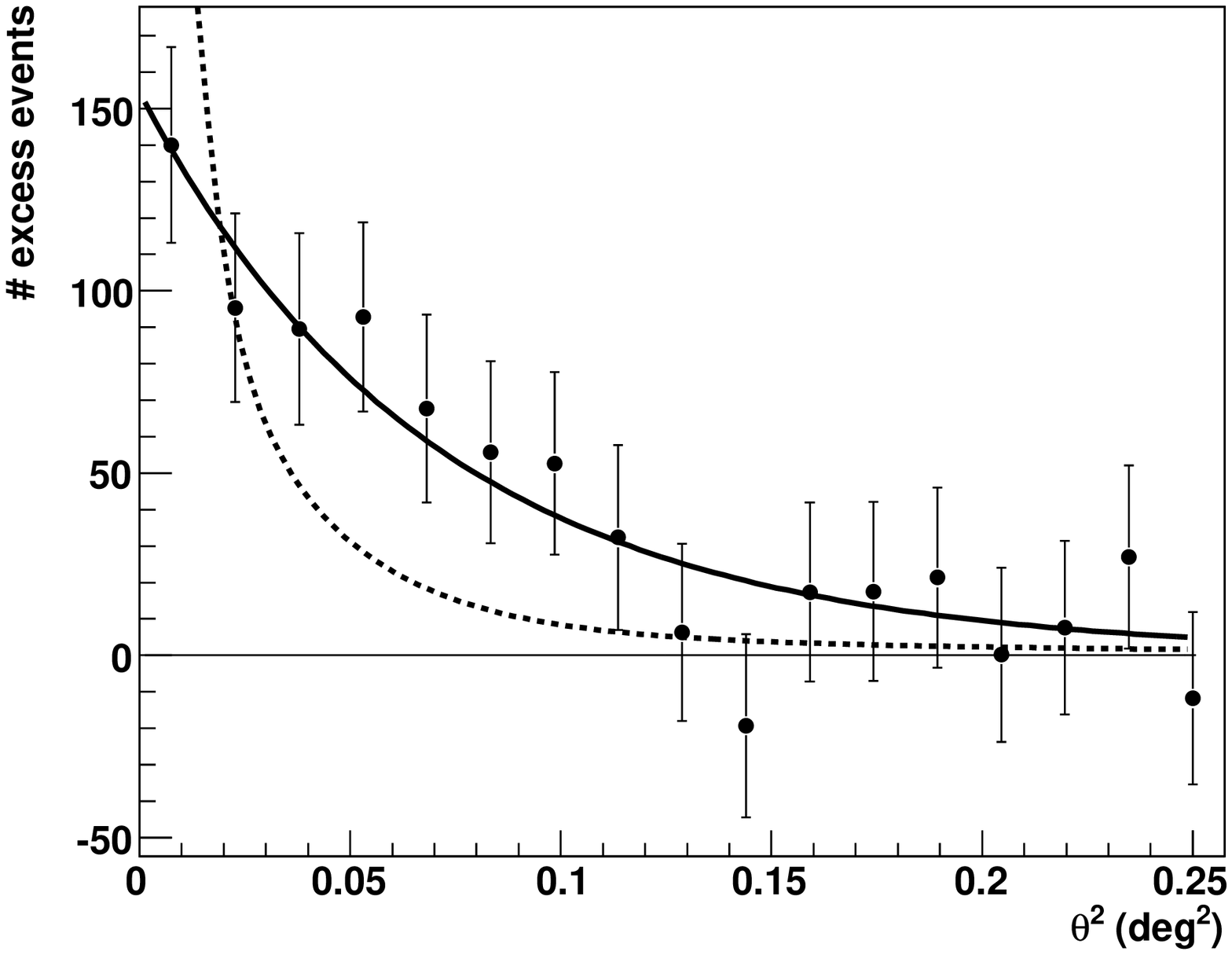} 
   \caption{Number of excess events versus the squared angular distance from the best fit position of the excess. 
            The dashed line shows the expectation for a point source derived from Monte Carlo data. The solid line is a 
			fit of the PSF folded with a Gaussian ($\sigma = 0.18^\circ \pm 0.02^\circ$).}
   \label{fig1}
\end{figure}

The differential energy spectrum for photons inside the corresponding 85\% containment radius of 0.39$^\circ$ is shown in Fig.~2. 
It can be described by a power law (dN/dE$= \Phi_0 \cdot (\mbox{E}/1\,\mbox{TeV})^{- \Gamma}$) 
with a photon index of $\Gamma=2.53 \pm 0.16_{\mathrm{stat}} \pm 0.1_{\mathrm{syst}}$ and a normalization at 1\,TeV of 
$\Phi_0 = (4.50 \pm 0.56_{\mathrm{stat}} \pm 0.90_{\mathrm{syst}}) \times 10^{-12}$\,TeV$^{-1}$\,cm$^{-2}$\,s$^{-1}$. 
The integral flux for the whole excess above the energy threshold of 380 GeV is (1.3 $\pm$ 0.3) $\times 10^{-11}$\,cm$^{-2}$\,s$^{-1}$. 
No significant flux variability could be detected in the data set. The fit of a constant function to the lightcurve binned 
in data segments of 28\,minutes has a chance probability of 0.14. The results were checked with independent analyses and 
found to be in good agreement.

\begin{figure}[ht]
   \centering
   \includegraphics[width=0.48\textwidth]{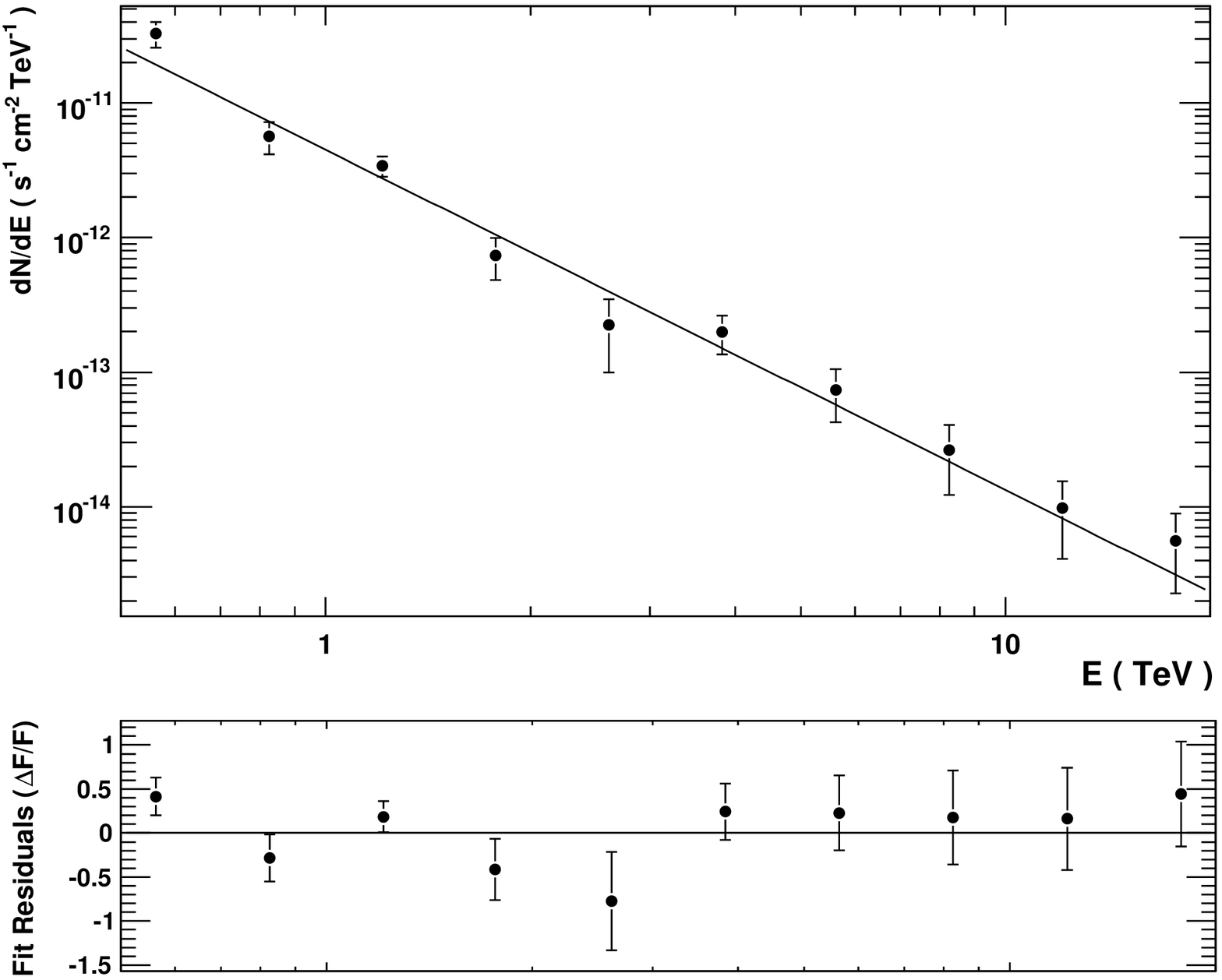}
   \caption{Differential energy spectrum and residuals to a single power-law fit of HESS~J1023--575 from photons 
            inside the 85\% containment radius (0.39$^\circ$) around the best fit position. The background is estimated with 
			background regions of the same size and distance from the camera center as the signal region.} 
      \label{fig2}
\end{figure}

\section*{HESS~J1023--575 in the context of $\gamma$-ray emission scenarios}

The detection of VHE gamma-ray emission from the Westerlund~2 region \cite{ref12} is proof for extreme high-energy particle 
acceleration associated with this young star forming region. With a projected angular size of submilliarcsecond 
scale, the WR~20a binary system, including its colliding wind zone, would appear as a point source for observations 
with the H.E.S.S. telescope array. Unless there are extreme differences in the spatial extent of the particle 
distributions producing radio, X-ray, and VHE gamma-ray emission, a \emph{colliding stellar wind scenario} for the WR~20a
binary faces the severe problem of accounting for the observed VHE source extension. At a nominal distance of 8.0~kpc, this 
source extension is equivalent to a diameter of 28~pc for the emission region, consistent in size with theoretical 
predictions of bubbles blown from massive stars into the ISM \cite{ref13}. The spatial extension found for HESS~J1023--575 
contradicts emission scenarios where the bulk of the gamma-rays are produced close to the massive stars. Alternatively, 
the emission could arise from \emph{collective effects of stellar winds} in the Westerlund~2 stellar cluster. Diffusive shock 
acceleration in cases where energetic particles experience multiple shocks can be considered for Westerlund~2. 
The stellar winds may provide a sufficiently dense target for high-energy particles, allowing the production of 
$\pi^0$-decay $\gamma$-rays via inelastic pp-interactions. Collective wind scenarios \cite{ref14, ref15} suggest that 
the spatial extent of the gamma-ray emission corresponds to the volume filled by the hot, shocked stellar winds, 
but HESS~J1023--575 substantially supersedes the boundary of Westerlund~2. Supershells, molecular clouds, and inhomogeneities embedded 
in the dense hot medium may serve as the targets for gamma-ray production in Cosmic Ray interactions. Such environments 
have been studied in the nonlinear theory of particle acceleration by large-scale MHD turbulence \cite{ref16}. \emph{Shocks and MHD 
turbulent motion inside a stellar bubble or superbubble} can efficiently transfer energy to cosmic rays if the particle acceleration time 
inside the hot bubble is much shorter than the bubble's expansion time. 

\begin{figure*}
\begin{center}
\includegraphics [width=0.95\textwidth]{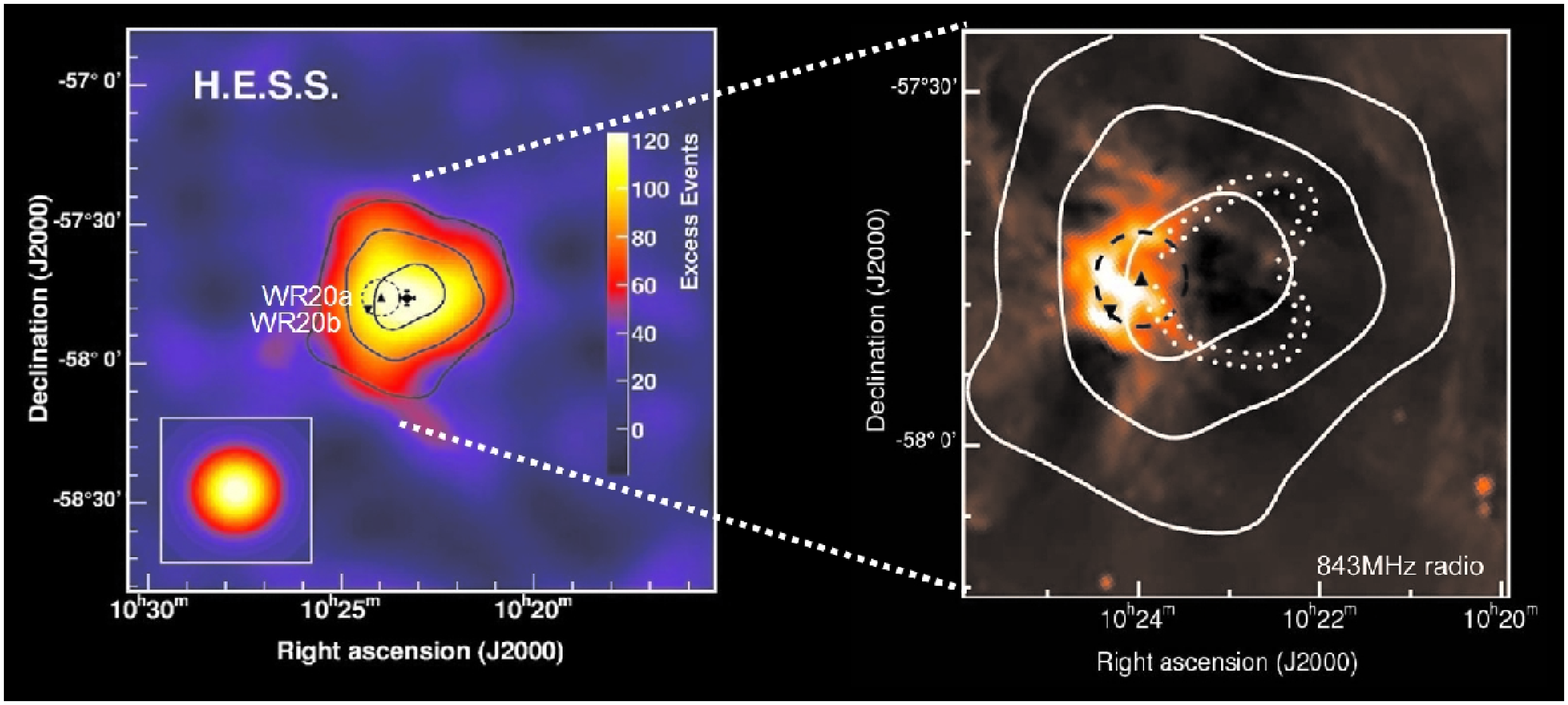}
\end{center}
\caption{
Left: H.E.S.S. $\gamma$-ray sky map of the Westerlund~2 region, smoothed to reduce the effect of statistical fluctuations. 
The inlay in the lower left corner shows how a point-like source would have been seen by H.E.S.S. WR~20a and WR~20b 
are marked as filled triangles, and the stellar cluster Westerlund~2 is represented by a dashed circle.    
Right: Significance contours of the $\gamma$-ray source HESS~J1023--575 (corresponding 5, 7 and 9$\sigma$ ), overlaid on a 
MOST radio image. The wind-blown bubble around WR~20a, and the blister to the west of it can be seen as depressions in 
the radio continuum. The blister is indicated by white dots as in \cite{ref4}, 
and appears to be compatible in direction and location with HESS~J1023--575.
}\label{fig3}
\end{figure*}

Finally, shock acceleration at the boundaries of the "blister" (Fig.~3 right) may enable particles to diffusively 
re-enter into the dense medium, thereby interacting in hadronic collisions and producing gamma-rays. A scenario as 
outlined in \cite{ref17} for a Supernova-driven expansion of particles into a low density medium may be applicable to the 
expanding stellar winds into the ambient medium. If one accepts such a scenario here, it might give the first 
observational support of gamma-ray emission due to diffusive shock acceleration from supersonic winds in a 
wind-blown bubble around WR~20a, or the ensemble of hot and massive OB stars from a superbubble in Westerlund~2, 
breaking out beyond the edge of a molecular cloud. Accordingly, one has to consider that such acceleration sites 
will also contribute to the observed flux of cosmic rays in our Galaxy \cite{ref18}.

Further observations with the H.E.S.S. telescope array will help to discriminate among the alternatives in the 
interpretation of HESS~J1023--575. However, the convincing association with a new type of astronomical object 
a massive HII region and its ionizing young stellar cluster profoundly distinguishes this new detection by 
the H.E.S.S. telescope array already from other source findings made during earlier Galactic Plane Scan observations.

\section*{Acknowledgements}
\scriptsize
The support of the Namibian authorities and of the University of Namibia
in facilitating the construction and operation of H.E.S.S. is gratefully
acknowledged, as is the support by the German Ministry for Education and
Research (BMBF), the Max Planck Society, the French Ministry for Research,
the CNRS-IN2P3 and the Astroparticle Interdisciplinary Programme of the
CNRS, the U.K. Science and Technology Facilities Council (STFC),
the IPNP of the Charles University, the Polish Ministry of Science and 
Higher Education, the South African Department of
Science and Technology and National Research Foundation, and by the
University of Namibia. We appreciate the excellent work of the technical
support staff in Berlin, Durham, Hamburg, Heidelberg, Palaiseau, Paris,
Saclay, and in Namibia in the construction and operation of the
equipment.

\normalsize
\bibliographystyle{plain}
{}

%%%%%%%%
%  24  %
%%%%%%%%

%The paper title
\title{HESS sources possibly associated with massive star clusters}

%Short title to print in the headers to the final publication (Not showed in this print).
\shorttitle{Hess sources possibly associated with massive star clusters}
%All paper authors
\authors{A. Marcowith$^{1}$, N. Komin$^{1}$, Y.A. Gallant$^{1}$, D. Horns$^{2}$, G. P\"uhlhofer$^{3}$, S. Schwemmer$^{3}$, O. Reimer$^{4}$
for the H.E.S.S. collaboration} 
%Short title to print in the headers to the final puplication (Not showed in this print).
\shortauthors{A. Marcowith et al}
%All the affiliations.
\afiliations{$^1$Laboratoire de Physique th\'eorique et astroparticules, universit\'e Montpellier II, CNRS/IN2P3, Montpellier, France 
\\ $^2$Landessternwarte Universit\"at Heidelberg, Germany \\ $^3$Institut f\"ur Theorische Physik, Lehrstuhl IV: Wletraum und Astrophysik,
Ruhr-Universit\"at Bochum, Germany \\ $^4$ Stanford University, HEPL \& KIPAC, CA 94305-4085, USA}
\email{Alexandre.Marcowith@lpta.in2p3.fr}

%The abstract.
\abstract{In view of the discovery of HESS J1023-575 (discussed in a separate presentation), we examine another very high energy (VHE) 
gamma-ray source possibly associated with massive star clusters. Particle acceleration in massive star forming regions 
can proceed at the interface of two interacting winds or result from a collective process; e.g. multiple shock acceleration or MHD 
turbulence. The gamma-ray emission can also take place at the edge of the superbubble blown by the winds and multiple supernova 
explosions. Non-thermal radiation from the shell structure then traces the interaction of energetic particles (ions and/or electrons) 
with the surrounding interstellar matter. In particular, HESS J1837-069 is spatially coincident with a recently discovered very massive 
star cluster. We discuss the VHE gamma-ray data resulting from H.E.S.S. observations on this or other possible such associations. 
We consider data in other wavelength domains, in particular in X-rays, and examine the available evidence that the VHE emission 
could originate in particles accelerated by the above-mentioned mechanisms in massive star clusters.}

\maketitle

\addcontentsline{toc}{section}{HESS sources possibly associated with massive star clusters}
\setcounter{figure}{0}
\setcounter{table}{0}
\setcounter{equation}{0}

\section*{Introduction}
High energy radiative processes are strongly connected with very energetic events as only a fraction of the free energy of the source 
is necessary to be injected into a small amount of relativistic particles. The Galactic cosmic-rays (GCRs) are probably 
connected with the explosion of supernov{\ae} (see \cite{ref1}). A majority of the supernova (SN) progenitors are associated 
with massive stars (the so-called core collapse supernova) and a majority of massive stars are born, live and die in groups: 
in massive star clusters and/or OB associations (both hereafter together called massive star forming regions or MSFR); 
see \cite{ref2}, \cite{ref3}, \cite{ref4}. The question of the production of energetic particles (EP), the non-thermal signatures 
and the contribution of MSFR to the GCR population then arises naturally. The acceleration and the propagation of EP in MSFR as 
well as their interaction with the ambient medium have been widely debated in the literature (\cite{ref5}, \cite{ref6}, \cite{ref7}, 
\cite{ref8}, \cite{ref2}). Observational probes of energetic events in MSFR have concentrated on the search of radiative 
signatures of supernova remnants in superbubbles (SB); e.g. the low density high temperature region blown up by the collective 
interaction of stellar winds and supernova explosions (see \cite{ref9}). Usually a firm detection have been proven to be difficult 
unless the shock is currently interacting with a molecular cloud (as it is probably the case of IC443 discussed by \cite{ref10} and 
recently by \cite{ref11}), or with the shell produced by the multiple wind system. 
The termination shock or wind-wind interaction shocks in massive star systems have been explored as sites of non-thermal radiation 
(\cite{ref12} and references therein). With its improved sensitivity (able to detect at 5 $\sigma$ level a source at 1$\%$ of the 
Crab in 25h), a large field of view ($5^o$) and a good angular resolution (about $0.1^o$) the H.E.S.S. (High Energy Stereoscopic System) 
telescope has the technical capacities to explore the faint and diffuse VHE gamma-ray emission that may be produced in MSFR. In this work, 
we examine different acceleration and radiation scenarii that should contribute to the high energy emission in MSFR (section \ref{S:Mech}). 
We discuss the possible association of the source HESS\,J1837-069 with a MSFR (section \ref{S:1809}). 

\section*{Particle acceleration and non-thermal radiation}
\label{S:Mech}
Several acceleration processes and radiation mechanisms could give rise to the TeV $\gamma$-ray emission detected from HESS\,J1023-575
(Reimer et al, these proceedings and \cite{ref13}), as well as in the source TeV J2032+4130 detected by HEGRA 
in the Cygnus OB2 SB \cite{ref14}. We discuss briefly here the most relevant acceleration and gamma-ray radiative mechanisms.\\ 
{\it Scenario 1: Massive star termination shock}: Several authors have considered the possibility of accelerating EP at 
the terminal shock of massive stars (\cite{ref15}, \cite{ref16}). The shock particle acceleration efficiency appears to be highly 
reduced by the magnetic field configuration; the toro\"{\i}dal component dominates at the termination shock radius $R_{\rm t}$ and 
is perpendicular to the shock normal. Unless a turbulent component randomizes the magnetic field orientation over a rotational period, 
the acceleration process is inoperant.
Even in the case of a stronger stellar magnetic field, the maximum energy EP can reach would hardly be above the threshold for neutral pion 
production. The result is also relatively insensitive to the modelling of the wind modulation which uses mean free path estimates strictly 
valid for protons with energies $\sim$ GeV in the solar wind. The re-acceleration of an ambient CR population has also been proven to 
be inefficient \cite{ref16}. It appears that the modulation factor (the ratio of advective and diffusive lengths) in the downstream 
region is probably so large that the shock region is shielded from the outer medium. Flare particles accelerated in the star 
atmosphere cool adiabatically in the strongly diverging wind flow and do not contribute to the EP at the shock front. Several issues 
seem however interesting to be investigated. Firstly, only proton or ion acceleration and radiation have been considered. A leptonic 
scenario (non-thermal bremsstrahlung or Inverse Compton) may contribute to a gamma-ray emission. Secondly, other turbulence regimes 
can lead to smaller and/or anisotropic mean free paths that are possibly more convenient for the particle transport.\\
{\it Scenario 2: Wind-wind interaction in massive binary systems}: As discussed above, non-thermal radio synchrotron emission 
from several massive star binaries supports the scenario of particle acceleration at the interface of two colliding winds. 
Eichler \& Usov (1993) \cite{ref17} demonstrated the possibility of an efficient particle acceleration at strong shocks created by the 
wind-wind collision. A more precise treatment of both acceleration and loss effects lead generally to a gamma-ray spectrum with a 
cut-off in the sub-TeV regime \cite{ref18}. Good knowledge of the viewing angle and the system parameters are also necessary to evaluate   
the $\gamma-\gamma$ absorption properly. Alternatively, VHE radiation could be produced by the interaction of a SNR shock 
wave and a stellar wind \cite{ref19}. The result is the production of two reflected shocks, one propagating inside the 
SNR, the second one inside the wind bubble and a converging flow which develops between the two shocks where particle acceleration 
can take place.\\ 
{\it Scenario 3: Collective wind scenario}: Gamma-rays produced by neutral pion decay can also result from collective interaction 
of winds, where the interaction region could serve both as accelerating region and target for the high energy hadrons, or just provide 
the target material alone. The former has been partly discussed above, the sources of radiation being in that case the termination shock of 
massive stars or the colliding winds. The second case requires an accelerator relatively close to the star 
cluster and has been considered by \cite{ref20}. There are several issues to the gamma-ray observability of these sources. Firstly, 
the opacity  from $\gamma-\gamma$ pair production in the stellar photon field should be low enough. Secondly, if the CR source is 
outside the star cluster where the interaction takes place, as discussed above, the wind modulation can prevent low energy particles 
to enter the wind regions. The threshold energy $E_{\rm lim}$ is very sensitive to the structure of the magnetic field, 
both to normalisation and energy dependence of the spatial diffusion coefficient and to the dominance of the parallel to the 
perpendicular diffusion coefficient. Depending on the strengths of these effects, $E_{\rm lim}$ can range from GeV to TeV energies. 
The modulation is treated similarly to the modelling in \cite{ref15}.\\
{\it Scenario 4:MHD turbulence and collective particle acceleration processes}: The SB turbulent model (\cite{ref7} and references 
therein) predicts a peak in the acceleration efficiency after a few times $10^5$ years after the ingnition of the turbulence (a network 
of MHD fluctuations and weak reflected shocks). The peak is reached at the maximum of turbulence conversion into non-thermal particles. 
The particles accelerated by the last or the aforementioned scenarii can interact with dense surrounding shells \cite{ref5}. In that 
case gamma-rays are expected to be produced not only from neutral pion decay but also by non-thermal Bremsstrahlung in dense surroundings 
shells or in molecular clouds. An accurate spectral measurement at lower energies by AGILE and GLAST would offer better constraints to the 
emitting process.  

\section*{A new possible association with a MSFR:HESS 1837-069}
\label{S:1809}
\begin{figure}
\begin{center}
\noindent
%\fbox{\hbox{\vbox{\hsize=50mm \hfill \vspace{50mm}}}}
%uncomment next line to include real image
\includegraphics [height=0.4\textwidth, width=0.5\textwidth]{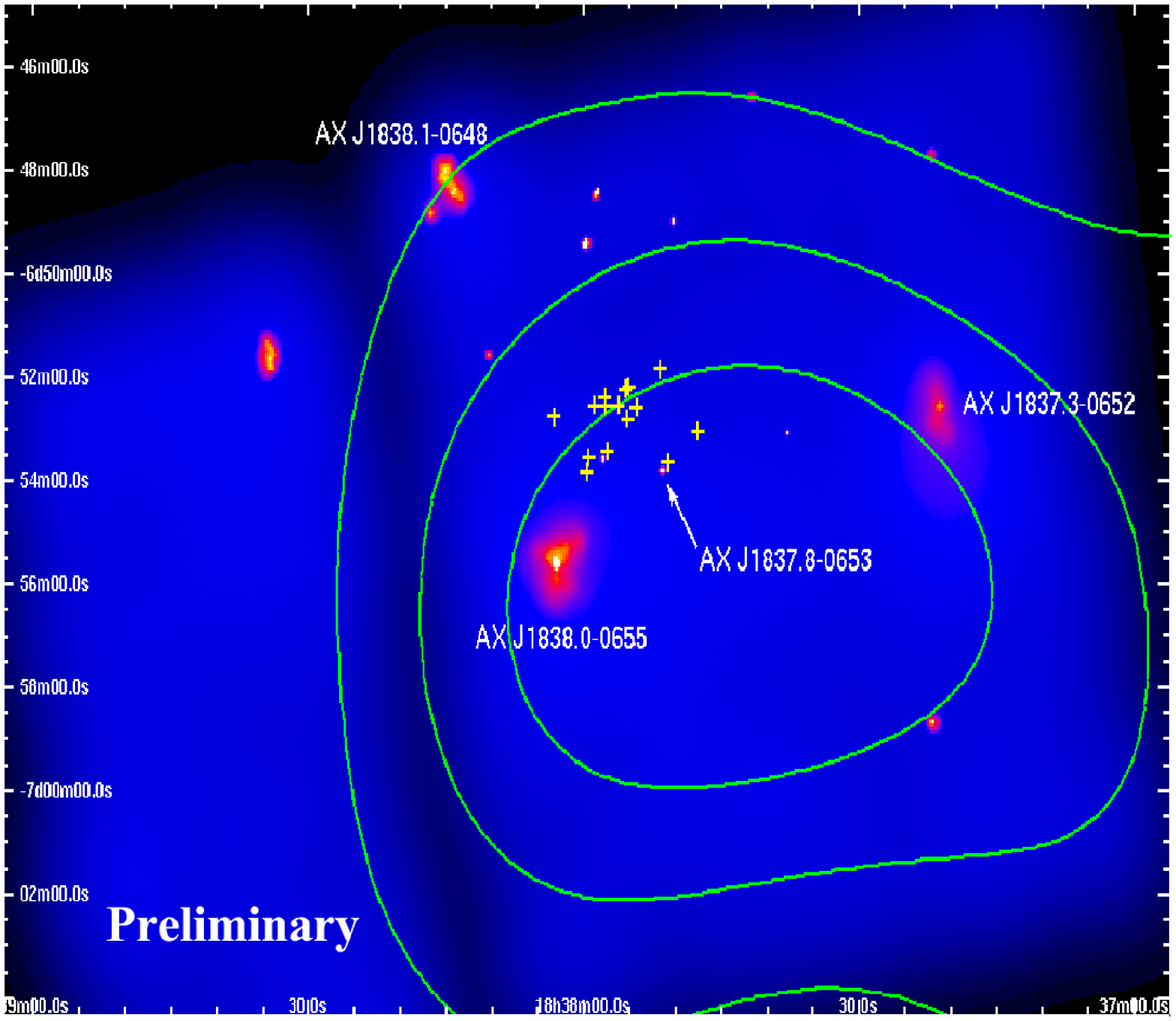}
\end{center}
\caption{Adaptively smoothed Chandra map (0.3-7 keV band - ObsId 6719 - 20 ks exposure) background-subtracted and
exposure corrected. H.E.S.S. excess contours are in green, RSG cluster stars as yellow crosses. Coordinates are J2000}
\label{fig1AM}
\end{figure}

HESS 1837-069 is an extended source discovered during the galactic plane scan survey \cite{ref21}. It shows a flux 
characterised by a power-law $I_0 (E/TeV)^{-\Gamma}$ with $I_0 \simeq 5\times10^{-12} \ \rm{TeV^{-1}} \ cm^{-2} \ s^{-1}$ 
and $\Gamma = 2.27 (\pm 0.06)$. The source has still no clear counterpart. At the edge (towards the north-east) of the peak
flux, \cite{ref22} reported on a cluster composed of 14 red super-giants (RSGs) using 2MASS data. The cluster is located at an 
estimated distance of 5.8 kpc with an age of $\sim$ 10 Myrs. In X-rays, a bright source AX J1838.0-0655 (with a 
photon index of $\Gamma=0.65$) was previously reported \cite{ref23} located at the south-East of the star cluster and close 
to the peak of the HESS source. Our Chandra observations (see figure \ref{fig1AM}) reveal a large number of previously unknown X-ray 
sources. Although diffuse X-ray emission is often associated with MSFRs, we find no evidence for extended emission coincident 
with the RSG star cluster. AX J1837.8-0653 appears as a point source coincident with a radio source (GPSR5 25.252-0139) itself at 
the center of elongated radio emission \cite{ref22}. Structure is resolved in the X-rays source AX J1838.10648 (coincident with the 
HII region W42) but this source appears too distant to power the H.E.S.S. source. Diffuse emission is apparent surrounding the 
hard X-ray source AX J1838.0-0655 (also detected by INTEGRAL \cite{ref24}).\\
Several possible scenarios can produce the gamma-ray radiation. The RSG stage is short compared to the main sequence 
phase (about one order of magnitude less). In this stage a massive star is known to have slow winds ($v_{\rm w} \sim 10-15$ km/s) and
mass losses a few times $10^{-6} M_{\odot}/\rm{yr}$. This aspect does not favor scenarios
relying on the wind activity (scenarios 1 to 3 above). Another important issue is that 
SN have probably already exploded (contrary to Westerlund 2) in the RSG cluster as suggested by \cite{ref22}. In the framework 
of an association between HESS\,J1837-069 and the RSG cluster, the scenarios of type 4 offer an interesting alternative. Further 
multiwavelength observations are necessary to confirm or reject them.

\section*{Perspectives and conclusions}
The high energy gamma-ray H.E.S.S. observations (as well as well multi-wavelength survey) are of prime importance to directly 
probe the occurrence of efficient particle acceleration processes in MSFR in connection with the origin of galactic cosmic rays.
Unfortunately, these regions are complex and associated with diverse environments (extended ionised regions, 
multiple shell structures, molecular clouds), all in themselves potential sites of particle acceleration and non-thermal radiation. 
Several scenarios of particle acceleration and radiation 
mechanisms in MSFR have been examined. They differ by the extension of the cluster and the acceleration / radiation zone, 
the massive star content, the impact of SN explosion over the cluster or the amount of dense material in the environment. 
The production of TeV gamma-rays is not systematic and requires favorable conditions: efficient conversion of free energy into 
turbulence or supersonic flows, sufficiently weak radiation losses, optically thin media. We discuss in some details the source 
HESS J1837-069. This VHE source is still unidentified and close to an exceptional cluster of RSG stars in the Galactic ridge. 
Among the scenarios we considered, an extended source associated with dense shells or a molecular cloud seems to be possible but 
further multi-wavelength observations are necessary to support this conclusion.

\section*{Acknowledgements}
\begin{small}
The support of the Namibian authorities and the University of Namibia in facilitating the construction and the operation of H.E.S.S.
is gratefully acknowledged, as is the support by the German Ministry for Education and Research (BMBF), the Max Planck Society, the French 
Ministry of Research, the CNRS-IN2P3 and the Astroparticle Interdisciplinary Programme of the CNRS, the U.K. Particle Physics and 
Astronomy Research Council (PPARC), the IPNP of the Charles University, the Polish Minisitry of Science and Higher Education, 
the South African Department of Science and Technology and National Research Foundation, and by the University of Namibia.We appreciate the 
excellent work of the technical support staff in Berlin, Durham, Hamburg, Heidelberg, Palaiseau, Paris, Saclay, and in Namibia in the 
construction and operation of the equipment.
\end{small}

\bibliographystyle{plain}

%%%%%%%%
%  25  %
%%%%%%%%

%The paper title
\title{H.E.S.S. Galactic Plane Survey unveils a Milagro Hotspot}
%Short title to print in the headers to the final publication (Not showed in this print).
\shorttitle{H.E.S.S. Galactic Plane Survey unveils a Milagro Hotspot}

%All paper authors
\authors{A. Djannati-Atai$^{1}$, E. O\~na-Wilhelmi$^{1}$,
M. Renaud$^{2}$ \& S.~Hoppe$^{2}$ for the H.E.S.S. Collaboration$^{3}$} 
%Short title to print in the headers to the final publication (Not shown in this print).
\shortauthors{Djannati-Atai et al.}
%All the affiliations.
\afiliations{$^1$APC, 11 Place Marcelin Berthelot, F-75231 Paris Cedex
05, France\\ $^2$Max-Plank-Institute fur Kernphysik, P.O. Box 103980,
D 69029 Heidelberg, Germany \\$^3$ \rm{ \texttt{www.mpi-hd.mpg.de/HESS}}}
\email{djannati@apc.univ-paris7.fr, emma@apc.univ-paris7.fr}

%The abstract.
\abstract{
We report here on a new VHE source, HESS~J1908+063, disovered during the extended
H.E.S.S. survey of the Galactic plane and which coincides with
the recently reported MILAGRO unidentified source MGRO~J1908+06.
The position, extension and spectrum measurements of the HESS
source are presented and compared to those of MGRO~J1908+06. Possible
counterparts at other wavelenghts are discussed.
For the first time one of the low-lattitude MILAGRO sources is
confirmed.

}

\maketitle

\addcontentsline{toc}{section}{H.E.S.S. Galactic Plane Survey unveils a Milagro Hotspot}
\setcounter{figure}{0}
\setcounter{table}{0}
\setcounter{equation}{0}

%Begin the section.
\section*{Introduction}
H.E.S.S. observations of the inner Galactic plane in the [$270^{\circ}$, $30^{\circ}$]
longitude range have revealed more than two dozens of  new VHE
sources, consisting of shell-type SNRs, pulsar wind nebulae,
X-ray binary systems, a putative young star cluster, etc, and yet
unidentified objects (see e.g. \cite{HESSScanII} and  \cite{HESSSurveyICRC07} in
these proceedings for a summary). 

The extended  H.E.S.S. survey in the
[$30^{\circ}$-$60^{\circ}$] longitude range performed between 2005 and
2007 overlaps with regions covered by the MILAGRO sky survey at longitudes
greater than $30^{\circ}$.
The latter experiment has recently reported \cite{MILAGRO} three
low-latitude sources including,  MGRO~J1908+06,
detected after seven years of operation (2358 days of data) at
8.3$\sigma$ (pre-trials) confidence level. 
MGRO~J1908+06, of which the extension remains unknown but bounded to
a maximum diameter of 2.6$^{\circ}$, is located near the 
galactic longitude $\sim 40^{\circ}$ and hence is covered by the
H.E.S.S. galactic plane survey. 

A new H.E.S.S. source, HESS~J1908+063, which coincides
with MGRO~J1908+06, is presented here. Its position, size and spectrum
are measured and compared to the MILAGRO source. 
Possible counterparts at other wavelengths are discussed in the light
of the H.E.S.S. measurements.

%%%%%%%%%%%%%%%%%%%%%%%%%%%%%%%%%%%%%%%%%%%%%%%%%%%%%%%%%%%%%%%%%%%%%%%%%%%%%%%%%
\section*{Observations, Analysis \& Results}
\label{results}
%%%%%%%%%%%%%%%%%%%%%%%%%%%%%%%%%%%%%%%%%%%%%%%%%%%%%%%%%%%%%%%%%%%%%%%%%%%%%%%%%
Results presented in this section should be
considered as preliminary.

Observations around HESS~J1908+063 were first performed during June 2005
and then from May to September 2006
as part of the extension of the Galactic plane survey in the range of
galactic longitude and latitude of 30$^{\circ}$ $<$~l~$<$60$^{\circ}$ and
$-3^{\circ}$~$<$~b~$<$3$^{\circ}$, respectively. Followup observations
were made during May and June 2007.  
In the available data-set the source is offset from the field of
view center, at different angular distances with an
average offset of 1.4$^{\circ}$. Observations for which the
source is offset by more than 2.5$^{\circ}$ were not considered for the
analysis. The total dead-time corrected and quality selected data-set
amounts to 14.9~hours with the zenith angle ranging from 30
to 46$^{\circ}$ and with a mean energy threshold  of $\sim$300~GeV. 

\begin{figure}[!t]% [!tp]
\begin{center}
\includegraphics[width=0.48\textwidth,angle=0,clip]{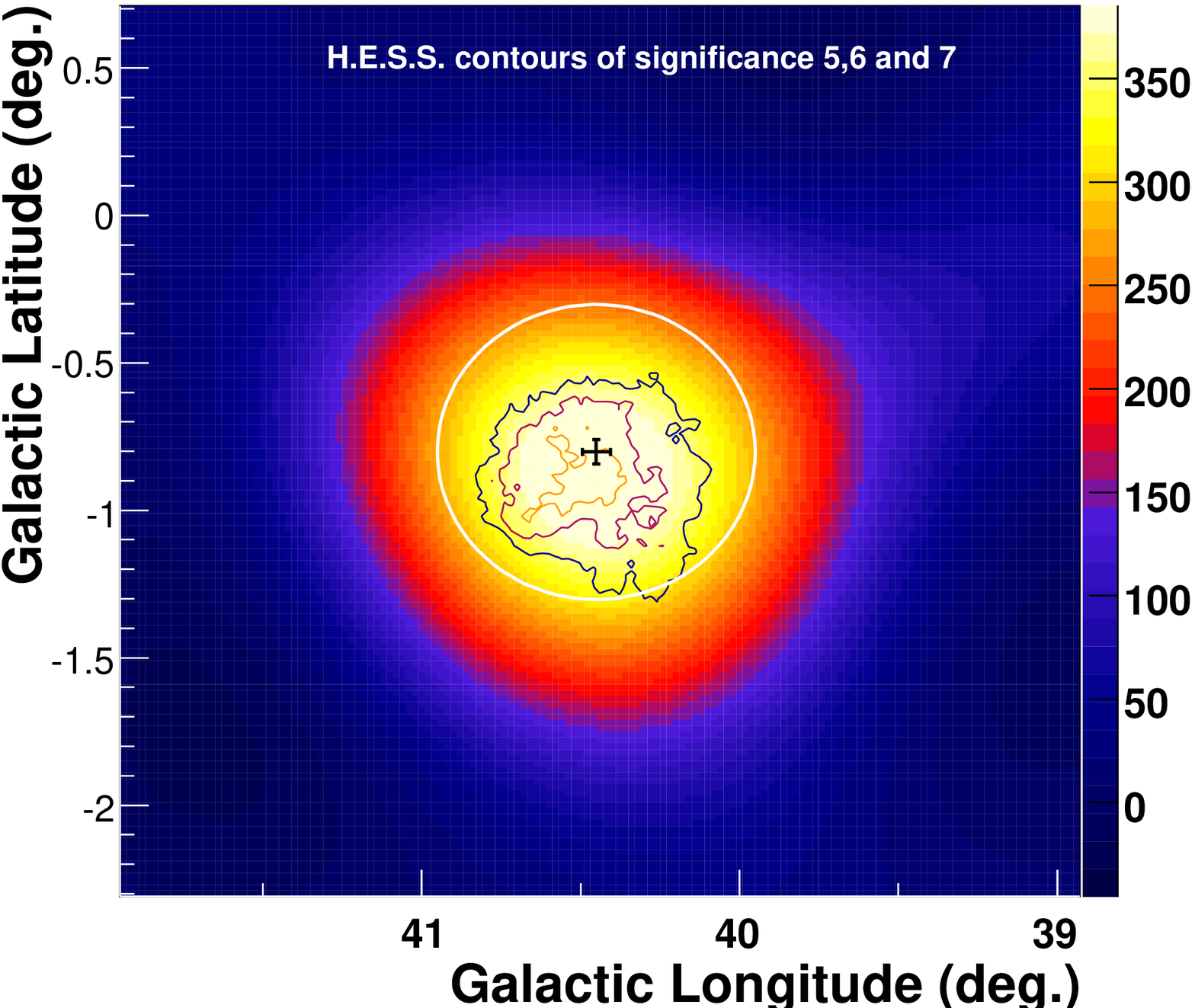}
\end{center}
\caption{ Smoothed excess map
($\sigma=0.5^{\circ}$) of the 1.5$^{\circ}\times1.5^{\circ}$ field of
view around the position of HESS~J1908+063. The contours show
the pre-trials significance levels for 5, 6 and 7$\sigma$, while the white
    circle shows the  $0.5^{\circ}$ integration radius used for the
    spectrum derivation. }
\label{skymap}
\end{figure}

After calibration, the standard H.E.S.S. event reconstruction scheme
was applied to the data \cite{HESSCrab}. In order to
reject the background of cosmic-ray showers, $\gamma$-ray like
events were selected using cuts on image shape scaled with their
expected values obtained from Monte Carlo simulations. As described in
\cite{HESSKooka}, two different sets of cuts, depending on the image
size, were applied. Cuts optimized for a hard spectrum and a weak
source with a rather tight cut on the image size of 200 p.e. (photoelectrons), 
which achieve a maximum signal-to-noise ratio, were applied to
study the morphology of the source, while for the spectral analysis,
the image size cut is loosened to 80 p.e. in order to cover the
maximum energy range. The background estimation
(described in \cite{HESSBack}) for each position in the
two-dimensional sky map is computed from a ring with an (apriori)
increased radius of $1.0^{\circ}$, as compared to the standard radius
of $0.5^{\circ}$, in order to deal with the large source diameter.
This radius yields four times a larger area for the background estimation
than the considered on-source region. Also events coming from known sources were
excluded to avoid contamination of the background. For the spectrum analysis, the
background is evaluated from positions in the field of view with the
same radius and same offset from the pointing direction as the source region.

Fig.~\ref{skymap} shows the Gaussian-smoothed excess map for a size
cut on the images above 200 p.e. The 
colored contours indicate the H.E.S.S. pre-trials significance contour
levels for 5,6 and 7 $\sigma$. HESS~J1908+063 was discovered first as a
hot-spot within the standard survey analysis scheme \cite{HESSScanII} and was
subsequently confirmed at 7.7 $\sigma$ (pre-trials). 
A conservative estimate of the trials yields a post-trials
significance of 5.7 $\sigma$.   

To evaluate the extension and the position of the source, 
the sky-map was fitted to a simple symmetrical two-dimensional Gaussian
function, convolved with the instrument PSF (point spread function).  
The best-fit position lies at
$l=40.45^{\circ}\pm0.06_{stat}^{\circ}\pm0.06_{sys}^{\circ}$ and
$b=-0.80^{\circ}\pm0.05_{sta}^{\circ}\pm0.06_{sys}^{\circ}$, while the intrinsic
extension derived is $\sigma_{src}=(0.21^{\circ}+
0.07_{sta}^{\circ}-0.05_{sta}^{\circ}$). As the shape of the source
seems to depart from a symmetrical Gaussian, these values should be taken
as first approximations.

\begin{figure}[!b] 
\begin{center}
\includegraphics[width=0.51\textwidth]{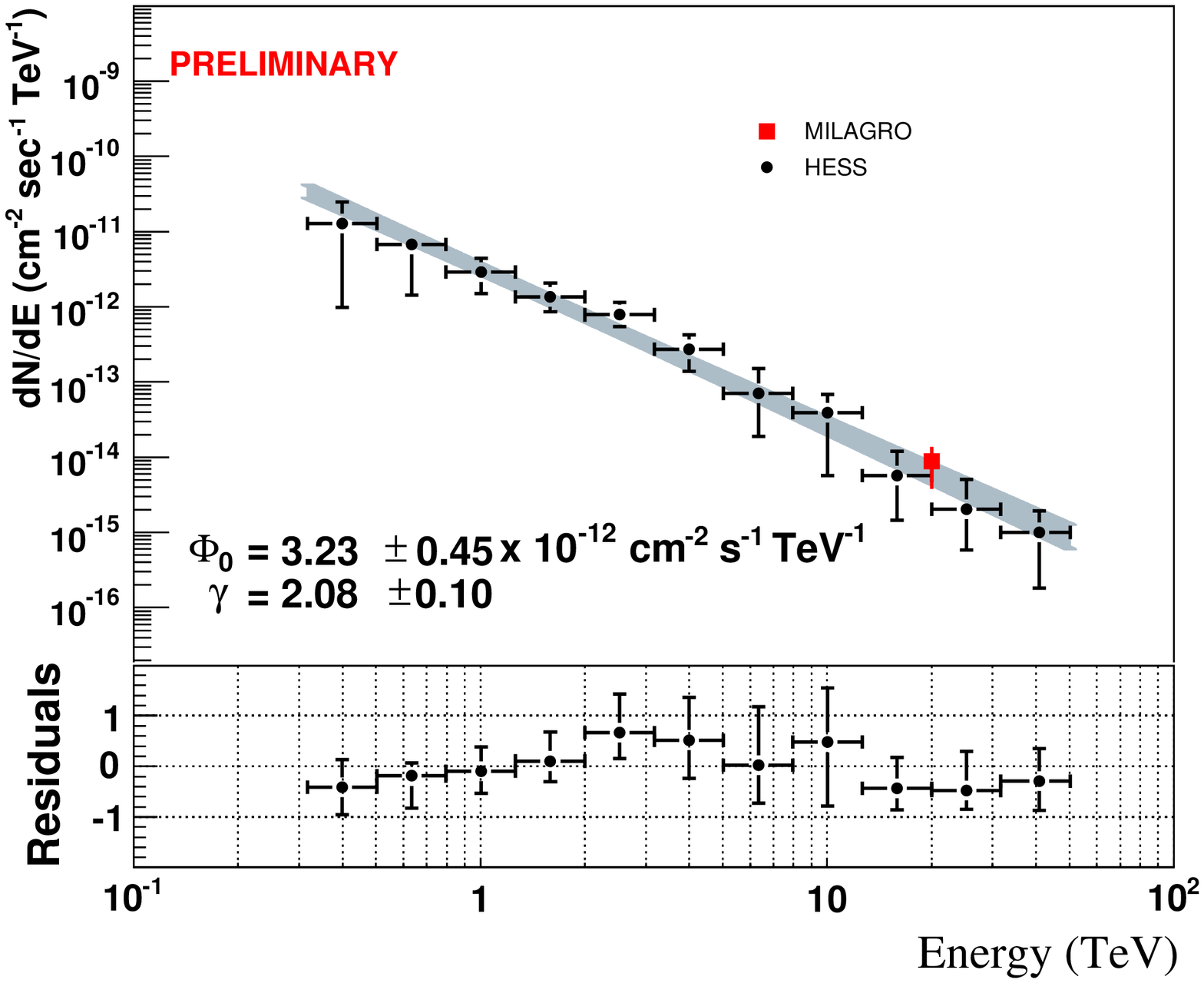}
\end{center}
\caption{Differential energy spectrum measured above 300~GeV for
HESS~J1908+063. The shaded area shows the 1 $\sigma$  
confidence region for the fit parameters. 
The differential flux of MGRO~J1908+06 at 20 TeV is shown in red. 
Fit residuals are given in the bottom panel.}  
\label{spectrum} 
\end{figure}

The differential energy spectrum was computed within an integration
radius of $0.5^{\circ}$ (corresponding to the FWHM of the source size
and shown as a white circle in Fig.~\ref{skymap}) 
centred on the best-fit position 
by means of a forward-folding maximum
likelihood fit \cite{CATSpectrum}. The spectrum is well fitted with a simple
power-law function (Fig.~\ref{spectrum}) with a hard photon index of
$2.08\pm0.10_{\rm stat}\pm 0.2_{\rm sys}$ 
and a differential flux at 1~TeV of ($3.23 \pm
0.45_{\rm stat} \pm 0.65_{\rm sys})\times 
10^{-12}$ cm$^{-2}$~s$^{-1}$. 
The integrated flux above 1~TeV corresponds to 14$\%$
of the Crab Nebula flux above that energy.

%%%%%%%%%%%%%%%%%%%%%%%%%%%%%%%%%%%%%%%%%%%%%%%%%%%%%%%%%%%%%%%%%%%%%%%%%%%%%%%%%
\section*{Comparison with MGRO1908+06 \& Search for Counterparts}
\label{comparison}
%%%%%%%%%%%%%%%%%%%%%%%%%%%%%%%%%%%%%%%%%%%%%%%%%%%%%%%%%%%%%%%%%%%%%%%%%%%%%%%%%

\begin{figure}[ht]
\begin{center}
\includegraphics*[width=0.5\textwidth,angle=0,clip]{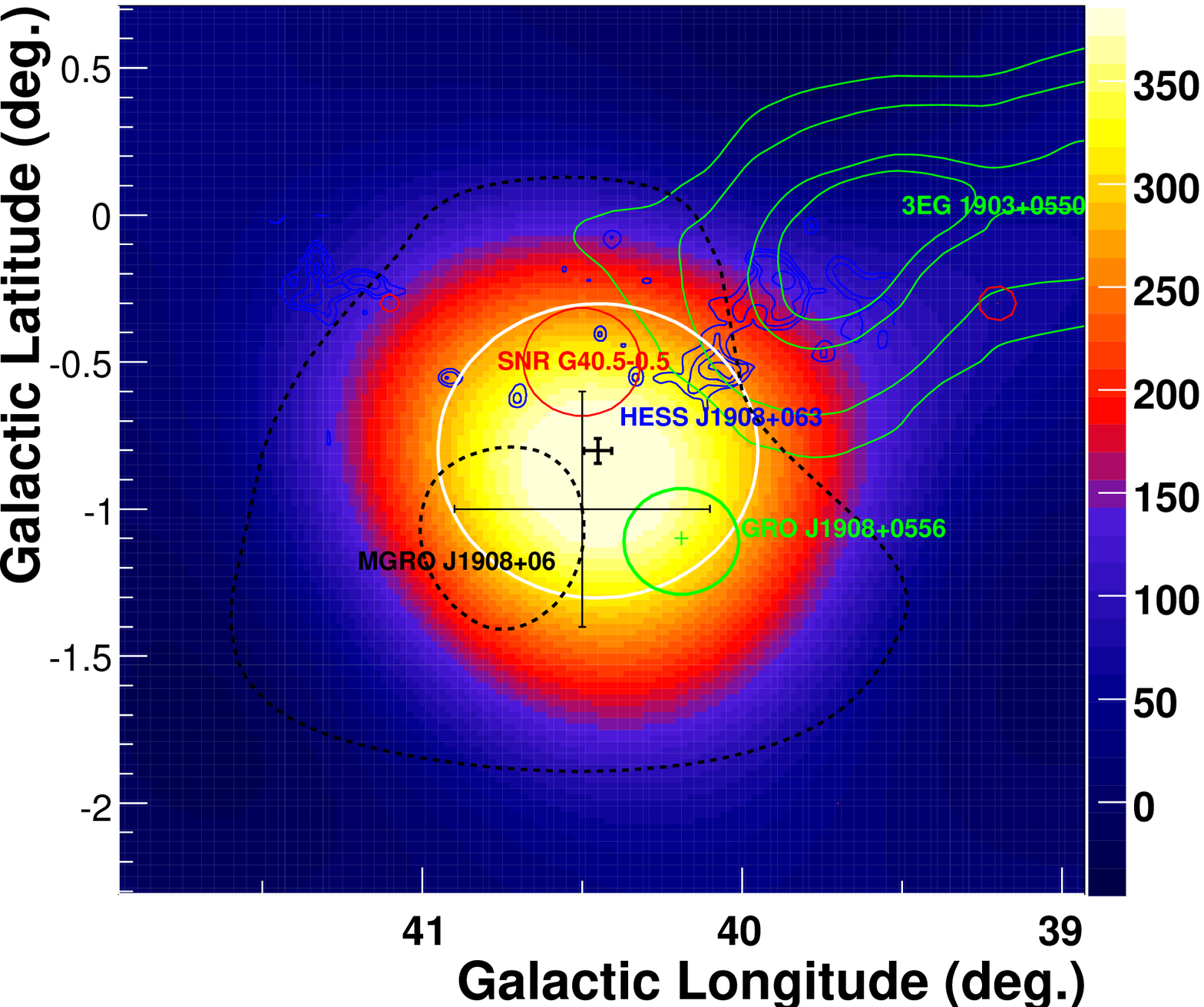}
\end{center}
\caption{ Multi-wavelength view of the 1.5$^{\circ}\times1.5^{\circ}$ field of
view around the position of HESS~J1908+063. The
    dotted black line shows the MILAGRO significance contours for 5 (inner) and
    8$\sigma$ (outer contour). The position of the EGRET GeV source
    GRO~J1908+0556 is marked with a green cross as well as the
    1$\sigma$ error in the position.  The 3EG 1903+0550 contours
    corresponding to 99, 95, 68 and 50$\%$ confidence levels are shown in
    green. The red circle marks the size and position of the radio-bright SNR
    G040.5-00.5. Contours in blue show the $^{13}$CO molecular cloud in the
    velocity range between (45,65) km/s.}
\label{skymapmwl}
\end{figure}

Fig.~\ref{skymapmwl} shows the $1.5^{\circ}\times1.5^{\circ}$ field of
view around the position of HESS~J1908+063 together with 
sources at other wavelengths including MGRO~J1908+06. The latter
source was discovered by the MILAGRO collaboration \cite{MILAGRO} 
after seven years of operation (2358 days of data) at the galactic
longitude and latitude of
$l=(40.4^{\circ}~\pm~0.1^{\circ}_{\rm stat}~\pm~0.3^{\circ}_{\rm sys}$) and  
$b=(-1.0^{\circ}~\pm~0.1^{\circ}_{\rm stat}~\pm 0.3^{\circ}_{\rm sys}$),
respectively.  The differential flux, at the median energy of 20~TeV, and
assuming a spectral index of -2.3, is at a level of  
(8.8$\pm$2.4$_{\rm stat}\pm$2.6$_{sys})\times
10^{-15}~{\rm TeV}^{-1}{\rm cm}^{-1}{\rm s}^{-1}$. MGRO~J1908+06 is reported to be 
both compatible with a point or extended source up to a diameter of
2.6$^{\circ}$. 

As clearly seen on Fig.~\ref{skymapmwl}, the positions of the two VHE sources
are fully compatible within errors. There is also a quite good
agreement between the differential flux at 
20~TeV of  MGRO~J1908+06 and the spectrum measured by HESS as shown on
Fig.~\ref{spectrum}. Given the larger integration radius of
$1.3^{\circ}$ for the MILAGRO source as compared to the $0.5^{\circ}$ radius
for HESS~J1908+063, the flux agreement implies the absence of any other
significant emission to the MILAGRO flux: the two sources can
consequently be identified to each other.

The better determination of the position of HESS~J1908+063 and the
measurement of its size and spectrum allow to search for counterparts
with stronger constraints.

At radio wavelengths, SNR~G40.5-0.5 \cite{Green} at
an estimated distance of 5.3~kpc overlaps with HESS~J1908+063. 
At EGRET energies, 3EG~J1903+0550, shown in green contours, lies close
to the SNR and has been suggested as possibly associated with it \cite{Sturner95}.
However G40.5-0.5 is not in exact coincidence with HESS~J1908+063
position and 3EG~J1903+0550 is only marginally overlapping with the
latter.  HEGRA observations of this region of
the sky \cite{HEGRAul} yielded an upper limit at 0.7~TeV at the SNR
position of 4.8$\%$ of the Crab Nebula flux.   
As this limit only applies for a point-like source it is not in
contradiction with the measurements reported here.

If the SNR is associated with the VHE source, the fact that the 22 
arc-min size of the shell is smaller than the FWHM of HESS~J1908+063
would contrast to previously discovered HESS sources identified
with shell-type VHE emitters, such as RX~J1713.7-3946
\cite{HESSG347} or RCW 86 reported at this conference
\cite{RCW86ICRC07}. The contribution of nearby unresolved sources or
interactions of accelerated cosmic rays with molecular matter in the vicinity of
the source could explain a larger size. However, for the latter
case, the position of the nearby $^{12}\rm CO$ cloud \cite{co} or
alternatively the $^{13}\rm CO$ contours (shown in blue on Fig.~\ref{skymapmwl}) do not favour
this scenario.

An analysis of the highest energy photons ($>$1~GeV) observed by EGRET
\cite{olaf97,lamb97} from this region shows a nearby and yet unidentified
source, GRO~J1908+0556/GEV J1907+0557. The positions of the two GeV
derivations are compatible within errors. GRO~J1908+0556, shown as a
green circle on Fig.~\ref{skymapmwl}, lies within a distance of less than
two times the EGRET 68\%  position measurement error to
HESS~J1908+063. A simple extrapolation of the H.E.S.S. spectrum to 
lower energies leads to a lower flux than that reported
for the EGRET source  ($6.33\times10^{-8}~{\rm cm}^{-1}{\rm s}^{-1}$). However
given the large PSF of EGRET even at GeV energies, other unresolved
sources can contribute to the flux measurement of GRO~J1908+0556. The
association of the HESS and MILAGRO sources to the GeV source is then
likely, although a coincidence by chance is not
excluded.

\section**{Summary}
In summary, a new source, HESS J1908+063 is reported above 300 GeV
at the level of 14$\%$ of the Crab Nebula flux and a post-trials
significance of 5.7~$\sigma$. The H.E.S.S.
source is extended, with a FWHM size of
0.5$^{\circ}$, and shows a hard spectrum with an index of 2.08$\pm$0.10. 
This detection confirms for the first time one of the low-latitude sources 
reported by the MILAGRO collaboration, MGRO~1908+062. 
A connection to the EGRET GeV source GRO~J1908+0556/GEV J1907+0557 at
lower energies remains possible. The association with SNR~G40.5-0.5 is not
excluded but the larger size of the TeV emission should then find an
explanation in terms of either contribution of unresolved sources or
interactions of ultra-relativistic particles with molecular matter in
the vicinity of the SNR. Deeper observations of this region with
Cherenkov telescopes and GLAST data would help the interpretation
of the detected VHE emission.

\section**{Acknowledgments}
The support of the Namibia authorities and of the University of Namibia
in facilitating the construction and operation of H.E.S.S. is gratefully
acknowledged, as is the support by the German Ministry for Education and
Research (BMBF), the Max Planck Society, the French Ministry for Research,
the CNRS-IN2P3 and the Astroparticle Interdisciplinary Programme of the
CNRS, the U.K. Particle Physics and Astronomy Research Council (PPARC),
the IPNP of the Charles University, the Polish Ministry of Science and
Higher Education, the South African Department of
Science and Technology and National Research Foundation, and by the
University of Namibia. We appreciate the excellent work of the technical
support staff in Berlin, Durham, Hamburg, Heidelberg, Palaiseau, Paris,
Saclay, and in Namibia in the construction and operation of the
equipment.

%This in the bibtex style, is ok.
\bibliographystyle{plain}

%%%%%%%%
%  26  %
%%%%%%%%

\title{Search for very high energy gamma-ray emission 
       from parts of the Gould belt with the H.E.S.S. 
       ground based Cherenkov telescopes}
%Short title to print in the headers to the final publication (Not showed in this print).
\shorttitle{H.E.S.S. observations of the Gould belt}
%All paper authors
\authors{D. Horns$^{1}$, G.P. Rowell$^{2}$, F. Aharonian$^{3,4}$,
S. Gabici$^{4}$, A. Santangelo$^{1}$, S. Schwarzburg$^{1}$ for the H.E.S.S. collaboration.}
%Short title to print in the headers to the final puplication (Not showed in this print).
\shortauthors{D. Horns et al. for the H.E.S.S. coll.}
%All the affiliations.
\afiliations{$^1$ Institute for Astronomy and Astrophysics, University Tuebingen\\ 
             $^2$  School of Chemistry and Physics, Adelaide, Australia\\
	     $^3$  Dublin Institute for Advanced Science, Dublin, Ireland \\
	     $^4$ Max-Planck Institute for nuclear physics, Heidelberg, Germany}
\email{horns@astro.uni-tuebingen.de}

%The abstract.
\abstract{The Gould belt, 
a well-known region of enhanced star formation in the solar
neighbourhood, is observed to be an expanding disk with a diameter of about
1~kpc  and a width of a few 100~pc.  Most of the nearby
OB stellar associations and molecular clouds are found to be
aligned with the Gould belt.  With the high star formation rate along the
Gould belt, the local supernova rate during the last few million years is
believed to be three to four times larger than the Galactic average. Under the
assumption that supernova remnants are efficient accelerators of cosmic rays,
the Gould belt and its environment should show an increased cosmic ray density
with respect to the Galactic average. The cosmic rays are expected to interact
with the dense molecular gas which results mainly in pi-meson production with
subsequent decay in gamma-rays and neutrinos. We have searched for gamma-ray
emission from various parts of the Gould belt
with the HESS Cherenkov telescopes. Results will be presented at the
conference.}

\maketitle

\addcontentsline{toc}{section}{Search for very high energy gamma-ray emission from parts of the Gould belt with the H.E.S.S. ground based Cherenkov telescopes}
\setcounter{figure}{0}
\setcounter{table}{0}
\setcounter{equation}{0}

\section*{Introduction}
The Gould belt is a local region with enhanced stellar formation and molecular clouds
within 0.5~kpc of the solar system (for a review see \cite{2001ASPC..243..667P}). A large fraction 
of young (spectral type O and B) massive stars in the solar vicinity 
are aligned with the Gould belt. However, late type stars have been 
associated to the Gould belt as well \cite{1998A&A..337..113G}.  The age of the stars ($\approx 50$~Myrs) 
tracing the Gould belt and the dynamical timescale derived from the expansion velocity 
of the belt like structure ($\approx 25$~Myrs, \cite{1999A&A..351..506C,2003A&A..404..519P}) 
constrains the age of the Gould belt within roughly a factor of 2 to be $25$--$50$~Myrs.\\ 
The nature of the event which triggered the star formation in the Gould belt is unclear, 
but various suggestions have been made including a cascade of supernova explosions or 
the impact of a high velocity cloud to the Galactic plane. An initial kinetic energy of $10^{52}$~ergs
(equivalent to 10 supernova explosions) is required to drive the Gould belt expansion \cite{2003A&A..404..519P}. 
\\
The supernova explosion rate of 75 to 95 Myr$^{-1}$~kpc$^{-2}$ implied by the age and 
stellar present population in the Gould belt 
is a factor of 3-5 larger than the expected local value of 20 Myr$^{-1}$~kpc$^{-2}$ \cite{2000A&A..364L..93G}.
The more massive molecular cloud complexes associated with the Gould belt (e.g. Taurus,
$\rho$-Oph, Lupus, Orion A, Orion B) are therefore good targets
to search for gamma-ray emission from the interaction of cosmic rays with the dense molecular clouds
(see e.g. \cite{1991Ap&SS.180..305A}).
\section*{Observations of regions in the Gould belt with the H.E.S.S. telescopes}
 The H.E.S.S. system of four imaging air Cherenkov telescopes is located at an altitude of 1\,800~m
 in the Khomas Highlands of Namibia. The telescope system is sensitive to Gamma-rays
 above 100~GeV. Parts of the Gould belt have been observed during dedicated observation
 runs as well as part of observations taken on dedicated targets which coincide with
 the Gould belt region. 

\section*{Results and Discussion}
 The final results of the analysis will be presented at the conference.
\section*{Acknowledgements}
The support of the Namibian authorities and of the University of Namibia
in facilitating the construction and operation of H.E.S.S. is gratefully
acknowledged, as is the support by the German Ministry for Education and
Research (BMBF), the Max Planck Society, the French Ministry for Research,
the CNRS-IN2P3 and the Astroparticle Interdisciplinary Programme of the
CNRS, the U.K. Science and Technology Facilities Council (STFC),
the IPNP of the Charles University, the Polish Ministry of Science and 
Higher Education, the South African Department of
Science and Technology and National Research Foundation, and by the
University of Namibia. We appreciate the excellent work of the technical
support staff in Berlin, Durham, Hamburg, Heidelberg, Palaiseau, Paris,
Saclay, and in Namibia in the construction and operation of the
equipment.

\bibliographystyle{plain}

%%%%%%%%
%  27  %
%%%%%%%%

%The paper title
\title{
Long-Term VHE Gamma-Ray Monitoring of \pks\ with H.E.S.S. and Multiwavelength measurements, 2002-2005}
%Short title to print in the headers to the final publication (Not showed in this print).
\shorttitle{Long-Term PKS \pks\ with H.E.S.S.}
%All paper authors
\authors{M. Punch$^{1}$, for the H.E.S.S. Collaboration$^{2}$.}
%Short title to print in the headers to the final puplication (Not showed in this print).
\shortauthors{M. Punch for the H.E.S.S. Collaboration}
%All the affiliations.
\afiliations{$^1$
Astroparticule et Cosmologie (APC), 
UMR 7164 (CNRS, Universit\'e Paris VII, CEA, Observatoire de Paris), Paris, France
%10, rue Alice Domon et Leonie Duquet, F-75205 Paris Cedex 13, France
$^2$ \rm{ \texttt{www.mpi-hd.mpg.de/HESS}} }
\email{punch@in2p3.fr}

%The abstract.
\abstract{
The high-frequency peaked BL Lac \pks, the lighthouse of the Southern hemisphere sky at VHE gamma-ray energies, has been followed by the H.E.S.S. array of atmospheric Cherenkov telescopes since the first light of the project, first with a single telescope in 2002, then with two \& three telescopes in 2003, and since 2004 with the full-sensitivity four-telescope array.  In this mode, a number of multi-wavelength campaigns have been performed with observations from the Rossi X-ray Timing Explorer (RXTE), Rotse (Optical), Spitzer (IR), James Clark Maxwell Telescope (JCMT, sub-mm) and others in both quiescent and active states, based on both fixed campaigns and triggers from H.E.S.S.  Here we present the results of this series of observations up to 2005 inclusive, together with the implications for source models of the spectral measurements and search for correlated variability with X-rays, Optical, and IR measurements.  The exceptional flare activity of 2006 is covered in a separate paper at this conference.
}

\maketitle

%%%%% Begin AGN %%%%%%
\addtocontents{toc}{\protect\contentsline {part}{\protect\large Active Galactic Nuclei (AGN)}{}}
\addcontentsline{toc}{section}{Long-Term VHE Gamma-Ray Monitoring of \pks\ with H.E.S.S. and Multiwavelength measurements, 2002-2005}
\setcounter{figure}{0}
\setcounter{table}{0}
\setcounter{equation}{0}

%Begin the section.
\section*{Introduction}
The H.E.S.S. experiment (High Energy Stereoscopic System, \cite{hinton}) has been in operation since 2002, though reaching full sensitivity in 2004 with the installation of the final element in the four-telescope array.  During all this time, a prime target of H.E.S.S. has been the high-frequency peaked BL Lac \pks, which is one of the brightest blazars in the Southern sky in the very-high-energy (VHE) $\gamma$-ray domain.
These observations have been coupled with multi-wavelength campaigns, both planned and resulting from Target of Opportunity (ToO) triggers by H.E.S.S. in the case of exceptional activity of this source, working with such instruments as the Rossi X-ray Timing Explorer (RXTE), the Chandra X-ray satellite, the X-ray telescope (XRT) and UV-optical telescope (UVOT) on board the SWIFT satellite, Spitzer (IR), James Clark Maxwell Telescope (JCMT, sub-mm), the Rapid Optical Transient Search Explorer (ROTSE-III), and the Nan\c{c}ay Radio Telescope (NRT).

The source \pks, at a red-shift of $z=0.116$, is an X-ray selected object of the BL Lac class, and has been the focus of studies in many wavelength ranges (see e.g. \cite{urry}) over the past 20 years.  It was first seen at VHE energies by the Durham Mark 6 telescope \cite{chadwick}, and this detection was confirmed by H.E.S.S. in the observations referred to here.  The multi-wavelength observations of this source have shown broad-band variability on timescales from years to minutes, as also seen in the VHE range with H.E.S.S.  This variability is presumed to be associated with a relativistic jet aligned close to the line of sight to the observer, allowing the processes in the jet to be probed by such studies. 

\comment{
\begin{table*}[t]
 \caption{Results, average flux above 1 TeV in units of $10^{12} \rm cm^{-2}s^{-1}TeV^{-1}$}
\begin{center}
\begin{tabular}{l|lrlccr}
\hline
Obs. Period		& Live-time	& Avg.	& Error		& Signif.	&	Excess & Telescope Status\\
{[}months--year{]}	& [hrs]		& flux	&&	($\sigma$)\\
\hline
07--02		& $\sim 4$	&15.6 &$\pm$2.1	&13  & Single dish, 1-telescope   \\
10--02		& $\sim 10$	&6.4  &$\pm$1.8	&8   & Faux-Stereo, 2-telescopes \\
vskip{10pt}
06--03		& $\sim 10$	&2.4  &$\pm$1.8	&8   & Stereo, 2-telescopes \\
07--03		& 		&1.8  &$\pm$1.8	&8   & Stereo, 2-telescopes \\
08--03		& 		&1.8  &$\pm$1.8	&8   & Stereo, 2-telescopes \\
09--03		& $\sim 50$	&2.4  &$\pm$1.8	&8   & Stereo, 2-telescopes \\
10:11--03	& 		&2.7  &$\pm$1.8	&8   & Stereo, 2-telescopes \\
vskip{10pt}
06--03		& $\sim 10$	&6.4  &$\pm$1.8	&8   & Stereo, 2-telescopes \\
\hline
\end{tabular}
\end{center}
\end{table*}}

\section*{Observations with H.E.S.S.}
\pks\ has been observed by H.E.S.S. since its inception, first with a single telescope in 2002, then with two \& three telescopes in 2003, and since 2004 with the full-sensitivity four-telescope array.
Figure \ref{fig1MP} shows a summary of the observations taken on this source, where in each case the average integral flux above $1\;$TeV is shown for each observation period, together with the significance of the detection above the threshold of the instrument in the given configuration.  All data here are analysed using the H.E.S.S. standard analysis procedures, for data passing run quality selection criteria, some previously reported in \cite{fontaine}.
The data for 2005/2006 are still undergoing analysis, and these preliminary results are shown for comparison, including the excess number of $\gamma$-rays above threshold for both standard and loose cuts. The 2006 data are not detailed on a month-by-month basis here.

The most striking conclusions which can be drawn from this series of observations is that the source is clearly variable on year-by-year and month-by-month timescales, and that it is detected in all months where a hour or more observation time was taken.  
It can also be seen that the activity of the source was very high in 2002, though it was observed with a low-sensitivity single-telescope configuration; this activity was greater on average than that in 2006.  However, as described in \cite{wystan} at this conference \& published in \cite{bigflare}, the large data-set in 2006 contains nights with exceptional activity, in one of which $\sim 3$ minute time-variability is seen.

\section*{Multi-wavelength campaign in 2003}

\begin{figure*}
\begin{center}
%\parbox{0.55\textwidth}{
\includegraphics [width=0.6\textwidth]{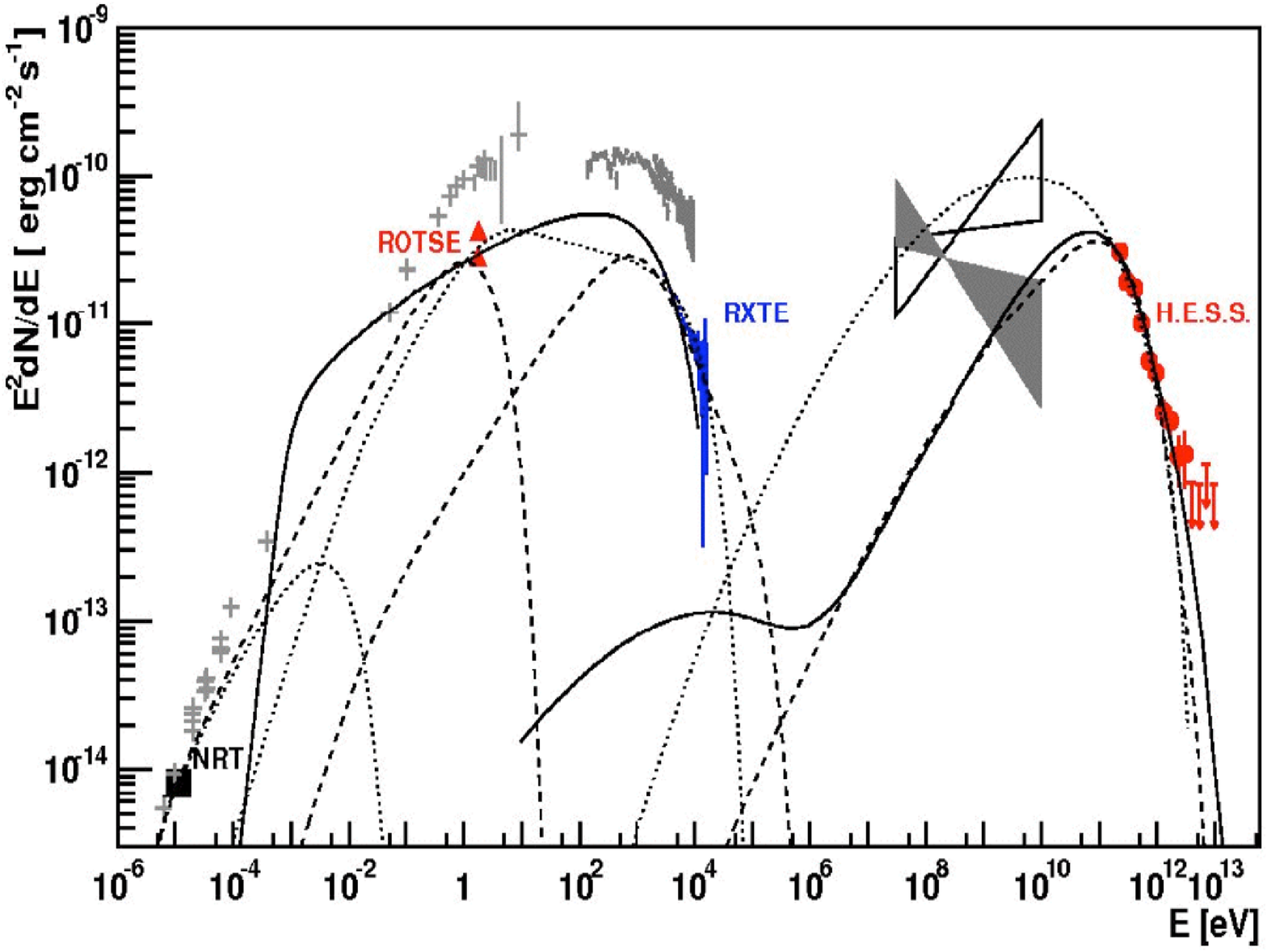}
%}
\end{center}
\vskip -20pt
%\parbox{0.45\textwidth}{
\caption{SED of \pks\ as measured in 2003. Only simultaneous data are labelled; non-contemporaneous data are shown in grey.
The H.E.S.S. spectrum is from Oct. and Nov. 2003 data (filled circles) as is the RXTE spectrum. The NRT radio point (filled square) is the average value for the observation during this period. The two triangles are the highest and lowest ROTSE measurements for the Oct.-Nov. observations. Archival SAX data show the high state observed in 1997 \cite{sax}. Archival EGRET data are from the third EGRET catalogue (shaded bowtie) and from a very high $\gamma$-state reported in \cite{egret} (open bowtie). The solid line is the hadronic blazar model described in the text, while the dotted and dashed lines are the leptonic models}
%}
\label{fig2MP}
\end{figure*}

On October 18th 2003, the highly-significant detections of this source with H.E.S.S. (in its 3-telescope stereo configuration) prompted the triggering of an RXTE ToO proposal, with a number of quasi-simultaneous observations being taken in conjunction with RXTE/PCA, ROTSE, and NRT.  From October 19th to November 26th, 2003, 
This campaign provided the first wide-band simultaneous spectrum on \pks, although the source appeared to be in a low state during these observations.  The Spectral Energy Distribution measured by the instruments in this campaign is shown in Figure \ref{fig2MP}.  The average spectral index measured with H.E.S.S. was $3.37\pm0.07_{\rm stat}$, similar to that measured at other epochs. The simultaneous data-points are fitted by several models.  
Due to the relatively distant red-shift of this source, the models must include absorption on the Extragalactic Background Light (EBL) in the $\mu$metre range.  
Two leptonic models were fitted:  single-zone Synchrotron Self-Compton (SSC) models in which the optical emission is either from the same population of electrons as the X-ray, or produced by an extra VLBI component, probably from the compact core, along with the radio.  These models imply values of EBL in the low range for reasonable values of fitted magnetic field and Doppler factor.  
One hadronic model was tested, the  Synchrotron Proton Blazar (SPB) model, which also could fit the data for a range of EBL values.
Note that subsequent observations of more distant blazars with H.E.S.S., 1ES$\;$1101$-$232 and H$\;$2356$-$309, favour the low range of EBL values also \cite{nature}.

\begin{figure}
\begin{center}
\includegraphics [width=0.46\textwidth]{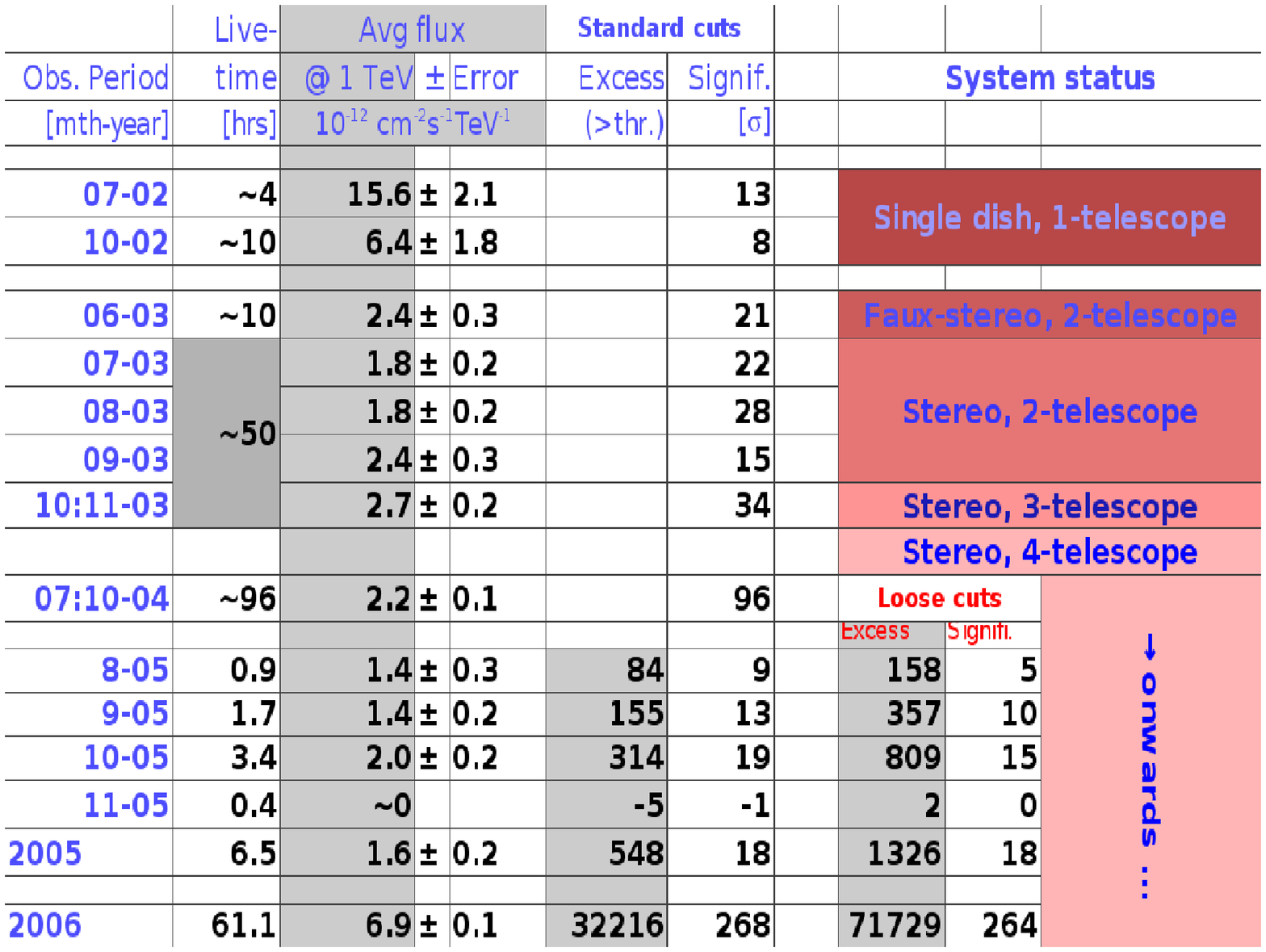}
\end{center}
\vskip -20pt
\caption{
Long-term monitoring of \pks\ with H.E.S.S., showing where available for each observation period, the observation time, average flux above 1 TeV in units of $10^{12} \rm cm^{-2}s^{-1}TeV^{-1}$, $\gamma$-ray excess above threshold, and significance.
}
\label{fig1MP}
\end{figure}

\begin{figure}[t]
\begin{center}
\includegraphics [width=0.46\textwidth]{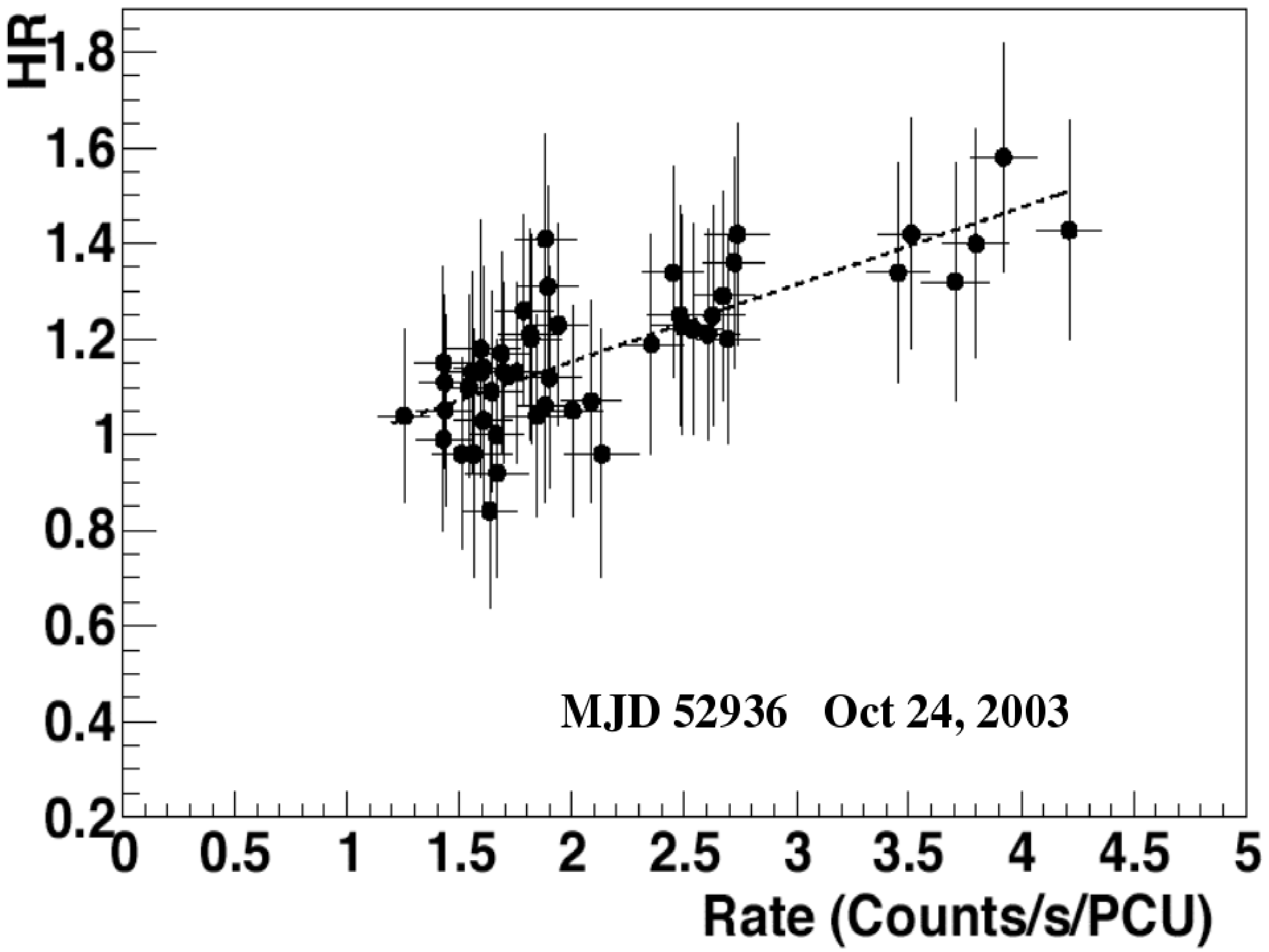}
\end{center}
\vskip -20pt
\caption{
Correlation within the RXTE measurements taken during the H.E.S.S. multi-wavelength campaign in 2003, where a clear correlation is seen with between the X-ray activity and the hardness ratio as defined in the text, with a correlation factor $r = 0.76\pm0.12$.
}
\label{fig3MP}
\end{figure}

In this campaign, no correlation was seen between the VHE $\gamma$-rays and the X-rays or the optical flux, or indeed between X-rays and optical.  However, within the X-ray band a correlation was seen between the hardness ratio (defined as ratio of the flux from 4--$11\;$keV to that from 1--$4\;$keV) and the rate, with a correlation of $r = 0.76 \pm 0.12$ (see Figure \ref{fig3MP}), implying that the spectrum becomes harder as the source brightens in X-rays.

This campaign is reported in greater detail in \cite{mwl2003}.

\section*{Multi-wavelength campaign in 2004}

In 2004, a planned monitoring programme on \pks, proposed by H.E.S.S., was carried out from July 14th to September 11th.  Observations were performed simultaneously with RXTE/PCA, ROTSE, and NRT.  The source in this case had become more active than in the previous year as seen by the X-ray data, and a preliminary analysis of the VHE data passing strict quality criteria show a strong positive correlation ($r = 0.71 \pm 0.05$) between the VHE $\gamma$-rays and the X-rays \cite{berrie}.  Compared to the previous campaign in 2003, then, it would appear that the correlation became apparent for the larger variability range seen in this years' data.

A number of nights of H.E.S.S. data from this campaign were affected by large-scale smokey haze; the correction of these data for loss of Cherenkov photons is covered in \cite{nolan} at this conference.  This should provide further data to be included for the VHE $\gamma$-ray / X-ray correlation.

\section*{Multi-wavelength campaign in 2006}

The exceptional activity in 2006 previously referred to prompted the publica\-tion of an Astronomers Telegram by H.E.S.S., and the observation of the source with RXTE, CHANDRA, SWIFT, and optical telescopes.  This campaign is still under analysis, but the much wider variability range shown (up to 15 Crab-level at peak) should provide ample testing-ground for multi-band correlations.

\begin{figure}[t]
\vskip -10pt
\begin{center}
\includegraphics [width=0.5\textwidth]{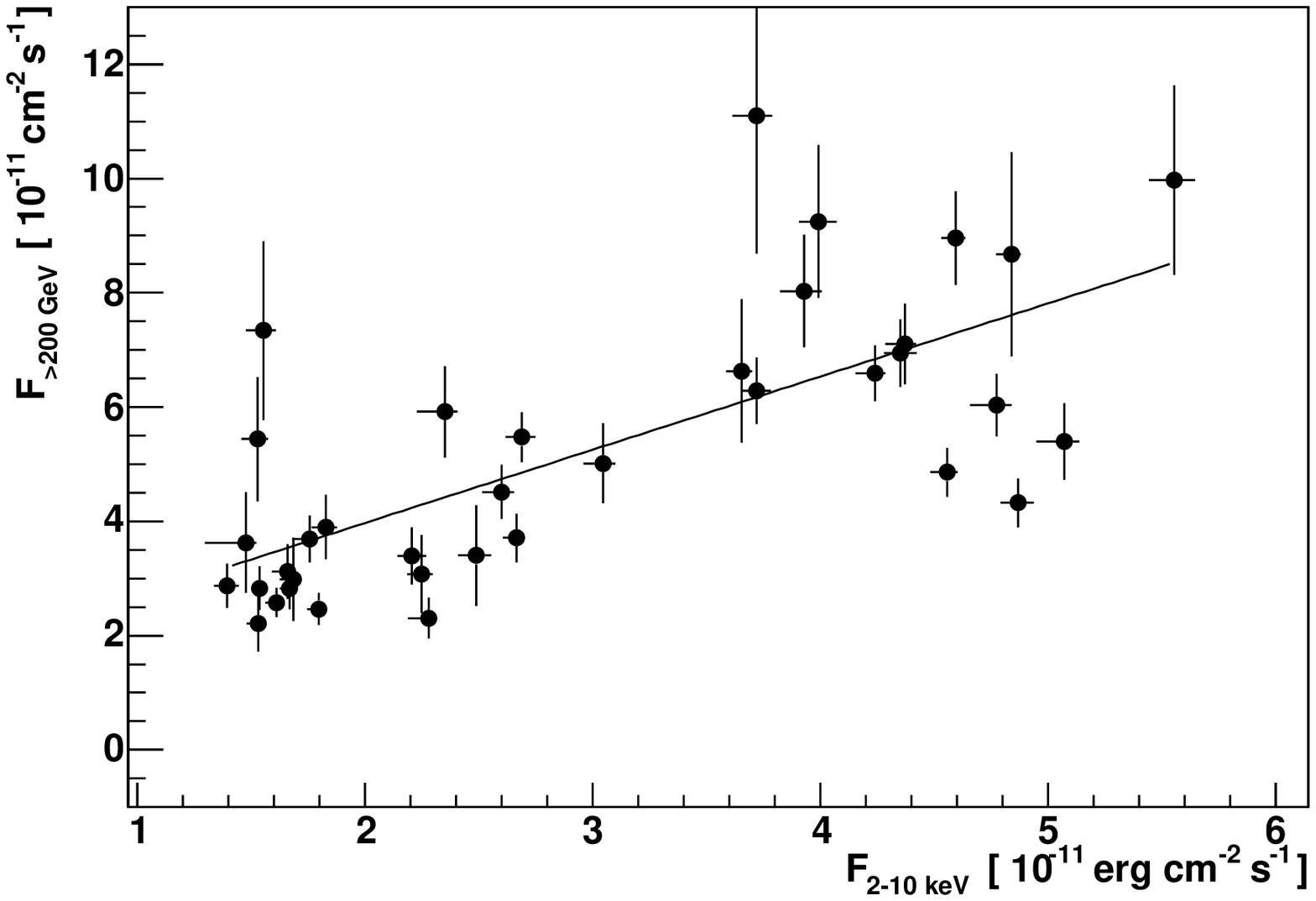}
\end{center}
\vskip -20pt
\caption{
Correlation between H.E.S.S. and RXTE for coincident measurements, in quality selected
runs during the X-ray flares of Aug. 2004 (44 data segments within
2 weeks); a close correlation is seen with a correlation factor
$r = 0.71\pm0.05$.
}
\label{fig4MP}
\end{figure}

\section*{Conclusions}

The observations of \pks\ with H.E.S.S. since the inception of the telescope has provided a relatively long baseline over which the activity of this brightest blazar in the Southern sky can be evaluated, and giving an estimate of its duty-cycle.  In VHE $\gamma$-rays alone, the activity is seen to vary over orders of magnitude, on time-scales of years, months, days, down to minutes, and very high activity was seen both in the first year of observation, 2002, and last year, in 2006.

Several multi-wavelength campaigns have been carried out, either planned or as target of opportunity.  In these campaigns, it has been seen that the X-ray spectrum became harder as the source increased in brightness (2003 campaign), and that inter-band (VHE $\gamma$-ray / X-ray) correlations became apparent only with a larger variability range (2004 campaign).  From the 2003 campaign, a multi-waveband SED has been produced which can be fitted by either leptonic or hadronic models, in the former case favouring a low range of Extragalactic Background Light.  The 2006 multi-wavelength campaign triggered by H.E.S.S. due to spectacular activity of the source is still under analysis but should yield further insights into the processes at work in this category of source.

This source, given its highly active and variable states and soft spectrum, should be a prime candidate for future GLAST/H.E.S.S. multi-wavelength campaigns.

\section*{Acknowledgements}

{
\small The support of the Namibian authorities and of the University of Namibia in facilitating the construction and operation of HESS is gratefully acknowledged, as is the support by the German Ministry for Education and Research (BMBF), the Max Planck Society, the French Ministry for Research, the CNRS-IN2P3 and the Astroparticle Interdisciplinary Programme of the CNRS, the U.K. Science and Technology Facilities Council (STFC), the IPNP of the Charles University, the Polish Ministry of Science and Higher Education, the South African Department of Science and Technology and National Research Foundation, and by the University of Namibia. We appreciate the excellent work of the technical support staff in Berlin, Durham, Hamburg, Heidelberg, Palaiseau, Paris, Saclay, and in Namibia in the construction and operation of the equipment.\par
}

%This is the reference to .bib file (Whitout .bib!)
%\bibliography{libros}
%This in the bibtex style, is ok.
\bibliographystyle{plain}
\bibliographystyle{plain}

%%%%%%%%
%  28  %
%%%%%%%%

%The paper title
\title{A Spectacular VHE Gamma-Ray Outburst from PKS 2155-304 in 2006}
%Short title to print in the headers to the final publication (Not showed in this print).
\shorttitle{A Spectacular VHE $\gamma$-ray Outburst from PKS\,2155$-$304}
%All paper authors
\authors{W.\,Benbow$^{1}$, 
C.\,Boisson$^{2}$, 
L.\,Costamante$^{1}$, 
O.\,de\,Jager$^{3}$,
G.\,Dubus$^{4}$,
D.\,Emmanoulopoulos$^{5}$, 
B.\,Giebels$^{6}$,
S.\,Pita$^{7}$,
M.\,Punch$^{7}$,
C.\,Raubenheimer$^{3}$,
M.\,Raue$^{8}$,
H.\,Sol$^{2}$,
and S.\,Wagner$^{5}$
for the HESS Collaboration
}
%Short title to print in the headers to the final puplication (Not showed in this print).
\shortauthors{W.\,Benbow et al.}
%All the affiliations.
\afiliations{$^1$ Max-Planck-Institut f\"ur Kernphysik, Heidelberg, Germany; 
$^2$ LUTH, UMR 8102 du CNRS, Observatoire de Paris, Section de Meudon, France; 
$^3$ Unit for Space Physics, North-West University, Potchefstroom, South Africa;
$^4$ Laboratoire d'Astrophysique de Grenoble, INSU/CNRS, Universit\'e Joseph Fourier, France;
$^5$ Landessternwarte, Heidelberg, Germany;
$^6$ LLR, CNRS/IN2P3, Ecole Polytechnique, Palaiseau, France;
$^7$ Astrophysique et Cosmologie (APC), Paris, France;
$^8$ Universit\"at Hamburg, Institut fuer Experimentalphysik, Germany
}
\email{Wystan.Benbow@mpi-hd.mpg.de}

%The abstract.
\abstract{

Since 2002 the VHE ($>$100 GeV) $\gamma$-ray flux of
the high-frequency peaked BL\,Lac PKS\,2155$-$304
has been monitored with the High Energy Stereoscopic System (HESS).  An extreme
$\gamma$-ray outburst was detected in the early hours of July 28, 2006
(MJD 53944).  The average flux above 200 GeV observed during this outburst is
$\sim$7 times the flux observed from the Crab Nebula above the same threshold.  
Peak fluxes are measured with one-minute time scale resolution at more than twice
this average value. Variability is seen up to $\sim$600 s in the
Fourier power spectrum, and well-resolved bursts varying on time
scales of $\sim$200 seconds are observed. There are no strong
indications for spectral variability within the data.  Assuming the
emission region has a size comparable to the Schwarzschild radius of a $\sim$$
10^9\,M_\odot$ black hole, Doppler factors greater than 100 are
required to accommodate the observed variability time scales.
}

\maketitle

\addcontentsline{toc}{section}{A Spectacular VHE Gamma-Ray Outburst from PKS 2155-304 in 2006}
\setcounter{figure}{0}
\setcounter{table}{0}
\setcounter{equation}{0}

%Begin the section.
\section*{Introduction}

In the Southern Hemisphere, PKS\,2155$-$304 (redshift $z=0.116$) is
generally the brightest blazar at VHE energies, and is probably the
best-studied at all wavelengths.  The VHE flux observed
\cite{HESS_2155A} from PKS\,2155$-$304 is typically of the order
$\sim$15\% of the Crab Nebula flux above 200 GeV.  The highest flux
previously measured in one night is approximately four times this
value and clear VHE-flux variability has been observed on daily time
scales. The most rapid flux variability measured for this source is
25\,min~\cite{HESS_2155B}, occurring at X-ray energies.  The fastest
variation published from any blazar, at any wavelength, is an event lasting
$\sim$800\,s where the X-ray flux from Mkn\,501 varied by 30\%
\cite{MKN501_dispute}\footnote{Xue \& Cui~\cite{MKN501_dispute}
also demonstrate that a 60\% X-ray flux increase in $\sim$200\,s observed
\cite{MKN501_flare} from Mkn\,501 is likely an artifact.}, 
while at VHE energies doubling time scales as fast as
$\sim$15 minutes have been observed from Mkn 421 \cite{Gaidos_Mkn421}.

As part of the normal HESS observation program
the flux from known VHE AGN is monitored regularly
to search for bright flares.  During the July 2006 dark period,
the average VHE flux observed by HESS from PKS\,2155$-$304 was
more than ten times its typical value.  In particular,
an extremely bright flare of PKS\,2155$-$304 was 
observed in the early hours of July 28, 2006 (MJD 53944).
This contribution focuses solely on this particular flare,
which is described in more detail in \cite{2155_letter}.

\section*{Results from MJD 53944}
\label{sect:results}

A total of three observation runs ($\sim$28 min each) were
taken on PKS\,2155$-$304 in the early hours of MJD 53944.  These data
entirely pass the standard HESS data-quality selection criteria,
yielding an exposure of 1.32\,h live time at a mean zenith angle of
13$^{\circ}$.  The analysis method is described in \cite{2155_letter}.
The observed excess is 11771 events (168$\sigma$), corresponding to a
rate of $\sim$2.5\,Hz.  This is the first time the
detected VHE  $\gamma$-ray rate has exceeded 1\,Hz.

\subsection*{Flux Variability}

\begin{figure}[t]
  \centering
  \includegraphics[width=0.45\textwidth]{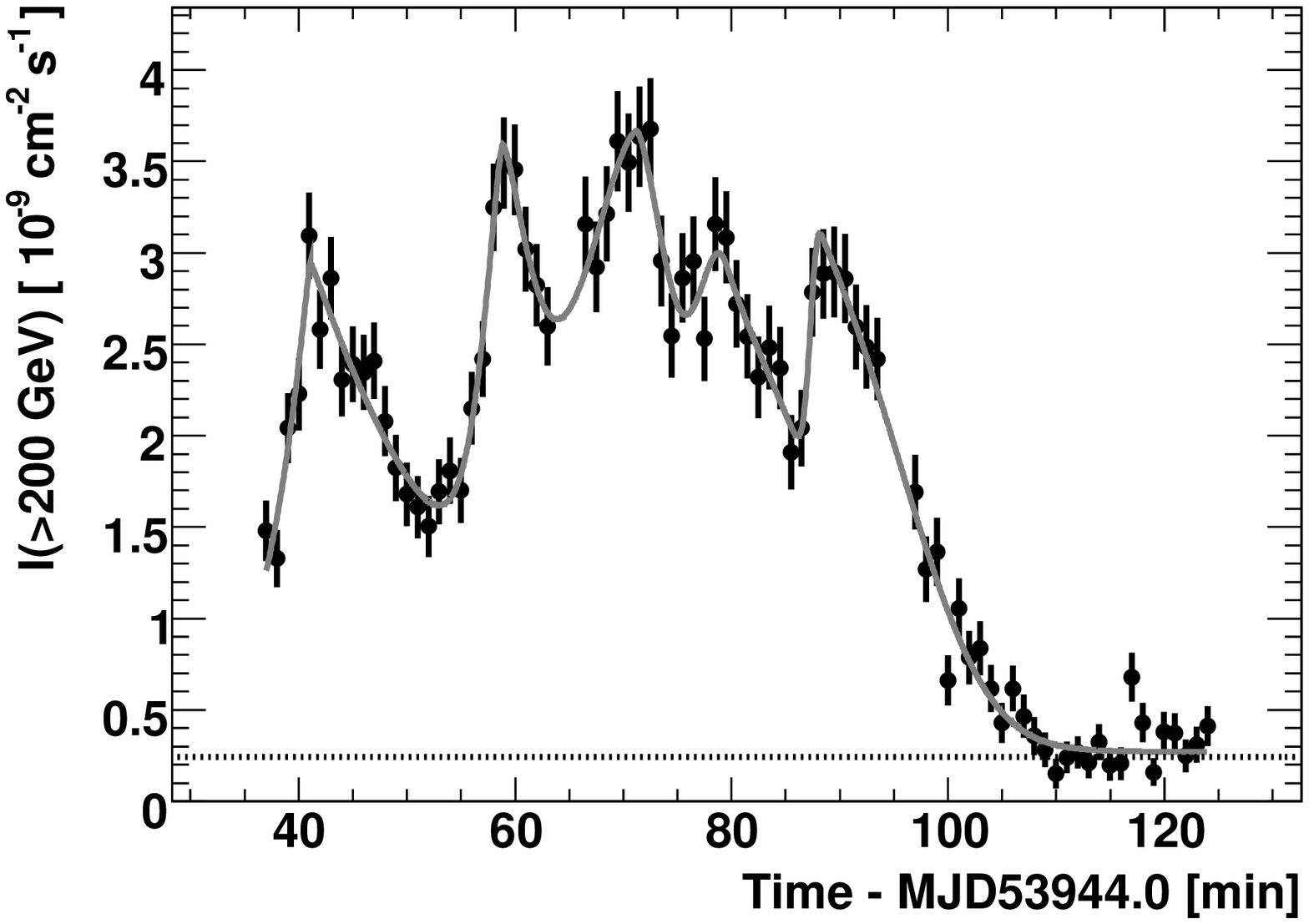}
  \caption{I($>$200 GeV) observed from PKS\,2155$-$304 binned
    in 1-minute intervals.  The horizontal line represents
    I($>$200 GeV) observed \cite{hess_crab}
    from the Crab Nebula.  The curve is the fit
    to these data of the superposition of five bursts (see text) and a
    constant flux.\label{flux_lc_1min}}
\end{figure}

\begin{figure}
  \centering
  \includegraphics[width=0.45\textwidth]{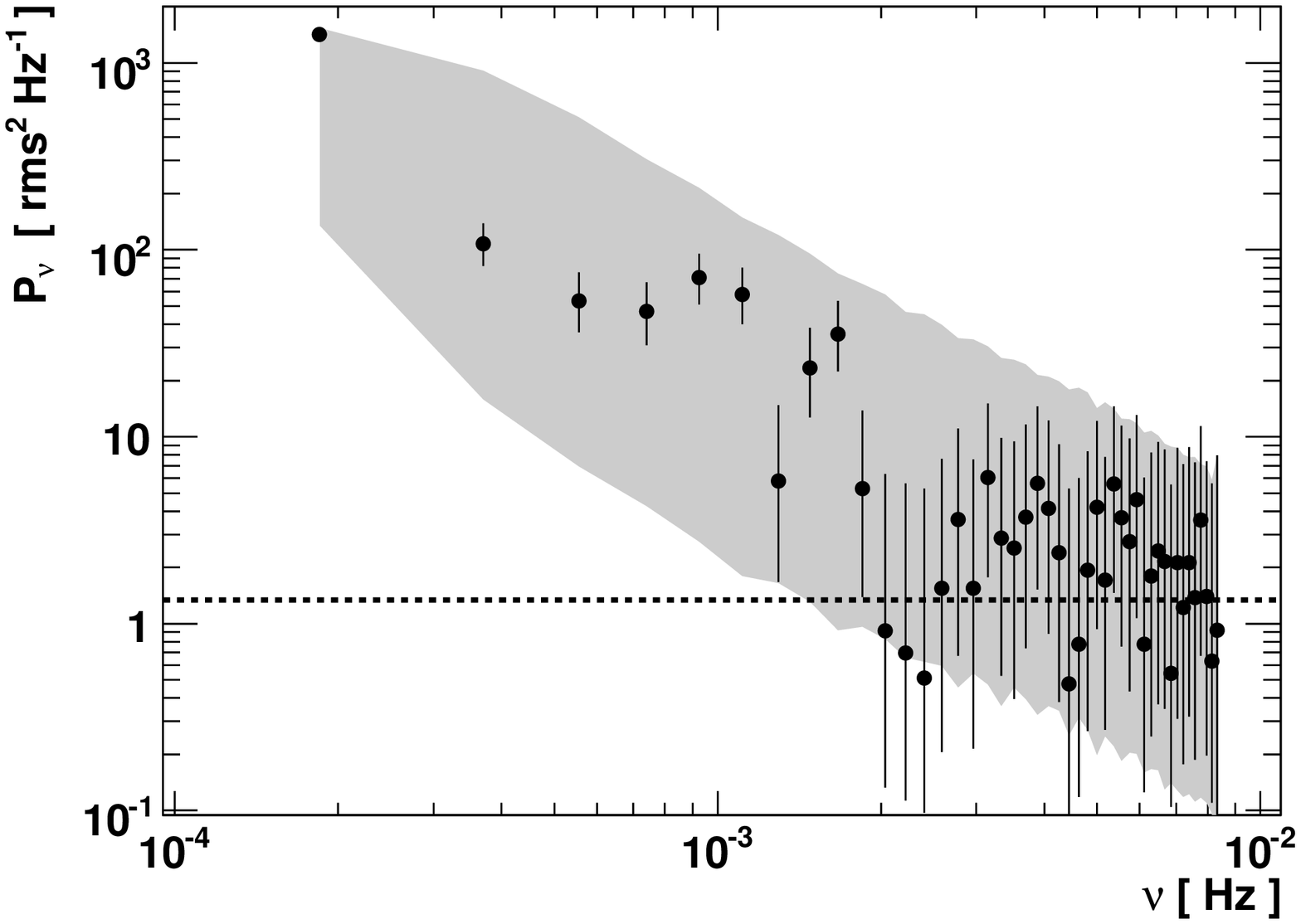}
  \caption{The Fourier power spectrum of the
    light curve and associated measurement error. The grey shaded area
    corresponds to the 90\% confidence interval for a light curve with a power-law
    Fourier spectrum $P_{\nu}\propto \nu^{-2}$. The horizontal line is the
    average noise level. \label{fourier_power}}
\end{figure}

The average integral flux above 200 GeV observed from PKS\,2155$-$304
is I($>$200 GeV) = 
(1.72$\pm$$0.05_{\rm stat}$$\pm$$0.34_{\rm syst}$)$\,\times\,$10$^{-9}$\,cm$^{-2}$\,s$^{-1}$, 
equivalent to $\sim$7 times the I($>$200 GeV) observed 
from the Crab Nebula (I$_{\mathrm{Crab}}$, \cite{hess_crab}).  Figure~\ref{flux_lc_1min}
shows I($>$200 GeV), binned in one-minute intervals, versus time.  The
fluxes in this light curve range from 0.65 I$_{\mathrm{Crab}}$ to 15.1
I$_{\mathrm{Crab}}$, and their fractional root mean square (rms)
variability amplitude \cite{rms_noise_ref} 
is F$_{\rm var}=0.58\pm0.03$.  This is $\sim$2
times higher than archival X-ray 
variability \cite{zhang1999,zhang2005}. 
The Fourier power spectrum
calculated from Figure~\ref{flux_lc_1min}
is shown in Figure~\ref{fourier_power}. 
There is power
significantly above the measurement noise level up to $1.6 \times
10^{-3}\,{\rm Hz}$ ($600\,{\rm s}$).  
The power spectrum derived from the data is
compatible with a light curve generated by a stochastic process with a
power-law Fourier spectrum of index -2. An index of -1 produces too
much power at high frequencies and is rejected. These power spectra are
remarkably similar to those derived in
X-rays \cite{zhang1999} from the same source.

Figure~\ref{flux_lc_1min} clearly contains substructures
with even shorter rise and decay time scales than found 
in the Fourier analysis.  Therefore, the light curve is considered as
consisting of a series of bursts, which is common for AGN and
$\gamma$-ray bursts (GRBs).  To characterize these bursts, 
the ``generalized Gaussian'' shape from
Norris et al.~\cite{norris} is used,
where the burst intensity is described by: ${\rm I}(t) = A \exp [
-(|t-t_{\rm max}|/\sigma_{\rm r,d})^\kappa]$, where $t_{\rm max}$ is
the time of the burst's maximum intensity (A); $\sigma_{\rm r}$ and
$\sigma_{\rm d}$ are the rise ($t<t_{\rm max}$) and decay ($t>t_{\rm
max}$) time constants, respectively; and $\kappa$ is a measure of the
burst's sharpness.  The rise and decay times, from half to maximum
amplitude, are $\tau_{r,d}=[\ln 2]^{1/\kappa}\sigma_{r,d}$. 
Five significant bursts were found with a peak finding tool based on 
a Markov chain algorithm \cite{mor02}. 
The data are well fit\footnote{All parameters are left free in the fit.}
by a function consisting of a superposition of an identical
number of bursts plus a constant signal.   
The best fit has a $\chi^2$ probability of
20\% and the fit parameters are shown in
Table~\ref{burst_info}. Interestingly, there is a marginal trend for
$\kappa$ to increase with subsequent bursts, 
making them less sharp, as the flare
progresses, which could imply the bursts are not stochastic.
The $\kappa$ values 
are close to the bulk of those found by Norris et al.~\cite{norris}, 
but the time scales measured here are two orders of
magnitude larger.

\begin{table}
\caption{The results of the best $\chi^2$ fit of the superposition
of five bursts and a constant to the
data shown in Figure~\ref{flux_lc_1min}\label{burst_info}.}
\centering
\begin{tabular}{cccc}
\\
\hline\hline
          \noalign{\smallskip}
$t_{\rm max}$  & $\tau_{\rm r}$& $\tau_{\rm d}$  & $\kappa$ \\
$[$min$]$ & [s] & [s] & \\
          \noalign{\smallskip}
          \hline
          \noalign{\smallskip}
	  41.0 & 173$\pm$28 & 610$\pm$129 & 1.07$\pm$0.20\\
	  58.8 & 116$\pm$53 & 178$\pm$146 & 1.43$\pm$0.83\\
	  71.3 & 404$\pm$219 & 269$\pm$158 & 1.59$\pm$0.42\\
	  79.5 & 178$\pm$55  & 657$\pm$268 & 2.01$\pm$0.87\\
	  88.3 & 67$\pm$44   & 620$\pm$75  & 2.44$\pm$0.41\\
          \noalign{\smallskip}
          \hline
\end{tabular}
\end{table}

During both the first two bursts there is clear doubling of the flux
within $\tau_{r}$.  Such doubling is sometimes used as a
characteristic time scale of flux variability.  For compatibility with
such estimators, the definition of doubling time, 
$T_2 = |{\rm I}_{ij} \Delta T / \Delta {\rm I}|$, from 
\cite{zhang1999} is also used\footnote{Only values of $T_2$ with 
less than 30\% uncertainty are considered.}.  Here, $\Delta T = T_j - T_i$, 
$\Delta {\rm I} = {\rm I}_j - {\rm I}_i$, $
{\rm {\rm I}}_{ij} = ({\rm I}_j + {\rm I}_i)/2$, with $T$ 
and I being the time and flux, respectively, of any pair
of points in the light curve. 
The fastest $T_2=224\pm60\,{\rm s}$ is
compatible with the fastest significant time scale found by the
Fourier transform. Averaging the five lowest $T_2$
values yields $330\pm40\,{\rm s}$.

\subsection*{Spectral Analysis}

\begin{figure}
   \centering
      \includegraphics[width=0.45\textwidth]{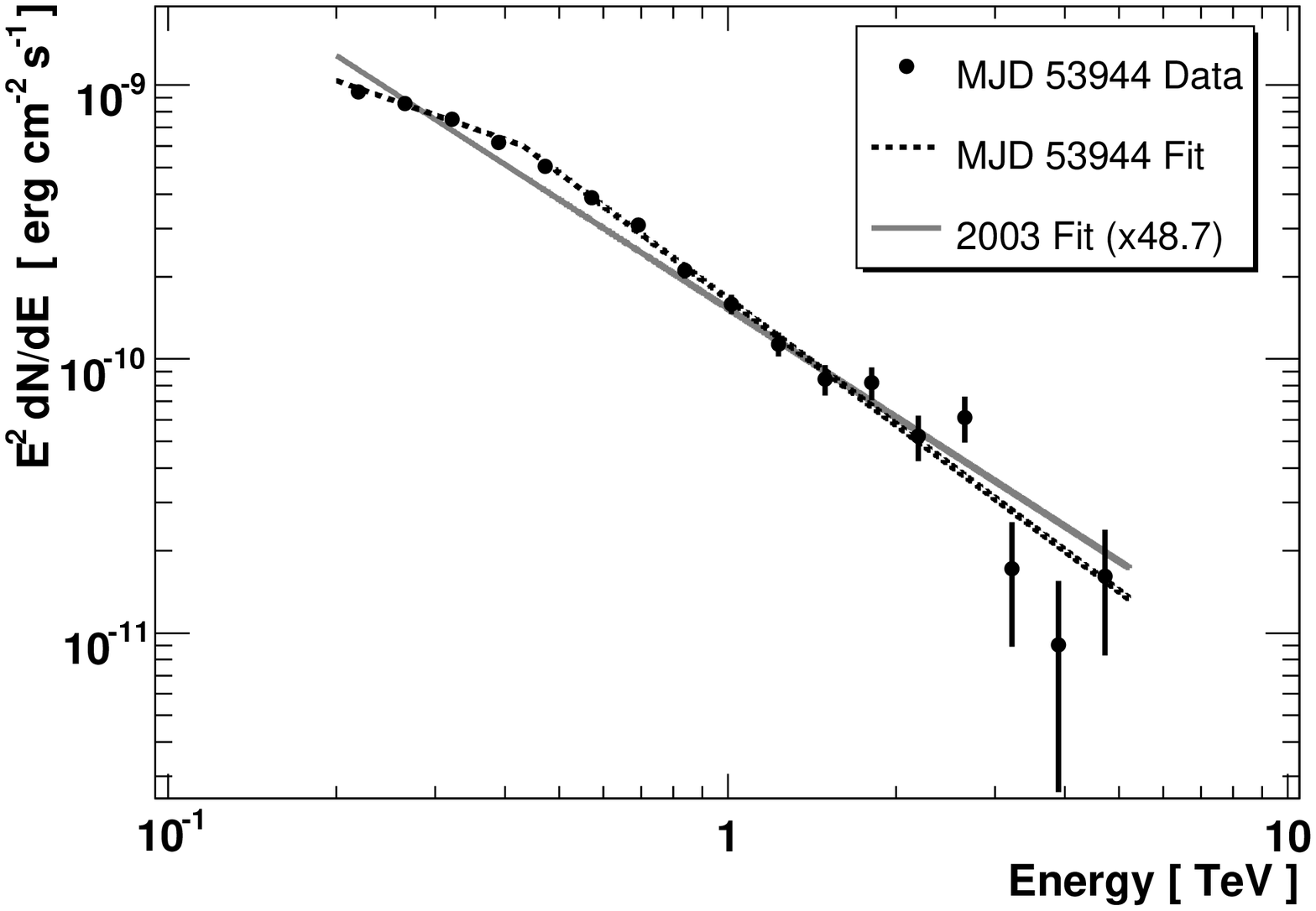}
\caption{The time-averaged spectrum observed
	from PKS\,2155$-$304 on MJD 53944.
	The dashed line is 
	the best $\chi^2$ fit of a broken power law to
	the data.  The solid line represents the fit to
	the time-averaged spectrum of PKS\,2155$-$304 
	from 2003 \cite{HESS_2155A} 
	scaled by 48.7.
\label{avg_spectrum}}
\end{figure}

Figure~\ref{avg_spectrum} shows the time-averaged photon 
spectrum for these data.  The data are well fit, $\chi^2=17.1$ for 13 
degrees of freedom (d.o.f.), by a broken power-law function with
$E_{\mathrm{B}}$\,=\,430$\pm$22$\pm$80 GeV,
$\Gamma_{1}$\,=\,2.71$\pm$0.06$\pm$0.10, 
and $\Gamma_{2}$\,=\,3.53$\pm$0.05$\pm$0.10. 
For each parameter, the two uncertainties are the 
statistical and systematic values, respectively.
The time-averaged spectrum ($\Gamma$\,=\,3.32) 
of PKS\,2155$-$304\ measured in 2003 \cite{HESS_2155A},
multiplied by the ratio (48.7) of I($>$200~GeV) from the 
respective data sets, is also shown in Figure~\ref{avg_spectrum}.
Despite a factor of $\sim$50 change in flux there 
is qualitatively little difference between the two spectra 
which is surprising. The lack of any strong ($\Delta \Gamma$\,$>$\,0.2)
temporal variability in the VHE spectrum within these data
(tested on time scales of 5, 10 and 28 minutes) is also surprising.

\section*{Discussion}

It is very likely that the electromagnetic emission in blazars 
is generated in jets that are beamed 
and Doppler-boosted toward the observer. 
Superluminal expansions observed with VLBI \cite{piner}
provide evidence for moderate Doppler boosting in PKS\,2155$-$304.
Causality implies that $\gamma$-ray variability on a 
time scale $t_{\rm var}$, with a Doppler factor\footnote{With 
$\delta$ defined in the standard
way as $[\Gamma(1-\beta\cos\theta)]^{-1}$, where $\Gamma$ is the bulk
Lorentz factor of the plasma in the jet, $\beta = v/c$, and $\theta$
is the angle to the line of sight.} ($\delta$), is related to the
radius ($R$) of the emission zone by $R \leq ct_{\rm
var}\delta/(1+z)$. Conservatively using the best-determined rise time 
(i.e. $\tau_r$ with the smallest error) 
from Table~\ref{burst_info} for $t_{\rm var} = 173\pm28\,{\rm s}$ 
limits the size of the emission region
to $R\delta^{-1} \leq 4.65 \times 10^{12}$ cm $\leq 0.31$ AU.

The jets of blazars are believed to be powered by accretion onto a 
supermassive black hole (SMBH).  Thus accretion/ejection 
properties are usually presumed to scale with the Schwarzschild radius 
$R_{\rm S}$ of the SMBH, where $R_{\rm S} = 2GM/c^2$, which
is the smallest, most-natural size of the system
(see, e.g., \cite{Blandford}).
Expressing the size $R$ of the $\gamma$-ray emitting region 
in terms of $R_{\rm S}$, the variability time
scale limits its mass by $M \leq (c^3 t_{\rm var}\delta/2G(1+z)) R_{\rm S}/R 
\sim 1.6\times10^7 M_\odot \delta R_{\rm S} / R$. 
The reported host galaxy luminosity $M_R=-24.4$
(Table 3 in \cite{kotilainen}) would imply a SMBH mass of order
1$-$2$\times 10^9M_\odot$ \cite{bettoni2003}, and therefore,
$\delta\geq 60-120\,R/R_{\rm S}$.  Emission regions of only a 
few $R_{\rm S}$ would
require values of $\delta$ much greater than those typically
derived for blazars ($\delta$$\sim$10) and come close
to those used for GRBs, which would be a challenge to understand.  

Although the choice of a $\sim$3 minute variability time scale in this article is 
conservative, it is still the fastest ever
seen in a blazar, at any wavelength, and is almost an order of
magnitude smaller than previously observed from this object.
The variability is a factor of five times faster than 
VHE variability previously measured from Mkn 421 \cite{Gaidos_Mkn421}
and comparable to that reported from Mkn\,501 \cite{MAGIC_501}.
However, in terms of the 
light-crossing time of the Schwarzschild
radius, $R_{\rm S}/c$, the variability of PKS\,2155$-$304 is
more constraining by 
another factor\footnote{These factors assume
black hole masses of 
$10^{8.22} M_\odot$ and $10^{8.62} M_\odot$ for Mkn\,421 
and Mkn\,501, respectively \cite{BH_Mass}.}
of $\approx 6-12$ for Mkn\,421, and a factor of $\approx 2.5-5$ 
for  Mkn\,501. 

The light curve presented here is strongly oversampled,
allowing for the first time in the VHE regime a detailed 
statistical analysis of a flare, which shows remarkable similarity to
other longer duration events at X-ray energies. More detailed discussion
of this outburst can be found in \cite{2155_letter}, and the
event continues to be investigated with other statistical
techniques. As the sensitivity of VHE instruments continues to improve, 
it is likely that similar extreme flaring episodes will be more commonly
detected in the future. Similar flares will
only strengthen the conclusion that
either very large Doppler factors can
be present in AGN jets, or that the observed variability is not
connected to the central black hole.

\section*{Acknowledgements}

The support of the Namibian authorities and of the University of Namibia
in facilitating the construction and operation of HESS is gratefully
acknowledged, as is the support by the German Ministry for Education and
Research (BMBF), the Max Planck Society, the French Ministry for Research,
the CNRS-IN2P3 and the Astroparticle Interdisciplinary Programme of the
CNRS, the U.K. Science and Technology Facilities Council (STFC),
the IPNP of the Charles University, the Polish Ministry of Science and
Higher Education, the South African Department of
Science and Technology and National Research Foundation, and by the
University of Namibia. We appreciate the excellent work of the technical
support staff in Berlin, Durham, Hamburg, Heidelberg, Palaiseau, Paris,
Saclay, and in Namibia in the construction and operation of the
equipment.

%This in the bibtex style, is ok.

%%%%%%%%
%  29  %
%%%%%%%%

\title{Active Atmospheric Calibration for H.E.S.S. Applied to PKS 2155-304}
\shorttitle{Active Atmospheric Calibration}
\authors{Nolan, S.J.$^1$, P\"uhlhofer G. $^2$, \& Chadwick, P.M. $^1$ for the H.E.S.S. Collaboration}
\afiliations{
$^1$ Physics Department, Durham University, South Road, Durham, County Durham, DH1 3LE, United Kingdom\\$^2$ Landessternwarte, Universit\"at Heidelberg, K\"onigstuhl, D 69117 Heidelberg, Germany}
\email{Email : s.j.nolan@dur.ac.uk}
\abstract{Using data derived from the H.E.S.S. telescope system and the LIDAR facility on site, a method of correcting for changing atmospheric quality based on reconstructed shower parameters is presented. The method was applied to data from the active galactic nucleus PKS 2155-304, taken during August and September 2004 when the quality of the atmosphere at  the site was highly variable. Corrected and uncorrected fluxes are shown, and the method is discussed as a first step towards a more complete atmospheric calibration.}
\maketitle

\addcontentsline{toc}{section}{Active Atmospheric Calibration for H.E.S.S. Applied to PKS 2155-304}
\setcounter{figure}{0}
\setcounter{table}{0}
\setcounter{equation}{0}

\section*{Introduction}Imaging Atmospheric Cherenkov Telescopes (IACTs) rely heavily on the atmosphere as their detecting medium. Although the atmosphere gives the telescope systems huge effective areas, daily variations in atmospheric quality can affect the system performance and lead, in the worst cases, to systematic bias in the estimated energy of a given event. Significant effort has been made in the past to take account of this problem by using the cosmic-ray background seen by the telescope on a given night to normalise the data \cite{ref1}. However, given a better understanding of the location of atmospheric aerosol populations from LIDAR measurements and via modelling of these populations, it is possible to determine an active atmospheric correction to the data. Herein, recent work on such a technique is discussed as applied to observations with the H.E.S.S. telescope array of the active galactic nucleus (AGN) PKS 2155-304, this work continues from that presented in an earlier proceedings \cite{ref2}.
\section*{Technique}The LIDAR system at the H.E.S.S. site works at a wavelength of 905 nm, and has an active range of 7.5 km. It is mounted on an alt-azimuth drive allowing on-source pointing during observations. During August and September 2004, a large population of  aerosols was seen by the LIDAR below 2 km above the site, concurrent with a significant drop in the H.E.S.S. array trigger-rate for cosmic-rays. This population was seen to vary on a night to night basis, but not within a given night. In order to simulate its effects, the atmospheric simulation code MODTRAN was used to generate optical depth tables for wavelengths in the range 200 to 750 nm and for successive heights above the site (which is 1.8 km above sea level). The aerosol desert model within MODTRAN introduces a layer of aerosols into the first 2 km above ground level, whose density is then increased as the wind speed parameter is increased. Thus optical depth tables were produced for the range of wind speeds from 0 m/s to 30 m/s. The wind speed therefore acts as a tuning parameter to match simultaneously cosmic-ray trigger-rate and image parameter distributions, and is not a reflection of the measured wind speed at the site. These tables were then applied to a set of CORSIKA cosmic-ray simulations at various zenith angles between 0 and 60 degrees and with a southern pointing, to best match the data taken on PKS 2155-304, and a cosmic-ray trigger-rate for each atmosphere was derived for the H.E.S.S. array based upon the spectra given in \cite{ref3}. By matching the trigger-rate from simulations and real data, taking into account zenith angle dependence effects and gain changes over the experiment lifetime, an atmospheric model can be selected, as discussed in \cite{ref2}. The real cosmic-ray trigger rate and that due to simulation for the PKS 2155-304 dataset discussed later are shown in figure \ref{fig1} for comparison. The figure clearly shows that the data can be separated into 3 classes corresponding to MODTRAN model wind speeds of 17.5, 20.0 and 22.5 m/s. \newline
\begin{figure*}[th]
\begin{center}
\includegraphics*[width=0.73\textwidth,height=0.30\textheight,clip]{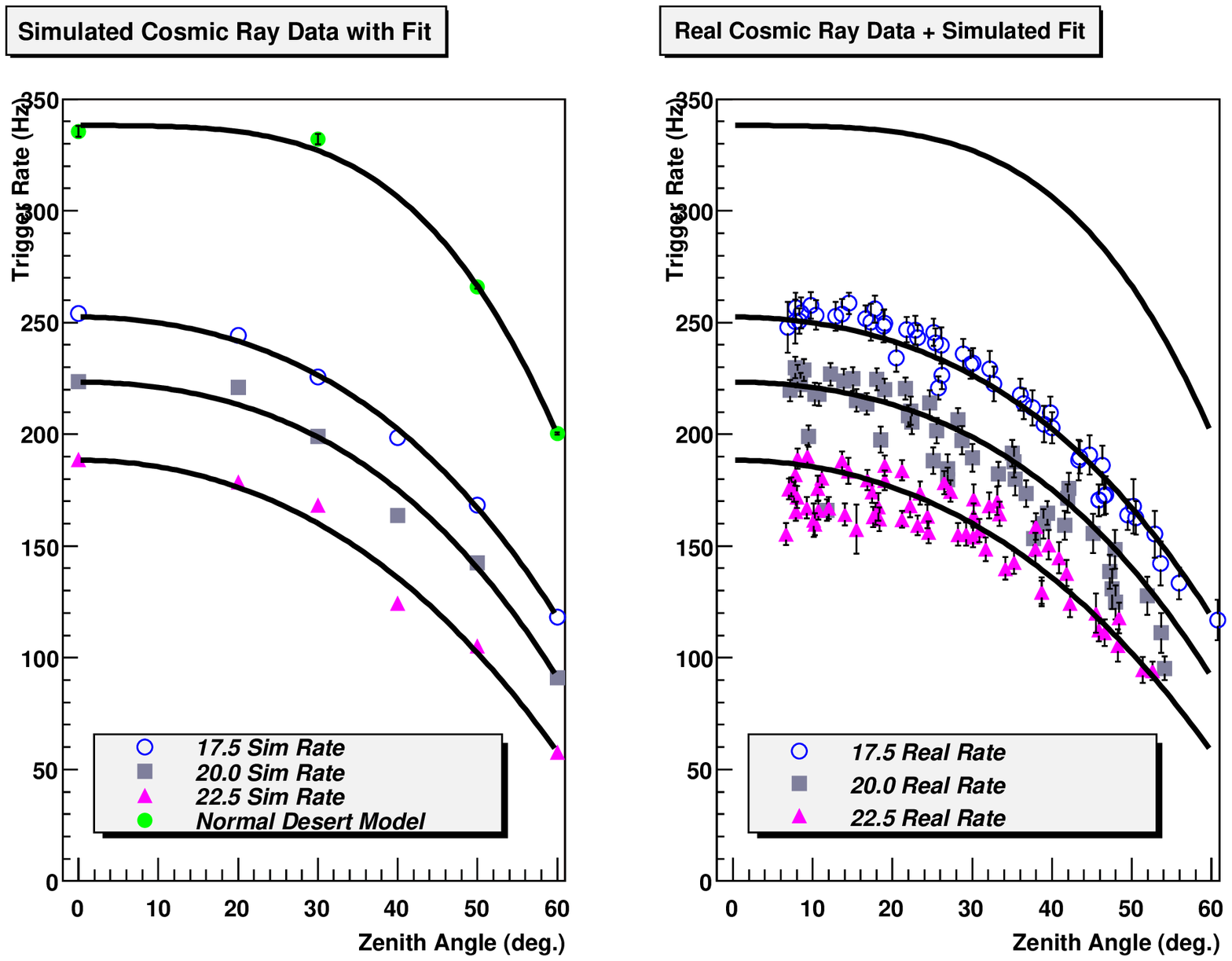}
\caption{\label {fig1} Simulated array trigger-rate for a spectrum of cosmic-rays \cite{ref3} for various atmospheric models with function fit in left panel, versus measured cosmic-ray trigger rates for the PKS 2155-304 2004 dataset in right hand panel. }
\end{center}
\end{figure*} In addition, as the LIDAR has a limited range and sensitivity, and to further confirm the choice of atmospheric models, a set of atmospheric models with aerosol densities at higher altitudes was simulated using MODTRAN. These simulated atmospheres represent conditions which could in principle also have occured during data-taking, as they result in similar cosmic-ray trigger rates as the low-level aerosol models.  As shown in figure \ref{fig2}, by comparing the reconstructed shower depth for gamma-rays between real-data and simulations, these models are shown to be considerably less favoured than the simple low-level aerosol models of 17.5, 20.0 and 22.5 m/s wind speed, which trigger-rate, image parameters, mean shower-depth and LIDAR data validate. \newline
\begin{figure*}[th]
\begin{center}
\includegraphics*[width=0.80\textwidth,height=0.24\textheight,clip]{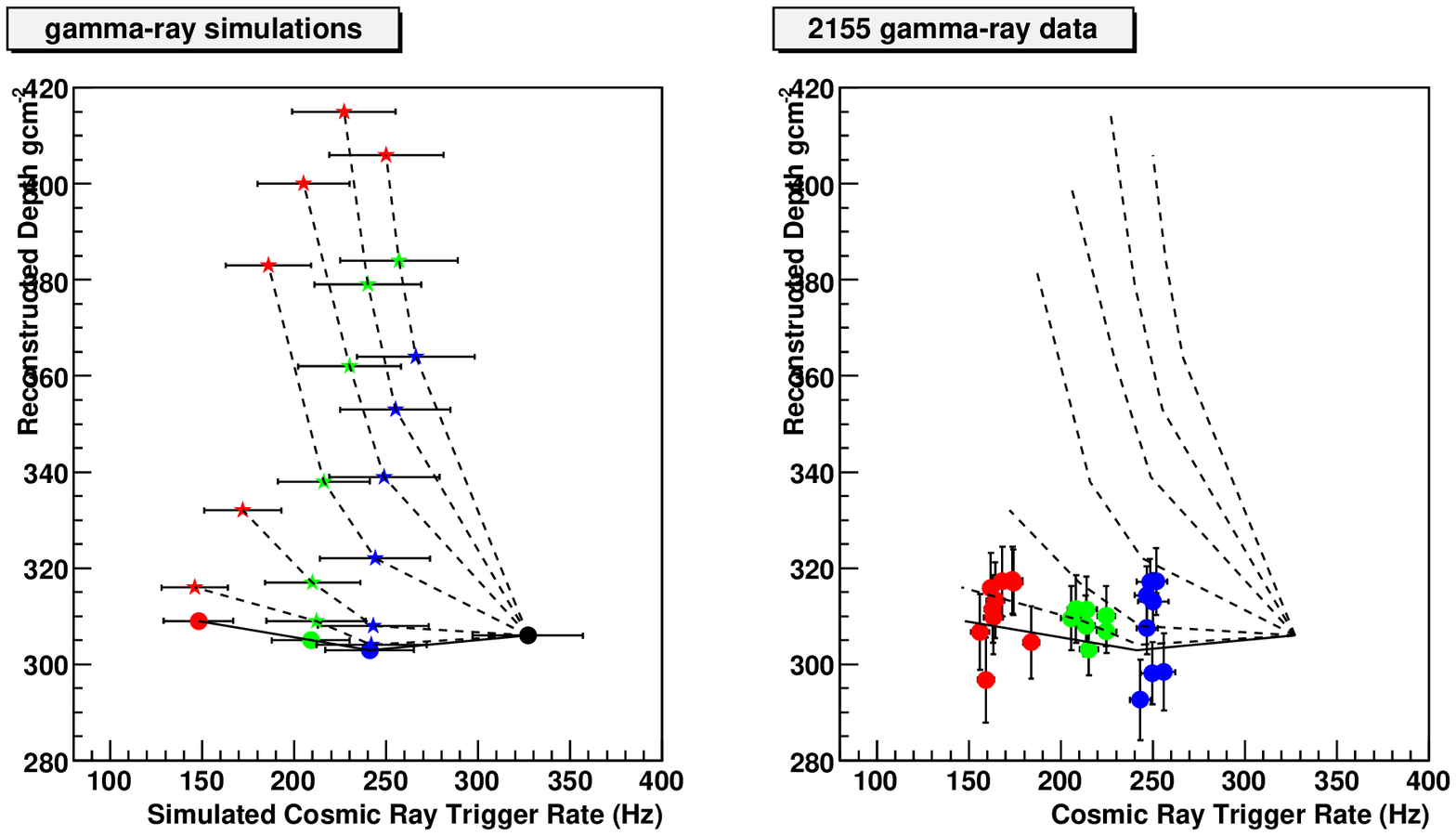}
\caption{\label {fig2} The left panel shows the mean of reconstructed depth (for a Gaussian fit) for gamma-ray shower simulations at 20 degrees zenith-angle versus telescope trigger-rate. The lower points (solid circles) show the results for the 17.5, 20.0 and 22.5 m/s wind speed models, with the other points showing show the result for atmospheres with increasing altitude of the aerosol contaminant layer, with lines connecting similar altitudes. These lines are reproduced on the right hand plot, which shows the preliminary real mean reconstructed depth for gamma-ray data on PKS 2155-304 taken during 2004 at zenith angles between 15 and 25 degrees, slightly scaled to match the results at 20 degrees.  The data shown no indication of high level aerosols, independently confirming the LIDAR results.} 
\end{center}
\end{figure*}The atmospheric model is then applied to a full set of CORSIKA gamma-ray simulations within a telescope simulation code. The simulations cover the zenith angle range of the observations, and produce lookup tables for image parameter cuts, energy and effective area, and these in turn are applied to the data using the standard H.E.S.S. analysis procedure \cite{ref4}. 
\section*{PKS 2155-304}
\begin{figure*}[th]
\begin{center}
\includegraphics[width=0.95\textwidth,height=0.3\textheight,clip]{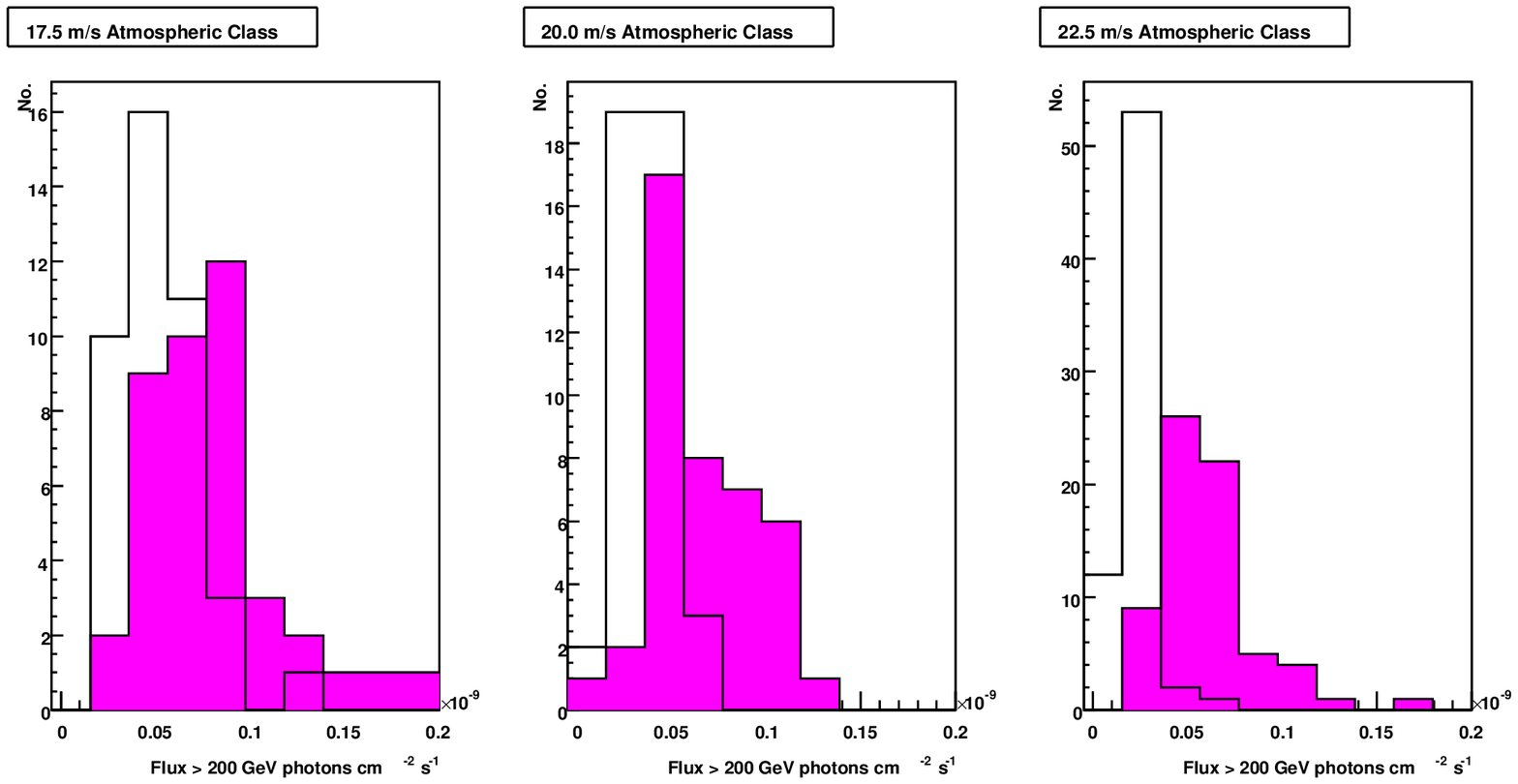}
\caption{\label{fig 3} The preliminary distribution of muon corrected integral flux for PKS 
2155-304 above 200 GeV derived from 28 minute runs is plotted before (open histograms) and 
after (filled histograms) the application of corrections for low-level dust. As noted each panel shows a subset of the data of differing atmospheric class.}
\end{center}
\end{figure*}

PKS 2155-304 is an AGN of the blazar class at a redshift of z$=0.116$. It was first detected in TeV gamma-rays by the Durham Mark 6 telescope \cite{ref5}, and has been observed from the earliest days of the H.E.S.S. experiment \cite{ref6}.   The data set from August and September 2004 is formed from 86 hours of four telescope observations. By combining flux data into atmospheric correction groups, figure \ref{fig 3} shows the results for corrected and non-corrected data in the form of a plot of the flux distribution derived on a run by run basis.  It is appears that in the data set considered here, as no run was taken under normal, clear atmospheric conditions, all runs are subject to systematically lowered detection rates, which if uncorrected may lead to significantly different results.
In addition, figure \ref{fig4} shows the spectra derived from this data. Without correction, significantly differing results are arrived at, with spectral index for a power-law fit differing by (at worst) $\Delta=0.7$, which is within errors marginally incompatible with a constant index. With correction all fit spectral indices agree well within errors.   
\begin{figure*}[hb]
\begin{center}
\includegraphics[width=0.95\textwidth,height=0.30\textheight,clip]{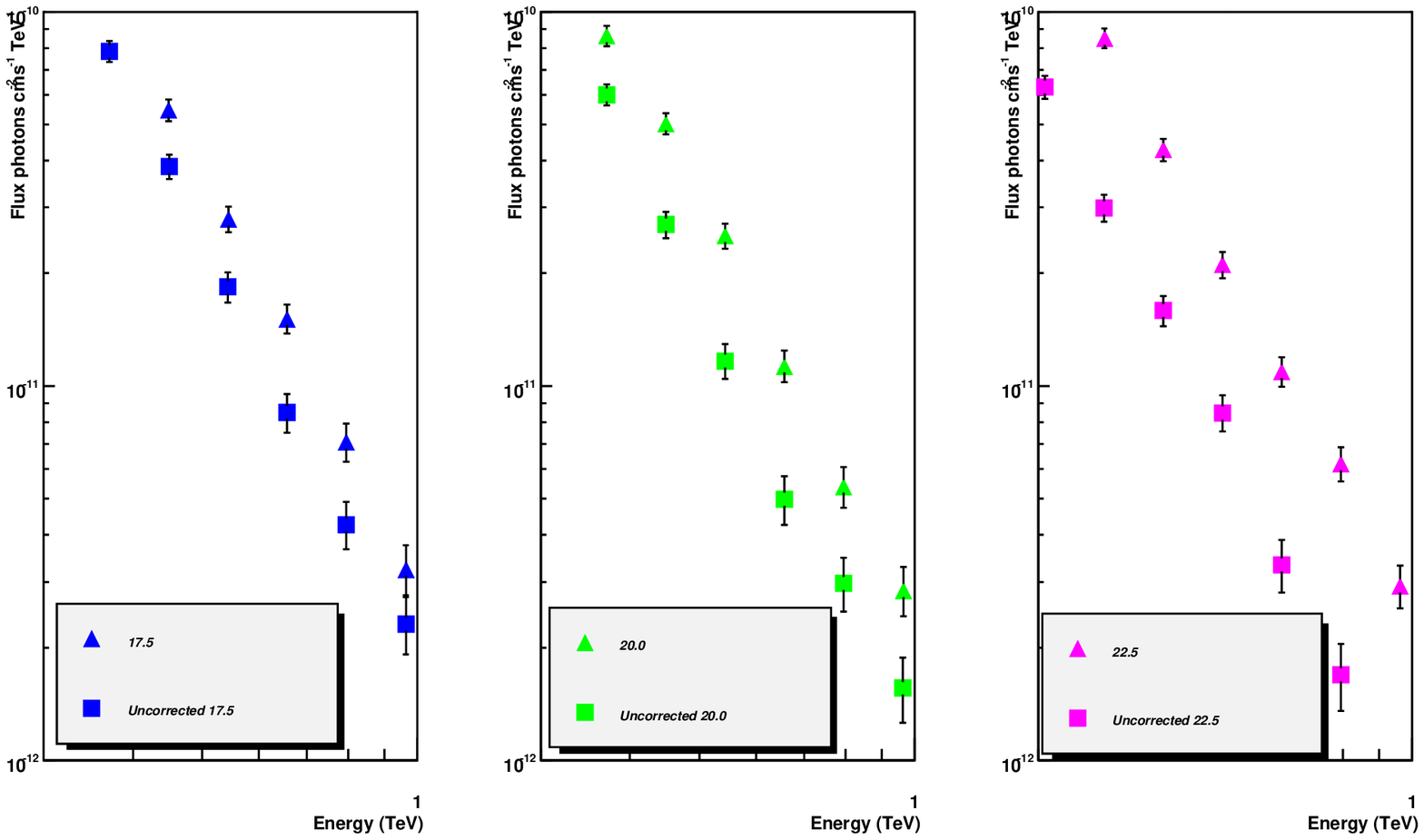}
\caption{\label{fig4}The preliminary uncorrected and corrected differential spectral for the 3 subsets of data is shown between 300 GeV and 1 TeV.  Above 1 TeV differences are negligible compared to statistical errors.  }
\end{center}
\end{figure*}
\section*{Conclusion}A new method for correcting for changes in low-level atmospheric quality is applied to the  variable source PKS 2155-304. The method, based on cosmic-ray trigger-rate, and LIDAR input, has allowed a corrected set of fluxes for PKS 2155-304 to be produced from data that would otherwise be unusable. This is particularly important as this data set forms part of a large multi-wavelength campaign so removing atmospheric biases is vital. To the lowest order, the effect on integral gamma-ray flux is seen to be proportional to the zenith- and time-corrected cosmic-ray trigger-rate. \newline
The current LIDAR system operates at a wavelength somewhat removed from typical Cherenkov photon wavelengths, and has a range which doesn't quite cover the maximum of shower development.  Two new LIDARs recently installed at the H.E.S.S. site operate at wavelengths closer to Cherenkov light and have a greater range, and will hopefully allow more straight forward correction. As has been shown, though, the comparison of real and simulated reconstructed shower depth under the application of different atmospheric models allows a coarse appreciation of atmospheric conditions, which is a useful check for the more accurate LIDAR dataset expected to be obtained soon.  

\bibliographystyle{plain}

%%%%%%%%
%  30  %
%%%%%%%%

%The paper title
\title{Discovery of fast variability of the TeV $\gamma$-ray flux from 
the radio galaxy M\,87 with H.E.S.S.}

%Short title to print in the headers to the final publication (Not showed in this print).
\shorttitle{Discovery of fast variable TeV $\gamma$-rays from M\,87 with 
H.E.S.S.}

%All paper authors
\authors{M. Beilicke$^{1}$, F. Aharonian$^{2}$, W. Benbow$^{2}$, 
G. Heinzelmann$^{1}$, D. Horns$^{3}$, O. Martineau-Huynh$^{4}$, M. 
Raue$^{1}$, J. Ripken$^{1}$, G. Rowell$^{5}$, H. Sol$^{6}$ for the 
H.E.S.S. collaboration}

%Short title to print in the headers to the final puplication (Not showed in this print).
\shortauthors{M. Beilicke and et al}

%All the affiliations.
\afiliations{$^1$ Institut f\"ur Experimentalphysik, Univ. of Hamburg, 
Luruper Chaussee 149, D-22761 Hamburg, Germany\\
$^2$ MPI-K Heidelberg, P.O. Box 103980, D-69029 Heidelberg, Germany\\
$^3$ Institut f\"ur Astronomie und Astrophysik, Universit\"at T\"ubingen, 
Sand 1, D-72076 T\"ubingen, Germany\\
$^4$ Laboratoire de Physique Nucleaire et de Hautes Energies, IN2P3/CNRS, 
Universites Paris VI \& VII, 4 Place Jussieu, F-75252 Paris Cedex 5, 
France\\
$^5$ School of Chemistry \& Physics, University of Adelaide 5005, 
Australia\\
$^6$ LUTH, UMR 8102 du CNRS, Observatoire de Paris, Section de Meudon, 
F-92195 Meudon Cedex, France}
\email{matthias.beilicke@desy.de}

%The abstract.
\abstract{The giant radio galaxy M\,87 was observed at GeV/TeV
$\gamma$-ray energies with the H.E.S.S. (High Energy Stereoscopic System)
Cherenkov telescopes in the years 2003--2006. The observations confirm
M\,87 as the first extragalactic TeV $\gamma$-ray source not of the blazar
type (first indications of a signal were reported by the HEGRA
collaboration earlier). The TeV $\gamma$-ray flux from M\,87 as measured
with H.E.S.S. was found to be variable on time-scales of years and
surprisingly also of days which strongly constrains the size of the
emission region. The results (position, energy spectrum and light curve)
as well as theoretical interpretations will be presented.}

\maketitle

\addcontentsline{toc}{section}{Discovery of fast variability of the TeV $\gamma$-ray flux from the radio galaxy M\,87 with H.E.S.S.}
\setcounter{figure}{0}
\setcounter{table}{0}
\setcounter{equation}{0}

%Begin the section.

%%%%%%%%%%%%%%%%%%%%%%%%%%%%%%%%%%%%%%%%%%%%
%% Introduction
%%%%%%%%%%%%%%%%%%%%%%%%%%%%%%%%%%%%%%%%%%%%
\section*{Introduction}

Observations of extragalactic objects at GeV/TeV $\gamma$-ray energies
play a key role in the understanding of non-thermal processes and emission
models of relativistic plasma jets. Meanwhile, more than $15$
extragalactic TeV $\gamma$-ray sources are established, whereas only the
giant FR\,I radio galaxy M\,87 \cite{HEGRA_M87_1, HEGRA_M87_2,
HESS_M87_ICRC} is not a blazar \footnote{Blazars are active galactic
nuclei (AGN) with their plasma jet pointing closely towards the observer's
line of sight (the energy and flux of the emitted photons are boosted due
to relativistic effects, making blazars detectable at TeV energies).},
making it an important object for the understanding of jet formation and
VHE $\gamma$-ray emission processes in AGN. M\,87 is well studied in
various wavelengths~-- allowing to constrain different system parameters,
as for example the black hole mass, the accretion rate, etc.

%==========================[ FIGURE
\begin{figure*}[th]
\begin{center}
\includegraphics [width=0.47\textwidth]{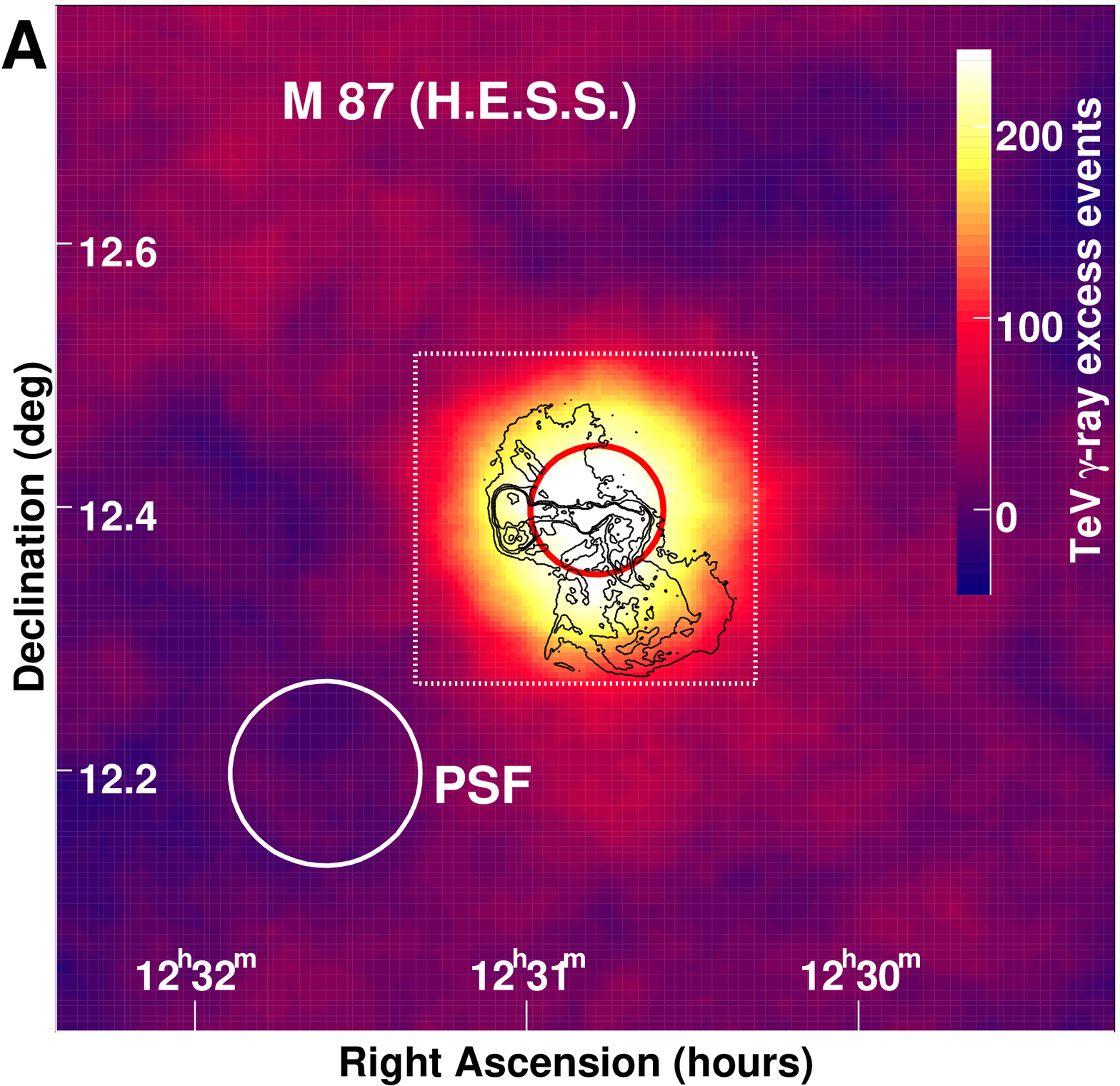} \hfill
\includegraphics [width=0.47\textwidth]{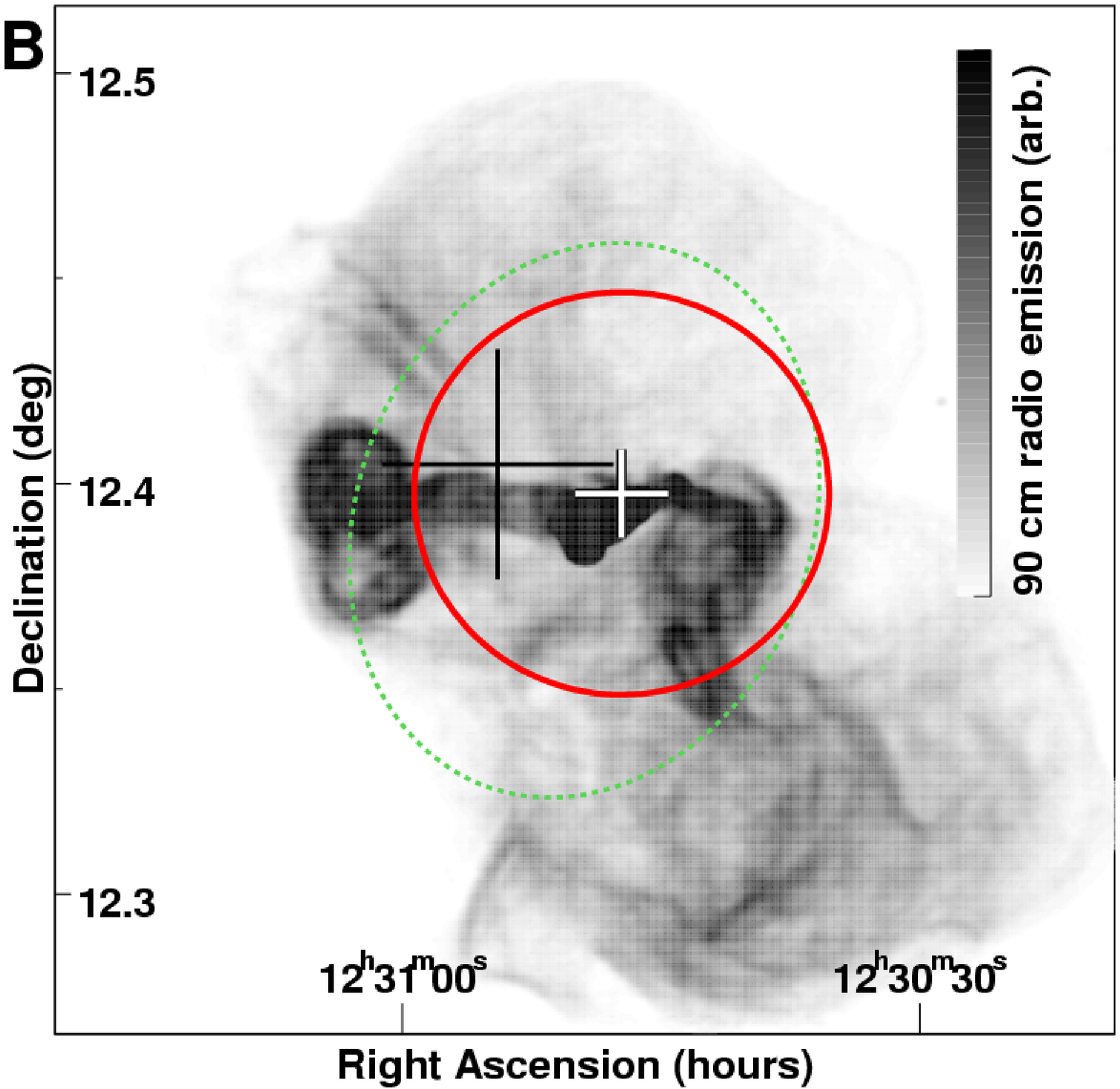}
\end{center}

\caption{{\bf Left:} Smoothed TeV $\gamma$-ray excess map and the upper
limit on the intrinsic source extension ($99.9\%$~c.l., red circle)
together with the $90 \, \rm{cm}$ radio contours adopted from
\cite{M87_Radio90cm} as well as the H.E.S.S. point spread function (PSF,
$r_{68}$). The white box indicates the cut-out of the right image. {\bf
Right:} The $90 \, \rm{cm}$ radio image \cite{M87_Radio90cm} showing
the large scale structure ($\sim 80 \, \rm{kpc}$ in diameter) of M\,87
together with the TeV position (white cross, including the statistical as
well as the $20''$ pointing uncertainty error) and again the extension
limit (circle). The black cross indicates the position of the excess
reported by HEGRA \cite{HEGRA_M87_1}. The green ellipse shows the 
extension of the M\,87 galaxy in the optical band.}

\label{Fig:M87SkyMap}

\end{figure*}

The H.E.S.S. collaboration operates an array of four Cherenkov telescopes
\cite{HESS_STATUS, HESS_HardCuts} situated in Namibia. The telescopes
measure cosmic $\gamma$-rays in the energy range between $100 \,
\mathrm{GeV}$ and several $10\, \mathrm{TeV}$ by recording the Cherenkov
light which is emitted from an air shower which develops when a very high
energy (VHE) particle (hadron or photon) enters the Earth's atmosphere.
The stereoscopic observation together with a corresponding hardware
trigger assures that an air shower is recorded by at least two of the four
telescopes, allowing for an angular and energy resolution per event of
$\delta \Theta < 0.1^{\circ}$ and $\Delta E / E \leq 15\%$, respectively,
as well as an improved cosmic ray (CR) background suppression.

%%%%%%%%%%%%%%%%%%%%%%%%%%%%%%%%%%%%%%%%%%%%
%% M87
%%%%%%%%%%%%%%%%%%%%%%%%%%%%%%%%%%%%%%%%%%%%
\section*{Variable TeV $\gamma$-ray emission from M\,87}

Radio-loud galaxies contain AGN with jets, but in contrast to blazars the
emission is not (strongly) Doppler boosted due to larger viewing angles
between the jet and the observer's line of sight. The radio-loud galaxy
M\,87 is located in the Virgo cluster of galaxies at a distance of $\sim16
\, \rm{Mpc}$ ($z = 0.0043$) and hosts a central black hole of $(3.2 \pm
0.9) \cdot 10^{9} \, \rm{M}_{\odot}$ \cite{M87_BH_Mass}. Due to its
proximity M\,87 is discussed as a possible source of the highest energy
($10^{20} \, \rm{eV}$) CRs \cite{M87_UHECR_2}. The $2 \, \rm{kpc}$ scale
plasma jet (inclination angle of $30^{\circ}$ \cite{M87_JetAngle}) is
spatially resolved in different wavelengths, ranging from radio, optical
to X-rays. Previously, evidence ($> 4 \, \sigma$) for $E > 730 \,
\rm{GeV}$ $\gamma$-ray emission from M\,87 in 1998/1999 was reported by
HEGRA \cite{HEGRA_M87_1, HEGRA_M87_2} and no significant emission above
$400 \, \rm{GeV}$ was observed by Whipple \cite{M87_Whipple} in 2000-2003.

M\,87 was observed by H.E.S.S. between 2003 and 2006 for a total of $89 \,
\rm{h}$ after data quality selection. Using hard event selection cuts
\cite{HESS_HardCuts} an excess of $243$ $\gamma$-ray events ($13 \,
\sigma$) was found in the whole data set. The position of the excess is
compatible with the nominal position of the nucleus of M\,87. With the
given angular resolution of H.E.S.S., the extension is consistent with a
point-like object with an upper limit for a Gaussian surface brightness
profile of $3'$ ($99.9\%$ c.l.), corresponding to a radial distance of $14
\, \rm{kpc}$ in M\,87, see Fig.~\ref{Fig:M87SkyMap}.

\begin{figure}
\begin{center}
\includegraphics [width=0.48\textwidth]{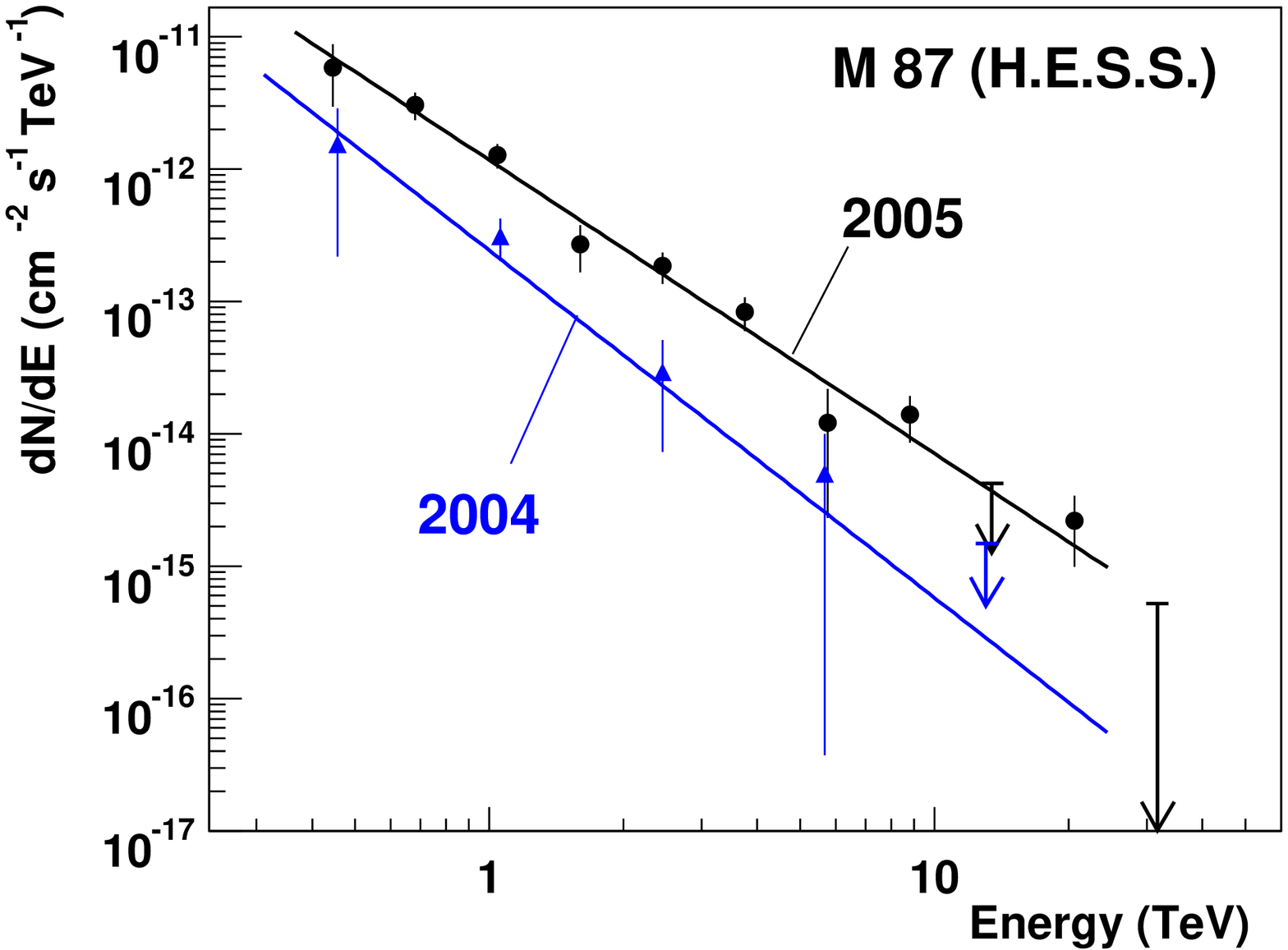}
\end{center}

\caption{Energy spectra of M\,87 (2004/05 data, using standard event
selection cuts \cite{HESS_HardCuts}) covering a range of $\sim 400 \,
\rm{GeV}$ to $\sim 10 \, \rm{TeV}$. Spectra for the 2003 and 2006
data sets could not be derived due to limited event statistics. The lines
show the fits of a power-law function.}

\label{Fig:M87Spectrum}

\end{figure}

The energy spectra for the 2004 and 2005 data sets are shown in
Fig.~\ref{Fig:M87Spectrum} (left) and are well fit by a power-law function
$\rm{d}N / \rm{d}E = I_{0} (E/1 \, \rm{TeV})^{-\Gamma}$ with
photon indices of $\Gamma = 2.62 \pm 0.35$ (2004) and $\Gamma = 2.22 \pm
0.15$ (2005). The systematic error on the photon index and flux
normalisation are estimated to be $\Delta \Gamma = 0.1$, and $\Delta I_{0}
/ I_{0} = 0.2$, respectively. The hard spectrum measured in 2005 (reaching
beyond an energy of $10 \, \rm{TeV}$) challenges hadronic as well as
leptonic VHE $\gamma$-ray emission models discussed for M\,87 in the
literature, see \cite{HESS_M87_Paper} and references therein.

\begin{figure*}[t]
\begin{center}
\includegraphics [width=0.7\textwidth]{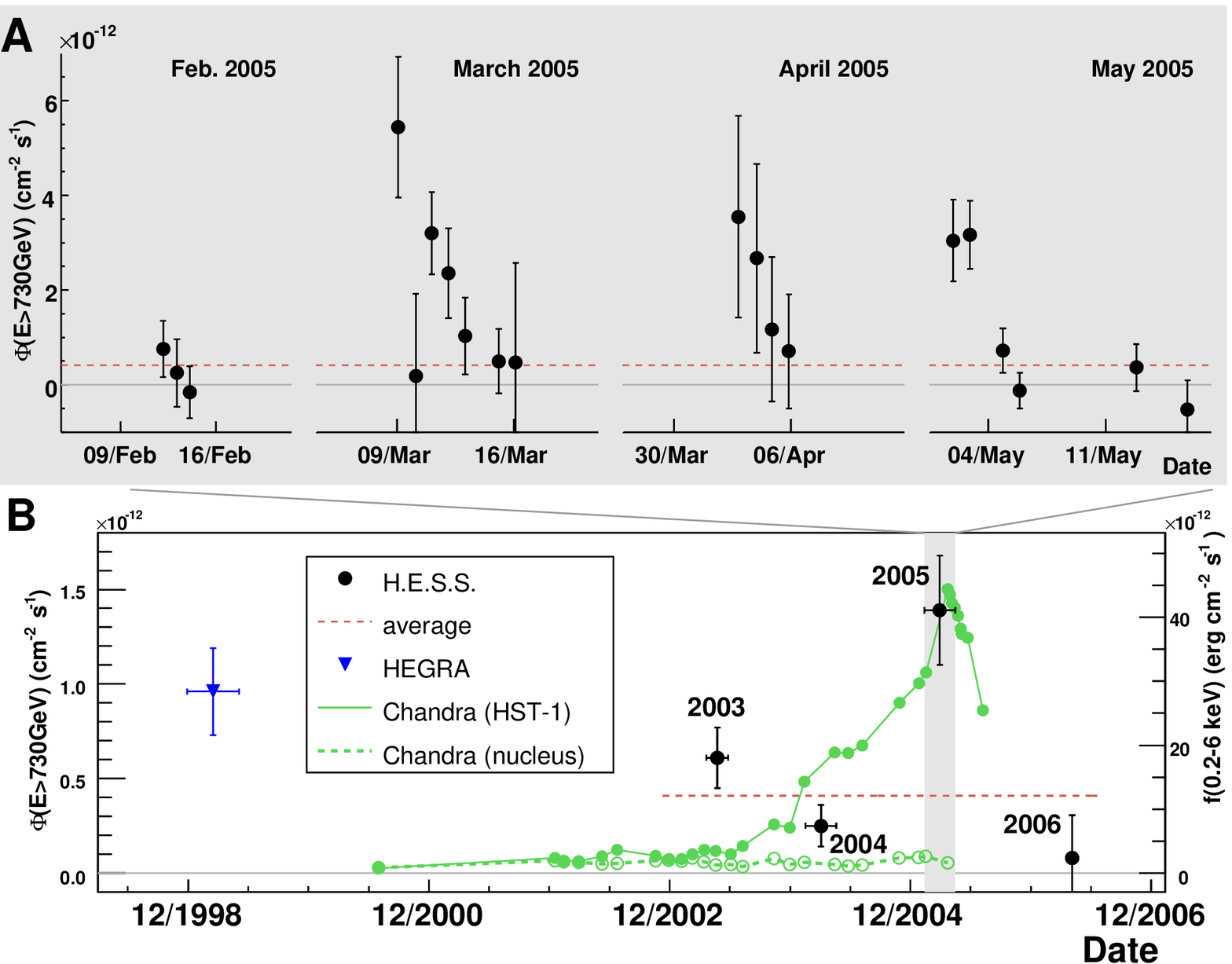}
\end{center}

\caption{Gamma-ray flux above $730 \, \rm{GeV}$. (B) The average flux
values for the years 2003 to 2006 as measured with H.E.S.S. together with
a fit of a constant function (red line). The flux reported by HEGRA is
also drawn. (A) The night-by-night fluxes for the four individual months
(February to May) of the high-state measurements in 2005 with significant
variability on (flux doubling) time-scales of days. The green points in
(B) correspond to the $0.2 - 6 \, \rm{keV}$ X-ray flux of the knot
HST-1 (solid, \cite{HST-1_Xrays}) and the nucleus (dashed, D.~Harris,
priv. comm.) as measured by Chandra; the lines are linear interpolations
of the flux points.}

\label{Fig:M87LC}

\end{figure*}

The integral $\gamma$-ray flux above $730 \, \rm{GeV}$ is shown in 
Fig.~\ref{Fig:M87LC} (right) for the years 2003--2006 with a statistical 
significance for variability on a yearly basis of $3.2 \, \sigma$. This is 
confirmed by a Kolmogorov test comparing the distribution of photon 
arrival times to the distribution of background arrival times yielding a 
significance for burst-like behaviour above $4 \, \sigma$. Surprisingly, 
variability on time-scales of days (flux doubling) was found in the high 
state data of 2005 (Fig.~\ref{Fig:M87LC}, upper right panel) with a 
statistical significance of more than $4 \, \sigma$. This is the fastest 
variability observed in any waveband from M\,87 and strongly constrains 
the size of the emission region of the TeV $\gamma$-rays to $R \leq c 
\cdot \Delta t \cdot \delta \approx 5 \times \delta \, R_{s}$, where 
$\delta$ is the relativistic Doppler factor of the source of $\gamma$ 
radiation and $R_{s} \approx 10^{15} \rm{cm}$ is the Schwarzschild radius 
of the supermassive black hole in M\,87. This very compact emission region 
excludes a variety of models for the emission of the TeV $\gamma$ 
radiation, i.e. CR interaction with matter in M\,87, radiation due to 
annihilation of dark matter particles, the kpc plasma jet or even 
individual knots of the jet, and leaves only a region very close to the 
central black hole as a reasonable production site of the TeV 
$\gamma$-rays with possibly novel mechanisms involved, see 
\cite{HESS_M87_Paper} for a more detailed discussion. For instance, the 
observed $\gamma$-ray flux may be explained by inverse Compton emission of 
ultrarelativistic electron-positron pairs which are produced in an 
electromegnetic cascade in the black hole magnetosphere 
\cite{M87_BH_Magnetosphere}.

M\,87 was monitored during the past years by the Chandra X-ray satellite,
see Fig.~\ref{Fig:M87LC} (right). The X-ray flux of the knot HST-1
(located very close to the nucleus) increased by a factor of $\sim 50$
between 2003 and 2005 \cite{HST-1_Xrays}, whereas the emission of the
nucleus itself remained rather constant. However, no unique correlation
between the X-ray and TeV $\gamma$-ray fluxes can be stated, since the
measurements were not performed simultaneously.

%==========================[ FIGURE

%==========================[ FIGURE

%%%%%%%%%%%%%%%%%%%%%%%%%%%%%%%%%%%%%%%%%%%%
%% Summary and Conclusion
%%%%%%%%%%%%%%%%%%%%%%%%%%%%%%%%%%%%%%%%%%%%
\section*{Summary and Conclusion}

H.E.S.S. confirmed the giant radio galaxy M\,87 as the first extragalactic
TeV $\gamma$-ray source which does not belong to the class of blazars. The
hard energy spectrum in 2005 challenges hadronic as well as leptonic
models. The surprisingly discovered variability of the TeV $\gamma$-ray
emission on short time-scales of days strongly constrains the size of the
emission region and excludes several models discussed for M\,87, leaving a
location close to the central black hole as reasonable production site of
the TeV $\gamma$-ray emission. Simultaneous multi-wavelengths observations
and observations in the MeV/GeV energy range with GLAST are of special
importance to estimate the position of the maximum of the VHE peak in the
SED and further constrain model parameters. Taken M\,87 as an established
TeV $\gamma$-ray emitting radio galaxy, one should also mention the FR\,I
radio galaxy Centaurus\,A (Cen\,A), which is located at an even closer
distance of $3.4 \, \rm{Mpc}$ ($z = 0.0018$) and shows a jet angle of
$\theta > 50^{\circ}$. Cen\,A is the only AGN not belonging to the class
of blazars which was detected in the GeV energy regime by EGRET
\cite{EGRET_CenA, EGRET_Catalogue}, making a detection with GLAST
promising. So far, no excess was found in the $\sim 5 \, \rm{h}$ of
H.E.S.S. data taken in 2004 and 2005 \cite{HESS_AGN}.

%%%%%%%%%%%%%%%%%%%%%%%%%%%%%%%%%%%%%%%%%%%%%%%%
%% BACKMATTER
%%%%%%%%%%%%%%%%%%%%%%%%%%%%%%%%%%%%%%%%%%%%%%%%
\section*{Acknowledgements}

The support of the Namibian authorities and of the University of Namibia
in facilitating the construction and operation of H.E.S.S. is gratefully
acknowledged, as is the support by the German Ministry for Education and
Research (BMBF), the Max Planck Society, the French Ministry for Research,
the CNRS-IN2P3 and the Astroparticle Interdisciplinary Programme of the
CNRS, the U.K. Particle Physics and Astronomy Research Council (PPARC),
the IPNP of the Charles University, the South African Department of
Science and Technology and National Research Foundation, and by the
University of Namibia. We thank D.~Harris for providing the Chandra X-ray
light curve of the M\,87 nucleus.

%%%%%%%%
%  31  %
%%%%%%%%

%The paper title
\title{Multiwavelength observations of PKS 2005-489 and H 2356-309 with HESS}
%Short title to print in the headers to the final publication (Not showed in this print).
\shorttitle{MWL observations of PKS 2005-489 and H 2356-309}
%All paper authors
\authors{Luigi Costamante$^{1}$, Wystan Benbow$^{1}$, Catherine Boisson$^{2}$, Santiago Pita$^{3}$, Helene
Sol$^{2}$, for the H.E.S.S. Collaboration}
%Short title to print in the headers to the final puplication (Not showed in this print).
%\shortauthors{L. Costamante et al.}
%All the affiliations.
\afiliations{$^1$ Max-Planck-Instituit f\"ur Kernphysik, PO box 103980, D69029 Heidelberg, Germany\\ 
$^2$ LUTH, UMR 8102 du CNRS, Observatoire de Paris, Section de Meudon, F-92195 Meudon Cedex, France\\
$^3$ APC, CNRS, Universite Paris 7 Denis Diderot, F-75205 Paris Cedex 13, France}
\email{Luigi.Costamante@mpi-hd.mpg.de}

%The abstract.
\abstract{
Very-high-energy (VHE; $>$100 GeV) $\gamma$-ray observations of PKS 2005-489 and H 2356-309 
were made with the High Energy Stereoscopic System (HESS) in 2005 and 2006. Previous 2004 data
have been reanalysed to correct for the degradation of the optical efficiency of the
HESS system. Both sources have been detected during all 3 years, at a level of 1-3\% of
the Crab flux. A total excess of $\sim$16$\sigma$ and $\sim$12$\sigma$, respectively, is accumulated. 
Significant flux variations are seen on a monthly basis for H 2356-309, 
and in 2006 for PKS 2005-489. The spectra confirm the previously reported values, 
in particular the hard spectrum of H 2356-309. Multiwavelength
observations performed with XMM and RXTE in 2004 and 2005 reveal remarkable flux (10x) and
spectral ($\Delta\Gamma$=0.7) variations for PKS 2005-489. 
Despite a $\sim$10$\times$ flux increase above 1 keV, no flux variation is seen at VHE, 
implying in an SSC scenario a corresponding decrease of the energy density of the seed photons 
for inverse Compton (IC) scattering, not observed in the SED. A possible explanation
is that a new component is emerging in the jet, whose electrons do not see the photons of the observed
synchrotron peak. The SED of both objects shows the potential for significantly higher VHE fluxes.}

\maketitle

\addcontentsline{toc}{section}{Multiwavelength observations of PKS 2005-489 and H 2356-309 with HESS}
\setcounter{figure}{0}
\setcounter{table}{0}
\setcounter{equation}{0}

%Begin the section.
\section*{Introduction}
\vspace*{-0.2cm}
The blazars PKS 2005-489 (z=0.071) and H 2356-309 (z=0.165) are two high-frequency-peaked BL Lac objects (HBL).
PKS 2005-489 is one of the brightest HBL in the southern hemisphere, and is characterized by very large
variability in the X-ray band \cite{1,2}. H 2356-309 is an {\it extreme} BL Lac \cite{3}, 
characterized by the synchrotron peak of the spectral energy distribution (SED) at energies above
a few keV. Both objects have been discovered by HESS as VHE sources in 2004 \cite{4,5}, 
though at a rather faint flux ($\sim$2-3\% Crab). Coordinated X-ray observations performed 
in the same epoch with XMM and RXTE revealed historically low fluxes, for both
objects. Since in HBLs the X-ray band usually samples the synchrotron emission of TeV electrons, which 
produce VHE photons by inverse Compton (IC) scattering of low energy photons, significantly higher VHE
fluxes can be expected. Monitoring observations in 2005 and 2006 were thus performed, both to increase
the event statistics and to catch flaring events. Further multiwavelength observations were also performed with
XMM (as pre-planned pointings due to the narrow overlap between HESS and XMM visibility windows) and
RXTE (as ToO). The main preliminary results on the average data are here reported.

\section*{HESS Results}
\vspace*{-0.2cm}
All data have been analyzed with the HESS standard analysis \cite{6,8}. 
For the spectral and flux determination the energy of each event event
is corrected \cite{6} for the absolute optical efficiency of the system 
using efficiencies determined from simulated and observed muons. 
This correction eliminates any potential long-term variations in the absolute energy scale
of the HESS analysis due to a changing optical throughput. 
The systematic error is $\sim$20\% on flux and $\sim$0.1 for the photon index. 
%The correction affects the flux of soft spectrum sources much more strongly than hard spectrum objects.

On PKS 2005-489, a total of 135.4 hours of observations were taken 
from 2004 through 2006. After data-quality selection, an exposure of
78.3 h livetime is obtained, at a mean zenith angle $36\gradi$. 
A point-like VHE $\gamma$-ray excess from PKS 2005-489 is detected 
each year, with an average flux of $\sim$2.8\% Crab. 
On a monthly basis (Fig. \ref{fig1}), there is indication of 
$\sim$3$\times$ flux variability in 2006. At shorter timescales, no
significant variability is detected, though comparable variations 
cannot be excluded. % given the low statistics. 
\begin{figure}[t]
\begin{center}
\includegraphics[width=0.48\textwidth]{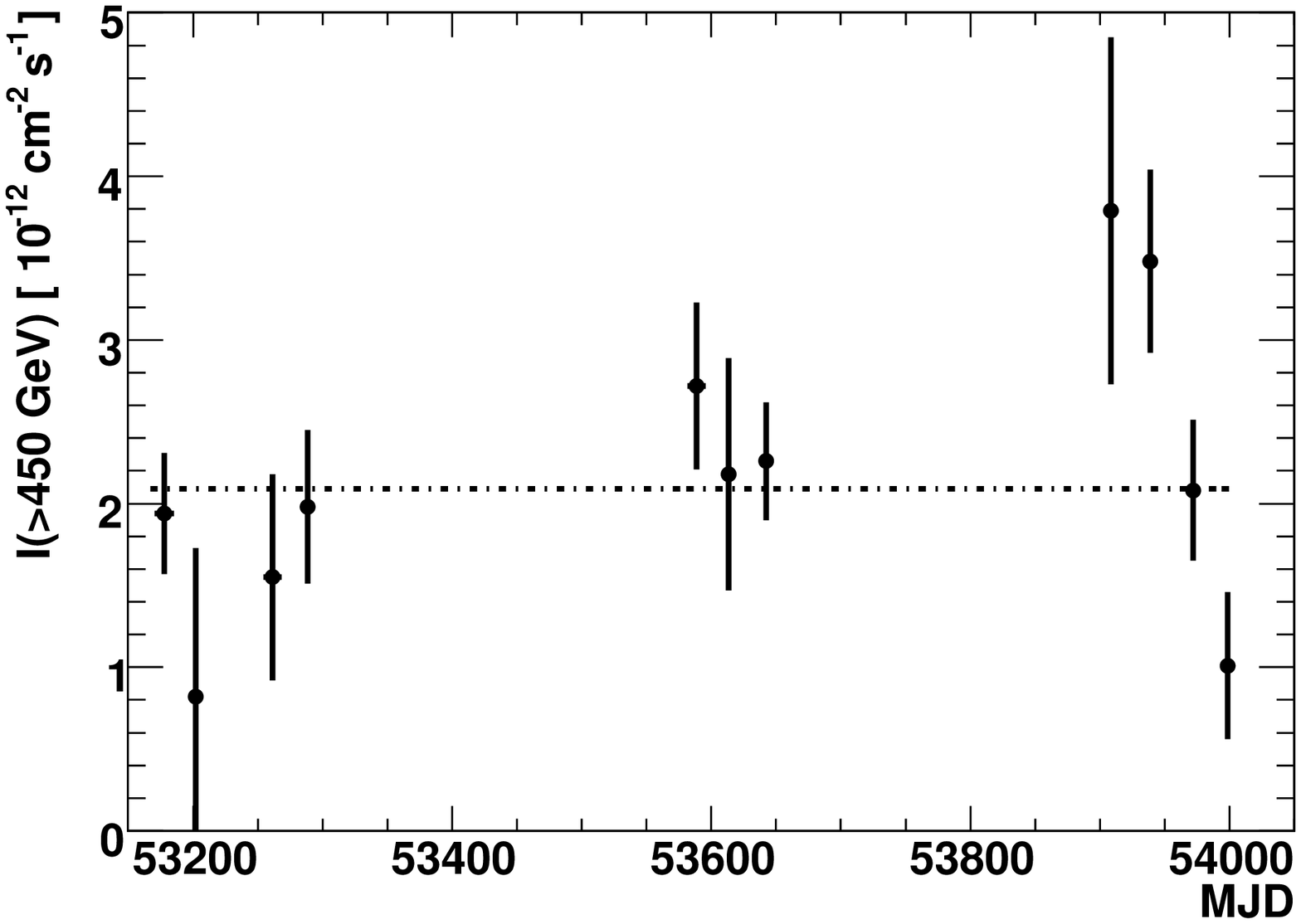}
\end{center}
\vspace{-0.8cm}
\caption{The integral flux ($>$450 GeV) measured by HESS from PKS
2005-489 in monthly bins. The 2004 values are $\sim$3 times
higher than previously published\cite{4} as all fluxes are corrected\cite{6} 
for degradation in the optical efficiency of the HESS system. 
Only the statistical errors are shown. The fluxes are calculated assuming
the time-average spectrum measured in the respective year (Table \ref{table1LC}).
Simultaneous X-ray observations were performed on MJD 53282, 
53608-53622 (RXTE) and 53641 (Table \ref{table2LC}, respectively). }
%53639 and 53641.}
\label{fig1}
\end{figure}
The annual VHE spectra measured are shown in Fig. \ref{fig2} and Table \ref{table1LC}. 
Among years, the flux below 1 TeV remains basically constant.
There is only a slight ($\sim$1.8$\sigma$) indication of hardening
between 2004 and 2006 spectra.
\begin{figure}[t]
\begin{center}
\includegraphics[width=0.48\textwidth]{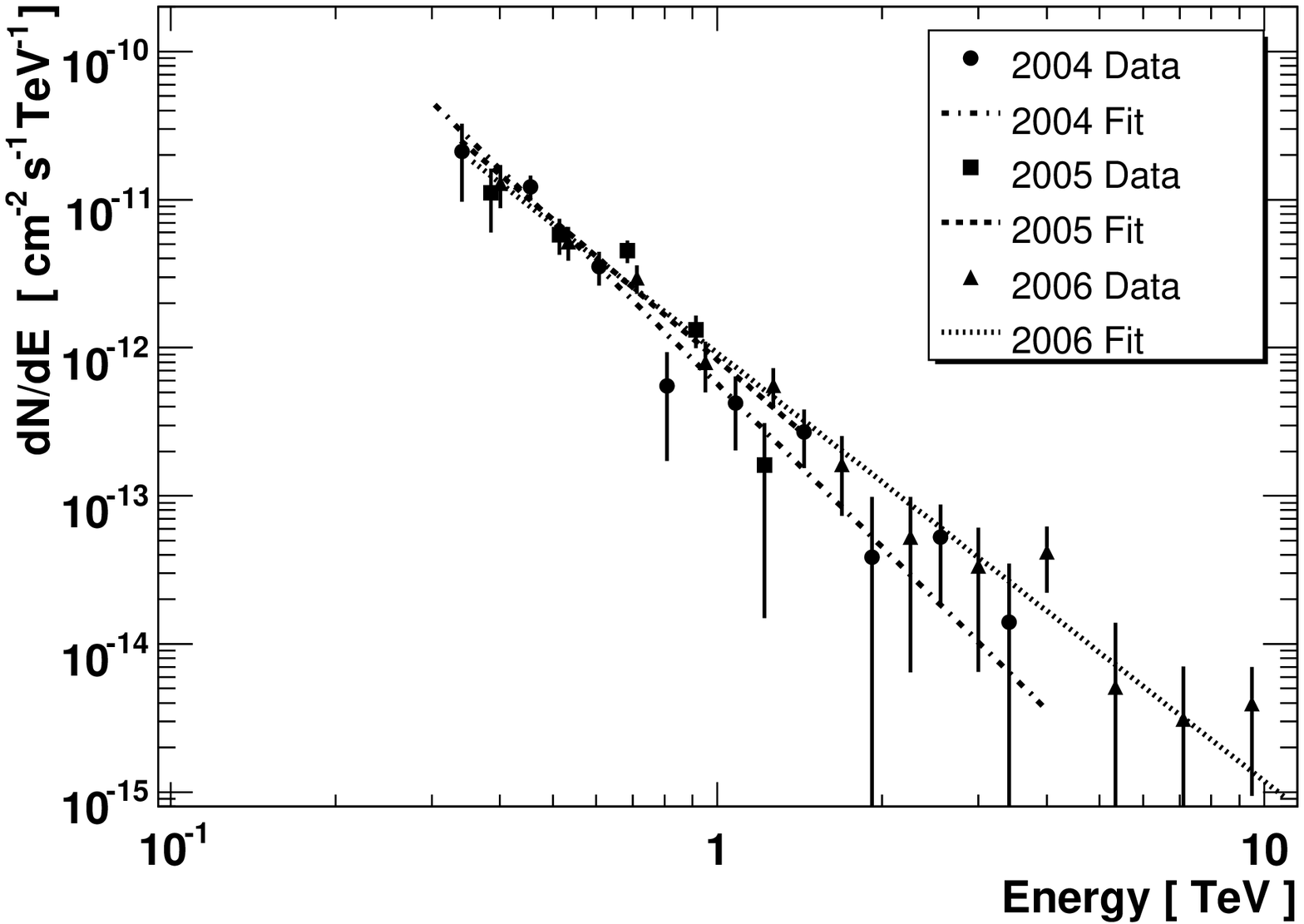}
\end{center}
\vspace{-0.8cm}
\caption{The annual VHE spectra observed from PKS 2005-489.
The lines represent the best fit of a power-law model to the observed data, 
with photon index $\Gamma$ reported in Table \ref{table1LC}.
}\label{fig2}
\end{figure}

\begin{table}[h!]
\caption{Results of the HESS observations. Shown are the epoch, exposure time, 
significance of the excess, integral flux (in units of $10^{-13}$ cm$^{-2}$ s$^{-1}$) 
and photon index for a single power-law fit. Only the statistical errors are shown.
}
\centering
\small
\begin{tabular}{c c c c c c}
\noalign{\smallskip}
\hline
\hline
\noalign{\smallskip}
Epoch & Time & Sign. & Int. Flux & $\Gamma$ \\
 & [h] &  [$\sigma$] &  [f.u.] &  \\
\noalign{\smallskip}
\hline
\hline
\noalign{\smallskip}
\multicolumn{3}{l}{\bf PKS 2005-489} &   ($>$450 GeV) & \\
\hline
\noalign{\smallskip}
2004 &  24.2   & 7.7  &   1.81$\pm$0.26 & 3.65$\pm$0.39 \\
2005 &  32.6   & 11.0 &   2.38$\pm$0.27 & 3.15$\pm$0.30  \\
2006 &  21.5   & 8.8  &   2.20$\pm$0.26 & 2.89$\pm$0.20  \\
\noalign{\smallskip}
\hline
\noalign{\smallskip}
Total & 78.3  &  15.9 & 2.08$\pm$0.15  & 3.18$\pm$0.16  \\
\noalign{\smallskip}
\hline
\hline
\noalign{\smallskip}
\multicolumn{3}{l}{\bf H 2356-309} &   ($>$200 GeV) &  \\
\hline
\noalign{\smallskip}
2004 &  39.9   &  9.6  & 5.97$\pm$0.61   & 2.97$\pm$0.19 \\
2005 &  46.7   &  5.9  & 3.28$\pm$0.65   & 2.99$\pm$0.39  \\
2006 &  23.2   &  5.1  & 3.49$\pm$0.82   & 3.43$\pm$0.41  \\
\noalign{\smallskip}			   
\hline
\noalign{\smallskip}
Total & 109.8  & 12.1  &  4.47$\pm$0.39 &  3.09$\pm$0.16 \\
\noalign{\smallskip}
\hline
\hline
\end{tabular}
\label{table1LC}
\end{table}

%\section*{h2356-309}
H 2356-309 has been observed by HESS for a total of 164 hours from 2004 through
2006. After data-quality selection, an exposure of 109.8 h is obtained, at mean
zenith angle $19\gradi$. Significant VHE emission is detected during each year, with clear
indications of variability on an annual and monthly timescale (probability of constant
flux $<$0.4\%). At shorter timescales no significant variability is detected, though
comparable variations cannot be excluded given the low statistics. 
Despite the variability, no significant spectral changes are observed (Table \ref{table1LC}).

For the discussion of the SED properties of these two HBL, 
all the HESS spectra have been corrected for $\gamma$-$\gamma$ absorption on the diffuse 
Extragalactic Background Light (EBL) with the P0.45 shape in \cite{7} 
(close to the level from galaxy counts).

\begin{figure}[t]
\begin{center}
\includegraphics[width=0.48\textwidth]{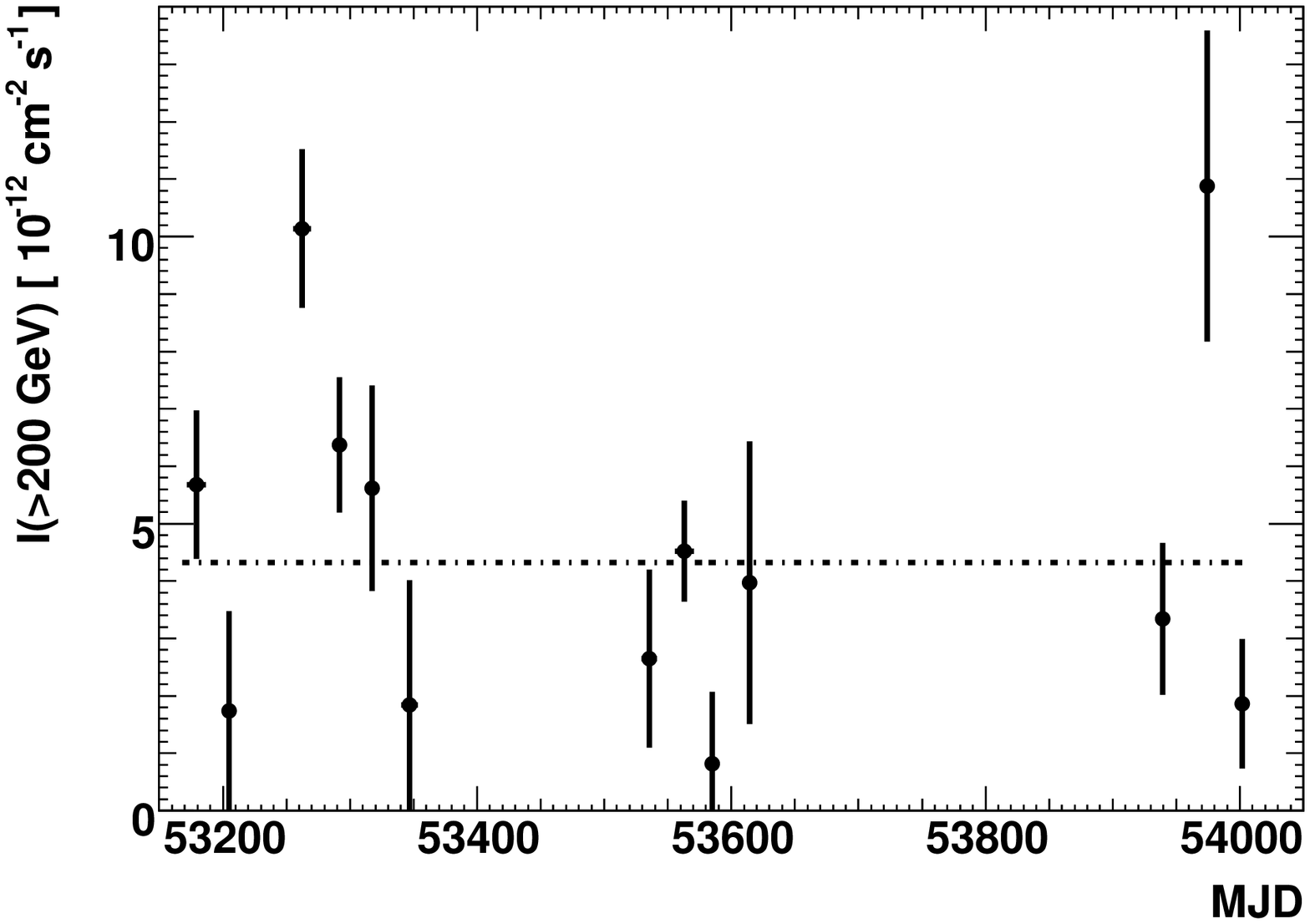}
\end{center}
\vspace{-0.8cm}
\caption{The integral flux ($>$200 GeV) measured by HESS from
H 2356-309, in monthly bins. All fluxes are corrected for the degradation 
of the HESS optical efficiency \cite{6}. Statistical errors only.
The fluxes are calculated assuming the time-average spectrum measured 
in the respective year (Table \ref{table1LC}). There is clear indication of variability
(probability of constant flux $<$0.4\%).
Simultaneous X-ray observations were performed on MJD 53320 (RXTE), 
53534 and 53536 (Table \ref{table2LC}, respectively).}
\label{fig3}
\end{figure}

\begin{figure}[t]
\begin{center}
\includegraphics[width=0.48\textwidth]{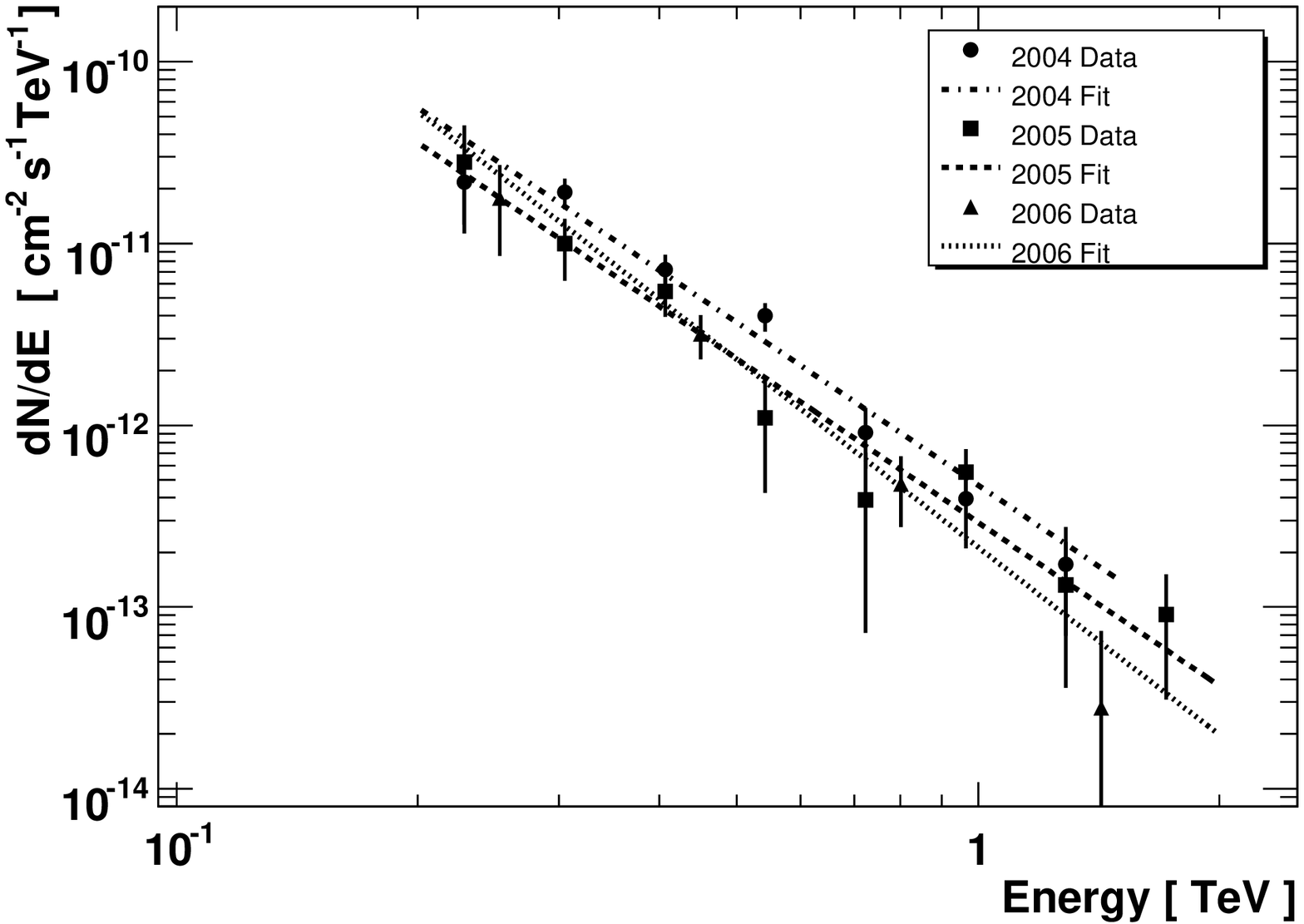}
\end{center}
\vspace{-0.8cm}
\caption{The annual VHE spectra observed from H 2356-309.
The lines represent the best fit of a power-law model to the observed data, 
with photon index $\Gamma$ reported in Table \ref{table2LC}.
No significant spectral variation is observed. 
An analysis of combined data for different flux levels is on-going.}
\label{fig4}
\end{figure}

\begin{table}[h!]
\caption{Best-fit parameters of the X-ray data. 
Single and broken power-law models (XMM: MOS+PN data).
Column density N$_{\rm H}$ fixed to galactic values, and modelled with Tbabs
using Wilms abundances. The errors are quoted at the 90\% confidence level.
Unabsorbed flux in units of erg cm$^{-2}$ s$^{-1}$ in the 2-10 KeV band.
%(Aharonian et al 2007, in preparation).
}
\centering
%\small
\footnotesize
\begin{tabular}{c c c c c c}
\noalign{\smallskip}
\hline
\hline
\noalign{\smallskip}
Instr.  & $\Gamma_1$ &  E$_{br}$ & $\Gamma_2$ &  Flux  \\
        &            &  [keV]  &            &  [f.u.]\\
\noalign{\smallskip}
\hline
\hline
\noalign{\smallskip}
\multicolumn{2}{l}{\bf PKS 2005-489} &    & \\
\hline
\noalign{\smallskip}
XMM  &    -  & -	  & 3.04$\pm$0.05 & 1.2E-12  \\
RXTE &    -  & -	  & 2.9$\pm$0.2   &  7.6E-12  \\
XMM  &    3.0$\pm$0.1 & 0.5 & 2.27$\pm$0.03 & 2.0E-11 \\
\noalign{\smallskip}
\hline
\hline
\noalign{\smallskip}
\multicolumn{2}{l}{\bf H 2356-309} &    &  \\
\hline
\noalign{\smallskip}
RXTE  &       -      &  -	&  2.43$\pm$0.25 & 9.7E-12 \\
XMM   &    2.00$\pm$0.05 & 1.0  &  2.34$\pm$0.03 & 7.2E-12 \\
XMM   &    1.92$\pm$0.06 & 0.9  &  2.23$\pm$0.03 & 8.7E-12 \\
\noalign{\smallskip}
\hline
\end{tabular}
\label{table2LC}
\end{table}

\section*{SED Changes in PKS\,2005$-$489}
\vspace*{-0.2cm}
Simultaneous X-ray observations were performed with XMM in Oct. 2004 and Sept. 2005, and with
RXTE in Aug-Sept 2005. No significant variability is observed within each data set, on any timescale.
\begin{figure}[t]
\begin{center}
\includegraphics[width=0.48\textwidth]{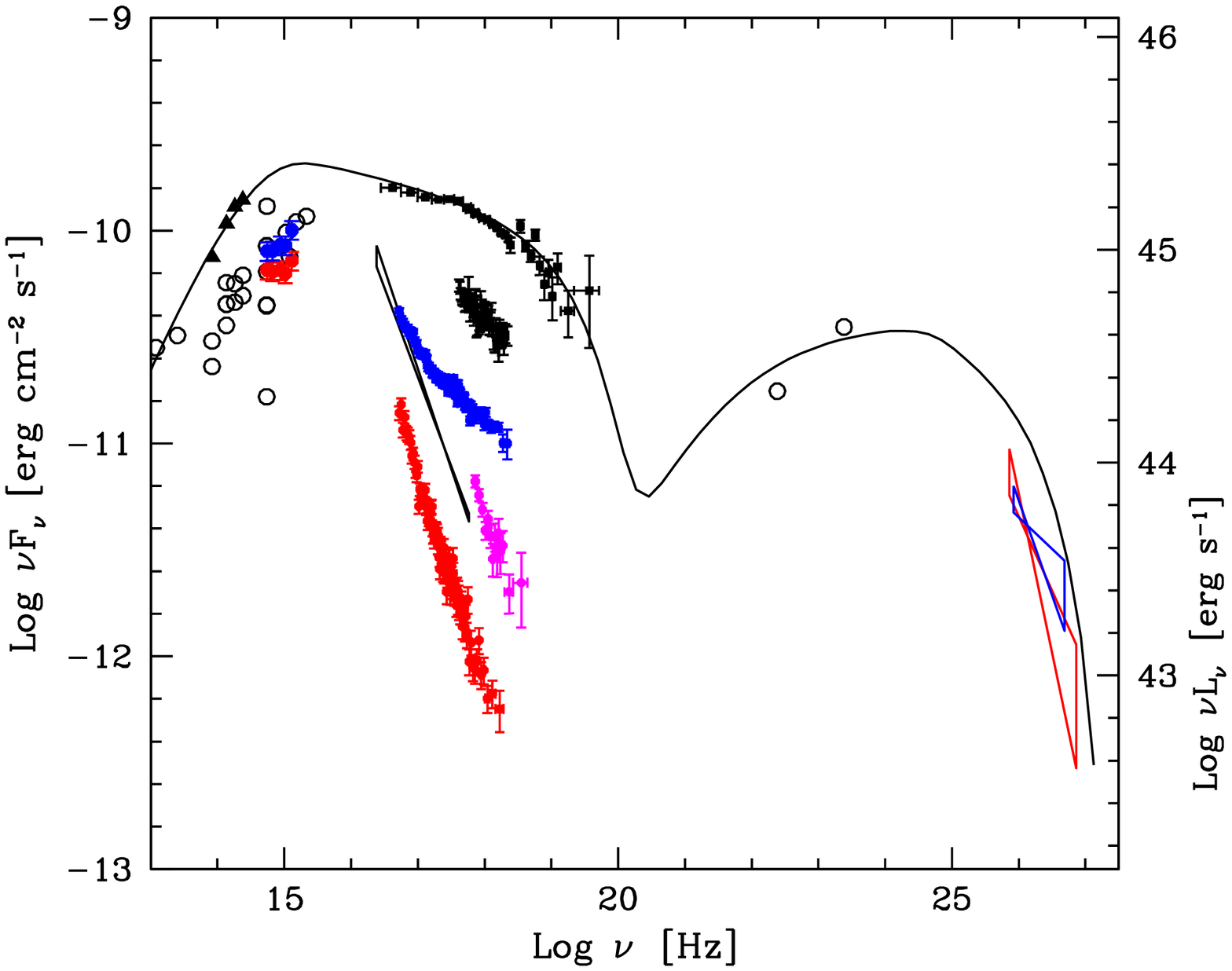}
\end{center}
\vspace{-0.8cm}
\caption{SED of PKS 2005-489. Black: historical data and
modelling of the strong 1998 flare \cite{2}.  
Corrected for EBL absorption (P0.45 curve in \cite{7}), 
the intrinsic VHE slopes obtained are $\Gamma_{int}$=$3.1\pm0.4$ (red, year 2004) 
and $\Gamma_{int}$=$2.6\pm0.3$ (blue, year 2005).
The Opt-UV fluxes from the Optical Monitor (OM) onboard XMM are
corrected for galactic extinction using the Cardelli et al. (1998)
curve. XMM data processed with SAS7.0. The hard ($\Gamma<$2) UV spectrum
indicated by the OM photometry locates the synchrotron peak
between the Far-UV and Soft X-ray range.
}\label{sed2005}
\end{figure}
From 2004 to 2005, the spectrum above 1 keV
hardens strongly ($\Delta\Gamma$=0.7), yielding a $\sim$10$\times$ flux increase.
The UV fluxes (close to the synchrotron peak) show a $\sim$30\% increase as well. 
In contrast, the VHE emission remains almost constant, with a spectrum
that suggests it can be produced by the same electrons emitting by synchrotron in the hard X-ray
band.  However, the VHE flux should have increased at least linearly with the X-ray flux between the
two epochs, if these electrons could upscatter by IC the observed synchrotron peak photons. 
For the VHE flux to remain constant, a corresponding decrease of the seed-photons energy density is
required. This suggests that a new jet component is emerging, physically separated from the main
emitting blob, and whose synchrotron peak emission remains at present hidden below the observed SED.

\begin{figure}[t]
\begin{center}
\includegraphics[width=0.48\textwidth]{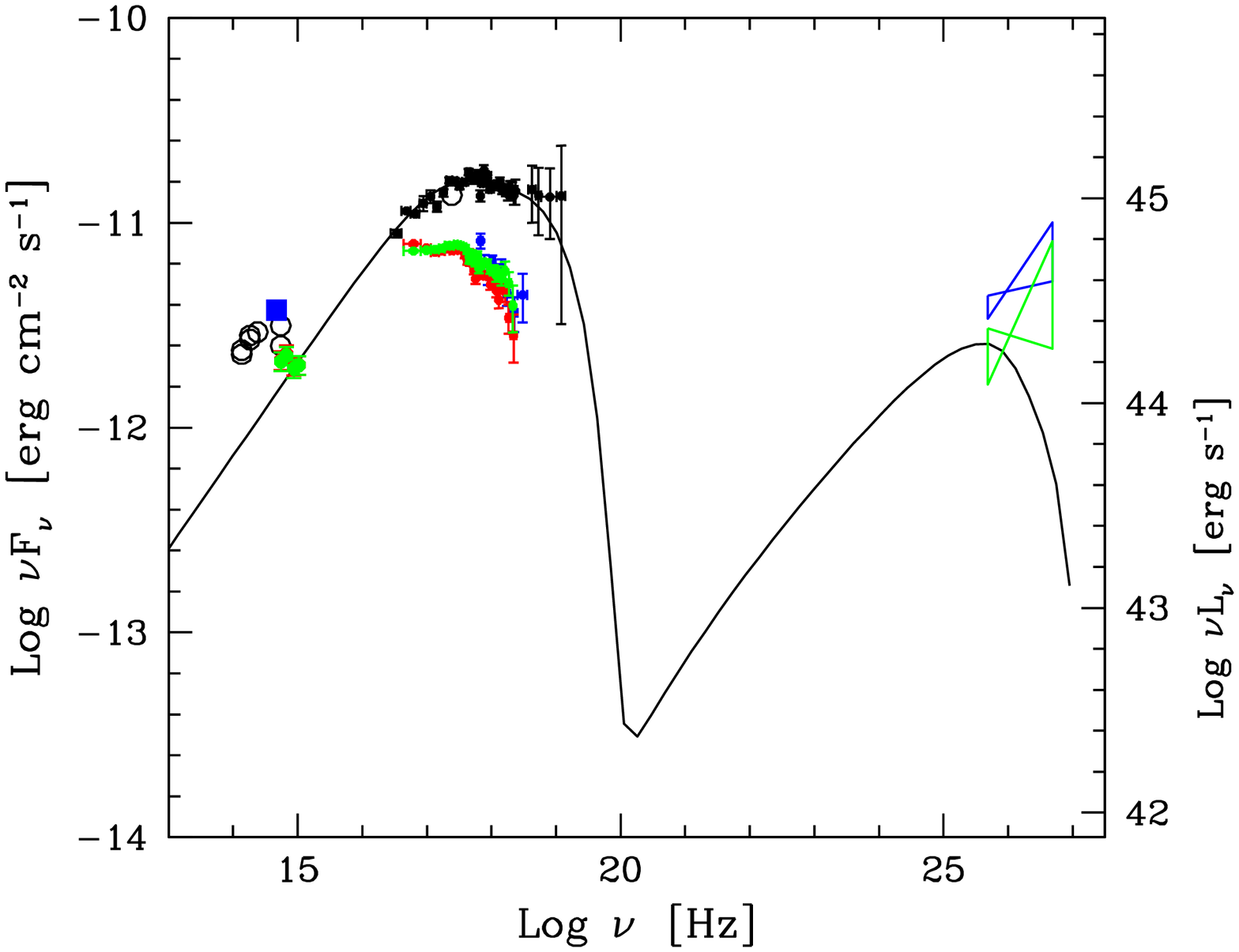}
\end{center}
\vspace{-0.8cm}
\caption{SED of H\,2356$-$309. Black: historical data and modelling\cite{3}.
After correction for EBL absorption, the intrinsic VHE slopes are $\Gamma_{int}$=$1.7\pm0.2$ 
(blue, year 2004) and $\Gamma_{int}$=$1.7\pm0.4$ (green, year 2005). 
Blue symbols: data discussed in \cite{5}. Optical fluxes from
ROTSE and XMM-OM . 
}\label{sed2356}
\end{figure}

\section*{SED Changes in H\,2356$-$309}
\vspace*{-0.2cm}
The X-ray flux and spectral properties appear to
be almost constant among these trhee epochs, at a flux level 
$\sim$3$\times$ lower than the \sax values (June 1998\cite{3}). 
The XMM spectra confirm the location of the synchrotron peak in
the X-ray band (at 1-2 KeV), as derived from the \sax data. 
The hard VHE spectra, now measured with better statistics, confirm the
constraints on the EBL previously obtained from the 2004 dataset\cite{7}. 
Once corrected for intergalactic $\gamma-\gamma$ absorption, the hard VHE spectrum
locates the IC peak of the SED above 1 TeV.

\section*{Conclusion} 
\vspace*{-0.2cm}
Observations performed by HESS from 2004 through 2006 have confirmed PKS 2005-489 and
H 2356-309 as VHE gamma-ray sources, at an average annual level of 1-3\% Crab. 
The VHE spectra confirm the very different SED properties of these two HBL, 
with very soft and hard intrinsic spectra respectively. 
Simultaneous observations with RXTE and XMM have confirmed the correlation between 
SED peak energies, with the higher synchrotron peak frequency observed in the object with the higher
IC peak energy. X-ray observations have also shown the objects to be in historically low states.
For PKS 2005-489, the overall SED evolution suggests that a new jet
component is emerging, with harder properties. Since PKS 2005-489 has historically
demonstrated a 100$\times$ dynamical range in the X-ray band, dramatically higher VHE fluxes 
($10^2-10^4\times$) can be expected in a leptonic scenario, unless counterbalanced by a strong 
($>$10$\times$) and simultaneous increase of the magnetic field. 
These results confirm the strong diagnostic potential
of coordinated Optical--X-ray--VHE observations. Further monitoring of these objects is highly
encouraged. 

\vspace{-0.3cm}
\section*{Acknowledgements}
\vspace{-0.4cm}
\begin{small}
The support of the Namibian authorities and of the University of Namibia
in facilitating the construction and operation of HESS is gratefully
acknowledged, as is the support by the German Ministry for Education and
Research (BMBF), the Max Planck Society, the French Ministry for Research,
the CNRS-IN2P3 and the Astroparticle Interdisciplinary Programme of the
CNRS, the U.K. Science and Technology Facilities Council (STFC),
the IPNP of the Charles University, the Polish Ministry of Science and 
Higher Education, the South African Department of
Science and Technology and National Research Foundation, and by the
University of Namibia. We appreciate the excellent work of the technical
support staff in Berlin, Durham, Hamburg, Heidelberg, Palaiseau, Paris,
Saclay, and in Namibia in the construction and operation of the
equipment.
\end{small}

%%%%%%%%
%  32  %
%%%%%%%%

%The paper title
\title{Multi-wavelength Observations of PG\,1553+113 with HESS}
%Short title to print in the headers to the final publication (Not showed in this print).
\shorttitle{Multi-wavelength Observations of PG\,1553+113}
%All paper authors
\authors{
W.\,Benbow$^{1}$, 
C.\,Boisson$^{2}$, 
R.\,B\"uhler$^{1}$,
and H.\,Sol$^{2}$
for the HESS Collaboration
}
%Short title to print in the headers to the final publication (Not showed in this print).
\shortauthors{W.\,Benbow et al.}
%All the affiliations.
\afiliations{$^1$ Max-Planck-Institut f\"ur Kernphysik, Heidelberg, Germany\\ 
$^2$ LUTH, UMR 8102 du CNRS, Observatoire de Paris, Section de Meudon, France\\
}
\email{Wystan.Benbow@mpi-hd.mpg.de}

%The abstract.
\abstract{
Very high energy (VHE; $>$100 GeV) $\gamma$-ray
observations of PG\,1553+113 were made with the 
High Energy Stereoscopic System (HESS) in 2005 and 2006.
A strong signal, $\sim$10 standard deviations, 
is detected by HESS during the 2 years of
observations (24.8 hours live time). 
The time-averaged energy spectrum, measured
between 225 GeV to $\sim$1.3 TeV,
is characterized
by a very steep power law (photon index of 
$\Gamma = 4.5\pm0.3_{\rm stat}\pm0.1_{\rm syst}$).
The integral flux above 300 GeV is $\sim$3.4\% of the Crab Nebula
flux and shows no evidence for any variations, on any time scale.
H+K (1.45$-$2.45$\mu$m) spectroscopy of PG\,1553+113 
was performed in March 2006 with SINFONI, an integral field 
spectrometer of the ESO Very Large Telescope (VLT) in Chile.
The redshift of PG\,1553+113 is still unknown, as
no absorption or emission lines were found.}

\maketitle

\addcontentsline{toc}{section}{Multi-wavelength Observations of PG\,1553+113 with HESS}
\setcounter{figure}{0}
\setcounter{table}{0}
\setcounter{equation}{0}

\section*{Introduction}

Evidence for VHE ($>$100 GeV) $\gamma$-ray emission from 
the high-frequency-peaked BL\,Lac object PG\,1553+113 was 
first reported by the HESS collaboration \cite{HESS_discovery} 
based on observations made in 2005. This detection was later
confirmed \cite{MAGIC_1553} with MAGIC observations 
in 2005 and 2006.   The measured VHE spectra are unusually soft 
(photon index $\Gamma$=4.0$\pm$0.6 and $\Gamma$=4.2$\pm$0.3 for the
HESS and MAGIC experiments, respectively) but the errors are large,
clearly requiring improved measurements before detailed 
interpretation of the complete SED is possible.
Further complicating any SED interpretation is the 
absorption of VHE photons \cite{EBL_effect3,EBL_effect2}
by pair-production on the Extragalactic Background Light (EBL).
This absorption, which is energy dependent and increases strongly 
with redshift, distorts the VHE energy spectra observed from 
distant objects. For a given redshift, the
effects of the EBL on the observed spectrum can be 
reasonably accounted for during SED modeling.
Unfortunately, the redshift of PG\,1553+113 
is unknown, despite many attempts to measure it
(see, e.g., \cite{Carangelo_03,no_lines}).

In 2005 and 2006, a total of 30.3 hours of HESS observations 
were taken on PG\,1553+113. The 2005 HESS observations are exactly the 
same as reported in \cite{HESS_discovery}.
The good-quality exposure is 24.8 hours live time.
The data are processed using the
standard HESS calibration \cite{calib_paper}
and analysis tools \cite{std_analysis}.
{\it Soft cuts} \cite{HESS_discovery}
are applied to select candidate $\gamma$-ray events,
resulting in an average post-analysis energy 
threshold of 300 GeV at the mean zenith angle of the observations, $37^{\circ}$. 

\section*{HESS Results}

   \begin{figure}[t]
   \centering
      \includegraphics[width=0.45\textwidth]{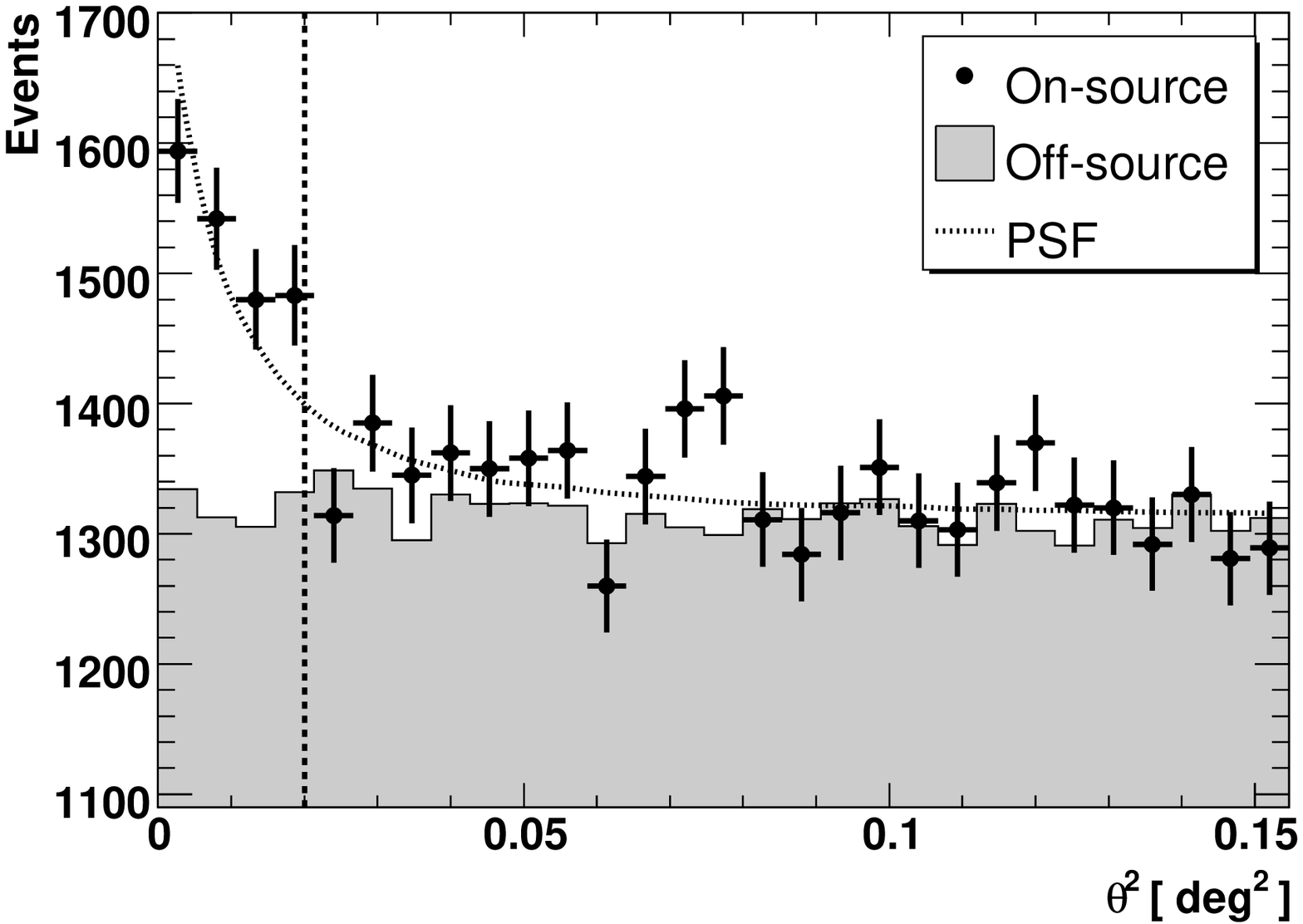}
      \caption{The distribution of $\theta^2$ for on-source 
        events (points) and
        normalized off-source events (shaded) from observations
        of PG\,1553+113.  The dashed curve represents
	the $\theta^2$ distribution expected for
	a point source of VHE $\gamma$-rays at $40^{\circ}$
	zenith angle with a photon index $\Gamma=4.5$.
	The vertical line represents the cut 
        on $\theta^2$ applied to the data.}
         \label{thtsq_plot}
   \end{figure}

  \begin{table}
      \caption{Shown are the excess, the significance of the excess, and 
        the integral flux above 300 GeV, from HESS observations
	of PG\,1553+113. The flux units are $10^{-12}$ cm$^{-2}$\,s$^{-1}$.
	The systematic error on the flux is 20\% and is not shown.}
         \label{results}
        \centering
         \begin{tabular}{c c c c c}
	   \\
            \hline\hline
            \noalign{\smallskip}
            Epoch & Time & Excess & Sig & I($>$300 GeV)\\
            & [h] & &  [$\sigma$] & [f.u.]\\
            \noalign{\smallskip}
            \hline
            \noalign{\smallskip}
            2005 & 7.6  & 249 & 6.0 & $5.44\pm1.23$\\
            2006 & 17.2 & 536 & 8.3 & $4.22\pm0.72$\\
            \noalign{\smallskip}
            \hline
            \noalign{\smallskip}
            Total & 24.8 & 785 & 10.2 & $4.56\pm0.62$\\             
            \noalign{\smallskip}
            \hline
       \end{tabular}
   \end{table}

A significant VHE $\gamma$-ray signal is detected 
in each year of HESS data taking.  The total observed
excess is 785 events, corresponding
to a statistical significance of 10.2 standard deviations ($\sigma$).
Table~\ref{results} shows the results of the HESS 
observations of PG\,1553+113. 
Figure~\ref{thtsq_plot} shows the on-source and normalized off-source
distributions of the square of the angular difference between
the reconstructed shower position and
the source position ($\theta^{2}$) for all observations. 
The background is approximately flat in $\theta^{2}$ as expected, and
there is a clear point-like excess of on-source events 
at small values of $\theta^{2}$, 
corresponding to the observed signal.
The peak of a two-dimensional Gaussian fit to a sky map 
of the observed excess is coincident with the position 
of PG\,1553+113.

\begin{figure}
   \centering
      \includegraphics[width=0.45\textwidth]{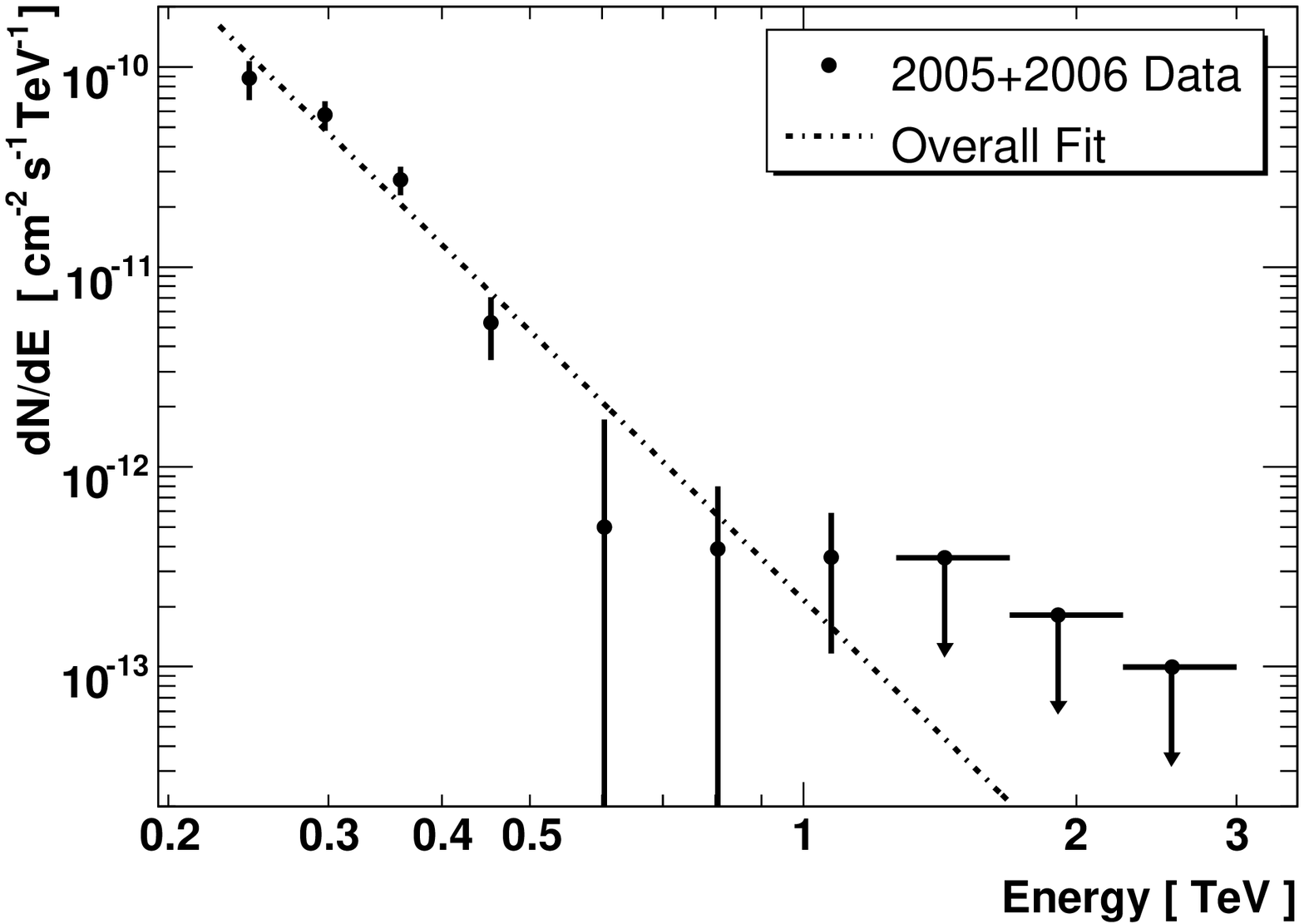}
      \caption{The overall VHE energy spectrum observed from PG\,1553+113. 
	The dashed line represents the best $\chi^2$ fit of a power law to
        the observed data.  The upper limits are at the 99\% confidence
	level \cite{UL_tech}. Only the statistical errors are shown.}
         \label{spectrum_plot}
   \end{figure}

The photon spectrum for the entire data set is shown 
in Figure~\ref{spectrum_plot}. These data are well fit
($\chi^2$ of 8.4 for 5 degrees of freedom) by
a power-law function, d$N$/d$E \sim E^{-\Gamma}$,
with a photon index  
$\Gamma=4.5\pm0.3_{\rm stat}\pm0.1_{\rm syst}$.
Fits of either a power law with an exponential cut-off
or a broken power law do not yield significantly
better $\chi^2$ values.  

\begin{figure}
   \centering
      \includegraphics[width=0.45\textwidth]{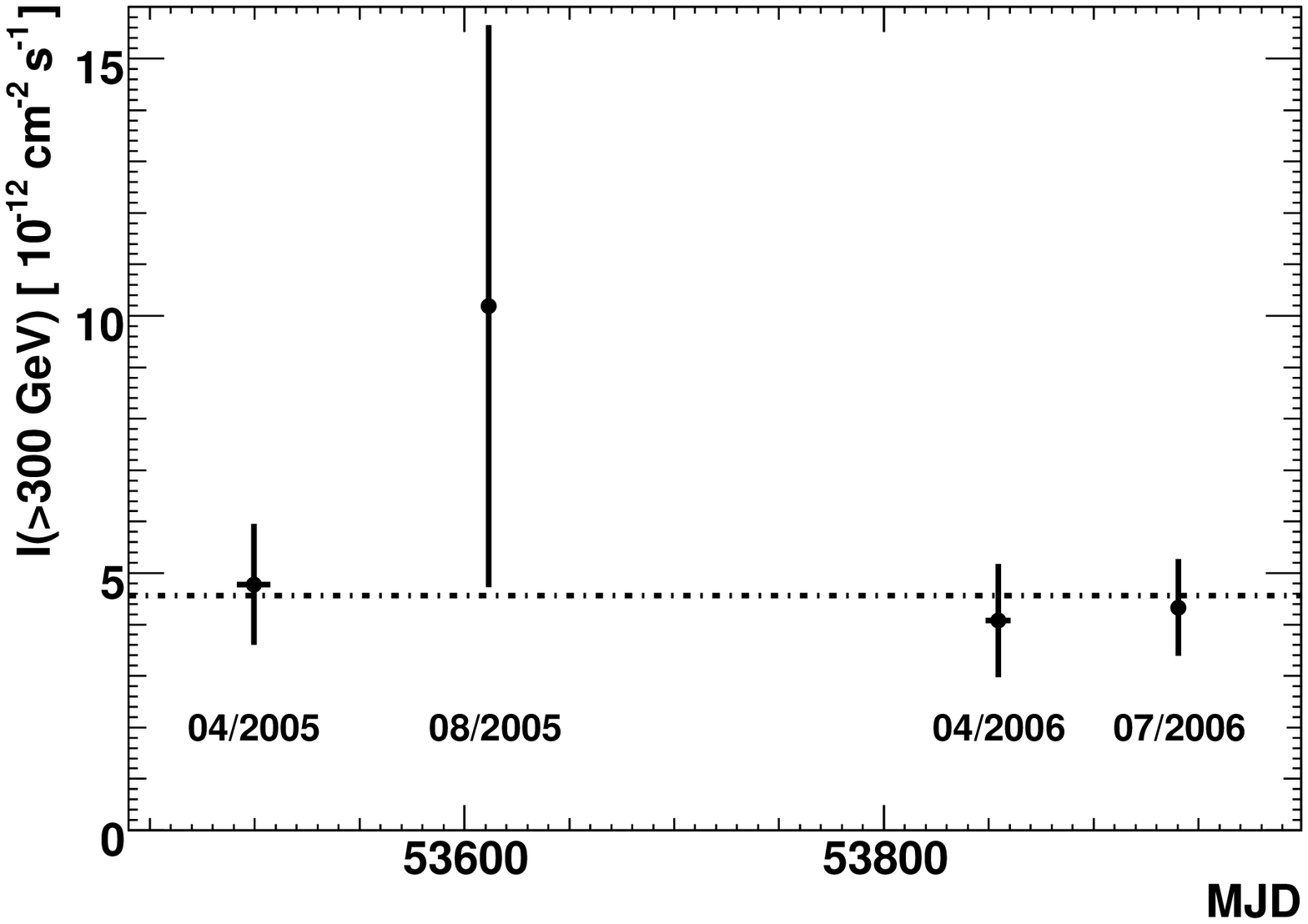}
      \caption{The integral flux, I($>$300 GeV), measured by HESS
from PG\,1553+113 during each dark period of observations.
The horizontal line represents the average flux for all 
the HESS observations. Only the statistical errors are shown.}
         \label{monthly_plot}
   \end{figure}

The observed integral flux above 300 GeV for the entire data set is
I($>$300 GeV) = $(4.6\pm0.6_{\rm stat}\pm0.9_{\rm syst}) \times 10^{-12}$ 
cm$^{-2}$\,s$^{-1}$.  This corresponds to $\sim$3.4\% of I($>$300 GeV)
determined from the HESS Crab Nebula spectrum \cite{HESS_crab}.
The integral flux, I($>$300 GeV), is shown in Table~\ref{results}
for each year of observations.   Figure \ref{monthly_plot} shows the flux 
measured for each dark period. There are no indications for flux variability 
on any time scale within the HESS data.
The data previously published \cite{HESS_discovery}
for HESS observations of PG\,1553+113 in 2005 were
not corrected for long-term changes in the optical sensitivity
of the instrument.  Relative to a virgin telescope, the total
optical throughput was decreased by 29\% in 2005 and 33\%
in 2006.  Correcting \cite{HESS_crab} for this decrease, 
using efficiencies determined from simulated and
observed muon events, increases the flux measured from the object.
The effect of this correction is larger for soft spectrum sources
than it is for hard spectrum sources. Due to the correction, 
the flux measured in 2005 is three times
higher than previously published.

On July 24 and 25, 2006, PG\,1553+113 
was observed by the Suzaku X-ray satellite (\cite{Suzaku_info},
Suzaku Observation Log: http://www.astro.isas.ac.jp/suzaku/index.html.en).
On these two dates HESS observed PG\,1553+113 for 3.1 hours live time,
resulting in a marginally significant excess of 101 events (3.9$\sigma$). 
The average flux measured on these two nights is I($>$300 GeV) = 
$(5.8\pm1.7_{\rm stat}\pm1.2_{\rm syst}) \times 10^{-12}$ 
cm$^{-2}$\,s$^{-1}$.  

\section*{SINFONI Near-IR Spectroscopy}

The determination of the redshift of an AGN is
generally based upon the detection of emission or
absorption lines in its spectrum.
In an attempt to detect absorption features from the host galaxy or
emission lines from the AGN,  H+K (1.50--2.40$\mu$m) spectroscopy 
of PG\,1553+113 was performed with SINFONI, an integral field 
spectrometer mounted at Yepun, Unit Telescope 4 of the 
ESO Very Large Telescope in Chile. The source was observed
on March 9, 2006 and March 15, 2006. The resulting images are spatially
unresolved and no underlying host galaxy is detected.
 
   \begin{figure}
   \centering
      \includegraphics[width=0.3\textwidth,angle=270]{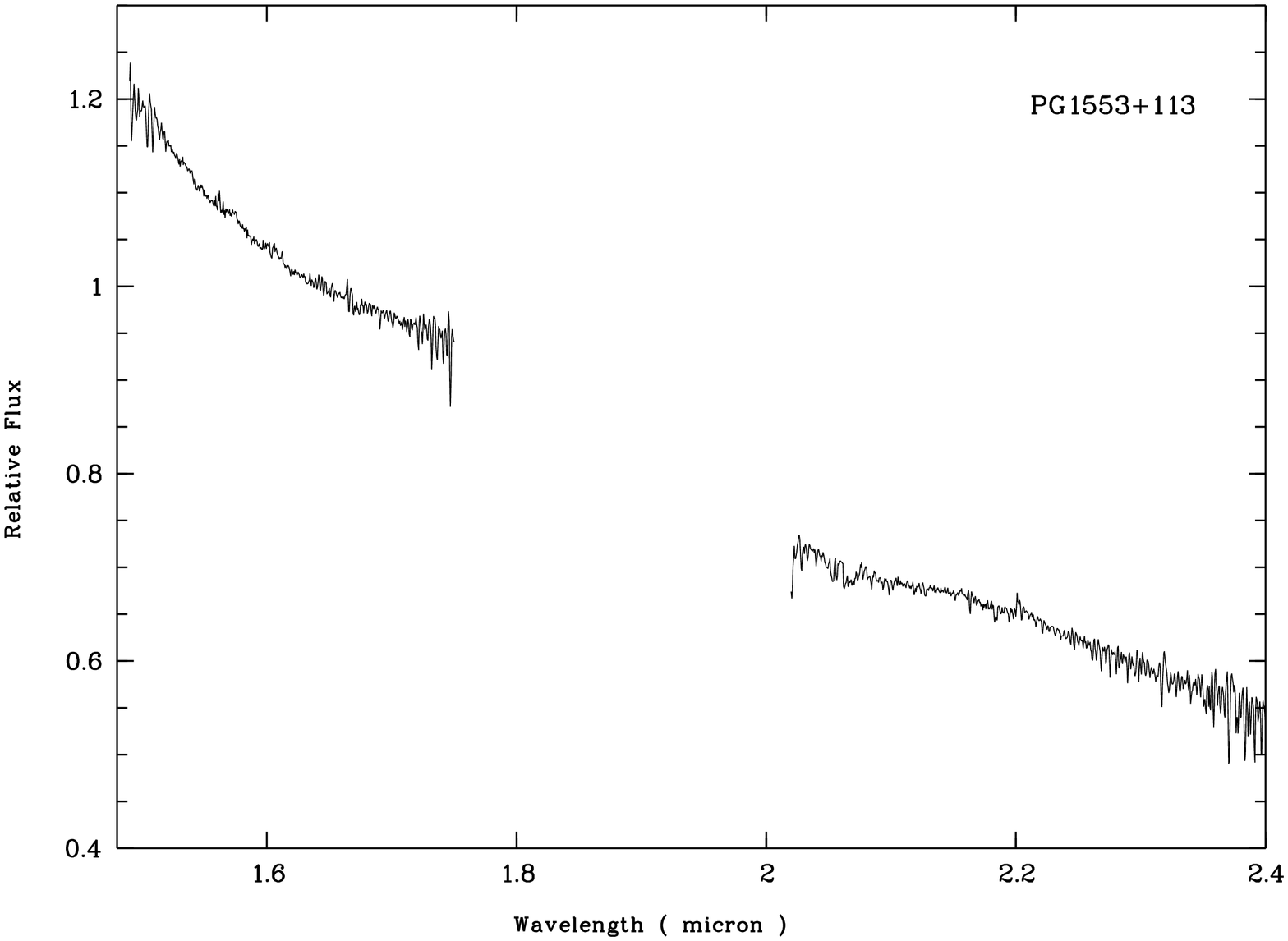}
      \caption{The measured H+K-band spectrum of PG\,1553+113
in relative flux units. The gap is due to the highly 
reduced atmospheric transmission between H and K bands. }
         \label{IR_spectrum}
   \end{figure}

The H+K-band spectrum of PG\,1553+113 measured here is shown in
Figure~\ref{IR_spectrum}. The signal-to-noise reach is $\sim$250 in
the H-band and $\sim$70 in the K-band.  The observed near-IR
spectrum is featureless apart from some residuals from the
atmospheric corrections. Thus in neither the broad-band images nor
in the spectrum are the influences of the gas of a host galaxy or
the AGN detected, even though PG\,1553+113 is bright in the IR.  As
a result, a redshift determination from these observations is not
possible.

\section*{Discussion}

\begin{figure}
   \centering
      \includegraphics[width=0.45\textwidth]{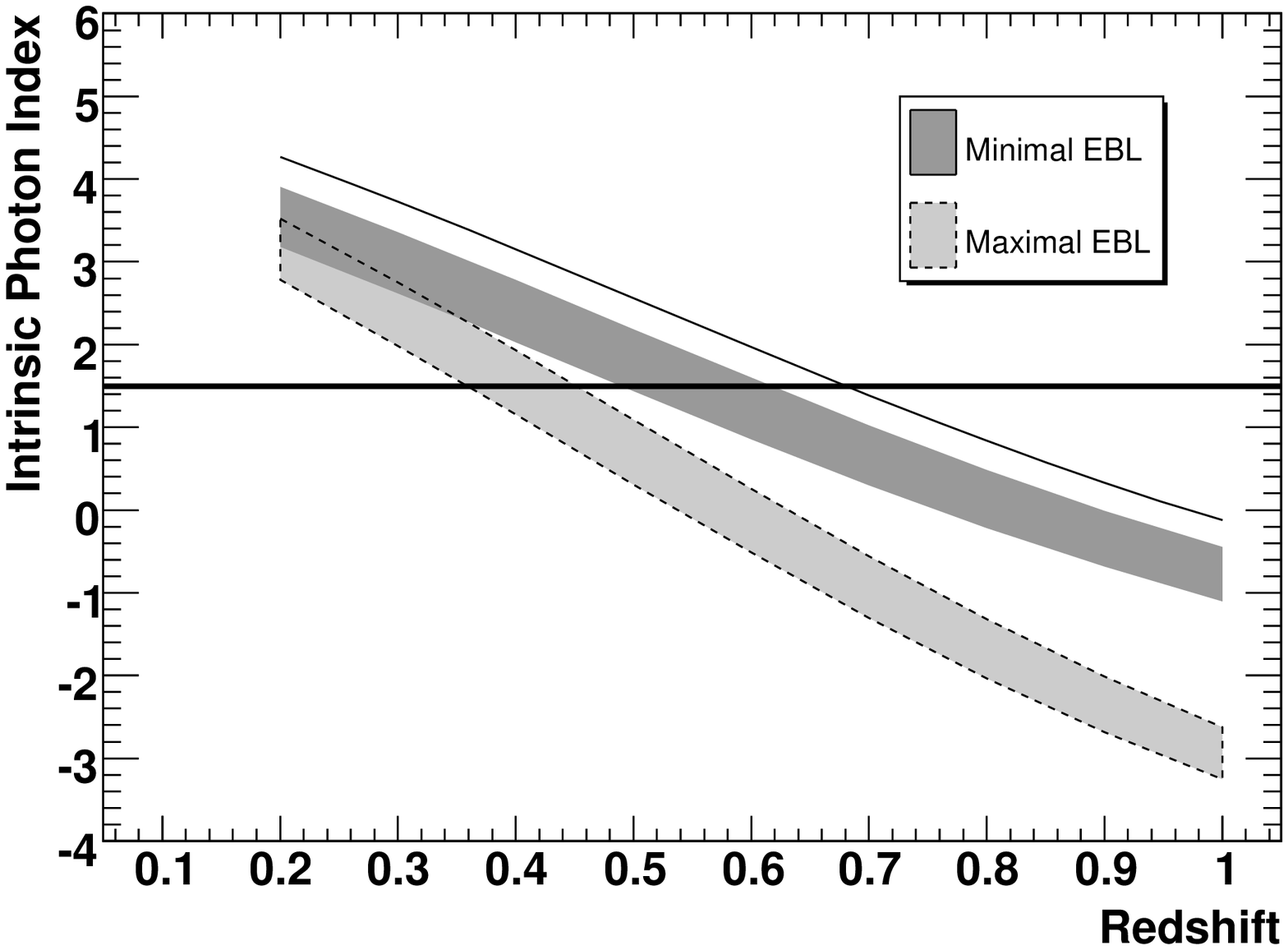}
      \caption{
The photon index $\Gamma_{\rm int}$ determined from a power-law fit to 
the intrinsic spectrum of PG 1553+113 (i.e. the H.E.S.S. data de-absorbed with an EBL model) 
for a range of redshifts. The contours reflect the 1$\sigma$ statistical uncertainty 
of the fits. The upper curve is the sum of $\Gamma_{\rm int}$ for the ¡ÈMinimal'' 
model and twice its statistical error.}
         \label{int_spec}
   \end{figure}

As the redshift of PG 1553+113 is likely $z>0.2$,
the observed VHE spectrum should be strongly affected by VHE $\gamma$-ray 
absorption on the EBL. If the redshift were known the spectrum intrinsic
to the source could be reconstructed assuming a model of the EBL density. 
However, the EBL SED is not well-determined. Using a {\it Maximal} 
EBL model at the level of the upper limits from \cite{Nature_EBL} 
or a {\it Minimal} model near the EBL lower limits 
from galaxy counts \cite{galaxy_counts} can yield a significantly different intrinsic spectrum.  
Figure~\ref{int_spec} shows the intrinsic spectrum versus redshift
for both the {\it Maximal} and {\it Minimal} EBL parameterizations.
Here, scaled models of \cite{Primack_EBL} are used, exactly as described 
in \cite{HESS_discovery}.  The redshift of the AGN can be limited 
using assumptions for the intrinsic spectrum. 
Assuming the intrinsic photon index is not harder than $\Gamma_{\rm int}=1.5$, 
a limit of $z<0.69$ is thus determined from the {\it Minimal} EBL model.

\section*{Conclusion}

With a $\sim$25 h data set, $\sim$3 times larger than previously
published \cite{HESS_discovery}, the HESS signal from 
PG\,1553+113 is now highly 
significant ($\sim$10$\sigma$). Thus the evidence 
for VHE emission from PG\,1553+113 
previously reported is now clearly verified.
However, the flux observed in 2005 is now $\sim$3 times higher 
than initially reported due to an improved calibration 
of the absolute energy scale of HESS, and agrees well with
the flux measured in 2005 by MAGIC \cite{MAGIC_1553}. 
The statistical error on the VHE photon index is still rather large
($\sim$0.3), primarily due to the extreme softness of the 
observed spectrum ($\Gamma=4.5$).  
The total HESS exposure on PG\,1553+113 is $\sim$25 hours.  Barring
a flaring episode, not yet seen in two years of observations,
a considerably larger total exposure ($\sim$100 hours) would be
required to significantly improve the spectral measurement.
However, the VHE flux from other AGN is known to vary dramatically
and even a factor of a few would reduce the observation
requirement considerably. Should such a VHE flare occur, 
not only will the error on the measured VHE spectrum be 
smaller, but the measured photon index may 
also be harder (see, e.g., \cite{VHE_hardening}).  Both effects would 
dramatically improve the redshift constraints and correspondingly
the accuracy of the source modeling. Therefore, the VHE flux
from PG\,1553$+$113 will continue to be monitored by HESS.
In addition, the soft VHE spectrum makes it an ideal target for
the lower-threshold HESS Phase-II \cite{HESSII} which should
make its first observations in 2009.

\section*{Acknowledgements}
The support of the Namibian authorities and of the University of Namibia
in facilitating the construction and operation of H.E.S.S. is gratefully
acknowledged, as is the support by the German Ministry for Education and
Research (BMBF), the Max Planck Society, the French Ministry for Research,
the CNRS-IN2P3 and the Astroparticle Interdisciplinary Programme of the
CNRS, the U.K. Science and Technology Facilities Council (STFC),
the IPNP of the Charles University, the Polish Ministry of Science and 
Higher Education, the South African Department of
Science and Technology and National Research Foundation, and by the
University of Namibia. We appreciate the excellent work of the technical
support staff in Berlin, Durham, Hamburg, Heidelberg, Palaiseau, Paris,
Saclay, and in Namibia in the construction and operation of the
equipment. Based on ESO-VLT SINFONI program 276.B-5036 observations.

%This in the bibtex style, is ok.
%\bibliographystyle{plain}

%%%%%%%%
%  33  %
%%%%%%%%

%The paper title
\title{Observations of 1ES\,1101$-$232 with H.E.S.S.\ and at lower frequencies:  
A hard spectrum blazar and constraints on the extragalactic background light
}
%Short title to print in the headers to the final publication (Not showed in this print).
\shorttitle{Observations of 1ES\,1101$-$232 with HESS}
%All paper authors
\authors{Gerd P{\"u}hlhofer$^{1}$, Wystan Benbow$^{2}$, Luigi Costamante$^{2}$, Helene Sol$^{3}$,
Catherine Boisson$^{3}$, Dimitrios Emmanoulopoulos$^{1}$, Stefan Wagner$^{1}$,
Dieter Horns$^{4}$, Berrie Giebels$^{5}$, for the H.E.S.S. collaboration.
}
%Short title to print in the headers to the final puplication (Not showed in this print).
\shortauthors{Gerd P{\"u}hlhofer et al., for the H.E.S.S. collaboration}
%All the affiliations.
\afiliations{
$^1$ Landessternwarte, Universit\"at Heidelberg, K\"onigstuhl, D 69117 Heidelberg, Germany\\
$^2$ Max-Planck-Institut f\"ur Kernphysik, P.O. Box 103980, D 69029 Heidelberg, Germany\\
$^3$ LUTH, UMR 8102 du CNRS, Observatoire de Paris, Section de Meudon, F-92195 Meudon Cedex, France\\
$^4$ Institut f\"ur Astronomie und Astrophysik, Universit\"at T\"ubingen, Sand 1, D 72076 T\"ubingen, Germany\\
$^5$ Laboratoire Leprince-Ringuet, IN2P3/CNRS, Ecole Polytechnique, F-91128 Palaiseau, France\\
}
\email{G.Puehlhofer@lsw.uni-heidelberg.de}
%The abstract.
\abstract{
VHE observations of the distant ($z$$=$0.186) blazar 1ES\,1101$-$232 with H.E.S.S. are used to constrain the
extragalactic background light (EBL) in the optical to near infrared band. As the EBL traces the galaxy formation
history of the universe, galaxy evolution models can be tested with the data. In order to measure the
EBL absorption effect on a blazar spectrum, we assume that usual constraints on the hardness of the intrinsic
blazar spectrum are not violated. We present an update of the VHE spectrum obtained with H.E.S.S. and the
multifrequency data that were taken simultaneously with the H.E.S.S. measurements. The data verify that the
broadband characteristics of 1ES\,1101-232 are similar to those of other, more nearby blazars, and strengthen the
assumptions that were used to derive the EBL upper limit.
%VHE observations of the distant ($z=0.186$) blazar 1ES\,1101$-$232 with H.E.S.S. can be used to constrain the
%extragalactic background light (EBL) in the optical to near infrared band. As the EBL traces the galaxy formation
%history of the universe, galaxy evolution models can therefore be tested with the data. In order to measure the
%EBL absorption effect on a blazar spectrum, we assume that usual constraints on the hardness of the intrinsic
%blazar spectrum are not violated. We present an update of the VHE spectrum obtained with HESS and the
%multifrequency data that were taken simultaneously with the HESS measurements. The data verify that the broadband
%characteristics of 1ES 1101-232 are similar to those of other, more nearby blazars, and strengthen the
%assumptions that were used to derive the EBL upper limit.
}

\maketitle

\addcontentsline{toc}{section}{Observations of 1ES\,1101$-$232 with H.E.S.S.\ and at lower frequencies: A hard spectrum blazar and constraints on the extragalactic background light}
\setcounter{figure}{0}
\setcounter{table}{0}
\setcounter{equation}{0}

%Begin the section.
\section*{Introduction} 

The detection of VHE emission from 1ES\,1101$-$232 with the H.E.S.S. Cherenkov telescopes has attracted particular attention for two reasons:
The object has been the farthest detected {\em VHE blazar} with confirmed redshift ($z$$=$0.186), and at the same time it has exhibited
% -- for a VHE blazar -- 
a hard spectrum, with a photon index of $\Gamma \approx 2.9$ between 0.2 and 4\,TeV. Both facts taken together
allowed to 
%draw a far-reaching conclusion for 
place a limit on
the density of the extragalactic background light (EBL) in the near-infrared band: under
the assumption of a normal-behaved intrinsic emission spectrum of 1ES\,1101$-$232, 
%it was shown that 
the energy density of the EBL at 
$1.5\,\mathrm{\mu m}$ has to be at or below $\nu F_{\nu}=14\,\mathrm{nWm^{-2}sr^{-1}}$ \cite{aharonian2006ebl}.

Here we present an update of the VHE spectrum and broadband data that were taken simultaneously to the H.E.S.S. data. We show that
the hard $\gamma$-ray spectrum of 1ES\,1101$-$232 is close to the borderline of what is possible to model using 
standard one-zone blazar emission scenarios. On the other hand, from the broadband data 
there is no evidence
% to claim 
that the emission characteristics of 1ES\,1101$-$232
are not in agreement with those of the remaining class of VHE blazars. A detailed description of the results 
can also be found in \cite{aharonian20071101}.

\section*{H.E.S.S.\ data and EBL limit}
\label{S:icrc0555_data}

\begin{figure*}
\begin{center}
%\vspace{-1cm}
\includegraphics [width=0.95\textwidth]{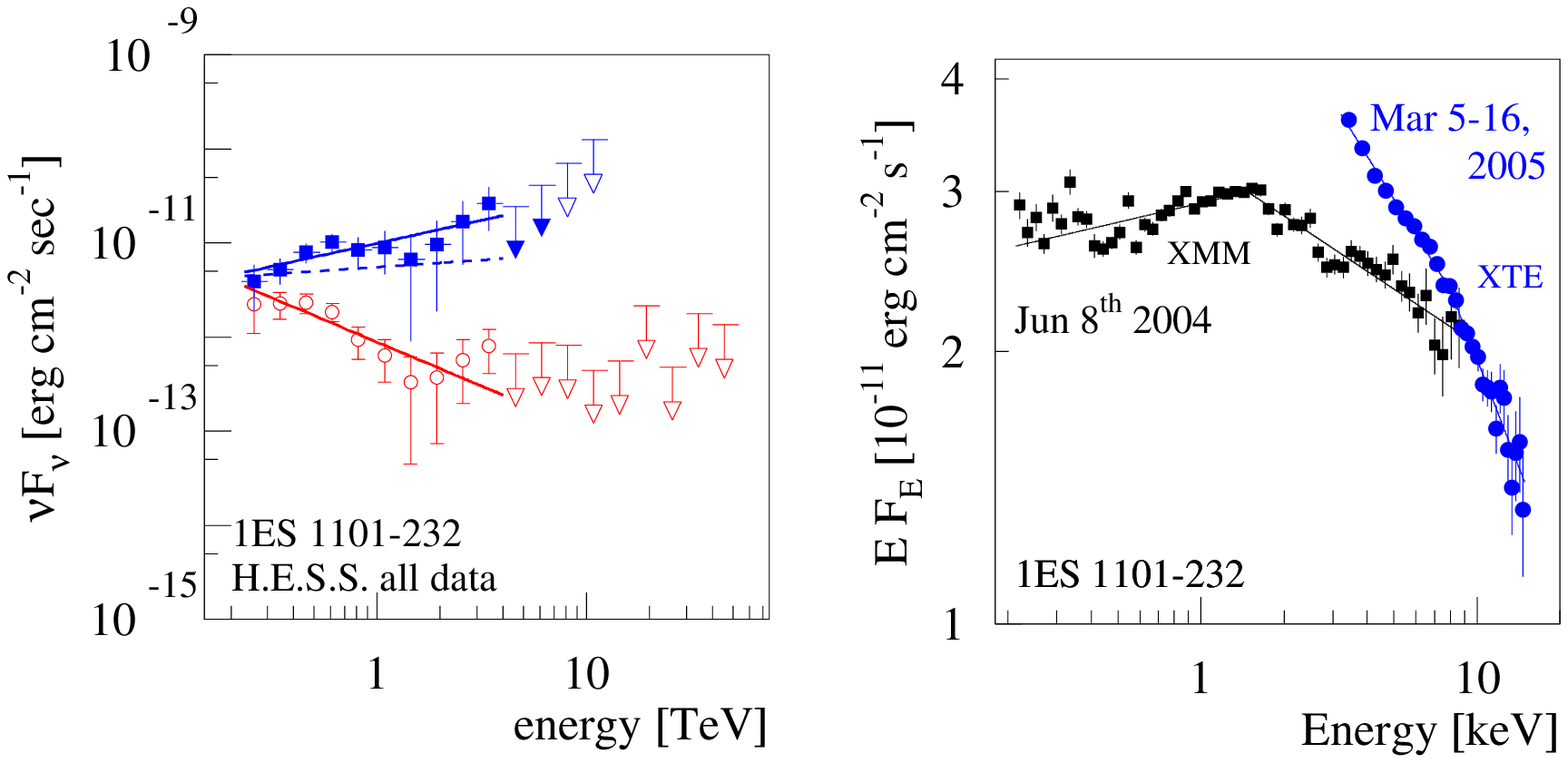}
%\vspace{-2cm}
\end{center}
\caption{{\bf Left panel:} 
VHE $\gamma$-ray spectrum of 1ES\,1101$-$232, from the total H.E.S.S data set of the years 2004 and 2005, in $\nu F_{\nu}$-representation.
%Spectral energy distribution from the same data.
The red, open circles denote the reconstructed flux as measured with H.E.S.S.
The spectrum after correction for {\em maximum} EBL absorption
%using an EBL model 
with $14\,\mathrm{nWm^{-2}sr^{-1}}$ at $1.5\,\mathrm{\mu m}$ 
is shown with
blue, filled circles. Upper limits in the deabsorbed spectrum above 7\,TeV are shown as open symbols only, 
because of strong EBL uncertainties at these high energies.
The solid lines denote power-law fits between 0.2 and 4\,TeV to the measured and deabsorbed spectra. The dashed line indicates the effect if
the EBL level used for deabsorption would be lowered to $10\,\mathrm{nWm^{-2}sr^{-1}}$ at $1.5\,\mathrm{\mu m}$.
%Extrapolations of the power laws to higher energies are shown as dotted lines.
{\bf Right panel:} 
X-ray spectra from observations taken simultaneously with the H.E.S.S. June 2004 (XMM-Newton) and March 2005 (RXTE) data.
}
\label{icrc0555_fig01}
%\vspace{-5cm}
\end{figure*}

\begin{table*} 
  \begin{center}
%\vspace{-4cm}
  \begin{tabular}{l|r||c||c|c} 
    \hline
                                                           & livetime & $\Gamma_{\mathrm{abs}}$  & $\Gamma_{\mathrm{deabs,max}}$       & $\Gamma_{\mathrm{deabs,min}}$ \\ 
    \hline
      EBL shape used for deabsorption                      &          &                          & $P0.45$                         & $P0.34$                         \\
      present day EBL shape                                &          &                          & $P0.55$                         & $P0.40$                         \\
      present day $\nu F\nu(1.5\,\mu\mathrm{m})$ &         &          & $14\,\mathrm{nWm^{-2}sr^{-1}}$  & $10\,\mathrm{nWm^{-2}sr^{-1}}$  \\
      fit energy range                                     &          & $0.23\textnormal{--}4.0\,\mathrm{TeV}$  & $0.23\textnormal{--}4.0\,\mathrm{TeV}$ & $0.23\textnormal{--}4.0\,\mathrm{TeV}$ \\
%      fit energy range                                    &          & \tiny{$0.23\textnormal{--}4.0\,\mathrm{TeV}$}  & \tiny{$0.23\textnormal{--}4.0\,\mathrm{TeV}$} & \tiny{$0.23\textnormal{--}4.0\,\mathrm{TeV}$} \\
    \hline
      All Data   & 42.7\,hrs & $ 2.94^{+0.20}_{-0.21}$  & $ 1.51^{+0.17}_{-0.19}$ & $ 1.85^{+0.18}_{-0.19}$ \\
      March 2005 & 31.6\,hrs & $ 2.94^{+0.21}_{-0.23}$  & $ 1.49^{+0.19}_{-0.20}$ & $ 1.84^{+0.20}_{-0.21}$ \\
      June 2004  & 8.4\,hrs  & $ 3.16^{+0.48}_{-0.61}$  & $ 1.70^{+0.47}_{-0.61}$ & $ 2.05^{+0.47}_{-0.61}$ \\
    \hline
  \end{tabular}
  \end{center}
\caption{
Photon indices from power-law fits to the VHE spectra of 1ES\,1101$-$232.
$\Gamma_{\mathrm{abs}}$ are from fits to the measured spectra. 
$\Gamma_{\mathrm{deabs,max}}$ are from fits to the deabsorbed spectra using the EBL shape $P0.45$, corresponding to the
{\em maximum} EBL level with the present day shape $P0.55$ ($\nu F_{\nu}(1.5\,\mathrm{\mu m}) = 14\,\mathrm{nWm^{-2}sr^{-1}}$) 
%$P0.45$ corresponds to the present day EBL level of $P0.55$, 
after scaling down by 15\% to take galaxy evolution effects into account.
$\Gamma_{\mathrm{deabs,min}}$ represents the result if the EBL level is lowered to 
$\nu F_{\nu}(1.5\,\mathrm{\mu m}) = 10\,\mathrm{nWm^{-2}sr^{-1}}$, at the level of the EBL lower limit from galaxy counts.
% as described in Section \ref{xxx}. 
$\Gamma_{\mathrm{abs}}$ and $\Gamma_{\mathrm{deabs,max}}$ correspond to the fits shown as solid lines in Fig.\,\ref{icrc0555_fig01},
the fit corresponding to $\Gamma_{\mathrm{deabs,min}}$ is shown as dashed line.
}
  \label{icrc0555_tab01}
\end{table*}

Fig.\,\ref{icrc0555_fig01} shows the spectrum of 1ES\,1101$-$232, derived from the total H.E.S.S. data set of the years 2004 and 2005. 
Compared to the analysis results used in \cite{aharonian2006ebl}, an improved energy calibration of the telescope system
was applied to the data
%, better taking into account the long-term optical sensitivity changes of the instrument
\cite{aharonian2006crab}.
For the given total data sample, 
this
energy scale recalibration yields a safe energy threshold of 225\,GeV 
(compared to 165\,GeV used in \cite{aharonian2006ebl})
and a flux normalisation increase of 27\% at 1\,TeV. 
After this correction, the systematic flux uncertainty is now estimated as 
20\% \cite{aharonian2006crab}. Reconstructed spectral indices were not affected significantly by these calibration updates,
the systematic error estimate for reconstructed photon indices is $\Delta \Gamma_{\mathrm{sys}} \sim 0.1$
\cite{aharonian2006ebl,aharonian2006crab}.
Lacking the intrinsic blazar spectrum,
% and -- to sufficient accuracy -- the absorbed spectrum at Earth,
one can
assume a power-law type intrinsic spectrum with $\Gamma_{\mathrm{deabs}}$ and 
%For 
a typical EBL model around $1.5\,\mu\mathrm{m}$.
Then EBL absorption simply leads to a softening of the VHE spectrum above 
$\sim 100\,\mathrm{GeV}$
of $\Delta \Gamma = \Gamma_{\mathrm{abs}} - \Gamma_{\mathrm{deabs}}$, 
where $\Gamma_{\mathrm{abs}}$ is the photon index of the measured 
%and  $\Gamma_{\mathrm{deabs}}$ that of the intrinsic blazar 
spectrum.
%$\Gamma_{\mathrm{deabs}} = \Gamma_{\mathrm{abs}} - \Delta\Gamma$, see \citeauthor*{aharonian2006ebl}; 
Using a template EBL spectrum ``$P$'' 
%adopted to the EBL model by \cite{primack2000ebl} 
and scaling it with a factor $p$ results in the
%simple 
relation 
$\Delta p = 0.34 \Delta \Gamma$ \cite{aharonian2006ebl}.
%within the range of scaling factors considered.

The EBL has a {\em lower limit} from galaxy counts \cite{madau2000galaxycounts}, with
about $10\,\mathrm{nWm^{-2}sr^{-1}}$ at $1.5\,\mathrm{\mu m}$, corresponding to $P0.40$ using the scaled EBL scheme.
Already with this smallest possible deabsorption, the VHE spectrum of 1ES\,1101$-$232 is harder than $\Gamma_{\mathrm{deabs}} = 2$
(see dashed line in Fig.\,\ref{icrc0555_fig01}), i.e.\ the 
intrinsic VHE power output peak of the source is above $\sim 3\,\mathrm{TeV}$ for any EBL level.

The intrinsic spectrum of VHE blazars is expected to be not harder than 1.5, i.e.\ $\Gamma_{\mathrm{deabs}} \ge 1.5$, 
taking the present observational and theoretical know\-ledge of VHE blazar spectra into account \cite{aharonian2006ebl}.
This translates into an {\em upper limit} of the 
%present day 
EBL of $P0.55$.
% using the scaled EBL scheme.
This {\em maximum} EBL has $14\,\mathrm{nWm^{-2}sr^{-1}}$ at $1.5\,\mathrm{\mu m}$.
This 
%value 
limit
is identical (within 1\%) using either the
spectrum used in \cite{aharonian2006ebl}, with $\Gamma_{\mathrm{abs}}=2.88$ between 0.16 - 3.3\,TeV,
or the recalibrated spectrum with $\Gamma_{\mathrm{abs}}=2.94$ between 0.23 - 4.0\,TeV.
We note that the limit at $1.5\,\mathrm{\mu m}$ is quite insensitive to the choice of the EBL parametrisation, see
\cite{aharonian2006ebl}, \cite{mazin2007eblblazars}.

This EBL upper limit is in conflict with models such as the
``fast evolution'' model by \cite{stecker2006ebl} and the ``best fit'' model by \cite{kneiske2004ebl}, with an EBL density of about
$\nu F_{\nu}(1.5\,\mathrm{\mu m}) \simeq 20\,\mathrm{nWm^{-2}sr^{-1}}$.
%As shown in \citet{stecker2006deltagamma}, the ``fast evolution'' EBL 
Such high EBL levels would lead to an intrinsic spectrum of 1ES\,1101$-$232 with $\Gamma_{\mathrm{deabs}} \sim 1$.
%but 
%such a spectrum 
Such a $\Gamma$
would isolate 1ES\,1101$-$232 in the class of VHE blazars.

\section*{Broadband data and SED}

\begin{figure*}[th]
\begin{center}
\includegraphics [width=0.8\textwidth]{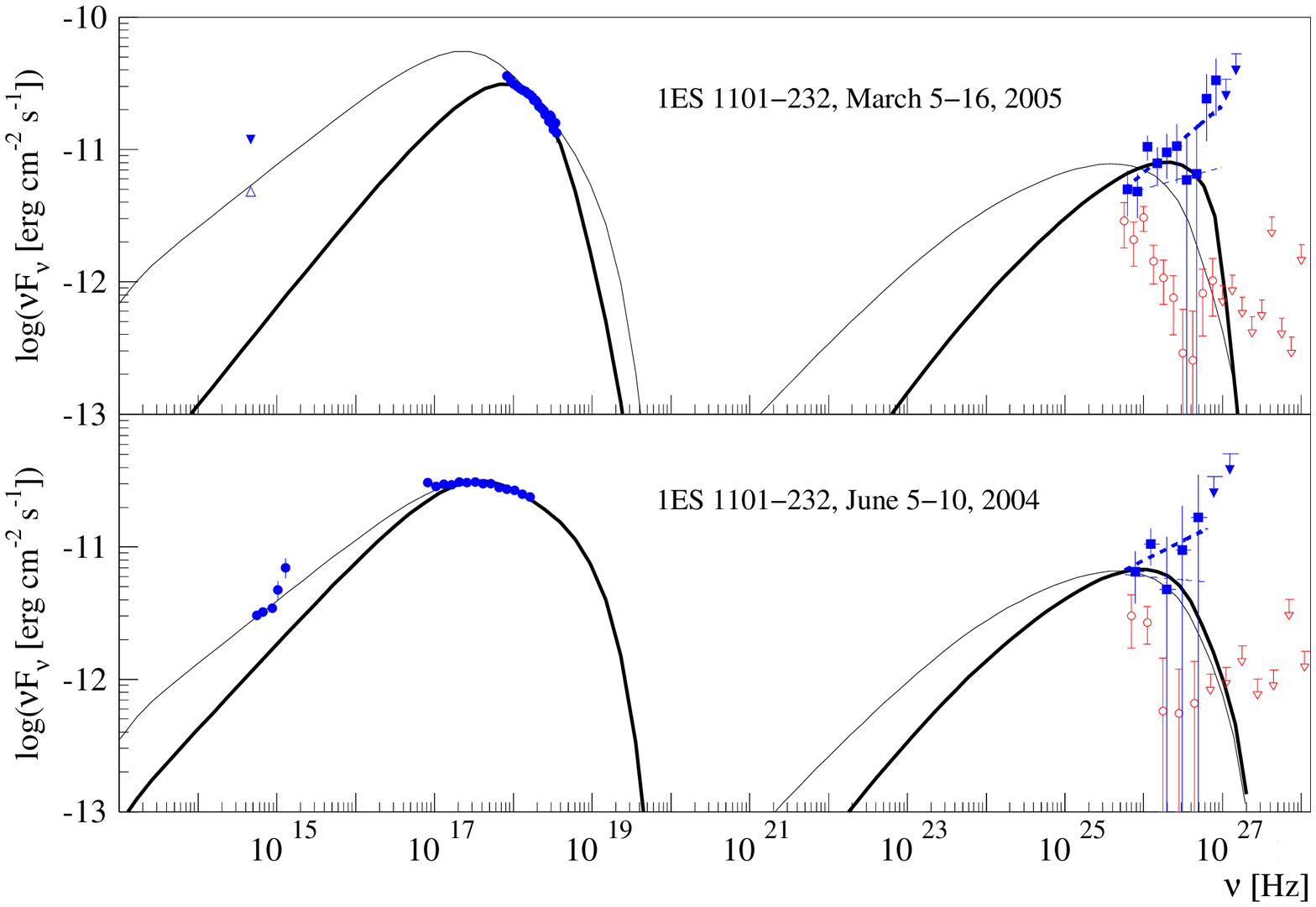}
\vspace{-0.5cm}
\end{center}
\caption{
Spectral energy distribution of 1ES\,1101$-$232. {\bf Upper panel:} Data from March 5-16, 2005. 
In the VHE band, the measured H.E.S.S. spectrum (red, open symbols) and the deabsorbed spectrum
using a maximum EBL level of $14\,\mathrm{nWm^{-2}sr^{-1}}$ at $1.5\,\mathrm{\mu m}$
% (see text) 
are shown.
%for better visibility, all 
%open symbols were slightly shifted to the left.
%, to 90\% of the respective original frequency.
The thick dashed line is a power-law fit to the deabsorbed data as plotted,
while 
the thin dashed 
line indicates the effect if the EBL is lowered to the minimum level of $10\,\mathrm{nWm^{-2}sr^{-1}}$.
X-ray data are from RXTE. In the optical band, an upper limit (filled triangle) and a tentative lower limit (open triangle)
from ROTSE 3c data are shown.
%, see Section \ref{SSS:rotse} for details. 
%The latter value corresponds to the minimum level of the EBL as derived from known resolved galaxy counts.
%Additionally, the results from power-law fits to 
%H.E.S.S. data are shown, after deabsorption of the data using 
%the maximum (thick line) and a minimum (thin line) EBL level.
% Thick and thin solid curves
% denote results from a single zone SSC model. The thick curves represent a model that was optimized to fit the
% H.E.S.S. and X-ray data, while the thin lines denote a model with an electron distribution $N_{\gamma} \propto \gamma^{-2}$ below the break.
{\bf Lower panel:} Data from June 2004. In the VHE band, 
H.E.S.S. data taken between June 5-10, 2004, are shown. 
%using the same procedure as described for the upper panel.
X-ray and optical data were derived from an XMM-Newton
pointing on June 8, 2004. 
% Solid curves denote results from a single zone emission model, also following the same prescription as for the upper panel.
}
\label{icrc0555_fig02}
\end{figure*}

Here we present truly simultaneous SEDs of 1ES\,1101$-$232. 
The data show no indication for 
%such
% effects
a 
%very hard 
$\Gamma$$\sim$$1$ type spectrum
in the synchrotron branch. 
%and 
%can 
%%in principle
%be understood in standard 
%%one-zone 
%emission models.
%
2004 H.E.S.S. observations were made in April and June (4 nights each). 
On June 8,
%$^{\mathrm{th}}$
2004, XMM-Newton X-ray and optical monitor 
%(OM)
data were obtained.
In March 2005, H.E.S.S., RXTE X-ray, and ROTSE 3c optical data were taken simultaneously during 11 nights.
% in a coordinated campaign.
The H.E.S.S. data did not reveal variability on any time scale. Also, the XMM-Newton data showed a constant flux.
The nightly averaged light curve of the RXTE data showed mild variations of 15\% (min-max), optical variations were below 10\%.
We note that the two X-ray spectra obtained in the two different years are quite different, see right panel of Fig.\,\ref{icrc0555_fig01}.

We therefore constructed two simultaneous SEDs, one for March 2005
% with 31.6\,hrs H.E.S.S. data, 
and one for June 2004, 
%comprising 8.4\,hrs of H.E.S.S. data from the June 2004 observations, 
see Fig.\,\ref{icrc0555_fig02}. 
The SED was modeled using 
a time-independent SSC model \cite{katarzynski2001ssc},
with a one-zone homogeneous, spherical emitting region 
%$R$ 
and a homogeneous magnetic field which propagates towards the observer. 
The high-energy electron distribution was modeled with a broken power law,
with particle energy index $p_{1}$ between Lorentz factors $\gamma_{\mathrm{min}}$ and $\gamma_{\mathrm{b}}$,
and $p_{2}$ between $\gamma_{\mathrm{b}}$ and $\gamma_{\mathrm{max}}$.
Thin lines correspond to models with $p_{1} = 2$ as expected from an uncooled, shock-accelerated particle distribution.
Thick lines are models with $p_{1} \simeq 1.5$, for instance from particle acceleration at strong shocks in a relativistic gas.

The $p_{1}$$=$$2$ models nicely reproduce the X-ray and optical data, but fail to fit the 2005 H.E.S.S. data. 
The $p_{1}$$=$$1.5$ model can marginally fit the 2005 H.E.S.S. data. 
An improved fit could be obtained if the EBL was below $\nu F_{\nu}=14\,\mathrm{nWm^{-2}sr^{-1}}$. 
Adopting the $p_{1}$$\simeq$$1.5$ models, one has to attribute the optical emission to a different emission zone 
(not modeled in Fig.\,\ref{icrc0555_fig02}), which
is viable because of the lack of correlated variability arguments. 

\section*{Conclusions}

The VHE SED of 1ES\,1101$-$232 peaks above 3\,TeV. 
%Despite this, 
Nevertheless,
a standard emission scenario can be used to essentially explain the broadband data of 1ES\,1101$-$232,
% can be understood in 
if the EBL used to
deabsorb the VHE data is at or below
$\nu F_{\nu}(1.5\,\mathrm{\mu m})=14\,\mathrm{nWm^{-2}sr^{-1}}$.
%$\nu F_{\nu}=14\,\mathrm{nWm^{-2}sr^{-1}}$. 
H.E.S.S. observations of other distant VHE blazars (H\,2356$-$309, 1ES\,0347$-$121) confirm the low EBL level as deduced from the 1ES\,1101$-$232
H.E.S.S. spectrum. The general ``hardness'' of the spectra of distant VHE blazars might be explained through 
the better VHE detectability of hard-spectrum sources.

\section*{Acknowledgements}
%\vspace{-0.5cm}
{\small
The support of the Namibian authorities and of the University of Namibia
in facilitating the construction and operation of H.E.S.S. is gratefully
acknowledged, as is the support by the German Ministry for Education and
Research (BMBF), the Max Planck Society, the French Ministry for Research,
the CNRS-IN2P3 and the Astroparticle Interdisciplinary Programme of the
CNRS, the U.K. Particle Physics and Astronomy Research Council (PPARC),
the IPNP of the Charles University, the South African Department of
Science and Technology and National Research Foundation, and by the
University of Namibia. We appreciate the excellent work of the technical
support staff in Berlin, Durham, Hamburg, Heidelberg, Palaiseau, Paris,
Saclay, and in Namibia in the construction and operation of the
equipment.
We thank the ROTSE collaboration for providing the ROTSE 3c optical data,
and L. Ostorero for help with the optical data analysis. 
}

\bibliographystyle{plain}

%%%%%%%%
%  34  %
%%%%%%%%

%The paper title
\title{Discovery of Two New TeV Blazars with the H.E.S.S. Cherenkov Telescope System}
%Short title to print in the headers to the final publication (Not showed in this print).
\shorttitle{Discovery of Two New TeV Blazars with the H.E.S.S. Cherenkov Telescope System}
%All paper authors
\authors{Martin Raue$^{1}$, Wystan Benbow$^{2}$, Luigi Costamante$^{2}$, and Dieter Horns$^{3}$\\ for the H.E.S.S. Collaboration}
%Short title to print in the headers to the final puplication (Not showed in this print).
\shortauthors{M. Raue for the H.E.S.S. Collaboration}
%All the affiliations.
\afiliations{$^1$Institut f\"ur Experimentalphysik, Universit\"at Hamburg, Luruper Chaussee 149, D-22761 Hamburg, $^2$MPI-K Heidelberg, P.O. Box 103980, D 69029 Heidelberg, Germany, $^3$Institut f. Astronomie und Astrophysik, Universit\"at T\"ubingen, Sand 1, D 72076 T\"ubingen, Germany}
\email{martin.raue@desy.de}

%The abstract.
\abstract{Since the new generation of imaging atmospheric-Cherenkov telescopes came online with the commissioning of the four telescopes of the H.E.S.S. experiment in 2004, the number of known extragalactic $\gamma$-ray emitters in the very high energy (VHE) domain has more than doubled. All of the sources detected so far are active galactic nuclei and all but one belong to the class of BL Lac objects. The emission process for VHE $\gamma$-rays in this class of objects is not fully understood and a large sample of sources and multi-wavelength data is needed to discriminate between different models. Furthermore, VHE photons from these distant sources are attenuated via pair production with the extragalactic photon field in the optical to infrared wavelength band (extragalactic background light, EBL), which contains cosmological information on the star and galaxy formation history. With assumptions about the source physics, limits on this photon field can be derived. We report the detection of VHE gamma-rays from the BL Lac 1ES 0229+200 (z = 0.14) and 1ES 0347-121 (z = 0.1880) with the H.E.S.S. Cherenkov telescope system. 1ES 0347-121 is among the most distant source detected in VHE gamma-rays to date.}

\maketitle

\addcontentsline{toc}{section}{Discovery of Two New TeV Blazars with the H.E.S.S. Cherenkov Telescope System}
\setcounter{figure}{0}
\setcounter{table}{0}
\setcounter{equation}{0}

%Begin the section.
%------------------------------------------------------------------------------------------------------------------------
% The H.E.S.S. Experiment
%------------------------------------------------------------------------------------------------------------------------

\section*{Introduction}

The H.E.S.S. collaboration operates an array of four large imaging atmospheric-Cherenkov telescopes (107\,m$^2$ mirror area per telescope; $\sim$5$^\circ$ field of view) located in the Southern hemisphere in Namibia (1800\,m a.s.l.) \cite{hinton04}. H.E.S.S. observes very high energy (VHE; $>$100\,GeV) $\gamma$-rays from many types of astrophysical objects. About one third of the H.E.S.S. observation time ($\sim$300 hours) is dedicated to study active galactic nuclei (AGN). In the following the detection of VHE $\gamma$-rays from two AGNs of the BL Lac class 1ES\,0347-121 and 1ES\,0229+200 is reported. Details on the discoveries and the implication for the extragalactic background light (EBL) are discussed in \cite{aharonian07} and \cite{aharonian07b}

%------------------------------------------------------------------------------------------------------------------------
% Methods
%------------------------------------------------------------------------------------------------------------------------

\section*{Methods}

H.E.S.S. takes data during dark moonless nights. The data are calibrated \cite{aharonian04b} and an image analysis provides properties of the primary particle like arrival direction, energy and particle type. The telescopes operate in coincidence mode to allow a stereoscopic event reconstruction \cite{funk04}. Usually the data is recorded in \textit{wobble mode} where the telescope point with and offset of 0.5$^\circ$ to the nominal source position to allow a simultaneous estimation of the background. The data presented here has been analyzed with a standard Hillas-type analysis as described in \cite{aharonian06}.
The signal is extracted from a circular region around the source position, the background (off-source data) is estimated using the {\it Reflected-Region} method \cite{berge07}

%------------------------------------------------------------------------------------------------------------------------
% 1ES 0347-121
%------------------------------------------------------------------------------------------------------------------------

\section*{1ES\,0347-121}

\begin{figure}[tb]
\centering
\includegraphics[width=0.45\textwidth]{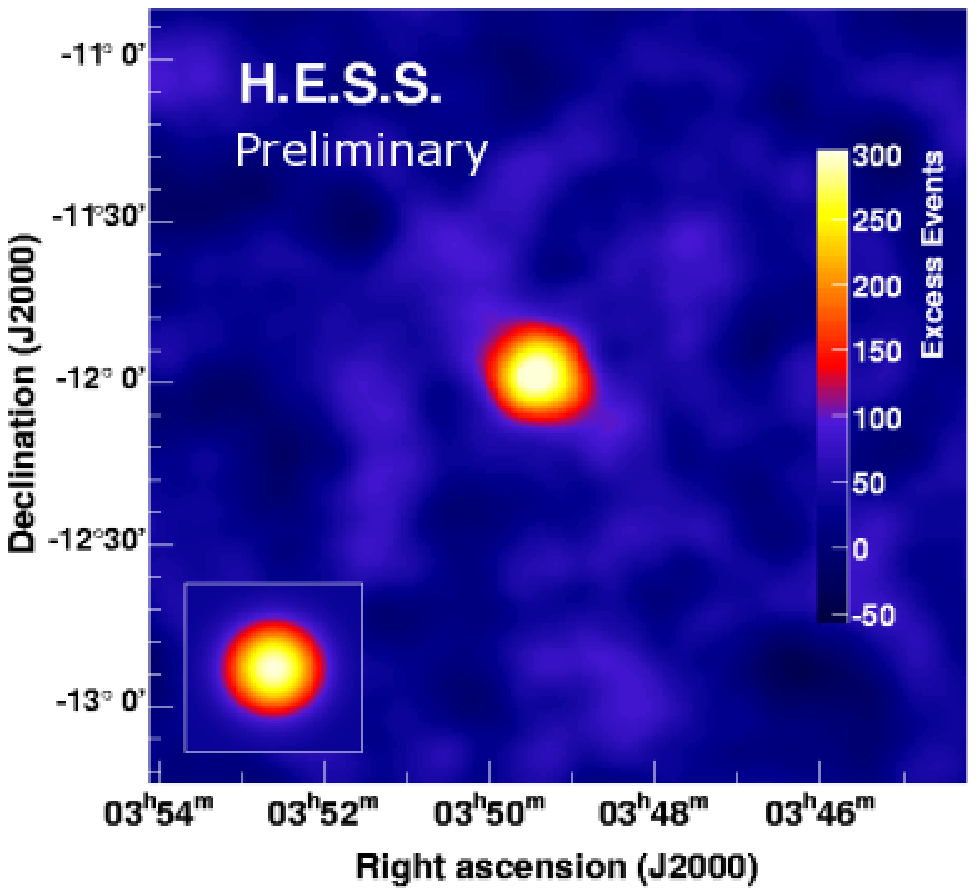}
\caption{Smoothed sky-map of excess events centered on the position of 1ES\,0347-121 (ring background). An excess of 327\,$\gamma$-ray candidates is detected corresponding to a statistical significance of 10.1 standard deviations.}
\label{fig1}
\end{figure}

\begin{figure}[tb]
\centering
\includegraphics[width=0.45\textwidth]{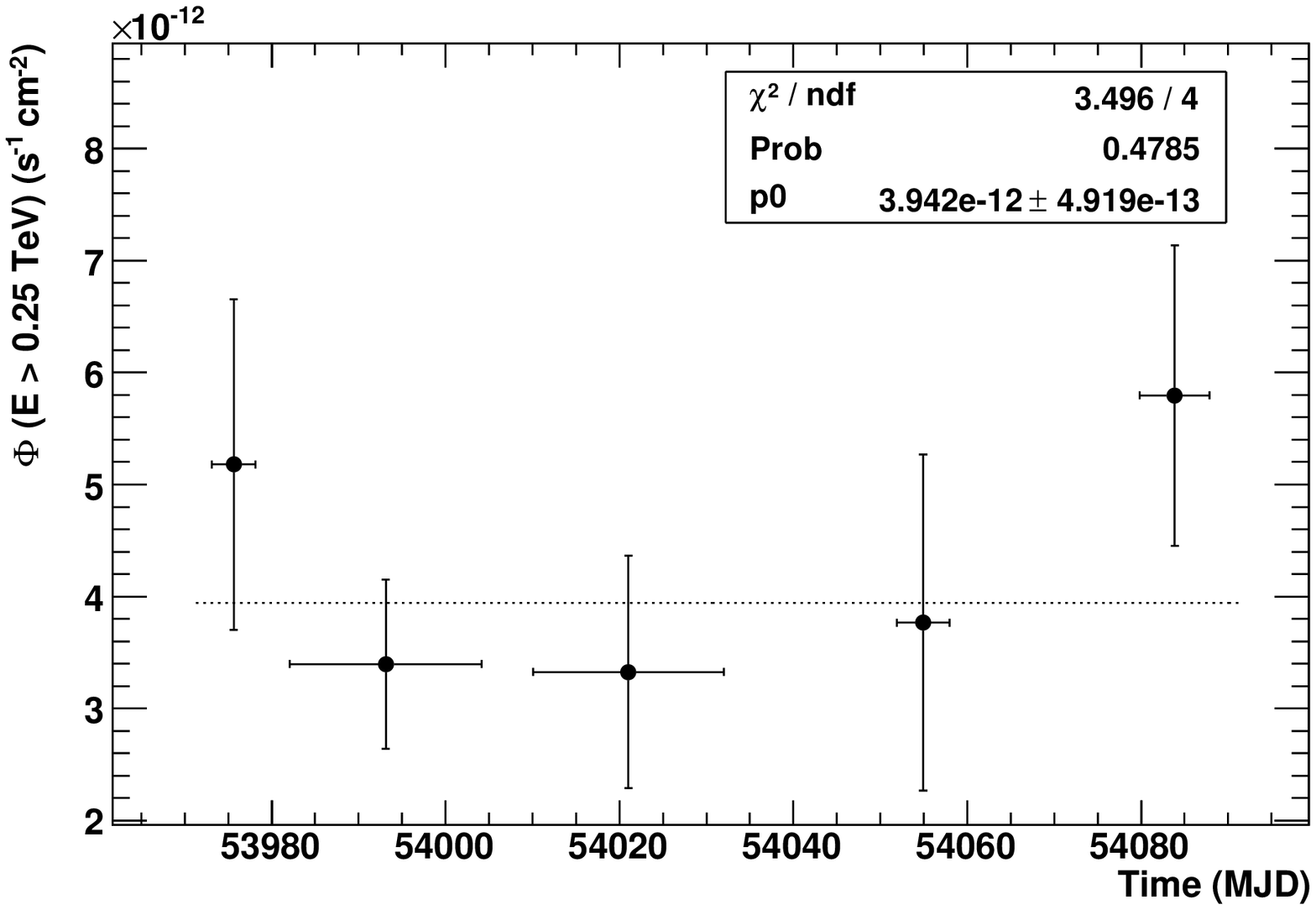}
\caption{Monthly averaged light-curve of 1ES\,0347-121. No significant variability is detected in the H.E.S.S. data-set. The dashed line is the fit of a constant function to the light-curve (parameters are given in the figure).}
\label{fig2}
\end{figure}

1ES\,0347-121 was discovered in the Einstein Slew Survey and was later classified as an BL Lac object \cite{elvis92,schachter93}. Located at a redshift of $z = 0.1880$ it harbours a super massive black hole of mass $\log(M_{\mathrm{BH}}/M_{\mathrm{Sun}}) = 8.02 \pm 0.11$ \cite{woo05}. \cite{stecker96} used simple physical considerations about the synchrotron and inverse Compton component of the spectrum to predict the flux above 0.3\,TeV of $3.8 \times 10^{-12}$\,cm$^{-2}$s$^{-1}$, which should easily be detectable with H.E.S.S. An upper limit on the integral flux above an energy threshold of 1.46\,TeV of $5.14 \times 10^{-12}$\,cm$^{-2}$s$^{-1}$ (0.56 Crab) has been reported by the HEGRA collaboration \cite{aharonian04}, considerably higher than the flux estimate from \cite{stecker96}.
 
The H.E.S.S. observations of 1ES\,0347-121 were carried out between August and December 2006. 25.4\,h (corrected for the detector deadtime) of good-quality data was recorded. The zenith angles of the observations ranged from 12 to 40$^\circ$, with a mean zenith angle of $\sim$19$^\circ$. The analysis energy threshold for the observation is $\sim$250\,GeV. 

An excess of 327\,$\gamma$-ray candidates was, found corresponding to a statistical significance of 10.1 standard deviations (Fig.~\ref{fig1}). The extension of the excess is compatible with a point-source whose position coincides with that of 1ES\,0347-121. The VHE flux is constant during the H.E.S.S. observation period (Fig.~\ref{fig2}).

The VHE energy spectrum, ranging from $\sim$250\,GeV to $>$3\,TeV, is well described by a power law with a photon index of $\Gamma \sim 3.1$. The integral flux above 250\,GeV corresponds to $\sim$2\% of the flux of the Crab Nebula above the same threshold. 

%------------------------------------------------------------------------------------------------------------------------
% 1ES 0229+200
%------------------------------------------------------------------------------------------------------------------------

\section*{1ES\,0229+200}

\begin{figure}[tb]
\centering
\includegraphics[width=0.5\textwidth]{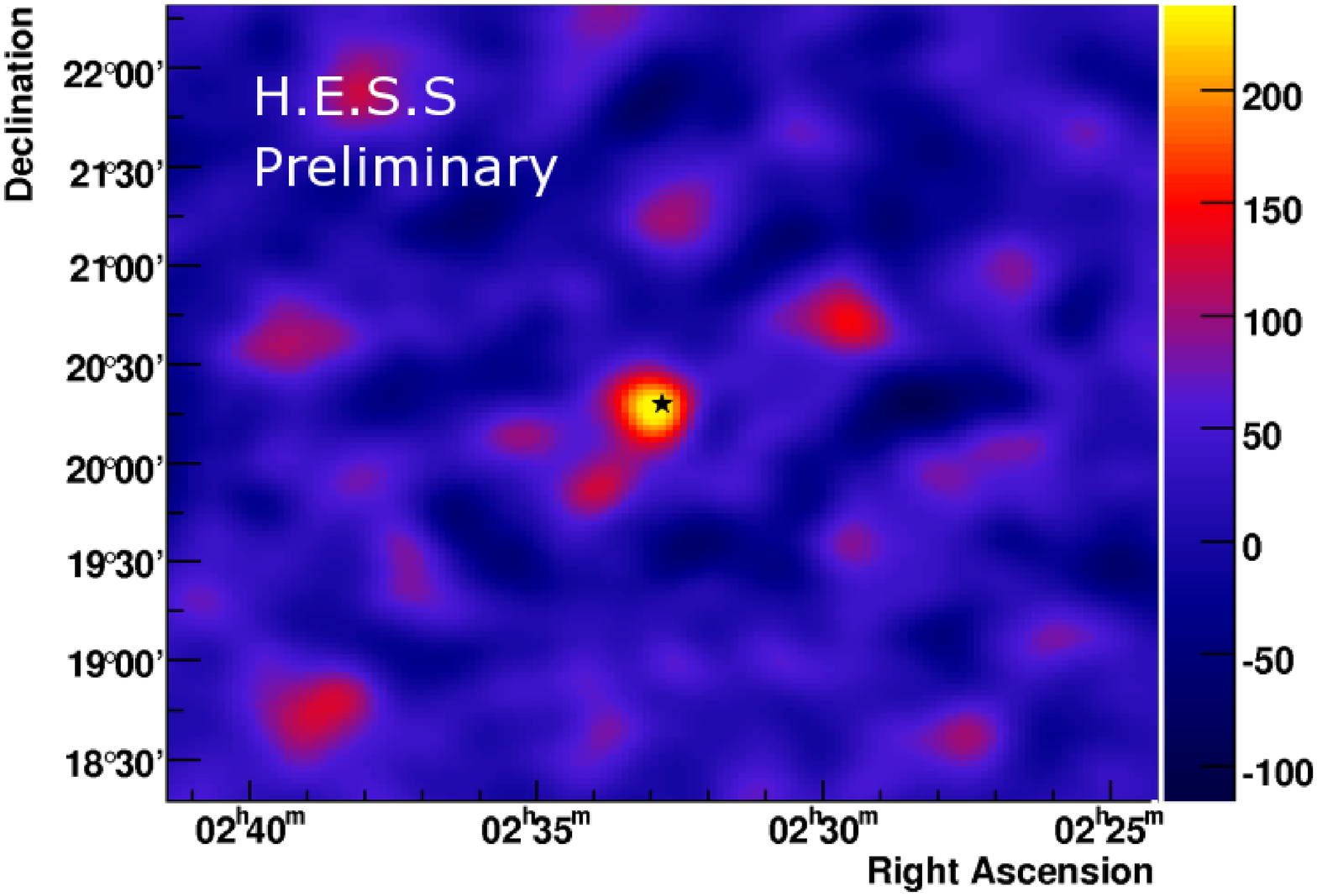}
\caption{Smoothed sky-map of excess events centered on the position of 1ES\,0229+200 (ring background). An excess of 261\,$\gamma$-ray candidates is detected corresponding to a statistical significance of 6.6 standard deviations.}
\label{fig3}
\end{figure}

\begin{figure}[tb]
\centering
\includegraphics[width=0.5\textwidth]{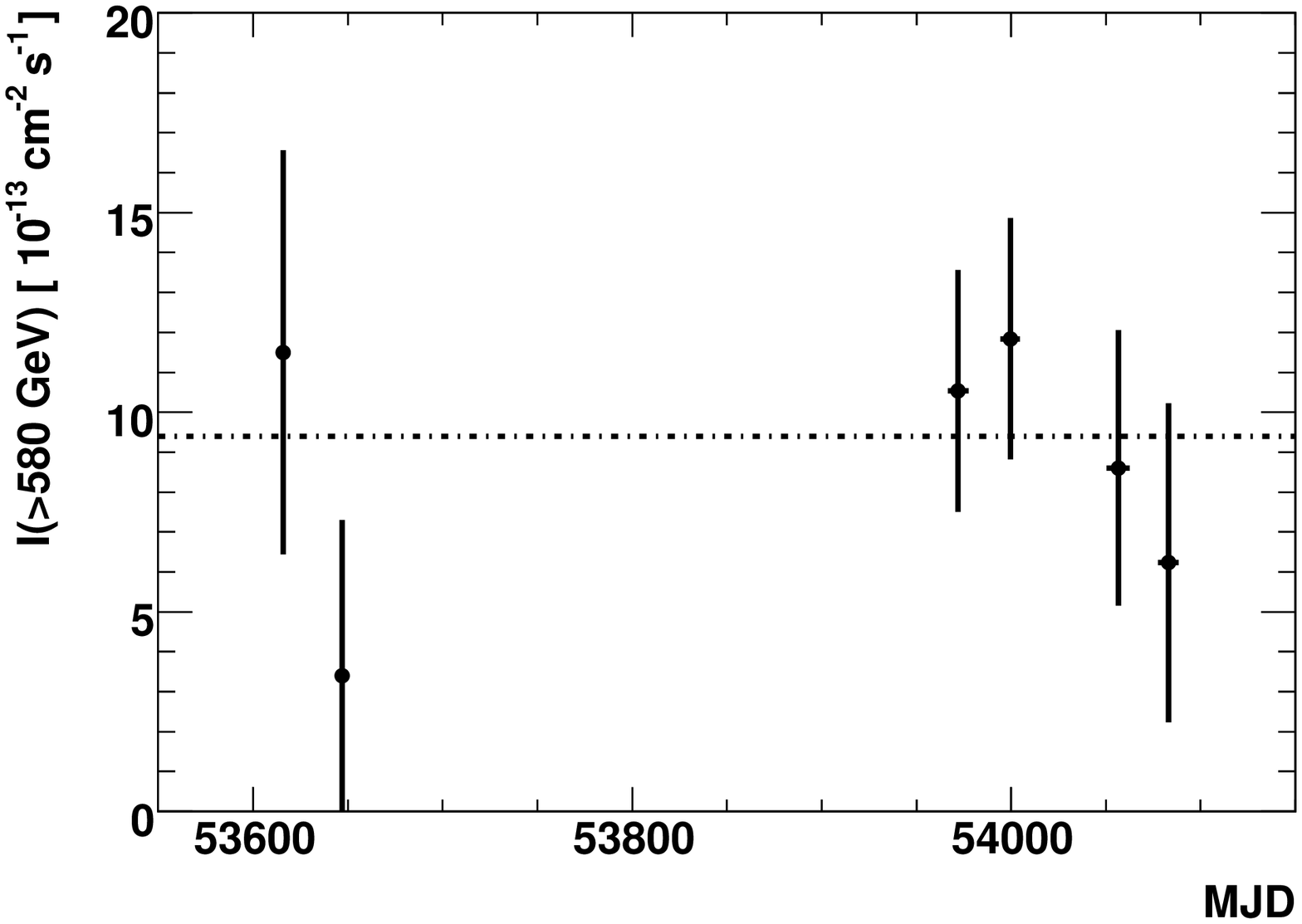}
\caption{Monthly averaged light-curve of 1ES\,0229+200. No significant variability is detected in the H.E.S.S. data-set. The dashed line corresponds to the average flux.}
\label{fig4}
\end{figure}

The active galactic nucleus (AGN) 
1ES\,0229+200 was initially discovered in the {\it Einstein} IPC Slew Survey
\cite{elvis92} and later identified as a BL Lac object \cite{schachter93}.  
It is now classified as a high-frequency peaked BL Lac (HBL) due to 
its X-ray-to-radio flux ratio \cite{giommi95}. 
The HBL is hosted \cite{falomo99} 
by an elliptical galaxy $M_R = -24.53$ located 
at a redshift of $z=0.14$ \cite{schachter93}.
Based on its spectral energy distribution (SED) and its relative
proximity, 1ES\,0229+200 was suggested \cite{costamente01,stecker96}
as a potential source of VHE $\gamma$-rays.
However, despite several attempts, it has not been 
previously detected in the VHE regime.
The HEGRA \cite{aharonian04}, Whipple \cite{delacalleperez03,horan04},
and Milagro \cite{williams05} collaborations have each reported 
upper limits on the flux from 1ES\,0229+200 during various epochs.
The most constraining upper limit (99.9\% confidence level) 
on the flux is I($>$410 GeV) $< 2.76\times 10^{-12}$ cm$^{-2}$\,s$^{-1}$, 
based on $\sim$1 hour of HESS observations made in 2004 \cite{aharonian05}.  

1ES\,0229+200 was observed with the H.E.S.S. array
for a total of 70.3 h (161 runs of $\sim$28 min each) in 2005 and 2006 \cite{aharonian07b}. After applying the standard HESS data-quality 
selection, 98 runs remain yielding an exposure of 41.8 h live time at 
a mean zenith angle $Z_{\textrm{mean}}=46^{\circ}$. The event-selection criteria are performed
using the {\it standard cuts} \cite{benbow05} resulting in a
post-analysis energy threshold of 580 GeV at $Z_{mean}$.
Results consistent with those presented below are also found using independent calibration and/or analysis chains.

A significant excess of 261 events (6.6$\sigma$)
from the direction of 1ES\,0229+200 is detected in the total data set (Fig.~\ref{fig3}).
The VHE flux is constant during the H.E.S.S. observation period (Fig.~\ref{fig4}) and the integral flux above 580\,GeV corresponds to $\sim$2\% of the flux of the Crab Nebula above the same threshold. The VHE energy spectrum, ranging from $\sim$500\,GeV to $\sim$10\,TeV, is well described by a hard power law with a photon index of $\Gamma \sim 2.5$. 

%------------------------------------------------------------------------------------------------------------------------
% Discussion & Conclusion
%------------------------------------------------------------------------------------------------------------------------

\section*{Discussion \& Conclusion}

BL Lac objects show highly variable emission in many wavelength bands. To understand and model the VHE emission processes in these objects contemporaneous multi-wavelength observations are necessary.

1ES\,0347-121 has been observed by the SWIFT satellite at X-rays and UV/optical wavelengths and by the ATOM optical telescope during the H.E.S.S. observation periods \cite{aharonian07}. The resulting broad-band spectral energy distribution (SED) can be described by a simple one-zone synchrotron-self-Compton model, though the data does not strongly constrain the model parameters. Due to the lack of contemporaneous X-ray data for 1ES\,0229+200 an accurate modeling of the source is not possible.

The VHE spectra from both sources enable the determination of stringent upper limit on the extragalactic background light (EBL) in the ultraviolet to near-IR wavelength region. These limit are reported in \cite{aharonian07} and \cite{aharonian07b}.

Future VHE monitoring of 1ES 0347-121 and 1ES\,0229+200, to search for high flux states (flares), is highly desirable. To derive stronger constraints on the emission models, further contemporaneous multiwavelength observations are also necessary. 

%------------------------------------------------------------------------------------------------------------------------
% Acknowledgments
%------------------------------------------------------------------------------------------------------------------------

\section*{Acknowledgements}

{\small
The support of the Namibian authorities and of the University of Namibia
in facilitating the construction and operation of H.E.S.S. is gratefully
acknowledged, as is the support by the German Ministry for Education and
Research (BMBF), the Max Planck Society, the French Ministry for Research,
the CNRS-IN2P3 and the Astroparticle Interdisciplinary Programme of the
CNRS, the U.K. Science and Technology Facilities Council (STFC),
the IPNP of the Charles University, the Polish Ministry of Science and 
Higher Education, the South African Department of
Science and Technology and National Research Foundation, and by the
University of Namibia. We appreciate the excellent work of the technical
support staff in Berlin, Durham, Hamburg, Heidelberg, Palaiseau, Paris,
Saclay, and in Namibia in the construction and operation of the
equipment.
}

%------------------------------------------------------------------------------------------------------------------------
% Bibliography
%------------------------------------------------------------------------------------------------------------------------
\bibliographystyle{plain}

%%%%%%%%
%  35  %
%%%%%%%%

%The paper title
\title{ Discovery of VHE gamma-rays from the BL Lac object PKS 0548-322 with H.E.S.S. }
%Short title to print in the headers to the final publication (Not showed in this print).
\shorttitle{Discovery of PKS0548-322}
%All paper authors
\authors{G. Superina$^{1}$, W. Benbow$^{2}$, T. Boutelier$^{3}$, G. Dubus$^{3}$, B. Giebels$^{1}$.}
%Short title to print in the headers to the final puplication (Not showed in this print).
\shortauthors{G. Superina and et al}
%All the affiliations
\afiliations{$^1$LLR, Ecole Polytechnique, CNRS-IN2P3, 91128 Palaiseau, France\\ $^2$Max-Planck-Institut f\"ur Kernphysik, P.O. 103980, Heidelberg, Germany\\$^3$LAOG, CNRS-INSU, 38400 Saint-Martin d'H\`eres, France}
\email{superina@poly.in2p3.fr}
\newcommand{\gev}{\ensuremath{\mathrm{\,Ge\kern -0.1em V}}}
\newcommand{\kev}{\ensuremath{\mathrm{\,ke\kern -0.1em V}}}
\def\cm   {\ensuremath{{\rm \,cm}}}
%The abstract.
\abstract{Observations and monitoring of active galactic nuclei (AGN) are a key part of the scientific observation programme of the High Energy Stereoscopic System (H.E.S.S). The instrument was used to search for very high energy (VHE: $>$ 100 \gev ) 
gamma rays coming from PKS 0548-322, a BL Lac object visible from the Southern Hemisphere. An excess of 
VHE gamma rays ($\sim$6$\sigma$) from the object is detected. The broad-band spectral energy distribution (SED), including the VHE spectrum ($\Gamma=2.8 \pm 0.3_{\rm{stat}}$) is presented.}

\maketitle

\addcontentsline{toc}{section}{Discovery of VHE gamma-rays from the BL Lac object PKS 0548-322 with H.E.S.S.}
\setcounter{figure}{0}
\setcounter{table}{0}
\setcounter{equation}{0}

%Begin the section.
\section*{Introduction}
PKS 0548-322, at a redshift of $z = 0.069$ is among the closest blazars of the Southern sky.  This extreme BL Lac is a promising source for VHE emission \cite{ref1} so H.E.S.S dedicated time ($\sim$ 45 hours) between 2004 and 2006 to its observation. The X-ray spectrum of this source has a rich history. The reported spectral indices show significant scatter (see \cite{ref5} and \cite{ref6}), with no well-defined correlation to the X-ray intensity. The X-ray fluxes and spectral indices of PKS 0548-322 show that the frequency of the synchrotron peak can vary from less than 1\ \kev\ to more than 20\ \kev, without any apparent correlation with the X-ray flux (\cite{ref7}). 

\section*{Analysis Technique and Results}
A total of 21.3 hours of good-quality data remain after application of 
quality-selection criteria and dead-time correction. The data are analyzed with the 3D-model method \cite{ref2}, where a model is used to reconstruct a 3D picture of the detected shower. For each detected shower, the direction, 
energy and 3D-width (used for gamma-hadron separation) are reconstructed. 

\begin{figure}
\begin{center}
\includegraphics [width=0.48\textwidth]{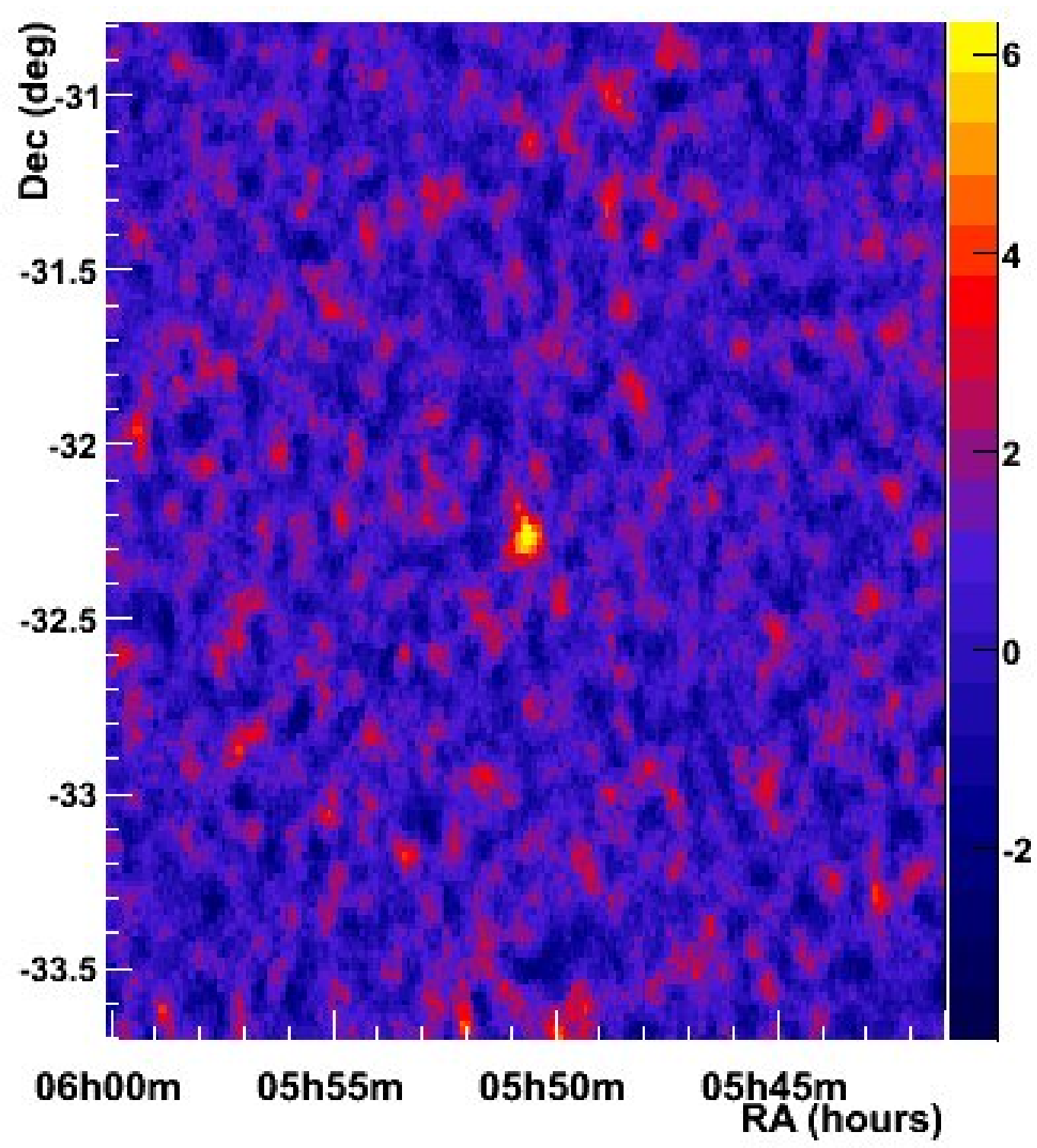}
\end{center}
\caption{Preliminary significance map centred on the position of the 
point-like source PKS 0548-322.}\label{icrcpks0548_fig1}
\end{figure}
\begin{figure}
\begin{center}
\includegraphics [width=0.48\textwidth]{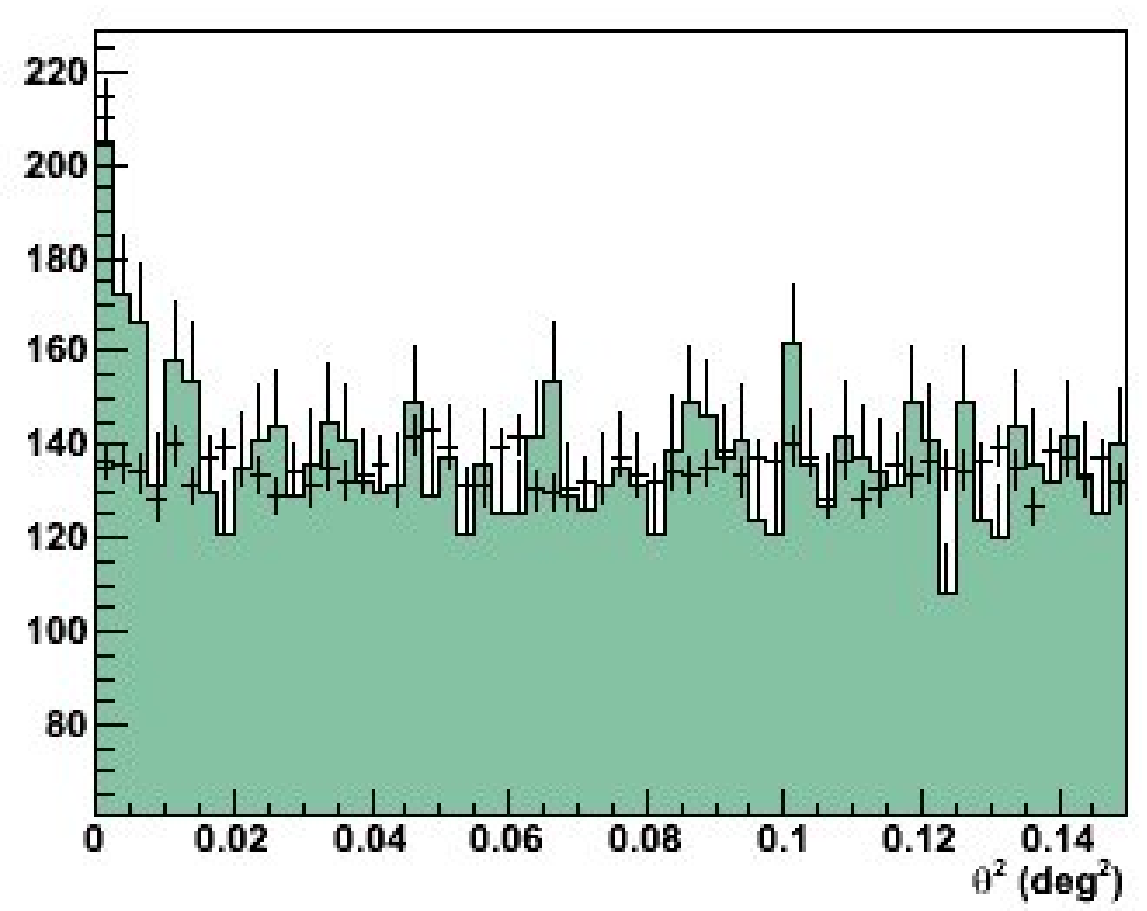}
\end{center}
\caption{Distribution of $\theta^{2}$ for on-source and normalized off-source events (preliminary). The excess is clearly visible in the region  $ \theta^{2} \leq 0.01$ $\deg^{2}$ corresponding to a statistical significance of 5.8 $\sigma$.}
\label{icrcpks0548_fig2}
\end{figure}
\begin{figure}
\begin{center}
\includegraphics [totalheight=0.4\textheight,width=0.48\textwidth] {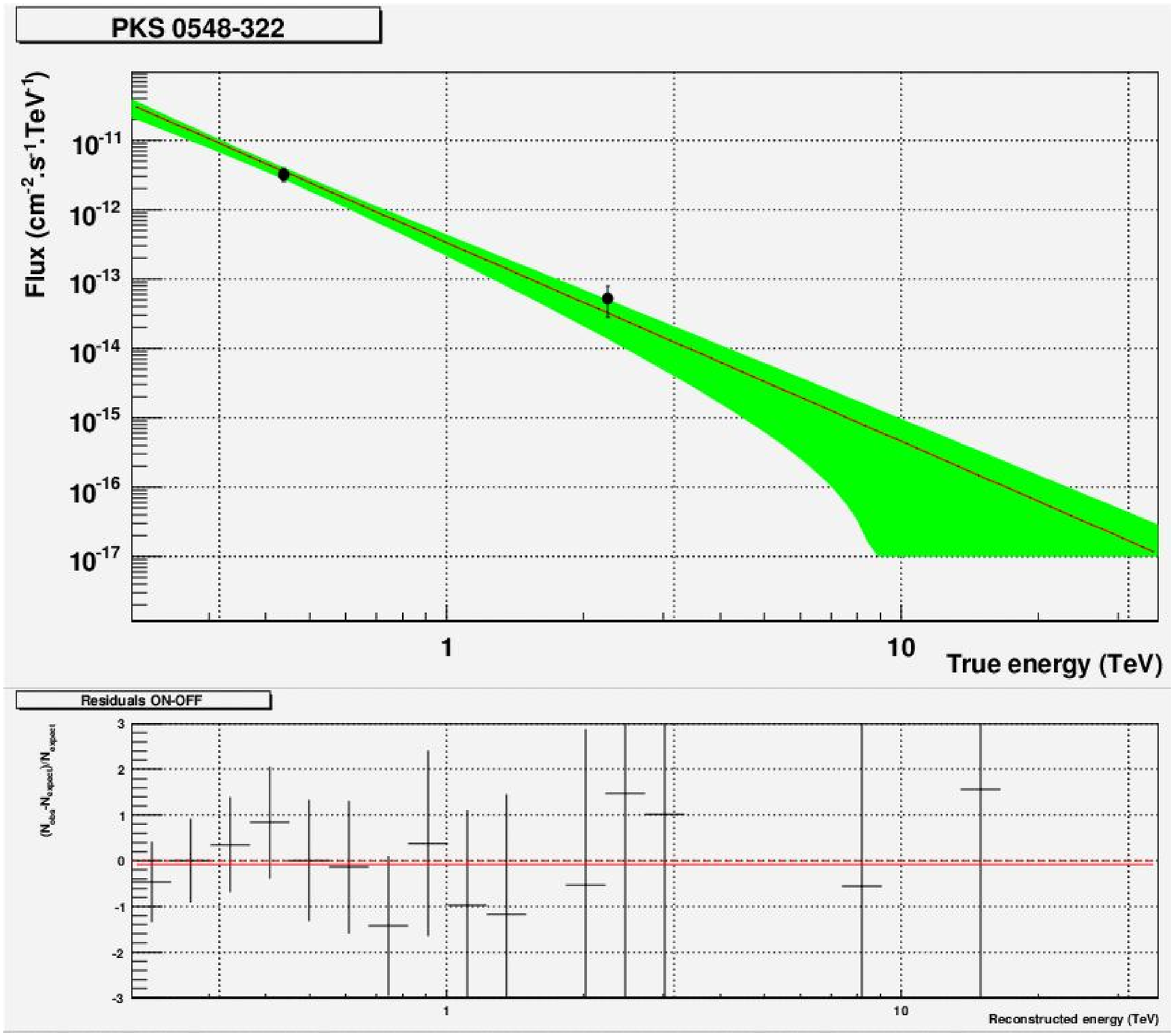}
\end{center}
\caption{VHE spectrum of PKS0548-322 (preliminary). The shaded region represents the 
1-sigma confidence bounds of 
the fitted spectrum (power-law hypothesis). }\label{icrcpks0548_fig3}
\end{figure}

Gamma-ray-like events are selected using cuts on image size, 3D-width, and telescope multiplicity. For this analysis, only events that triggered at least three telescopes are kept, in order to have better gamma-hadron separation.

The significance map of PKS 0548-322 is shown in Figure 1. The distribution of squared angular distance from the source is given in Figure 2. 
The analysis yields an excess of 181 gamma rays, corresponding to a significance of 5.8 standard deviations.
The differential energy spectrum is presented in Figure 3. For a simple power-law hypothesis, the likelihood maximization yields a spectral index of $2.8\pm0.3_{\rm{stat}}\pm0.1_{\rm{sys}}$ and an integral flux above the energy threshold:
(I $>$\ 200 \gev )= $(3.3\pm0.7)×10^{-12}\rm{cm}^{-2}\rm{s}^{-1}$. This correspond to $\sim$1.4\% of the HESS Crab Nebula flux \cite{ref4} above the same threshold. No evidence for flux variability is seen in these data.

\section*{Spectral Energy Distribution}
A simple homogeneous one-zone synchrotron self Compton (SSC) model (code from G.Dubus) is used to characterize the SED of PKS 0548-322 :\\
- the low-frequency emission, extending up to X-ray energies, is most likely due to synchrotron radiation of high-energy electrons.\\
- the VHE emission is believed to be produced through Compton upscattering of seed photons by the same population of relativistic electrons.\\
This model is adapted to the data of a soft state (X-ray data from BeppoSAX and RXTE) as well as of a hard state (BeppoSAX only). The HESS data derived from the spectrum (circles  -opened: observed data, filled: intrinsic spectrum, i.e with the effects of EBL absorption removed using the upper limit of
 \cite{ref3}) are well fitted, simultaneously, with either of these two sets of archival X-ray data.
\begin{figure*}[th]
\begin{center}
\includegraphics [width=0.70\textwidth]{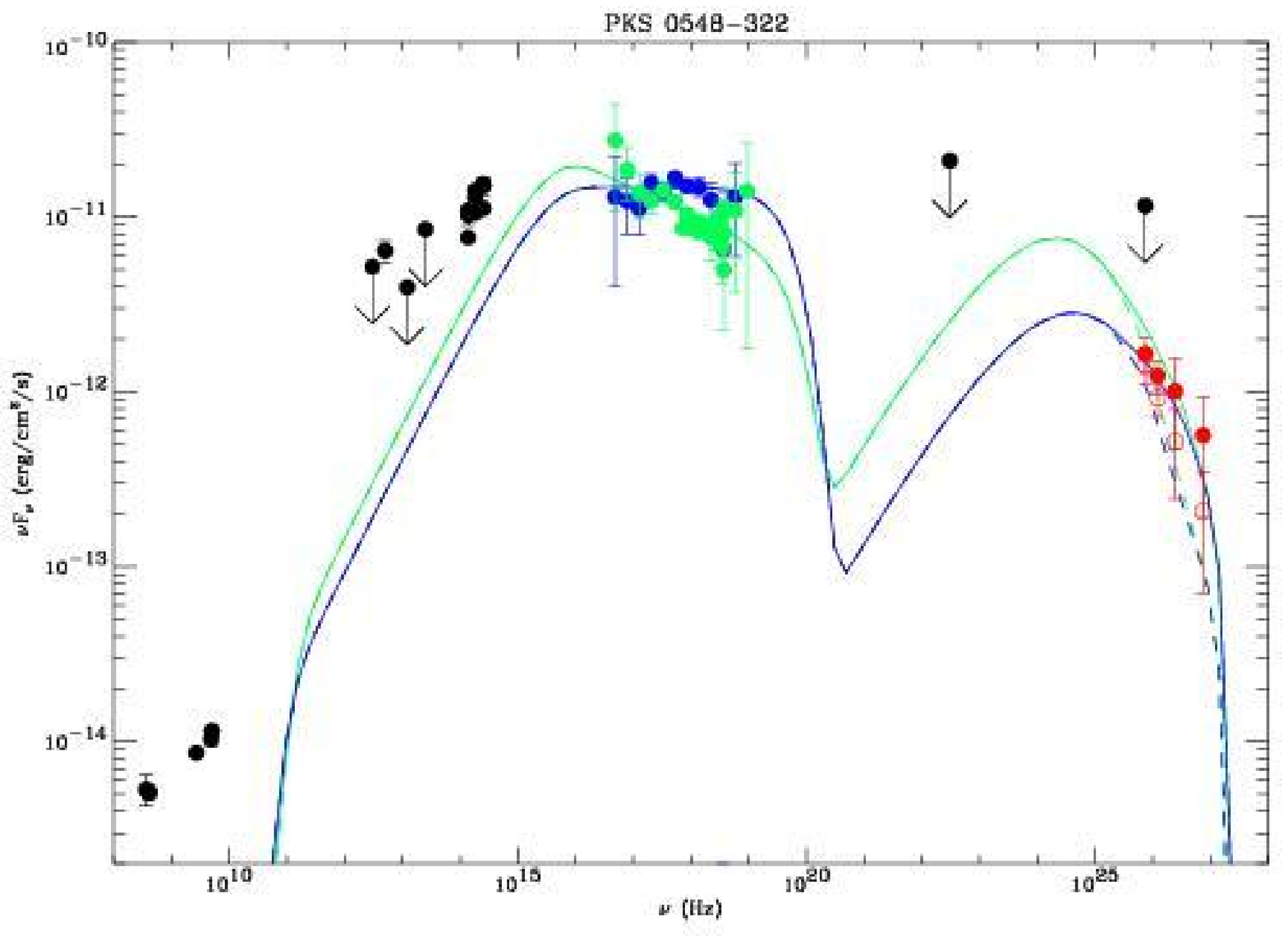}
\end{center}
\caption{Preliminary spectral energy distribution of PKS 0548-322, modelled for both a soft and a hard X-ray state. Using a model with a reasonable set of parameters provides a satisfactory fits to the archival X-ray and VHE data. The emitting region is characterised by a Doppler factor $\delta=20$, a magnetic filed B = 0.6 G and a region blob size of R$\sim2\times10^{15}\cm$.  }\label{icrcpks0548_fig4}
\end{figure*}

\section*{Conclusions}
Observations performed by H.E.S.S from 2004 up to 2006 have established PKS 0548-322 as a VHE gamma-ray source. It is among the closest TeV blazars discovered.
Given its X-ray behaviour, the source is a very interesting object for understanding its VHE emission and discriminating between different models. Quasi-simultaneous Swift data were taken during the 2006 H.E.S.S. observations and will later be used in a more detailed SED study.
\section*{Acknowledgements}
The support of the Namibian authorities and of the University of Namibia in facilitating the construction 
and operation of H.E.S.S. is gratefully acknowledged, as is the support by the German Ministry for Education
and Research (BMBF), the Max Planck Society, the French Ministry for Research, the CNRS-IN2P3 and the 
Astroparticle Interdisciplinary Programme of the CNRS, the U.K. Science and Technology Facilities Council 
(STFC), the IPNP of the Charles University, the Polish Ministry of Science and Higher Education, the South 
African Department of Science and Technology and National Research Foundation, and by the University of 
Namibia. We appreciate the excellent work of the technical support staff in Berlin, Durham, Hamburg, 
Heidelberg, Palaiseau, Paris, Saclay, and in Namibia in the construction and operation of the equipment.

%This is the reference to .bib file (Whitout .bib!)

%This in the bibtex style, is ok.
\bibliographystyle{plain}

%%%%%%%%
%  36  %
%%%%%%%%

%The paper title
\title{Upper Limits from HESS Observations of AGN in 2005-2007}

%Short title to print in the headers to the final publication (Not showed in this print).
\shorttitle{Upper Limits from HESS Observations of AGN}
%All paper authors
\authors{W.\,Benbow$^{1}$ and R.\,B\"uhler$^{1}$ for the HESS Collaboration}
%Short title to print in the headers to the final publication (Not showed in this print).
\shortauthors{W.\,Benbow et al.}
%All the affiliations.
\afiliations{$^1$ Max-Planck-Institut f\"ur Kernphysik, Heidelberg, Germany}
\email{Wystan.Benbow@mpi-hd.mpg.de}

%The abstract.
\abstract{
Very high energy (VHE; $>$100 GeV) observations of a sample of 
selected active galactic nuclei (AGN) were performed
between January 2005 and April 2007 with the High Energy Stereoscopic System (HESS),
an array of imaging atmospheric-Cherenkov telescopes.
Significant detections are reported elsewhere for many of these objects.  
Here, integral flux upper limits for twelve candidate
very-high-energy (VHE; $>$100 GeV)  gamma-ray emitters are presented.
In addition, results from HESS observations of four known VHE-bright AGN
are given although no significant signal is measured.  For three of
these AGN (1ES\,1101$-$232, 1ES\,1218+304, and Mkn\,501) simultaneous 
data were taken with the Suzaku X-ray satellite.}

\maketitle

\addcontentsline{toc}{section}{Upper Limits from HESS Observations of AGN in 2005-2007}
\setcounter{figure}{0}
\setcounter{table}{0}
\setcounter{equation}{0}

%Begin the section.
\section*{Introduction}

The HESS array \cite{HESS_jim} of four imaging atmospheric-Cherenkov 
telescopes located in Namibia is used to search for
VHE $\gamma$-ray emission from various
classes of astrophysical objects.
Approximately 300 hours per year, $\sim$30\% of the total
observation budget, are dedicated to
HESS studies of AGN. These observations are
divided between monitoring the flux of
known VHE-bright AGN and searching for new VHE sources. 
For the monitoring observations, an 
AGN is typically observed for a few hours, distributed
over several nights, a month for $\sim$3 months, with
the hopes of detecting a bright flaring episode (see, e.g., \cite{2155_flare}).
In the discovery part of the AGN program, a candidate from 
a large, diverse sample of relatively nearby AGN is typically
observed for $\sim$10 hours.  If any of these observations 
show an indication for a signal (e.g., an excess with significance
more than $\sim$3 standard deviations), a deeper 
exposure is promptly scheduled to increase the overall significance
of the detection and to allow for a spectral measurement.

The targets of HESS AGN observations are primarily blazars, 
a class which includes both BL\,Lac objects and
Flat Spectrum Radio Quasars (FSRQ). The spectral energy
distributions (SEDs) of these objects are generally
characterized by two peaks: a lower-energy one in the optical to X-ray
regime, and another which potentially extends to
$\gamma$-ray energies. Based on their SEDs, 
BL\,Lacs are generally categorized into groups that are either 
low (LBL), intermediate (IBL), or high-frequency-peaked (HBL).
An overwhelming majority of VHE-emitting AGN
are HBL, therefore these objects
are the primary targets of HESS AGN observation
program.  However, prominent examples of different
types of AGN are also observed with the hopes of 
detecting new AGN classes.  These include
radio-loud objects such as Fanaroff-Riley (FR) galaxies
and narrow line Seyfert (NLS) galaxies, and radio-weak objects
like typical Seyfert (Sy) galaxies,  all of which come
in several types (generally I or II).

For all the following results, the HESS standard analysis \cite{std_analysis} is used.
All upper limits are given at the 99\% confidence level \cite{UL_tech}.
The flux quantities are calculated assuming a power-law spectrum
with photon index $\Gamma$=3.0, with the exception of those for 1ES\,1101$-$232
where  $\Gamma$=2.94, as measured \cite{Gerd_1101} in 2004-05, is chosen.
The reported values change by less than $\sim$10\% when a 
different photon index (i.e. $\Gamma$ between 2.5 and 3.5) 
is assumed.  The effects of changes in the absolute optical efficiency
of HESS are corrected for using efficiencies 
determined from simulated and observed muons \cite{HESS_crab}. 
The systematic error on all flux quantities is estimated to
be $\sim$20\%.

\section*{Limits from Discovery Observations}

   \begin{table*}[ht]
      \caption{The candidate AGN in groups of blazars and non-blazars. 
The asterisk denotes the four candidates detected by the EGRET satellite 
\cite{EGRET_catalog}. The redshift ($z$), total good-quality
live time (T), mean zenith angle of observation (Z$_{\mathrm{obs}}$),
the observed excess and significance (S) are shown.
Integral flux upper limits above the energy threshold of the observations (E$_{\mathrm{th}}$), 
and the corresponding percentage of the HESS Crab Nebula flux \cite{HESS_crab} above the same
threshold, are also shown.  The flux units are $10^{-12}$ cm$^{-2}$ s$^{-1}$. The $\dagger$
represents the six upper limits which are the most constraining ever reported for the object.}
         \label{results} 
        \centering
         \begin{tabular}{c c c c c c c c c c}
            \hline\hline
            \noalign{\smallskip}
	     Object & $z$ & Type & T & Z$_{\mathrm{obs}}$ & Excess & S & E$_{\mathrm{th}}$ & I($>$E$_{\mathrm{th}}$) & Crab \\
             & &  & [hrs] & [$^{\circ}$] & & [$\sigma$] & [GeV] & [f.u.] & \%  \\
            \noalign{\smallskip}
            \hline
            \noalign{\smallskip}
	{\it Blazar} \\
            \noalign{\smallskip}
             III\,Zw\,2           & 0.0893 & FSRQ & 1.7 & 37 & 12 & 1.4 & 420 & 5.36$^{\dagger}$ & 6.4 \\ 
	     BWE\,0210+116$^{*}$  & 0.250  & LBL & 6.0 & 43 & $-$13 & $-$0.9 & 530 & 0.72$^{\dagger}$ & 1.2\\
	     1ES\,0323+022        & 0.147  & HBL & 7.2 & 27 & 13 & 0.7 & 300 & 2.52 & 1.9\\
	     PKS 0521$-$365$^{*}$ & 0.0553 & LBL & 3.1 & 26 & 11 & 0.8 & 310 & 5.40$^{\dagger}$ & 4.2\\
	     3C\,279$^{*}$	  & 0.536 & FSRQ & 2.0 & 26 & 5 & 0.5 & 300 & 3.98$^{\dagger}$ & 2.9\\
             RBS\,1888            & 0.226 & HBL	& 2.4 & 15 & 30 & 2.2 & 240 & 9.26 & 4.9\\
             PKS\,2316$-$423      & 0.055 & IBL	& 4.1 & 20 & 29 & 1.6 & 270 & 4.74 & 3.0\\
 	     1ES\,2343$-$151      & 0.226 & IBL	& 8.6 & 17 & $-$16 & $-$0.6 & 230 & 2.45$^{\dagger}$ & 1.2\\
\\	{\it Non-blazar} \\
            \noalign{\smallskip}
             NGC\,1068		& 0.00379 & Sy II & 1.8 & 29 & 9 & 1.1 & 330 & 5.76 & 4.9\\
             Pictor\,A		& 0.0342 & FR II & 7.9 & 31 & $-$23 & $-$1.1 & 320 & 2.45 & 2.0\\
	     PKS\,0558$-$504	& 0.137 & NLS I & 8.3 & 28 & $-$14 & $-$0.7 & 310 & 2.38$^{\dagger}$ & 1.8\\
             NGC\,7469		& 0.0164 & Sy I	& 3.4 & 34 & $-$14 & $-$1.3 & 330 & 1.38 & 1.2\\
          \noalign{\smallskip}
            \hline
       \end{tabular}
   \end{table*}

Twenty-nine AGN were observed by HESS from January 2005 through April 2007.
Some of these objects were previously shown by HESS
to emit VHE $\gamma$-rays, and the discoveries of VHE emission
from others are reported elsewhere. Of the remaining AGN
with non-zero good-quality exposure,
twelve show no indication of any VHE emission.  
As many of the HBL observed by HESS have been detected, 
the twelve candidates discussed in this section are largely not HBL.
Table~\ref{results} shows these AGN, their
redshift and AGN type, as well as details of their observations.
The mean good-quality exposure for 
the candidates is 4.7 hours live time at a
mean zenith angle of 28$^{\circ}$. In 5 hours of observations, 
the sensitivity of HESS \cite{std_analysis}
enables a 5$\sigma$ detection of an $\sim$2\% Crab Nebula flux 
source at 20$^{\circ}$ zenith angle.

\begin{figure}
\begin{center}
\noindent
\includegraphics [width=0.5\textwidth]{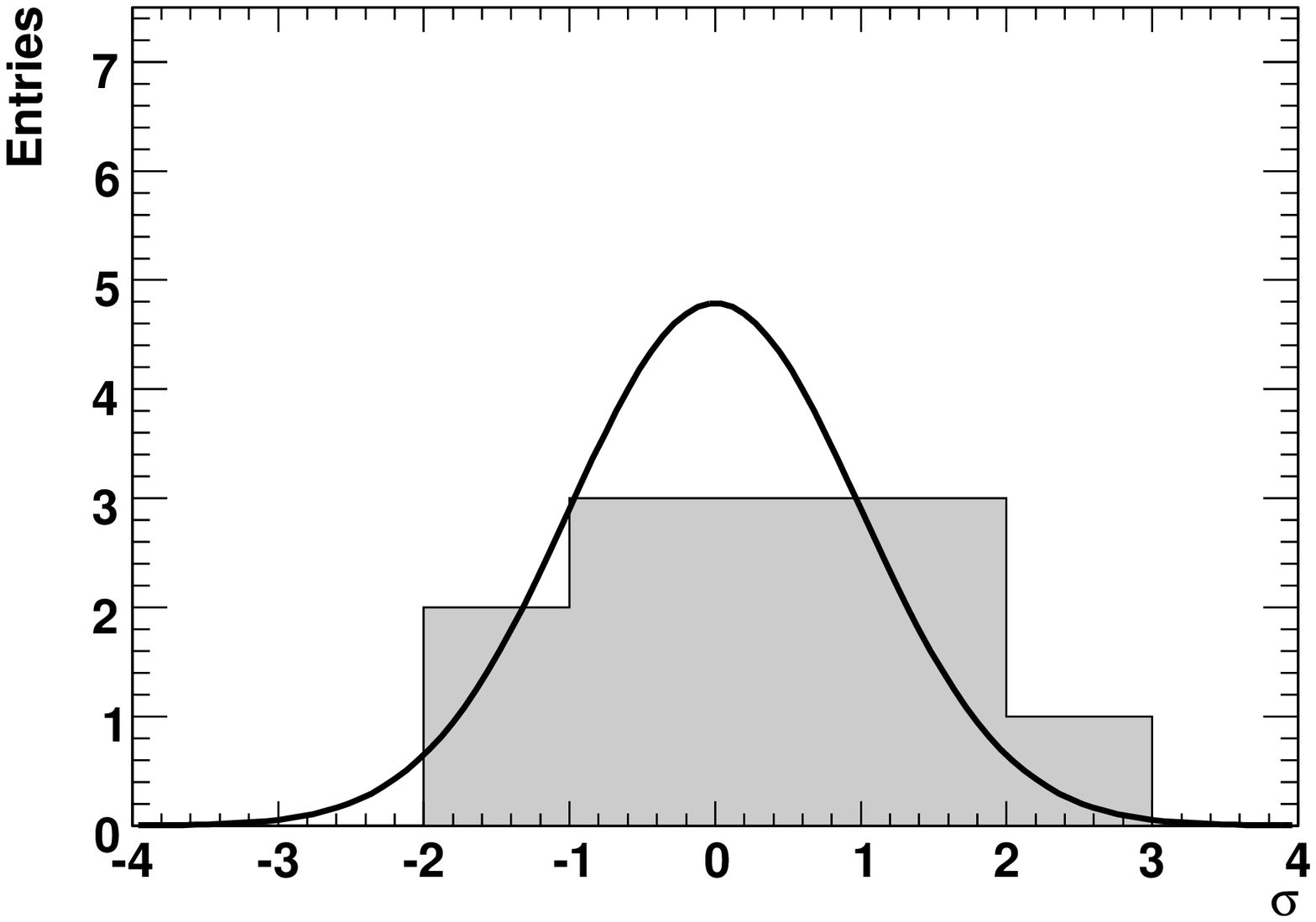}
\end{center}
      \caption{Distribution of the significance observed 
from the twelve candidate AGN.  The curve
represents a Gaussian distribution with zero mean 
and a standard deviation of one.}
         \label{AGN_sigma}
\end{figure}

As mentioned previously, no significant excess of VHE $\gamma$-rays is found 
from any of these twelve AGN in the given exposure time. 
Figure~\ref{AGN_sigma} shows the distribution 
of the significance observed from the direction of each AGN.  
The measured excess, corresponding significance and resulting
integral flux limits are given in Table~\ref{results} 
for each AGN.  Six of the upper limits are the most constraining
ever reported from these objects, and 
the other six limits are only surpassed by those from
HESS observations in 2004 \cite{HESS_AGN_UL}.
Combining the excess from all twelve candidates only 
yields a total of 29 events 
and a statistical significance of
1.1$\sigma$.  No significant excess is found in a search for serendipitous 
source discoveries in the HESS field-of-view centered on each of the 
AGN.  Further, as the nightly flux from each target is well-fit by a
constant, no evidence for VHE flares is found from any of the twelve AGN.

\section*{Low Altitude HESS Observations}

Three northern AGN, known to emit VHE $\gamma$-rays,
were briefly (good-quality live time $<$2.2 h) 
observed at low altitudes with HESS. 
At such altitudes the threshold of HESS is higher 
and the sensitivity is reduced.  However,
observations at low altitudes sample
the VHE spectrum at much higher energies than
the typical near-zenith observations made with
Cherenkov-telescope arrays.  Simultaneous measurements
of the same northern target with HESS and a Northen Hemisphere
instrument enable both the determination of the object's spectrum
over several orders of magnitude in energy, as well as
cross-calibration between the instruments 
(see, e.g., \cite{HESS_MAGIC_421}).
For two of these targets (1ES\,1218+304 and Mkn\,501)
simultaneous observations
were successfully performed by the MAGIC VHE telescope
and the Suzaku X-ray satellite \cite{Suzaku_info}.

HESS observed Mkn\,421 on April 12, 2005.  
The good-quality exposure is 0.9 h live time at a 
mean zenith angle of $63^{\circ}$.
A marginally significant excess (28 events, 3.5$\sigma$)
is found.  The corresponding integral flux
above the 2.1 TeV analysis threshold is  
I($>$2.1 TeV) = $(3.1\pm1.0_{\rm stat}) \times 10^{-12}$
cm$^{-2}$\,s$^{-1}$, or 45\% of the HESS Crab Nebula flux above
the same threshold. 

The HESS observations of 1ES\,1218+304 on May 19, 2006
yield a good-quality data set of 1.8 h live time at a 
mean zenith angle of $56^{\circ}$.
The resulting excess is not significant (9 events, 
1.2$\sigma$). The upper limit on the 
integral flux above the 1.0 TeV analysis threshold is  
I($>$1.0 TeV) $ < 3.9 \times 10^{-12}$
cm$^{-2}$\,s$^{-1}$.  This corresponds to
17\% of the HESS Crab Nebula flux 
above the same threshold.

HESS observations of Mkn\,501 occurred on July 18, 2006.  
All data pass the standard quality-selection
criteria, yielding an exposure of 2.2 h live time at a 
mean zenith angle of $64^{\circ}$.
Mkn\,501 is not detected by HESS
as the resulting excess is $-9$ events ($-0.8$$\sigma$).
The upper limit on the integral flux above the 2.5 TeV analysis threshold is  
I($>$2.5 TeV) $ < 1.1 \times 10^{-12}$
cm$^{-2}$\,s$^{-1}$, or 22\% of the 
HESS Crab Nebula flux above the same threshold.

\section*{VHE Monitoring of 1ES\,1101$-$232\label{1101_Sect}}
\begin{figure}
\begin{center}
\noindent
\includegraphics [width=0.45\textwidth]{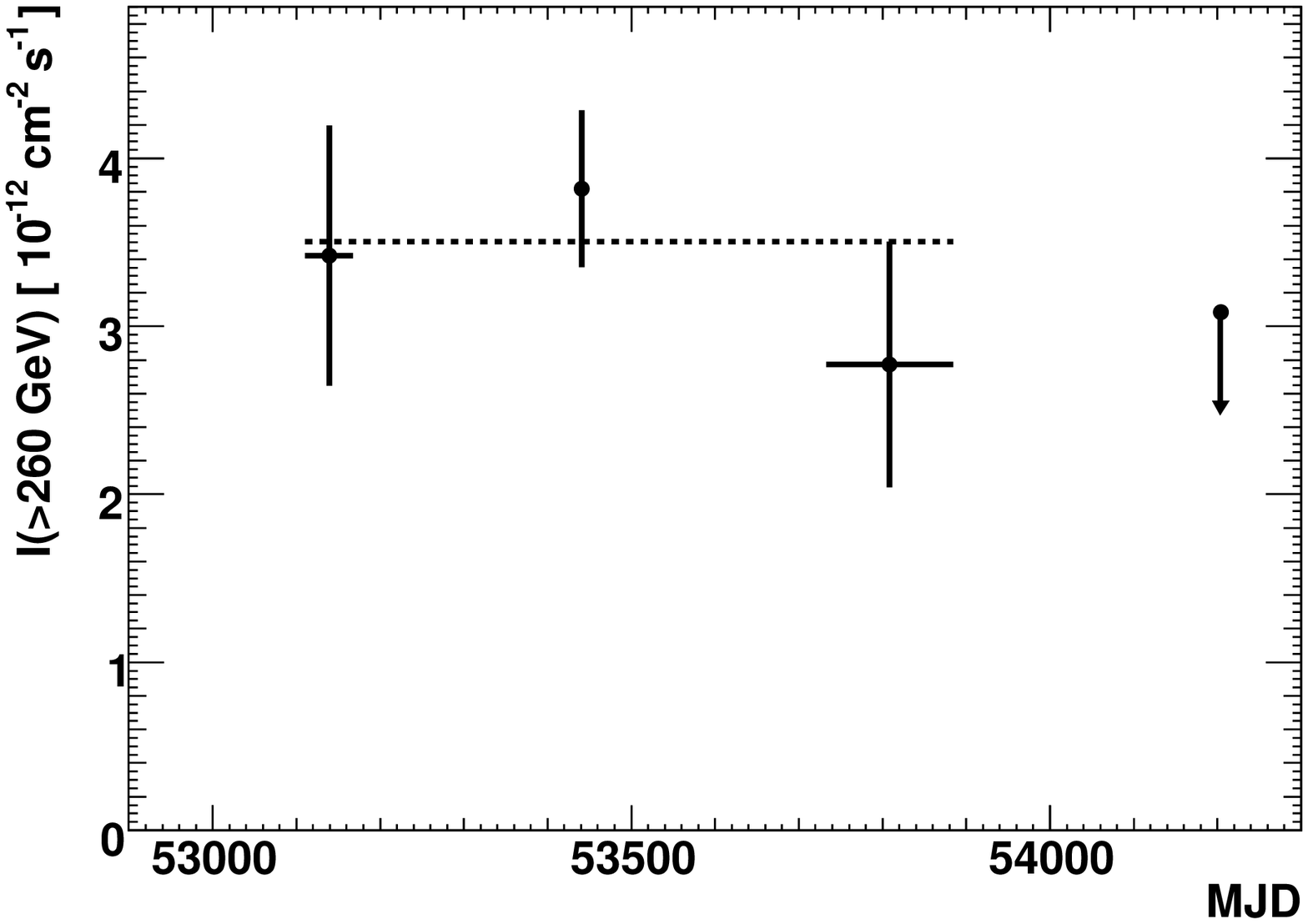}
\end{center}
      \caption{The annual light curve, I($>$260 GeV), 
from HESS measurements of 1ES\,1101$-$232. 
The upper limit in 2007 is at the 99.9\% confidence level.
The 2004 and 2005 data are published elsewhere \cite{Gerd_1101}.  
The actual observation dates are shown by the x-error bars.  
The dashed line is the average flux measured from 2004-2006. }
         \label{1101_lc}
\end{figure}

1ES\,1101$-$232 was discovered by HESS \cite{Nature_EBL,Gerd_1101} 
to emit VHE $\gamma$-rays during observations in 2004-2005.
As part of a campaign to monitor its VHE flux, it was re-observed 
for a total (good-quality observations) of 18.3 h in 2006-07.
A marginally significant excess (117 events, 3.6$\sigma$) 
is detected from 1ES\,1101$-$232
in the 2006 observations (13.7 h), and the object is not detected 
(16 events,  0.9$\sigma$) in 2007.   As can be seen from Figure~\ref{1101_lc}
the upper limit from 2007 falls below the average flux measured by
HESS from 2004-2006.  Some of the 2006 
HESS data (4.3 h) are simultaneous with Suzaku X-ray 
observations.  In these data, the blazar is again marginally
detected (51 events,  0.9$\sigma$) and the corresponding flux
is I($>$260 GeV) $ < (3.2\pm1.4_{\rm stat}) \times 10^{-12}$
cm$^{-2}$\,s$^{-1}$.

\section*{Discussion \& Conclusions}

One of the defining characteristics of AGN is their extreme
variability. The VHE flux from any of these AGN may increase 
significantly during future flaring episodes 
(see, e.g., \cite{2155_flare}) and could potentially
exceed the limits presented here.  In
addition, accurate modeling of the SED 
requires that the state of the source is accounted for.  
Therefore, in the absence of contemporaneous observations
at lower energies,  it is recommended that these results 
be conservatively interpreted as limits on,
or measurements of, the steady-component or quiescent flux from the AGN.
Clearly, the simultaneous Suzaku X-ray data 
from Mkn\,501, 1ES\,1218+304, and 1ES\,1101$-$232, make the
HESS results from these objects particularly useful.
Finally, interpretation of the SED of an AGN not only requires accounting
for the state of the source, but also the redshift and energy dependent
absorption \cite{EBL_1} of VHE photons on the 
Extragalactic Background Light (EBL),
which is potentially large \cite{Nature_EBL,EBL_2} for some of these sources.

With the detection of ten VHE AGN, including the discovery of
seven, the HESS AGN observation program has been highly successful. 
However, despite more than five years of operations, the observation 
program is not complete as many proposed candidates 
have either not yet been observed
or only have a fraction of their intended exposure.  
Therefore, the prospects of finding additional VHE-bright AGN 
with HESS are still excellent.

\section*{Acknowledgements}

The support of the Namibian authorities and of the University of Namibia
in facilitating the construction and operation of H.E.S.S. is gratefully
acknowledged, as is the support by the German Ministry for Education and
Research (BMBF), the Max Planck Society, the French Ministry for Research,
the CNRS-IN2P3 and the Astroparticle Interdisciplinary Programme of the
CNRS, the U.K. Science and Technology Facilities Council (STFC),
the IPNP of the Charles University, the Polish Ministry of Science and 
Higher Education, the South African Department of
Science and Technology and National Research Foundation, and by the
University of Namibia. We appreciate the excellent work of the technical
support staff in Berlin, Durham, Hamburg, Heidelberg, Palaiseau, Paris,
Saclay, and in Namibia in the construction and operation of the
equipment.

%This in the bibtex style, is ok.
%\bibliographystyle{plain}

%%%%%%%%
%  37  %
%%%%%%%%

%The paper title
\title{A search for Very High Energy \gr emission from Passive Super-massive Black Holes}
%Short title to print in the headers to the final publication (Not showed in this print).
\shorttitle{A search for VHE emission from passive SMBH}

%All paper authors
\authors{G.Pedaletti$^{1,2}$, S. Wagner$^1$, W. Benbow$^3$, for the \hess collaboration}
%Short title to print in the headers to the final publication (Not shown in this print).
\shortauthors{Pedaletti et al.}
%All the affiliations.
\afiliations{$^1$Landessternwarte, Universit\"{a}t Heidelberg, K\"{o}nigstuhl, 69117 Heidelberg, Germany\\ 
$^2$ Max-Planck-Institut f\"{u}r Astronomie, K\"{o}nigstuhl, 69117 Heidelberg, Germany \\$^3$Max-Planck-Institut f\"{u}r Kernphysik, PO Box 103980, 69029 Heidelberg, Germany}
\email{gpedalet@lsw.uni-heidelberg.de}

%The abstract.
\abstract{
Jets of Active Galactic Nuclei (AGN) are established emitters of very high energy (VHE; $>$100 GeV) $\gamma$-rays. VHE radiation is also expected to be emitted from the vicinity of super-massive black holes (SMBH), irrespective of their activity state. Accreting SMBH rotate and generate a dipolar magnetic field. In the magnetosphere of the spinning black hole, acceleration of particles can take place in the field gaps. VHE emission from these particles is feasible via leptonic or hadronic processes. Therefore quiescent systems, where the lack of a strong photon field allows the VHE emission to escape, are candidates for emission.  The \hess experiment has observed the passive SMBH in the nearby galaxy NGC 1399. No VHE \gr signal is observed from the galactic nucleus. Constraints set by the NGC 1399 observations are discussed in the context of different mechanisms for the production of VHE \gr emission.}

\maketitle

\addcontentsline{toc}{section}{A search for Very High Energy \gr emission from Passive Super-massive Black Holes}
\setcounter{figure}{0}
\setcounter{table}{0}
\setcounter{equation}{0}

\section*{Introduction}

Spheroidal systems (such as elliptical galaxies, lenticular galaxies and early-type spiral galaxies with bulges) are commonly believed to host in the central region super-massive black holes with masses in the range $\mbhï»= 10^6 - 10^9 M_\odot$ \cite{rich}. During the early stages of galaxy evolution these SMBH accrete matter at high rates and are observed as bright QSOs. The radiative output at low energy (e.g. optical) decays from redshift z$>$3 to z=0 by almost 2 orders of magnitude. Therefore, the majority of SMBH in the local universe are not embedded in dense radiation fields. This enables VHE \grs to escape from the nuclear region without suffering from strong absorption via $\gamma$-photon pair absorption.
Several models \cite{gal-cen,lev,neron,slane} are proposed for the production of VHE \grs emission from these passive AGN. In all cases a large mass of the central object is the most important characteristic for generating a high VHE flux. \hess has already observed nine nearby galaxies whose black hole mass is measured \cite{mag,pel}. Only the case of NGC 1399 is considered here. Constraints on the physical parameters of the system (e.g. the magnetic field \textbf{B}) are derived using several of the aforementioned models.
\section*{Acceleration Mechanism}
If the central black hole is accreting matter from a disk that also carries magnetic flux, it will develop a magnetosphere similar to those surrounding neutron stars. If the charge density is not too high in the magnetosphere of the spinning black hole, it is possible to have a non-zero component of the electric field \textbf{E} parallel to the magnetic field \textbf{B}. In this configuration field gaps are created, where acceleration of particles can take place \cite{slane}.

Various methods can be used to estimate the magnetic field B. For example, B is estimated:
\begin{itemize}
 \item assuming equipartition 
\begin{equation}\label{eqn:magn}
\frac{B^2}{8\pi} = \frac{1}{2} \rho(r_0)v^2_r(r_0),
\end{equation}
where $\rho$ is the mass density and $v_r$ is the radial infall velocity of the accreting matter (both being a function of $r_0$, the distance to the inner edge of the disk);
 \item from the angular momentum as in \cite{bick}
$$ B = 3.1\times10^3 \frac{\dot{m}^{1/2}}{M_{10}^{1/2}}\left(\frac{r}{r_\mathrm{g}}\right)^{-5/4} \textrm{Gauss,}$$
where $\dot{m}$ is the mass accretion rate in units of the Eddington mass accretion rate, $r_\mathrm{g}$ is the gravitational radius of the black hole and $M_{10}=(\mbhrm/10^{10}M_{\odot})$.
\end{itemize}

In the model of \cite{slane} protons accelerated in the outer part of the black hole magnetosphere will collide with other protons present in the accretion disk producing pions some of which decay into VHE \grs.
The available power is
\begin{equation}\label{eq:power}
 W_{\mathrm{max}} \sim 10^{27} \left(\mbh\right)^2 \left(B_4\right)^2 \textrm{ ergs s}^{-1}, 
\end{equation}

where $B_4=(B/10^4 \textrm{ Gauss})$.
Here it is assumed that the magnetic energy density is in equipartition with the accretion energy density, which depends on various properties of the accretion disk (see Eq.~\ref{eqn:magn}). 

In other models \cite{gal-cen,lev,neron} VHE \grs originate from electromagnetic processes such as synchrotron or curvature emission. Following the analogous arguments given in \cite{gal-cen} for the Galactic Center, synchrotron emission is not feasible due to a cut-off for protons and electrons at $\epsilon_{\gamma,\mathrm{max}} \simeq 0.3$ TeV and $\epsilon_{\gamma,\mathrm{max}} \simeq 0.16$ GeV respectively. These cut-offs are independent of the magnetic field strength. 
The energy of curvature photons (when curvature losses are the dominant ones) does not depend on the mass of the particle, so it is the same for electron or proton originated photons. The emission spectrum from curvature radiation can extend up to VHE energies, with a cut-off at:
\begin{equation}\label{eq:cutoff}
 E_{\mathrm{max}}\simeq14\left(M_{10}\right)^{1/2}\left(B_4\right)^{3/4} \textrm{TeV.}
\end{equation}

\section*{VHE Observation of NGC 1399}

The giant elliptical galaxy NGC 1399 is located in the central region of the Fornax cluster at a distance of $20.3$ Mpc. An SMBH of $ \mbh = 1.06\times10^9 M_\odot$ resides in the central region. 
The nucleus of this galaxy is well known for its low emissivity at all wavelengths \cite{oconn}. Considering also the visibility of candidate sources for \hessns, NGC 1399 therefore emerged as the best candidate for this study.

\begin{figure}[htbp]
\centering
  \includegraphics[width=0.5\textwidth]{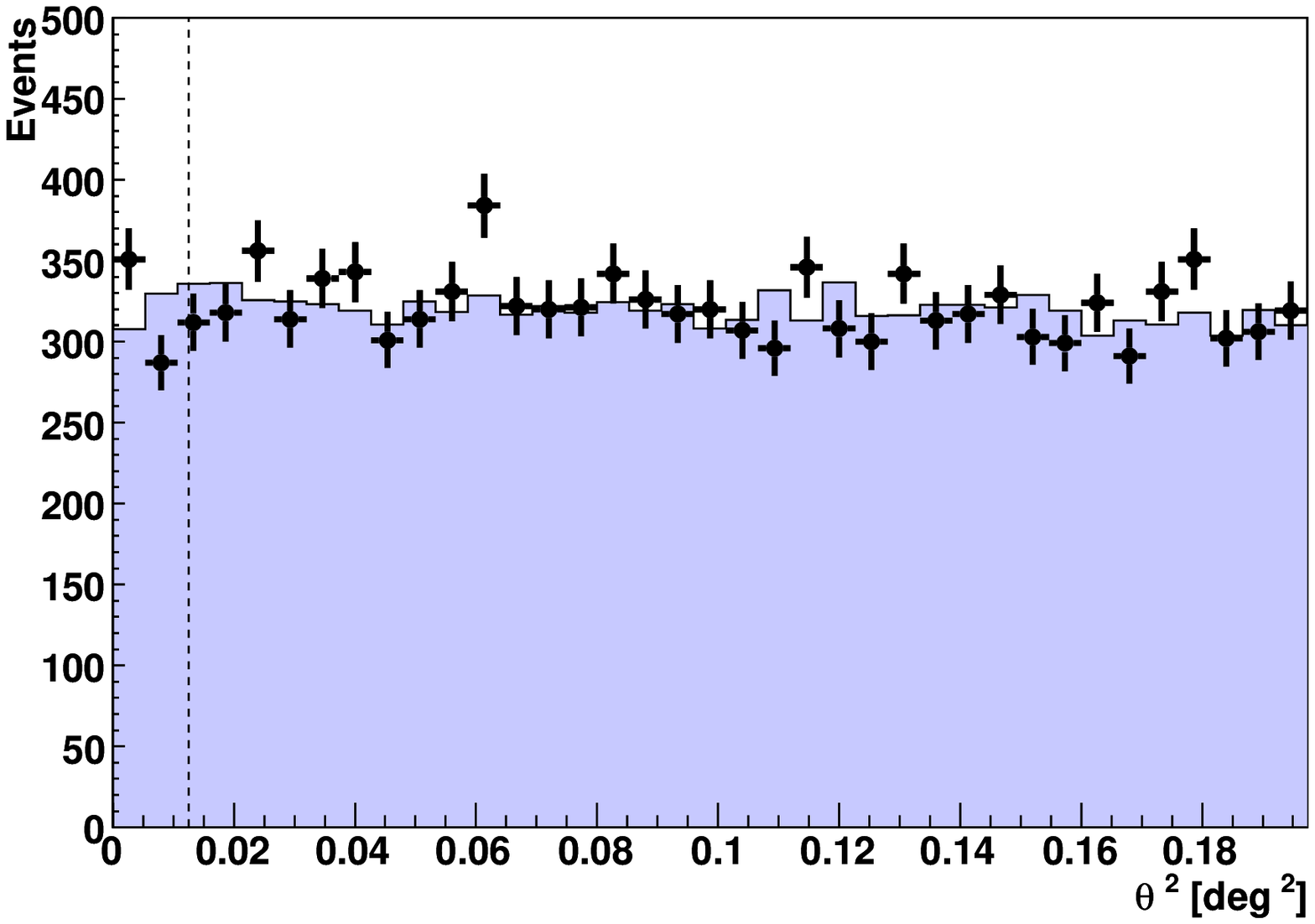}
  \caption{\small Distribution of squared angular distance from NGC 1399 for gamma-ray-like events in the ON region (dots) and in the OFF region (filled area, normalized). The dotted line represents the cut for point-like sources. Preliminary.  \normalsize}
 \label{fig:thetasq}
\end{figure}

NGC 1399 was observed with the \hess array of imaging atmospheric-Cherenkov telescopes for a total of 22.4 h (53 runs of $\sim$28 min each). After applying the standard \hess data-quality selection criteria a total of 13.9 hours live time remain. The mean zenith angle is $Z_{\mathrm{mean}} = 22^\circ$. 
The data were reduced using the standard analysis tools and selection cuts \cite{benb} and the Reflected-Region method \cite{berge} for the estimation of the background. This leads to a post-analysis threshold of 200 GeV at $Z_{\mathrm{mean}}$. No significant excess (-29 events, -1$\sigma$) is detected from NGC 1399 (see Fig. \ref{fig:thetasq} and Fig.~\ref{fig:skymap}). Results are consistent with independent analysis in the collaboration.

Assuming a photon index of $\Gamma$=2.6, the upper limit (99$\%$ confidence level; \cite{fc}) on the integral flux above 200 GeV is:
$$I \left(>200\textrm{GeV}\right) < 2.3 \times 10^{-12} \textrm{ cm}^{-2}\textrm{s}^{-1},$$
or ~1\% of the Crab Nebula flux.
\begin{figure}[htbp]
\centering
  \includegraphics[width=0.5\textwidth]{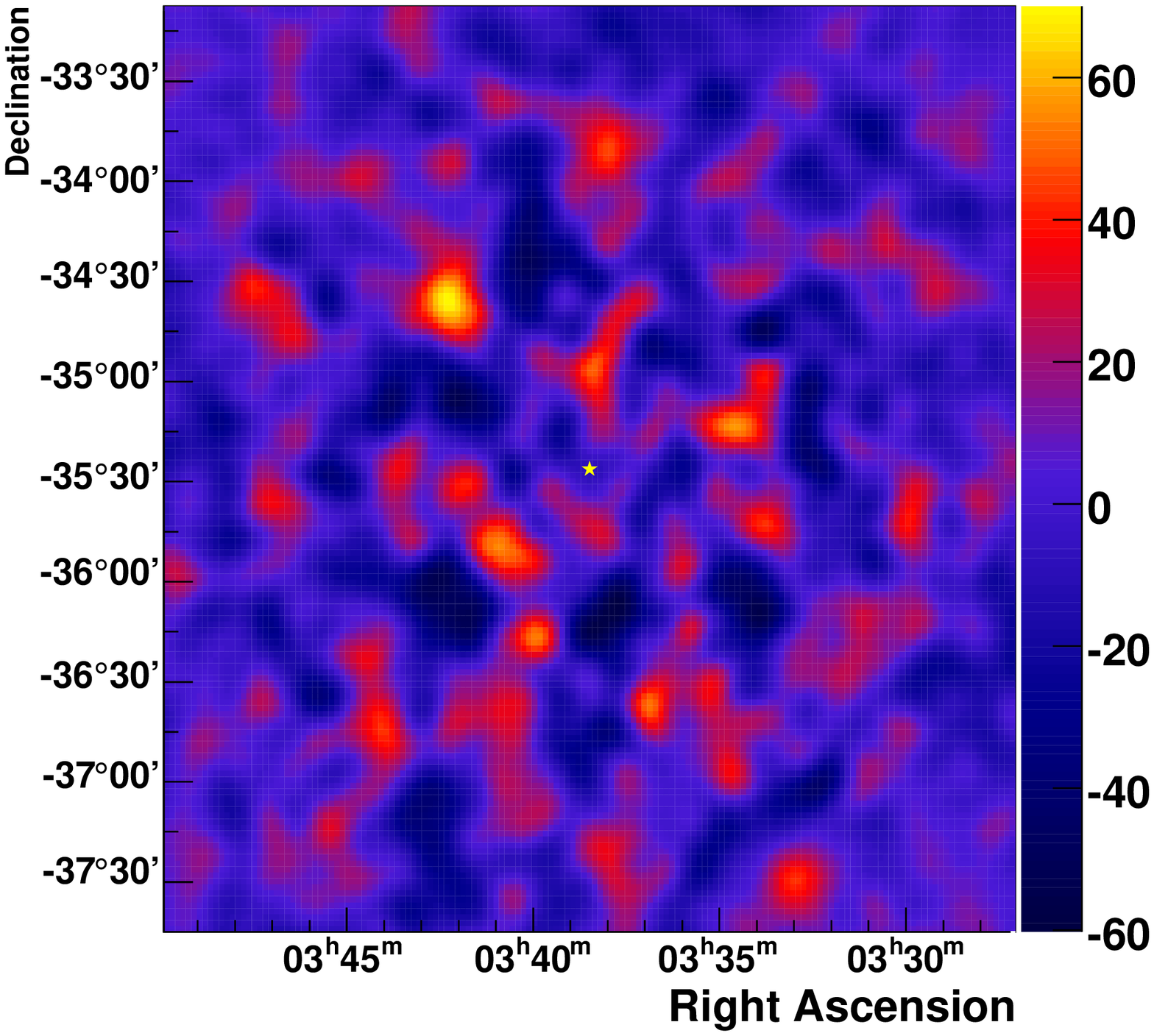}
  \caption{\small The smoothed (smoothing radius r=0.09Ë) VHE excess in the region centered on NGC 1399. The yellow star indicates the position of the optical centre of NGC 1399. Preliminary.\normalsize}
 \label{fig:skymap}
\end{figure}
\section*{Constraints from  NGC 1399 Observations}

As can be seen from the spectral energy distribution (SED) of NGC 1399 in Fig.~\ref{fig:SED}, the VHE fraction of its total energy budget is potentially not-negligible. The \hess limit on the isotropic VHE \gr luminosity is:
$$L_\gamma < 9.6 \times 10^{40} \textrm{ erg s}^{-1}.$$
Here it is assumed that the \gr emission originates solely from the nucleus, even though the entire galaxy is point-like considering the angular resolution of \hess
\begin{figure}[htbp]
\centering
  \includegraphics[width=0.5\textwidth]{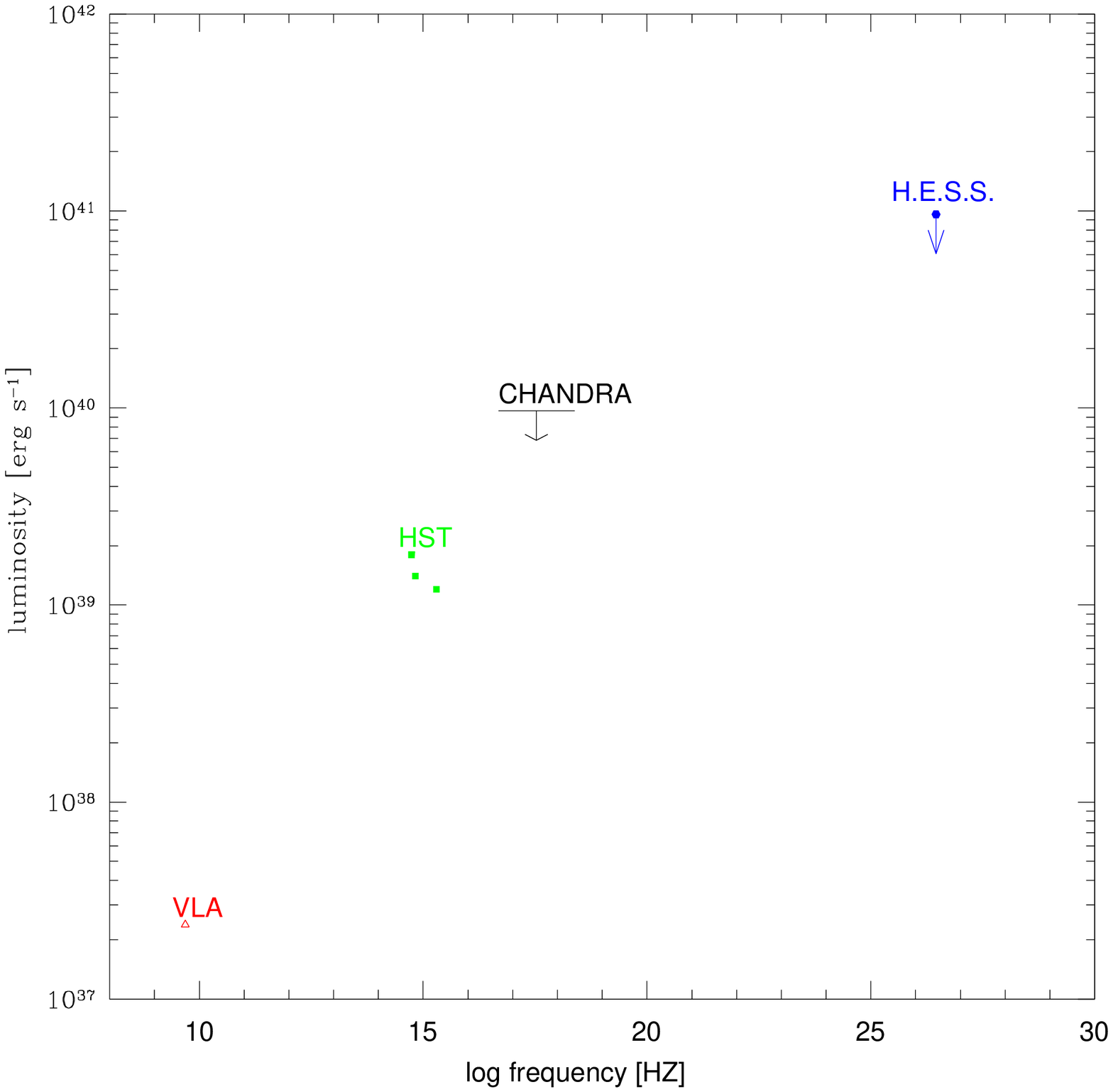}
  \caption{\small The SED of NGC 1399. All the data are for the core region. The archival points are VLA radio data (red triangles; \cite{sadl}), HST optical data (green squares; \cite{oconn}), and Chandra X-ray upper limits (solid line; \cite{lowen}). The blue dot is the  H.E.S.S. upper limit derived from the 2005 observations. Preliminary. \normalsize}
 \label{fig:SED}
\end{figure}

In the case of NGC 1399 photon-photon pair absorption would not hide any possible VHE emission. The cross section $\sigma_{\gamma\gamma}$  of this process depends on the product of the energies of the colliding photons. In the case of VHE photons, the most effective interaction is with background photons of energy:
$$\epsilon_{\mathrm{IR}} \approx \left(E/\mathrm{1TeV}\right)^{-1} \textrm{ eV.} $$

The optical depth resulting from this absorption, in a source of luminosity $L$ and radius $R$, reads:

\begin{eqnarray}
\tau\left(E,R_{\mathrm{IR}}\right) & = & \frac{L_{\mathrm{IR}}\sigma_{\gamma\gamma}}{4\pi R_{\mathrm{IR}} \epsilon_{\mathrm{IR}}}
\nonumber\\
& \simeq & 1 \left[\frac{L_{\mathrm{IR}}\left(\epsilon \right)}{10^{-7} L_{\mathrm{Edd}}}\right] \left[\frac{R_\mathrm{S}}{R_{\mathrm{IR}}} \right]\left[\frac{E}{\textrm{1 TeV}}\right],
\nonumber
\end{eqnarray}
where $R_\mathrm{S}$ is the Schwarzschild radius of the black hole and $L_{\mathrm{Edd}}$ is the Eddington luminosity.
In the system here presented, the visibility of a 200 GeV photon requires  $L_{\mathrm{IR}} <7.9 \times 10^{40} \textrm{ ergs s}^{-1}$, a condition that seems to be satisfied.

In the p-p interaction scenario, assuming that all the available power (Eq. \ref{eq:power}) will be radiated in the VHE domain, the following limit for the magnetic field is obtained from the H.E.S.S. result:
$$ B < 92.6 \textrm{ Gauss.}$$

In order to maintain gaps in the magnetosphere, as is essential for particle acceleration, pair production should be avoided. Translating this condition into an upper limit for the magnetic field yields: 
$$B < 3.6 \times 10^4 \left(M_{10}\right)^{-2/7} = 6.8\times10^4 \textrm{ Gauss.}$$

Therefore the H.E.S.S. NGC 1399 data allow plausible values of the magnetic field.
Considering the production of a 1 TeV photon via curvature emission (Eq. \ref{eq:cutoff}) requires in the case of NGC 1399:
$$B=1.3\times10^3 \textrm{ Gauss.} $$
The non-detection of NGC 1399 does not constrain the magnetic field. In all the aforementioned scenarios, hadronic and/or leptonic, no clear constraints on the magnetic field are derived.

\section*{Conclusions}
VHE emission from passive SMBH is plausible either via leptonic or hadronic processes. In order to detect this emission the giant elliptical galaxy NGC 1399 was observed by H.E.S.S. in 2005. NGC 1399 is not detected in these observations. The corresponding upper limit does not allow a firm estimation of the circumnuclear magnetic field. 

%\clearpage

\subsubsection*{Acknowledgments}
%\tiny 
The support of the Namibian authorities and of the University of Namibia in facilitating the construction and operation of H.E.S.S. is gratefully acknowledged, as is the support by the German Ministry for Education and Research (BMBF), the Max Planck Society, the French Ministry for Research, the CNRS-IN2P3 and the Astroparticle Interdisciplinary Programme of the CNRS, the U.K. Science and Technology Facilities Council (STFC), the IPNP of the Charles University, the Polish Ministry of Science and Higher Education, the South African Department of Science and Technology and National Research Foundation, and by the University of Namibia. We appreciate the excellent work of the technical support staff in Berlin, Durham, Hamburg, Heidelberg, Palaiseau, Paris, Saclay, and in Namibia in the construction and operation of the equipment. \\This work has been supported by the International Max Planck Research School (IMPRS) for Astronomy \& Cosmic Physics at the University of Heidelberg.
%\normalsize

%This is the reference to .bib file (Without .bib!)

%This in the bibtex style, is ok.
\bibliographystyle{plain}

%%%%%%%%
%  38  %
%%%%%%%%

%The paper title
\title{Wide-range multiwavelength observations of northern TeV blazars with MAGIC/HESS, Suzaku and KVA}

%Short title to print in the headers to the final publication (Not showed in this print).
\shorttitle{MWL observations for Northern TeV blazars}
%All paper authors
\authors{M.~Hayashida$^{1,*}$, S.~R\"ugamer$^{2}$, D.~Mazin$^{3}$, R.~Firpo$^{3}$, K.~Mannheim$^{2}$, F.~Tavecchio$^4$, M.~Teshima$^{1}$ on behalf of the MAGIC collaboration.\\
D.~Horns$^{5}$, L.~Costamante$^{6}$, S.~Schwarzburg$^{5}$, S.~Wagner$^{7}$ on behalf of the HESS collaboration.\\
T.~Takahashi$^{8}$, J.~Kataoka$^{9}$, G.~Madejski$^{10}$, R~Sato$^{8}$, M.~Ushio$^{8}$ for the Suzaku team
}

%Short title to print in the headers to the final puplication (Not showed in this print).
\shortauthors{M. Hayashida et al}
%All the affiliations.
\afiliations{$^1$Max-Planck-Institut f\"ur Physik, D-80805 M\"unchen, Germany, $^2$Universit\"at W\"urzburg, D-97074 W\"urzburg, Germany, $^3$Institut de F\'\i sica d'Altes Energies, Edifici Cn., E-08193 Barcelona, Spain, 
$^4$INAF/Osservatorio Astronomico di Brera, Merate, Italy,  
$^5$Institut f\"ur Astronomie und Astrophysik Eberhard Karls Universit\"at, D-72076 T\"ubingen, Germany, 
$^6$Max-Planck-Institut f\"ur Kernphysik, D-69029 Heidelberg, Germany, $^7$Landessternwarte, Universit\"at Heidelberg, K\"onigstuhl, D-69117 Heidelberg, Germany, $^8$Institute of Space and Astronautical Science/JAXA, Kanagawa, 229-8510, Japan, $^9$Tokyo Institute of Technology, Tokyo, 152-8551, Japan, $^{10}$Stanford Linear Accelerator Center, Stanford, CA, 943099-4349, USA}
\email{mahaya@mppmu.mpg.de}

%The abstract.
\abstract{We conducted multiwavelength observations of the northern TeV blazars, Mkn501 and Mkn421, employing the ground-based
$\gamma$-ray telescopes MAGIC and HESS, the Suzaku X-ray satellite and the KVA optical telescope. 
%The sources were clearly detected in all energy ranges.
The observations for Mkn501 were performed in July 2006. The source showed one of the lowest fluxes both in very high energy (VHE) $\gamma$-ray and X-ray. No significant flux variability could be found in the VHE band while an overall increase of about 50\% on a 1-day time scale could be seen in the light curve of the X-ray flux. A one-zone synchrotron self-Compton model can well describe our simultaneous spectral data of the VHE $\gamma$-ray and the X-ray emissions of Mkn501 in the quiescent state. The simultaneous observations of Mkn421 were carried out in April 2006.  The source was clearly detected in all observations and showed a high state of activity both in VHE $\gamma$-ray and X-ray. 
}

\maketitle

\addcontentsline{toc}{section}{Wide-range multiwavelength observations of northern TeV blazars with MAGIC/HESS, Suzaku and KVA}
\setcounter{figure}{0}
\setcounter{table}{0}
\setcounter{equation}{0}

%Begin the section.
\section*{Introduction}
Blazars, a sub-class of active galactic nuclei (AGNs) characterized by small angles between the jet axis and the line of sight, can provide excellent opportunities for studying particle acceleration mechanism in the jet.
One of the most successful models for the emission mechanism in the jet for TeV blazars is the synchrotron self-Compton (SSC) models~\cite{CandC02}, in which the radiation is originated from relativistic electrons.
Models based on the acceleration of hadrons can also sufficiently describe the observed emission~\cite{Man93}.
Blazars often show strong flux variability. Hence, simultaneous multiwavelength observations over a wide-energy range in different states are essential to studying the evolution of physical conditions and the shock mechanism in the jet~[e.g.]\cite{Mas97}.
However, most of the previous simultaneous multiwavelength observations could only be conducted during flaring states due to the low sensitivity of the participating $\gamma$-ray telescopes.

Multiwavelength campaigns for several northern TeV blazars were coordinated in 2006. A new generation of Imaging Atmospheric Cherenkov Telescopes (IACTs) for very high energy (VHE; $>$100 GeV) $\gamma$-ray, the MAGIC and HESS telescopes, the Suzaku X-ray satellite and the KVA optical telescope were involved in those campaigns. In this paper we present the observational results of the campaigns for Mkn501 and Mkn421.

\section*{Involved Instruments}
\subsection*{MAGIC}
The MAGIC telescope is an IACT with a 17-m diameter dish, located on the Canary Island of La Palma (28.2$^{\circ}$~N, 17.8$^{\circ}$~W, 2225 m\,a.s.l.).  The telescope is operating at a $\gamma$-ray trigger threshold of $\sim50$\,GeV and a spectral threshold of $\sim$100\,GeV.
%representing the lowest energy threshold among IACTs. 
The telescope parameters and performance are described in detail in~\cite{Crab}.

\subsection*{HESS}
The HESS array consists of four IACTs, each with a tessellated 13-m diameter mirror, located in the Khomas highlands in Namibia (23.3$^{\circ}$~S, 16.5$^{\circ}$~E, 1800 m\, a.s.l.).  Due to large zenith angle for northern objects, the observations with HESS array are sensitive to an energy range shifted towards higher energies. The telescope parameters and performance are described in detail in~\cite{Hin04}.

\subsection*{Suzaku}
The joint Japanese-US satellite Suzaku~\cite{Mit07}, launched successfully into orbit on 10 July 2005, 
has four X-ray Imaging Spectrometers (XIS) and a separate Hard X-ray Detector (HXD). The XIS are sensitive in the 0.2-10 keV band with CCDs. The HXD's silicon PIN diode array is the most sensitive detector in the 10-70 keV band thanks to the good noise shielding.
Its high sensitivity both in the soft and hard X-rays makes it an excellent instrument for studying the synchrotron component of TeV blazar emission, making the detection of the X-ray peak position feasible.

\subsection*{KVA}
KVA\footnote{more information at http://tur3.tur.iac.es/} is a 35-cm optical telescope also situated on La Palma. 
%KVA can be operated remotely via internet e.g.\,from Finland. 
Selected blazars are regularly observed with KVA as a part of the Tuorla Observatory blazar monitoring program. The KVA telescope can be fully committed to monitoring the target sources during the multiwavelength campaign.

\section*{Markarian 501}
Mkn501 (\textit{z} = 0.034) is the second established TeV blazar~\cite{Qui96}. In 1997 this source went into a state of surprisingly high activity and strong variability, becoming 10 times brighter than the Crab Nebula in the TeV range~\cite{Aha99}. In 1998-1999 the mean flux dropped by an order of magnitude~\cite{Aha01}. Recently, rapid flux variability with flux-doubling times down to 2 minutes has been reported~\cite{Mkn501}.

\subsection*{VHE $\gamma$-ray observations} 
Mkn501 was observed for 10.5 h with the MAGIC telescope in the night of 18th, 19th and 20th of July, 2006.
The observations were performed in the so-called wobble mode~\cite{Dau97}, where the object is observed with an $0.4^{\circ}$ offset from the camera center.
After the data selection by the quality and zenith angle ($< 35^{\circ}$),
the remaining data of 9.1 h were analyzed using the MAGIC standard analysis chain. The detailed information can be found in~\cite{Crab, Mkn421}. 
%Using the DISP method~\cite{DISP} for the reconstruction of the shower direction, a final cut on the $\theta^{2}$ parameter (the squared angular distance between the nominal source position and the reconstructed $\gamma$-ray direction) was applied.
An observed excess signal 
%below $\theta^{2} <  0.03\,{\rm deg}^{2}$ 
corresponding to 13.4 $\sigma$ excess was found.

The source was also observed with the HESS array during the campaign. The data analysis is currently ongoing.

\subsection*{X-ray observations}
The X-ray observation window of Suzaku was between 53934.789 and 53935.727 in MJD time (from 18th to 19th of July, 2006). The net exposure times after screening are 35 ksec in both XIS and HXD detectors.

%_____________________________________________________________
%                 A figure as large as the width of the column
%-------------------------------------------------------------
  \begin{figure}[h]
  \centering
 \includegraphics[width=7cm,clip]{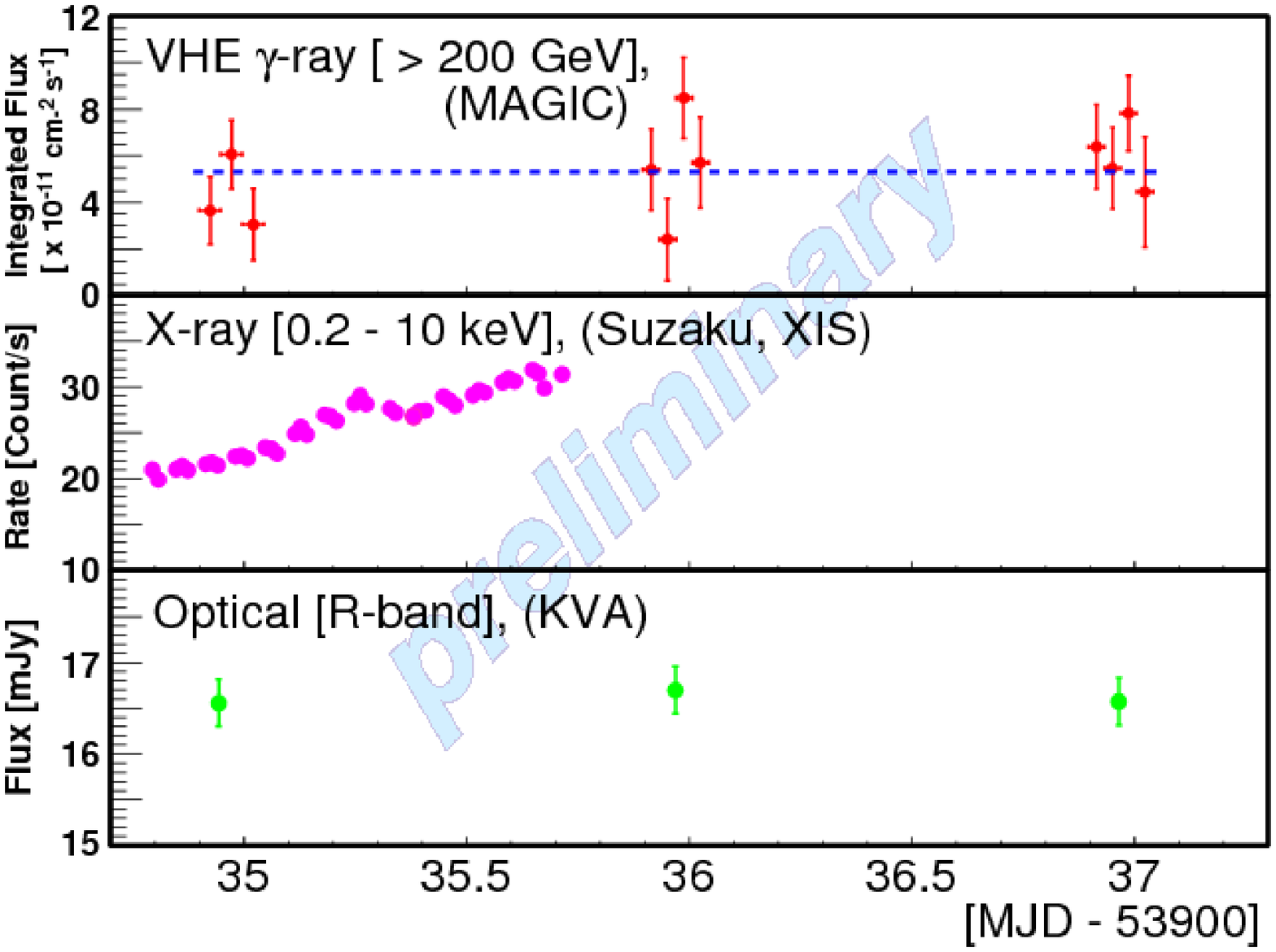}
   \caption{Light curves of Mkn501 in different energy bands during the campaign in 2006. \textbf{Top:} VHE $\gamma$-ray flux measured by the MAGIC telescope with 1-hour binning. A dotted horizontal line represents the average flux. \textbf{Middle:} The X-ray count rate recorded by the Suzaku XIS detectors. \textbf{Bottom:} Measured Optical R-band flux by KVA.}
   \label{MWLC}
   \end{figure}

%--------

\subsection*{Light curves}
Figure~\ref{MWLC} shows the light curves in various energy bands.
The source was rather quiet during the campaign. The fluxes in VHE $\gamma$-ray and optical R-band were consistent with constant levels, while the X-ray count rate was growing during the observations with an overall increase of about 50\% on a 1-day time scale between the beginning and the end of the observations.

The average integrated flux above 200 GeV is $(5.3\pm0.5)\times 10^{-11}\ {\rm cm}^{-2}\ {\rm s}^{-1}$ $ (\chi^{2}/\rm{dof} = 12.7/10)$, which corresponds to about 27\% of the Crab Nebula flux as measured by the MAGIC telescope~\cite{Crab}.

%_____________________________________________________________
%                 A figure as large as the width of the column
%-------------------------------------------------------------
  \begin{figure}[h]
  \centering
 \includegraphics[width=7cm, clip]{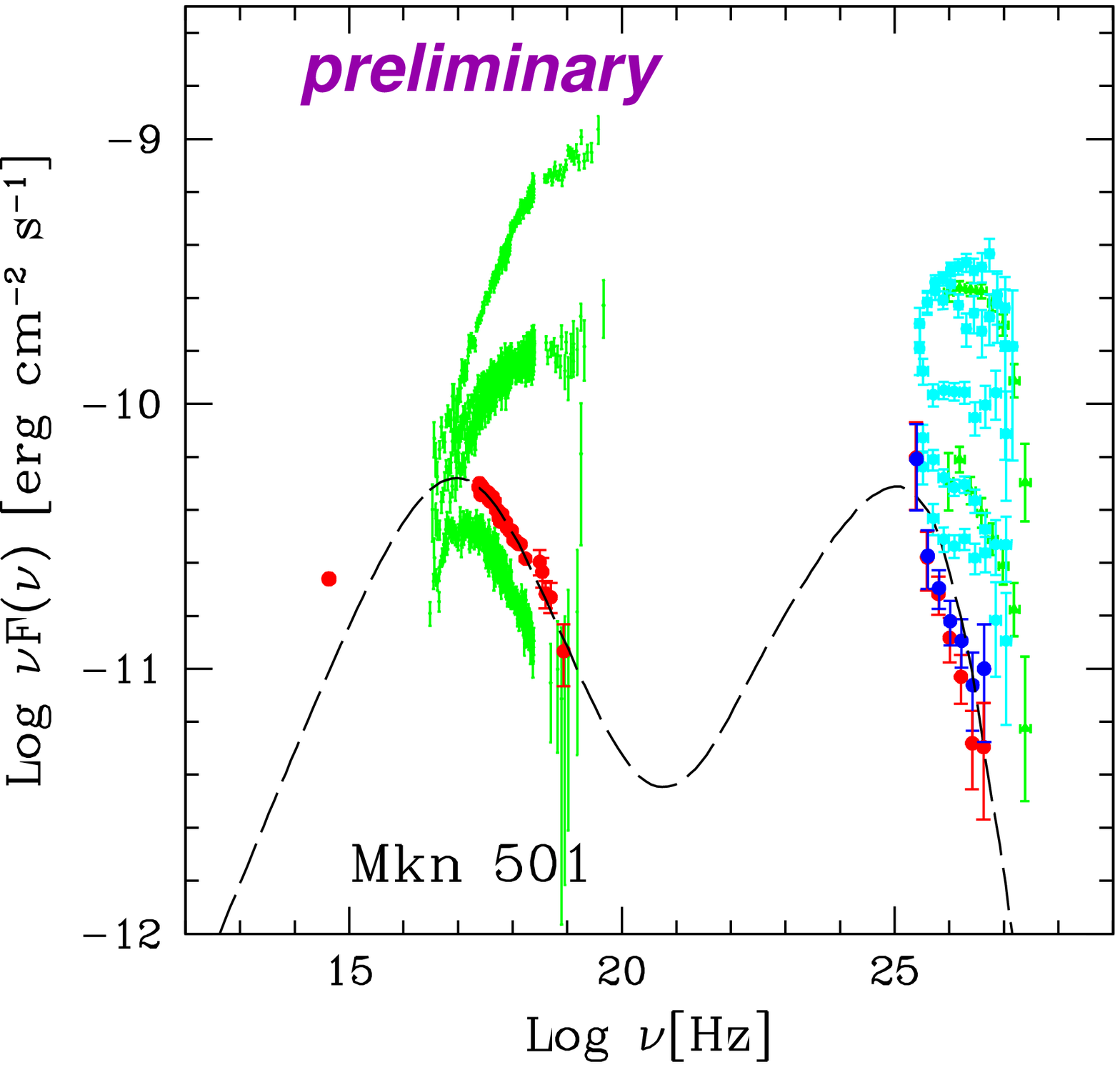}
   \caption{SED of Mkn501. The simultaneous spectral data in optical (KVA), X-ray (Suzaku) and VHE $\gamma$-ray (MAGIC) in the 2006 campaign are shown by red points. Blue points represent a "de-absorbed" spectrum corrected for the EBL absorption. The line describes the best fit to this campaign data with a one-zone SSC model developed by~\cite{Tav98, Tav01}. See text for the fit parameters. For comparison, green points denote historical data taken by BeppoSAX for X-rays data from 1997 to 1999 and  by CAT in the VHE $\gamma$-ray range in 1997~\cite{Tav01}. Flux levels from the MAGIC observations in 2005 are also denoted by cyan points~\cite{Mkn501}. 
   }
   \label{SSC}
   \end{figure}

%_____________________________________________________________
%
%\clearpage

\subsection*{Spectra}
The spectrum in the VHE band is well described by a simple power law from 85 GeV to 2 TeV with ${dN/dE} = (1.24 \pm 0.11) \times 10^{-10} \left({E}/{300\ {\rm GeV}}\right)^{-2.85\pm0.14}$ [${\rm TeV}^{-1}\ {\rm s}^{-1}\ {\rm cm}^{-2}$]. 
The flux level and the photon index of the measured spectrum are comparable to those in the lowest state among 2005 MAGIC observations (${dN/dE} = (1.36 \pm 0.21) \times 10^{-10} \left({E}/{300\ {\rm GeV}}\right)^{-2.73\pm0.29}$ \cite{Mkn501} ) for this object. 

\subsection*{Spectral Modeling}
Figure~\ref{SSC} shows the spectral energy distribution (SED) of Mkn501 with data of this multiwavelength campaign and some historical data~\cite{Mkn501,Tav01}.
The "de-absorbed" data in blue points at the VHE band were corrected for the extra-galactic background light (EBL) absorption using the "Best" fit model of~\cite{Kne04}. In optical, the host galaxy contributions $(12.0\pm0.3)$ [mJy]~\cite{Nil07} has already been subtracted. 
Since systematic errors at soft X-ray energies are still under investigation, we use only X-ray data above 1 keV for the model fit.

Assuming a uniform injection of the electrons throughout a homogeneous emission region,
%and all emissions in optical, X-ray and VHE $\gamma$-ray are generated in the same region,
we applied a one-zone SSC model for the campaign data to estimate the physical parameters of the emitting region using the code developed by \cite{Tav98, Tav01}.
Briefly, a spherical shape (blob) is adopted for the emission region with radius $R$, filled with a tangled magnetic field with intensity $B$ .
An electron distribution is described by a smoothed broken power-law energy distribution with slopes $s_1$ from $\gamma _{\rm min}$ to the break energy $\gamma _{\rm b}$ and $s_2$ up to the limit of $\gamma _{\rm max}$ and with a normalization factor of $K$. The relativistic effects are taken into account by the Doppler factor $\delta$.

$R$ is selected to be $1\times 10^{15}$ cm, which has been adopted in~\cite{Mkn501} for the SED during the rapid flare observed with MAGIC in 2005.
%which is equivalent to a previous study~\cite{Tav01}. 
Since no cut off can be seen both in the X-ray and VHE $\gamma$-ray spectra, $\gamma _{\rm min}$ and $\gamma _{\rm max}$ are fixed at 1 and $10^7$, respectively.
The best fit was achieved with the following parameters: $\delta=20$, $B=0.26$ G, $K=1\times 10^{5}\,{\rm cm}^{-3}$, $\gamma _{\rm b}=6.7\times 10^4$, $s_1=2$ and $s_2=4$.
This one-zone SSC model can reproduce well the obtained X-ray and VHE $\gamma$-ray fluxes in the low state of activity during the campaign. However, it is apparent that the model underestimates the optical flux. This can be explained by the assumption that the emission from radio to UV has another origin than the high energy emission. This interpretation has already been applied to previous SEDs of Mkn501 by~\cite{Katar01}.
%The Doppler factor ($delta$=28) is rather high but still compatible with the previous results ($delta$=8~36)~\cite{}. Together with $R$ and the 1 day time variability ($t_{var}$) in our X-ray data, they are consistent with $R < ct_{var}\delta$.
Compared to historical measurements our data show one of the lowest states both in X-ray and VHE $\gamma$-ray. 
The historical data suggest a correlation between the peak positions and the source luminosity~\cite{Tav01}. Following~\cite{Mas97, Tav01} we can assume this feature is related to the evolution of $\gamma _{\rm b}$. In that framework we can find our derived value of $\gamma _{\rm b}$ to be small compared to
previous observation results~[e.g.]\cite{Mkn501, Katar01, Pia98, Tav01}. More details on the analysis and the results will be found in~\cite{MWL501}.

\section*{Markarian 421}
Mkn421 (\textit{z} = 0.030) is the closest known source and the first extragalactic one detected in the TeV energy range using IACTs~\cite{Pun92}, and it is one of the best studied TeV $\gamma$-ray blazar.

\subsection*{Observations and Results}
The multiwavelength observations for this source were conducted in the night of April 28th, 2006. The X-ray observations with Suzaku were carried out between 53853.267 and 53854.271 in MJD time.
The source was observed by the MAGIC telescope for 3.8 h and by the HESS array for 1.5 h. Both observations were performed during the Suzaku pointing to the source. Clear detections of signals can be found in all observations. 
%Figure~\ref{Mkn421SSC} shows the SED of Mkn421 with data taken during this campaign and some previous observation results. 
The X-ray spectrum is extracted from the data which are exactly coincident with the HESS or MAGIC observation times.
The spectrum in the X-ray band as well as in the VHE $\gamma$-ray band indicates that the source showed a high state and a rather stable activity during the simultaneous observations. Details of the observational results of this campaign will be discussed in~\cite{Tak07}.

\section*{Acknowledgments}
The MAGIC collaboration thanks the IAC for the excellent working conditions at the ORM. The MAGIC project is supported by the German BMBF and MPG, the Italian INFN, the Spanish CICYT, the Swiss ETH and the Polish MNiI.

%This is the reference to .bib file (Whitout .bib!)

%This in the bibtex style, is ok.
\bibliographystyle{plain}

%%%%%%%%
%  39  %
%%%%%%%%

%The paper title
\title{Simultaneous observation of GRB060602B with the H.E.S.S. Air Cherenkov array}
%Short title to print in the headers to the final publication (Not showed in this print).
\shorttitle{Simultaneous observation of GRB060602B with H.E.S.S.}

%All paper authors
\authors{Pak-Hin Tam$^{1}$, Konrad Bernl\"ohr$^{2}$, Paula Chadwick$^{3}$, Jim Hinton$^{1,2,4}$, Dalibor Nedbal$^{2}$, Gerd P\"uhlhofer$^{1}$, Stefan Wagner$^{1}$ for the H.E.S.S. collaboration}
%Short title to print in the headers to the final publication (Not shown in this print).
\shortauthors{P.-H. Tam et al. for the H.E.S.S. collaboration}
%All the affiliations.
\afiliations{
$^1$Landessternwarte, Universit\"{a}t Heidelberg, K\"{o}nigstuhl, D 69117 Heidelberg, Germany\\
$^2$Max-Planck-Institut f\"{u}r Kernphysik, P.O. Box 103980, D 69029 Heidelberg, Germany\\
$^3$University of Durham, Department of Physics, South Road, Durham DH1 3LE, U.K.\\
$^4$School of Physics \& Astronomy, University of Leeds, Leeds LS2 9JT, U.K. }
\email{phtam@lsw.uni-heidelberg.de}

%The abstract.
\abstract{
    On June 2, 2006, the {\it Swift} Burst Alert Telescope (BAT) triggered a bursting event in the 15-350 keV energy band. The burst position was being observed with the H.E.S.S. array of IACTs before the burst, throughout the duration of the burst, and after the burst. In particular, the burst position accidentally fell in the f.o.v. of the H.E.S.S. camera when the burst occurred. This is the first completely simultaneous observation of a soft gamma-ray bursting event with an IACT instrument. A search for VHE gamma-rays coincident with the burst event as well as that during the afterglow period was performed. No signal was found during the period covered by the H.E.S.S. observation. The {\it Swift} X-ray Telescope, which started observation 83~seconds after the BAT trigger, detected an X-ray counterpart of the event. No optical/infrared counterpart was found. Due to the very soft BAT spectrum (photon index $\Gamma\approx5$) of the burst compared to other {\it Swift} GRBs and its proximity to the galactic center, the burst might have been caused by a galactic X-ray burster (e.g. a low-mass X-ray binary). However, the possibility of it being a cosmological GRB cannot be ruled out. Since the nature of the event is still unclear, we discuss the implications according to the two different bursting scenarios.
}

\maketitle

%%%%% Begin GRB %%%%%%
\addtocontents{toc}{\protect\contentsline {part}{\protect\large Gamma-Ray Bursts (GRB)}{}}
\addcontentsline{toc}{section}{Simultaneous observation of GRB060602B with the H.E.S.S. Air Cherenkov array}
\setcounter{figure}{0}
\setcounter{table}{0}
\setcounter{equation}{0}

%Begin the section.
\section*{Introduction}

    The temporally and spatially unpredictable, and fast-fading nature of gamma-ray bursts (GRBs)
    makes it operationally rather difficult to study the prompt phase of GRBs simultaneously
    in other wavelengths. There are two operational techniques currently
    employed in the very-high-energy (VHE; $100\,\mathrm{GeV}-100\,\mathrm{TeV}$) $\gamma$-ray regime: (1)
    To slew quickly to the GRB position provided by a burst alert from
    satellites. This technique is suitable for IACTs such as H.E.S.S. which have a field of view (f.o.v.)
    of several degrees. However, there is always a delay in time for IACT operating in this GRB-follow-up
    mode, as long as the GRB position lies outside the camera f.o.v. at the GRB onset.
    The MAGIC telescope, operating in this mode, was able to slew to the position
    of GRB~050713a, 40 seconds after the GRB onset, while the prompt keV emission was still active. A total of 37-minute observation was made and no evidence of emission above 175 GeV was obtained \cite{albert06}; (2) To observe a large part of the sky continually, at the expense of a much lower sensitivity at the GRB position than the IACT technique. This technique is used, e.g. by the water Cherenkov detector Milagro \cite{Milagro}.

    Here we report on the first completely simultaneous observation of a
    soft $\gamma$-ray bursting event (a possible GRB) with H.E.S.S., an IACT
    instrument. The burst position accidentally fell in the f.o.v. (albeit with a large offset from the center of the f.o.v.)
    of the H.E.S.S. camera when the burst occurred.

\section*{GRB 060602B}

    At 23:54:33 UT on 2 June, 2006 (denoted by $t_0$), the Burst Alert Telescope (BAT)
    on board \emph{Swift}, which operates in the 15-350 keV energy band,
    triggered a bursting event. This event was announced as a gamma-ray burst (GRB) and designated GRB~060602B (e.g. \cite{jalinek5198,kubanek5199,schady5200}). The refined BAT
    position was (RA,Dec) = ($17\mathrm{^h}49\mathrm{^m}28\mathrm{^s}.2, -28\mathrm{^o}7'15".5$) (J2000, \cite{palmer5208}). The BAT light curve showed a single-peaked structure lasting from $t_0-1.0\,\mathrm{s}$ to $t_0+12.0\,\mathrm{s}$.
    The peak was the strongest in the 15-25~keV energy band among other bands and was absent above 50~keV.
    $T_{90}$ (defined as the duration when 90\% of the total fluence of the
    15-350~keV was emitted) is $9.0 \pm 2\,\mathrm{sec}$.
    The time-averaged energy spectrum from $t_0-1.1\,\mathrm{s}$ to $t_o+8.8\,\mathrm{s}$ can best be fit by a simple power law (PL), with a photon index of $5.02 \pm 0.52$ (one of the softest among \emph{Swift} GRBs).
    The fluence in the 15-150 keV band was $(1.8 \pm 0.2) \times 10^{-7} \mathrm{ergs\,cm}^{-2}$
    \cite{palmer5208}.

    Although no counterpart was found in the optical/IR bands \cite{kubanek5199,khamitov5205, blustin5207,melandri5229}, an X-ray counterpart was found using \emph{Swift}/XRT and was fading as a PL with an index $1.05\pm0.07$ for at least 11 hours, after a small rise before
    $t_0+200\,\mathrm{s}$ \cite{beardmore5209}.

The time-averaged X-ray energy spectrum during $100\mathrm{s}-11.4\mathrm{ks}$ after $t_0$ can be fit equally well by an absorbed PL or an absorbed blackbody (BB) model. A photon index of
$2.6^{+1.0}_{-0.8}$ and a hydrogen column density of $3.8^{+2.1}_{-1.4}\times 10^{22}
\mathrm{cm}^{-2}$ were obtained for the PL model. For an absorbed BB
model, a temperature of $0.90^{+0.23}_{-0.17}$~keV and a column density of
$1.8^{+1.2}_{-0.8} \times 10^{22} \mathrm{cm}^{-2}$ were obtained.

    The galactic coordinates of the source were (lon, lat) $=$ ($1.15^\circ$, $-0.30^\circ$). This position has raised up a possibility of the
    event originating from a galactic source (eg. an X-ray burster). The fact that the BAT spectrum was one of the softest \emph{Swift} GRBs could be used to argue for this scenario \cite{palmer5208}.

\section*{The H.E.S.S. Observations}

The H.E.S.S. array is a
system of four 13m-diameter imaging atmospheric Cherenkov telescopes
located in the Khomas Highland of Namibia.
The system has a point source sensitivity above 100 GeV of
$\sim3.0\times~10^{-13} \mathrm{cm}^{-2} \mathrm{s}^{-1}$ ($1\%$ of
the flux from the Crab nebula) for a $5 \sigma $ detection in a 25~h
observation.
 Its has a f.o.v. of radius $\sim~2.6^\circ$ (at this radius the relative gamma-ray acceptance is about 10\% of that at the targeting position of the telescopes), thus enabling it to
detect serendipitous sources, as have been demonstrated in the galactic scan
survey \cite{aha05a}.

The position of GRB~060602B was being observed with H.E.S.S. before
the burst, throughout the duration of the burst, and after the
burst. The source offsets from the observation positions were large (up to $\sim~2.8^\circ$)
in the beginning. %The observation pattern is shown in
%Table~\ref{obspattern}. The zenith angles (ZA) and source offsets
%from the observation position are shown for each observation period.
A total of 4.9 hours of observation was obtained during the night of
2-3 June, 2006. This includes 1.7hr \emph{pre-burst}, 9s \emph{prompt}, and 3.2hr \emph{afterglow} epochs.
An additional 4.7 hours of observation at the burst position was
obtained during the next 3 nights. All data were taken in good
weather conditions and good hardware status. The observations were
taken with the source position placed at different offsets relative
to the center of the f.o.v. of the telescopes. This is because most
observational runs were not dedicated to the position
of the event GRB~060602B.

%A total of 4.9 hours of observation was obtained during the night of
%2-3 June, 2006. Both non-GRB-dedicated observations (which covered
%the \emph{pre-burst} (1.7 hrs), \emph{prompt} (9s), as well as part
%of the \emph{afterglow} epochs, 1.3 hrs) and target-of-oppurtunity
%(ToO) observations under the H.E.S.S. GRB observation program (which includes
%the subsequent 1.9 hrs of the \emph{afterglow} epoch) are included. An
%additional 4.7 hours of observation at the burst position was
%obtained during the next 3 nights. All data were taken in good
%weather conditions and good hardware status. The observations were
%taken with the source position placed at different offsets relative
%to the centre of the f.o.v. of the camera. This is because most
%observational runs were not dedicated observations of the position
%of the event GRB~060602B.

The source position fell into the H.E.S.S. f.o.v. during other H.E.S.S. observations (which targeted objects like Sgr A*) as well. This allows us to investigate a possible long-term emission of the source at VHE energies. It is of particular interest in the X-ray burst (XRB) scenario, which we will discuss later.

%                                                One column figure
%----------------------------------------------------------- simultaneous
\begin{figure*}[th]
\begin{center}
\includegraphics [width=0.70\textwidth]{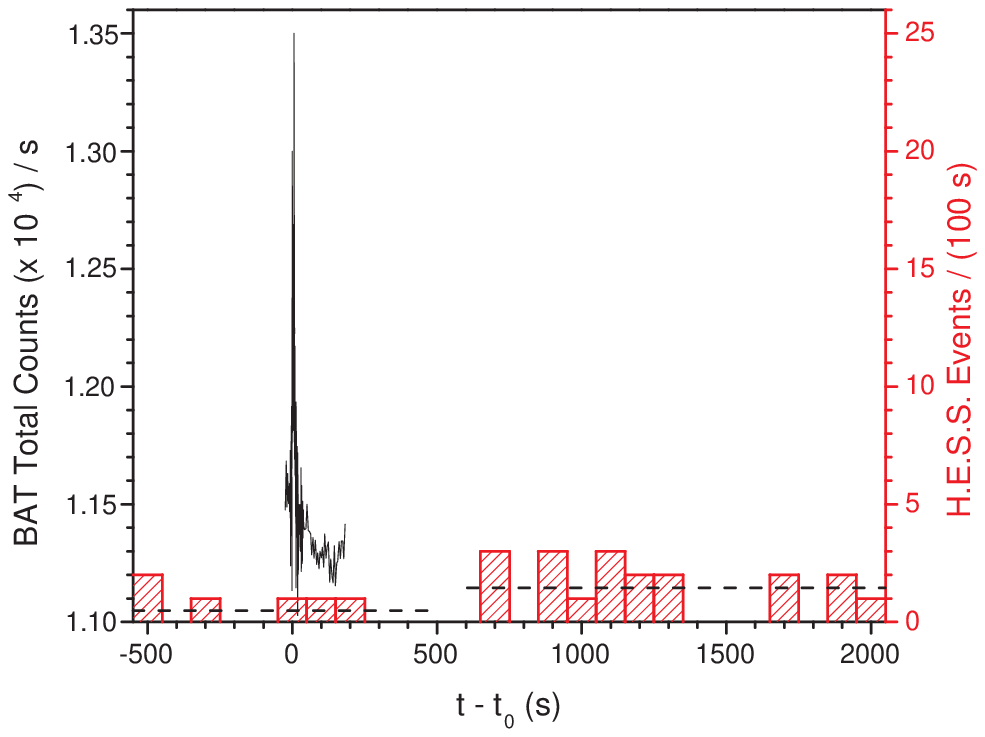}
\caption{Shaded, red histogram: Events observed with H.E.S.S. within a circular region of radius $\theta_{\rm cut}=0.32^\circ$ (for $t < t_0+500$s) and $\theta_{\rm cut}=0.2^\circ$ (for $t > t_0+600$s) centered at the \emph{Swift}/XRT position in 100-second
bins. The dashed horizontal lines indicate the expected number of background events in
the circular regions, using the reflected-region background model~\cite{berge07}. The difference in event rates is due to the different offsets, zenith angles, and $\theta_{\rm cut}$ for different periods. The gap between 500s and 600s is due to the transition of observation runs. Solid, black curve: \emph{Swift}/BAT light curve in the 15-150 keV band. }
\end{center}
\label{simultaneous}
\end{figure*}

\section*{Data Analysis}

Calibration of data, event reconstruction and rejection of the cosmic-ray background (i.e. $\gamma$-ray event selection criteria) were performed as described in~\cite{aha06a}. On-source data were taken from a circular region of radius $\theta_{\rm cut}$ centered at the source.
    %Unlike cosmic-ray particles which are coming isotropically from the sky, any gamma-ray photons are expected to come from positions around the known source position. Applying a $\theta$-cut thus helps lower the background level. However, the point spread function from an source-offset position of 2.9 degrees is considerably more extended than that for a normal offset. This is due to geometrical considerations such as fewer air-shower images available in direction reconstruction and that images used are only from directions which are always to one side of the source position, thus worsening the accuracy of the direction reconstruction.
The background was then estimated using the reflected-region model as described in \cite{berge07}, which makes use of the off-source data obtained during the same observation period.
    %The relative acceptance of gamma-ray photons at an offset of 2.9 degrees from the observation position is about 0.05
    %    that of the nominal one (i.e. at an offset of 0.5 degree), according to Monte-Carlo
    %    simulations of air showers. %Effective areas at 3.0 degree source offsets at zenith angles 0 and 20 were derived.

    Two sets of analysis cuts were applied to search for a signal. These include standard cuts (as described in \cite{aha06a}) and soft cuts (with lower-energy thresholds, as described in \cite{aha06b}). Furthermore, for the periods with large offsets, we also used larger $\theta_{\rm cut}$'s to match the larger point spread functions (PSFs).
%The standard cuts are most sensitive for a 10\%
%     Crab flux source with photon index $\Gamma=2.6$. The soft cuts are optimized for sources with steep spectra ($\Gamma=5.0$) and weak flux (1\% Crab in $> 200$ GeV), thus having the best sensitivity at lower energies.
%The latter is useful for a cosmological source like a GRB since extra-galactic background light (EBL) would greatly soften the VHE radiation from the source if there was any.
%The energy thresholds ($E_{\rm th}$) used in the standard cut analysis for each observational period are shown in Table~\ref{obspattern}.

Figure~1 shows the independent events observed
within a circular region of radius $\theta_{\rm cut}=0.32^\circ$ (for $t < t_0+500$s) and $\theta_{\rm cut}=0.2^\circ$ (for $t_0+600\mathrm{s} < t < t_0+2050\mathrm{s}$) centered
at the source. The $\theta_{\rm cut}$ values were chosen to match the different PSFs for observations with different source offsets.% On top of it is the
%\emph{Swift} light curve in 15-150 keV band.

%----------------------------------------------------------- light curve (HESS Upper limits, logscale)
%   \begin{figure}
%   \centering
%   \includegraphics[width=0.48\textwidth]{HESS_ULfg2.eps}
%      \caption{The 99\% c.l. flux upper limits~\cite{feldman98} at energies $>1.1\,\mathrm{TeV}$ derived from H.E.S.S. observations during the prompt and afterglow epochs.}
%         \label{lightcurve}
%   \end{figure}

\section*{Results}

No evidence of excess events as observed with H.E.S.S.
in any time before, during, or after the soft-gamma-ray bursting
event is seen. There was no hint of radiation in the next 3 nights as well.
%The 99\% confidence level (c.l.) upper limits using the method of Feldman and Cousins \cite{feldman98}
%for every observation run using standard cuts are included in
%Table~\ref{obspattern}.
%We show in Figure~\ref{lightcurve} the 99\% confidence level (c.l.) upper limits using the method of Feldman \& Cousins~\cite{feldman98}, at VHE energies $>1.1$~TeV derived from H.E.S.S. observations during the prompt and afterglow epochs.

%The nightly and run-by-run (for the first
%night) light curves are shown in Figure~\ref{lightcurve} and
%\ref{lightcurve_2} respectively. All flux values shown in the
%figures are consistent with zero.

We also looked into the H.E.S.S. observations from 2004 to 2006
containing the XRT position in the f.o.v. Again no emission was found and the 99\% confidence level (c.l.) upper limits using the method of Feldman \& Cousins~\cite{feldman98} derived
from 128.3 hours of observation using standard cuts is $1.06\times
10^{-12}\,\mathrm{photons\,cm^{-2}\,s^{-1}}$ above 200 GeV (i.e. 0.5\%
of the Crab flux).

\section*{Discussions}

The nature of the bursting event that occurred at 23:54:33 UT on 2
June, 2006 is unclear. %The empirical properties (fluence, time
%duration, softening of its spectrum) of the event are consistent
%with it being either a GRB or an XRB.
It was firstly announced as a GRB by \cite{jalinek5198} and several other teams including \cite{kubanek5199,schady5200}. Thus the event has been designated GRB~060602B. However, its very soft spectrum (photon index $\Gamma\approx5$) in the 15-150 keV band % (for comparison, the mean
%of a large sample of other \emph{Swift} GRBs is 1.68)
and its galactic position support a galactic origin of the event, as first suggested in~\cite{palmer5208}. %% (LMXBs),
%%believed to be the progenitors of XRBs, concentrate toward the
%%galactic bulge of our galaxy.
A faint XMM-Newton source was detected
on 23 September 2000 in the proximity of the \emph{Swift}/XRT
position, as first noticed by \cite{halpern5210}. Whether this source is related to the bursting event on 2
June, 2006 is unclear.
%It is
%listed in the second XMMSSC-XMM-Newton Serendipitous Source Catalog
%as having a flux of $2.31\times10^{-13}
%\mathrm{erg\,cm^{-2}\,s^{-1}}$ in the 0.2-12 keV range.
However, the possibility of the event as a cosmological GRB is not ruled out.
Since the results are still not conclusive, we briefly
discuss the implications of the H.E.S.S. observations according to
these two scenarios.

\subsection*{The gamma-ray burst scenario}

If GRB~060602B were a cosmological GRB, it would be a GRB with a very steep keV spectrum and happened to be to the direction next to the galactic center.
Our observation with H.E.S.S. would be the first strictly simultaneous observation of a GRB with an IACT
instrument during its prompt phase. %This also means that the dense regions near the galactic centre
%locate on the path of GRB radiation. Any optical counterpart would
%be absorbed (consistent with no optical emission detected). %, while
%the keV radiation of the GRB would not be affected.

%As the redshift of the progenitor is unknown under this scenario, we
%could not set the intrinsic upper limits to the VHE flux.

%For a GRB with an assumed redshift $z= 0.1$, the 9-sec and 3-hour H.E.S.S. upper
%limits would translate into intrinsic fluence limits of $2.5\times
%10^{-7}\,\mathrm{erg\,cm^{-2}}$ and $9.6\times 10^{-8}\,\mathrm{erg\,cm^{-2}}$ respectively. Here the EBL absorption was applied assuming the P0.45 EBL model used in
%\cite{aha06c}.

High-energy components from GRBs have been predicted (e.g. \cite{wang01,peer05}) and
observed (e.g. in the case of GRB~940217 up to $\sim18$~GeV
\cite{hurley94}) in the prompt and afterglow phases. The H.E.S.S. data can be used to constrain the
level of such a high-energy component of GRB~060602B at VHE energies.

\subsection*{The X-ray burst scenario}

If GRB~060602B were a galactic XRB, it would be the first
simultaneous observation of an XRB with an IACT instrument. Its
galactic position suggests that it is quite near the galactic center
and thus probably at a distance of $\sim 8$~kpc. Any optical
emission would have been absorbed.

XRBs have only been detected from low-mass-X-ray binaries \cite{lewin95}. Early claims of persistent detections in the VHE regime of this class of objects were not confirmed with more sensitive experiments (see e.g. \cite{weekes92}).
%(see e.g. \cite{reynolds93}.
Our long-term upper limit of the burst position is among the most constraining ones. Possible models which suggest continual VHE emission from X-ray binaries were proposed \cite{cheng91} (see also \cite{moskalenko95} for a review).

%__________________________________________________________________

\section*{Conclusions}

For the first time, a strictly simultaneous observation of a
soft $\gamma$-ray bursting event with an IACT instrument without time
delay were obtained on 2 June, 2006.

The burst position was being observed with H.E.S.S. at VHE energies
before the burst, throughout the duration of the burst, and after
the burst. This was also the first time a soft
$\gamma$-ray bursting event was observed with an IACT instrument before its onset. A search for TeV
signal coincident with the burst event as well as that during the
afterglow period was performed. No signal has been found during the
period covered by the H.E.S.S. observation. The data analysis is still ongoing and final results will be published elsewhere \cite{aha07}.

%\section*{Acknowledgements}
%The support of the Namibian authorities and of the University of Namibia
%in facilitating the construction and operation of H.E.S.S. is gratefully
%acknowledged, as is the support by the German Ministry for Education and
%Research (BMBF), the Max Planck Society, the French Ministry for Research,
%the CNRS-IN2P3 and the Astroparticle Interdisciplinary Programme of the
%CNRS, the U.K. Science and Technology Facilities Council (STFC),
%the IPNP of the Charles University, the Polish Ministry of Science and
%Higher Education, the South African Department of
%Science and Technology and National Research Foundation, and by the
%University of Namibia. We appreciate the excellent work of the technical
%support staff in Berlin, Durham, Hamburg, Heidelberg, Palaiseau, Paris,
%Saclay, and in Namibia in the construction and operation of the
%equipment.
%This is the reference to .bib file (Whitout .bib!)
%\bibliography{libros}
%This in the bibtex style, is ok.
\bibliographystyle{plain}

%%%%%%%%
%  40  %
%%%%%%%%

%The paper title
\title{Gamma-ray burst observations with the H.E.S.S. Air Cherenkov array}
\shorttitle{GRB Observations with H.E.S.S.}

%All paper authors
\authors{Pak-Hin Tam$^{1}$, Paula Chadwick$^{2}$, Yves Gallant$^{3}$, Dieter Horns$^{4}$, Gerd P\"uhlhofer$^{1}$, Gavin Rowell$^{5}$, Stefan Wagner$^{1}$ for the H.E.S.S. collaboration}
%Short title to print in the headers to the final publication (Not shown in this print).
\shortauthors{P.-H. Tam et al. for the H.E.S.S. collaboration}
%All the affiliations.
\afiliations{
$^1$Landessternwarte, Universit\"{a}t Heidelberg, K\"{o}nigstuhl, D 69117 Heidelberg, Germany\\
$^2$University of Durham, Department of Physics, South Road, Durham DH1 3LE, U.K. \\
$^3$Laboratoire de Physique Th\'eorique et Astroparticules, IN2P3/CNRS, Universit\'e
Montpellier II, CC 70, Place Eug\`ene Bataillon, F-34095 Montpellier Cedex 5, France\\
$^4$Institut f\"ur Astronomie und Astrophysik, Universit\"at T\"ubingen, Sand 1, D 72076 T\"ubingen, Germany\\
$^5$School of Chemistry \& Physics, University of Adelaide, Adelaide 5005, Australia}
\email{phtam@lsw.uni-heidelberg.de}

%The abstract.
\abstract{
    Gamma-ray bursts (GRBs) are among the potential very-high-energy (VHE) gamma-ray and cosmic-ray sources. Particles are accelerated to highly-relativistic speeds. This might give rise to emission of VHE gamma-ray and/or cosmic-ray particles with ultra-high energy $>10^{19}$~eV. Despite its generally fast-fading behavior seen in many longer wavebands, the time evolution of any VHE radiation is still not clear. In order to probe the largely unexplored VHE spectra of GRBs, a GRB observing program has been set up by the H.E.S.S. collaboration. With the high sensitivity of the H.E.S.S. array and given favorable observational conditions, VHE flux levels predicted by GRB models are within reach. Extra-galactic background light (EBL) absorption is considered in cases where redshifts of the GRBs are reported. We present the H.E.S.S. VHE gamma-ray observations of and results from some of the reported GRB positions during the past few years, including recent observations in early 2007.
}

\maketitle

\addcontentsline{toc}{section}{Gamma-ray burst observations with the H.E.S.S. Air Cherenkov array}
\setcounter{figure}{0}
\setcounter{table}{0}
\setcounter{equation}{0}

%Begin the section.
\section*{Introduction}

Gamma-ray bursts (GRBs), being established as originating from highly-relativistic ejecta, are among potential VHE gamma-ray and cosmic-ray sources. First detected in late 1960s, the origin of GRBs is still not well understood, especially compared with other objects such as pulsars and quasars which were also detected in the same era but the origins of which are much more understood nowadays than GRBs. The main reason is that to obtain multi-wavelength information other than gamma-rays proves to be operationally very difficult. Two breakthroughs in understanding GRBs include the establishment of the isotropic spatial distribution by the BATSE experiment on board CGRO in the 1990s and the discovery of longer-wavelength counterparts after the launch of BeppoSAX in Feb 1997. Since then not only GRBs are confirmed to be of cosmological origin, but the theories of GRBs and their modeling have drawn a lot of attentions from the astrophysical community.

After the highly variable radiation seen in keV-MeV gamma-ray energies (known as the prompt GRB phase), fast-fading behavior is generally seen in counterparts in longer wavebands (known as the afterglow phase). It is believed that during the GRBs, particles are accelerated to highly-relativistic speeds. Highly-relativistic particles might give rise to emission of VHE gamma-ray and/or cosmic-ray particles with ultra-high energy $> 10^{19}$eV (see e.g. \cite{waxman06}). Although VHE emission from GRBs during the prompt GRB phase or the afterglow phase is predicted, its flux level and temporal behavior is not clear, given that no emission in this energy range has ever been detected unambiguously by now. With the high sensitivity of the H.E.S.S. array and given favorable observational conditions, VHE flux levels predicted by GRB models are within reach.

\section*{VHE emission from GRBs}

The highest energy radiation from GRBs ever detected firmly by any instrument was a $\sim18$~GeV photon coming from GRB~940217, detected using EGRET about 1.5 hour after the onset of the GRB~\cite{hurley94}. From the theoretical point of view, photons with energies up to $\sim10$ TeV from GRBs are expected (for review, see e.g. \cite{zhang04} and references therein). In one case considered by \cite{zhang01} where electron IC emission dominates, an energy flux of about $5\times10^{-12}\mathrm{erg}\,\mathrm{cm}^{-2}\mathrm{s}^{-1}$ at 1 TeV one day after GRB onset is predicted\footnote{The attenuation factor due to EBL absorption is about 0.1 at 1 TeV for a GRB with $z=0.15$ ~\cite{aha06a}.}, if one assumes a redshift of 0.15. This is within H.E.S.S. detection limit. The detection of VHE photons (and its quantity) or upper limits (in cases of null detections) could be used to constrain GRB properties, eg. bulk Lorentz factor and ambient density~\cite{peer05,wang05}.

Currently, the most sensitive detectors in the VHE gamma-ray regime are air Cherenkov systems, including H.E.S.S., MAGIC, VERITAS, and Whipple. While still no detection using any of the instruments is established, results on upper limits have been reported by the MAGIC collaboration \cite{albert07} and the Whipple collaboration \cite{horan07}. In general, their results are consistent with power-law extrapolation of the keV spectra obtained with satellite data.

At cosmological distances, one has to take into account the absorption of VHE gamma-ray by extragalactic background light (EBL). For low-redshift GRBs and sub-TeV energies, the attenuation is less significant. Moreover, there is evidence from distant blazar spectra that the Universe is more transparent for VHE gamma-ray than previously thought~\cite{aha06a}. Thus, current air Cherenkov systems are able to observe out to $z\sim1$ at $\sim100$~GeV.

\section*{H.E.S.S. GRB observing program}

The H.E.S.S. array is a system of four 13m-diameter Imaging Atmospheric Cherenkov Telescopes (IACTs) located in the Khomas Highland of Namibia~\cite{hinton04}. Since the completion of the whole array in late 2003, H.E.S.S. has proven to be very successful in VHE gamma-ray astronomy, thus opening a new era in astronomy in this observational window. The array is one of the most sensitive VHE gamma-ray detectors. For a point source with integral flux $\sim1.4\times10^{-11}$~ph~cm$^{-2}$s$^{-1}$ above 1~TeV and spectral index 2.6, only a 2-hour H.E.S.S. observation is required for a 5-$\sigma$ detection.

We have been observing GRBs since early 2003. At the beginning of 2005, a GRB coordination team was set up and since then our GRB observation program has been fully established. An automated program is running on site to keep the shift crew alerted of any new detected GRBs in real-time.

We have followed on-board GRB triggers distributed by {\it Swift}, as well as triggers from INTEGRAL and HETE II confirmed by ground-based analysis. Upon the reception of a GRB Coordinates Network (GCN) notice from one of these satellites (with good indications that the source is a genuine GRB), we will observe the burst position as soon as possible, limited to $\mathrm{ZA} < 45^\circ$ (to ensure a reasonably low energy threshold) and HESS dark time\footnote{H.E.S.S. observations are taken in darkness and when the moon is below the horizon. The fraction of H.E.S.S. dark time is about $0.2$.}. We start observing the burst position up to 24 hours after the burst time. The nominal observation time is 2 hours (within the above observational constraints). If there are tentative positive signals indicated by a quick analysis, further observations will be carried out accordingly.

In Table~\ref{GRBtable}, we show 17 GRBs which we have observed using H.E.S.S. during the period from March 2003 to April 2007. All GRB data shown were taken in good weather conditions and good hardware status, while data taken in non-optimal situations are not shown and not used in further analyzes. For each burst, the start observation time, live-time of the observation and the mean zenith angle (ZA) are presented.

\section*{Data Analysis and Results}

Calibration of data, the event reconstruction and rejection of the cosmic-ray background (i.e. gamma-ray event selection criteria) were performed as described in~\cite{aha06b}. Except for the case of GRB~030329, where a different analysis cut was used because only two telescopes were operating, standard analysis cuts as described in~\cite{benbow05} were applied to each GRB to search for any possible signal. The background was then estimated using the reflected-region model as described in \cite{berge07}. The energy threshold ($E_{\rm th}$) after analysis cuts of each GRB observation (mainly depending on the zenith angle) is shown in Table~\ref{GRBtable}.

No evidence of excess events for any GRB observed using H.E.S.S. was seen. The 99.9\%
confidence level (c.l.) integral photon flux upper limits above $E_{\rm th}$ using the method of~\cite{feldman98}
for each GRB are included in Table~\ref{GRBtable}. For GRBs with reported redshifts $<1$, the EBL-corrected values using the P0.45 model as in \cite{aha06a} are also shown. For GRBs with redshifts $>1$, the attenuation of VHE gamma-ray by EBL is large, thus the EBL-corrected values are much higher (which are not shown here).

\begin{table*}
\caption{GRBs observed with H.E.S.S. from March 2003 to April 2007. All GRB data shown were taken in good weather conditions and good hardware status. For each burst, start observation time, live-time, mean zenith angle (ZA), energy threshold ($E_{\rm th}$) and 99.9\% c.l. upper limit (UL) of the observation are presented.}
\label{GRBtable}
\begin{center}
\begin{tabular}{lcccrcc}
    \hline
      GRB   & Observation starts  & live time & mean ZA & $E_{\rm th}$ & Flux ULs$^\mathrm{\star}$ & redshift \\
            & after GRB onset     & (hrs)     & (deg)   & (GeV)        & (ph cm$^{-2}$ s$^{-1}$) & \\
    \hline
    070429A & 64 min              & 0.5      & 21      & 310     & $1.06\times10^{-12}$& --             \\
    070419B & 15.1 h              & 1.0      & 48      & 750     & $3.89\times10^{-12}$& --             \\
    070209  & 15.4 h              & 1.0      & 41      & 660     & $4.49\times10^{-12}(1.70\times10^{-9})$  & 0.314$^\mathrm{\dagger}$ \\
    060526  & 4.7 h               & 1.9      & 25      & 200     & $5.90\times10^{-12}$& 3.21~\cite{berger5170}  \\
    060505  & 19.4 h              & 2.0      & 42      & 450     & $6.29\times10^{-12}(1.55\times10^{-11})$ & 0.089~\cite{ofek06} \\
    060403  & 13.6 h              & 0.9      & 39      & 310     & $9.37\times10^{-12}$& --             \\
    050922C & 52 min              & 0.7      & 23      & 200     & $1.22\times10^{-11}$& 2.199~\cite{delia4044}      \\
    050801  & 16 min              & 0.5      & 43      & 370     & $3.40\times10^{-12}$& 1.56$^\mathrm{\ddagger}$       \\
    050726  & 10.8 h              & 2.0      & 40      & 400     & $4.22\times10^{-12}$& --             \\
    050607  & 14.8 h              & 1.5      & 37      & 290     & $5.39\times10^{-12}$& --             \\
    050509C & 21 h                & 1.0      & 22      & 220     & $1.08\times10^{-11}$& --             \\
    050209  & 20.2 h              & 2.5      & 48      & 520     & $3.32\times10^{-12}$& --             \\
    041211  & 9.5 h               & 2.0      & 46      & 420     & $4.00\times10^{-12}$& --             \\
    041006  & 4.7 h               & 1.4      & 27      & 220     & $1.01\times10^{-11}(2.69\times10^{-8})$  & 0.716~\cite{soderberg06}\\
    040425  & 26 h                & 0.4      & 28      & 230     & $2.37\times10^{-11}$& --             \\
    030821  & 18 h                & 1.0      & 28      & 290     & $1.52\times10^{-11}$& --             \\
    030329  & 11.5 d              & 0.5      & 60      & 1400    & $2.58\times10^{-12}(5.59\times10^{-11})$ & 0.169~\cite{stanek03}  \\
    \hline
\end{tabular}
\end{center}

\vspace*{.6cm}
\noindent
$^\mathrm{\star}$ Flux upper limits ($>E_{\rm th}$) are at the 99.9\% c.l. using the method given in~\cite{feldman98}. For GRBs with redshifts~$<0.5$, the EBL-corrected values in the energy range [$E_{\rm th}$,$10$~TeV] are also given in the brackets. For GRB041006, the EBL-corrected value in the energy range [$E_{\rm th}$,$0.7$~TeV] is given. \\
$^\mathrm{\dagger}$ Redshift of a host galaxy candidate found by~\cite{berger6101}. \\
$^\mathrm{\ddagger}$ Photometric redshift according to~\cite{pasquale07}.
\end{table*}

\section*{Acknowledgements}
The support of the Namibian authorities and of the University of Namibia
in facilitating the construction and operation of H.E.S.S. is gratefully
acknowledged, as is the support by the German Ministry for Education and
Research (BMBF), the Max Planck Society, the French Ministry for Research,
the CNRS-IN2P3 and the Astroparticle Interdisciplinary Programme of the
CNRS, the U.K. Science and Technology Facilities Council (STFC),
the IPNP of the Charles University, the Polish Ministry of Science and
Higher Education, the South African Department of
Science and Technology and National Research Foundation, and by the
University of Namibia. We appreciate the excellent work of the technical
support staff in Berlin, Durham, Hamburg, Heidelberg, Palaiseau, Paris,
Saclay, and in Namibia in the construction and operation of the
equipment.
%\bibliography{libros}
%This in the bibtex style, is ok.
\bibliographystyle{plain}

%%%%%%%%
%  41  %
%%%%%%%%

%The paper title
\title{Search for a Dark Matter annihilation signal from the Sagittarius dwarf galaxy with H.E.S.S.}
%Short title to print in the headers to the final publication (Not showed in this print).
\shorttitle{Search for a Dark Matter annihilation signal from the
Sagittarius dwarf galaxy}
%All paper authors
\authors{E. Moulin$^{1}$, C. Farnier$^{2}$, J.-F. Glicenstein$^{1}$, A. Jacholkowska$^{2}$, L. Rolland$^{3}$, M. Vivier$^{1}$, for the H.E.S.S. Collaboration}
%Short title to print in the headers to the final puplication (Not showed in this print).
\shortauthors{E. Moulin and et al}
%All the affiliations.
\afiliations{$^1$DAPNIA/DSM/CEA, CE Saclay, F-91191 Gif-sur-Yvette Cedex, France\\
$^2$ Laboratoire de Physique Th\'eorique et Astroparticules,
Universit\'e Montpellier II, CC 70, Place Eug\`ene Bataillon, F-34095 Montpellier Cedex 5, France\\
$^3$ Laboratoire d'Annecy-le-Vieux de Physique des Particules,
IN2P3/CNRS, 9 Chemin de Bellevue, BP 110 F-74941 Annecy-le-Vieux
Cedex, France} \email{emmanuel.moulin@cea.fr}

%The abstract.
\abstract{Dwarf Spheroidal galaxies are amongst the best targets
to search for a Dark Matter (DM) annihilation signal. The
annihilation of WIMPs in the center of Sagittarius dwarf
spheroidal (Sgr dSph) galaxy would produce high energy
$\gamma$-rays in the final state. Observations carried out with
the H.E.S.S. array of Imaging Atmospheric Cherenkov telescopes are
presented. A careful modelling of the Dark Matter halo profile of
Sgr dwarf was performed using latest measurements on its
structural parameters. Constraints on the velocity-weighted cross
section of Dark Matter particles are derived in the framework of
Supersymmetric and Kaluza-Klein models.
%Observations of the Sagittarius dwarf spheroidal (Sgr dSph) galaxy
%were carried out with the H.E.S.S. array of four imaging air
%Cherenkov telescopes. A total of 11 hours of high quality data are
%available after data selection. There is no evidence for a very
%high energy $\gamma$-ray signal above the energy threshold at the
%target position. A 95\% C.L. flux limit of $\rm 3.6 \times
%10^{-12} cm^{-2}s^{-1}$ above 250 GeV has been derived.
%Constraints on the velocity-weighted cross section $\rm \langle
%\sigma v \rangle$ are calculated in the framework of Dark Matter
%particle annihilation using realistic models for the Dark Matter
%halo profile of Sagittarius dwarf galaxy. A 95\% C.L. exclusion
%limit on $\rm \langle \sigma v \rangle$ of the order of $\rm 2
%\times 10^{-25} cm^3s^{-1}$  is obtained for a core profile in the
%100 GeV - 1 TeV neutralino mass range.
}

\maketitle

%%%%% Begin DM %%%%%%
\addtocontents{toc}{\protect\contentsline {part}{\protect\large Dark Matter}{}}
\addcontentsline{toc}{section}{Search for a Dark Matter annihilation signal from the Sagittarius dwarf galaxy with H.E.S.S.}
\setcounter{figure}{0}
\setcounter{table}{0}
\setcounter{equation}{0}

%Begin the section.
\section*{Introduction}
Astrophysical and cosmological observations provide a substantial
body of evidences for the existence of Cold Dark Matter (CDM)
although its nature remains still unknown. It is commonly assumed
that CDM is composed of yet undiscovered non-baryonic particles
for which plausible candidates are Weakly Interacting Massive
Particles (WIMPs). In most theories, candidates for CDM are
predicted in theories beyond the Standard Model of particle
physics~\cite{bertone}. The annihilation of WIMPs into
$\gamma$-rays may lead to detectable very high energy (VHE, E $>$
100 GeV) $\gamma$-ray fluxes above background via continuum
emission from the hadronization and decay of the cascading
annihilation products, predominantly from $\pi^0$'s generated in
the quark jets. Among the best-motivated CDM candidates are the
lightest neutralino $\tilde{\chi}$ provided by R-parity conserving
supersymmetric extensions of the Standard Model~\cite{jungman},
and
the lightest Kaluza-Klein particle (LKP)~\cite{servant} %provided
%by the
in universal extra dimension theories which is most often the
first KK
mode of the hypercharge gauge boson, $\tilde{B}^{(1)}$.
% is the
%best motivated.
% hadronization
%of gauge bosons and heavy quarks, or $\gamma$-ray lines through
%loop-induced processes.
%The annihilation of DM particles can generate $\gamma$-ray fluxes
%through different processes depending on the particle physics
%scenarios.Generally, WIMP annihilations will produce a continuum
%of $\gamma$-rays with energies up to the WIMP mass issued from the
%hadronization and decay of the cascading annihilation products,
%predominantly from $\pi^0$'s generated in the quark jets. Among
%the best motivated CDM candidates are the lightest neutralino
%$\tilde{\chi}$ provided by R-parity conserving supersymmetric
%extensions of the Standard Model~\cite{jungman}, and the lightest
%Kaluza-Klein particle (LKP)~\cite{servant} provided by the
%universal extra dimension (UED) theories in which the first KK
%mode of the hypercharge gauge boson, $\tilde{B}^{(1)}$ is the best
%motivated.\\
The H.E.S.S. array of Imaging Atmospheric Cherenkov Telescopes
(IACTs), designed for high sensitivity measurements in the 100 GeV
- 10 TeV energy regime, is a suitable instrument to
detect VHE $\gamma$-rays and investigate their possible origin.\\
Dwarf Spheroidal galaxies such as Sagittarius or Canis Major,
discovered recently in the Local Group, are among the most extreme
DM-dominated environments. Indeed, measurements of roughly
constant radial velocity dispersion of stars imply large mass to
luminosity ratios~\cite{wilkinson}.
%Nearby dwarfs are ideal
%astrophysical probes of the nature of DM as they usually consist
%of a stellar population with no hot or warm gas, no cosmic ray
%population and little dust.
The core of the Sgr dSph at l=5.6$^{\circ}$ and b=-14$^{\circ}$ in
galactic coordinates at a distance of about 24 kpc from the
Sun~\cite{majewski}. Latest velocity dispersion measurements on M
giant stars with 2MASS yields a mass to light ratio of about
25~\cite{majewski2}.
%The Sgr
%dSph core is positioned behind the bulge of Milky Way but outside
%the Galactic plane, thus reduced foreground  $\gamma$-ray
%contaminations are expected.
The luminous density profile of Sgr
dSph has two components~\cite{monaco}. The compact component,
namely the core, is characterized by a size of about 3 pc FWHM,
which corresponds to a point-like region for H.E.S.S. This is the
DM annihilation region from which $\gamma$-ray signal may be
expected. A diffuse component is well fitted by a King model with
a characteristic size of 1.6 kpc.\\
We present in this paper the observations of the Sgr dSph galaxy
by the H.E.S.S. array of Imaging Atmospheric Cherenkov Telescopes.
A careful modeling of the Dark Matter halo using the latest
measurements on the structural parameters of Sagittarius is
presented to derive constraints on the WIMP velocity-weighted
annihilation rate.

\section*{Search for VHE $\gamma$-rays from observations of Sagittarius dwarf by H.E.S.S.}
H.E.S.S. (High Energy Stereoscopic System)
%is an array of four
%imaging atmospheric Cherenkov telescopes located in the Khomas
%Highlands of Namibia at an altitude of 1800 m a.s.l. Each
%telescope consists of an optical reflector of about 107 m$^2$
%effective area which collects the Cherenkov light emitted by the
%charged particles composing the electromagnetic shower initiated
%by the interaction of the primary $\gamma$-ray in the Earth's
%upper atmosphere. The light is focused on a 960 fast
%photomultiplier (PMT) camera~\cite{vincent}. %The total field of
%%view of the H.E.S.S. instrument is 5$^{\circ}$ in diameter.
%The stereoscopy technique used in the imaging atmospheric
%Cherenkov telescopes allows for accurate reconstruction of the
%direction and energy of the primary $\gamma$-rays as well as an
%efficient rejection of the background induced by cosmic ray
%interactions~\cite{trigger}. The energy threshold of H.E.S.S. at
%zenith before selection cuts is 160 GeV due to the degradation of
%the optical performance in 2006. The point source sensitivity is
%better than $\rm 2\times 10^{-13} cm^{-2}s^{-1}$ above 1 TeV for a
%5$\sigma$ detection in 25 hours~\cite{crabe}.\\
has observed the Sgr dSph in June 2006 with zenith angles ranging
from 7$^{\circ}$ to 43$^{\circ}$ around an average value of
19$^{\circ}$. A total of 11 hours of high quality data are
available for the analysis after standard selection cuts. After
calibration of the raw shower images from PMT
signals~\cite{aharonian0}, two independent reconstruction
techniques were combined to select $\gamma$-ray events and
reconstruct their direction and energy. The first one uses the
Hillas moment method~\cite{aharonian1}. The second analysis
referred hereafter as ``Model Analysis'', is based on the
pixel-per-pixel comparison of the shower image with a template
generated by a semi-analytical shower development
%The event
%reconstruction relies on a maximum likelihood method which uses
%available pixels in the camera, without requirement for an image
%cleaning
model~\cite{denauroi0,denauroi1}.
%The reconstructed shower
%parameter (energy, impact, direction and primary interaction
%point) are obtained as a product of the minimization procedure.
 The separation between $\gamma$ candidates and hadrons is done
 using a combination of the Model goodness-of-fit
 parameter~\cite{denauroi1} and the Hillas mean scaled width and length
 parameters, which results in an improved background
 rejection~\cite{crabe}.
%Standard cuts on the width and the length of Hillas ellipses
%combined with the goodness-of-fit are used to suppress the
%hadronic background~\cite{aharonian1}.
An additional cut on the primary interaction depth is used to
improve background rejection.\\
%Both methods yields typical energy resolution of 15\% above energy
%threshold.
%The analysis yields a typical energy resolution of 15\%
%above energy threshold and the angular resolution at the 68\%
%containment radius is found to be better than 0.06$^{\circ}$
%per $\gamma$-ray.\\
The on-source signal is defined by integrating all the events with
angular position $\theta$ in a circle around the target position
with a radius of $\theta_{cut}$. The target position is chosen
according to the photometric measurements of the Sgr dSph luminous
cusp showing that the position of the center corresponds to the
center of the globular cluster M 54~\cite{monaco1}. The target
position is thus found to be $\rm (RA = 18^h55^m59.9s, Dec =
-30d28'59.9'')$ in equatorial coordinates (J2000.0).
%or $\rm (l =
%5^{\circ}41'12.9'',b = -14^{\circ}16'29.8'')$ in Galactic
%coordinates.
The signal coming from Sgr dSph is expected to come from a region
of 1.5 pc, about 30'', much smaller the H.E.S.S. point spread
function (PSF). A $\theta_{cut}$ value of 0.14$^{\circ}$ suitable
for a point-like source was therefore used in the analysis. In
case of a Navarro-Frenk-White (NFW) density profile~\cite{nfw} for
which $\rho$ follows r$^{-1}$ or a cored profile~\cite{evans}
folded with the point spread function (PSF) of H.E.S.S., the
integration region allows to retrieve a significant fraction of
the expected signal. See Table~\ref{tab:table}.\\
%A cut on the image size of 60
%photoelectrons is used to obtain a good sensitivity for weak
%sources. In order to reduce systematic effects which affect images
%close to the edges of the camera, only events reconstructed within
%a maximum distance of 2.5$^{\circ}$ from the camera center are
%used for this analysis. The excess sky map is obtained by the
%subtraction of a background model on the $\gamma$-ray candidate
%sky distribution. The background level is estimated using the
%ring-background method~\cite{puhlhofer} where the background rate
%is calculated from the integration of $\gamma$-like events falling
%in an annulus around the center of the camera with identical
%observation conditions and acceptances than that used for the
%on-source region, which allows
%an estimate of the background on every sky position.\\
No significant $\gamma$-ray excess is detected in the sky map. We
thus derived the 95\% confidence level upper limit on the observed
number of $\gamma$-rays: $\rm N_{\gamma}^{95\%\, C.L.}$. The limit
is computed knowing the numbers of events in the signal and
background regions above the energy of 250 GeV
%in the signal region
%$\rm N_{ON} = 437$, in the background region $\rm N_{OFF} = 4270$,
%and the ratio of the off-source time to the on-source time
%$\rm\alpha = 10.1$. We
using the Feldman \& Cousins method~\cite{feldman} and we obtain:
%\begin{equation}
$\rm N_{\gamma}^{95\%\,C.L.} = 56$.
%\end{equation}
Given the acceptance of the detector for the observations of the
Sgr dSph, a 95\% confidence level upper limit on the $\gamma$-ray
flux is also derived:
\begin{eqnarray}
\Phi_{\gamma}({\rm E_{\gamma} > 250\,GeV}) < 3.6 \times 10^{-12}
\,{\rm cm^{-2}s^{-1}}\, \nonumber \\(95\%\,C.L.)
\end{eqnarray}

\section*{Predictions of $\gamma$-ray from Dark Matter annihilations}
\begin{figure}[!bp]
\begin{center}
\includegraphics [width=0.48\textwidth]{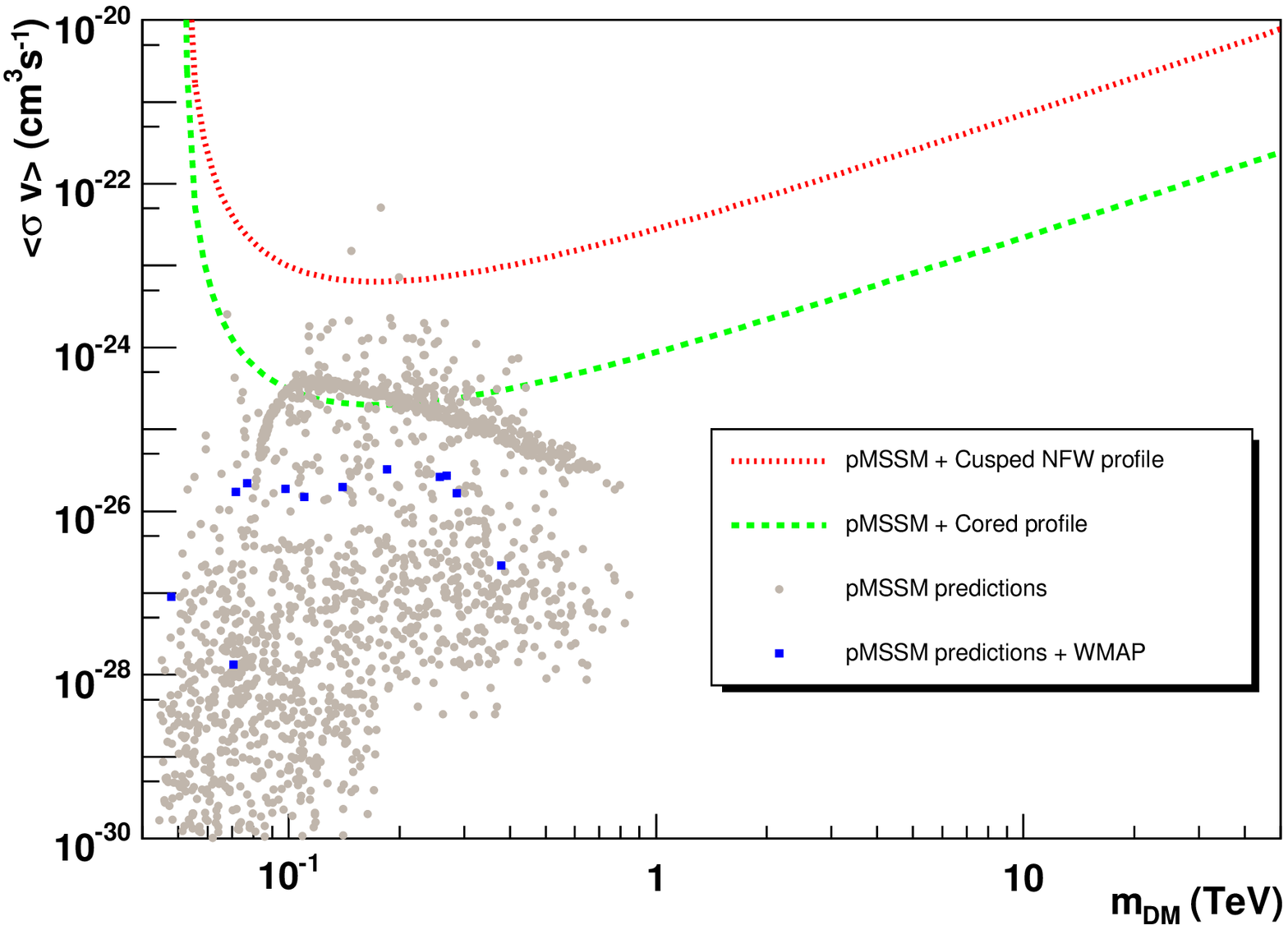}
\end{center}
\caption{Upper limits at 95\% C.L. on $\langle \sigma v \rangle$
versus the DM particle mass in the case of a cusped NFW (red
dotted line) and a cored (green dashed line) DM halo profiles
respectively. The pMSSM parameter space was explored with DarkSUSY
4.1~\cite{darksusy}, each point on the plot corresponding to a
specific model (grey point). Amongst these models, those
satisfying in addition the WMAP constraints on the CDM relic
density are overlaid as blue square. The limits in case of
neutralino dark matter from pMSSM are derived using the
parametrisation from~\cite{bergstrom} for a higgsino-type
neutralino annihilation $\gamma$ profiles.}\label{fig:sigmav}
\end{figure}

The $\gamma$-ray flux from  annihilations of DM particles of mass
$m_{DM}$ accumulating in a spherical DM halo can be expressed in
the form:
\begin{equation}\label{equ4}
\frac{d\Phi(\Delta\Omega,E_{\gamma})}{dE_{\gamma}}\,=\frac{1}{4\pi}\,\underbrace{\frac{\langle
\sigma
v\rangle}{m^2_{DM}}\,\frac{dN_{\gamma}}{dE_{\gamma}}}_{Particle\,
Physics}\,\times\,\underbrace{\bar{J}(\Delta\Omega)\Delta\Omega}_{Astrophysics}
\end{equation}
as a product of %characterized by
a particle physics component with an astrophysics component. The
particle physics part contains $\langle \sigma v\rangle$, the
velocity-weighted annihilation cross section, and
$dN_{\gamma}/dE_{\gamma}$,  the differential $\gamma$-ray spectrum
summed over the whole final states with their corresponding
branching ratios.
\begin{table*}[!htp]
\begin{center}
\begin{tabular}{|c||c|c|c|}
\hline
Halo type&Parameters&$\rm \bar{J}$&Fraction of signal \\
&&$\rm (10^{24} GeV^{2}cm^{-5})$&in $\rm\Delta\Omega = 2 \times 10^{-5}$sr\\
%&$r_c/r_s$&$v_a$/A&$\bar{J}$\\
\hline \hline
%Cusped halo ($\rm \gamma = 1$)&r$_s$ = 0.2 kpc&$\rm 2.2$& \\
Cusped NFW halo &r$_s$ = 0.2 kpc&$\rm 2.2$& 93.6\%\\
&$\rm A = \rm 3.3\times 10^7 M_{\odot}$&& \\
\hline
Cored halo&r$_c$ = 1.5 pc&$\rm 75.0$&99.9\%\\
&v$_a$ = 13.4 $\rm km\ s^{-1}$&&\\
\hline
\end{tabular}
\end{center}
\caption{\label{tab:table}Structural parameters for a cusped NFW
($\rm r_s,A$) and a cored ($\rm r_c,v_a$) DM halo model,
respectively. The values of the solid-angle-averaged l.o.s
integrated squared DM distribution are reported in both cases for
the solid angle integration region $\rm \Delta \Omega = 2\times
10^{-5} sr$.}
\end{table*}
The astrophysical part corresponds to the line-of-sight-integrated
squared density of the DM distribution J, averaged over the
instrument solid angle integration region for H.E.S.S. ($\rm
\Delta\Omega=2\times 10^{-5}$sr):
\begin{eqnarray}
\label{eqnj}
%J\,=\,\int_{l.o.s}\rho^2(r[s])ds \nonumber \\
\bar{J}(\Delta\Omega)\,=\,\frac{1}{\Delta\Omega}\int_{\Delta\Omega}
{\rm PSF}*\int_{l.o.s}\rho^2(r[s])ds\,d\Omega
\end{eqnarray}
where PSF is the point spread function of H.E.S.S.\\
The mass distribution of the DM halo of Sgr dwarf has been
described by plausible models taking into account the best
available measurements of the Sgr dwarf galactic structure
parameters. We have used two widely different models. The first
has a NFW cusped profile~\cite{nfw} with the mass density given
by:
\begin{equation}
\rho_{NFW}(r)\,=\,\frac{A}{r(r+r_s)^{2}}
\end{equation}
with A the normalization factor and $r_{s}$ the scale radius taken
from~\cite{evans}. Using Eq.~\ref{eqnj}, the value of $\bar{J}$
obtained with this model is reported in Table~\ref{tab:table}.
%In the spirit of reference~\cite{evans},
We have also studied a core-type halo model  as in~\cite{evans}
characterized by the mass density:
\begin{equation}
\rho_{core}(r)\,=\,\frac{v_a^2}{4\pi G}\frac{3 r_c^2
+r^2}{(r_c^2+r^2)^2}
\end{equation}
where $r_c$ is the core radius and $v_a$ a velocity scale.
 However, we have tried to update the $v_{a}$ and $r_{c}$
values which were used in~\cite{evans}. By inserting in the Jeans
equation the luminosity profile of the Sgr dwarf core of the form:
\begin{equation}
\nu (r)= \frac{\nu_{0}{r_c}^{2\alpha}}{(r_c^2+r^2)^{\alpha}}
\end{equation}
we estimated from the data of \cite{monaco1} $\alpha = 2.69\pm
0.10$ and $ r_c = 1.5\ \mbox{pc}.$ Note that the value of $r_{c}$
is only an upper limit. The value of the central velocity
dispersion of Sgr Dwarf is
 $\sigma = 8.2\,\pm\, 0.3\,\rm km\,s^{-1}$~\cite{zaggia}.
We have assumed that the velocity dispersion is independent of
position. The value of $v_{a}$ is then given by $v_{a} =
\sqrt{\alpha}\,\sigma = 13.4\,\rm km\, s^{-1}$. The cored model
gives a very large value of $\bar{J},$ which is reported in
Table~\ref{tab:table}. The third column of Table~\ref{tab:table}
gives the amount of signal expected in the solid angle integration
region $\rm \Delta\Omega = 2\times 10^{-5}$sr.
Fig.~\ref{fig:sigmav} shows the limits in the case of a cored
(green dashed line) and cusped NFW (red dotted line) profile using
the value of $\bar{J}$ computed above.
%The differential
%$\gamma$ spectrum is parametrized using the expression given
%in~\cite{bergstrom} for a higgsino-type neutralino.
Predictions for phenomenological MSSM (pMSSM) models are displayed
(grey points) as well as those satisfying in addition the WMAP
constraints on the CDM relic density (blue points). The SUSY
models are calculated with DarkSUSY4.1~\cite{darksusy}.
% in pMSSM
%framework and characterized by a basic set of independent
%parameters  : the higgsino mass parameter $\mu$, the gaugino mass
%parameter M$_2$, the CP-odd Higgs mass M$_A$, the common scalar
%mass m$_0$ , the trilinear couplings A$_{t,b}$ and the Higgs
%vacuum expectation value ratio $\tan \beta$. The set of parameters
%for a given model is randomly chosen in a parameter region
%encompassing a large class of pMSSM models, as described in
%Tab.~\ref{tab:table2}.
In the case of a cusped NFW profile, the
H.E.S.S. observations do not establish severe constraints on the
velocity-weighted cross section. For a cored profile, due to a
higher central density,
stronger constraints are derived and some pMSSM models
%yielding neutralino masses in the 100 - 400 GeV range,
can be excluded in the upper part of the pMSSM scanned region. \\
In the case of KK dark matter, the differential $\gamma$ spectrum
is parametrized using Pythia~\cite{pythia} simulations and
branching ratios from~\cite{servant}. Predictions for the
velocity-weighted cross section of B$^{(1)}$ dark matter particle
are performed using the formula given in~\cite{sigmaKK}. In this
case, the expression for $\langle \sigma v \rangle$ depends
\begin{figure}[!hb]
\begin{center}
\includegraphics [width=0.48\textwidth]{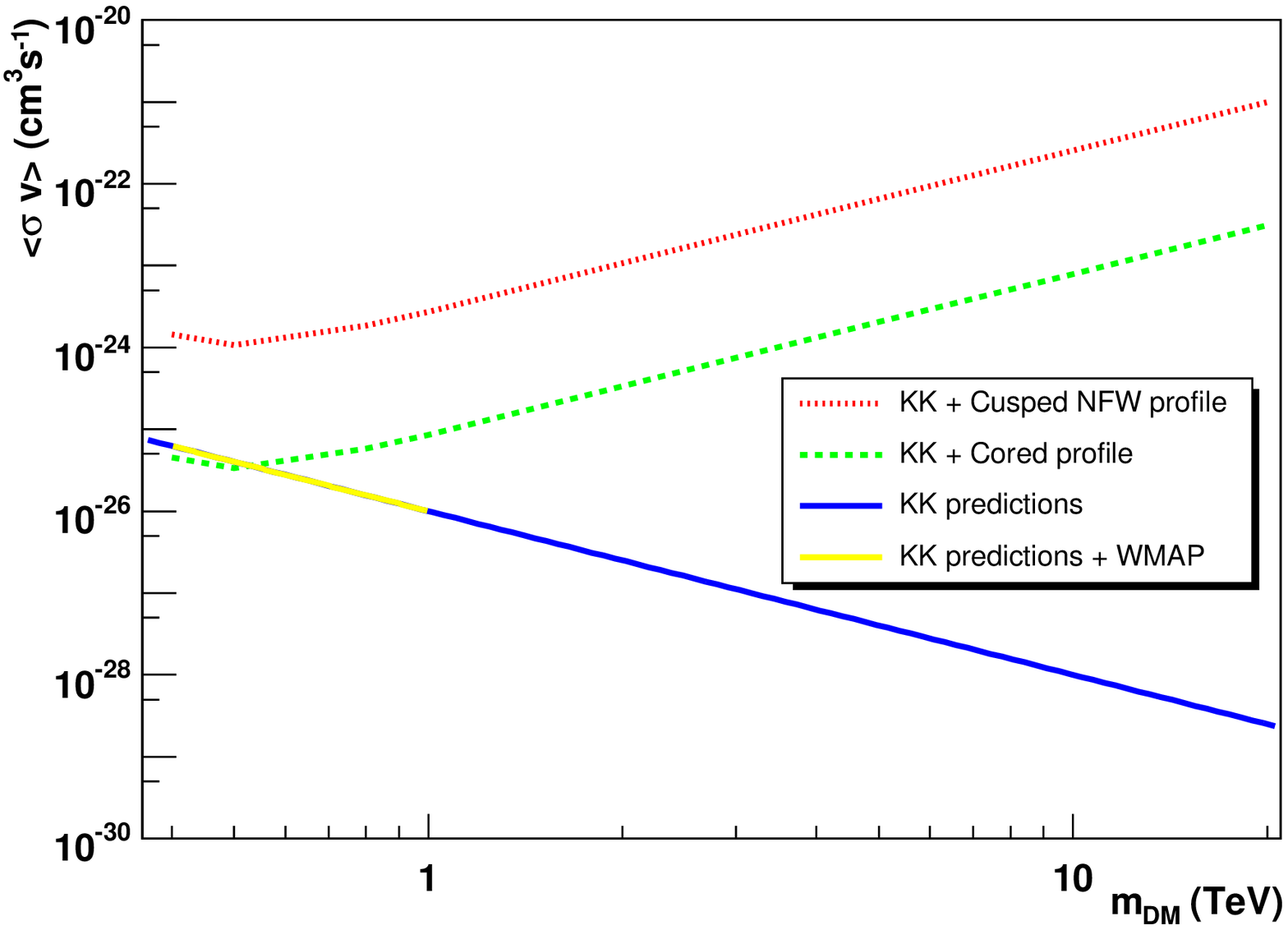}
\end{center}
\caption{Upper limits at 95\% C.L. on $\langle \sigma v \rangle$
versus the DM particle mass in the  B$^{(1)}$ Kaluza-Klein
scenarios for a cusped NFW (red dotted line) and a cored (green
dashed line) DM halo profiles respectively. The blue line
corresponds to Kaluza-Klein models~\cite{servant}. Overlaid
(yellow line) are the KK models satisfying WMAP constraints on the
CDM relic density.}\label{fig:sigmav2}
\end{figure}
analytically on the $\rm \tilde{B^{(1)}}$ mass square.
Fig.~\ref{fig:sigmav2} shows the sensitivity of H.E.S.S. in the
case of Kaluza-Klein models where the hypercharge boson B$^{(1)}$
is the LKP, for a cored (green solid line) and a cusped NFW (red
solid line) profile respectively using the value of $\rm \bar{J}$
computed in section 3.2. With a NFW profile, no Kaluza-Klein
models can be tested. In the case of a cored model, some models
providing a LKP relic density compatible with WMAP contraints can
be excluded. From the sensitivity of H.E.S.S., we exclude
B$^{(1)}$ masses lying in the range 300 - 500 GeV.
%derive a lower limit on the B$^{(1)}$
%mass of 500 GeV.

\section*{Conclusions}
The observations of Sgr dSph with H.E.S.S. reveal no significant
$\gamma$-ray excess at the nominal target position. The
Sagittarius dwarf DM halo profile has been modeled using latest
measurements of its structure parameters. Constraints have been
derived on the velocity-weighted cross section of the DM particle
in the framework of supersymmetric and Kaluza-Klein models.

%\section*{Acknowledgements}
%The support of the Namibian authorities and of the University of
%Namibia in facilitating the construction and operation of H.E.S.S.
%is gratefully acknowledged, as is the support by the German
%Ministry for Education and Research (BMBF), the Max Planck
%Society, the French Ministry for Research, the CNRS-IN2P3 and the
%Astroparticle Interdisciplinary Programme of the CNRS, the U.K.
%Particle Physics and Astronomy Research Council (PPARC), the IPNP
%of the Charles University, the South African Department of Science
%and Technology and National Research Foundation, and by the
%University of Namibia. We appreciate the excellent work of the
%technical support staff in Berlin, Durham, Hamburg, Heildelberg,
%Palaiseau, Paris, Saclay, and in Namibia in the construction and
%operation of the equipment.
%\nocite{ref4} \nocite{ref5}
%\nocite{ref6} \nocite{ref7}
%This is the reference to .bib file (Whitout .bib!)

%This in the bibtex style, is ok.
\bibliographystyle{plain}

%%%%%%%%
%  42  %
%%%%%%%%

\title{The \boldmath{$\gamma$}-radiation from the Galactic
  Center observed by H.E.S.S. and the possible dark matter
  interpretation} 
\shorttitle{Very high energy $\gamma$-radiation from
  Sgr A+ in terms of dark matter annihilation}

\authors{J. Ripken$^{1}$, G. Heinzelmann$^{1}$, J. F. Glicenstein$^{2}$, J. Hinton$^{3}$, D. Horns$^{4}$, L. Rolland$^{5}$, on behalf of
the H.E.S.S. collaboration}
\shortauthors{J. Ripken et al for the H.E.S.S. collaboration}
\afiliations{$^{1}$Institut f\"ur Experimentalphysik, Univ. Hamburg, Luruper Chaussee 149, D-22761 Hamburg, Germany \\
  $^{2}$DAPNIA/DSM/CEA, CE Sacley, F-91191 Gif-sur-Yvette, Cedex, France \\
  $^{3}$Max-Planck-Institut f\"ur Kernphysik, Saupfercheckweg 1, D-69117 Heidelberg, Germany \\ 
  $^{4}$Institut f\"ur Astronomie and Astrophysik, Eberhard Karls Universit\"at T\"ubingen, Sand 1, D-72076 T\"ubingen \\
  $^{5}$LPNHE, IN2P3/CNRS, Universit\'es Paris VI \& VII, 4 Place Jussieu, F-75252 Paris Cedex 5, France}

\email{ripkenj@mail.desy.de}

\abstract{With the H.E.S.S. system of four Cherenkov telescopes a
  signal of very high energy $\gamma$-radiation from the direction of
  the Galactic center has been detected. The interpretation of the
  signal due to dark matter annihilation is discussed and limits on
  the annihilation cross sections and density profiles are given.}

\maketitle

\addcontentsline{toc}{section}{The \boldmath{$\gamma$}-radiation from the Galactic Center observed by H.E.S.S. and the possible dark matter interpretation}
\setcounter{figure}{0}
\setcounter{table}{0}
\setcounter{equation}{0}

\section*{Introduction}
\vspace{-0.3cm} The nature of the dark matter particles is still one
of the outstanding problems of astrophysics. A possible candidate for
a dark matter particle is provided by the $R$-parity conserving
supersymmetric extension of the standard model of particle physics
e.g. the neutraline $\chi$. Another possible candidate is predicted by
Kaluza Klein (KK) theories, the $B^{(1)}$ \cite{bib_kksugg}. Both
particles are neutral, stable, and could naturally match the measured
matter density. Besides direct measurements of dark matter in
underground experiments, indirect detections via measurement of the
secondary particles produced in the self-annihilation in deep
gravitational potential wells has been suggested. One product of the
self-annihilation of dark matter particles is high energy
$\gamma$-radiation. Regions of high mass accumulations such as the
Galactic center (GC) could produce a detectable very high energy (VHE)
$\gamma$-ray flux \cite{bib_bergfirst}. With the H.E.S.S. Cherenkov
telescope array \cite{bib_HESS} the Galactic Center has been observed
in 2003 and 2004 \cite{bib_HESSobs}. High energy $\gamma$-radiation
has been observed with high significance. This radiation has been
investigated in the framework of dark matter annihilation
\cite{bib_horns, bib_icrc, bib_dmint}.

\begin{figure*}[t]
\begin{center}
\includegraphics[width=0.8\linewidth]{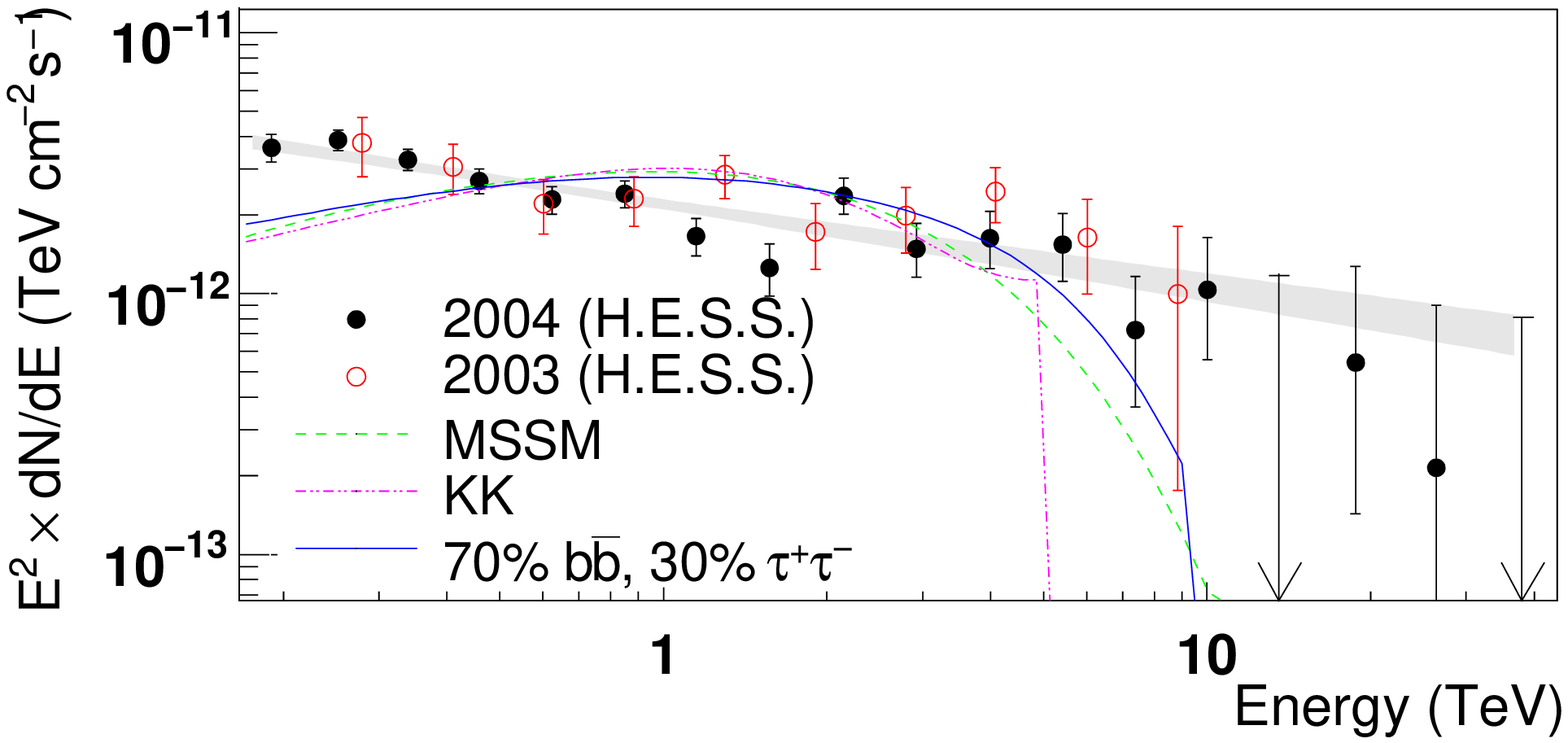}
\end{center}
\caption{(Color online) Spectral energy density $E^2\times 
  \mathrm{d}N/\mathrm{d}E$ of $\gamma$-rays from the GC source, for
  the 2004 data (full points) and 2003 data (open
  points).  Upper limits are 95\% CL.  The shaded area shows the
  power-law fit $\mathrm{d}N/\mathrm{d}E \sim E^{-\Gamma}$.  The
  dashed line illustrates a typical spectrum of neutralino DM
  annihilation for best fit neutralino masses of 14~TeV. The dotted
  line shows the distribution predicted for KK DM with a mass of
  5~TeV.  The solid line gives the spectrum of a 10~TeV DM particle
  annihilating into $\tau^+\tau^-$ (30\%) and $b\bar{b}$ (70\%).}
\label{spec_ex}
\end{figure*}

\vspace{-0.4cm}
\section*{\boldmath{$\gamma$}-rays from dark matter annihilation}
\vspace{-0.3cm} Since the $\chi$ and the $B^{(1)}$ are Majorana
particles, they can annihilate producing photons with energies up to
the particle masses.  Whereas the direct production of photons leading
to monoenergetic $\gamma$-rays are loop suppressed most of the high
energy photons are produced in decays of secondaries from the
annihilation processes. These photons have a continuous energy
spectrum up to the mass of the dark matter particle and are not easily
distinguished from other astrophysical processes for VHE
$\gamma$-rays.  The calculation of the $\gamma$-ray flux leads to the
formula
\begin{eqnarray}
\Phi(E) &=& 2.8 \cdot 10^{-10} \, \rm{cm}^{-2} \rm{s}^{-1} \nonumber \\ 
& & \times \frac{dN_{\gamma}}{dE} \Bigl{(} \frac{\langle \sigma v \rangle}{\rm{pb} \, c} \Bigr{)}
\Bigl{(} \frac{100 \, \rm{GeV}}{m_{\rm{DM}}} \Bigr{)}^{2} \nonumber \\
& & \times \bar{J}(\Delta \Omega) \Delta \Omega \nonumber \\
\bar{J}(\Delta \Omega) \Delta \Omega &=& \frac{1}{(0.3 \, \rm{GeV}/\rm{cm}^{3})^{2} \cdot 8.5 \, \rm{kpc}} \nonumber \\
& & \times \ \ \int d \Omega \ 
\int_{\rm{los}} \! dl \ \varrho^{2}
\label{eq_anni}
\end{eqnarray}
where $\langle \sigma v \rangle$ denotes the mean of the annihilation
cross section multiplied with the velocity of the particles and
$dN_{\gamma}/dE$ the photon energy spectrum per annihilation. The
solid angle $\Delta \Omega$ denotes the resolution of the detector or
the investigated solid angle. $\rho$ denotes the dark matter density
along the line of sight (los).

\begin{figure}[t]
\begin{center}
\includegraphics[width=\linewidth]{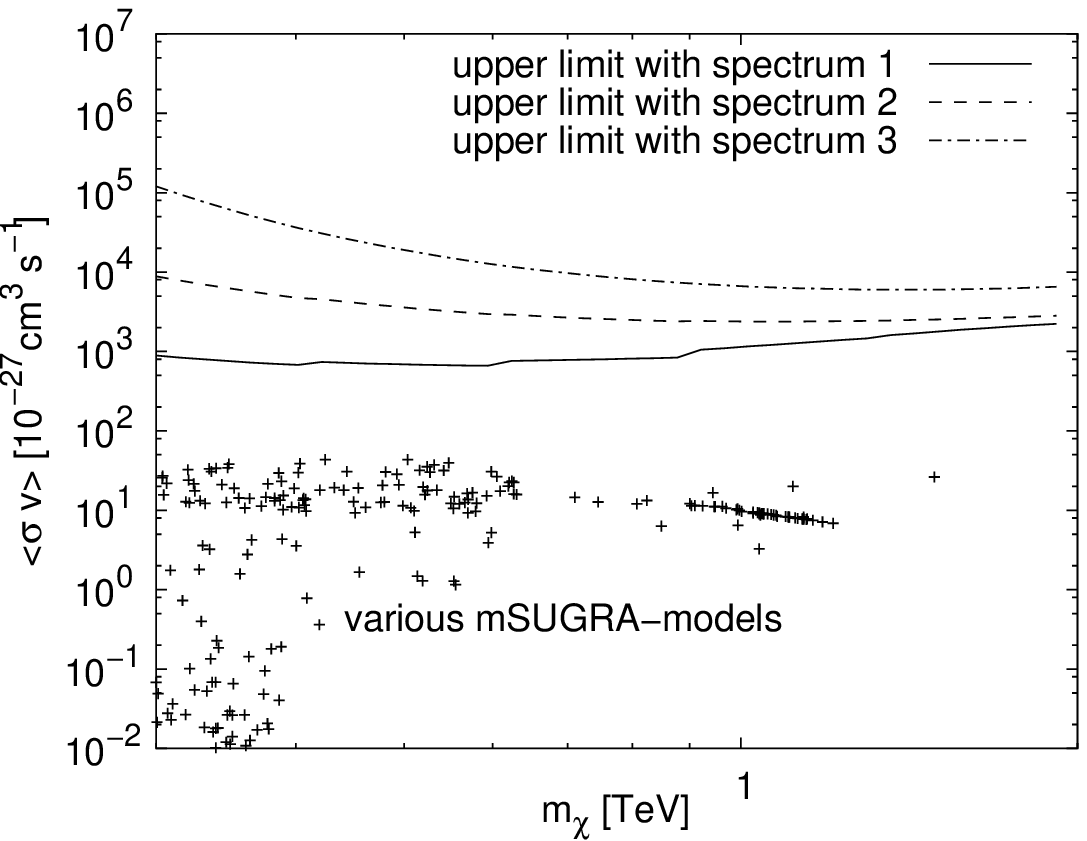}
\includegraphics[width=\linewidth]{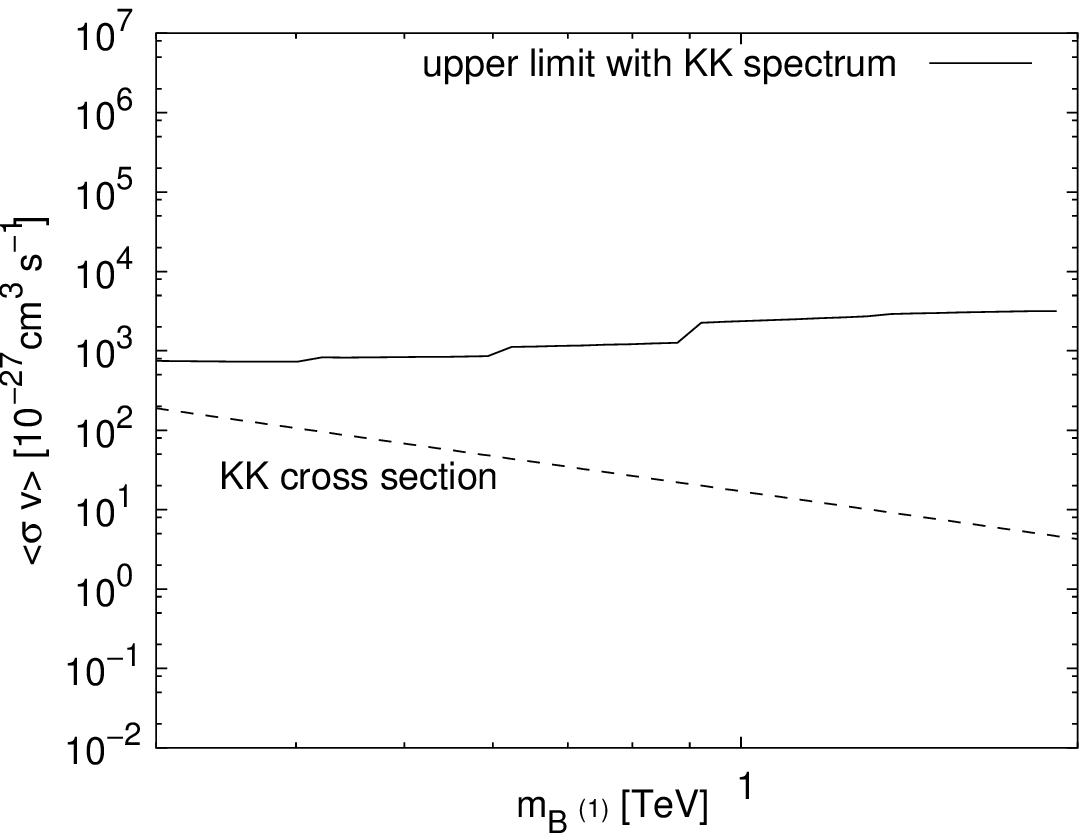}
\end{center}
\caption{Upper limits ($99 \%$ CL) on the annihilation cross section for neutralinos 
  (upper panel) and KK particle (lower panel) as function of the dark
  matter particle mass assuming an NFW profile.}
\label{partspec1}
\end{figure}

\begin{figure}[t]
\begin{center}
\includegraphics[width=\linewidth]{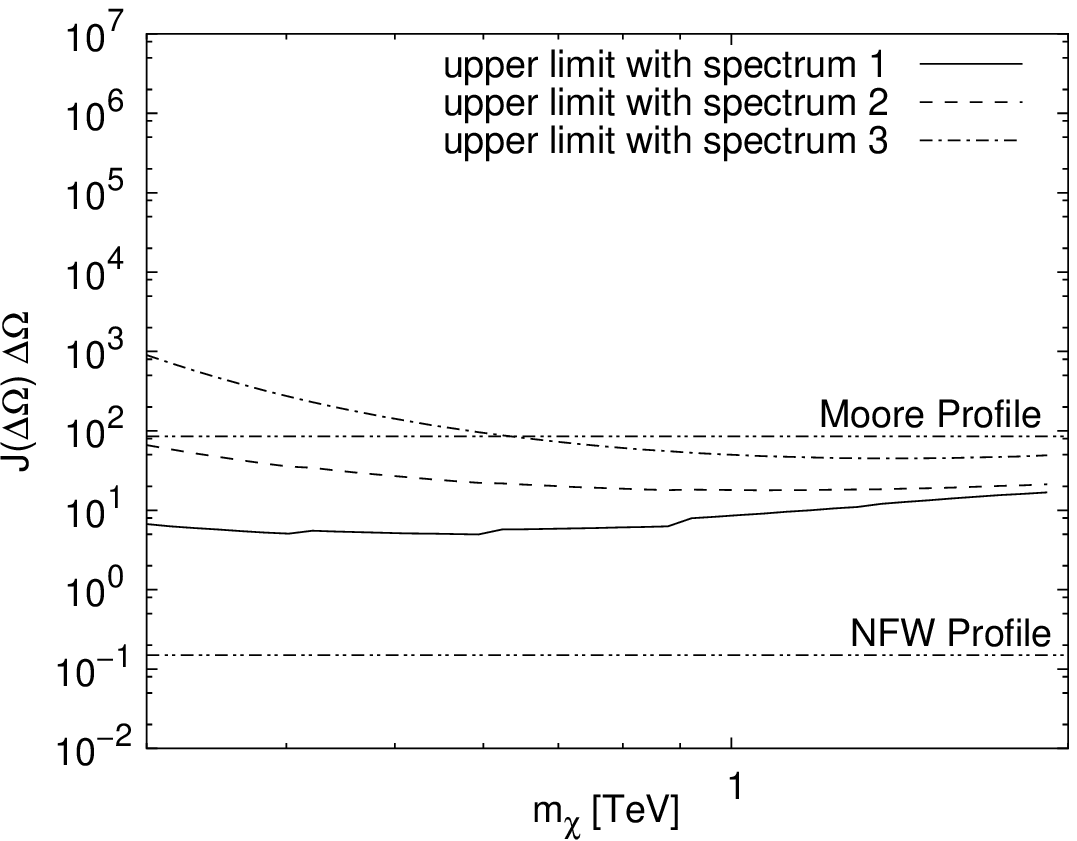}
\includegraphics[width=\linewidth]{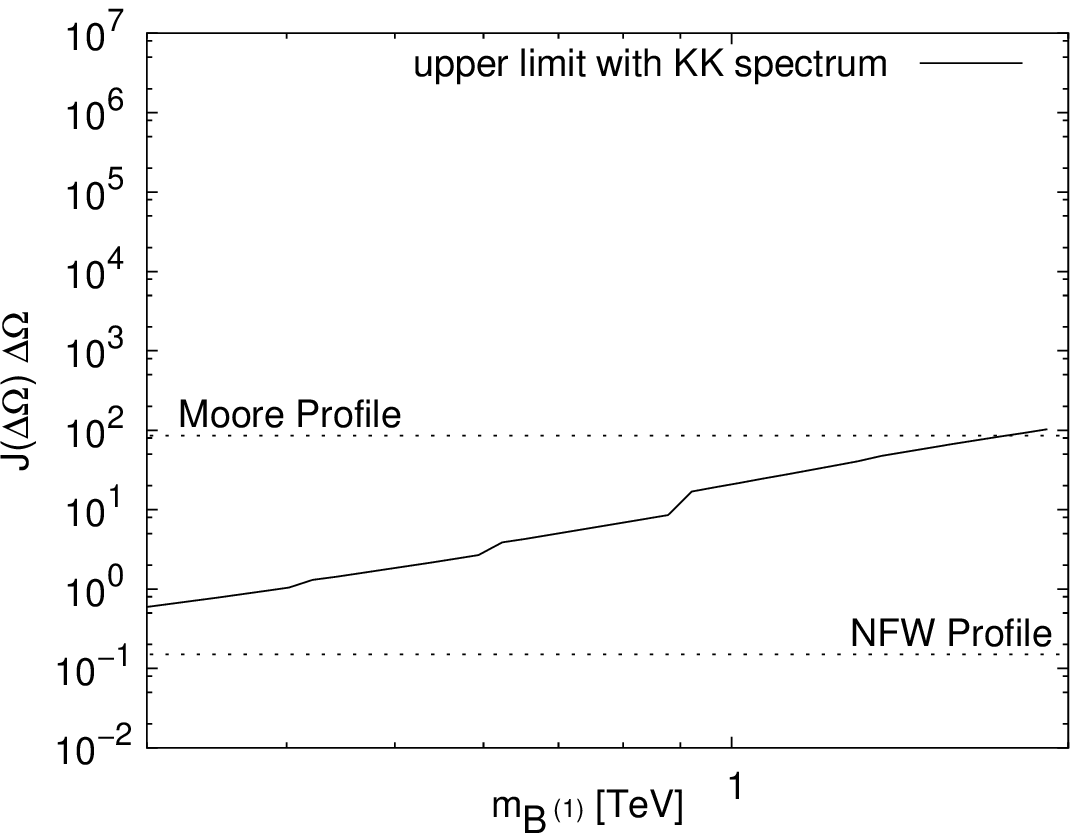}
\end{center}
\caption{Upper limits on $\bar{J}(\Delta \Omega) \Delta \Omega$ assuming a cross
  section for neutralinos (upper panel) and KK particles (lower panel)
  as function of the dark matter particle mass.}
\label{partspec2}
\end{figure}

\vspace{-0.4cm}
\section*{The center of our Galaxy}
\vspace{-0.3cm} The GC region containing the supermassive black hole
Sgr A*, was observed with the H.E.S.S. telescopes in the years 2003
and 2004.  High energy $\gamma$-radiation above $200 \, \rm{GeV}$ was
detected with a significance of $38$ standard deviations (in 2004)
without indications for variability. Further data have been collected
in 2005 and 2006 not yet included in this analysis.

The central region of our Galaxy is due to its proximity ($\approx
8.5$ \ kpc) and the expectation of high mass concentration a target
for the indirect search for dark matter. The scaling factor
$\bar{J}(\Delta \Omega) \Delta \Omega$ depends strongly on the not
well known density profile. We consider the results of
$N$-body-simulations from Navarro, Frenk and White (NFW) predicting
$\rho(r)\propto r^{-1}$ \cite{bib_NFW} and Moore et al. predicting
$\rho(r)\propto r^{-1.5}$ \cite{bib_Moore}.

The measured spectrum can be described by a powerlaw
\begin{equation}
\Phi(E) = \Phi_{0} \Bigl{(} \frac{E}{1 \ \rm{TeV}} \Bigr{)}^{-\Gamma}
\end{equation}
with an index of $\Gamma = 2.25 \pm 0.04_{\rm{stat}} \pm
0.10_{\rm{syst}}$. The integral flux above $1 \ \rm{TeV}$
is $(1.87 \pm 0.10_{\rm{stat}} \pm 0.30_{\rm{syst}})
\cdot 10^{-12} \ \rm{cm}^{-2} \rm{s}^{-1}$.

In the following, two different assumptions are used to derive
conclusions on dark matter annihilation from the observed radiation:
\begin{enumerate}
\item The flux results solely from dark matter annihilation as
  discussed in \cite{bib_horns} and \cite{bib_kk_full} exploring the
  consistency of the required mass density and cross section with
  other observations. This hypothesis explores the mass and cross
  section of the dark matter particle and the density profile of the
  dark matter halo in the central region of our Galaxy.
\label{hypa}
\item Only a part of the signal originates from dark matter
  annihilation, whereas the remaining part is produced by other
  processes and sources. This hypothesis can constrain either particle
  properties assuming a density profile of the factor $\bar{J}(\Delta
  \Omega) \Delta \Omega$ assuming in turn a range of cross sections.
\label{hypb}
\end{enumerate}

\vspace{-0.2cm}
\subsection*{Hypothesis \ref{hypa}: 100 \% dark matter annihilation radiation}
\vspace{-0.1cm}
\textbf{Density profile:} The density profile of the dark matter in
the inner part of the halo can be approximated by $\varrho \sim
r^{-\alpha}$. Instead of the integration over the solid angle $\Delta
\Omega$ in equation \ref{eq_anni} we convolute the line of sight
integral with the point spread function of the detector (the H.E.S.S.
experiment). In the GC region are molecular clouds which
lead to a diffuse radiation of $\gamma$-radiation \cite{bib_mole}
which has been substracted for this investigation. The remaining signal
from the Galactic Center is compatible with a point source. For $\alpha$
a lower limit of $1.2$ with a confidence level of $95 \%$ was
derived \cite{bib_dmint}. This is not compatible with an NFW profile, but
with a Moore profile.

\textbf{Energy spectrum:} The energy spectrum measured by H.E.S.S. 
reaches up to more than $10 \, \rm{TeV}$. This would require a very
massive dark matter particle close to the unitarity limit
\cite{bib_unit}. The favored mass range of the dark matter particles
is below $1 \, \rm{TeV}$ \cite{bib_elliswmap} in the considered
models, but such high masses, not violating the unitarity limit,
cannot be ruled out completely.

In Figure \ref{spec_ex} the spectral energy distribution
measured by H.E.S.S. is shown together with fits of a neutralino
annihilation spectrum parameterized according to \cite{bib_horns} and
a $B^{(1)}$ annihilation spectrum from \cite{bib_kk_full}. Clearly,
the expected curvature of the predicted energy spectra is not matching
the data which is in reasonable agreement with a power law type
function. With nonminimal SUSY models flatter spectra can be obtained,
but they also don't fit the measured spectrum well. Hypothesis \ref{hypa}
can therefore be ruled out.

\vspace{-0.2cm}
\subsection*{Hypothesis \ref{hypb}: Background and dark matter annihilation radiation}
\vspace{-0.1cm} In the GC region other processes may produce
$\gamma$-radiation above $100 \, \rm{GeV}$. This may result in a
$\gamma$-ray background to a hypothetical annihilation. The strength
of the annihilation radiation (equation \ref{eq_anni}) for a given
particle mass $m_{\rm{DM}}$ is proportional to $A = \langle
\sigma v \rangle \cdot J(\Delta \Omega) \Delta \Omega$.  Fitting the
assumed background (a power law) plus the fixed annihilation component
we get a function $\chi^{2}(A)$, which provides the upper limits on
$A$. With this limits we can produce upper limits either on the cross
section $\langle \sigma v \rangle$ of the annihilation by assuming a
density profile or on $\bar{J}(\Delta \Omega) \Delta \Omega$ with a
fixed cross section. In the Figures \ref{partspec1} and
\ref{partspec2} these upper limits are shown as functions of the
particle mass for neutralino dark matter and for KK dark matter. In
the supersymmetric scenario the number of photons per annihilation
depends on the parameter set used. Three spectra which are
encompassing all probabilities are used. The mSUGRA model cross
sections are calculated with DarkSUSY 4.1 \cite{bib_ds41}. The KK
annihilation spectrum and its cross section is described in
\cite{bib_kk_full}. With an NFW profile no cross section neither from
supersymmetric models (calculated with DarkSUSY 4.1 \cite{bib_ds41})
nor with KK dark matter can be ruled out. With a mean cross section
for neutralinos or the expected cross section for KK particles a
profile suggested by Moore can be ruled out for all considered
neutralino masses.

\vspace{-0.4cm}
\section*{Conclusion}
\vspace{-0.3cm}
We have investigated wether part or all of the high energy
$\gamma$-radiation from the GC observed by H.E.S.S. could be
attributed to dark matter annihilation. The energy spectrum can't be
reconciled by either a neutralino annihilation spectrum or with a
spectrum produced by KK dark matter only. Considering an additional
background we can exclude a large line of sight integral of the squared
density. For an NFW profile still no model, neither mSUGRA nor KK dark
matter, can be ruled out.

\textbf{Acknowledgements:}
{
The support of the Namibian authorities and of the University of Namibia
in facilitating the construction and operation of H.E.S.S. is gratefully
acknowledged, as is the support by the German Ministry for Education and
Research (BMBF), the Max Planck Society, the French Ministry for Research,
the CNRS-IN2P3 and the Astroparticle Interdisciplinary Programme of the
CNRS, the U.K. Particle Physics and Astronomy Research Council (PPARC),
the IPNP of the Charles University, the South African Department of
Science and Technology and National Research Foundation, and by the
University of Namibia. We appreciate the excellent work of the technical
support staff in Berlin, Durham, Hamburg, Heidelberg, Palaiseau, Paris,
Saclay, and in Namibia in the construction and operation of the equipment.
}

%%%%%%%%
%  43  %
%%%%%%%%

\title{Energy spectrum of cosmic iron nuclei measured by H.E.S.S.}

\authors{R.~B\"uhler$^{1}$ for the H.E.S.S. collaboration}

\afiliations{$^1$Max-Planck-Institut f\"ur Kernphysik, P.O. Box 103980, D 69029 Heidelberg, Germany}
\email{rolf.buehler@mpi-hd.mpg.de}

%The abstract.
\abstract{
A recently proposed novel technique for the detection of cosmic rays with arrays of \emph{Imaging Atmospheric Cherenkov Telescopes} is applied to data from the High Energy Stereoscopic System (H.E.S.S.). The method relies on the ground based detection of Cherenkov light emitted from the primary particle prior to its first interaction in the atmosphere. The charge of the primary particle (Z) can be estimated from the intensity of this light, since it is proportional to Z$^2$. Using H.E.S.S. data, an energy spectrum for cosmic-ray iron nuclei in the energy range 13--200 TeV is derived. The reconstructed spectrum is consistent with previous direct measurements and is the most precise measurement so far in this energy range.}

\maketitle

%%%%% Begin CR %%%%%%
\addtocontents{toc}{\protect\contentsline {part}{\protect\large Cosmic Rays}{}}
\addcontentsline{toc}{section}{Energy spectrum of cosmic iron nuclei measured by H.E.S.S.}
\setcounter{figure}{0}
\setcounter{table}{0}
\setcounter{equation}{0}

\section*{Introduction}
At present the best measurements of elemental composition of cosmic rays in the energy range 1 GeV to 0.5 PeV come from long duration balloon flights \cite{RUNJOB}. Because of the decreasing flux of cosmic rays and the limited collection area of these experiments ($\approx$1 m$^2$), it is hard to extend such measurements to higher energies. A further improvement in the accuracy and energy range of composition measurements of cosmic rays could however provide crucial information about the acceleration mechanism and propagation of these particles.

In 2001, Kieda et al. \cite{KIEDA} proposed a new method for the measurement of cosmic rays with \emph{\textbf{I}maging \textbf{A}tmospheric \textbf{C}herenkov \textbf{T}elescopes} (IACTs). The measurement from the ground takes advantage of the huge detection area ($\approx$$10^5$ m$^2$) of IACTs, in principle enabling the extension of spectral and composition measurements up to $\sim$ 1 PeV. Here we review this technique and describe its application to data from the \emph{High Energy Stereoscopic System} (H.E.S.S.). We present the measurement of the iron spectrum and give an outlook on future applications of this method (a more detailed description of the analysis and the results can be found in \cite{IRON}).

\section*{Technique}
\begin{figure}
  \begin{center}
    \includegraphics[width=7cm]{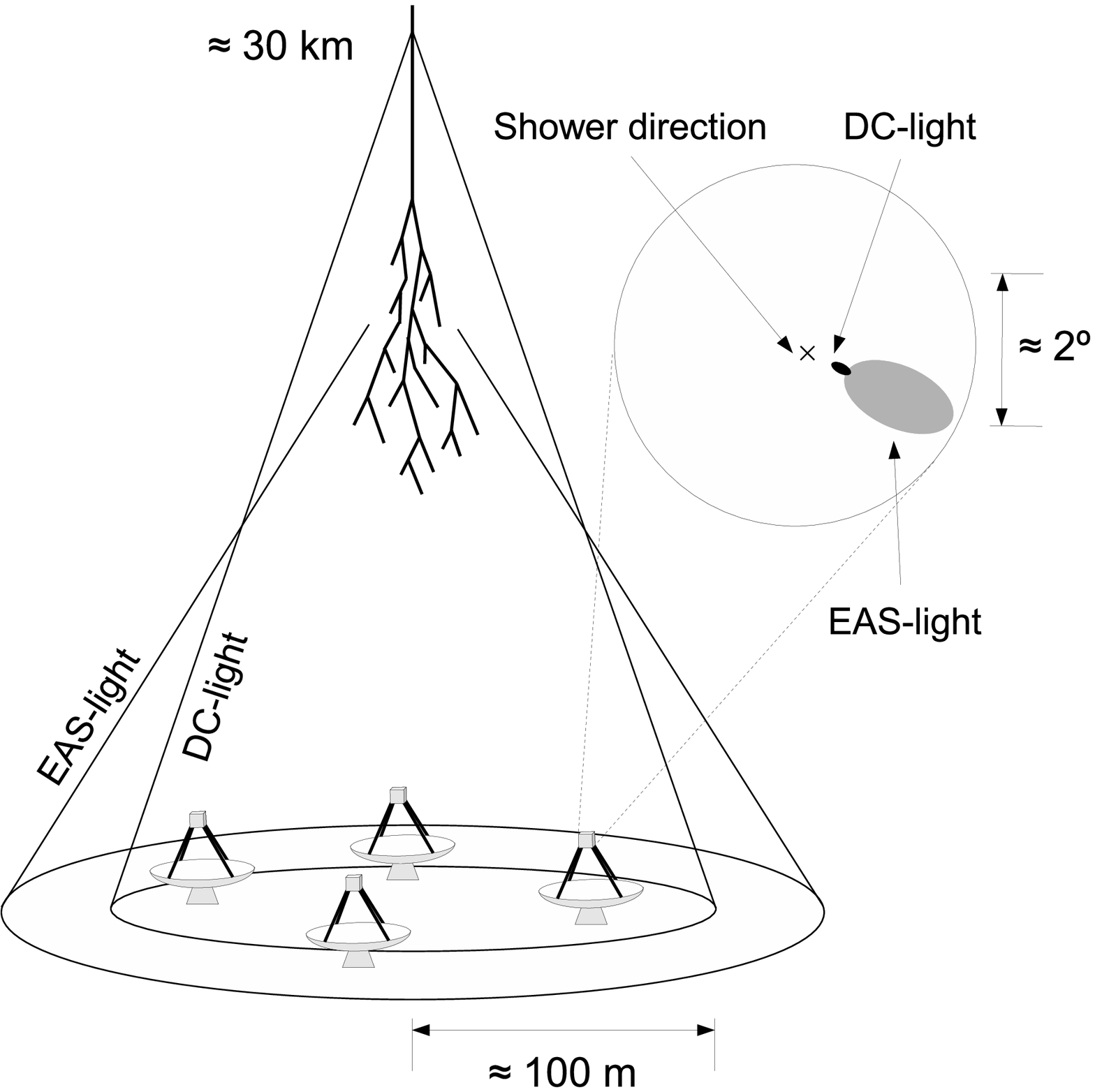}
  \end{center}
  \caption{Schematic representation of the Cherenkov emission from a cosmic-ray primary particle and the light distribution on the ground and in the camera plane of an IACT.}
  \label{sketch}
\end{figure}
When cosmic rays enter the atmosphere they emit Cherenkov light ( so called \emph{\textbf{D}irect \textbf{C}herenkov Light} ) above an element-dependent energy threshold. The Cherenkov angle increases with the density of the surrounding medium. The emission angle of the DC-light therefore increases with increasing depth of the primary particle in the atmosphere, creating a light cone on the ground with a radius of roughly 100 m (see Fig. \ref{sketch}).  At a typical height of 30 km the particle interacts and a particle cascade is induced (Extensive Air Shower, EAS). The Cherenkov light from these secondary particles creates a second, wider, light cone on the ground.

The intensity of the DC-light is proportional to the square of the charge Z of the emitting particle, and can therefore be used to identify the primary particle. The challenge for detecting DC-light is to distinguish it from the $\approx$100 times brighter EAS-light background. Because the DC-light is emitted higher in the atmosphere, it is emitted at a smaller angle than the EAS-light, and is therefore imaged closer to the shower direction in the camera plane. A typical  emission angle for DC-light is $0.15^\circ$ to $0.3^\circ$, whereas most of the EAS-light is emitted at angles greater $0.4^\circ$ from the direction of the primary particle. The H.E.S.S. Cherenkov cameras, with pixel sizes of $0.16^\circ$, are therefore able to resolve the DC-emission as a single bright pixel between the reconstructed shower direction  and the \emph{center of gravity} (cog) of the EAS-image in the camera plane (Fig. \ref{sketch}). 

The energy range over which this technique can be applied depends on the charge of the primary particle \cite{KIEDA}. At lower energies the limiting factor is that the primary particle momentum must exceed the Cherenkov threshold. At very high energies, the EAS-light outshines the DC-light, making the detection of the latter impossible. The reason for this is that the intensity of the EAS-light increases approximately linearly with energy, whereas the amount of emitted DC-photons remains basically constant above a certain energy. Because of their large atomic number and high flux compared to other heavy elements, iron nuclei are well suited for DC-light detection. The lower energy threshold for the detection of these nuclei is $\sim 10$ TeV. 

\section*{H.E.S.S. Data}

A total of 357 hours live time of H.E.S.S. data were considered for the analysis. The camera images were calibrated and the particle energy and direction were reconstructed using the standard H.E.S.S. analysis \cite{CALIB,CRAB}. Afterwards, the DC-Light is identified as a single high intensity pixel in the aforementioned angular area of the camera (Fig. \ref{sketch}). Once a DC-light pixel is found in a camera image, the DC-light intensity $I_{\rm{DC}}$ is reconstructed by subtracting the mean intensity of the neighboring pixels $I_{\rm{neighb. pixels}}$ from the DC-pixel intensity:
\begin{equation}
I_{\rm{DC}}=I_{\rm{DC-pixel}} - <I_{\rm{neighb. pixels}}>
\end{equation}

In total 1899 events with DC-light detection in at least two camera images were found in the data-set (events with DC-light detection in only one camera image were not considered in the analysis to minimize systematic uncertainties \cite{IRON}). The elemental composition of these events is estimated using the $Z$ dependence of the DC-light intensity. The reconstructed charge $Z^{*}$ is defined as:   
\begin{equation}
Z^* = d(E,\theta) \sqrt{I_{\rm{DC}}} ,
\end{equation}
where $\theta$ is the zenith angle, E the reconstructed energy of the particle and $d(E,\theta)$ is a factor that normalizes the mean of the $Z^*$ distribution from iron simulations to the atomic number $Z$ of iron. The fraction $k_{\rm{Fe}}$ of iron events among the data is then measured by a two-component model fit to the $Z^*$ distribution of the data. The first component of this model is the $Z^*$ distribution of simulated iron nuclei. The second component is a sum of the $Z^*$ distribution of lighter nuclei. The relative composition of the lighter charge bands (= all except the iron band, defined as Z$>$24) is kept fixed to a reference composition, so that $k_{\rm{Fe}}$ is the single free parameter of the fit. The reference composition is taken from the elemental flux compilations given in \cite{HOERANDEL,WIEBEL}. The flux errors of these compilations are afterwards propagated into the fit result. 

\section*{Iron Flux}
\begin{figure*}[th]
  \begin{center}
    \includegraphics[width=13cm]{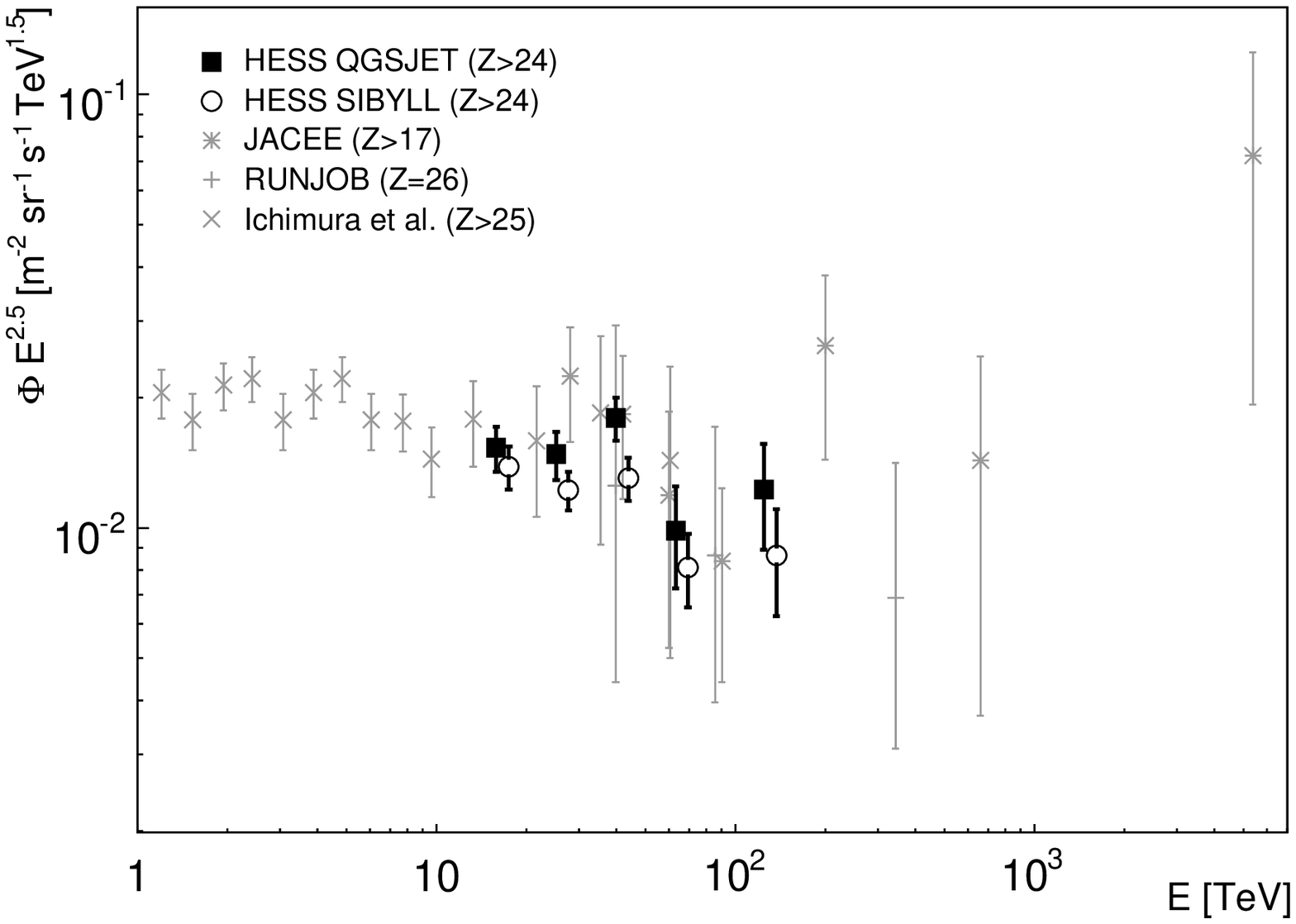}
  \end{center}
  \caption{Differential iron energy spectrum measured with H.E.S.S. for the hadronic models QGSJET and SIBYLL multiplied by E$^{2.5}$ for better visibility of structures. The  spectral points for both models are measured for the same energies. For better visibility  the SIBYLL points were shifted 10\% upwards in energy. The error bars show the statistical errors.   The systematic flux error in each bin is 20\%. The measurements from balloon experiments with data points at the highest energies are shown for comparison \cite{JACEE,RUNJOB,ICHIMURA}. When comparing the measurements one should bear in mind that the experiments have different charge thresholds for their definition of the iron band (see legend). }
  \label{figspectrum}
\end{figure*}

The iron fraction in the data was measured in five energy bins. The differential iron flux $\phi(E)$ can then be estimated in each bin as:
\begin{equation}
\phi(E)  = \frac{N_{\rm{DC}}(E)}{A_{\rm{eff}}(E)\cdot \Delta E \cdot  t}\cdot k_{\rm{Fe}} ,
\end{equation}
where $N_{\rm{DC}}(E)$ is the number of detected DC-events in the energy interval from $E$ to $E+\Delta E$, $t$ is the total live-time of the data-set and  $A_{\rm{eff}}$ is the mean effective area times the field of view of the detector, averaged over the zenith angle of the observations, taking into account the efficiency of selection cuts.
 
$A_{\rm{eff}}$ is derived from atmospheric shower simulations of iron nuclei. These simulations rely on the detailed modeling of the hadronic interaction in EAS-showers at energies that are not accessible to current particle accelerators. To assess the systematic errors arising from uncertainties in these interactions, the analysis is performed with simulations based on two independent hadronic interaction models, SIBYLL 2.1 \cite{SIBYLL} and QGSJET 01f \cite{QGSJET}.

The resulting iron energy spectrum is shown in Fig. \ref{figspectrum} for both hadronic models, together with the highest energy baloon measurements.  The derived spectra agree well with these measurements for both models and are well fit by a power law $\phi(E) =\phi_0 (\frac{E}{\rm{TeV}})^{-\gamma}$. The best fit values for the SIBYLL spectrum are given by
$\phi_0 = (0.029 \pm 0.011)$ m$^{-2}$sr$^{-1}$ TeV$^{-1}$ and 
$\gamma = 2.76\pm0.11$  with an 
$\chi^2$/ndf of 3.0/3. For the QGSJET spectrum the best fit values are 
$\phi_0 = (0.022 \pm 0.009)$ m$^{-2}$sr$^{-1}$ TeV$^{-1}$ and 
$\gamma = 2.62 \pm 0.11 $  with 
$\chi^2$/ndf of 5.3/3. 

The difference between the two measured spectra gives an estimate of the systematic error introduced due to hadronic modeling. We note that, despite this uncertainty, the presented measurement is the most precise so far in its energy range. The result confirms the flux measurements from balloon experiments with an independent technique, giving confidence in both results.

\section*{Outlook}

Future improvements of the DC-light technique could extend the energy range of the measurement to an energy of $\sim 1$ PeV. Besides larger statistics, this extension requires additional separation power of the DC-light  from the EAS-light.  As shown in \cite{KIEDA}, additional separation power can be achieved using the time structure of the DC-light, since it arrives with a typical delay of 4 ns with respect to the EAS light. This fact could not be exploited in the analysis presented  because the H.E.S.S. data used here were taken with the standard integration window of 16 ns. However, current and planned Cherenkov telescopes, which routinely store pulse timing information \cite{MAGIC,VERITAS}, may take advantage of this characteristic. 

Due to the strong dependence of the DC-light yield on the charge of the primary particle, the DC-light technique has great potential for composition measurements. The limiting factor in the charge resolution is currently the accuracy of shower reconstruction. However, future IACT's, with pixels of smaller angular size and more nearby telescopes could provide the needed reconstruction accuracy \cite{IRON}. 

\section*{Acknowledgments}
The support of the Namibian authorities and of the University of Namibia
in facilitating the construction and operation of H.E.S.S. is gratefully
acknowledged, as is the support by the German Ministry for Education and
Research (BMBF), the Max Planck Society, the French Ministry for Research,
the CNRS-IN2P3 and the Astroparticle Interdisciplinary Programme of the
CNRS, the U.K. Science and Technology Facilities Council (STFC),
the IPNP of the Charles University, the Polish Ministry of Science and 
Higher Education, the South African Department of
Science and Technology and National Research Foundation, and by the
University of Namibia. We appreciate the excellent work of the technical
support staff in Berlin, Durham, Hamburg, Heidelberg, Palaiseau, Paris,
Saclay, and in Namibia in the construction and operation of the
equipment.

\small
%This is the reference to .bib file (Whitout .bib!)

%This in the bibtex style, is ok.
\bibliographystyle{plain}
\normalsize

%%%%%%%%
%  44  %
%%%%%%%%

%The paper title
\title{Measurement of Cosmic Ray Electrons with H.E.S.S.}
%Short title to print in the headers to the final publication (Not showed in this print).
\shorttitle{Measurement of Cosmic Ray Electrons with H.E.S.S.}

%All paper authors
\authors{Kathrin Egberts$^{1}$, Christopher van Eldik$^{1}$, Jim Hinton$^{2}$ for the H.E.S.S. collaboration}
%Short title to print in the headers to the final publication (Not shown in this print).
\shortauthors{K. Egberts$^{1}$ et al.}
%All the affiliations.
\afiliations{$^1$Max-Planck-Institut f\"ur Kernphysik, P.O. Box 103980, D 69029
Heidelberg, Germany\\ $^2$School of Physics \& Astronomy, University of Leeds, Leeds LS2 9JT, UK }
\email{kathrin.egberts@mpi-hd.mpg.de}

%The abstract.
\abstract{Due to energy losses in the interstellar medium, cosmic ray electrons at TeV energies carry information on local (within a few hundred parsecs) accelerators. However, measurements of the spectrum of the cosmic ray electrons beyond 1 TeV are extremely difficult due to the rapidly declining flux and the much more numerous background of nucleonic cosmic rays. The very large collection area of Cherenkov telescope arrays makes them promising instruments with which to measure these high energy electrons. While Cherenkov telescopes solve the problem of low fluxes of cosmic ray electrons in the TeV range, they still have to deal with the problem of distinguishing electrons from the nucleonic background. Here we report on first results towards a measurement of the cosmic ray electron spectrum with the High Energy Stereoscopic System (H.E.S.S.). The improved background supression that is needed for such a measurement is achieved by an event classification with the ``Random Forest'' algorithm based on decision
trees. }

\maketitle

\addcontentsline{toc}{section}{Measurement of Cosmic Ray Electrons with H.E.S.S.}
\setcounter{figure}{0}
\setcounter{table}{0}
\setcounter{equation}{0}

%Begin the section.
\section*{Introduction}
\begin{figure}[ht]
\begin{center}
\includegraphics [width=6.4cm]{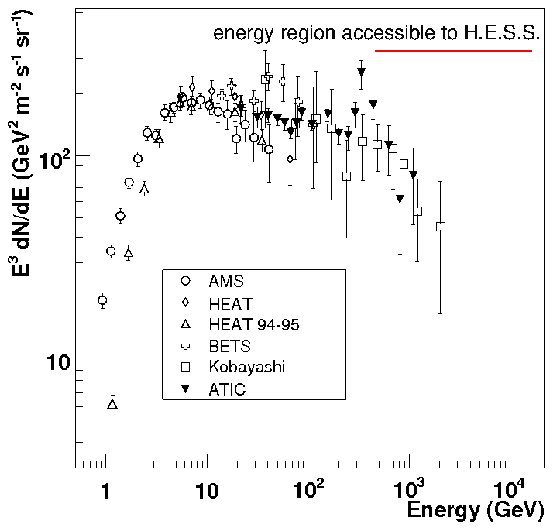}
%[width=0.52\textwidth]{espec.eps}
  \vspace{-4mm}
\caption{The electron spectrum measured with balloon and satellite experiments. Data are taken from \cite{ams}\cite{HEAT}\cite{HEAT2}\cite{ATIC}\cite{BETS}. The energy region accessible to H.E.S.S. with sufficient statistics, assuming a spectral index of 3.3, is indicated by the red bar. Systematic effects may limit the results.}\label{Spec.eps}
\end{center}
\end{figure}

Cosmic ray electrons are with about 1\% of the flux in the GeV range a small but peculiar fraction of cosmic rays.
In contrast to hadronic cosmic rays, they lose their energy rapidly via inverse Compton scattering and synchrotron radiation leading to a steep spectrum following a power law $dN/dE \propto E^{-\Gamma}$ with spectral index $\Gamma \approx 3.3$, which is observed in the GeV range by various balloon and satellite experiments as shown in Fig. \ref{Spec.eps}. Furthermore, at high energies, the energy of the electron limits its lifetime, $t \propto 1/E$, and hence its propagation distance. Therefore, at TeV energies, distinct features of single nearby sources can be expected in the electron spectrum \cite{tevatron}\cite{Kobayashi}.
At these energies, however, no measurements exist, as the rapidly declining electron flux calls for larger detector areas than balloon and satellite experiments can provide. 
Thus, %das muss praegnant!!!
while interesting theoretical predictions for the TeV range of the spectrum exist, it has been impossible so far to measure.\newline
J. Nishimura proposed an alternative approach using imaging atmospheric Cherenkov telescopes (IACTs) \cite{Idea}. They use the earth's atmosphere as detector and therefore provide an order of $10^5$ larger collection areas. Designed for the measurement of $\gamma$-rays, they can be used to study cosmic ray electrons, which, like $\gamma$-rays, produce electromagnetic showers. The High Energy Stereoscopic System (H.E.S.S.) is an array of four imaging atmospheric Cherenkov telescopes in the Khomas highlands in Namibia \cite{HESS}. Its sensitivity and its large field of view make such a measurement of cosmic ray electrons in the TeV range now possible.

\section*{Measuring cosmic ray electrons with H.E.S.S.}
\begin{figure}[ht]
\begin{center}
\includegraphics[width=6.2cm]{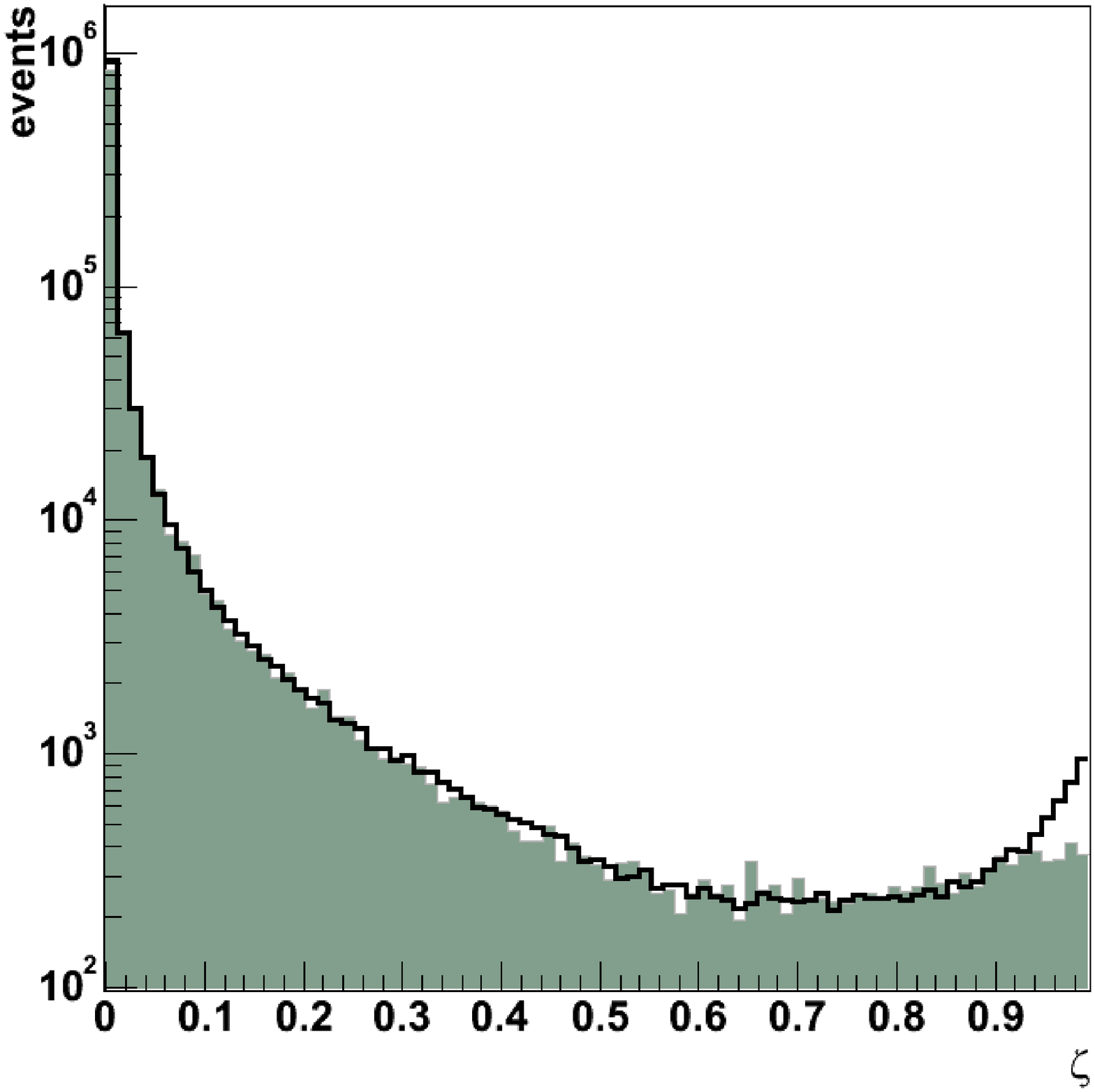}
  \vspace{-4mm}
\caption{The distribution of the Random Forest output $\zeta$ for data with energy 0.4-1.0 TeV(solid line) and background simulation consisting of contributions from protons, helium, nitrogen, silicon and iron. A clear excess of electron events can be seen at higher values of $\zeta$ that cannot be explained by background simulations.}\label{fig1.eps}
\end{center}
\end{figure}

\begin{figure}
\begin{center}
\includegraphics[width=6.2cm]{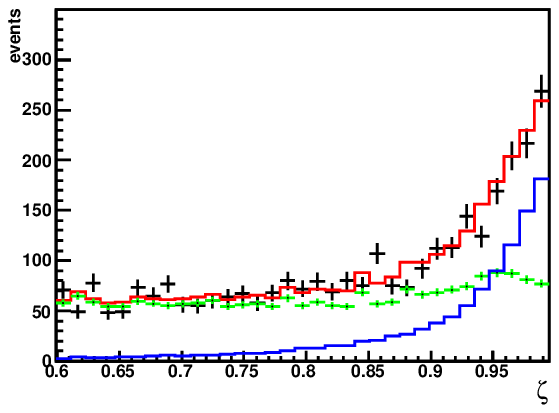}
  \vspace{-4mm}
\caption{The distribution of $\zeta$ in the signal region of $\zeta>0.6$ for data (black) with a reconstructed energy between 0.7 and 1 TeV and the fitted proton (green) and electron (blue) simulations in this energy range. The best fit model of electrons and protons is shown in red.}\label{fig2.eps}
\end{center}
\end{figure}

%data used: extragalactic fields, excluding circle of 0.4 degree radius around any $\gamma$-ray source
For the analysis of cosmic ray electrons, all data that were taken by the complete four telescope array, targeting extragalactic fields to avoid contamination of diffuse $\gamma$-ray emission from the galactic plane, were used. Any known or potential $\gamma$-ray source was excluded.\newline
While the big advantage of IACTs is their large collection area, the challenge that is posed by such a cosmic ray electron measurement is the discrimination of the electrons from the much more numerous hadronic background.
%random forest:
Therefore, a sophisticated machine learning algorithm was chosen to separate electron and hadron events.
The  ``Random Forest'' program \cite{Forest}\cite{Bock} is based on decision trees and was trained using Monte Carlo simulations of electrons and off-source data. The input parameters for the training contain camera image information like the width and length of the elliptical image scaled to the expected width and length defined by simulations, and intensity information. Only those events are used that triggered all four telescopes in order to assure that only the best measured events are chosen for analysis.
As the input parameters are partially energy dependent, an improved performance was achieved by training in five energy bands. The zenith angle dependence was taken into account by restricting the data set to zenith angles smaller than $28^{\circ}$, thus approximately matching the simulations at $20^{\circ}$.\newline
The Random Forest converts the input parameters into an output $\zeta$ between zero and one, denoting the electron-likeness of the respective event. A large $\zeta$ represents an electron-like event, while $\zeta=0$ stands for background events.
The Random Forest method allows for an improved background rejection shown in Fig. \ref{fig1.eps}.
Only about $0.5-2.0 \%$, depending on energy, of all proton events passing the four-telescope cut end up in the electron signal region of $\zeta>0.6$. This large suppression of hadronic background events makes a measurement of cosmic ray electrons possible in the first place.
In Fig. \ref{fig1.eps} the $\zeta$ distribution of data is shown together with a Monte Carlo simulation of the background, showing the good agreement at low values of $\zeta$ and a clear signal of an electromagnetic showers at $\zeta=1$.\newline 
%Background estimation:
In order to estimate the remaining background and extract the number of electrons from the data, simulated electrons and protons are fitted to the data in the $\zeta$ distribution. The simulations are produced using CORSIKA \cite{CORSIKA} with SIBYLL as hadronic interaction model \cite{SIBYLL}. Modeling the background with simulated protons only is possible because heavier nuclei contained in the hadronic background %due to their even more hadronic nature 
show an even better classification power than protons and therefore, background in the signal region is completely dominated by protons. 
This method of fitting electron and proton simulations to the data is demonstrated in Fig. \ref{fig2.eps} exemplarily for the energy between 0.7 and 1.0 TeV. A good match between data and electron-proton combination is observed.    

This method has the potential to extend the spectrum of cosmic ray electrons to energies of several TeV. While statistical errors are small compared to direct measurements, systematic effects have to be taken into account.
Prime source of systematic uncertainties is the usage of proton simulations to model the data. To quantify this effect, two different hadronic interaction models, SIBYLL \cite{SIBYLL} and QGSJET \cite{QGSJET} are compared.
Additionally, confusion with $\gamma$-rays might occur as $\gamma$-rays and electrons produce very similar air showers in the atmosphere. They can be separated on a statistical basis by the height of their shower maximum $X_{max}$, which occurs half a radiation length higher in the atmosphere for electrons than for $\gamma$-rays. As $X_{max}$ is not measured precisely enough, a contribution of extragalactic $\gamma$-rays can experimentally not be excluded. However, theoretical predictions for diffuse extragalactic $\gamma$-ray flux lie far below the electron flux \cite{coppi}.\newline
Ongoing work concerns the in-depth study of systematic errors and model-dependence of the results with the goal to derive a reliable electron spectrum.

\section*{Conclusion}

For the first time, cosmic ray electrons have been measured with IACTs. A clear signal of electrons can be seen in the H.E.S.S. data. Therefore, IACTs seem to be able to extend the measured spectrum of cosmic ray electrons into the TeV range, where the shape of the spectrum is completely unknown, but expected to give information on the existence of nearby cosmic ray accelerators.

%This in the bibtex style, is ok.
\bibliographystyle{plain}

%%%%%%%%
%  45  %
%%%%%%%%

%The paper title
\title{H.E.S.S. observations of galaxy clusters}
%Short title to print in the headers to the final publication (Not showed in this print).
\shorttitle{H.E.S.S. observations of galaxy clusters}
%All paper authors
\authors{W. Domainko$^{1}$, W. Benbow$^{1}$, J. A. Hinton$^{2}$, O. Martineau-Huynh$^{3}$, M. de Naurois$^{3}$, D. Nedbal$^{4}$, G. Pedaletti$^{5}$, G. Rowell$^{6}$ for the H.E.S.S. collaboration.}
%Short title to print in the headers to the final puplication (Not showed in this print).
\shortauthors{W. Domainko et al}
%All the affiliations.
\afiliations{$^1$Max-Planck-Institut f\"ur Kernphysik, Heidelberg, Germany\\ $^2$School of Physics \& Astronomy University of Leeds, UK\\ $^3$Laboratoire de Physique Nucl\'eaire et de Hautes Energies, Universit\'es Paris VI \& VII, France\\ $^4$Institute of Particle and Nuclear Physics, Charles University, Prague, Czech Republic\\ $^5$Landessternwarte, Universit\"at Heidelberg, Germany\\ $^6$School of Chemistry \& Physics, University of Adelaide, Australia}
\email{wilfried.domainko@mpi-hd.mpg.de}

%The abstract.
\abstract{Clusters of galaxies, the largest gravitationally bound objects in the
universe, are expected to contain a significant population of hadronic
and leptonic cosmic rays. Potential sources for these particles are
merger and accretion shocks, starburst driven galactic winds and radio
galaxies. Furthermore, since galaxy clusters confine cosmic ray protons
up to energies of at least 1 PeV for a time longer than the
Hubble time they act as storehouses and accumulate all the hadronic
particles which are accelerated within them.  Consequently clusters of galaxies are potential sources of VHE ($>$ 100 GeV) gamma rays. Motivated by these considerations, promising galaxy clusters are observed with the H.E.S.S. experiment as part of an ongoing campaign. Here, upper limits for the VHE gamma ray emission for the Abell 496 and Coma cluster systems are reported.}

\maketitle

%%%%% Begin EEGO %%%%%%
\addtocontents{toc}{\protect\contentsline {part}{\protect\large Extended Extragalactic Objects}{}}
\addcontentsline{toc}{section}{H.E.S.S. observations of galaxy clusters}
\setcounter{figure}{0}
\setcounter{table}{0}
\setcounter{equation}{0}

%Begin the section.
\section*{Introduction}

Galaxy clusters are the largest non-thermal sources in the universe. Radio \cite{giovannini00}, \cite{feretti04} and hard X-ray \cite{rephaeli02}, \cite{fusco04} observations show the presence of accelerated electrons in these systems. It is understood that hadronic cosmic rays accelerated within the cluster volume will be confined there (with energies of up to 10$^{15}$ eV) for timescales longer than the Hubble time \cite{voelk96}, \cite{berezinsky97}. Hence clusters of galaxies act as storehouses for such particles, and therefore a large component of cosmic rays is expected in these systems.

Several sources of cosmic rays can be found in galaxy clusters. Accretion and merger shocks driven by large-scale structure formation have the ability to accelerate cosmic rays \cite{colafrancesco00}, \cite{loeb00}, \cite{ryu03}. Supernova remnant shocks and galactic winds can also produce high-energy particles \cite{voelk96}. Additionally AGN outbursts can distribute non-thermal particles in the cluster volume \cite{ensslin97}, \cite{aharonian02}, \cite{hinton07}.

Due to the expected large component of non-thermal particles, galaxy clusters are potential sources for gamma-ray emission (see \cite{blasi07} for a recent review). Various processes can lead to the production of gamma-ray radiation in these objects. Inelastic collisions between cosmic ray protons and thermal nuclei from the intra-cluster medium (ICM) will lead to gamma-ray emission through $\pi^0$-decay \cite{dennison80}, \cite{voelk96}. Electrons with sufficiently high energies can up-scatter cosmic microwave background (CMB) photons to the gamma-ray range in inverse Compton processes \cite{atoyan00}, \cite{gabici03}, \cite{gabici04}. 

Despite the arguments for potential gamma-ray emission given above, no galaxy cluster has firmly been established as a source of high-energy and very high-energy electromagnetic radiation \cite{reimer03}, \cite{perkins06}.

\section*{The H.E.S.S. experiment}

The H.E.S.S. experiment is an array of imaging atmospheric Cherenkov telescopes located in the Khomas highlands, Namibia \cite{hinton04}. It observes in the VHE gamma-ray regime and has a field of view of $\sim$5$^\circ$. Due to the large field of view it is possible to detect extended sources such as supernova remnants \cite{aharonian04}, \cite{aharonian07}. Galaxy clusters are expected to feature extended VHE gamma-ray emission. The H.E.S.S. experiment is well suited to search for such a signal (see e.g. \cite{aharonian04}). 

\section*{Targets}

\subsection*{Abell 496}

\begin{figure*}
\begin{center}
%\noindent \fbox{\hbox{\vbox{\hsize=130mm \hfill \vspace{50mm}}}}
\includegraphics [width=0.99\textwidth]{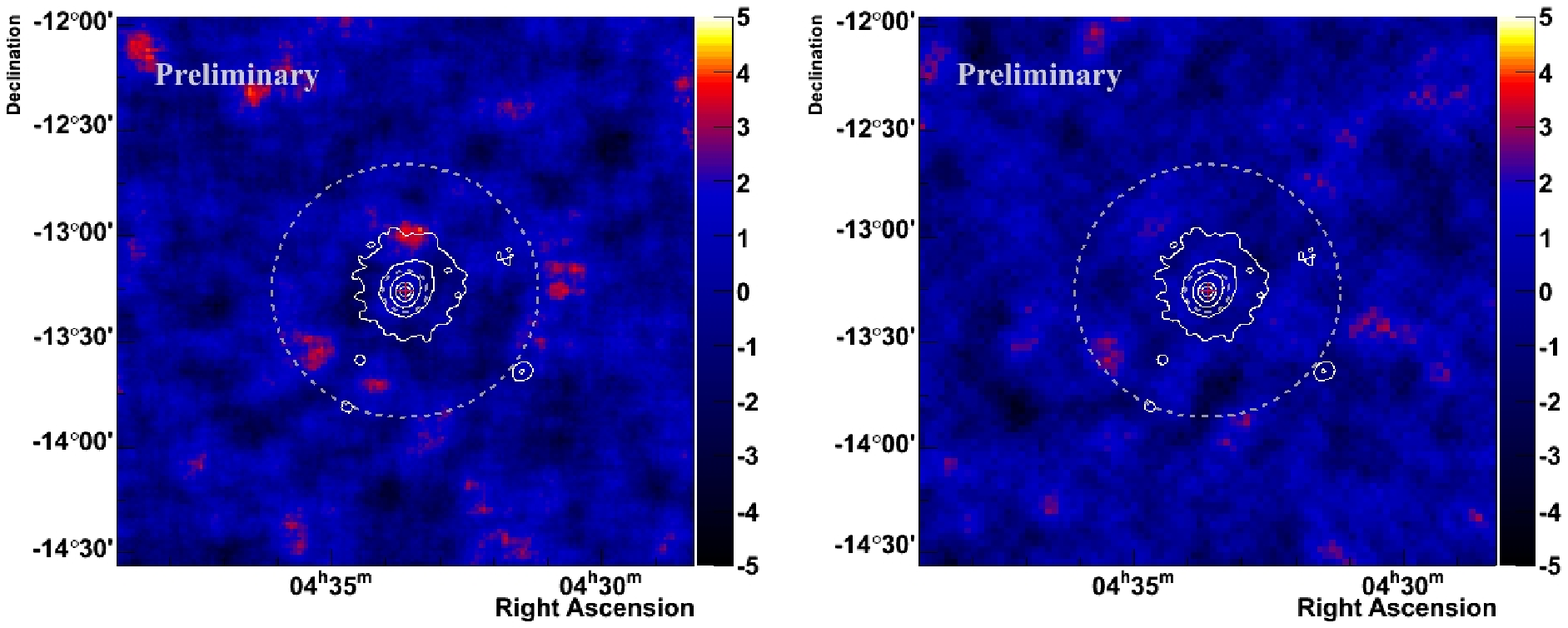}
\end{center}
\caption{Significance map of the cluster Abell 496 seen by H.E.S.S. with standard cuts (left panel) and hard cuts (right panel). No signal is detected from this region of the sky. The dashed circles show the two size cuts and the white contours correspond to \textit{ROSAT} X-ray contours \cite{durret00}.}\label{abell}
\end{figure*}

Abell 496 is a nearby (z = 0.033), relaxed  cluster of galaxies with a mean temperature of 4.7 keV. It features a cooling core at its center \cite{markevitch99}. It is located in the Southern Hemisphere \cite{boehringer04} and is therefore well suited for observations with H.E.S.S. Data taking was performed during moonless nights in the time period from October to December 2005, and in October 2006. In total 23.4 hours of data were taken, with 15.9 hours passing standard data-quality selection (live time 14.6 hours). The mean zenith angle is 27.6$^\circ$ which results in an energy threshold of 0.31 TeV for standard cuts and 0.57 TeV for hard cuts. H.E.S.S. standard data analysis (described in \cite{benbow05}) was performed using different geometrical size cuts to account for the extended nature of the target. No significant excess of VHE gamma-ray emission is found at the position of Abell 496 (see Fig. \ref{abell}). Upper limits for this object for two different size cuts are derived. All upper limits are obtained following \cite{feldman98} assuming a power law spectral index of -2.1 and are given at the 99.9\% confidence level. The first radial size cut, 0.1$^\circ$, is applied to test gamma-ray emission associated with the high density core region of the cluster. This is of particular interest for a hadronic scenario, since the gamma-ray emission should be enhanced in regions with higher density of target material. In this region an upper limit of F$_\mathrm{UL}$($>$0.31 TeV) = $1.0 \times 10^{-12}$ ph cm$^{-2}$ s$^{-1}$ (0.8\% Crab flux) is determined. A radial size cut of 0.6$^\circ$ is also applied, which covers the entire cluster \cite{reiprich02}. For this extended region an upper limit of F$_\mathrm{UL}$($>$0.57 TeV) = $2.4 \times 10^{-12}$ ph cm$^{-2}$ s$^{-1}$ (4.5\% Crab flux) is found. It should be noted that the H.E.S.S. upper limits scale approximately with $r/r_0$ with $r$ and $r_0$ being geometrical size cuts and this relation can be used to convert the presented upper limits to other sizes.

\subsection*{Coma cluster}

\begin{figure*}
\begin{center}
%\noindent \fbox{\hbox{\vbox{\hsize=130mm \hfill \vspace{50mm}}}}
\includegraphics [width=0.99\textwidth]{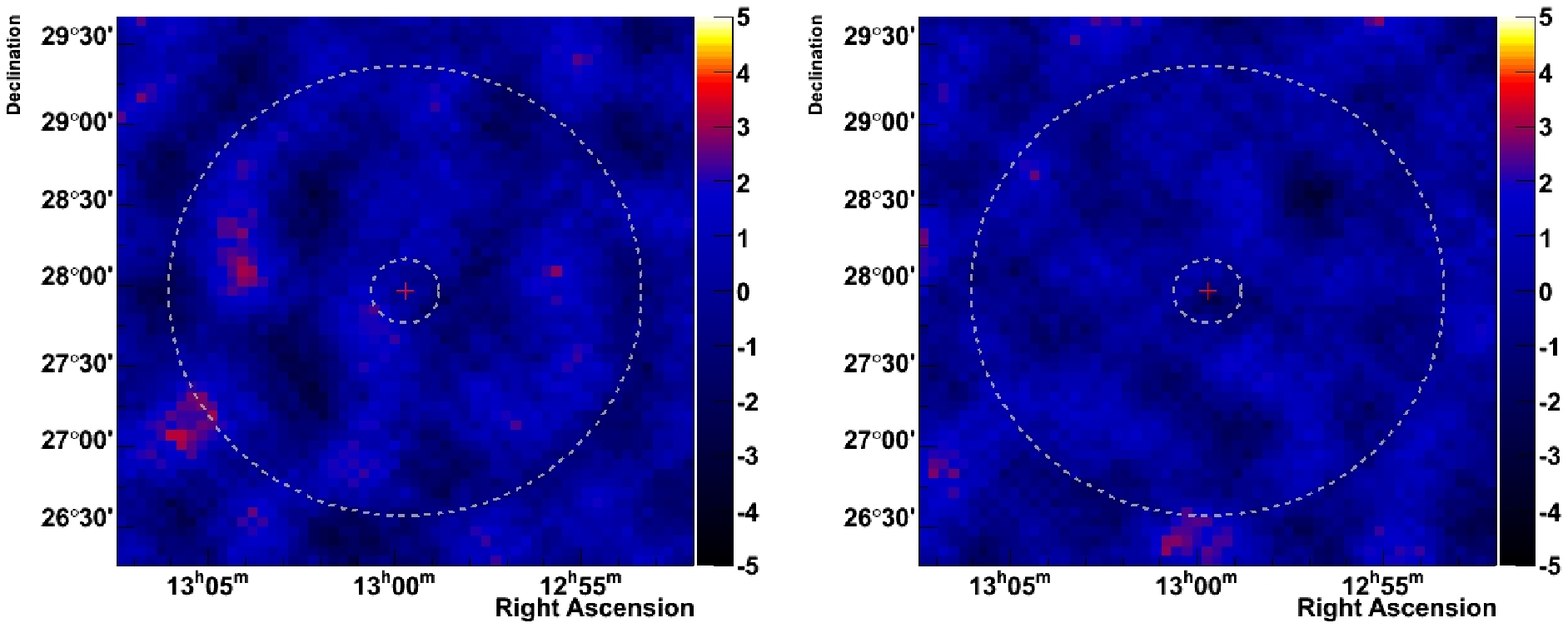}
\end{center}
\caption{Significance map with standard cuts (left) and hard cuts (right) of the Coma cluster. No signal is found in the data. The two dashed circles correspond to the two size cuts.}\label{coma}
\end{figure*}

The Coma cluster is a prominent hot (T = 8.25 keV, \cite{arnaud01}), nearby (z = 0.023) galaxy cluster which shows a merger signature in the X-ray gas (\cite{neumann03}). It features a hard X-ray excess (\cite{rephaeli02}, \cite{fusco04} but see \cite{rossetti04} for a different interpretation)  and a radio halo \cite{giovannini93}. The Coma cluster is often considered as a "standard cluster", and, due to the wealth of data on this object, it is very important for theoretical interpretations. It is located in the Northern Hemisphere which makes it less accessible for H.E.S.S. This cluster was observed during moonless nights in April and May 2006. 7.9 hours of good data were obtained resulting in 7.3 hours live time. The mean zenith angle of these observation is 53.5$^\circ$ which results in an energy threshold of 1.0 TeV for standard cuts and 2.0 TeV for hard cuts. No significant signal is found in these observations using various geometrical size cuts (see Fig. \ref{coma}). Upper limits on the VHE gamma ray emission of the Coma cluster for the core region and for the entire cluster are derived. Applying a radial size cut of 0.2$^\circ$ (core region) an upper limit of F$_\mathrm{UL}$($>$1.0 TeV) = $8.3 \times 10^{-13}$ ph cm$^{-2}$ s$^{-1}$ (3.7\% Crab flux) is found. For the entire cluster, with a radial size cut of 1.4$^\circ$, an upper limit of F$_\mathrm
{UL}$($>$2.0 TeV) = $4.8 \times 10^{-12}$ ph cm$^{-2}$ s$^{-1}$ (65.6\% Crab flux) is obtained. For the latter, very extended analysis, only data with a live time of 6.4 hours are used due to an insufficient number of OFF-source runs with such a large zenith angle for the background estimation. 

\section*{Summary \& outlook}

Clusters of galaxies are the most massive gravitationally bound structures in the universe and as such they are believed to be representatives for the universe as a whole. Therefore they are important tools for cosmology. The detection of gamma-ray emission from these objects will give important information about structure formation and supernova activity over the entire history of the universe. No gamma-ray excess has been found with H.E.S.S. from any cluster, with observation times in the range of 10 $-$ 20 hours. As a next step, one promising galaxy cluster will be given a very deep H.E.S.S. exposure of at least 50 hours.

\section*{Acknowledgments}

The support of the Namibian authorities and of the University of Namibia
in facilitating the construction and operation of H.E.S.S. is gratefully
acknowledged, as is the support by the German Ministry for Education and
Research (BMBF), the Max Planck Society, the French Ministry for Research,
the CNRS-IN2P3 and the Astroparticle Interdisciplinary Programme of the
CNRS, the U.K. Science and Technology Facilities Council (STFC),
the IPNP of the Charles University, the Polish Ministry of Science and 
Higher Education, the South African Department of
Science and Technology and National Research Foundation, and by the
University of Namibia. We appreciate the excellent work of the technical
support staff in Berlin, Durham, Hamburg, Heidelberg, Palaiseau, Paris,
Saclay, and in Namibia in the construction and operation of the
equipment.

%%%%%%%%
%  46  %
%%%%%%%%

%The paper title
\title{VHE $\gamma$-ray observations of starburst galaxies with H.E.S.S.}
%Short title to print in the headers to the final publication (Not showed in this print).
\shorttitle{VHE gamma-ray observations of starburst galaxies with H.E.S.S.}
%All paper authors
\authors{D. Nedbal$^1$, W. Benbow$^2$, K. Bernl\"{o}hr$^{2,3}$, J. Hinton$^4$, M.
Lemoine-Goumard$^5$ for the H.E.S.S. collaboration.}
%Short title to print in the headers to the final puplication (Not showed in this print).
\shortauthors{D. Nedbal et al}
%All the affiliations.
\afiliations{$^1$Institute of Particle and Nuclear Physics, Charles
  University, V Holesovickach 2, 180 00, Prague 8, Czech Republic\\
  $^2$ Max-Planck-Institut f\"ur Kernphysik, P.O. Box 103980, D 69029
Heidelberg, Germany\\
$^3$ Institut f\"ur Physik, Humboldt-Universit\"at zu Berlin, Newtonstr. 15,
D 12489 Berlin, Germany\\
$^4$ School of Physics and Astronomy, University of Leeds, U.K.\\
$^5$ CENBG - CNRS - IN2P3, France}
\email{nedbal@ipnp.troja.mff.cuni.cz}

%The abstract.
\abstract{Starburst galaxies are characterized by extremely high
  star-formation rates and, as a consequence, very high supernova
  rates. These rates, as well as the gas density, are orders of
  magnitude higher than in the Milky Way. Starburst galaxies contain
  both a high cosmic-ray flux and high density of target material for
  proton-proton and inverse-Compton interactions. These objects are
  therefore viable candidates for observable levels of VHE
  $\gamma$-ray emission. Nearby starburst galaxies, such as NGC 253
  and M83, allow a study of general processes during galaxy formation
  and evolution of high redshift galaxies. These two galaxies were
  observed with H.E.S.S. stereoscopic array of atmospheric-Cherenkov
  telescopes. Upper limits from these observations are presented here.}

\maketitle

\addcontentsline{toc}{section}{VHE $\gamma$-ray observations of starburst galaxies with H.E.S.S.}
\setcounter{figure}{0}
\setcounter{table}{0}
\setcounter{equation}{0}

%Begin the section.
\section*{Introduction}
Starburst galaxies are galaxies with very intense star formation rate in
the central region. The starburst activity could be triggered by
 mechanisms such as a galaxy merger, causing compression of the
interstellar gas. Consequent high star formation is responsible for
a high supernovae explosion rate in this region. Supernovae then
provide shocks necessary for accelerating cosmic-ray particles.

Detectable levels of VHE $\gamma$-ray emission are predicted for
starburst galaxies such as M82 \cite{Voelk1996} and NGC 253
\cite{Aharonian:2005de}, \cite{Torres2005}. Gas
densities in the starburst regions reach values 100 times larger than
in the Milky Way. Such high densities provide enough target material
for inelastic proton-proton collisions of cosmic-ray
protons. Resulting $\pi^0$ mesons decay into high energetic
$\gamma$-ray photons, that can be detected by earthbound Cherenkov
imaging telescopes.

Despite encouraging predictions no significant detection of VHE
$\gamma$-rays has been confirmed from starburst galaxies so
far.

Results of M83 observations and re-observations of NGC 253 with the
full array of four H.E.S.S. telescopes are presented here.

\section*{H.E.S.S. experiment}
The H.E.S.S. Collaboration operates an array of four imaging
atmospheric-Cherenkov telescopes (IACTs), located in Namibia. Each
telescope has 107 m$^2$ mirror area and a camera consisting of 960
pixels providing a wide field-of-view of 5$^\mathrm{o}$. Images of $\gamma$-ray
showers are analysed in order to reconstruct the direction and energy
of the primary $\gamma$ photon. Background suppression is based on a
system of cuts on the image parameters (e.g. width). All results
presented here are obtained using the standard Hillas-based
H.E.S.S. analysis. For further details see
\cite{Aharonian:2006pe}. All upper limits are calculated assuming a
power-law spectrum with the photon index $\Gamma=2.1$ of the
$\gamma$ rays, at the 99.9\% confidence level (CL) \cite{Feldman:1998}. Not
included in the upper limits is a 20\% systematic error on flux.

\section*{Observed targets}
\subsection*{NGC 253}
NGC 253 is the closest (D $\sim$ 2.5 Mpc \cite{vaucouleurs78}) starburst
galaxy viewed edge-on. It is considered to be, along with M 82, an
archetypal starburst galaxy \cite{Rieke1980}.

The object is very well studied in all wavelengths from radio to
X-rays. Chandra X-ray observations \cite{Weaver2002} show, that the
central region of size $\sim$100 pc exhibits very intense starburst
activity. The central region is surrounded by a torus of dense gas
collimating a galactic wind driven by the starburst
\cite{McCarthy1987}.

NGC 253 is suggested as a possible source of detectable VHE $\gamma$-ray
flux (see for example \cite{Torres2005}).

Several instruments have already published upper limits on
$\gamma$-ray emission from NGC 253. CANGAROO III published results of
12.5 hours on-source observations \cite{itoh2007}. The
outcome was a 2$\sigma$ upper limit of $3.4\times 10^{-12}$
ph. cm$^{-2}$ s$^{-1}$ above 450 GeV, corresponding to 4.5\% of Crab
flux. H.E.S.S. observed NGC 253 for 24.6 hours with 2 telescopes and
for 3.4 hours using 3 telescopes. The result was an upper limit of
1.9$\times 10^{-12}$ ph. cm$^{-2}$ s$^{-1}$ above 300 GeV (1.4\% of
Crab flux using Crab spectrum of \cite{Aharonian:2006pe}) at 99\%
CL. \cite{Aharonian:2005de}. Upper limits were also
determined by the HEGRA experiment \cite{hegra2007} at 11\% Crab units
at 99\% CL. EGRET set an upper limit at lower energies of $F(>100
\mathrm{MeV}) < 3.4\times 10^{-8}$ ph. cm$^{-2}$ s$^{-1}$ \cite{egret1999}.

H.E.S.S. observed NGC 253 in several campaigns during 2003 and
2005. The total live time used for analysis (after applying current
data selection criteria) is 21.3 hours taken with only two telescopes
and 14.2 hours performed using the full array of four telescopes.

\begin{figure}[htbp]
\includegraphics[width=8cm]{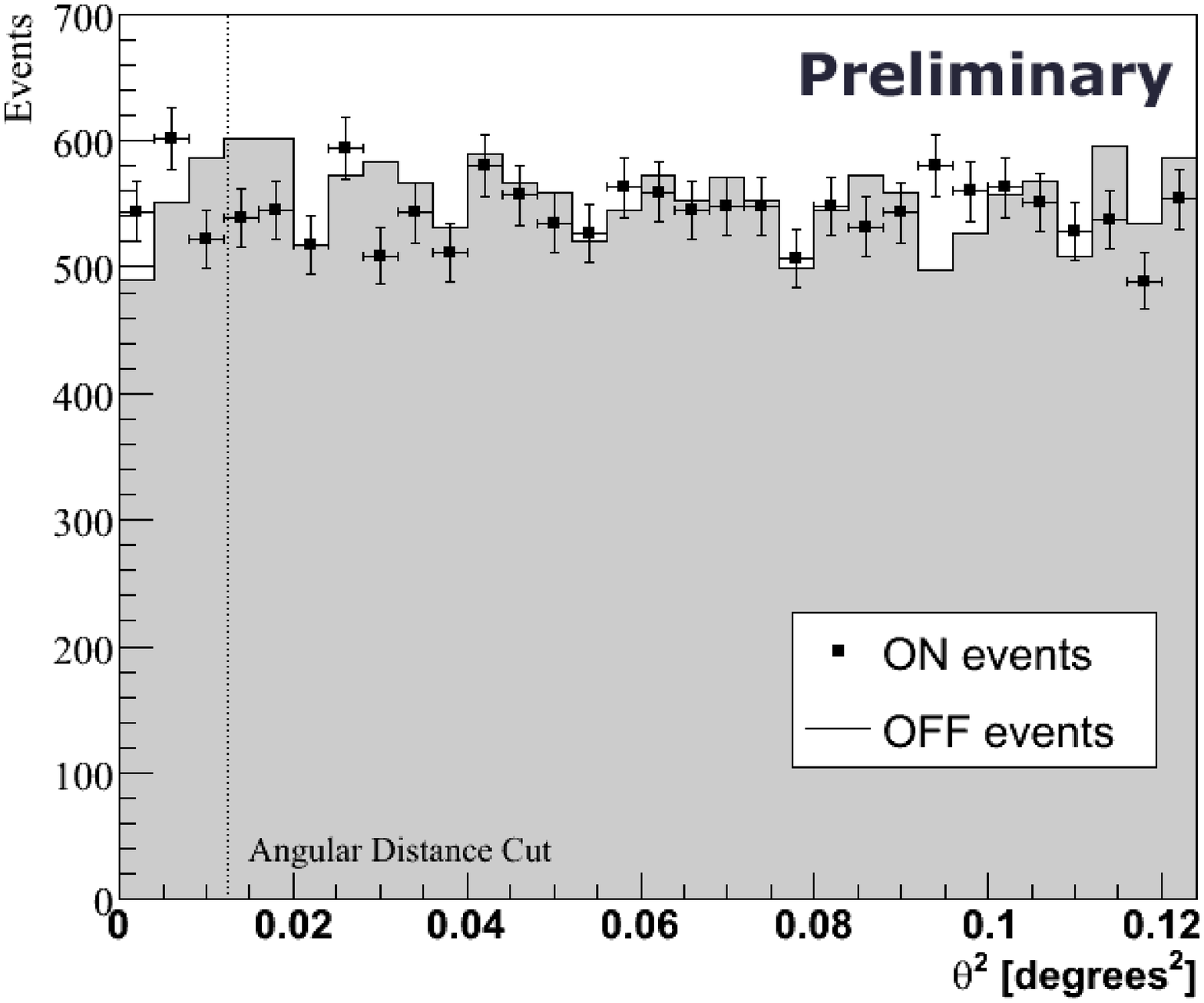}
\caption{Angular distribution of on-source and off-source events of the
  H.E.S.S. observations of NGC 253}\label{ngc253_theta}
\end{figure}

No VHE significant signal is found. The angular distribution of
on-source and off-source events shows no hint of a $\gamma$-ray excess (see
Figure \ref{ngc253_theta}) from the NGC 253 region. The 99.9\% upper limit
on the integral flux is $I_{\mathrm{ul}}(>270 GeV) = 1.2 \times
10^{-12}$ ph. cm$^{-2}$ s$^{-1}$, corresponding to 0.7\% of the Crab
Nebula flux. The upper limits are considerably lower than all the
previously published limits \cite{Aharonian:2005de}, \cite{itoh2007}. 

\begin{figure*}[htbp]
\includegraphics[width=16cm, height=7.5cm]{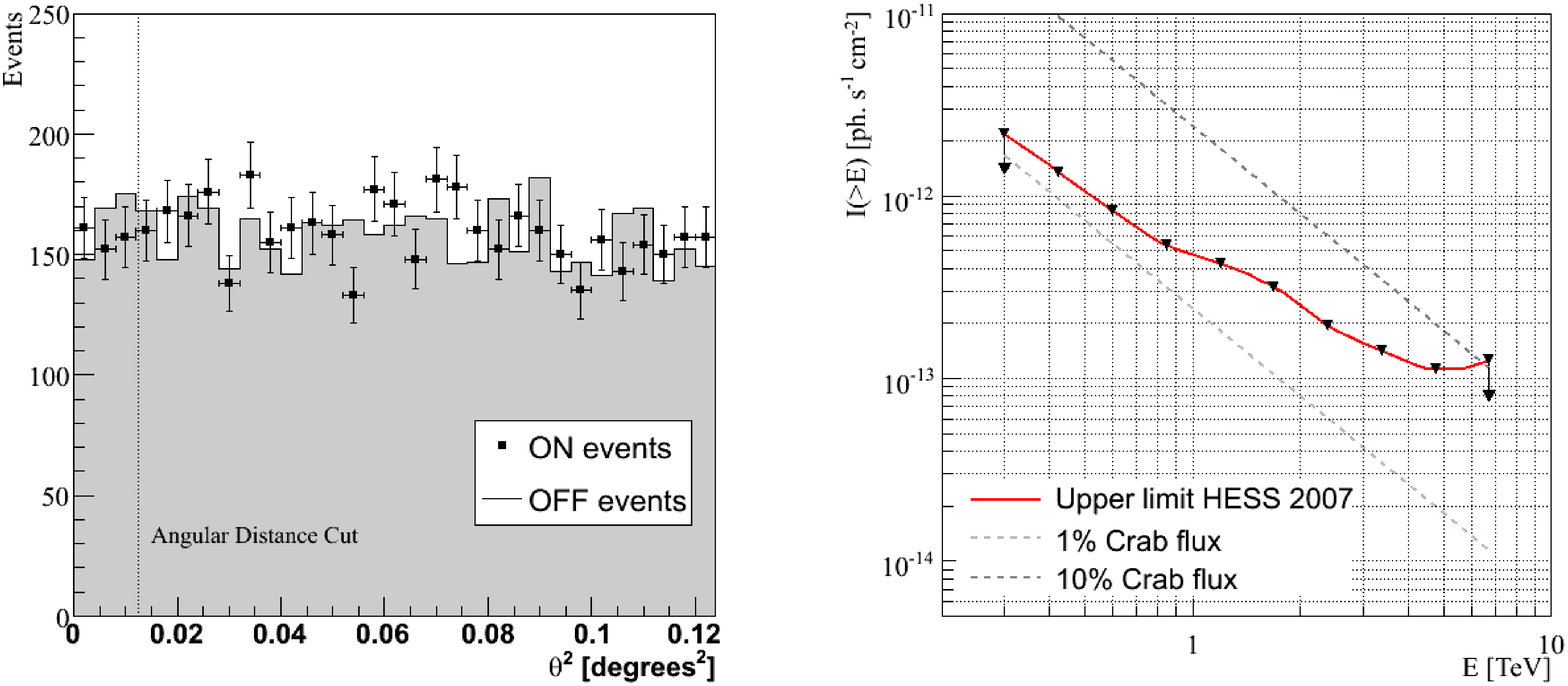}
\caption{Left panel: Angular distribution of on-source and off-source
  events for M83. Right panel: Upper limit on the M83 integral flux
  above the energy, E. The H.E.S.S. results are compared to 1\% and 10\% of
  Crab integral flux according to \cite{Aharonian:2006pe}.}\label{fig_m83}
\end{figure*}

\subsection*{M 83}
M83 is a nearly face-on spiral galaxy with high metallicity around
twice the solar value \cite{bresolin2002}. The central $\sim$300 pc is
shown to host a starburst nucleus \cite{harris2001}.  M83 is slightly
further away (D $\sim$ 4.5 Mpc \cite{thim2003}) and the FIR flux is
also lower than in the case of NGC 253. This makes M83 a slightly
worse VHE candidate with respect to NGC 253.  On the other hand, M83
does not show any sign of a galactic wind, contrary to NGC 253. The
convective losses of high-energy cosmic rays may thus be considerably
lower than in the case of NGC 253, implying a detectable $\gamma$-ray
flux as well.

M83 was observed by H.E.S.S. in July 2006. The total live time used
for analysis is 8.2 hours of good-quality data. No significant VHE
$\gamma$-ray emission is found. The angular distribution of on-source and
off-source events (see Figure \ref{fig_m83}) does not show any
hint of an excess. The resulting upper limit on the integral flux is
$I(> 330 GeV) = 2.2 \times 10^{-12}$ cm$^{-2}$ s$^{-1}$ at the 99.9\%
confidence level. These are the first VHE $\gamma$-ray upper limits to
be presented for M83. The integral flux upper limits are also shown in the
Figure \ref{fig_m83}.

In addition, no other significant signal is found in the field-of-view
of M83.

\section*{Summary and conclusions}
No significant VHE $\gamma$-ray signal was found in the
H.E.S.S. follow-up observations of NGC 253 or in the observations of
M83. Upper limits from these observations are presented. These limits
are already very close to the theoretical predictions. Since there is
only limited space for uncertainties in the flux-prediction models,
the detection of starburst galaxies by the current or future
generation of $\gamma$-ray experiments seems to be
inevitable. Starburst galaxies thus remain very good candidates for
future observations.

\section*{Acknowledgements} %UPDATE!!!
The support of the Namibian authorities and of the University of Namibia
in facilitating the construction and operation of H.E.S.S. is gratefully
acknowledged, as is the support by the German Ministry for Education and
Research (BMBF), the Max Planck Society, the French Ministry for Research,
the CNRS-IN2P3 and the Astroparticle Interdisciplinary Programme of the
CNRS, the U.K. Science and Technology Facilities Council (STFC),
the IPNP of the Charles University, the Polish Ministry of Science and 
Higher Education, the South African Department of
Science and Technology and National Research Foundation, and by the
University of Namibia. We appreciate the excellent work of the technical
support staff in Berlin, Durham, Hamburg, Heidelberg, Palaiseau, Paris,
Saclay, and in Namibia in the construction and operation of the
equipment.
\bibliographystyle{plain}

%%%%%%%%%%%
% The End %
%%%%%%%%%%%
\end{document}